\documentclass[twoside,a4paper]{report}

\usepackage[pdftex]{graphicx}
\graphicspath{{Figures/}{../Figures/}}
\usepackage{subfiles}
\providecommand{\main}{.}
\usepackage[backend=bibtex]{biblatex}
\makeatletter
    \newrobustcmd*{\nobibliography}{%
      \@ifnextchar[
        {\blx@nobibliography}
        {\blx@nobibliography[]}}
    \def\blx@nobibliography[#1]{}
    \appto{\skip@preamble}{\let\printbibliography\nobibliography}
\makeatother

\makeatletter
    \newrobustcmd*{\notableofcontents}{%
      \@ifnextchar[
        {\blx@notableofcontents}
        {\blx@notableofcontents[]}}
    \def\blx@notableofcontents[#1]{}
    \appto{\skip@preamble}{\let\tableofcontents\notableofcontents}
\makeatother

\usepackage[english]{babel}

\usepackage[T3,OT2,T1]{fontenc}
\usepackage[noenc]{tipa}

\usepackage{amsfonts,amsmath,amssymb,eucal,mathrsfs,amsthm}
\usepackage{enumerate}
\usepackage{framed}
\usepackage{mathdots}
\usepackage[pdftex]{color}
\usepackage{multicol}
\usepackage[Sonny]{fncychap} 

\usepackage{tikz}
\usetikzlibrary{matrix,arrows,decorations.pathmorphing}
\usetikzlibrary{calc}

\definecolor{BlueViolet}{rgb}{0,0.56,0.72}
\definecolor{darkdelion}{rgb}{0.8,0.48,0}
\definecolor{cyan}{rgb}{0,0.56,0.72}
\definecolor{dandelion}{rgb}{0.8,0.48,0}

\newcounter{dummy} \numberwithin{dummy}{section}

\theoremstyle{plain}
\newtheorem{tm}[dummy]{Theorem}
\newtheorem*{tm*}{Theorem}
\newtheorem{pp}[dummy]{Proposition}
\newtheorem*{pp*}{Proposition}
\newtheorem{lm}[dummy]{Lemma}
\newtheorem*{lm*}{Lemma}
\newtheorem{cl}[dummy]{Corollary}

\newtheorem*{lm1}{Lemma 10.1.2}
\newtheorem*{lm2}{Lemma 10.1.8}

\theoremstyle{definition}
\newtheorem{df}[dummy]{Definition}
\newtheorem{dfpp}[dummy]{Definition/Proposition}
\newtheorem{ej}[dummy]{Example}

\theoremstyle{remark}
\newtheorem{rk}[dummy]{Remark}

\newcommand{\bigslant}[2]{{\raisebox{.2em}{$#1$}\left/\raisebox{-.2em}{$#2$}\right.}}
\newcommand{\restrict}[2]{{\raisebox{.2em}{$#1$}\left|\raisebox{-.2em}{$#2$}\right.}}
\newcommand{\overbar}[1]{\mkern 1.5mu\overline{\mkern-1.5mu#1\mkern-1.5mu}\mkern 1.5mu}

\newcommand{\longleftmapsto}{\longleftarrow\joinrel\mapstochar}
\newcommand{\defeq}{\mathrel{\rlap{\raisebox{0.3ex}{$\cdot$}}\raisebox{-0.3ex}{$\cdot$}}                    =}
\DeclareMathOperator{\dec}{dec}
\DeclareMathOperator{\dia}{\scriptstyle \Diamond}

\newcommand{\catname}[1]{{\normalfont\textbf{#1}}}

\newcommand{\Cat}{\catname{C}}
\newcommand{\Diff}{\catname{Diff}}
\newcommand{\iMan}{\catname{iMfld}}
\newcommand{\elMan}{\catname{elMfld}}
\newcommand{\blMan}{\catname{blMfld}}
\newcommand{\lMan}{\catname{lMfld}}
\newcommand{\Fr}{\catname{Fr}}
\newcommand{\Man}{\catname{Mfld}}

\newcommand{\VBun}{\catname{VBun}}
\newcommand{\Set}{\catname{Set}}
\newcommand{\Sh}{\catname{Sh}}

\newcommand{\Top}{\catname{Top}}
\newcommand{\Vect}{\catname{Vec}}

\def \A {\mathcal{A}}

\def \C {\mathscr{C}^{\infty}}
\def \D {\mathcal{D}}
\def \E {\mathcal{E}}
\def \F {\mathcal{F}}
\def \G {\mathcal{G}}
\def\Linf{\mbox{$L_{\infty}$}-algebra}
\def\Linfoid{\mbox{$L_{\infty}$}-algebroid}

\def\ihom{\mbox{$\underline{\textrm{Hom}}$}}
\def \M {\mathcal{M}}
\def \N {\mathcal{N}}
\def \RE {\mathbb{R}}

\def \Tg {\rm{T}}
\def \V {\mathcal{V}}
\def \W {\mathscr{W}}

\def \Oml {\Omega_{{\textrm{loc}}}^{p,q}(\E \times M)}
\def \S {\mathbb{S}}

\newcommand{\dem}{\vspace{2.5ex}{\bf Proof. }}
\def \qed {\hfill $\blacksquare$\\}

\usepackage{sectsty}
\allsectionsfont{\color{BlueViolet} \sffamily}
\partfont{\color{darkdelion} \sffamily}

\usepackage{fancyhdr}
 
\author{N\'estor Le\'on Delgado \thanks{Within the doctoral work under the supervision of Christian Blohmann at the Max Planck Institute for Mathematics in Bonn.}}
\title{\sf {\Huge \textcolor{darkdelion}{Lagrangian field theories:}} \\ \textcolor{BlueViolet}{ind/pro-approach and \Linf { of} local observables}}
\date{Version \today}

\usepackage[pdftex]{hyperref}

\begin{document}


\thispagestyle{empty}
\pagenumbering{gobble}
\begin{centering}
\phantom{a}

\vspace{2em}

{\bf
{\huge Lagrangian field theories: \\ ind/pro-approach and \boldmath{\Linf} \\
\vspace{0.25em} { of} local observables}

\vspace{6em}

{\Large 
Dissertation\\
\vspace{0.5em}
zur \\
\vspace{0.5em}
Erlangung des Doktorgrades (Dr. rer. nat.) \\
\vspace{0.5em}
der \\
\vspace{0.5em}
Mathematisch-Naturwissenschaftlichen Fakult\"{a}t \\
\vspace{0.5em}
der \\
\vspace{0.5em}
Rheinischen Friedrich-Wilhelms-Universit\"{a}t Bonn\\
}

\vspace{5em}

{\Large 
vorgelegt von}\\
\vspace{0.9em}
{\LARGE N\'estor Le\'on Delgado} \\
\vspace{0.5em}
{\Large aus \\
\vspace{0.5em}
San Crist\'obal de La Laguna, Teneriffa, Spanien\\
}

\vspace{9em}

{\large Bonn, December 2017}

}

\end{centering}

\newpage
\thispagestyle{empty}
\begin{centering}

Angefertig mit Genehmigung der Mathematisch-Naturwissenschaftlichen Fakult\"{a}t der Rheinischen Friedrich-Wilhelms-Universit\"{a}t Bonn

\end{centering}

\vspace{10em}

1. Gutachter: Prof. Dr. Peter Teichner

2. Gutachter: Prof. Dr. Werner Ballmann

Tag der Promotion: 4. Mai 2018

Erscheinungsjahr: 2018

\newpage
\thispagestyle{empty}

\pagenumbering{roman}



\chapter*{\color{darkdelion} Abstract}

\begin{quotation}

Field theories in Physics and in particular their quantization have been a continuous source of mathematical challenge during decades. The Lagrangian formalism point of view is to start by a Lagrangian: a function on the fields (sections of a smooth fiber bundle) and on the derivatives of the fields, valued in densities of the base manifold. We then arrive at Lagrangian field theories. The work by Deligne and Freed \cite{DEL} is a very good exposition to this formalism. The key property is that the Lagrangian depends only on finitely many derivatives of the fields, i.e., on a finite jet bundle. Fixing a finite jet degree is not a convenient solution since all the usual machinery in differential geometry does not leave the jet degree invariant (differentiation, for example).\\

The book by Anderson on the infinite jet bundle \cite{AND} is the most exhaustive example of the theory necessary to include all jet degrees. The work by Anderson is still very inspirational, and new results keep on being published to extend his work and bring it to both a more modern mathematical language, but also closer to the physical motivations of field theories. The works of G\"{u}neysu and Pflaum \cite{GP} and Khavkine and Schreiber \cite{KS} are good examples for this current development.\\

The approach on this thesis is to talk about the infinite jet bundle as a pro-finite dimensional smooth manifold and, similarly, to view forms on it as ind-differential forms. This idea, already present in the work of Blohmann \cite{B}, has been explored in some depth. Part I of the thesis introduces the relevant notions to talk about the infinite jet bundle as a pro-finite smooth manifold. Part II focuses on comparing this approach to others, including different topological and Fr\'echet structures on the infinite jet bundle.\\

The Lagrangian, nevertheless, is not a form on the infinite jet bundle, but a local form on the space of fields and the base manifold. The adjective ``local'' roughly means that the corresponding structure on the space of fields times the base manifolds has been pulled back from the corresponding infinite jet bundle. This is the way the theory is studied by Deligne and Freed \cite{DEL}. Updating their work to fit into the language of ind- and pro-categories has been a motivation and a source of results for the present thesis. Part IV is a review of results related to the bicomplex of local forms and Lagrangian field theories based on that categorical approach. At that point we do not only follow Deligne and Freed but also Zuckerman's paper \cite{ZUC}. Later in Chapter 13 we review Noether's second theorem (from her paper \cite{NOE}) using the bicomplex of local forms.\\

Local forms are one of the pieces coming from a category which is in many ways similar to that of smooth manifolds. A category in which objects are spaces of smooth sections times the base manifold and whose objects are maps descending to a finite jet bundle: local maps. As a matter of fact, it is important to be a bit less strict and also to consider maps between pairs in which the base manifolds are not the same (think for example of restricting a set of solutions of a differential equation to its boundary, or extending the boundary data to solutions of the equation in the whole manifold). The resulting category, called the category of insular manifolds is defined in Part III.\\

The adjective insular suggest that locality is not enough. The Cartan distribution, extensively treated in the Russian literature on Lagrangian field theories (for example Vinogradov \cite{VINO}), plays a fundamental role in defining that category. The corresponding notions of tangent bundles, vector fields and differential forms in the insular world are studied in Part IV; while that of Lie groups, Lie algebras, and \Linf s are presented in Part V. Finally, Part VI focuses on multisymplectic structures in the insular manifold category.\\

The language of ind- and pro-categories is particularly well suited for the problems in Lagrangian field theory. A categorical approach to local maps and forms is not only natural, but also lacking in the analysis of similar problems as in the paper of G\"{u}neysu and Pflaum \cite{GP} or when talking about pseudogroups as in the PhD thesis of Yudilevich \cite{ORI}. Moreover, other approaches to quantum field theories, such as that of topological quantum field theories are based on categories and functors. In order to be able to relate all those different approaches, it was necessary to go in the direction of creating the category of insular manifolds.\\

Deformation quantization has been one of the most relevant tools to the study of field theories in the past decades. From the paper of Kontsevich \cite{KON} to the work of Costello and Gwilliam \cite{CG2}, we can see how producing \Linf s out of a field theory is the first step towards having a deformation functor and towards considering deformation quantization of field theories. The ind/pro-categorical approach is also helpful when talking about \Linf s, since the different characterizations of an {\Linf} in the finite dimensional world also hold in pro-finite dimensional manifolds. Chapter 14 focuses in defining \Linf s in this world and Part VI defines the {\Linf} of local observables on a field theory. This {\Linf} has the virtue to depend only on the cohomology of the Lagrangian. Moreover, the brackets are naturally antisymmetric and multilinear, without the need of introducing dual fields, ghosts, or any other extra fields to resolve any kind of symmetries. This is an advantage in comparison with the similar structures coming out of the BV-BRST formulation of field theories, such as the one by Costello and Gwilliam.\\

Rogers in his PhD thesis \cite{ROG}, shows that given any finite dimensional multisymplectic manifold, one gets an {\Linf} structure on its Hamiltonian forms. That results also holds in the category of insular manifolds. As a matter of fact, the {\Linf} of local observables is the {\Linf} of Hamiltonian forms coming from certain local pre-multisymplectic form. The generalization of Rogers' result to our setting and the study of the relation of the {\Linf} of local observables to other Lie algebra-like structures often used in field theories can be found in Part VI.\\

That local pre-multisymplectic form, called the Poincar\'e-Cartan form associated to a Lagrangian, comes from the cohomological study of the Lagrangian in the variational bicomplex (which was done by Zuckerman). That form has been studied for certain field theories by Gotay, Isenberg, Marsden, Montgomery, \'{S}niatycki, and Yasskin (usually known as GiMmsy) \cite{GIMMSY} or by de Le\'on, Mart\'{i}n de Diego, and Santamar\'{i}a-Merino \cite{dL}. Both studies seem to disregard the fact that the form comes from the study of the variational bicomplex and focus only in finding such forms explicitly in local coordinates for various examples of field theories. The study presented in Part VI can be thought of as a coordinate free, cohomological version of the work of GiMmsy, de Le\'on, Mart\'{i}n de Diego, and Santamar\'{i}a-Merino which holds for every finite jet degree.\\

Finally, this thesis opens new directions for further research. The study of Lie pseudogroups, using similar techniques to the ones developed here such as the category of insular manifolds, could bring potentially very interesting results. Lie pseudogroups have been proven to model correctly symmetries in field theories and a theory of reduction is available in that context. Yudilevich's work brings some light into this problem.
Following with reduction, multisymplectic reduction via Hamiltonian moment maps (as Defined in Part VI) is a natural way of using the Poincar\'e-Cartan form to get rid of the symmetries in field theories in order to get well-posedness for the equations of motion. The work of Blohmann and Weinstein follows this idea. And last but not least: quantization. Once the {\Linf} of local observables has been defined it is natural to consider deformation functors induced by it. It is very relevant to mention that the Maurer-Cartan elements of the {\Linf} here presented are trivial for degree reasons. Modifying the {\Linf} to allow higher Hamiltonian multi-vector fields (such as the ones in N.L.D. \cite{YO}) but most importantly, Hamiltonian $0$-vector fields (functions such that $f \omega = d \alpha$ for some $\alpha$) will help solving this problem. As a matter of fact, the quantization of such Maurer-Cartan elements seems to be related to the second quantization operators derived from the fields. Comparing those deformation functors with the ones from Costello and Gwilliam, and more generally comparing the {\Linf} of local observables with others available in the literature (see for example the work of Schiavina \cite{CS}) are potential applications of the work here presented.\\

\end{quotation}

\tableofcontents


\chapter*{\color{darkdelion} Acknowledgments}

I would like to begin thanking my three advisers. Christian Blohmann, who has been my real adviser, besides not appearing as an official one due to administrative reasons. Also Peter Teichner and Werner Ballmann for having agreed to be the official advisers and acted as such, especially in the final period of the thesis. Equally, I would like to thank the Max-Planck-Gesellschaft for having provided me with economic support and the Bonn International Graduate School of Mathematics. In particular  I would like to thank Karen Bingel. Her work, dedication and personality have been one of the biggest highlights of this adventure in Bonn.\\

At a mathematical level I would like to thank several mathematicians who have provided me with very important insights and sources of knowledge. I have to start by Igor Khavkine, who has answer always so willingly and intelligently every question I have posed him. I have shared long conversations with Pavel Mn\"ev, Michele Schiavina and Alessandro Valentino which have affected my research and me as a person in always a positive way. I thank also the rest of the <<Z\"{u}rich Crew>>: Ricardo Miguel Campos, Vincent Schlegel and Konstantin Wernli: all our mathematical adventures have been very fun (especially in Morocco). Owen Gwilliam and Claudia Scheimbauer: thank you for your wonderful lecture on derived deformation theory and for being such great mathematicians, full of enthusiasm and always willing to share your knowledge. I also need to mention David Carchedi, Honglei Lang, Dmitry Roytenberg, Ori Yudilevich and Marco Zambon for all their comments and for always showing passion for the kind of mathematics they do.\\

Coming back to Michele Schiavina, I would also like to thank him for all the discussions while my stay at Berkeley. Thanks to Alan Weinstein for having, not only invited me to Berkeley, but offered to share his time and mathematical knowledge with me during my month at the Mathematical Department.\\

I could not finish without thanking all my friends in Bonn, without them nothing would have been the same. The Max-Planckers, the Fat-Flat and Esmee te Winkel: thank you for such wonderful times. And thanks to my family for their continuous support throughout the years.


\newpage
\begin{centering}
\phantom{a}
\vspace{12em}

{\bf \color{darkdelion}
{\huge Lagrangian field theories: \\ ind/pro-approach and \boldmath{\Linf} \\
\vspace{0.25em} { of} local observables}
}

\vspace{6em}

{\Large A tale of insular maps, whose islands are full of \\
sunflower fields and water wells.
}

\end{centering}

\newpage
\pagenumbering{arabic}


\newpage
\part{The Variational Bicomplex: an ind-/pro-categorical approach}

\newpage
\tableofcontents

\chapter*{\color{darkdelion} The Variational Bicomplex: an ind-/pro-categorical approach}

In the framework of Lagrangian field theory the space of fields is given by sections of a smooth fiber bundle over a smooth manifold $M$. This is, we have $\pi \colon E \rightarrow M$ a smooth fiber bundle and its set of smooth sections is denoted by $\E \defeq  \Gamma^{\infty}(M, E)$. Typically, the action depends on the field and a finite number of derivatives of the field. This is expressed in our setting by requiring the Lagrangian to factor through some finite jet bundle of $E$. The theory deals with insertion of vector fields, de Rham differentials and similar differential geometry tools. Some of these operators applied to the Lagrangian change the degree of the jet bundle that the resulting object factors through. This is why it is not possible to fix a finite jet degree, but it is necessary to work with the infinite jet bundle $J^{\infty} E$.\\


There are various ways to work with the infinite jet bundle. The approach in this thesis is to use ind- a pro-categories. In this way, the infinite jet bundle is the pro-finite smooth manifold given by the sequence of finite dimensional jet bundles. Moreover, smooth functions, differential forms, or derivations are to be considered as part of an ind-algebra or to be ind-derivations, respectively.\\

This first part represents the starting point of the thesis. It defines and states some of the main constructions and results about the variational bicomplex in the ind-/pro-categorical language. The up short is that all the usual formulas (known as Cartan calculus) hold in the infinite dimensional setting provided we work with ind-/pro-structures. It also present some comparison results relating Fr\'echet smooth maps and pro-finite smooth maps. This is a first step into dealing with the Fr\'echet structures on the space of fields versus the local structure.\\

The main result besides the development of the ind-/pro-language for the variational bicomplex is about jet prolongations. Jet prolonged pro-smooth maps are a particular kind of maps between two infinite jet bundles that can be recovered from a single map between two associated finite jet bundles. In Section \ref{jpg} we extend the class of pro-smooth maps that admit a unique jet prolongation to include maps covering a submersion which can vary from section to section.\\

{\it This part is designed to provide a reference for an ind-/pro-categorical approach to the infinite jet bundle. It gives a general overview of the basics of the variational bicomplex, restating some results in this other language. It also proves the existence of pro-smooth jet prolongations for a larger class of maps than the ones studied so far. Throughout the Part, we compare pro-smooth manifolds to the non-equivalent notion with the same name recently published by G\"uneysu and Pflaum \cite{GP}. The main references in this part are Anderson \cite{AND}, Chetverikov \cite{CHE}, and Saunders \cite{SAU}.}


\newpage
\chapter{The infinite jet bundle}

The infinite jet bundle is to be thought of as a projective limit of the finite dimensional jet bundles. Since this limit does not exist in the category of smooth finite dimensional manifolds, the most natural way to deal with it is to consider the associated category of pro-smooth manifolds (where these limits are added formally).\\

This chapter defines the finite and the infinite jet bundles associated to a smooth fiber bundle using the language of ind- and pro-categories. 
We explore this approach further by showing that the set of pro-finite smooth maps from $J^{\infty} E$ to $\RE$ can be given the structure of a an ind-algebra (formal colimit of algebras): this leads to the definition of the ind-algebra of smooth functions on the infinite jet bundle.\\

The first two sections consist of definitions taken from different sources in the literature and of examples of pro-smooth maps. The most prominent example is related to the study of pullbacks in pro-smooth manifolds, where we show that the fiber product of two infinite jet bundles is isomorphic to the infinite jet bundle of the fiber product (Proposition \ref{cqc}).\\

Section \ref{jif} focuses on the comparison of the pro-smooth approach to the Fr\'echet manifold approach followed by other authors. The infinite jet bundle is a Fr\'echet manifold. We specialize a result by Dodson-Galanis-Vassiliou to show that Fr\'echet spaces of certain kind are sequential pro-finite normed vector spaces. Similarly for the morphisms: morphisms of pro-finite spaces are morphisms of the corresponding Fr\'echet spaces, also called smooth.  Pro-smooth maps are actually smooth.\\

{\it 
The main references in this chapter are Dodson-Galanis-Vassiliou \cite{DGV}, Saunders \cite{SAU}, and SGA 4.1, expos\'e i \cite{SGA}.}


\section{Jet bundles}\label{11}

{\it In this section we introduce ind- and pro-categories, fixing our approach to work with infinite jet bundles. Finite jet bundles are defined, as well as the pro-smooth infinite jet bundle together with all the structure maps. The main references here are Saunders \cite{SAU} and SGA 4.1, expos\'e i \cite{SGA}.}\\

We consider a fiber bundle in the category of smooth manifolds $\pi \colon E \rightarrow M$. We denote the set of smooth sections of $\pi$ by $\E = \Gamma^{\infty} (M, E)$. We are interested in classes of local sections in the following sense: two sections around a certain point are equivalent if their partial derivatives agree up to a finite degree. We need a formal definition, extracted from Saunders \cite{SAU}.

\begin{df}[Space of $k$-jets of local sections]\label{jetspa}
Given a fiber bundle $\pi \colon E \rightarrow M$ and an integer $k$ we say that two local sections of $\pi$ around a point $x \in M$ have the same $k$-jet at $x$ if their partial derivatives up to order $k$ agree at $x$, in some chart around $x$. $J_x^k E$ denotes the set of such equivalent classes of local sections around $x$. 

\end{df}

We denote the class $[(\varphi, x)]_k$ also by $j_x^ k \varphi$. Observe that the definition does not depend on the coordinate chart chosen (this is clear for $k=1$ using the chain rule, and for the general statement see Saunders \cite[Lemma 6.2.1]{SAU}). Observe also, that $\varphi$ and $\varphi^{\prime}$ have the same $k$-th jet at $x$ if and only if $\, \restrict{T^k(\varphi)}{T^k_x M} = \restrict{T^k(\varphi^{\prime})}{T^k_x M}$ where $T^k \varphi$ is the iterated tangent map:
$$T^k \varphi= T(T( \cdots T \varphi) \cdots ) \colon T(T( \cdots T M) \cdots ) \rightarrow T(T( \cdots T E) \cdots ).$$

\begin{dfpp}[Bundle of $k$-jets of local sections]\label{jet}
Given a fiber bundle $\pi \colon E \rightarrow M$ and an integer $k$ we denote the set $J^k E \defeq  \{ j_x^k \varphi \in J_x^k M \colon x \in M\}$. It is a smooth bundle over $E$ and hence over $M$.
\begin{align*}
	\pi_k \colon J^k E  & \longrightarrow M \\
	j_x^k(\varphi) & \longmapsto x.
\end{align*}
\end{dfpp}

The previous definition/proposition follows again Saunders \cite{SAU}. Given a global section, we can evaluate its $k$-th jet:
\begin{align*}
	j^k \colon \E \times M & \longrightarrow J^k  E \\
	(\varphi, x) & \longmapsto  j_x^k(\varphi).
\end{align*}

\begin{rk} The bundle of $k$-jets is defined not only for global but for {\bf local sections}. Given an open subset of $U$ of $M$, any local section $\varphi \, \in \Gamma^{\infty}(U, \restrict{E}{\pi^{-1}U})$ defines an element in $J^k E$. This induces a map $j^k (U) \colon \E(U) \rightarrow J^k E$. In some cases local sections cannot be extended to global ones. In other words, sometimes $j^k$ is not surjective. When the sheaf $\E$ is soft, $j^k$ is surjective. (A sheaf of sections is soft when it is possible to extend local sections defined over closed subsets to global ones).
\end{rk}

The space of $0$-jets is just $E$ and $j^0$ is the evaluation map. The space of sections $\Gamma^{\infty}(M, J^k E)$ will be denoted by $\mathcal{J}^k \E$. There are canonical projections between different jet bundles:
\begin{align*}
    \pi_k^l \colon J^k E & \longrightarrow J^l E  \textrm{ for } k \geq l \\
    j_x^k(\varphi) & \longmapsto j_x^l(\varphi).
\end{align*}

These maps are smooth fiber bundles, sequences of affine bundle projections $\pi_k^{k+1}$ (see Saunders \cite{SAU}). All the bundle maps are compatible with the family of projections $\{\pi_k^l\}$. From one side $\pi_k^l \circ \pi_i^k \colon J^i E \rightarrow J^l E$, also $\pi_l \circ \pi_i^l = \pi_i \colon J^i E \rightarrow M$ and $\pi_i^k \circ j^i = j^k \colon \E \times M \rightarrow J^k E$. This is summarized in the following commutative diagram in which $i \geqslant k \geqslant l$:

\begin{center}
\begin{tikzpicture}[description/.style={fill=white,inner sep=2pt}]
\matrix (m) [matrix of math nodes, row sep=2em,
column sep=2em, text height=1.5ex, text depth=0.25ex]
{ \E \times M & & J^k E && J^l E && M \\
  &&& J^i E &&& \\};
\path[->,font=\scriptsize]
(m-1-1) edge node[auto] [swap]{$j^i$} (m-2-4)
(m-2-4) edge node[auto] [swap]{$\pi_i^k$} (m-1-3)
(m-2-4) edge node[auto] {$\pi_i^l$} (m-1-5)
(m-2-4) edge node[auto] [swap]{$\pi_i$} (m-1-7)
(m-1-1) edge node[auto] {$j^k$} (m-1-3)
(m-1-3) edge node[auto] {$\pi_k^l$} (m-1-5)
(m-1-5) edge node[auto] {$\pi_l$} (m-1-7);
\end{tikzpicture}
\end{center}

\begin{ej}\label{ene} The identity $\textrm{id} \colon M \rightarrow M$ is a trivial fiber bundle over $M$ with typical fiber the singleton (it is the trivial vector bundle of dimension $0$). Any local section $\varphi \colon U \rightarrow M$ is such that $\textrm{id} \circ \varphi = \varphi = \textrm{id}_U$. This means that there is only one section of the bundle and it is globally defined: the identity. The space of $k$-th jets is simply $M \cong \bigslant{\{(\textrm{id},x) \colon x \in \, M \}}{\sim}$. In this way $J^k M = M$ and $\pi_k^l = \pi_k = \textrm{id}$ for every $k$ and $l$.  
\end{ej}

Definition \ref{jet} can be extended to the case where the partial derivatives of two sections agree at any order. In that case, the sections are said to have the same $\infty$-jet. The limit of the sequence 
$$E = J^0 E  \longleftarrow J^1 E  \longleftarrow J^2 E  \longleftarrow J^3 E  \longleftarrow \cdots $$
\noindent{does not exist in the category of smooth fiber bundles in general.}

Our chosen way to get around this is to work with the ind- and pro-categories associated to a category (we follow the description given in the monograph SGA 4.1, expos\'e i, \cite{SGA}).

\begin{df}[Ind-category]
Let $\Cat$ be a category. The ind-category given by $\Cat$ which will be denoted by $\textrm{Ind}(\Cat)$ has as objects $X \colon \mathcal{I} \rightarrow \Cat$ functors, where $\mathcal{I}$ is an essentially small filtered category. Morphisms between two of objects $X \colon \mathcal{I} \rightarrow \Cat$ and $Y \colon \mathcal{J} \rightarrow \Cat$ are given by 
$$\textrm{Hom}_{\textrm{Ind}(\Cat)}(X, Y) \defeq  \lim_{i \in \mathcal{I}} \underset{j \in \mathcal{J}}{\textrm{colim}} \, \, \textrm{Hom}_{\Cat}(X_i, Y_j).$$
\end{df}

We want to emphasize that there is no essentially small filtered category fixed in the construction. Every ind-object is a functor from an, every time possibly different, essentially small filtered category $\mathcal{I}$.

\begin{df}[Pro-category]
Let $\Cat$ be a category. We define the pro-category given by $\Cat$ by $\textrm{Pro}(\Cat) = \left( \textrm{Ind}(\Cat^{\textrm{op}}) \right)^{\textrm{op}}$.
\end{df}

To be precise, objects in $\textrm{Pro}(\Cat)$ are functors $X \colon \mathcal{I} \rightarrow \Cat$,  where $\mathcal{I}$ is an essentially small cofiltered category. Morphisms between two of objects $X \colon \mathcal{I} \rightarrow \Cat$ and $Y \colon \mathcal{J} \rightarrow \Cat$ are given by 
$$\textrm{Hom}_{\textrm{Pro}(\Cat)}(X, Y) \defeq  \lim_{j \in \mathcal{J}} \underset{i \in \mathcal{I}}{\textrm{colim}} \, \, \textrm{Hom}_{\Cat}(X_i, Y_j).$$

At this point we are only going to use pro-categories, but later when talking about algebras of functions we will need ind-categories as well.

When the category $\Cat$ is $\Man$, that is finite dimensional manifolds, we talk about pro-finite dimensional manifolds, or pro-smooth manifolds for short. G\"uneysu and Pflaum \cite{GP} also use the term {\it pro-finite dimensional manifold} but with a different meaning. First of all, they only consider objects indexed by $\mathbb{N}$. But even when restricted to the same objects, the morphisms are not the same. Morphisms of pro-smooth manifolds between their {\it pro-finite dimensional manifolds} are called {\it local} in their work. We will explain what they call {\it morphisms of pro-finite dimensional manifolds} in the following section.

We are ready to consider the collection of $\infty$-jets as the pro-object given by the sequence of finite jet bundles, $(J^k E, \pi_k^l)$ in the category of smooth manifolds $\Man$ (this follows Blohmann \cite{B}):

\begin{df}[$J^{\infty}  E $]
Given a fiber bundle $\pi \colon E \rightarrow M$, the space of $\infty$-jets of $\pi$ is the pro-object in the category of finite dimensional smooth manifolds given by $E = J^0 E  \leftarrow J^1 E  \leftarrow J^2 E  \leftarrow \cdots $. We denote it by $J^{\infty} E$.
\end{df}

The term space in the previous definition is used in a vague way. From now on, pro-smooth manifolds will sometimes be called spaces to make the reading lighter. Observe that this is partially justified by the fact that the forgetful functor from $\Man$ to $\Top$ induces a forgetful functor from $\textrm{Pro}(\Man)$ to $\textrm{Pro}(\Top)$. As a matter of fact, given any functor $F \colon \Cat \rightarrow \Cat^{\prime}$, we get induced functors 
$\textrm{Pro}(F) \colon \textrm{Pro}(\Cat) \rightarrow \textrm{Pro}(\Cat^{\prime})$ and 
$\textrm{Pro}(F) \colon \textrm{Ind}(\Cat^{\prime}) \rightarrow \textrm{Ind}(\Cat)$. In the case in which the functor $F$ is contravariant we have instead
$\textrm{Pro}(F) \colon \textrm{Pro}(\Cat) \rightarrow \textrm{Ind}(\Cat^{\prime})$ and 
$\textrm{Pro}(F) \colon \textrm{Ind}(\Cat^{\prime}) \rightarrow \textrm{Pro}(\Cat)$.

\begin{rk}\label{functs} Informally, we can think of ind- and pro-categories as an extension of the original ones where we have added some of the missing limits and colimits respectively; and we have done it accordingly to the properties of limits and colimits with respect to morphisms. We only add filtered colimits and cofiltered limits and define the morphisms the same way they will be defined in the category of presheaves. Recall that a filtered colimit is a colimit over a diagram from a filtered category, for example a sequential colimit:

$$A_0 \longrightarrow A_1  \longrightarrow A_2  \longrightarrow A_3 \cdots \longrightarrow ``\underset{i \in \mathbb{N}}{\textrm{colim }}" A_i$$

Consider the case in which two such colimits $\underset{i \in \mathcal{I}}{\textrm{colim }} A_i$ and $\underset{j \in \mathcal{J}}{\textrm{colim }} B_j$ do exist in the category $\Cat$. If the functor $\Cat \rightarrow \textrm{Ind}(\Cat)$ is fully-faithful and if the elements of $\Cat$ are compact in $\mathrm{Ind}(\Cat)$ (i.e. mapping out of elements of $\Cat$ commutes with filtered colimits), the morphims between the colimits in $\Cat$ are as follows
\begin{eqnarray*}
	\begin{split}
		\mathrm{Hom}_{\Cat}(A,B) &= \mathrm{Hom}_{\Cat}(\underset{i \in \mathcal{I}}{\textrm{colim }} A_i, \underset{j \in \mathcal{J}}{\textrm{colim }} B_j) \\
		& \cong \lim_{i \in \mathcal{I}} \mathrm{Hom}_{\Cat}(A_i, \underset{j \in \mathcal{J}}{\textrm{colim }} B_j) \\
		& \cong \lim_{i \in \mathcal{I}} \underset{j \in \mathcal{J}}{\textrm{colim }} \mathrm{Hom}_{\Cat}(A_i, B_j) \\
		& \cong \lim_{i \in \mathcal{I}} \underset{j \in \mathcal{J}}{\textrm{colim }} \mathrm{Hom}_{\Cat}(A_i, B_j).
	\end{split}
\end{eqnarray*}

All the previous assumptions are satisfied if the filtered category indexing the diagram is essentially small (SGA 4.1, expos\'e i, \cite{SGA} and Mac Lane \cite{MAC} are good references at this point). This is one way of remembering the formula for the morphisms between pro-objects. Another way is to compute the morphisms as pre-sheaves. Given $A \colon \mathcal{I} \rightarrow \Cat$, consider the pre-sheaf $\widehat{A} \defeq  \underset{i \in \mathcal{I}}{\textrm{colim }} y(A_i)$ where $y$ denotes the Yoneda embedding. Do the same for $\widehat{B}$. Now we can repeat the calculation above in pre-sheaves to see that the morphisms of ind-objects are actually calculated in pre-sheaves
\begin{eqnarray*}
	\begin{split}
		\mathrm{Hom}_{\Set^{\Cat^{\textrm{op}}}}(\widehat{A},\widehat{B}) &= \mathrm{Hom}_{\Set^{\Cat^{\textrm{op}}}}(\underset{i \in \mathcal{I}}{\textrm{colim }} y(A_i), \widehat{B}) \\
		& =\lim_{i \in \mathcal{I}} \mathrm{Hom}_{\Set^{\Cat^{\textrm{op}}}}(y(A_i), \widehat{B})  \\
		& \cong \lim_{i \in \mathcal{I}} \widehat{B}(A_i) \\
		& = \lim_{i \in \mathcal{I}} \underset{j \in \mathcal{J}}{\textrm{colim }} \mathrm{Hom}_{\Cat}(A_i, B_j) \\
		& = \mathrm{Hom}_{\textrm{Iro}(\Cat)}(A, B).
	\end{split}
\end{eqnarray*}

The statement here is that the map from $\textrm{Ind}(\Cat) \rightarrow \Set^{\Cat^{\textrm{op}}}$ is fully faithful. 
\end{rk}

In Appendix \ref{app1} we give a brief summary of the key computations to work with morphisms of pro-objects. But, the fact that $\pi_j^k$, $\pi_k$ and $j^k$ behave well with respect to $\{\pi_k^l\}$ allows us to define the corresponding maps from or to $J^{\infty} E$. Observe what the spaces of morphisms to which they belong are:
\begin{eqnarray*}
\begin{split}
\textrm{Hom}_{\textrm{Pro}(\Man)}(J^{\infty} E, J^l E) &= \underset{k \in \mathbb{N}}{\textrm{colim }} \textrm{Hom}_{\Man}(J^k E, J^l E) = \underset{k \in \mathbb{N}}{\textrm{colim }} \C(J^k E, J^l E), \\
\textrm{Hom}_{\textrm{Pro}(\Man)}(J^{\infty} E, M) &= \underset{k \in \mathbb{N}}{\textrm{colim }} \textrm{Hom}_{\Man}(J^k E, M) = \underset{k \in \mathbb{N}}{\textrm{colim }} \C(J^k E, M)  \textrm{ and}\\
\textrm{Hom}_{\textrm{Pro}(\Fr)}(\E \times M, J^{\infty} E) &\defeq  \lim_{k \in \mathbb{N}} \textrm{Hom}_{\Fr}(\E \times M, J^k E).
\end{split}
\end{eqnarray*} 

Observe that $\E \times M$ is not a finite dimensional smooth manifold, but rather a Fr\'echet manifold (this will be further discussed in Part II). Until then we can think of the maps appearing in the last line to be morphisms of pro-sets and of sets instead. $\Fr$ denotes the category of Fr\'echet manifolds.

The following maps are examples of pro-finite smooth maps: $\pi_{\infty}^l$, $\pi_{\infty}$ and $j^{\infty}$.
\begin{eqnarray*}
\begin{split}
\pi_{\infty}^l \colon J^{\infty} E & \longrightarrow J^l E,\\
\pi_{\infty} \colon J^{\infty} E & \longrightarrow M \textrm{  and}\\
j^{\infty} \colon \E \times M & \longrightarrow J^{\infty} E.
\end{split}
\end{eqnarray*} 
\noindent{These are given by the colimits associated to $\textrm{id} \colon J^l E \rightarrow J^l E$, $\pi = \pi_0 \colon J^0 E = E \rightarrow M$ and the limit of the family of maps $\{j^k \colon \E \times M \rightarrow J^k E\}_{k\in \mathbb{N}}$ respectively.}

Observe that we could have equally defined $J^{\infty} E$ in the pro-category of smooth bundles over $M$ instead of on the one of smooth manifolds and later construct the map $\pi_{\infty}$. This does not matter at this point, but it gives more flexibility to construct maps between infinite jet bundles that do not cover a smooth map between the bases. We will explore this idea further in the following sections.

We have a commutative diagram involving all the maps related to $J^{\infty} E$ which is basically the one we had for $J^i E$, $J^k E$ and $J^l E$ in which we replace $i$ by $\infty$. The following commutative diagram expresses all the information stated in the previous paragraphs ($i > k > l$). The diagram takes place in $\textrm{Pro}(\Fr)$.

\begin{center}
\begin{tikzpicture}[description/.style={fill=white,inner sep=2pt}]
\matrix (m) [matrix of math nodes, row sep=2em,
column sep=2em, text height=1.5ex, text depth=0.25ex]
{ &&& J^{\infty} E &&&\\
 \E \times M & & J^k E && J^l E && M \\
  &&& J^i E &&& \\};
\path[->,font=\scriptsize]
(m-2-1) edge node[auto] {$j^{\infty}$} (m-1-4)
(m-1-4) edge node[auto] [below, right]{$\pi_{\infty}^k$} (m-2-3)
(m-1-4) edge node[auto] [below, left]{$\pi_{\infty}^l$} (m-2-5)
(m-1-4) edge node[auto] {$\pi_{\infty}$} (m-2-7)
(m-2-1) edge node[auto] [swap]{$j^i$} (m-3-4)
(m-3-4) edge node[auto] [swap]{$\pi_i^k$} (m-2-3)
(m-3-4) edge node[auto] {$\pi_i^l$} (m-2-5)
(m-3-4) edge node[auto] [swap]{$\pi_i$} (m-2-7)
(m-2-1) edge node[auto] {\phantom{m} $j^k$} (m-2-3)
(m-2-3) edge node[auto] {$\pi_k^l$} (m-2-5)
(m-2-5) edge node[auto] {$\pi_l$ \, \,} (m-2-7);
\end{tikzpicture}
\end{center}


\section{Pro-smooth maps between infinite jet bundles}\label{13}

{\it This section provides explicit formulas for pro-finite smooth maps between infinite jet bundles. It focuses on the definition of the ind-algebra of smooth functions, showing that the set of pro-smooth functions on $J^{\infty} E$ can be given the structure of an ind-algebra. We work with different examples, including how infinite jet bundles behave with respect to fiber products. Then we show that $J^{\infty}(E \times_M F) \cong J^{\infty} E \times_M J^{\infty} F$. The main reference in this section is Saunders \cite{SAU}.}\\

Given two smooth fiber bundles $\pi \colon E \rightarrow M$ and $\rho \colon F \rightarrow N$ we are going to give a better insight into the pro-finite smooth maps between $J^{\infty} E$ and $J^{\infty} F$. For details we refer to Appendix \ref{app1}. Let 

$$f^{\infty} \in \, \textrm{Hom}_{\textrm{Pro}(\Cat)}(J^{\infty} E, J^{\infty} F) \defeq  \lim_{l \in \mathbb{N}} \underset{k \in \mathbb{N}}{\textrm{colim}} \, \, \textrm{Hom}_{\Cat}(J^k E, J^l F).$$

This means that we have a limit of smooth maps $f^l \colon J^{k(l)} E \rightarrow J^l F$ satisfying certain compatibility relations, summarized in the diagram below being commutative for all $l \in \mathbb{N}$:

\begin{center}
\begin{tikzpicture}[description/.style={fill=white,inner sep=2pt}]
\matrix (m) [matrix of math nodes, row sep=3em,
column sep=3.5em, text height=1.5ex, text depth=0.25ex]
{J^{k(l+1)} E & J^{l+1} F\\
 J^{k(l)} E & J^{l} F\\};
\path[->,font=\scriptsize]
(m-1-1) edge node[auto] {$f^{l+1}$} (m-1-2)
(m-2-1) edge node[auto] {$f^{l}$} (m-2-2)
(m-1-1) edge node[auto] {$\pi_{k(l+1)}^{k(l)}$} (m-2-1)
(m-1-2) edge node[auto] {$\rho_{l+1}^{l}$} (m-2-2);
\end{tikzpicture}
\end{center}

Observe that we have implicitly assumed that we can take $k$ as a function to be non-decreasing, see Appendix \ref{app1} for more details. If that were not the case, for instance $k(l) > k(l+1)$, we can always take $k^{\prime}(l) \defeq  k(l+1)$ and $f^l \defeq  \rho_{l+1}^l \circ f^{l+1}$.

We can summarize all the information in the following diagram (for convenience $k(0)$ is simply denoted by $k$):

\begin{center}
\begin{tikzpicture}[description/.style={fill=white,inner sep=2pt}]
\matrix (m) [matrix of math nodes, row sep=3em,
column sep=3.5em, text height=1.5ex, text depth=0.25ex]
{J^{\infty} E & J^{\infty} F\\
 J^{k(l)} E & J^{l} F\\
 J^{k} E & F\\};
\path[->,font=\scriptsize]
(m-1-1) edge node[auto] {$f^{\infty}$} (m-1-2)
(m-2-1) edge node[auto] {$f^{l}$} (m-2-2)
(m-3-1) edge node[auto] {$f^{0}$} (m-3-2)
(m-1-1) edge node[auto] {$\pi_{\infty}^{k(l)}$} (m-2-1)
(m-1-2) edge node[auto] {$\rho_{\infty}^{l}$} (m-2-2)
(m-2-1) edge node[auto] {$\pi_{k(l)}^{k}$} (m-3-1)
(m-2-2) edge node[auto] {$\rho_{l}^{0}$} (m-3-2);
\end{tikzpicture}
\end{center}

The map $f^{\infty}$ could have been represented by any other family $\{\widetilde{f}^l\}$ for a different $\widetilde{k}$. In that case, the two families represent the same $f^{\infty}$ if for every $l$ the following diagram commutes (we assume $k(l) \geqslant \widetilde{k}(l)$):

\begin{center}
\begin{tikzpicture}[description/.style={fill=white,inner sep=2pt}]
\matrix (m) [matrix of math nodes, row sep=3em,
column sep=3.5em, text height=1.5ex, text depth=0.25ex]
{J^{k(l)} E & J^{l} F\\
 J^{\widetilde{k}(l)} E &  \\};
\path[->,font=\scriptsize]
(m-1-1) edge node[auto] {$f^{l}$} (m-1-2)
(m-2-1) edge node[auto] {$\widetilde{f}^{l}$} (m-1-2)
(m-1-1) edge node[left] {$\pi_{k(l)}^{\widetilde{k}(l)}$} (m-2-1);
\end{tikzpicture}
\end{center}

Let us look at examples of maps of this kind, in particular when one of the two bundles is the trivial bundle $M \rightarrow M$.

{\noindent {\it Example }\ref{ene} {\it continued.} Consider the trivial bundle $M \rightarrow M$. We are going to show that $J^{\infty} M$ is isomorphic to $M$ (considered as the pro-object $M \colon \{*\} \rightarrow \Cat$ sending $*$ to $M$). Let $f \colon M \rightarrow J^{\infty} M$ be the morphism in the pro-category given by $f^l = \textrm{id} \colon M \rightarrow J^l M \cong M$ and let $g \colon J^{\infty} M \rightarrow M$ be given by $g^{\{*\}}=\textrm{id} \colon J^{0} M \rightarrow M$.}\\

From one side $g \circ f \colon M \rightarrow M$ is given by $(g \circ f)^{\{*\}} = g^0 \circ f^{\{*\}} = \textrm{id} \colon M \rightarrow M$. This shows that $g \circ f$ is the identity on $M$.\\

From the other side $f \circ g \colon J^{\infty} M \rightarrow J^{\infty} M$ is given by $(f \circ g)^l = f^{\{*\}} \circ g^l = \textrm{id}$ as a map from $M \cong J^0 M$ to $J^l M$. This represents the same maps as $\textrm{id} \colon J^{\infty} M \rightarrow J^{\infty} M$ because the following diagram (trivially) commutes:
\begin{center}
\begin{tikzpicture}[description/.style={fill=white,inner sep=2pt}]
\matrix (m) [matrix of math nodes, row sep=3em,
column sep=3.5em, text height=1.5ex, text depth=0.25ex]
{J^{l} M \cong M& J^{l} M \cong M\\
 M &  \\};
\path[->,font=\scriptsize]
(m-1-1) edge node[auto] {id} (m-1-2)
(m-2-1) edge node[auto] {id} (m-1-2)
(m-1-1) edge node[left] {id} (m-2-1);
\end{tikzpicture}
\end{center}

\begin{ej} Let $E = M \stackrel{\scriptscriptstyle{\pi = \textrm{id}}}{\rightarrow} M$ and $f^{\infty} \colon J^{\infty} M \rightarrow J^{\infty} F$ be a map in the pro-category of smooth finite dimensional manifolds. In this case, the function $k$ can be chosen to be identically $0$ (we can just take the composition with $\textrm{id} = \pi_{k^{\prime}}^0$). $f^{\infty}$ is just a family of maps $f^l \colon M \rightarrow J^l F$ such that the following family of diagrams (for each $l$) commute:
\begin{center}
\begin{tikzpicture}[description/.style={fill=white,inner sep=2pt}]
\matrix (m) [matrix of math nodes, row sep=1.5em,
column sep=2.5em, text height=1.5ex, text depth=0.25ex]
{  & J^{l+1} F\\
 M & \\
   & J^{l} F\\};
\path[->,font=\scriptsize]
(m-2-1) edge node[auto] {$f^{l+1}$} (m-1-2)
(m-2-1) edge node[auto] {$f^{l}$} (m-3-2)
(m-1-2) edge node[auto] {$\rho_{l+1}^l$} (m-3-2);
\end{tikzpicture}
\end{center}
In other words,
$$\textrm{Hom}_{\textrm{Pro}(\Cat)}(J^{\infty} M, J^{\infty} F) \cong \textrm{Hom}_{\textrm{Pro}(\Cat)}(M, J^{\infty} F) =  \lim_{l \in \mathbb{N}} \, \textrm{Hom}_{\Cat}(M, J^l F).$$
This is conquered just by using the isomorphism between $M$ and $J^{\infty} M$ explained in the previous example. Pro-smooth sections of $\pi_{\infty} \colon J^{\infty} E \rightarrow M$ are of this kind.
\end{ej}

\begin{ej}\label{123} In the other direction, let $F = N \stackrel{\scriptscriptstyle{\rho = \textrm{id}}}{\rightarrow} N$ and $f^{\infty} \colon J^{\infty} E \rightarrow J^{\infty} F \cong N$ be a map in the pro-category of smooth finite dimensional manifolds. In this case, the function $k$ can be chosen to be constant $k = k(0)$. $f^{\infty}$ is given just by a single map $f^0 \colon J^k E \rightarrow N$, where all the other maps $f^l$ are just $f^0$ again. In other words, using that $J^{\infty} N \cong N$: $$\textrm{Hom}_{\textrm{Pro}(\Cat)}(J^{\infty} E, J^{\infty} N) \cong \textrm{Hom}_{\textrm{Pro}(\Cat)}(J^{\infty} E, N) \defeq  \underset{k \in \mathbb{N}}{\textrm{colim}} \, \, \textrm{Hom}_{\Cat}(J^k E, N).$$
\end{ej}

We can apply the previous example to the case in which $N = \RE$. $$\textrm{Hom}_{\textrm{Pro}(\Cat)}(J^{\infty} E, \RE) = \underset{k \in \mathbb{N}}{\textrm{colim}} \, \, \textrm{Hom}_{\Cat}(J^k E, \RE).$$

We would like to call this $\C(J^{\infty} E)$ but observe there is another approach to define such a thing which actually endows $\C(J^{\infty} E)$ with a richer structure: for every natural number $k$, we denote by $\C(J^k E)$ the algebra of smooth functions of $J^k E$. If $k > l$, $\pi_k^l \colon J^k E \rightarrow J^l E$ induces a map $(\pi_k^l)^* \colon \C(J^l E) \rightarrow \C(J^k E)$: this is the pullback of $\pi_k^l$ or simply pre-composition with $\pi_k^l$. Following Blohmann \cite{B} we have:

\begin{df}[$\C(J^{\infty} E) $]\label{cije}
	Given a fiber bundle $\pi \colon E \rightarrow M$, the space of smooth functions on $J^{\infty} E$ is defined as the ind-object in the category of algebras given by the diagram $$\C(E) = \C(J^0 E) \longrightarrow \C(J^1 E) \longrightarrow \C(J^2 E) \longrightarrow \C(J^3 E) \longrightarrow \cdots .$$
\end{df}

Observe that this is an immediate consequence of the comment about applying contravariant functors to pro- and  ind-categories made before Remark \ref{functs}.
The limit exists in the category of algebras but the interesting feature is that we want to consider ind-algebra maps instead. Observe that the multiplication is an ind-finite map. We will comment this in further detail when talking about derivations and ind-finite derivations of this algebra.

\begin{rk}\label{gune} A pro-smooth function on $J^{\infty} E$ has globally bounded jet degree. By this we mean that it only depends globally on $J^{k} E$ for some finite $k$. It is also possible to talk about locally bounded pro-smooth functions: in order to do so we consider $J^{\infty} E$ as the projective limit in topological spaces of $\{ J^k E \}$ (this will be done in Part II). A map from $J^{\infty} E$ to $\RE$ is said to have locally bounded jet degree if for all $\chi \in J^{\infty} E$ there exists an open neighborhood of $\chi$ such that the restriction of the map to that neighborhood depends on a finite jet bundle. This is precisely the way in which the algebra of smooth functions on $J^{\infty} E$ is defined by G\"uneysu and Pflaum \cite{GP}, where they, unfortunately, call $J^{\infty} E$ with this smooth structure a pro-finite dimensional manifold. That contradicts the language of ind-/pro-categories, but it is an interesting concept on its own. As a matter of fact, as we will see in the following Chapter, their morphisms are smooth maps with the standard Fr\'echet structure on the projective limit.
\end{rk}

We finish this section with a different example, which does not involve the trivial bundle of $M$ over itself. In this case we want to show that two infinite jet bundles are isomorphic. This result will be used when talking about local Lie algebra or group actions in Lagrangian field theories.

\begin{ej}\label{fpb} Consider two smooth fiber bundles over $M$ the same base manifold: $\pi \colon E \rightarrow M$ and $\rho \colon F \rightarrow F$. We can construct the fiber product bundle $E \times_M F$ which is given by the following pullback:
\begin{center}
\begin{tikzpicture}[description/.style={fill=white,inner sep=2pt}]
\matrix (m) [matrix of math nodes, row sep=2.5em,
column sep=2.5em, text height=1.5ex, text depth=0.25ex]
{ E \times_M F & F\\
  E & M \\};
\path[->,font=\scriptsize]
(m-1-1) edge node[auto] {} (m-1-2)
(m-1-1) edge node[auto] {} (m-2-1)
(m-2-1) edge node[auto] {$\pi$} (m-2-2)
(m-1-2) edge node[auto] {$\rho$} (m-2-2);
\begin{scope}[shift=($(m-1-1)!.5!(m-2-2)$)]
\draw +(-.3,0) -- +(0,0)  -- +(0,.3);
\fill +(-.15,.15) circle (.05);
\end{scope}
\end{tikzpicture}
\end{center}

In the case the bundles were vector bundles, this is called the Whitney sum of the bundles. But the situation is more general. Still, the fiber at each point is given by the products of the fibers.

The space of sections $\Gamma^{\infty}(M, E \times_M F)$ is simply $\E \times \F$. In the case of vector bundles, this fact is usually written as $\Gamma(M, E \oplus F) = \E \oplus \F$. The finite jet bundles are of a very specific kind: $J^k (E \times_M F) \cong J^k E \times_M J^k F$, the isomorphism between the two is the one induced by the map $(\E \times \F) (U) \rightarrow \E(U) \times \F(U)$ for every $U$ open in $M$. We denote this isomorphism by $$s_k \colon J^k (E \times_M F) \rightarrow J^k (E) \times_M J^k(F).$$

We can also take the pullback of $J^{\infty} E$ and $J^{\infty} F$ in the category of pro-finite smooth manifolds:

\begin{center}
\begin{tikzpicture}[description/.style={fill=white,inner sep=2pt}]
\matrix (m) [matrix of math nodes, row sep=2.5em,
column sep=2.5em, text height=1.5ex, text depth=0.25ex]
{ J^{\infty} E \times_M J^{\infty} F & J^{\infty} F\\
  J^{\infty} E & M \\};
\path[->,font=\scriptsize]
(m-1-1) edge node[auto] {} (m-1-2)
(m-1-1) edge node[auto] {} (m-2-1)
(m-2-1) edge node[auto] {$\pi_{\infty}$} (m-2-2)
(m-1-2) edge node[auto] {$\rho_{\infty}$} (m-2-2);
\begin{scope}[shift=($(m-1-1)!.5!(m-2-2)$)]
\draw +(-.3,0) -- +(0,0)  -- +(0,.3);
\fill +(-.15,.15) circle (.05);
\end{scope}
\end{tikzpicture}
\end{center}

$J^{\infty} E \times_M J^{\infty} F$ is the pro-object given by the functor $\mathbb{N} \times \mathbb{N} \rightarrow \Cat$ sending $(k,l)$ to $J^k E \times_M J^{l} F$. To be precise about why is this the pullback, consider a morphism in the index category $(k,l) \mapsto (k^{\prime}, l^{\prime})$. We get induced maps from $J^k E \rightarrow J^{k^{\prime}} E$ to $J^l F \rightarrow J^{l^{\prime}} F$ which also give maps $J^k E \times_M J^{l} F \rightarrow J^{k^{\prime}} E \times_M J^{l^{\prime}} F$:

\begin{center}
	\begin{tikzpicture}[description/.style={fill=white,inner sep=2pt}]
	\matrix (m) [matrix of math nodes, row sep=1.2em,
	column sep=1em, text height=1.5ex, text depth=0.25ex]
	{ J^{k} E \times_M J^{l} F & & J^l F &  \\
		& J^{k^{\prime}} E \times_M J^{l^{\prime}} F & & J^{l^{\prime}} F \\
		J^{k} E &  & & \\
		& J^{k^{\prime}} M & & M \\};
	\path[->,font=\scriptsize]
	(m-1-1) edge node[auto] {} (m-1-3)
	(m-1-1) edge node[auto] {} (m-3-1)
	(m-2-2) edge node[auto] {} (m-2-4)
	(m-2-2) edge node[auto] {} (m-4-2)
	(m-3-1) edge node[auto] {$\pi_k^{k^{\prime}}$} (m-4-2)
	(m-1-3) edge node[auto] {$\rho_l^{l^{\prime}}$} (m-2-4)
	(m-4-2) edge node[auto] {$\pi_{k^{\prime}}$} (m-4-4)
	(m-2-4) edge node[auto] {$\rho_{l^{\prime}}$} (m-4-4);
	\path[->,font=\scriptsize]
	(m-1-1) edge[dashed] node[right=2em,above=-0.4em] {$\pi_k^{k^{\prime}} \times_M \rho_l^{l^{\prime}}$} (m-2-2);
	\begin{scope}[shift=($(m-2-2)!.5!(m-4-4)$)]
	\draw +(-.3,0) -- +(0,0)  -- +(0,.3);
	\fill +(-.15,.15) circle (.05);
	\end{scope}
	\end{tikzpicture}
\end{center}

It is straight-forward to verify that this object is indeed the pullback of the corresponding diagram for the infinite jet bundles.

The content of this example is to prove the following Proposition:

\begin{pp}\label{cqc} Given two smooth fiber bundles $E, \, F  \rightarrow M$, one has in the category of pro-finite dimensional smooth manifolds that $J^{\infty}(E \times_M F) \cong J^{\infty} E \times_M J^{\infty} F$.
\end{pp}

\dem
We consider two maps
\begin{eqnarray*}
f \colon J^{\infty} E \times_M J^{\infty} F & \longrightarrow & J^{\infty}(E \times_M F) \textrm{ and} \\
g \colon J^{\infty}(E \times_M F) & \longrightarrow & J^{\infty} E \times_M J^{\infty} F.
\end{eqnarray*}

The first one is given by $$f^k \defeq  s_k^{-1} \colon J^k E \times_M J^k F \rightarrow J^{k}(E \times_M F).$$ Observe that here we are implicitly using final functors to calculate the colimit in $\Delta(\mathbb{N})$ instead of in $\mathbb{N} \times \mathbb{N}$ (we have avoided to talk about final functors for simplicity). ($\Delta$ denotes the diagonal map.)

The second one is $$g^{k,l} \defeq  (\pi_{\max (k,l)}^k \times \rho_{\max (k,l)}^l) \circ s_{\max (k,l)} \colon J^{\max (k,l)} (E \times_M F) \rightarrow J^k E \times_M J^l F.$$
Now we can verify that these maps actually are compatible with the structures of $J^{\infty} E \times_M J^{\infty} F$ and $J^{\infty}(E \times_M F)$.

Let us look at the composition $f \circ g \colon J^{\infty}(E \times_M F) \rightarrow J^{\infty}(E \times_M F)$. This map is given by $(f \circ g)^k = f^k \circ g^{k,k} = s_k^{-1} \circ s_k \colon J^k (E \times_M F) \rightarrow J^k (E \times_M F)$, which is precisely the identity on $J^k (E \times_M F)$.

On the other direction, $g \circ f \colon J^{\infty}(E) \times_M J^{\infty}(F) \rightarrow J^{\infty}(E) \times_M J^{\infty}(F)$ is given by 
\begin{eqnarray*}
(g \circ f)^{k,l} &=& g^{k,l} \circ f^{\max (k,l)} = (\pi_{\max (k,l)}^k \times \rho_{\max (k,l)}^l) \circ s_{\max (k,l)} \circ s_{\max (k,l)}^{-1} \\
&=& \pi_{\max (k,l)}^k \times \rho_{\max (k,l)}^l
\end{eqnarray*}

 We need to see that this map represents the identity on $J^{\infty}(E) \times_M J^{\infty}(F)$. But it is enough to look at the following diagram:

\begin{center}
\begin{tikzpicture}[description/.style={fill=white,inner sep=2pt}]
\matrix (m) [matrix of math nodes, row sep=2.5em,
column sep=2.5em, text height=1.5ex, text depth=0.25ex]
{ J^{\max (k,l)} (E) \times_M J^{\max (k,l)}(F) &  \\
  &  J^k (E) \times_M J^l (F) \\
  J^k (E) \times_M J^l (F) &  \\ };
\path[->,font=\scriptsize]
(m-1-1) edge node[left=-0.25em] {$\pi_{\max (k,l)}^k \times \rho_{\max (k,l)}^l$} (m-3-1)
(m-1-1) edge node[right=3.25em,above=-0.5em] {$\pi_{\max (k,l)}^k \times \rho_{\max (k,l)}^l$} (m-2-2)
(m-3-1) edge node[auto] {id} (m-2-2);
\end{tikzpicture}
\end{center}

This shows that $J^{\infty}(E \times_M F)$ is isomorphic to $J^{\infty} E \times_M J^{\infty} F$. \qed

\end{ej}


\section{Fr\'echet manifold structure on \boldmath{\mbox{$J^{\infty} E$} } }\label{jif}

{\it Another common approach taken when working with the infinite jet bundle is to consider a Fr\'echet manifold structure on it (this is done by Saunders \cite{SAU}, for instance). In this section we show that pro-finite smooth maps are Fr\'echet-smooth and give a comparison between some Fr\'echet spaces and pro-finite normed spaces. In particular, we study the infinite jet bundle as a Fr\'echet manifold and state the smoothness of the structure maps. The principal references in this chapter are Saunders \cite{SAU} and Schaefer \cite{SCH}.}\\

In short, a Fr\'echet space is a vector space $V$ equipped with a Hausdorff topology coming from a countable family of seminorms $\{ | \cdot |_n \}_{n \in \mathbb{N}}$. For details we refer to Appendix \ref{app2}.

Since our first example of Fr\'echet structures is going to be a sequential (indexed by $\mathbb{N}$) projective limit of finite dimensional normed spaces ($J^{\infty} E$) we will focus on that case. The following results are inspired on the ones found in the book by Dodson, Galanis, and Vassiliou in  \cite{DGV}:

\begin{lm}\label{lmfs} Sequential pro-finite dimensional normed spaces are Fr\'echet.
\end{lm}

\dem Consider $V$ the projective limit of $\{f_n^m \colon V_n \rightarrow V_m \}_{n, m \in \mathbb{N}, \, n > m}$, denote by $f_{\infty}^n$ the associated map from $V$ to $V_n$ for each $n \in \mathbb{N}$. V is a vector space and the maps $\{f_{\infty}^n\}_{n \in \mathbb{N}}$ are linear. Given any $n \in \mathbb{N}$, the norm $| \cdot |_n$ on $V_n$ induces a seminorm on $V$ via $| \cdot |_n = |\cdot |_n \circ f_{\infty}^n$. Explicitly:
\begin{enumerate}
	\item $| v |_n = |f_{\infty}^n(v)|_n \geqslant 0$ for all $v \in V$.
	\item $| v+ w |_n = | f_{\infty}^n(v+ w) |_n = | f_{\infty}^n(v) + f_{\infty}^n(w) |_n \leqslant | f_{\infty}^n(v) |_n + | f_{\infty}^n(w) |_n = | v |_n + | w |_n$ for all $v$ and $w$ in $V$.
	\item $| a \cdot v |_n = | f_{\infty}^n(a \cdot v) |_n = | a \cdot f_{\infty}^n( v) |_n = |a| \cdot | f_{\infty}^n(v) |_n = |a| \cdot | v |_n$ for all $v \in V$ and all $a \in \RE$.
\end{enumerate}
The corresponding locally convex topological vector space with respect to that family of seminorms is metrizable (see Proposition \ref{haumet} or Schaefer \cite{SCH}) and it is also Hausdorff since
$$v = 0 \Longleftrightarrow f_{\infty}^n(v) = 0 \, \forall n \in \mathbb{N} \Longleftrightarrow |f_{\infty}^n(v)|_n = 0 \, \forall n \in \mathbb{N} \Longleftrightarrow | v |_n = 0 \, \forall n \in \mathbb{N}.$$
The vector space $V$ is hence a locally convex topological space which is Hausdorff and metrizable. Finite dimensional vector spaces are complete and projective limits of topological vector spaces which are complete are complete (this is done for instance by Schaefer \cite[II 5.3]{SCH}). This proves the claim. 
\qed

\begin{ej}\label{rinf} $\RE^{\infty}$ as the sequential limit of all the normed vector spaces $\RE^n$ is a Fr\'echet space.
\end{ej}

A more general result is also proven in more detail by Schaefer \cite[II 5.4]{SCH}. Inspecting that proof we can get some insight into the converse of this statement: which Fr\'echet spaces are sequential pro-finite dimensional normed spaces.

\begin{cl}\label{chus} A Fr\'echet space $(V, \{ | \cdot |_n \}_{n \in \mathbb{N}})$ such that $\, \bigslant{V}{\textrm{ker} \, | \cdot |_n}$ is finite dimensional for any $n \in \mathbb{N}$ is isomorphic to a sequential pro-finite dimensional normed space. 
\end{cl}

\dem Let $(V, \{ | \cdot |_n \}_{n \in \mathbb{N}})$ be such a Fr\'echet space and consider for every $n \in \mathbb{N}$ a complement $V_n$ of $ \textrm{ker} \, | \cdot |_n $ in $V$. They are finite dimensional by assumption. The seminorm $|\cdot |_n$ on $V$ gives rise to a norm on $V^n$ for every $n$.

The following step is to consider a reordering of the subspaces so that the norms are compatible. In order to do that consider $W^1 \defeq  V^1$ and $W^{n+1}$ the vector space generated by $W^n \cup V^{n+1}$ for higher $n$. Define a norm on $W^{n+1}$ inductively to be $| \cdot |_{n}^{\prime}$ in $W^n \cap V^{n+1}$ and $| \cdot |_{n+1}$ in the complementary subspace. The change in the norm happens in a finite dimensional vector space, and hence the topology on $V$ with respect to the $W^n$'s agrees with the one induced by the $V^n$'s. Under this conditions, we can form $W$ the Fr\'echet space colimit of the sequence of the $W^n$ as in Lemma \ref{lmfs}.

The claim is that $W$ and $V$ are isomorphic. The map $V \rightarrow W$ sending $v$ to each of its projections is linear and injective since $V$ was Hausdorff. The proof that it is also surjective follows from the aforementioned result from Schaefer \cite[II 5.4]{SCH} that we have included as Theorem \ref{probanach} in the appendices.
\qed

Fr\'echet spaces have the advantage that one can talk about smooth maps between them (see Appendix \ref{app2}). For our purposes it is enough to understand how smooth maps to and from $\RE^{\infty}$ work. We have the following helpful results by Saunders \cite{SAU}.

\begin{lm}[Saunders {\cite[Lemma 7.1.8]{SAU}}]\label{saun1}
	Let $U \subset V$ be an open subset of a Fr\'echet space. $f \colon U \rightarrow \RE^{\infty}$ is smooth if and only if $\textrm{pr}_{\RE^n} \circ f \colon U \rightarrow \RE^n$ is smooth for all $n \in \mathbb{N}$. 
\end{lm}

\begin{lm}[Saunders, \cite{SAU}]\label{saun2}
	A map $f \colon \RE^{\infty} \rightarrow \RE$ is smooth if at every point only a finite number of its partial derivatives do not vanish. 
\end{lm}

At this point we can include the definition of a smooth manifold with charts in Fr\'echet spaces. (See for instance Schaefer \cite{SCH})

\begin{df}[Fr\'echet manifold] A Hausdorff topological space $\M$ is a Fr\'echet manifold if it is provided with an atlas of homeomorphisms to open sets of a Fr\'echet space $V$ such that the transition functions are smooth in the sense of Definition \ref{FrSm}.
\end{df}

We fix a smooth fiber bundle $\pi \colon E \rightarrow M$. The associated infinite jet bundle $J^{\infty} E$ is from the topological point of view a fiber bundle over $M$ whose fiber is a projective limit of vector spaces. Choosing Riemannian structures on the finite jets, we can see that the fiber is a sequential projective limit of finite dimensional normed vector spaces, and hence a Fr\'echet space. It is then natural to think that $J^{\infty} E$ could be given the structure of a Fr\'echet manifold, perhaps with values in $\RE^{\infty}$ as in Example \ref{rinf}. We follow Saunders to introduce the Fr\'echet manifold structure on the infinite jet bundle:

\begin{df}\label{above}
Let $E \rightarrow M$ be a smooth fiber bundle, let $\{U_{a}\}_{a \in A}$ be a cover of $E$ by trivial coordinate charts and consider the induced cover on $J^{\infty} E$ given by $\{\left( \pi_{\infty}^0 \right)^{-1} \left( U_{a} \right)\}_{a \in A}$. If we choose coordinate systems on each $J^{k} E$ which are compatible with the projections $\pi_l^k$ (as it will be done in Definition \ref{coordinates}), then the induced maps $u_{a}^{\infty} \colon \left( \pi_{\infty}^0 \right)^{-1} \left( U_{a} \right) \rightarrow \RE^{\infty}$ are coordinates in $J^{\infty} E$.
\end{df}

That set of trivializations define a smooth structure on $J^{\infty} E$.

\begin{pp}[Saunders {\cite[Proposition 7.2.4]{SAU}}]\label{saunp} The infinite jet bundle with the maps introduced in Definition \ref{above}, $\left( J^{\infty} E, \{u_{a}^{\infty} \colon \left( \pi_{\infty}^0 \right)^{-1} \left( U_{a} \right) \rightarrow \RE^{\infty}\}_{a \in A} \right)$, is a Fr\'echet manifold.
\end{pp}

There is a statement about the topologies in Proposition \ref{saunp} which we would like to make explicit: the charts $u_{a}^{\infty} \colon \left( \pi_{\infty}^0 \right)^{-1} \left( U_{a} \right) \rightarrow \RE^{\infty}$ are homeomorphisms, and hence {\it the Fr\'echet manifold topology on $J^{\infty} E$ coincides with the limit topology.} This can be seen in the following diagram which commutes for every finite $k$:
\begin{center}
\begin{tikzpicture}[description/.style={fill=white,inner sep=2pt}]
\matrix (m) [matrix of math nodes, row sep=3em,
column sep=3.5em, text height=1.5ex, text depth=0.25ex]
{\left( \pi_{\infty}^0 \right)^{-1} \left( U_{a} \right) & \RE^{\infty} \\
  \left( \pi_{k}^0 \right)^{-1} \left( U_{a} \right) & \RE^{\textrm{dim}(J^k E)} \\};
\path[->,font=\scriptsize]
(m-1-1) edge node[auto] {$u_{a}^{\infty}$} (m-1-2)
(m-1-1) edge node[auto] {$\pi_{\infty}^{k}$} (m-2-1)
(m-1-2) edge node[auto] {$u_{a}^{k}$} (m-2-2)
(m-2-1) edge node[auto] [swap]{$\textrm{pr}_{\RE^{\textrm{dim}(J^k E)}}$} (m-2-2);
\end{tikzpicture}
\end{center}
The bottom map is a coordinate chart for $J^k E$. Observe that the topology on $J^{\infty} E$ is the sequential limit of the $J^k E$ on the left; on the other side, the topology on $\RE^{\infty}$ is the sequential limit of the $\RE^{\textrm{dim}(J^k E)}$ as a consequence of Lemma \ref{lmfs}.

Using Lemmas \ref{saun1} and \ref{saun2} we get to the following corollary, which is a well know fact in the study of infinite jet bundles.

\begin{cl}\label{prosmosmo} Let $J^{\infty} E$ and $J^{\infty} F$ be the infinite jet bundles of some smooth fiber bundles. Any pro-finite smooth morphism $f \colon J^{\infty} E \rightarrow J^{\infty} F$ is Fr\'echet smooth.
\end{cl}

Observe that $\RE$ can be regarded as an infinite jet bundle. In that case, we know that pro-smooth maps from $J^{\infty} E$ to $\RE$ descend to a map from a finite jet bundle. On the other hand, smooth maps from $J^{\infty} E$ to $\RE$ descend {\it locally} to maps from finite jet bundles (this follows from Lemma \ref{saun2}). In this way we see that the converse of the previous corollary does not hold in general. We will have to wait until we get a partial converse result to this one (see Corollary \ref{chocho}). Smooth functions on $J^{\infty} E$ are precisely the maps considered by G\"uneysu and Pflaum \cite{GP}, that we have already discussed in Remark \ref{gune}. Khavkine and Schreiber work with a smaller category than that of Fr\'echet spaces where these maps live: the category of locally pro-manifolds. That terminology is very clarifying. (See for example their work \cite{KS}.)

\dem A map $f \colon J^{\infty} E \rightarrow J^{\infty} F$ is smooth if and only if it is smooth when trivialized, for any trivializing chart. We pick any of them:
\begin{center}
\begin{tikzpicture}[description/.style={fill=white,inner sep=2pt}]
\matrix (m) [matrix of math nodes, row sep=3em,
column sep=2.5em, text height=1.5ex, text depth=0.25ex]
{ \RE^{\infty} & J^{\infty} E & J^{\infty} F & \RE^{\infty} \\ };
\path[->,font=\scriptsize]
(m-1-1) edge node[auto] {} (m-1-2)
(m-1-2) edge node[auto] {} (m-1-3)
(m-1-3) edge node[auto] {} (m-1-4);
\end{tikzpicture}
\end{center}
By Lemma \ref{saun1} such a map is smooth if and only if it is smooth after any projection to $\RE^n$. The way $J^{\infty} F$ is constructed, this is equivalent to the following composition to be smooth for all $l \in \mathbb{N}$:
\begin{center}
\begin{tikzpicture}[description/.style={fill=white,inner sep=2pt}]
\matrix (m) [matrix of math nodes, row sep=3em,
column sep=2.5em, text height=1.5ex, text depth=0.25ex]
{ \RE^{\infty} & J^{\infty} E & J^{\infty} F & \RE^{\infty} \\
& & J^l F & \RE^{\textrm{dim}(J^{l} F)} \\};
\path[->,font=\scriptsize]
(m-1-1) edge node[auto] {} (m-1-2)
(m-1-2) edge node[auto] {} (m-1-3)
(m-1-3) edge node[auto] {} (m-1-4)
(m-1-3) edge node[auto] {} (m-2-3)
(m-1-4) edge node[auto] {} (m-2-4)
(m-2-3) edge node[auto] {} (m-2-4);
\end{tikzpicture}
\end{center}
Since by assumption $f$ is a pro-finite smooth map we can complete the diagram on the lower row:
\begin{center}
\begin{tikzpicture}[description/.style={fill=white,inner sep=2pt}]
\matrix (m) [matrix of math nodes, row sep=3em,
column sep=2.5em, text height=1.5ex, text depth=0.25ex]
{ \RE^{\infty} & J^{\infty} E & J^{\infty} F & \RE^{\infty} \\
\RE^{\textrm{dim}(J^{k(l)} E)} & J^{k(l)} E & J^l F & \RE^{\textrm{dim}(J^{l} F)} \\};
\path[->,font=\scriptsize]
(m-1-1) edge node[auto] {} (m-1-2)
(m-1-2) edge node[auto] {} (m-1-3)
(m-1-3) edge node[auto] {} (m-1-4)
(m-1-3) edge node[auto] {} (m-2-3)
(m-1-4) edge node[auto] {} (m-2-4)
(m-2-3) edge node[auto] {} (m-2-4)
(m-1-1) edge node[auto] {} (m-2-1)
(m-1-2) edge node[auto] {} (m-2-2)
(m-2-1) edge node[auto] {} (m-2-2)
(m-2-2) edge node[auto] {} (m-2-3);
\end{tikzpicture}
\end{center}
Now, the map $\RE^{\infty} \rightarrow \RE^{\textrm{dim}(J^{l} F)}$ is smooth if and only if it is smooth on each component of the map, and those by Lemma \ref{saun2} are smooth if at every point only finitely many partial derivatives do not vanish at every point. In this case we have an even stronger result. All the components of the map at all points depend only on finitely many entries (the $\textrm{dim}(J^{k(l)} E)$ first ones), so that the original map $f$ is smooth.
\qed

Using the descriptions of smooth maps involving $\RE^{\infty}$ (Lemmas \ref{saun1} and \ref{saun2}) it is possible to prove that all the structure maps involving the infinite jet bundle are smooth. The following results can be found in the book by Saunders \cite{SAU} (results $7.2.5$ to $7.2.9$):

\begin{pp}[Saunders \cite{SAU}]\label{proprosmo} Let $\pi \colon E \rightarrow M$ be a smooth fiber bundle with associated jet bundles $J^k E$ for all $k \in \mathbb{N} \cup \{\infty\}$ (the manifold structure on $J^{\infty} E$ is the one in Proposition \ref{saunp}). Then the following statements hold:
\begin{itemize}
	\item $\pi_{\infty}^k \colon J^{\infty} E \rightarrow J^k E$ is a smooth fiber bundle for each $k \in \mathbb{N}$.
	\item $\pi_{\infty} \colon J^{\infty} E \rightarrow M$ is a smooth fiber bundle.
	\item $j^{\infty} \varphi \colon M \rightarrow J^{\infty} E$ is smooth for every $\varphi$ smooth section $\varphi$ of $\pi$.
\end{itemize}
\end{pp}


\newpage
\chapter{The Cartan Distribution}

The pro-finite smooth structure on infinite jet bundles gives plenty of flexibility to construct the usual tools in differential geometry. We have already seen in the last chapter how to define the ind-algebra of pro-smooth functions and now it is the turn of ind-differential forms and vector fields.\\ 

We will be able to talk about pullbacks, insertion of vector fields in ind-differential forms, Lie derivatives, bracket of vector fields, and so on. The fundamental relations between these terms, usually known as Cartan calculus, also holds in this setting.\\

The ind-/pro-categorical approach has the advantage that calculations are easier than in the Fr\'echet manifold setting. This is so because we have conditions to insure that all structures descend to the corresponding ones in a finite jet degree. But another advantage is that often obscure and complicated results known for the infinite jet bundle can be stated in a more concrete manner using the ind-/pro-categorical approach. An example of this is is Proposition \ref{vfeq} in Section \ref{vf} which states that ind-derivations of $\C(J^{\infty}E)$ are in one to one correspondence with pro-smooth sections of the pro-finite tangent bundle $T (J^{\infty}E) \rightarrow J^{\infty}E$.\\

The infinite jet bundle has been studied independently of the ind-/pro-categorical approach, and almost everything that can be said about it has already been published. In this chapter we have wanted to compile the relevant results and notations for the rest of the document. But, most importantly, we have presented them as an ind-/pro-categorical way of defining vector fields and differential forms on the infinite jet bundle.\\

The Cartan distribution is an intuitive geometric feature on infinite jet bundles. The image of the infinite jet prolongations of local sections of a bundle ($j^{\infty} \varphi (U)$) define a local distribution on the infinite jet bundle known as the Cartan distribution. The first useful consequence of the study of the Cartan distribution is that it allows us to split the ind-complex of differential forms into a vertical and a horizontal part giving rise to the variational bicomplex. This bicomplex (studied by Anderson \cite{AND} in great detail) is the key underlying structure underneath the complex of local forms, central in Lagrangian field theory.\\

The associated co-distribution (the annihilator of the Cartan distribution) is usually called the contact ideal. Maps and vector fields preserving the contact ideal have special features: they are jet prolongations of their lowest representatives $J^k E \rightarrow F$ (where $F = TE$ in the case of vector fields). We first review the known results concerning jet prolongations along a diffeomorphism: their existence (here as Proposition \ref{pro}, originally by Saunders \cite{SAU}) and their factorization properties (here Proposition \ref{chet}, by Chetverikov \cite{CHE}). Vector fields preserving the contact ideal split into a vertical and a horizontal component called evolutionary and total (here Proposition \ref{last}, by Anderson \cite{AND}).\\

The original work in this chapter is concentrated in the study of a more general class of maps that admit pro-finite smooth prolongations. Maps in that class give rise to unique jet prolongations that preserve the contact ideal. These maps are not only pro-smooth but also smooth. These will be an important tool when talking about insular maps and Lagrangian field theories.\\

{\it 
The main references in this chapter are Anderson \cite{AND}, Chetverikov \cite{CHE}, and Saunders \cite{SAU}.}


\section{Differential forms on \boldmath{\mbox{$J^{\infty} E$}} }\label{dforms}
{\it In this section we continue introducing ind-/pro-structures related to the infinite jet bundle. We introduce the pro-finite tangent bundle of the infinite jet bundle and the ind-complex of differential forms on it. This complex splits into a horizontal and a vertical part, giving rise to a bicomplex called the variational bicomplex. We give explicit coordinate expressions for the horizontal and vertical differentials. We also introduce the Cartan distribution, a very intuitive geometric feature of the infinite jet bundles which is important in future results. In particular, since pullbacks of ind-differential forms are well defined, we use the Cartan distribution to prove a result related to pullback of sections in the infinite jet bundle. This section is a recollection of definitions that provide a basis for the rest of the thesis. The main references are Anderson \cite{AND}; Dodson, Galanis, and Vassiliou \cite{DGV}; and Chetverikov \cite{CHE}.}\\

The tangent space at a point of the infinite jet bundle can be defined in different equivalent ways. We will begin with the pro-smooth definition. 

\begin{df}\label{tje}[Tangent bundle of the infinite jet bundle]
Let $E \rightarrow M$ be a smooth fiber bundle. Let $\chi \in J^{\infty} E$ be a point in the associated infinite jet bundle. $T_{\chi} (J^{\infty} E)$ is defined as the pro-object in the category of vector spaces given by $\left\{ \left( T_{\pi_{\infty}^{k}(\chi)} J^k E, T \pi_{k}^l \right) \right\}$.
The tangent bundle $T (J^{\infty} E) = \bigcup_{\chi \in J^{\infty} E} T_{\chi} (J^{\infty} E)$ is a pro-finite smooth manifold modeled on $\left\{ \left( T J^k E, T \pi_{k}^l \right) \right\}$. The base-point projection map $$\textrm{pr}_{J^{\infty} E} \colon T (J^{\infty} E) \rightarrow J^{\infty} E$$ is pro-smooth and can be represented by $\left\{ \textrm{pr}_{J^{\infty} E}^l = \textrm{pr}_{J^{l} E} \right\}_{l \in \mathbb{N}}$.
\end{df} 

The infinite jet bundle is a Fr\'echet manifold and as such it has an associated Fr\'echet tangent bundle. $T (J^{\infty} E) \defeq  \bigslant{\C(\RE, J^{\infty} E)}{\sim}$ where $c \sim \tilde{c}$ if and only if $c(0) = \tilde{c}(0)$ and $D(\varphi \circ c)(0,1) = D(\varphi \circ \tilde{c})(0,1)$ for all $\varphi$ chart around $c(0)$, where $D$ is  the G\^{a}teaux derivative as in Definition \ref{gtd}. This is the approach followed by Dodson, Galanis, and Vassiliou \cite{DGV}. The tangent space at a point $\chi \in J^{\infty} E$ is then the subset given by the curves passing through $\chi$ at time zero. It can be given the structure of a Fr\'echet space with finite dimensional cokernels of the seminorms (and hence a sequential pro-finite dimensional normed spaces by Corollary \ref{chus}). It is not difficult to convince oneself that this space is actually the same as $T_{\chi} J^{\infty} E$ as in Definition \ref{tje}, a proof of such fact can be found in the book by Dodson, Galanis, and Vassiliou, \cite[Proposition 3.2.2]{DGV}.

Moreover, we can also talk about derivations at $\chi \in J^{\infty} E$ of the algebra of pro-smooth functions on $J^{\infty} E$. Anderson shows \cite{AND} that that approach is also equivalent to Definition \ref{tje}.  Global derivations are not the correct way of approaching vector fields on the infinite jet bundle. We should consider ind-derivations instead. We will define vector fields only a bit later (see in Chapter \ref{vf}), since the theory is a bit more involved than the dual of ind-differential forms on $J^{\infty} E$. 

\begin{df}[The ind-complex of differential forms, $\Omega^{\bullet}(J^{\infty} E, d) $]\label{ldf}
Given a fiber bundle $\pi \colon E \rightarrow M$, the space of differential forms on $J^{\infty} E$ is defined as the ind-object in the category of differential complexes given by the diagram
$$\Omega^{\bullet}(E) = \Omega^{\bullet}(J^0 E) \rightarrow \Omega^{\bullet}(J^1 E) \rightarrow \Omega^{\bullet}(J^2 E) \rightarrow \cdots.$$
It is denoted by $\Omega^{\bullet}(J^{\infty} E)$.
\end{df}

Observe that since $\Omega^{\bullet}(J^{\infty} E)$ is an ind-object, ind-morphisms from $\RE$ to $\Omega^{\bullet}(J^{\infty} E)$ can be thought as elements of $\Omega^{\bullet}(J^{\infty} E)$. Those are given precisely by some $\omega^k \in \Omega(J^k E)$.

\begin{rk}
	Once again, we want to refer to the work of G\"uneysu and Pflaum \cite{GP} as in Remark \ref{gune}. They consider locally bounded jet degrees, both for referring to the smooth locally ringed structure on $T (J^{\infty} E)$ and for differential forms on it. They refer to the ind-differential forms as local forms.
\end{rk}

\begin{rk}  Ind-differential complexes are not only ind-vector spaces but the differential also plays a role. In this case, it allows us to define exterior differentiation of forms in all jet degrees. At every finite jet-bundle we have $d_l = d \colon \Omega^{\bullet}(J^l E) \rightarrow \Omega^{\bullet+1}(J^l E)$. This can be interpreted as a morphism of ind-graded vector spaces $d \colon \Omega^{\bullet}(J^{\infty}  E ) \rightarrow \Omega^{\bullet}(J^{\infty}  E )$ given by $\{d_l\}_{l \in \mathbb{N}}$. This map squares to zero in the sense that given any $\omega$ in $\Omega^{\bullet}(J^{\infty} E )$ (that is, a map $\mathbb{R} \rightarrow \Omega^{\bullet}(J^{\infty} E )$ or equivalently $\omega_k \in \Omega^{\bullet}(J^k E)$) the ind-form $d \circ d (\omega)$ is the zero form.  
\end{rk}

The tensor product of two differential complexes $(Q_1, d_1)$ and $(Q_2, d_2)$ is again a differential complex with differential $d(\alpha_1 \otimes \alpha_2) \defeq d_1(\alpha_1) \otimes \alpha_2 +(-1)^{|\alpha_1|} \alpha_1 \otimes d_2(\alpha_2)$. We can hence talk about the ind-differential complex $\Omega^{\bullet}(J^{\infty}  E) \otimes \Omega^{\bullet}(J^{\infty}  E)$ which will be indexed by $\mathbb{N} \times \mathbb{N}$.

\begin{pp} The ind-differential forms $\Omega^{\bullet}(J^{\infty}  E)$ are equipped with an ind-morphism $\wedge \colon \Omega^{\bullet}(J^{\infty}  E) \otimes \Omega^{\bullet}(J^{\infty}  E) \rightarrow \Omega^{\bullet}(J^{\infty}  E)$ called the wedge product, given by the maps $\{ \wedge_{k,l} \colon \Omega^{\bullet}(J^k  E) \otimes \Omega^{\bullet}(J^l  E) \rightarrow \Omega^{\bullet}(J^{\textrm{max}(k,l)}  E) \}_{(k,l)}$. Those are defined using pullbacks: $\wedge_{k,l} = \wedge \circ ((\pi_{k}^{\textrm{max}(k,l)})^* \otimes (\pi_{l}^{\textrm{max}(k,l)})^*)$. 
\end{pp}

Observe that this in particular means that the differential $d$ is a derivation of the product. In other words, when applied to two ind-differential forms $\alpha_1$ and $\alpha_2$ we have that $d(\alpha_1 \wedge \alpha_2) = d_1(\alpha_1) \wedge \alpha_2 + (-1)^{|\alpha_1|} \alpha_1 \wedge d_2(\alpha_2)$. Again we are to understand this as applied to $\mathbb{R}$-points of $\Omega^{\bullet}(J^{\infty}  E)$.

Recall that elements of $\Omega^{\bullet}(J^{\infty} E)$ are given by some $\omega^k \in \Omega^{\bullet}(J^k E)$. We distinguish between order and degree of a differential form $\omega \in \Omega(J^{\infty} E)$. The smallest $k$ such that $\omega$ can be represented by $\omega^k \in \Omega^{\bullet}(J^k E)$ is called the {\it order} of $\omega$. The form is of {\it degree} $p$ if it is an element of the ind-object
$$\Omega^p (J^{\infty} E) = \Omega^p(E) \rightarrow \Omega^p(J^1 E) \rightarrow \cdots.$$
The exterior derivative has order zero and degree $+1$.

Once again we want to point out that this definition is compatible with other equivalent ones: forms in Fr\'echet manifolds or sections of the limit of $(\wedge^p (J^k E))^*$. This can be found in the books by Anderson \cite{AND} and Dodson, Galanis, and Vassiliou \cite{DGV}. Since we focus on the pro/ind-categorical approach, we will stop talking about these comparisons in general, unless something remarkable happens. Using those equivalences, $\omega_{\chi}$ is a multilinear totally antisymmetric function on $T_{\chi} J^{\infty} E$ where $\omega \in \Omega^{\bullet}(J^{\infty} E)$ and $\chi \in J^{\infty} E$.

We are going to use {\it local coordinates} in the infinite jet bundle to be able to talk about the splitting of the differential into a horizontal and a vertical direction. What follows can be found in the book by Anderson, \cite{AND}.

Given a smooth fiber bundle $\pi \colon E \rightarrow M$, the associated finite jet bundles are smooth bundles over $M$. We fix a chart around a point in $E$: $(x^1, \ldots , x^m, u^1, \ldots , u^n)$ where $(x^1, \ldots , x^m)$ are coordinates on the base $M$ and $(u^1, \ldots , u^n)$ are coordinates on the fiber. We interpret those as maps in $\C(E)$.

\begin{df}[Anderson \cite{AND}]\label{coordinates}
On $J^k E$ a system of coordinates is given by $(x_i, u_I^{\alpha})$ where $i$ runs from $1$ to $m$, $\alpha$ runs from $1$ to $n$ and $I$ runs over all multi-indices of length at most $k$ with values between $1$ and $m$:
\begin{equation*}
\begin{split}
	x^i (j^{k}(\varphi, x)) & \defeq  x^i(x) \textrm{ and } \\
  u_I^{\alpha} (j^{k}(\varphi, x)) & \defeq  {\left. \frac{\partial^{|I|}(u^{\alpha} \circ \varphi)}{\partial x^I} \right|}_x . \\
\end{split}
\end{equation*}
\end{df}

All those functions are smooth in a certain neighborhood of $(\varphi, x) \in \E \times M$ and they are also well defined in $J^k E$, by definition of the finite jet space. Even more, they locally determine $j^k(\varphi, x)$. This is because $j^{k}(\varphi, x) = j^{k}(\widetilde{\varphi}, \widetilde{x})$ if and only if $x = \widetilde{x}$ (which happens only if $x^i(x)=x^i(\widetilde{x})$ for all $i$) and $\partial^{|I|} \varphi (x) = \partial^{|I|} \widetilde{\varphi} (x)$ for all multi-indices of length at most $k$ (which happens if an only if ${\left. \frac{\partial^{|I|}(u^{\alpha} \circ \varphi)}{\partial x^I} \right|}_x = {\left. \frac{\partial^{|I|}(u^{\alpha} \circ \widetilde{\varphi})}{\partial x^I} \right|}_x$ for all $\alpha$ and all multi-indices of length at most $k$).

Observe that the maps $\{x_i\}$ and $\{ u_I^{\alpha} \}$ are compatible with the projections $\{\pi_k^l\}$. Fix $k \geqslant l$, $I$ of length at most $l$, $i$, and $\alpha$:
\begin{equation*}
\begin{split}
	x^i \circ \pi_k^l (j^{k}(\varphi, x)) & = x^i(j^{k}(\varphi, x)) = x^i(x) = x^i (j^{k}(\varphi, x)) \textrm{ and}\\
  u_I^{\alpha} \circ \pi_k^l (j^{k}(\varphi, x)) & = u_I^{\alpha} (j^{l}(\varphi, x)) = {\left. \frac{\partial^{|I|}(u^{\alpha} \circ \varphi)}{\partial x^I} \right|}_x = u_I^{\alpha} (j^{k}(\varphi, x)).
\end{split}
\end{equation*}

%
%
%
In these coordinates the horizontal subspace is spanned by the coordinates $x^i$. We could then define the horizontal differential by differentiating with respect to these coordinates. But observe that taking the derivative of a function with respect to $x^i$ should heuristically be done by taking into consideration that each of the functions $u_I^{\alpha}$ also depends on $x^i$.

\begin{df}[$d_H$ and $d_V$] Given an ind-smooth function $f \in \C(J^{\infty} E)$, the horizontal differential of $f$ is defined using the one forms $\{ d x^i \}$:

$$d_H f \defeq  \frac{\partial f}{\partial x^i} dx^i + \frac{l_1! \cdots l_m! }{k!} \frac{\partial f}{\partial u_I^{\alpha}} u_{I, i}^{\alpha} dx^i,$$
where $k = |I|$ and $l_j$ is the number of occurrences of $j$ in $I$. The horizontal differential extends to all forms as a derivation of the wedge product. The vertical differential is defined as the difference between the other two: $d_V \defeq  d - d_H$.  
\end{df}

We would like to introduce some new notation. Anderson uses \cite{AND} the operators 
$$\partial_{\alpha}^I \defeq \frac{l_1! \cdots l_m! }{k!} \frac{\partial }{\partial u_I^{\alpha}},$$
so that $d_H f = \frac{\partial f}{\partial x^i} dx^i + u_{I, i}^{\alpha} \partial_{\alpha}^I f dx^i$.  Denoting
\begin{equation}\label{di}
D_i \defeq  \frac{\partial}{\partial x^i} + u_{I, i}^{\alpha} \partial_{\alpha}^I.
\end{equation}

With this notation we have \footnote{Iterated partial differentials are elements of the symmetric algebra generated by the partial differentials. There are two ways of considering the symmetric algebra as invariants or coinvariants of the symmetric action on the tensor algebra generated by them. The appearance of the factorial coefficients is due to mixing these two choices. We have decided to not fix the convention used by Anderson so that the two equations in \ref{ventiuno} have the same structure. They both use $D_i$ and $\partial_{\alpha}^I$ which avoid fractions.}
\begin{eqnarray}\label{ventiuno}
d_H f &=& D_i f \, dx^i \nonumber \\
d_V f & =& \partial_{\alpha}^I f \, d_V u_I^{\alpha}.
\end{eqnarray}

Moreover $d_H \circ \pi_{\infty}^* = \pi_{\infty}^* \circ d_M$ where $d_M$ is the de Rham differential on $M$. The local forms $d_V u_I^{\alpha} = d u_I^{\alpha} - u_{I,i}^{\alpha} dx^i$ generate a differential ideal $\mathtt{C}$. Using the definitions we can see that in particular $d_H x^i = d x^i$ and $d_H(u_I^{\alpha}) = u_{I, i}^{\alpha} dx^i$. This suggest that the horizontal and vertical differentials respect the contact ideal. As a matter of fact, any ind-differential form splits into vertical and horizontal parts. First we need to define vertical vectors:

\begin{df}[Vertical vector] Considering $\pi_{\infty} \colon J^{\infty} E \rightarrow M$ as a fiber bundle (even a pro-smooth fiber bundle) we define $V (J^{\infty} E) \defeq \ker(T \pi_{\infty})$. To be precise, at $\chi \in J^{\infty} E$, 
$$V_{\chi}(J^{\infty} E) \defeq  \{ X_{\chi} \in T_{\chi} J^{\infty} E \colon T \pi_{k} (T \pi_{\infty}^k X_{\chi}) = 0 \in T_{\pi_{\infty}(\chi)} M \textrm{ for all } k \}.$$
\end{df}

Horizontal forms will be the ones annihilated by horizontal vectors and vertical forms the ones spanned by $\mathtt{C}$. The definitions are the following:

\begin{df}[Horizontal and vertical forms] Given natural numbers $p$, $r$, and $s$ we define $(p,s)$-horizontal forms to be:
\begin{eqnarray*}
\Omega_H^{p,s} (J^{\infty} E) &\defeq & \\
\{ \omega \in \Omega^p (J^{\infty} E) & \colon & \omega_{\chi}(X_1, \cdots, X_{p-s+1}, -) = 0 \, \forall X_i \in T_{\chi} J^{\infty} E \textrm{ vertical} \forall \chi \textrm{ in } J^{\infty} E\}.
\end{eqnarray*}
On the other hand, $(p,r)$-vertical forms are defined as:
$$\Omega_V^{p,r} (J^{\infty} E) \defeq  \{ \omega \in \Omega^p (J^{\infty} E) \colon \omega = \alpha_1 \wedge \cdots \wedge \alpha_r \wedge \widetilde{\omega} \textrm{ where } \alpha_1 , \cdots , \alpha_r \in \mathtt{C} \}.$$ 
\end{df}

$(p,1)$-horizontal forms are simply called horizontal. And $(p,p)$-vertical forms, vertical. With this notation we can clearly see that $d_H \Omega_H^{p,s} (J^{\infty} E) \subset \Omega_H^{p+1,s} (J^{\infty} E)$ and $d_H \Omega_V^{p,r} (J^{\infty} E) \subset \Omega_V^{p+1,r} (J^{\infty} E)$. It is also immediate to see that the horizontal differential of horizontal forms is again horizontal. But there is much more to it:

\begin{df}[Variational bicomplex \cite{AND}]\label{JVB} 
Let $E \rightarrow M$ be a smooth fiber bundle and let $J^{\infty} E$ denote its infinite jet bundle. The variational bicomplex associated to $E \rightarrow M$ is $(\Omega^{r,s}(J^{\infty} E), d_H, d_V)$ where $$\Omega^{r,s}(J^{\infty} E) \defeq  \Omega_V^{s+r,r} (J^{\infty} E) \cap \Omega_H^{s+r,s} (J^{\infty} E).$$
\end{df}

The total space of the variational bicomplex is that of $\Omega(J^{\infty} E)$. A form $\omega$ in $\Omega^{\bullet}(J^{\infty} E)$ is of bidegree $(r,s)$ if and only if it is locally a finite sum of forms of the type:
$$ \omega_{{\alpha}_1, \ldots, {\alpha}_r; i_1, \ldots, i_{s}}^{I_1, \ldots, I_r} d_V u_{I_1}^{{\alpha}_1} \wedge \cdots \wedge d_V u_{I_r}^{{\alpha}_r} \wedge d x^{i_1} \wedge \cdots \wedge d x^{i_{s}}, $$
{\noindent where each $\omega_{{\alpha}_1, \ldots, {\alpha}_r; i_1, \ldots, i_{s}}^{I_1, \ldots, I_r}$ is an element of $\C(J^{\infty} E)$.}\\

The following observations can be made about the variational bicomplex:
\begin{itemize}
	\item The exterior derivative splits between the two differentials and they anticommute: $d = d_V + d_H$, $d_H^2 = 0$, $d_V^2 = 0$ and $d_H \circ d_V = -d_V \circ d_H$.
	
	\item Using all the previous equations and the fact that $d_H(u_I^{\alpha}) = u_{I, i}^{\alpha} dx^i$ we can write:
$d_H(d_H x^i) = 0$, $d_V(d_H x^i) = 0$, $d_V(d_V u_I^{\alpha}) = 0$ but $d_V (d_H u_I^{\alpha}) = d_V u_{I, i}^{\alpha} dx^i $. 

	\item For a diagram describing the bicomplex we refer to Part IV where we study the pullback of this complex via $j^{\infty}$ on $\E \times M$.

	\item The cohomology of this complex has been extensively studied. In terms of bounds of the jet degree, the locally finite bounded case was studied by Takens \cite{TAK}, the globally finite order by Bauderon \cite{BAU} and Anderson \cite{AND}. For a non-bounded version (something more general than the variational bicomplex) we refer to the paper by Giachetta, Mangiarotti and Sardanashvily \cite{GMS}. 
	
	\item In Part \ref{lft} we will explore some of the results of the cohomology of the variational bicomplex. We will focus in the study of the interior Euler operator which is one of the main tools in the study of that cohomology. 
\end{itemize} 

As a matter of fact, for most of what is done in this thesis, the relevant complex is not that of ind-differential forms on $J^{\infty} E$, but that of $M$-twisted forms. Loosely speaking, we want vertical forms valued in densities on $M$ rather than valued in differential forms on $M$. We refer to Appendix \ref{tf} for the details. 


\subsection{The contact ideal}

There is an extra ingredient involved into the definition of the variational bicomplex that we have avoided to talk about so far for simplicity of the argument. Nevertheless, it is an essential piece into understanding what is special about the infinite jet bundle in comparison to other pro-finite smooth manifolds. This elephant in the room is no other than the Cartan distribution. 

Jet bundles (finite or infinite) come naturally equipped with a local foliation and a distribution. We know that every point $\chi \in J^k E$ can be represented by $j_{x_0} \varphi$ for some local section $\varphi \in \E(U)$ of $\pi \colon E \rightarrow M$ containing ${x_0} \in U$, $U$ open. Using the map $j^k (U)$ defined in Chapter 1, we get $j^k \varphi \colon U \rightarrow J^k E$. We can then consider $j^k \varphi (U) \defeq  \{ j_x^k \varphi \colon x \in U \} \subset J^k E$. The union of all such sets defines a local foliation on $J^{k} E$ which is called the local Cartan foliation. The associated distribution is called the Cartan distribution:

\begin{df}[Cartan Distribution, Chetverikov \cite{CHE}]\label{cd} Given a smooth fiber bundle $\pi \colon E \rightarrow M$ and $\chi \in J^k E$ ($k$ finite), the $k$-th Cartan distribution at $\chi$ is defined to be
$$C_{\chi}^k \defeq  \bigcup_{j_{\pi_k(\chi)}^{k+1} \varphi \in (\pi_{k+1}^{k})^{-1}(\chi)} T j^k \varphi \left( T_{\pi_k (\chi)} U \right).$$
The $\infty$-Cartan distribution at $\chi \in J^{\infty} E$, $C_{\chi}^{\infty}$ is given by the pro-object indexed by the limit of the finite distributions.
\end{df}

Observe that the fact that $T j^k \varphi$ is well defined in the previous equation is due to the fact that we know the $(k+1)$-th jet of $\varphi$ at $\pi_k(\chi)$ and jets where equivalence classes of iterated tangent maps.

\begin{rk} It is important to observe that the term ``distribution'' here is used in a more general setting that usual. $C_{\chi}^k$ is not a linear subspace of $T_{\chi} J^k E$ but rather a union of linear subspaces. This can be easily illustrated in the following example. Consider the trivial one dimensional vector bundle over $\RE$, $\RE \times \RE \rightarrow \RE$, whose space of sections is simple smooth functions on $\RE$. $C_{(x,y)}^0$ is the whole tangent space without the vertical axis.
\end{rk}

The Cartan distribution has various interesting applications and has been studied in the Russian literature in depth (see Vinogradov \cite{VINO} for example). We would like to start by pointing out some applications related to differential forms in the infinite jet bundle.

Any pro-smooth map $f^{\infty} \colon J^{\infty} E \rightarrow J^{\infty} F$ can {\it pull-back} differential forms using representatives. Let $\{f^l\}$ represent $f$ and $\omega_l$ represent the form $\omega$ in $\Omega^{\bullet}(J^{\infty} F)$. Then $(f^l)^* \omega_l \in \Omega(J^{k(l)} E)$ is a differential form which represents up to isomorphism a unique differential form $(f^{\infty})^* \omega$  in $\Omega^{\bullet}(J^{\infty} E)$ independently of all the representatives chosen.

\begin{dfpp}[Contact ideal, Anderson \cite{AND}]
Given a smooth fiber bundle $\pi \colon E \rightarrow M$, the contact ideal $\mathtt{C} \subset \Omega^{\bullet}(J^{\infty} E)$ is given by forms vanishing along the local Cartan foliation:
$$\mathtt{C} = \{\omega \in \Omega^{\bullet}(J^{\infty} E) \colon (j^{\infty} \varphi)^*\omega = 0 \in \Omega^{\bullet}(U) \, \forall (\varphi, U) \textrm{ local section of } \pi\}.$$
\end{dfpp}

To be precise, if $\omega$ is of order $k$ and it is represented by $\omega_k \in \Omega^{\bullet}(J^k E)$, by $(j^{\infty} \varphi)^*\omega$ we mean the differential form represented by $(j^{k} \varphi)^*\omega_k$. The condition that it equals zero is independent of the choice of the representative $\omega_k$. The contact ideal can also be defined for $J^k E$ for finite $k$ in the same way. We adopt the notation $\mathtt{C}(J^k E)$ in case it is necessary to distinguish between different $k$'s or different $E$'s.

The contact ideal is sometimes called the Cartan co-distribution. It is clear that it is a differentiable ideal in $\Omega^{\bullet}(J^{\infty} E)$. Observe that we have used the same letter $\mathtt{C}$ to denote the contact ideal and the ideal generated by $d_V u_I^j$. This is not a coincidence, both ideals agree and that is the content of the Proposition above (for a proof, see Anderson \cite{AND}):
$$ (j^{\infty} \varphi)^* d_V u_I^{\alpha} = (j^{\infty} \varphi)^* (d u_I^{\alpha} + u_{I, i}^{\alpha} d x^i ) = \frac{\partial u_I^{\alpha} \circ \varphi}{\partial x^i} dx^i - \left( u_{I,i}^{\alpha} \circ \varphi \right) dx^i = 0.$$

The contact ideal has very interesting features on its own. In order to point out one of the basic results related to the contact ideal we need to introduce the {\it prolongation of sections}. This uses once again the maps $j^k (U)$ defined in Chapter 1: given any local section $\varphi \in \E(U)$ we can construct a local section $j^k \varphi \in \Gamma^{\infty}(U, \restrict{J^k E}{\pi_k^{-1}(U)})$ for all $k$ (even $k = \infty$ taking the corresponding colimit). The map sending $\varphi$ to $j^k \varphi$ is nothing else but the induced map $j^k$ on $\E(U)$:
\begin{center}
    \begin{tabular}{rcl}
    $j^k \colon \E(U)$ & $\longrightarrow$ & $\mathcal{J}^k \E(U) \defeq  \Gamma^{\infty}(U, \restrict{J^k E}{\pi_k^{-1}(U)})$ \\
    $\varphi$ & $\longmapsto$ & $\left[ x \mapsto [(\varphi, x)] \, \right].$ \\
    \end{tabular}
\end{center}

Any prolonged section is called a {\it holonomic section}. We focus on the case $k=\infty$. Not all sections $\Xi$ of $J^{\infty} E \rightarrow M$ are prolongations of sections $\varphi$ of $E \rightarrow M$. As a matter of fact, those are precisely the ones that pull back the contact ideal to zero:

\begin{lm}[Anderson \cite{AND}]\label{XI} Let $E \rightarrow M$ be a smooth fiber bundle with associated infinite jet bundle $J^{\infty} E$. A local section $\Xi$ of $\pi_{\infty} \colon J^{\infty} E \rightarrow M$ is holonomic (that is, it is the infinite prolongation of a local section $\varphi$ of $E \rightarrow M$) if and only if $\Xi^* \mathtt{C} = 0$.
\end{lm}

The proof is clear but illuminating. $(\Xi)^*(d_V u_{I}^{\alpha}) = 0$ if and only if $\frac{\partial (u_I^{\alpha} \circ \Xi)}{\partial x^i} = u_{I,i}^{\alpha} \circ \Xi$ for all indices, so that $\Xi$ can be constructed inductively from $u_i \circ \Xi$. Now defining $\varphi \defeq  \pi_{\infty}^0 \circ \Xi$ it is clear that $\Xi = j^{\infty} \varphi$. The proof also works when considering smooth sections instead of pro-smooth sections since it relies on the coordinates and those are the same due to Proposition \ref{saunp}.\\

Pro-smooth maps preserving the contact ideal are extremely interesting. Intuitively, a map preserving the contact ideal, i.e. the Cartan co-distribution also preserves the Cartan distribution and the local Cartan foliation. That means it sends infinite prolongations of sections to infinite prolongations of sections.  This seems tremendously natural and one could decide to study maps $f^{\infty} \colon J^{\infty} E \rightarrow J^{\infty} F$ that preserve the contact ideal. This has been done extensively in the Russian literature and it is also covered in the book by Anderson \cite{AND}. We develop this intuition in detail in the following sections.


\section{Jet prolongations}\label{jpg}

{\it Maps between finite jet bundles induce, under certain assumptions, pro-finite maps between the corresponding infinite jet bundles. We first review the known results concerning jet prolongations including the factorization theorem of Chetverikov that asserts that any pro-finite smooth function factors through 3 maps which are jet prolongations. The original work in this chapter is concentrated in the study of a more general class of maps that admit pro-finite smooth prolongations. Maps in that class give rise to unique jet prolongations that preserve the contact ideal. The starting points of this section are the texts of Chetverikov \cite{CHE} and Saunders \cite{SAU}.}\\

Infinite jet prolongations are pro-smooth maps $f^{\infty} = j^{\infty} f^0$ where all higher $f^{l \geqslant 1}$ are obtained from $f^0$. In the literature (Kock \cite{K}, Anderson \cite{AND}, Saunders \cite{SAU}) jet prolongations are always studied in the setting in which all maps are vector bundles over a diffeomorphism. In this section we have extended the set of such maps for which infinite jet prolongations exist to include maps between pairs in which the base manifolds are not the same. The motivation to do so is physical, we would like to include restrictions of a set of solutions of a differential equation to their boundaries, or conversely, extending the boundary data to solutions of the equation in the whole manifold.

As a motivation for jet prolongations, we begin by describing jet prolongations in an easy case, that is treated by Anderson \cite{AND} and Saunders \cite{SAU}. Consider a bundle map over a diffeomorphism $\tau \colon M \rightarrow M$:
\begin{center}
\begin{tikzpicture}[description/.style={fill=white,inner sep=2pt}]
\matrix (m) [matrix of math nodes, row sep=2.5em,
column sep=2.5em, text height=1.5ex, text depth=0.25ex]
{E & F\\
 M & M\\};
\path[->,font=\scriptsize]
(m-1-1) edge node[auto] {$f^0$} (m-1-2)
(m-2-1) edge node[auto] {$\tau$} (m-2-2)
(m-1-1) edge node[auto] {$\pi$} (m-2-1)
(m-1-2) edge node[auto] {$\rho$} (m-2-2);
\end{tikzpicture}
\end{center}

Given any local section of $E$, $\varphi \colon U \subset M \rightarrow E$ we can construct a local section of $F$ simply by following $\tau^{-1}$ and then $f^0$: thus we have a map $\E \rightarrow \F$ given by $\varphi \mapsto f^0 \circ \varphi$ (it is a section because $\rho \circ f^0 \circ \varphi \circ \tau^{-1} = \tau \circ \pi \circ \varphi \circ \tau^{-1} =  \tau \circ \tau^{-1} = \textrm{id}_M$).

Using the chain rule and the inverse function theorem, we can see that in order to know the $l$-th jet of $f^0 \circ \varphi \circ \tau^{-1}$ at $\tau(x)$ it is enough to know the $l$-th jet of $\varphi$ at $x$ for all $(\varphi, x) \in \E \times M$.

We can hence define the {\it Lie-jet prolongation}: 
\begin{center}
    \begin{tabular}{rcl}
    $j^l f^0 \colon J^l E$ & $\longrightarrow$ & $J^l F$ \\
    $[(\varphi ,x)]$ & $\longmapsto$ & $[(f^0 \circ \varphi \circ \tau^{-1}, \tau(x))]$ \\
    \end{tabular}
\end{center}
    
Explicitly for the first derivatives, taking local coordinates $\{x^i\}$ in $M$ and $\{ u^{\alpha} \}$ in $E$ as in Definition \ref{coordinates}):
$$\left. \frac{\partial (f^0 \circ \varphi \circ \tau^{-1})}{\partial x^i} \right|_{\tau(x)} =  \left. \frac{\partial f^0}{\partial u^{\alpha}} \right|_{\varphi(x)} \cdot \left. \frac{\partial \varphi_{\alpha}}{\partial x^j} \right|_{x} \cdot \left. \frac{\partial \tau_j^{-1}}{\partial x^i} \right|_{\tau(x)} = \left. \frac{\partial f^0}{\partial u^{\alpha}} \right|_{\varphi(x)} \cdot \left. \frac{\partial \varphi_{\alpha}}{\partial x^j} \right|_{x} \cdot \left( \left. \frac{\partial \tau}{\partial x} \right|_{x} \right)_{(i,j)}^{-1}.$$

The prefix {\it Lie} here comes from the Russian school in which such maps are called Lie-transformations (see Chetverikov \cite{CHE} for example). 

By staring at the previous construction it is obvious that there are three natural ways of generalizing this result: 

\begin{enumerate}
	\item The first one is to consider maps $f^0 \colon J^k E \rightarrow F$ rather than $E \rightarrow F$. This is done by Kock \cite{K} and it is also used by Saunders \cite{SAU} without explicitly stating any result. One has to deal with holonomic jets since $J^l(J^k E) \neq J^{k+l} E$. 
	\item But we do not need $\tau$ to be a diffeomorphism. It is enough that $\tau$ is a submersion and instead of $\tau^{-1}$ we take any local section around $\tau(x)$ passing through $x$. The trick with the derivatives and the inverse function theorem still applies.
	\item We also do not need $f^0$ to be a bundle map over $\tau$. We only need that for a fixed $\varphi$, $f^0 \circ j^k \varphi \colon U \rightarrow F$ is a bundle map over $\tau_{\varphi} \defeq  \rho \circ f^0 \circ j^k \varphi$ (which is true) and that $\tau_{\varphi}$ is a submersion.
\end{enumerate}

First of all, we finish the literature review by introducing holonomic jets and the result by Kock about jet prolongations which was mentioned in the previous enumeration.

If we can replace $E$ by $J^k E$ in all the formulas above, we only need to change $\pi$ by $\pi_k$ and to do some adjustments. Observe that the map we get so far using the same procedure, goes from $J^l (J^k E)$ to $J^l F$. We want to start from $J^{n} E$ instead (for some value of $n$). The spaces $J^l (J^k E)$ and $J^{k+l} E$ are not the same. A dimension count shows that the first is larger than the second. If the dimension of $M$ is $m > 1$ and the rank of $E$ is $e$, by looking at the local coordinates, we see that for $k=l=1$ the rank of the first bundle is $e(m+1)^2$ while for the other it is $e(m+1)\frac{m+2}{2}$ (this is so because partial derivatives commute). We conclude that $J^1 (J^1 E)$ is strictly larger than $J^2 E$.

But still, $J^{k+l} E$ sits inside of $J^l (J^k E)$, via the following map:
\begin{center}
    \begin{tabular}{rcl}
    $\iota_{l,k} \colon J^{k+l} E$ & $\longrightarrow$ & $J^l (J^k E) $ \\
    $[(\varphi, x)]_{k+l}$ & $\longmapsto$ & $[(j^k \varphi, x)]_l.$ \\
    \end{tabular}
\end{center}

We have denoted the equivalence classes in $J^{k+l}(-)$ and $J^l(-)$ by $[-]_{k+l}$ and $[-]_l$ respectively in order to avoid confusion. As we said earlier $j^k \varphi$ is a section of $J^k E$ so that we can take its equivalence class in $J^l (J^k E)$. This map is well defined and it is a bundle map (see Saunders \cite[5.2.1]{SAU}). The elements in $\iota_{l,k}(J^{k+l} E)$ are called {\it holonomic jets}.

We can use the maps $\iota_{l,k}$ to interpret the jet prolongation of a map $J^k E \rightarrow F$ not as a map $J^l (J^k E) \rightarrow J^l F$ but rather as a more convenient map $J^{k+l} E \rightarrow J^l F$. Now it is clear that if we apply $\iota_{l,k}$ and later the jet prolongation along $\tau$ we get a new jet prolongation fitting our new situation. This notion includes the previous one.

Given $f^0 \colon J^k E \rightarrow F$ a bundle map covering a diffeomorphism $\tau$ and $l$ a non-negative integer, we  define the {\it holonomic-jet prolongation} of $f^0$ as:
\begin{center}
    \begin{tabular}{rcl}
    $j^l f^0 \colon J^{k+l} E$ & $\longrightarrow$ & $J^l F$ \\
    $[(\varphi ,x)]$ & $\longmapsto$ & $[(f^0 \circ j^k \varphi \circ \tau^{-1}, \tau(x))]$ \\
    \end{tabular}
\end{center}

\begin{pp}\label{pro}
Let $\pi \colon E \rightarrow M$ and $\rho \colon F \rightarrow M$ be two smooth fiber bundles. Any $f \colon J^{k} E \rightarrow F$ bundle map covering a diffeomorphism $\tau$ induces a smooth and pro-smooth bundle map $j^{\infty} f \colon J^ {\infty} E \rightarrow J^{\infty} F$ covering $\tau$ in the category of pro-finite smooth manifolds.
\begin{center}
\begin{tikzpicture}[description/.style={fill=white,inner sep=2pt}]
\matrix (m) [matrix of math nodes, row sep=2.5em,
column sep=2.5em, text height=1.5ex, text depth=0.25ex]
{J^{\infty} E & J^{\infty} F \\
 M & M \\};
\path[->,font=\scriptsize]
(m-1-1) edge node[auto] {$j^{\infty} f^0$} (m-1-2)
(m-2-1) edge node[auto] {$\tau$} (m-2-2)
(m-1-1) edge node[auto] {$\pi_{\infty}$} (m-2-1)
(m-1-2) edge node[auto] {$\rho_{\infty}$} (m-2-2);
\end{tikzpicture}
\end{center}
\end{pp}

Saunders \cite[Proposition 7.2.10]{SAU} proves that holonomic-jet prolongation are smooth. The previous proposition simply says that holonomic-jet prolongations are pro-smooth (as a consequence of their definition above). We simply need to apply Corollary \ref{prosmosmo} to conclude that the map is also smooth without using the result of Saunders.

Holonomic-jet prolongations of bundle maps covering the identity will be extremely common in the theory. Thus, we want to give an explicit formula for the prolongations (define $D_{i_1 , \cdots , i_n} \defeq  D_{i_1} \cdots D_{i_n}$, recall the meaning of $D_i$ as in Equation \ref{di}):

\begin{eqnarray}\label{proI}
(j^{\infty}f^0)_i &=& (f^0)_i = x^i  \nonumber \\
(j^{\infty}f^0)_{\alpha} &=& (f^0)_{\alpha} \nonumber \\
(j^{\infty}f^0)_{\alpha}^I &=& D_I (j^{\infty} f^0)_{\alpha} = D_I (f^0)_{\alpha}.
\end{eqnarray}

As a matter of fact, Lie- and holonomic-jet prolonged maps give a great understanding of pro-smooth maps preserving the contact ideal. We mention here the classical result by Chetverikov about bundle maps over a diffeomorphism. Generalizations to a more abstract setting are possible.

\begin{df}[$\mathtt{C}$-transformation, Chetverikov \cite{CHE}] Consider two smooth fiber bundles over the same base manifold $\pi_1 \colon E_1 \rightarrow M$ and $\pi_2 \colon E_2 \rightarrow M$. A map $f \colon J^{\infty} E_1 \rightarrow J^{\infty} E_2$ is called a $\mathtt{C}$-transformation if
	\begin{enumerate}
		\item $f$ is a diffeomorphism and
		\item $f$ preserves the contact ideal: $f^* (\mathtt{C}(J^{\infty} E_2)) \subset \mathtt{C}(J^{\infty} E_1)$.
	\end{enumerate}
\end{df}

Chetverikov actually defines $\mathtt{C}$-transformations as continuous maps preserving the algebra of pro-smooth functions. He then argues that these maps are pro-smooth. This is indeed the case: continuous maps are pro-continuous since $\Top$  has all limits. Each representative $f^l \colon J^{k(l)} E \rightarrow J^l E$ preserves the algebra of smooth functions of those finite dimensional manifolds, hence it is smooth. That shows that the original map is indeed pro-smooth. 

The main result \cite{CHE} by Chetverikov is that $\mathtt{C}$-transformations factorize locally as jet prolongations:

\begin{tm}[Chetverikov {\cite[Theorem 2]{CHE}}]\label{chet} Given $f \colon J^{\infty} E_1 \rightarrow J^{\infty} E_2$ a $\mathtt{C}$-transformation and $\chi \in J^{\infty} E_1$ generic (for some notion of generic points). There exist $U_i \subset J^{\infty} E_i$, for $i \in \{1, 2\}$ open neighborhoods of $\chi$ and $f(\chi)$ respectively; there exist $g_i^{\infty} \colon U_i \rightarrow J^{\infty} F_i$ for $i \in \{1, 2\}$ and $f^{\infty} \colon J^{\infty} F_1 \rightarrow J^{\infty} F_2$ such that:
	\begin{itemize}
		\item $f^{\infty}$ factors through the other maps, that is, the following diagram commutes
		\begin{center}
			\begin{tikzpicture}[description/.style={fill=white,inner sep=2pt}]
			\matrix (m) [matrix of math nodes, row sep=2.5em,
			column sep=2.5em, text height=1.5ex, text depth=0.25ex]
			{U_1 & U_2\\
				J^{\infty} F_1 & J^{\infty} F_2 \\};
			\path[->,font=\scriptsize]
			(m-1-1) edge node[auto] {$\restrict{f}{U_1}$} (m-1-2)
			(m-2-1) edge node[auto] {$f^{\infty}$} (m-2-2)
			(m-1-1) edge node[left] {$g_1^{\infty}$} (m-2-1)
			(m-1-2) edge node[auto] {$g_2^{\infty}$} (m-2-2);
			\end{tikzpicture}
		\end{center}
		\item $g_i^{\infty}$ is a Lie-prolongation of $g_i^0 \colon E_i \rightarrow F_i$ both for $i=1$ and $i=2$.
		\item $f^{\infty}$ is a holonomic-prolongation of $f^0 \colon J^k F_1 \rightarrow F_2$.
		\item $g_1^{\infty}$, $g_2^{\infty}$ and $f^{\infty}$ are pro-smooth.
		\item The set of generic points is an open everywhere dense set in $J^{\infty} E_1$.
	\end{itemize}
\end{tm}

The second to last item is unnecessary since it follows from the previous ones and Proposition \ref{pro}.

We finish this section with an example of jet prolongations that relates the maps that we already know with this new technique.

\begin{ej}
	Let $\pi \colon E \rightarrow M$ be a fiber bundle. We can think of it as a bundle map from $(E, \pi)$ to $(M, \textrm{id})$ over the identity.
	\begin{center}
		\begin{tikzpicture}[description/.style={fill=white,inner sep=2pt}]
		\matrix (m) [matrix of math nodes, row sep=2em,
		column sep=2.5em, text height=1.5ex, text depth=0.25ex]
		{E & M \\
			M & M \\};
		\path[->,font=\scriptsize]
		(m-1-1) edge node[auto] {$\pi$} (m-1-2)
		(m-2-1) edge node[auto] {id} (m-2-2)
		(m-1-1) edge node[auto] {$\pi$} (m-2-1)
		(m-1-2) edge node[auto] {id} (m-2-2);
		\end{tikzpicture}
	\end{center}
	We can take the $k$-th jet prolongation of $\pi$ to get a map $j^k \pi \colon J^k E \rightarrow J^k M \cong M$ sending $[(\varphi, x)]$ to $[(\textrm{id}, x)] \cong \{x\}$. We have recovered the bundle projection, so that $j^k \pi = \pi_k \colon J^k E \rightarrow M$.\\
	
	If we fix a local section, we can interpret it as a bundle map over the identity again.
	\begin{center}
		\begin{tikzpicture}[description/.style={fill=white,inner sep=2pt}]
		\matrix (m) [matrix of math nodes, row sep=2em,
		column sep=2.5em, text height=1.5ex, text depth=0.25ex]
		{U & \restrict{E}{\pi^{-1}(U)} \\
			U & U \\};
		\path[->,font=\scriptsize]
		(m-1-1) edge node[auto] {$\varphi$} (m-1-2)
		(m-2-1) edge node[auto] {id} (m-2-2)
		(m-1-2) edge node[auto] {$\pi$} (m-2-2)
		(m-1-1) edge node[auto] {id} (m-2-1);
		\end{tikzpicture}
	\end{center}
	Its $k$-th jet prolongation is a map $j^k \varphi \colon U \rightarrow J^k E$ sending $x \cong [(\textrm{id}, x)]$ to $[(\varphi, x)] \in J^k E$. It is obvious that $\pi_k \left( j^k \varphi \right) = \textrm{id}_U$ so that $j^k \varphi$ is a local section of $J^k E$. We have recovered the map $j^k \colon \E(U) \rightarrow \mathcal{J}^k \E (U)$ which we defined before. 
\end{ej}


\subsection{General case}

Following the observations made when treating Lie-jet prolongations, we can consider the following more general case:

\begin{df}[Jet prolongation]\label{PRO1}
Let $\pi \colon E \rightarrow M$ and $\rho \colon F \rightarrow N$ be two smooth fiber bundles. Let $f \colon J^k E \rightarrow F$ be a map such that:
\begin{enumerate}
\item For all local section $\varphi \in \E(U)$ the map $\tau_{\varphi} \defeq  \rho \circ f \circ j^k \varphi \colon U \rightarrow N$ is a submersion.
\end{enumerate}
For each $\chi \in j^k \varphi (U)$ we take a local section $s_{\chi}$ of $\tau_{\varphi}$ passing through $\pi_{\infty}(\chi)$. Consider $s$ a family of such $s_{\chi}$. We define the set theoretic infinite jet prolongation of $f$ associated to $s$ to be the map $j_s^{\infty} f$ represented for each $l$ non-negative integer by
\begin{center}
    \begin{tabular}{rcl}
    $j_s^l f \colon J^{k+l} E$ & $\longrightarrow$ & $J^l F$ \\
    $[(\varphi ,x)]$ & $\longmapsto$ & $[(f \circ j^k \varphi \circ s_{\chi}, \tau_{\varphi}(x))]$ \\
    \end{tabular}
\end{center}
\end{df}

\begin{rk}
	The map $j_s^l f$ is not smooth a priori, hence even if they commute with the bundle maps $\pi_{k}^{k^{\prime}}$ and $\rho_{l}^{l^{\prime}}$, the map $j_s^{\infty} f$ is not pro-smooth.
	
	A way of solving this difficulty is the following. Since $\tau_{\varphi}$ is a submersion for every $\varphi$, one can show that $J^k E \rightarrow N$ is also a submersion. Taking families of sections of this map instead simplifies the calculations to show that $j_s^l f$ is smooth. But as soon as we want this sections to define sections of the $\tau_{\varphi}$ via $\pi_k$ we notice that we need $s$ to preserve the local Cartan foliation. This approach is explored in what follows.
	
	Nevertheless, there is another way to work this problem out and it is to choose families of sections encoded in a pro-smooth map $\iota = s \colon J^{\infty} E \rightarrow J^{\infty}(M \times N)$ where $M \times N$ is viewed as the trivial bundle over $N$. In this way the pro-smoothness of $j_{\iota}^{\infty} f$ is ensured.
\end{rk}

Another question related to jet prolongations is how unique $j_s^{\infty} f$ actually is. Consider a pro-finite smooth map $f^{\infty}$ between infinite jet bundles:

\begin{center}
\begin{tikzpicture}[description/.style={fill=white,inner sep=2pt}]
\matrix (m) [matrix of math nodes, row sep=3em,
column sep=3.5em, text height=1.5ex, text depth=0.25ex]
{J^{\infty} E & J^{\infty} F\\
 J^{k(l)} E & J^{l} F\\
 J^{k} E & F\\};
\path[->,font=\scriptsize]
(m-1-1) edge node[auto] {$f^{\infty}$} (m-1-2)
(m-2-1) edge node[auto] {$f^{l}$} (m-2-2)
(m-3-1) edge node[auto] {$f^{0}$} (m-3-2)
(m-1-1) edge node[auto] {$\pi_{\infty}^{k(l)}$} (m-2-1)
(m-1-2) edge node[auto] {$\rho_{\infty}^{l}$} (m-2-2)
(m-2-1) edge node[auto] {$\pi_{k(l)}^{k}$} (m-3-1)
(m-2-2) edge node[auto] {$\rho_{l}^{0}$} (m-3-2);
\end{tikzpicture}
\end{center}

If $f^0$ is under the assumptions of Theorem \ref{chet}, we can construct the $\infty$-jet prolongation of $f^0 \colon E \rightarrow F$. We have not discussed if such a map $j_s^{\infty} f^0$ covering $f^0$ is unique or even whether or not $f^{\infty} = j_s^{\infty} f^0$ for some choice of sections.

This question is answered for Lie-jet prolongations (i.e. prolongations of bundle maps $f^0 \colon E \rightarrow F$ along a diffeomorphism) by Anderson \cite[Proposition 1.6]{AND}). The key observation is that the map $j_s^{\infty} f^0$, the Lie-jet prolongation of $f^0$ is the only one map covering $f^0$ if it preserves the Cartan co-distribution. Translating this fact to the full generality of jet prolongations given by Definition \ref{PRO1} is possible.

\begin{pp}\label{PR}
Let $\pi \colon E \rightarrow M$ and $\rho \colon F \rightarrow N$ be two smooth fiber bundles. Let $f^{\infty} \colon J^{\infty} E \rightarrow F$ be a pro-smooth map such that:
\begin{enumerate}
\item For every local section $\varphi \in \E(U)$ the map $\tau_{\varphi} \defeq  \rho_{\infty} \circ f^{\infty} \circ j^{\infty} \varphi \colon U \rightarrow N$ is a submersion.
\item $f^* (\mathtt{C}(J^{\infty} F)) \subset \mathtt{C}(J^{\infty} E)$.
\end{enumerate}
Given $\{f^l\}$ a representative of $f^{\infty}$, infinite jet prolongations of $f^0 \colon J^k E \rightarrow F$ in the sense of Definition \ref{PRO1} exist, they are pro-smooth and $f^{\infty} = j_s^{\infty} f^0$ for every choice of sections $s$. That is:
\begin{center}
    \begin{tabular}{rcl}
    $f^{\infty} \colon J^{k+l} E$ & $\longrightarrow$ & $J^l F$ \\
    $j_x^{\infty} \varphi$ & $\longmapsto$ & $j_{\rho \circ f^0 \circ j^{k} \varphi (x)}^{\infty} (f^0 \circ j^k \varphi \circ s_{\chi})$. \\
    \end{tabular}
\end{center}
\end{pp}

\dem Fix $\{f^l\}$ a representative of $f^{\infty}$. Let $\varphi \in \E(U)$ be a local section of $\pi$. Since $f^0 \circ \pi_{\infty}^k = \rho_{\infty}^0 \circ f^{\infty}$ we have that 
$$\tau_{\varphi} \defeq  \rho_{\infty} \circ f^{\infty} \circ j^{\infty} \varphi = \rho \circ f^0 \circ \pi_{\infty}^k \circ j^{\infty} \varphi = \rho \circ f^0 \circ  j^k \varphi$$
{\noindent is a submersion and thus $f^0$ is in the hypothesis of Definition \ref{PRO1}. Hence, infinite jet prolongations of $f^0$ do exist. Fix $\chi \in \left( j^{\infty} \varphi \right) (U)$ and take $s_{\chi}$ a local section of $\tau_{\varphi}$ around $V \subset N$ passing through $\pi_{\infty}(\chi)$.}
\begin{equation}\label{aaa}
j_s^{\infty} f^0 (\chi) = j_{\rho \circ f^0 \circ j^{k} \varphi (\pi_{\infty}(\chi))}^{\infty} (f^0 \circ j^k \varphi \circ s_{\chi}).
\end{equation}

We can construct the map $f^{\infty} \circ j^{\infty} \varphi \circ s_{\chi} \colon V \rightarrow J^{\infty} F$. It is a pro-smooth section of $\rho_{\infty}$ since 
$$\rho_{\infty} \circ \left( f^{\infty} \circ j^{\infty} \varphi \circ s_{\chi} \right) = \left( \rho_{\infty} \circ f^{\infty} \circ j^{\infty} \varphi \right) \circ s_{\chi} = \tau_{\varphi} \circ s_{\chi} = \textrm{id}_V.$$

We are going to pullback to $V \subset N$ a form $\omega$ in $\mathtt{C}(J^{\infty} F)$.

$$\left( f^{\infty} \circ j^{\infty} \varphi \circ s_{\chi} \right)^* \omega = (s_{\chi})^* \circ (j^{\infty} \varphi)^* \circ (f^{\infty})^* \omega = 0$$

This is so because since $f^{\infty}$ preserves the contact ideal, $(f^{\infty})^* \omega$ is in $\mathtt{C}(J^{\infty} E)$ and hence, by definition $(j^{\infty} \varphi)^* (f^{\infty})^* \omega = 0$.

Applying Lemma \ref{XI} we get that there exists $\psi_{\chi} \in \F(V)$ such that 
\begin{equation}
f^{\infty} \circ j^{\infty} \varphi \circ s_{\chi} = j^{\infty} \psi_{\chi}.
\end{equation}

By further composing with $\rho_{\infty}^0$ on both sides of the equation we get: 
\begin{equation}\label{yo}
f^0 \circ j^k \varphi \circ s_{\chi} = \psi_{\chi},
\end{equation}
{\noindent and hence $j_s^{\infty} f^0 (\chi) =[(\psi_{\chi}, \rho_{\infty} \circ f^{\infty} (\chi))]$.}

On the other hand, 
\begin{equation}\label{bbb}
f^{\infty}(\chi) = f^{\infty} \circ j^{\infty} \varphi \, (s_{\chi}(\rho_{\infty} \circ f^{\infty}(\chi))) = j^{\infty} \psi_{\chi} (\rho_{\infty} \circ f^{\infty}(\chi)).
\end{equation}

Comparing equations \ref{bbb} and \ref{yo} we get that $j_s^{\infty} f^0 = f^{\infty}$ and in particular that $j^{\infty} f^0$ is pro-smooth. \qed\\

\begin{rk}\label{genius}
In the notation of the proof, different choices of $s_{\chi}$ give rise to different $\psi_{\chi}$, but their infinite jets at $\rho(f^{\infty}(\chi))$ agree.

Even though the proof uses the local Cartan foliation, we do not know a priori if $f^{\infty}$ preserves the foliation, that is, if $f^{\infty} j^{\infty} \varphi (U) = \bigcup_{\alpha} j^{\infty} \psi_{\alpha} (V)$ for every local section.

There are particularly well behaved choices of $s_{\chi}$ where we can actually show that $f^{\infty}$ preserves the local Cartan foliation. Take $x_0 \in U$ and $s_0 \defeq  s_{j_{x_0}^{\infty} \varphi}$ a local section at $V$ passing through $x_0$. For all $x \in s_{j_{x_0}^{\infty} \varphi} (V)$ define $s_{j_{x}^{\infty} \varphi} \defeq  s_0$.  It is a section of $\tau_{\varphi}$ and it passes through the relevant point. By the previous procedure $s_0$ defines a local section $\psi_0$ in $\F(V)$. In that case $f^{\infty} j^{\infty} \varphi (s_{0}(V)) = j^{\infty}(\psi_{s_0})(V)$ and by repeating the process for every point in $U$ we get that (shrinking $U$ if necessary) $f^{\infty} j^{\infty} \varphi (U) = \bigcup_{\alpha} j^{\infty} \psi_{\alpha} (V)$, i.e. $f^{\infty}$  preserves the local Cartan foliation.
\end{rk}

We can get a final conclusion that serves as a partial converse of Proposition \ref{prosmosmo}:

\begin{cl}\label{chocho}
Let $\pi \colon E \rightarrow M$ and $\rho \colon F \rightarrow N$ be two smooth fiber bundles. Let $f^{\infty} \colon J^{\infty} E \rightarrow J^{\infty} F$ be a smooth map such that:
\begin{enumerate}
\item For all local section $\varphi \in \E(U)$ the map $\tau_{\varphi} \defeq  \rho_{\infty} \circ f^{\infty} \circ j^{\infty} \varphi \colon U \rightarrow N$ is a submersion.
\item $f^* (\mathtt{C}(J^{\infty} F)) \subset \mathtt{C}(J^{\infty} E)$.
\item It covers a smooth map $f^0 \colon J^k E \rightarrow F$.
\end{enumerate}
Then $f^{\infty}$ is pro-smooth.
\end{cl}

\dem The proof of Proposition \ref{PR} only uses the fact that $f^{\infty}$ is pro-smooth when referring to Lemma \ref{XI}. We have remarked after the statement of that Lemma that it also holds when $f^{\infty}$ is smooth. Otherwise, in order to apply Proposition \ref{PR} we only need the extra condition that $f^{\infty}$ covers $f^0$. We can thus apply that Proposition to get the desired result.
\qed

\section{Vector Fields on \boldmath{\mbox{$J^{\infty} E$}} }\label{vf}
{\it The next and last piece of information related to the infinite jet bundle is about vector fields. There is a correspondence between pro-smooth sections of the tangent bundle and ind-derivations of the space of smooth functions: those are called vector fields. The usual formulas known as Cartan calculus apply to the infinite jet bundle. Vector fields preserving the contact ideal are once again prolongable and they actually split into a vertical and a horizontal component called evolutionary and total. This section provides a translation into the ind/pro-categorical language of the results about generalized vector fields by Anderson \cite{AND} and Olver \cite{OLV}.}\\

Recall from the previous chapter that the tangent bundle of the infinite jet bundle, $T( J^{\infty} E) = \bigcup_{\chi \in J^{\infty} E} T_{\chi} (J^{\infty} E)$ is a pro-smooth manifold modeled on $\left\{ \left( T (J^k E), T \pi_{k}^l \right) \right\}$ and that the base-point projection map $\textrm{pr}_{J^{\infty} E} \colon T (J^{\infty} E) \rightarrow J^{\infty} E$ is pro-smooth and can be represented by $\left\{ \textrm{pr}_{J^{\infty} E}^l = \textrm{pr}_{J^{l} E} \right\}$. We can talk about pro-smooth sections of that pro-smooth fiber bundle, those do deserve to be called vector fields on $J^{\infty} E$:

\begin{df}[Vector field on the infinite jet bundle] Given a smooth fiber bundle $\pi \colon E \rightarrow M$, a vector field on $J^{\infty} E$ is a pro-smooth map $X \colon J^ {\infty} E \rightarrow T(J^{\infty} E)$ such that $\textrm{pr}_{J^{\infty} E} \circ X = \textrm{id}_{J^{\infty} E}$.
\end{df} 

Since vector fields are pro-finite smooth maps, they are represented by families of maps $\{X^l \colon J^{k(l)}  E \rightarrow T (J^l E) \}$ such that:
\begin{itemize}
	\item They are compatible, i.e. $T \pi_l^{l^{\prime}} \circ X^l = X^{l^{\prime}} \circ \pi_{k(l)}^{k(l^{\prime})}$.
	\item They are sections, i.e. $\textrm{pr}_{J^l E} \circ X^l = \pi_{k(l)}^l$.
\end{itemize}

\begin{center}
	\begin{tikzpicture}[description/.style={fill=white,inner sep=2pt}]
	\matrix (m) [matrix of math nodes, row sep=3em,
	column sep=3.5em, text height=1.5ex, text depth=0.25ex]
	{	J^{k(l)} E & T (J^{l} E)\\
		J^{k} E & TE\\};
	\path[->,font=\scriptsize]
	(m-1-1) edge node[auto] {$X^{l}$} (m-1-2)
	(m-2-1) edge node[auto] {$X^{0}$} (m-2-2)
	(m-1-1) edge node[auto] {$\pi_{k(l)}^{k}$} (m-2-1)
	(m-1-2) edge node[auto] {$T \pi_{l}^{0}$} (m-2-2);
	\end{tikzpicture}
\end{center}

This is the definition of a compatible generalized vector field given by Anderson \cite{AND}. A vector field which is representable by $\{X^l\}$ where $k(l)$ is taken minimum is said to be of order $(k(0), k(1), \ldots )$. G\"uneysu and Pflaum \cite{GP} call such vector fields local.

A different approach is to view vector fields as derivations of the algebra of pro-smooth functions $\C(J^{\infty} E)$. Following the ind/pro-categorical spirit, we should consider ind-derivations. Let us be precise about this:

The product in $\C(J^{\infty} E)$ is an ind-endomorphism of $\C(J^{\infty} E)$. For every $(l, k) \in \mathbb{N} \times \mathbb{N}$ the map 
$$(\pi_{\max (l, k) }^{l})^* \cdot_{\C(J^{\max (l, k) } E)} (\pi_{\max (l, k) }^{k})^* \colon \C(J^{l} E) \times \C(J^{k} E) \rightarrow \C(J^{\max (l, k) } E)$$
{\noindent represents the product $\cdot \colon \C(J^{\infty} E) \times \C(J^{\infty} E) \rightarrow \C(J^{\infty} E)$ since it commutes with all the structure maps $\{ (\pi_{l}^{l^{\prime}})^* \}$.}

\begin{df}[Ind-derivation] An ind-linear map $D \colon \C(J^{\infty} E) \rightarrow \C(J^{\infty} E)$ such that $D(f \cdot g) = D(f) \cdot g + f \cdot D(g)$ for all $f$ and $g$ in $\C(J^{\infty} E)$ is called an ind-derivation.
\end{df}

Observe that all ind-derivations are derivations of the algebra product, but not the other way around. This is so, because ind-derivations are ind-morphisms and hence given an in-derivation $D$, for every $l \in \mathbb{N}$ there exists $k(l) \in \mathbb{N}$ such that $D$ can be represented by maps $D_l$:

\begin{center}
	\begin{tikzpicture}[description/.style={fill=white,inner sep=2pt}]
	\matrix (m) [matrix of math nodes, row sep=3em,
	column sep=3.5em, text height=1.5ex, text depth=0.25ex]
	{\C(J^{\infty} E) & \C( J^{\infty} E)\\
	\C( J^{l} E) & \C( J^{k(l)} E)\\};
	\path[->,font=\scriptsize]
	(m-1-1) edge node[auto] {$D$} (m-1-2)
	(m-2-1) edge node[auto] {$D_{l}$} (m-2-2)
	(m-2-1) edge node[auto] {$(\pi_{\infty}^{l})^*$} (m-1-1)
	(m-2-2) edge node[auto] {$(\pi_{\infty}^{k(l)})^*$} (m-1-2);
	\end{tikzpicture}
\end{center}

In that case $D$ is said to have order $(k(0), k(1), \ldots)$. Not all derivations are of this kind. Anderson shows \cite[Proposition 1.3]{AND} that, in the case in which $M$ and $E$ are compact, all derivations are ind-derivations.

Using this new interpretation of Anderson's result in terms of ind-derivations, Proposition 1.5 in the book by Anderson, \cite{AND} now becomes the interesting (and trivial) result:

\begin{pp}[Anderson, {\cite[Proposition 1.5]{AND}}]\label{vfeq} Let $X$ be a vector field on $J^{\infty} E$ of order $(k(0), k(1), \ldots )$ and $f$ be a pro-smooth function represented by $f_l \colon J^l E \rightarrow \RE$. $X(f)$ is defined to be the pro-smooth function on $J^{\infty} E$ represented by $X^l (f_l) \colon J^{k(l)} E \rightarrow \RE$ (where $X^l (f_l)$ is the projection to the fiber of $T f_l \circ X^l$). The associated map $\widehat{X} \colon \C(J^{\infty} E) \rightarrow \C(J^{\infty} E)$ is an ind-derivation of the pro-smooth algebra of functions of order $(k(0), k(1), \ldots )$.

Conversely, given $D$ an ind-derivation of $\C(J^{\infty} E)$ there exists a unique vector field $X$ such that $\widehat{X} = D$.
\end{pp}

G\"uneysu and Pflaum also notice \cite[Theorem 3.2.6]{GP} this result, but they are more interested in the derivations of the algebra of smooth functions with locally bounded jet degree as in Definition \ref{gune}.

We denote the Lie algebra of vector fields on $J^{\infty} E$ by $(\mathfrak{X}(J^{\infty} E), [-,-])$, where $[X,Y]$ is the unique vector field such that $\widehat{[X, Y]} = [\widehat{X}, \widehat{Y}]$.

\begin{rk} It is very important to remark that we do not take the most general of the possible definitions of vector fields, but the one that allows as the most flexibility to work with. This also happens to be the one coming from the ind-/pro-object approach.
\end{rk}

The next step is to consider insertion and Lie derivation of vector fields on ind-differential forms: From now on, all the definitions and results are taken from Anderson \cite{AND}. Our only input is to write all formulas in a more appropriate coordinate system.

\begin{df}[Insertion, action and Lie derivative]\label{CC2}
Let $\pi \colon E \rightarrow M$ be a smooth fiber bundle. Let $\omega$ be a differential form of degree $p$ and order $l$ in $\Omega^{\bullet}(J^{\infty} E)$ and let $\omega_l \in \Omega^p(J^l E)$ be a representative of $\omega$. Let $X$ be a vector field of order $(k(0), k(1), \ldots )$. Let $f$ be a pro-smooth function on $J^{\infty} E$.
\begin{itemize}
	\item The insertion of $X$ into $\omega$, denoted by $\iota_X \omega$ is the differential form of degree $p-1$ and order $k(l)$ represented by $\iota_{X^l}\omega_l$.
	\item $X(f)$ is defined to be pro-smooth function $\iota_X (df) = \widehat{X}(f)$.
	\item The Lie derivative of $\omega$ along $X$, denoted by $\mathcal{L}_X \omega$ is defined to be the degree $p$ form: $\mathcal{L}_X \omega (X_1, \ldots , X_p) = X (\omega (X_1, \ldots , X_p)) -(-1)^i \omega ([X,X_i], X_1, \ldots , X_p)$.
\end{itemize}
\end{df}

The total derivative $d$, the insertion operators and the Lie derivatives are all graded endomorphisms of $\Omega^{\bullet}(J^{\infty} E)$ (as ind-differential complexes). All such ind-endo\-morphisms form a Lie algebra  with bracket $[f,g] \defeq f \circ g - (-1)^{|f|\cdot|g|} g \circ f$, since composition and addition of ind-endomorphisms are well behaved. From all the finite dimensional pieces of $\Omega^{\bullet}(J^{\infty} E)$ we see that the commutation relations known as Cartan calculus also hold in the infinite jet bundle:

\begin{pp}[Cartan calculus]\label{CC}
The total derivative $d$, the insertion operators $\iota_X$ and the Lie derivatives $\mathcal{L}_X$ (for each vector field $X$) form a closed subalgebra of the Lie algebra of graded endomorphisms of $\Omega^{\bullet}(J^{\infty} E)$ with only the following non-zero commutators:
\begin{itemize}
	\item $\mathcal{L}_X \omega = \iota_X d \omega + d \iota_X \omega = [d, \iota_X]$.
	\item $[\mathcal{L}_X, \iota_{X^{\prime}}] = \iota_{[X, X^{\prime}]}$.
	\item $[\mathcal{L}_X, \mathcal{L}_{X^{\prime}}] = \mathcal{L}_{[X, X^{\prime}]}$.
\end{itemize}
\end{pp}

\dem The first, and most important of the items is proven by Anderson \cite[Proposition 1.16]{AND}. For the other two statements, we have found no reference in the literature for the infinite jet bundle. Nevertheless, the second statement clearly follows from the definition of the insertion operator in terms of the insertion of a vector filed on a form on a finite dimensional manifold (Definition \ref{CC2}) and the corresponding formula for finite dimensional manifolds (for instance Cannas da Silva \cite[18.3]{DS}). For the third point, we simply use the two previous statements and the graded Jacobi identity or refer again to the finite dimensional case (again Cannas da Silva \cite[18.3]{DS}).
\qed

\begin{rk}[Local coordinates]\label{IMPO}

The vector field $X$ is given in local coordinated by $X = \widetilde{X}_i \frac{\partial}{\partial x^i} + \widetilde{X}_{\alpha}^I \partial_{\alpha}^I$ where $X_i$ and $X_{\alpha}^I$ are pro-smooth functions on $J^{\infty} E$. Observe that these coordinates are the dual ones to $dx^i$ and $d u_I^{\alpha}$, but we are interested in finding coordinates dual to $d_H x^i$ and $d_V u_I^{\alpha}$ instead. In that case we can see that $\iota_{\frac{\partial}{\partial x^i}} d_V u_I^{\alpha} = - u_{I,i}^{\alpha}$ which is clearly non-zero. Nevertheless, the dual vector fields to $dx^i$ and $d u_I^{\alpha}$ are $D_i$ and $\partial_{\alpha}^I$ (this is already hinted in equation \ref{ventiuno}): the only non-trivial equation to verify is
$$\iota_{D_i} d_V u_I^{\alpha} = -u_{I,i}^{\alpha} + u_{I,i}^{\alpha} = 0.$$
Using these coordinates, we can write $X = X_i D_i + X_{\alpha}^I \partial_{\alpha}^I$. Now we have to know what is the relation between the local functions $\widetilde{X}_i$ and $\widetilde{X}_{\alpha}^I$ and $X_i$ and $X_{\alpha}^I$. Since the insertion operator is linear over local functions we have:
$$\widetilde{X}_j =  \iota_{\widetilde{X}_i \frac{\partial}{\partial x^i} + \widetilde{X}_{\alpha}^I \partial_{\alpha}^I} dx^j = \iota_X dx^j = \iota_{X_i D_i + X_{\alpha}^I \partial_{\alpha}^I} dx^j = X_j.$$
Now for the vertical components:
$$\widetilde{X}_{\beta}^J = \iota_{\widetilde{X}_i \frac{\partial}{\partial x^i} + \widetilde{X}_{\alpha}^I \partial_{\alpha}^I} d u_J^{\beta} = \iota_X d u_J^{\beta} = \iota_{X_i D_i + X_{\alpha}^I \partial_{\alpha}^I} d u_J^{\beta}  = \sum_{j} X_j u_{J,j}^{\beta} + X_{\beta}^J.$$
Reorganizing these equations we have that
\begin{eqnarray}\label{ventibis}
X_j & = & \widetilde{X}_j \nonumber \\
X_{\alpha}^I & = & \widetilde{X}_{\alpha}^I - \sum_{j} \widetilde{X}_j u_{I,j}^{\alpha}
\end{eqnarray}
These coordinates without tildes will be called {\it variational coordinates} to suggest that they are dual to the variational bicomplex coordinates.
We can use the variational coordinates to, using Equation \ref{ventiuno}, express the action of a pro-smooth vector field on a pro-smooth function: 
\begin{equation}\label{ventidos}
X(f)= \iota_X df = X_i D_i f + X_{\alpha}^I \partial_{\alpha}^I f.
\end{equation}
\end{rk}

We need a different definition of ``preserving the contact ideal'' for a vector field since the usual one fails. Recall that we can represent a vector field in the following way:

\begin{center}
\begin{tikzpicture}[description/.style={fill=white,inner sep=2pt}]
\matrix (m) [matrix of math nodes, row sep=3em,
column sep=3.5em, text height=1.5ex, text depth=0.25ex]
{J^{\infty} E & T (J^{\infty} E)\\
 J^{k(l)} E & T (J^{l} E)\\
 J^{k} E & TE\\};
\path[->,font=\scriptsize]
(m-1-1) edge node[auto] {$X^{\infty}$} (m-1-2)
(m-2-1) edge node[auto] {$X^{l}$} (m-2-2)
(m-3-1) edge node[auto] {$X^{0}$} (m-3-2)
(m-1-1) edge node[auto] {$\pi_{\infty}^{k(l)}$} (m-2-1)
(m-1-2) edge node[auto] {$T \pi_{\infty}^{l}$} (m-2-2)
(m-2-1) edge node[auto] {$\pi_{k(l)}^{k}$} (m-3-1)
(m-2-2) edge node[auto] {$T \pi_{l}^{0}$} (m-3-2);
\end{tikzpicture}
\end{center}

$T (J^{\infty} E)$ is not $J^{\infty} F$ for any other $F$, but it is a different kind of pro-object. Nevertheless, now that we know how the Lie derivative is defined, we can consider vector fields $X$ such that $\mathcal{L}_X \mathtt{C} \subset \mathtt{C}$. This is the correct notion of Cartan-preserving vector field. The property that $X^{\infty}$ can be recovered from $X^0$ for such vector fields still holds. The proof can be found in the books by Anderson \cite{AND} or Olver \cite{OLV}.

\begin{pp}\label{achus} Let $\pi \colon E \rightarrow M$ be a smooth fiber bundle over $M$. If $X \in \mathfrak{X}(J^{\infty} E)$ preserves the contact ideal in the sense that $\mathcal{L}_X \mathtt{C} \subset \mathtt{C}$ then $X$ is the jet prolongation of $X^0 \colon J^{k(0)} E \rightarrow TE$, denoted $\textrm{pr}(X^0)$. Conversely, $X$ is the unique vector field covering $X^0$ and preserving the contact ideal.

In local coordinates, if $X^0 = X_i D_i + X_{\alpha} \partial_{\alpha}$, then $\textrm{pr}X =: X^0 + \sum_{|I|\geqslant 1} \textrm{pr}X_{\alpha}^I \partial_{\alpha}^I$ and the formula for the extra components is:
\begin{equation}\label{pr}
\textrm{pr}X_{\alpha}^I = D_I(X_{\alpha}).
\end{equation}
\end{pp}

Observe that the two equations for a prolongation are very similar. For a holonomic-jet prolongation we had that $(j^{\infty} f^0)_{\alpha}^I = D_I (f^0)_{\alpha}$ (equation \ref{proI}) and for a prolongation of a vector field we have $\textrm{pr}X_{\alpha}^I = D_I(X_{\alpha})$ (equation \ref{pr}). It is important to keep in mind that the meaning if the indices is different in each case. In the prolongation of a map $(j^{\infty} f^0)_{\alpha}^I$ denotes the coordinate in $u_{I}^{\alpha}$, while in the prolongation of a vector field $\textrm{pr}X_{\alpha}^I$ means the component of $\textrm{pr} \, X$ with respect to the change of basis given by equation \ref{ventibis}. If we were to express equation \ref{pr} in the standard coordinates (the dual to $d x^i$ and $d u_{I}^{\alpha}$) we would get the following formula (which is the one Anderson proves \cite{AND}):
\begin{equation}\label{pr2}
\textrm{pr}\widetilde{X}_{\alpha}^I = D_I(\widetilde{X}_{\alpha} - u_i^{\alpha} \widetilde{X}_i) + u_{I, i}^{\alpha} \widetilde{X}_i.
\end{equation}

From our choice of coordinates, it is clear that a vector field preserving the contact ideal splits into two non-interacting parts: 
$$X = \left(X_i D_i \right) + \left(D_I(X_{\alpha}) \partial_{\alpha}^I  \right).$$
The first part is horizontal (usually called total) and the second one vertical (usually called evolutionary). This is an alternative proof to the following fact that can be found in the article of Ibragimov and Robert Anderson \cite[Theorem 4]{IA}. The result also appears in book by Ian Anderson \cite[Proposition 1.20]{AND}, but with weaker hypothesis.

\begin{pp}[Ibragimov and Anderson {\cite[Theorem 4]{IA}}] \label{last}
Every vector field $X \in \mathfrak{X}(J^{\infty} E)$ preserving the contact ideal is decomposable. This means, $X = \textrm{pr} (\xi) + \textrm{tot} (\nu)$ where $\nu = (\pi_{\infty})_* X \defeq  T \pi_{\infty} \circ X$ and $\xi = \left( X - \textrm{tot} (\nu) \right)^0$.
\end{pp}

We want to give explicit definitions of evolutionary and total vector fields for the purpose of future references (we follow the presentation from Anderson \cite{AND}, but Olver does the same \cite{OLV}):

\begin{df}[Evolutionary and total vector fields]\label{evtot} Let $\pi \colon E \rightarrow M$ be a smooth fiber bundle. Let $VE \defeq \ker T \pi \subset T E$ denote the subbundle of vertical vectors in $TE$.

A pro-smooth map $\xi \colon J^{\infty} E \rightarrow VE$ such that $\textrm{pr}_E \circ \xi = \pi_{\infty}^0$ gives rise to a unique contact ideal preserving vector field $\textrm{pr}(\xi) \in \mathfrak{X}(J^{\infty} E)$. All such vector fields are called {\it evolutionary}. 

A pro-smooth map $\nu \colon J^{\infty} E \rightarrow TM$ such that $\textrm{pr}_M \circ \nu = \pi_{\infty}$ defines via the formula $(\pi)_* \nu \defeq  T \pi \circ \nu \colon J^{\infty} E \rightarrow TE$ which gives rise to a unique contact ideal preserving vector field which annihilates all vertical forms $\textrm{tot}(\nu) \in \mathfrak{X}(J^{\infty} E)$. All such vector fields are called {\it total}.

A vector field $X \in \mathfrak{X}(J^{\infty} E)$ is called {\it decomposable} if it is the sum of an evolutionary and a total vector field.
\end{df}

Observe that $(\pi)_* \nu \defeq  T \pi \circ \nu$ has coordinates $\nu_i \frac{\partial}{\partial x^i} + u_i^{\alpha} \nu_i \partial_{\alpha}^i$. That means that with respect to the variational coordinates $(\pi)_* \nu = \nu_i D_i$ alone, since $((\pi)_* \nu)_{\alpha} = u_i^{\alpha} \nu_i - u_i^{\alpha} \nu_i = 0$. As an example $\textrm{tot}(\frac{\partial}{\partial x^j}) = D_j$. For evolutionary vector fields $X_i = 0$ so that $\widetilde{X}_{\alpha}^I = {X}_{\alpha}^I$.\\

We want to finish this section with some remarks about the Lie bracket of contact-ideal preserving vector fields and their behavior with respect to the splitting into vertical and horizontal parts. A coordinate free version of these results can be found in the book by Anderson \cite[Proposition 1.21]{AND}:

\begin{pp}\label{closed} The bracket of two contact-ideal preserving vector fields preserves again the contact ideal. The bracket of two evolutionary ones is again evolutionary and the bracket between two total ones is again total. In local coordinates:
\begin{eqnarray*}
	{[}X, Y{]}_i &=& X_j D_j Y_i + X_{\alpha}^I \partial_{\alpha}^I Y_i - Y_j D_j X_i - Y_{\alpha}^I \partial_{\alpha}^I X_i \\
	{[}X, Y{]}_{\alpha} &=& X_i Y_{\alpha}^i + X_{\beta}^J \partial_{\beta}^J Y_{\alpha} - Y_i X_{\alpha}^i - Y_{\beta}^J \partial_{\beta}^J X_{\alpha} \\
	{[}X, Y{]}_{\alpha}^I &=& D_I {[}X, Y{]}_{\alpha} \\
	&&\\
	{[}\textrm{tot}(X), \textrm{tot}(Y){]}_i &=& X_j D_j Y_i - Y_j D_j X_i \\
	&&\\
	{[}\textrm{ev}(X), \textrm{ev}(Y){]}_{\alpha}  &=&  X_{\beta}^J \partial_{\beta}^J Y_{\alpha} - Y_{\beta}^J \partial_{\beta}^J X_{\alpha} \\
	{[}\textrm{ev}(X), \textrm{ev}(Y){]}_{\alpha}^I &=& D_I {[}\textrm{ev}(X), \textrm{ev}(Y){]}_{\alpha}
\end{eqnarray*}
\end{pp}

We want to point out that the Lie bracket of an evolutionary vector field and a total vector field is not total, contrary to what one could interpret from one of the items in the aforementioned Proposition in Anderson's book. We want to be explicit about this, since it is a general source of confusion:
\begin{eqnarray*}
	{[}\textrm{tot}(X), \textrm{ev}(Y){]}_i &=& - Y_{\alpha}^I \partial_{\alpha}^I X_i \\
	{[}\textrm{tot}(X), \textrm{ev}(Y){]}_{\alpha} &=& X_i Y_{\alpha}^i \\
	{[}\textrm{tot}(X), \textrm{ev}(Y){]}_{\alpha}^I &=& D_I {[}\textrm{tot}(X), \textrm{ev}(Y){]}_{\alpha}
\end{eqnarray*}

\dem Let $X$ and $Y$ be two vector fields preserving the contact ideal. Using Cartan calculus (Proposition \ref{CC}) $$\mathcal{L}_{[X,Y]} \mathtt{C} =[\mathcal{L}_X, \mathcal{L}_Y] \mathtt{C} = \mathcal{L}_X ( \mathcal{L}_Y \mathtt{C}) - \mathcal{L}_Y ( \mathcal{L}_X \mathtt{C}) \subset \mathtt{C}.$$
The computation in local coordinates of the bracket uses simply equation \ref{ventidos} and the Cartan calculus. Finally, we adjust $D_i X_{\alpha}$ by $X_{\alpha}^i$ using the equation for the prolongation from Proposition \ref{achus}.
\qed

At this point Anderson also explains how the insertion of evolutionary vector fields behaves with respect to the Lie algebra from Proposition \ref{CC} (Cartan calculus). At this moment, this is not necessary and we refer the reader to Section \ref{lill} where we discuss the matter for the bicomplex of local forms.

\printbibliography


\setcounter{part}{1}
\setcounter{chapter}{2}


\part{Local maps and smooth maps involving \boldmath{\mbox{$\E \times M$}} }

\newpage
\tableofcontents

\chapter*{\color{darkdelion} Local maps and smooth maps involving \boldmath{\mbox{$\E \times M$}}}

In Lagrangian field theory, the objects are products of the kind $\E \times M$ where $\E$ is the space of smooth sections of a fiber bundle over $M$. There is a category having these as objects: the category of local manifolds. A morphism between two local manifolds is required to cover a pro-smooth map between the associated infinite jet bundles. By pulling back the ind-algebra of functions to the infinite jet bundle we are able to talk about local functions. Local maps respect the space of local functions.\\

The approach to maps in Lagrangian field theory is that they should descend to a map between pro-finite dimensional manifolds. There are other angles to it: both $J^{\infty} E$ and $\E \times M$ can be given topological and even smooth structures. Asking for continuous or smooth maps between the infinite jet bundles instead of pro-finite smooth maps is possible. Yet another possibility: we could work with continuous or smooth maps directly on $\E \times M$ instead of working with local maps.\\

From the topological perspective, $J^{\infty} E$ is a topological space with respect to the limit topology. Pro-finite smooth maps are continuous with respect to that topology. In the same direction,  $\E \times M$ can be topologized in several ways including the compact-open topology or the Whitney $\C$-topology. Jet evaluations are continuous with respect to the Whitney $\C$-topology and open with respect to the compact-open topology. For local maps of easy kinds, such as $f_{\F} \times \textrm{id}_M \colon \E \times M \rightarrow \F \times M$, we can prove that they are also continuous.\\

$J^{\infty} E$ and $\E$ are infinite dimensional manifolds modeled on Fr\'echet spaces. Jet evaluations are always smooth. In general, there is no stronger or weaker notion among smooth maps and local maps. The topologies on $\E$ coming from the Fr\'echet manifold structures are finer than the previous ones, so that some of the topological results still hold in this setting.\\

{\it This explores local, topological and smooth structures on $\E$ and $\E \times M$. It focuses in the comparison between continuous and smooth maps and pro-smooth and local maps. In particular it shows that for some topologies, jet evaluations are open, and hence a larger category of local manifolds can be defined. It also shows that under certain assumptions, local maps are smooth. The references in this part are Blohmann \cite{B}; Hamilton \cite{HAM}; and Kriegl and Michor \cite{KM}.}


\newpage
\chapter{Local maps and functions}\label{ldc}


\section{ Locality}

{\it In this section we define the algebra of local functions on $\E \times M$ and local maps between $\E \times M$ and $\F \times N$. We show that the pullback of a smooth function along a local maps is again local. We give some insight into the theory of local maps, pointing out that the jet evaluations are not surjective and that the next natural step into the theory is to study topological structures on $\E \times M$. We follow the ideas from Deligne and Freed \cite{DEL} and Blohmann \cite{B}.}\\
 
We fix a smooth fiber bundle $\pi \colon E \longrightarrow M$ with space of sections $\E$. In field theory, one views the Lagrangian, the Euler-Lagrange term or the conserved currents as maps on $\E$ and variations of $\E$ valued in forms on $M$. Moreover, it is said that they depend on finitely many derivatives of the field and of the variations. It is then suggested that we do not work with $\Omega^{\bullet}(\E \times M)$, nor with $\Omega^{\bullet}(J^{\infty} E)$, but with the pullback of $\Omega^{\bullet}(J^{\infty} E)$ by $j^{\infty}$. The first, and easiest, step is to take the pullback of a pro-smooth functions:

\begin{df}[Local function]\label{locfun}
Given a fiber bundle $\pi \colon E \longrightarrow M$, the space $\C_{\mathrm{loc}}(\E \times M) \defeq (j^\infty)^* \left(\C(J^{\infty} E) \right) \subset \textrm{Hom}_{\Set}(\E \times M, \RE)$ is called the space of local smooth functions, or simply local functions, on $\E \times M$.
\end{df}

A map $f \colon \E \times M \longrightarrow \RE$ is local if and only if it descends to a finite jet; that is if there exists a $k\in \mathbb{N}$ and a map $f_k \colon J^k E \longrightarrow \mathbb{R}$ such that the following diagram commutes: 
\begin{center}
\begin{tikzpicture}[description/.style={fill=white,inner sep=2pt}]
\matrix (m) [matrix of math nodes, row sep=2em,
column sep=2.5em, text height=1.5ex, text depth=0.25ex]
{\E \times M & \RE \\
  J^k E & \\};
\path[->,font=\scriptsize]
(m-1-1) edge node[auto] {$f$} (m-1-2)
(m-1-1) edge node[auto] {$j^k$} (m-2-1)
(m-2-1) edge node[auto] [swap]{$f_k$} (m-1-2);
\end{tikzpicture}
\end{center}

The concept of locality in field theory has been used extensively in the literature but it has not been investigated much on their own. The previous Definition is inspired in a similar one by Deligne and Freed \cite{DEL}, although they consider directly $M$-twisted local forms (we have decided to treat the $M$-twisted case later since we need some more results in order to prove basic properties about $M$-twisted local functions).

\begin{pp}\label{subal} $\C_{\textrm{loc}}(\E \times M)$ is a sub-algebra of $\textrm{Hom}_{\Set}(\E \times M, \RE)$. 
\end{pp}

\dem The proof relies on the fact that $\C(J^{\infty} E)$ is an ind-algebra. In other words if $f$ and $g$ are local functions, being the pullback of $f^{\infty}$ and $g^{\infty}$ respectively. Then $f+g$, $f \cdot g$ and $t f$ ($t \in \mathbb{R}$) are the pullbacks of $f^{\infty}+g^{\infty}$, $f^{\infty} \cdot g^{\infty}$ and $t f^{\infty}$ respectively. 
The algebra commutation relations follow from those on $\textrm{Hom}_{\Set}(\E \times M, \RE)$. To be precise, we can take representatives $f=(j^k)^*f_k$, $g=(j^l)^*g_l$, $k \geqslant l$ for $k, l \in \mathbb{N}$, $f_k \in \C(J^k E)$ and $g_l \in \C(J^l E)$. Now let $g_k \defeq g_l \circ \pi_k^l$, we have that $g = (j^k)^* g_k$ and $g+f = (j^k)^*(f_k + g_k)$, $g \cdot f = (j^k)^*(f_k \cdot g_k)$ and $t f = (j^k)^*t f_k$.
\qed

We now consider $\pi \colon E \longrightarrow M$ and $\rho \colon F \longrightarrow N$ two smooth fiber bundles over possibly different base manifolds. The associated spaces of smooth sections are denoted by $\E$ and $\F$ respectively. As we have mentioned, the relevant algebras of functions on $\E \times M$ and $\F \times N$ are the local ones (Definition \ref{locfun}). Not every map between $\E \times M$ and $\F \times N$ induces a map from $\C_{\textrm{loc}}(\F \times N)$ to $\C_{\textrm{loc}}(\E \times M)$, but the ones descending to a map between the corresponding infinite jet bundles do. The following definition can be found in the work of Blohmann \cite{B}.

\begin{df}[Local map]\label{locmap} A map $f \colon \E \times M \rightarrow \F \times N$ is local if it descends to a pro-finite smooth map between the associated infinite jet bundles.
\begin{center}
\begin{tikzpicture}[description/.style={fill=white,inner sep=2pt}]
\matrix (m) [matrix of math nodes, row sep=3em,
column sep=2.5em, text height=1.5ex, text depth=0.25ex]
{\E \times M & \F \times N \\
  J^{\infty} E & J^{\infty} F \\};
\path[->,font=\scriptsize]
(m-1-1) edge node[auto] {$f$} (m-1-2)
(m-1-1) edge node[auto] {$j_{E}^{\infty}$} (m-2-1)
(m-1-2) edge node[auto] {$j_{F}^{\infty}$} (m-2-2)
(m-2-1) edge node[auto] [swap]{$f^{\infty}$} (m-2-2);
\end{tikzpicture}
\end{center}
\end{df}

The diagram takes place in the category of sets since $\E$ is only a set so far. The map between the infinite jet bundles is required to be a morphism in the category of pro-finite dimensional smooth manifolds.

\begin{ej}\label{secti} Consider two smooth fiber bundles over the same manifold $E \rightarrow M$ and $F \rightarrow M$. Given a bundle map over the identity $f \colon E \rightarrow F$ we can consider the infinite jet prolongation of such map $j^{\infty} f \colon J^{\infty} E \rightarrow J^{\infty} F$ which is simply given by $[(\varphi, x)] \mapsto [(f \circ \varphi ,x)]$ (it is indeed a Lie-jet prolongation and it is pro-smooth by Proposition \ref{pro}). The induced map $f_* \times \textrm{id}_{M} \colon \E \times M \rightarrow \F \times M$ given by $(\varphi, x) \mapsto (f \circ \varphi, x)$ is clearly local.
\end{ej}

We want to point out that the maps as in the previous example are very special: they are products. In principle, given a local map $f \colon \E \times M \rightarrow \F \times N$ we do not necessarily have a splitting into a product $f_{\F} \times f_N$ but we can only talk about the components $f = (f_{\F}, f_N)$ where $f_{\F} \colon \E \times M \rightarrow \F$ and $f_{N} \colon \E \times M \rightarrow N$. This is also the case when the base manifold is the same. Local maps in which $f_{M=N} = \textrm{pr}_{M}$ are called local maps {\it covering the identity} and local maps in which $f = f_{\F} \times \textrm{id}_M$ are called local maps {\it along the identity}. The maps in the previous example are local maps along the identity. In Part III we will inspect with further detail local maps along the identity and give precise definitions.

Looking back at Definition \ref{locmap}, we can compose two squares and get another square of the same kind. This induces a category structure on local maps between the product spaces. The category structure is simply pulled back from the one on pro-finite smooth manifolds.

\begin{df}[Category of local manifolds]\label{map} We define the category of local manifolds $\lMan$ with objects given by products $\E \times M$, where $\E$ denotes the space of smooth sections of some smooth fiber bundle $E \rightarrow M$. Morphisms are given by local maps in the sense of Definition \ref{locmap}.
\end{df}

\begin{center}
\begin{tikzpicture}[description/.style={fill=white,inner sep=2pt}]
\matrix (m) [matrix of math nodes, row sep=3em,
column sep=2.5em, text height=1.5ex, text depth=0.25ex]
{\E \times M & \F \times N  & \G \times P\\
  J^{\infty} E & J^{\infty} F & J^{\infty} G\\};
\path[->,font=\scriptsize]
(m-1-1) edge node[auto] {$f$} (m-1-2)
(m-1-1) edge node[auto] {$j_{E}^{\infty}$} (m-2-1)
(m-1-2) edge node[auto] {$j_{F}^{\infty}$} (m-2-2)
(m-2-1) edge node[auto] [swap]{$f^{\infty}$} (m-2-2)
(m-1-2) edge node[auto] {$g$} (m-1-3)
(m-1-3) edge node[auto] {$j_{G}^{\infty}$} (m-2-3)
(m-2-2) edge node[auto] [swap]{$g^{\infty}$} (m-2-3);
\end{tikzpicture}
\end{center}

\begin{ej}\label{aa} Given any smooth manifold $M$ we can view it as a bundle over a point $M \rightarrow \{*\}$. Its space of sections $\Gamma(\{*\}, M)$ is isomorphic to $M$ and all the tangent bundles are again isomorphic to $M$: $J^k M \cong M$ for all $k$. Given a smooth map $f \colon M \rightarrow N$ we can view it as a map $M \cong \Gamma(\{*\}, M) \times \{*\} \rightarrow N \cong \Gamma(\{*\}, N) \times \{*\}$ which is trivially local. As a matter of fact, all the local maps $M \rightarrow N$ come in this way:
\begin{center}
\begin{tikzpicture}[description/.style={fill=white,inner sep=2pt}]
\matrix (m) [matrix of math nodes, row sep=3.5em,
column sep=3.5em, text height=1.5ex, text depth=0.25ex]
{M & N \\
 M & N \\};
\path[->,font=\scriptsize]
(m-1-1) edge node[auto] {$f$} (m-1-2)
(m-1-1) edge node[auto] [swap]{$\textrm{id}_M = j^{k(l)}$} (m-2-1)
(m-1-2) edge node[auto] {$\textrm{id}_N = j^l$} (m-2-2)
(m-2-1) edge node[auto] [swap]{$f=f^{k(l)}$} (m-2-2);
\end{tikzpicture}
\end{center}
We have shown that $\Man \rightarrow \lMan$ is a fully faithful functor.
\end{ej}

\begin{lm}\label{lol} Given a smooth fiber bundle $E \rightarrow M$, local functions are in one to one correspondence to local maps $\E \times M \rightarrow \RE$.
\end{lm}

\dem We view $\RE$ as a line bundle over a point as in Example \ref{aa}. Since $J^l \RE \cong \RE$ for all $l$, a local map from $\E \times M$ to $\RE$ is indexed with constant k: $k(l)=k(0)$ for all $l$. It is hence equivalent to a local function:
\begin{center}
\begin{tikzpicture}[description/.style={fill=white,inner sep=2pt}]
\matrix (m) [matrix of math nodes, row sep=3.5em,
column sep=3.5em, text height=1.5ex, text depth=0.25ex]
{\E \times M & \RE \\
 J^k E & \RE \\};
\path[->,font=\scriptsize]
(m-1-1) edge node[auto] {$f$} (m-1-2)
(m-1-1) edge node[auto] [swap]{$j^{k=k(l)}$} (m-2-1)
(m-1-2) edge node[auto] {$\textrm{id}_\RE = j^l$} (m-2-2)
(m-2-1) edge node[auto] [swap]{$f^{k}$} (m-2-2);
\end{tikzpicture}
\end{center}
\qed

Following this point of view, pullbacks of local functions are well defined in the category of local maps:

\begin{dfpp}\label{pull}
Given a local function $g \in \C_{\textrm{loc}}(\F \times N)$ and a local map $f \colon \E \times M \rightarrow \F \times N$, the pullback map: $f^*(g) \defeq g \circ f \in \textrm{Hom}_{\Set}(\E \times M, \RE)$ is again local. 
\end{dfpp}

The proof follows from Lemma \ref{lol} and the fact that $\lMan$ is a category.

\subsection{Jet evaluations are not surjective}

At this point we could follow the flow of definitions and results for the pro-smooth manifold $J^{\infty} E$ and have a verbatim discussion about $\E \times M$. This will include discussing the Cartan distribution, local forms, local vector fields and smooth structures on $\E \times M$. We will not do this in this order this time around.

For instance, we are ready to introduce local analogues of the tangent and the cotangent bundles in $\lMan$ as well as talking about sections of those bundles to get local vector fields, local $1$-forms, and even local higher forms. This amounts simply to give a definition of the tangent motivated to what happens at the level of the infinite jets. We refer to Chapters \ref{lmlvf} and \ref{blf}.

Nevertheless we want to point out that there is a fundamental difference between pro-smooth maps $f^{\infty}  \colon  J^{\infty} E \rightarrow J^{\infty} F$ and the local analogues $f \colon \E \times M \rightarrow \F \times N$: the general lack of surjectivity of $j^k \colon \E \times M \rightarrow J^{\infty} E$. Even if $(f,f^{\infty})$ is a local map, and $f^{\infty}$ preserves the Cartan distribution, it is unrealistic to think that there is a unique map $f^0 \colon J^k E \rightarrow F$ covered by $\E \times M \rightarrow \F \times N$ from which we can recover $f$. There is simply too much space outside of $j^{k}(\E \times M)$ in $J^k E$ in some cases (think of the two extrema: if $\E$ is soft, all jet evaluations are surjective but if $E \rightarrow M$ has no global sections $j^k(\E \times M)= \emptyset$).

What we could do is to try to replace $J^k E$ by $j^k(\E \times M)$, thus avoiding the problem. But observe that $j^{\infty}(\E \times M)$ is in principle not a pro-smooth manifold since we do not know whether or not $j^k(\E \times M)$ is a smooth manifold. As a matter of fact, the jet evaluations are open maps with respect to some topologies on $\E \times M$. Hence, $j^{\infty}(\E \times M)$ is indeed a pro-smooth manifold and we can talk about maps $f$ factoring in the following way:

\begin{center}
\begin{tikzpicture}[description/.style={fill=white,inner sep=2pt}]
\matrix (m) [matrix of math nodes, row sep=3em,
column sep=2.5em, text height=1.5ex, text depth=0.25ex]
{ \E \times M & \F \times M \\
  j^{\infty}(\E \times M)  & j^{\infty}(\F \times M) \\};
\path[->,font=\scriptsize]
(m-1-1) edge node[auto] {$f$} (m-1-2)
(m-1-1) edge node[auto] {$j^{\infty}$} (m-2-1)
(m-2-1) edge node[auto] {$f^{\infty}$} (m-2-2)
(m-1-2) edge node[auto] {$j^{\infty}$} (m-2-2);
\end{tikzpicture}
\end{center}

We should deal with this problem even before we try to develop a theory of Cartan preserving local maps (as it will be done in Chapter \ref{ins}). This brings us to study first the topological structures and then the smooth (Fr\'echet) structures on $\E \times M$. The comparison between local maps and smooth maps in that context is not so simple as in the $J^{\infty} E$ case and it will involve the Cartan-preserving property.


\newpage
\chapter{Topologies on \boldmath{\mbox{$\E \times M$}}}\label{top}

We have identified in the previous section a relevant category in Lagrangian field theory: that of local manifolds. Its objects are given by $\E \times M$ where $\E$ is the space of smooth section of a smooth fiber bundle over $M$; and its morphisms are local maps. Local maps are defined using finite and infinite jet evaluations.\\

$\E \times M$ can be topologized in different ways. From our point of view the interesting topologies are those in which the jet evaluations are continuous or open. Pro-smooth maps are continuous and then, under certain assumptions we can prove that local maps are also continuous with respect to some of those valuable topologies on $\E \times M$. To be precise, that result holds for the Whitney $\C$-topology ($WO^{\infty}$-) and the $CO^{\infty}$-topologies.\\

Among the topologies here studied, we also have the compact-open ($CO$-) topology, which is useful since jet evaluations are open with respect to this topology. The same goes for the Whitney-open ($WO$-) topology. This solves the problem pointed out at the end of the previous section that the lack of surjectivity of the jet evaluations poses to the theory of local manifolds. \\

{\it This chapter introduces several topologies on $\E \times M$ and studies whether the jet evaluations are continuous or open with respect to them. It also shows that local maps along a continuous surjection are continuous with respect to the $CO^{\infty}$- and $WO^{\infty}$-topologies. The main references in this chapter are Kriegl and Michor \cite{KM} and Whitney \cite{WHI}.}


\section{Whitney's extension theorem}

{\it In this section we provide a version in terms of local maps of a very relevant theorem in this for this chapter: Whitney's extension theorem. The main reference is Whitney in  \cite{WHI}.}\\

There is a very relevant result in the literature that we will be using extensively in this chapter: Whitney's extension theorem. In short, it states that given  a smooth function in a compact subset of $\RE^n$, it can be extended to a smooth function in the whole of $\RE^n$.

\begin{tm}[Whitney's extension theorem, Whitney \cite{WHI}]\label{WET}
A smooth function of class $k \in \mathbb{N} \cup \{ \infty \}$ in the sense of Whitney on $K \subset \RE^n$ compact is a collection of continuous functions $\{ f_{I} \colon K \rightarrow \RE \}_{|I| \leqslant k}$ (where $I$ is a multi-index in \{1, \ldots, n\}) such that for every $m \leqslant k$, the remainders
$$R_{I}^m (y, x) \defeq f_I(x) - \sum_{|J| \leqslant m - |I|} \frac{1}{J!} f_{I+J}(y) (x-y)^J$$ defined for all $|I| \leqslant m$ and all $x, y \in K$ satisfy that
$$\lim_{x,y \in K, |x-y| \rightarrow 0} \frac{| R_{I}^m (y, x) |}{ |x-y|^{m-|I|}} = 0.$$

Any smooth function of class $k \in \mathbb{N}$ in the sense of Whitney on $K \subset \RE^n$ can be extended as a $\mathscr{C}^k$-smooth function on $\RE^n$.  This means that there exists a smooth function $f \in \mathscr{C}^k(\RE^n)$ such that for any multi-index $I$, $|I| \leqslant k$ we have that $$\frac{\partial^I f}{\partial x^I}\restrict{\frac{}{}}{_K} = \restrict{f_I}{_K}.$$
\end{tm}

We will apply a version of Whitney's extension theorem in our setting of multi-valued jets: 

\begin{cl}\label{WETc} Let $\pi \colon E \rightarrow M$ be a smooth fiber bundle. Consider a smooth section $\varphi \in \E = \Gamma^{\infty}(M, E)$ and any jet $\chi \in J^{\infty} E$ such that $\pi_{\infty}^0 (\chi)$ lies in a trivializing neighborhood of $\varphi (\pi_{\infty}(\chi))$ in $E$. Then there exists a smooth section $\varphi_{\chi} \in \E$ with the jet $\chi$ at $\pi_{\infty}(\chi)$ and coinciding with $\varphi$ outside a compact neighborhood of $\pi_{\infty}(\chi)$.
\end{cl}

\dem Consider a coordinate system trivializing $\pi$ around $(x, \varphi(x))$ where $x$ denotes $\pi_{\infty}(\chi)$. In the image of the chart for $x$, which can be taken equal to $\RE^n$ and to be centered around $x$, we consider a compact annulus $A$ around $0$ and we call $K \defeq A \cup \{0\}$ which is a compact subset of $\RE^n$. Now, given any direction in the fiber $\RE^e$ (which is trivial using the coordinate system) we can consider the smooth function in the sense of Whitney (Theorem \ref{WET}) given by the Taylor coefficients of $\varphi$ in $A$ and $\chi$ in $0$ (after the trivialization and the restriction to that given direction and using the fact $\chi$ can actually be realized in that neighborhood by assumption). These functions satisfy indeed the conditions since $A$ is disjoint from $\{0\}$ and $\varphi$ is smooth in $A$. Applying Whitney's extension theorem \ref{WET}, we get a smooth function from $\RE^n$ to $\RE^e$ which can be glued to $\varphi$ since they agree on $A$ to form a smooth section $\varphi_{\chi} \in \E$ with $\chi$ as infinite jet at $x$ .
\qed


\section{\boldmath{\mbox{$CO$}}- and \boldmath{\mbox{$WO$}}-topologies}

{\it This section introduces 3 topologies on $\E \times M$ and studies whether the finite jet evaluations are continuous or open with respect to these topologies. For the first one, the jet evaluations are continuous, but only open in the case in which $\E$ is soft. For the other two, the products with the compact-open- and Whitney-open- ($CO$- and $WO$-) topologies on $\E$, the finite jet evaluations are always open. This leads us to the definition of a new category of local maps. As a reference for the topologies on $\E$ we follow Kriegl and Michor \cite{KM}.}\\

We fix a smooth fiber bundle $E \rightarrow M$ with space of smooth sections $\E$. Recall that the main point of this chapter is to compare continuous maps to local maps. The first step is to understand the topology on $J^{\infty} E$ and compare pro-finite smooth maps to continuous maps. $J^{\infty} E$ is a topological space with respect to the projective limit topology, that is, the one that makes each of the maps $\pi_{\infty}^k \colon J^{\infty} E \rightarrow J^{k} E$ continuous. Using that pro-finite smooth maps are smooth from Corollary \ref{chus} we can state the following result:

\begin{lm}\label{pfsac} Pro-finite smooth maps $J^{\infty} E \rightarrow J^{\infty} F$ are continuous with respect to the projective limit topologies on the infinite jet bundles.
\end{lm}

In the other direction, there are clearly continuous maps which are not pro-finite smooth. As an example of such a map consider simply $\RE \cong J^{\infty} \left(\RE \rightarrow \{*\}\right) \rightarrow \RE$ given by a continuous but not smooth function such as the absolute value.

The interesting topologies on $\E \times M$ from our point of view, are those in which $j^{\infty} \colon \E \times M \rightarrow  J^{\infty} E$ is continuous and in some cases open. The easiest way to achieve this is to topologize $\E \times M$ via the initial topology of the infinite jet evaluation, precisely the one making that map continuous.

\begin{df}[$j^{\infty}$-topology]
The initial topology on $\E \times M$ via $j^{\infty} \colon \E \times M \rightarrow J^{\infty} E$ is called the $j^{\infty}$-topology.
\end{df}

Clearly, the infinite jet and the finite jet evaluations are continuous with this topology: $j^k = \pi_{\infty}^k \circ j^{\infty} \colon \E \times M \rightarrow J^{\infty} E \rightarrow J^{k} E$ for each $k \in \mathbb{N}$ since the topology on $J^{\infty} E$ is the initial topology with respect to all the finite jets. We have also wished for these maps to be open under certain assumptions: these would be the space of smooth sections being soft.

\begin{lm}\label{infop} If  $\E$ is soft, $ j^{\infty} \colon \E \times M \rightarrow J^{\infty} E$ is an open map with respect to the $j^{\infty}$-topology.
\end{lm} 

\begin{lm}\label{midop} $\pi_{\infty}^k \colon J^{\infty} E \rightarrow J^k E$ is an open map for each $k \in \mathbb{N}$.
\end{lm}

\begin{cl}\label{kop} If  $\E$ is soft, $ j^{k} \colon \E \times M \rightarrow J^{k} E$ is an open map with respect to the $j^{\infty}$-topology for each $k \in \mathbb{N}$.
\end{cl}

{\bf Proof of Corollary \ref{kop}. } Assuming the two lemmas,
$\pi_{\infty}^{k}$ is open from Lemma \ref{midop} and $j^{\infty}$ is an open map in the case $\E$ is soft from Lemma \ref{infop}. Since the composition of open maps is open, $j^k = \pi_{\infty}^k \circ j^{\infty}$ is open for each $k \in \mathbb{N}$.
\qed

Lemma \ref{infop} is immediate: the fact that $\E$ is soft implies that $j^{\infty}$ is surjective. Any surjective map defining an initial topology is open. Let $f \colon X \rightarrow Y$ be a surjection where $X$ is topologized via the initial topology. All open sets in $X$ are of the kind $f^{-1}(V)$ where $V$ is open in $Y$. $V = f (f^{-1} (V))$ since $f$ is surjective, so that $f$ is open.\\

{\bf Proof of Lemma \ref{midop}. }
We want to show that $\pi_{\infty}^k \colon J^{\infty} E \rightarrow J^k E$ is an open map for each $k \in \mathbb{N}$. We fix such a $k$ and $U \subset J^{\infty} E$ open. We will show that for each $x \in U$ we can find an open neighborhood of $\pi_{\infty}^k (x)$ in $J^k E$ fully contained in $\pi_{\infty}^k (U)$. We fix such an $x$. Since the topology on $J^{\infty} E$ is the initial one with respect to all the finite jets we are assured the existence of $n \in \mathbb{N}$, $k_{i} \in \mathbb{N}$ and $U^{k_i} \subset J^{k_i} E$ open for each $i \in \{1, \ldots, n\}$ such that $$ x \in \left( \bigcap_{i = 1}^n \left(\pi_{\infty}^{k_i}\right)^{-1} (U^{k_i}) \right) \subset U.$$ We will construct $V$ an open neighborhood of $\pi_{\infty}^k (x)$ in $\pi_{\infty}^k (U)$.
\begin{itemize}
	\item {\it Case $n=1$.} Call $l = k_1$.
		\begin{itemize}
			\item If $k \leqslant l$ take $V = \pi_{l}^k (U^l)$, it is open since the map $\pi_l^k$ is a fiber bundle (see for example Saunders \cite{SAU}). $$\pi_{\infty}^k (x) = \pi_{l}^k \circ \pi_{\infty}^l (x) \subset \pi_{l}^k \circ \pi_{\infty}^l \left( \left(\pi_{\infty}^l\right)^{-1} (U^l) \right) = \pi_{l}^k  (U^l) = V.$$ Given $\pi_{l}^l (y)$ with $y \in U^l$ there is $x^{\prime} \in U$ such that $\pi_{\infty}^l (x^{\prime}) = y$ and hence $\pi_{\infty}^k (x^{\prime}) = \pi_l^k (\pi_{\infty}^l (x^{\prime})) = \pi_l^k y$ so that $\pi_{\infty}^k (x) \in V = \pi_l^k (U^l) \subset \pi_{\infty}^k U$.
			\item If $k \geqslant l$ take $V = \left(\pi_{k}^l\right)^{-1} (U^l)$, it is open since $\pi_k^l$ is continuous. Now, $x \in \left(\pi_{\infty}^l \right)^{-1} (U^l) = \left(\pi_{\infty}^k \right)^{-1} \circ \left(\pi_{k}^l \right)^{-1} (U^l) = \left(\pi_{\infty}^k \right)^{-1} (V)$ so that its image $\pi_{\infty}^k(x) \in \pi_{\infty}^k  \left( \pi_{\infty}^k \right)^{-1} (V) = V$ since $\pi_{\infty}^k$ is surjective.  Even more, because $\left(\pi_{\infty}^{k}\right)^{-1}(V) = \left(\pi_{\infty}^{l}\right)^{-1}(U^l) \subset U $ we get that $V \subset \pi_{\infty}^k (U)$. This shows the result, in other words, $\pi_{\infty}^k (x) \in V \subset  \pi_{\infty}^k U$.
		\end{itemize}
	\item {\it Case $n > 1$.} Call $l \defeq \max_{1 \leqslant i \leqslant n}{k_i}$, $U^i \defeq \left( \pi_{l}^{k_i} \right)^{-1} (U^{k_i})$. In this case $$ x \in \left( \bigcap_{i = 1}^n \left(\pi_{\infty}^{k_i}\right)^{-1} (U^{k_i}) \right)  =  \left( \bigcap_{i = 1}^n  \left(\pi_{\infty}^{l}\right)^{-1} (U^i) \right) = \left(\pi_{\infty}^{l}\right)^{-1} \left( \bigcap_{i = 1}^n  U^i \right) \subset U$$ and we are reduced to the previous case where now $U^l = \left( \bigcap_{i = 1}^n  U^i \right)$ is open since all $\pi_{l}^{k_i}$ are continuous and the intersection is finite.
\end{itemize}

Explicitly, we have constructed $V$ open in $J^k E$ such that $\pi_{\infty}^k (x) \in V \subset \pi_{\infty}^k (U)$ showing that $\pi_{\infty}^k$ is open where
\[
 V =
  \begin{cases} 
      \hfill \pi_l^k  \left( \bigcap_{i = 1}^n  \left( \pi_{l}^{k_i} \right)^{-1} (U^{k_i}) \right)   \hfill & \text{ if $k \leqslant l \defeq \max_{1 \leqslant i \leqslant n}{k_i}$} \\
      \hfill \left(\pi_l^k\right)^{-1}  \left( \bigcap_{i = 1}^n  \left( \pi_{l}^{k_i} \right)^{-1} (U^{k_i}) \right)   \hfill & \text{ if $k \geqslant l \defeq \max_{1 \leqslant i \leqslant n}{k_i}$.} \\
  \end{cases}
\]
\qed

The $j^{\infty}$-topology on $\E \times M$ has great advantages as we have seen in the previous results. On the other hand, it is not clear whether or not it is compatible with the projections to the $\E$ and the $M$ factors. On the one hand, $\E \times M \rightarrow M$ is continuous with respect to the $j^{\infty}$-topology since it is the composition of two continuous maps $j^{0}$ and $\pi$; it is even open if $\E$ is soft as a consequence of Corollary \ref{kop}. On the other hand, there are different topologies on $\E$ that can be considered: for example those as subspaces of $\mathscr{C}^{0}(M, E)$. On $\mathscr{C}^{0}(M, E)$ there are two interesting topologies sometimes called weak and strong topologies on the space of smooth maps. The subspace topology on $\E$ with respect to these topologies have some interesting features from our point of view.

\begin{df}[$CO$-topology, following Kriegl and Michor \cite{KM}]
Given two topological spaces $X$ and $Y$, the compact-open topology, weak- or $CO$-topology, on $\mathscr{C}^0(X,Y)$ is given by the sub-base $\{CO(K, V) \colon K \subset X \textrm{ compact, } V \subset Y \textrm{ open}\}$ where $$CO(K, V)\defeq \{f \in \mathscr{C}^0(X,Y) \colon f(K) \subset V\}.$$
The compact-open topology on $\E=\Gamma^{\infty}(M, E)$ is the topology as a subspace of $\mathscr{C}^0(M,E)$. We denote $CO_{\E}(K,V)\defeq CO(K,V) \cap \E$. 
\end{df}

\begin{df}[$WO$-topology, following Kriegl and Michor \cite{KM}]
Given two topological spaces $X$ and $Y$, the Whitney-open topology, strong- or $WO$-topology, on $\mathscr{C}^0(X,Y)$ is given by the base $\{WO(V) \colon V \subset Y \textrm{ open}\}$ where $WO(V)\defeq \{f \in \mathscr{C}^0(X,Y) \colon f(X) \subset V\}$.\\
The Whitney-open topology on $\E=\Gamma^{\infty}(M, E)$ is the topology as a subspace of $\mathscr{C}^0(M,E)$. We denote $WO_{\E}(V)\defeq WO(V) \cap \E$. 
\end{df}

If $X$ is compact, the two topologies agree, but in case it is not, the Whitney-open topology is strictly finer than the compact-open one (this result can be found in the book by Kriegl and Michor, \cite{KM}).

The compact-open topology on $\E$ is not compatible with the $j^{\infty}$-topology on $\E \times M$, to be precise we have the following badly behaved properties:

\begin{pp}\label{noco} Let $E \rightarrow M$ be a smooth fiber bundle with non-empty space of smooth sections $\E$. Then
\begin{enumerate}
\item $\textrm{pr}_{\E} \colon \E \times M \rightarrow \E$ is not continuous for the $j^{\infty}$- and the $CO$-topologies respectively.
\item $\textrm{id} \colon \E \times M \rightarrow \E \times M$ is not continuous for the $j^{\infty}$-topology on the left and the product topology with the $CO$-topology on the right.
\item $\textrm{id} \colon \E \times M \rightarrow \E \times M$ is not open for the $j^{\infty}$-topology on the left and the product topology with the $CO$-topology on the right.
\end{enumerate}
\end{pp}

The results $2$ and $3$ in the above proposition say that the $j^{\infty}$-topology and the product topology on $\E \times M$ with respect to the $CO$-topology on $\E$ are not comparable.

\dem
\begin{enumerate}
\item Consider $K\subset M$ containing at least two points $x$ and $x^{\prime}$ in some trivializing open set and $V \subset E$ not containing a whole connected component of the fiber along $x^{\prime}$. Take $\varphi \in CO_{\E}(K,V)$. Given any $U \subset J^{\infty} E$ open such that $(\varphi, x) \in (j^{\infty})^{-1}(U)$ we can choose $\varphi^{\prime}$ with the same infinite jet at $x$ as $\varphi$, but with arbitrary value at $x^{\prime}$, even outside of $V$ (applying Whitney's extension theorem in our setting, Corollary \ref{WETc}). This shows that $j_x^{\infty} \varphi^{\prime} \in U$ but $(\varphi^{\prime}, x) \notin \textrm{pr}_{\E}(CO_{\E}(K,V))$ since $\varphi^{\prime}(x^{\prime}) \notin V$.
\item We can simply take $CO_{\E}(K,V) \subset \E$ such that $\textrm{pr}_{\E}^{-1}(CO_{\E}(K,V))$ is not open in $\E \times M$ in the $j^{\infty}$-topology and consider the open set $CO_{\E}(K,V) \times M$ in the product topology on the right. It is such that the preimage along the identity is precisely $\textrm{pr}_{\E}^{-1}(CO_{\E}(K,V))$ which is not open in the $j^{\infty}$-topology.
\item Fix a point $e$ in $E$ such that there exists a global section passing through that point. Given any neighborhoods of $e$ and $\pi(e)$, we can find a local section passing through $e$ with arbitrarily high derivatives close enough to $\pi(e)$ taking values in the given neighborhoods because of Whitney's extension theorem in our setting (Corollary \ref{WETc}). This means that any open set in the product topology with the $CO$-topology on $\E \times M$ contains elements with arbitrarily high jets close to any given point. That shows that there are open sets for the $j^{\infty}$-topology which are not open for the product with the $CO$-topology.
\qed
\end{enumerate}

The Whitney-open and the compact-open topology have, on the other side a very interesting property: jet evaluations are open maps with respect to these topologies. In some cases, it will be interesting to work with local maps defined on an open subset $\mathcal{W}$ of $\E \times M$ and to consider $j^{\infty}(\mathcal{W})$ as a pro-finite smooth manifold. In the case in which jet evaluations are open maps, $j^k(\mathcal{W})$ is an open submanifold of $J^k E$ and thus $j^{\infty} (\mathcal{W})$ is indeed a pro-finite dimensional smooth manifold.

\begin{pp}\label{open} Given a smooth fiber bundle $E \rightarrow M$ with smooth space of sections $\E$, the finite jet evaluations $j^k \colon \E \times M \rightarrow J^k E$ are open with respect to the $WO$-topology (and hence with respect to the $CO$-topology).
\end{pp}

\dem Given $V \subset E$ open and $U \subset M$ open, consider $(\varphi, x) \in WO_{\E}(V) \times U$. Consider a trivialization of the bundle around $(x, \varphi(x))$ which is fully contained in $V$ (which is possible since $\varphi(x) \in \varphi(M) \subset V$). In that trivializing neighborhood we can choose an open ball around the image of $x$ which does not fill the image of the chart. Now for any $y$ in the ball, we can apply Whitney's extension theorem in our setting (Corollary \ref{WETc}) to give rise to a smooth section $\varphi^{\prime}$ of $E$ with any given $k$-th jet at $y$. This shows that $j^k$ is an open map.
\qed

Proposition \ref{open} allows us to work with local maps defined on open subsets of $\E \times M$ with respect to the product topology with the Whitney-open topology on $\E$ and hence with respect to the compact-open topology as well.

\begin{dfpp}[Local map on an open subset]\label{locmapex}
Let $E \rightarrow M$ and $F \rightarrow N$ be two smooth fiber bundles with spaces of smooth sections $\E$ and $\F$ endowed with the Whitney-open topology. Let $\mathcal{V}$ and $\mathcal{W}$ be open subsets of $\E \times M$ and $\F \times N$ respectively.
A map $f \colon \mathcal{V} \subset \E \times M \rightarrow \mathcal{W} \subset \F \times N$ is called local if it descends to a pro-finite smooth map  $f^{\infty} \colon j^{\infty} (\mathcal{V}) \rightarrow j^{\infty}(\mathcal{W})$.
\begin{center}
\begin{tikzpicture}[description/.style={fill=white,inner sep=2pt}]
\matrix (m) [matrix of math nodes, row sep=3em,
column sep=2.5em, text height=1.5ex, text depth=0.25ex]
{\mathcal{V} \subset \E \times M & \mathcal{W} \subset \F \times N \\
  j^{\infty}\left( \mathcal{V} \right) & j^{\infty} \left( \mathcal{W} \right) \\};
\path[->,font=\scriptsize]
(m-1-1) edge node[auto] {$f$} (m-1-2)
(m-1-1) edge node[auto] {$j_{E}^{\infty}$} (m-2-1)
(m-1-2) edge node[auto] {$j_{F}^{\infty}$} (m-2-2)
(m-2-1) edge node[auto] [swap]{$f^{\infty}$} (m-2-2);
\end{tikzpicture}
\end{center}
\end{dfpp}


\dem The only necessary thing to check is that $j^{\infty}(\mathcal{V})$ and $j^{\infty}(\mathcal{W})$ are well defined pro-finite dimensional manifolds. This is the case as a consequence of Proposition \ref{open}.
\qed

\begin{df}[Category of extended local manifolds]\label{elMan}
The category of extended local manifolds $\elMan$ has as objects Whitney-open open subsets of $\E \times M$ where $E \rightarrow M$ is a smooth fiber bundle with space of smooth sections $\E$; and local maps in the sense of Definition/Proposition \ref{locmapex} as morphisms.
\end{df}

\begin{rk} Observe that $\lMan$ is a full subcategory of $\elMan$ but not a faithful one unless $\E$ is soft. This is not the end of the discussion about categories modeling local maps: we will consider even a subcategory $\elMan$ later when dealing with maps preserving the Cartan distribution.
\end{rk}


\section{\boldmath{\mbox{$CO^{\infty}$}}- and \boldmath{\mbox{$WO^{\infty}$}}-topologies on $\E$}

{\it In this section we introduce the $CO^{\infty}$ and $WO^{\infty}$-topologies on $\E$. Local maps are continuous with respect to these topologies under certain hypothesis. The extra assumption is that the map is a product where the second component is a continuous surjection, namely $f = f_{\F} \times f_ N \colon \E \times M \rightarrow \F \times N$ and $f_N$ is continuous and surjective. The main reference keeps being the book of Kriegl and Michor \cite{KM}. }\\

The $j^{\infty}$-topology on $\E \times M$,  the compact Whitney-open and the compact-open topologies on $\E$ are convenient to work with for our purposes as we have seen through some results in previous section. On the other hand, they are not well behaved with respect to the projection to the $\E$ factor (Proposition \ref{noco}).

Another point of view, the one followed in Kriegl and Michor \cite{KM}, is that of considering initial topologies on $\E$ with respect to the infinite jet prolongation:
\begin{center}
    \begin{tabular}{rcl}
    $J^{\infty} \colon \E$ & $\longrightarrow$ & $\mathscr{C}^0(M, J^{\infty}  E) $ \\
    $\varphi$ & $\longmapsto$ & $j^{\infty} \varphi,$ \\
    \end{tabular}
\end{center}
where $\mathscr{C}^0(M, J^{\infty} E)$ is endowed with some topology.

\begin{df}[$CO^{\infty}$- and $WO^{\infty}$-topologies] The initial topologies on $\E$ with respect to $J^{\infty}$ coming from the compact-open- and the Whitney-open-topologies on $\mathscr{C}^0(M, J^{\infty} E)$ are called the $CO^{\infty}$- and the $WO^{\infty}$-topologies respectively.
\end{df}

The topologies such that the evaluation $\textrm{ev} \colon \mathscr{C}^0(M, J^{\infty} E) \times M \rightarrow J^{\infty} E$ is continuous are interesting for us. In that way $j^{\infty} = \textrm{ev} \circ (J^{\infty} \times \textrm{id}_M) \colon \E \times M \rightarrow J^{\infty} E$ would be continuous with respect to that new topology. 

\begin{pp}\label{evcon}
Let $X$ and $Y$ be two topological spaces. If $X$ is Hausdorff and locally compact, then $\textrm{ev} \colon \mathscr{C}^0(X, Y) \times X \rightarrow Y$ is continuous with respect to the $CO$-topology and the $WO$-topologies on $\mathscr{C}^{0}(X, Y)$.
\end{pp}

\begin{cl}\label{ary}
Let $E \rightarrow M$ be a smooth fiber bundle with space of sections $\E$. Then
\begin{enumerate}
	\item $j^k \colon \E \times M \rightarrow J^k E$ is continuous for each $k \in \mathbb{N} \cup \{\infty\}$.
	\item $\textrm{pr}_{\E} \colon \E \times M \rightarrow \E$ and $\textrm{pr}_{M} \colon \E \times M \rightarrow M$ are continuous and open maps.
\end{enumerate}
Both statements hold with respect to either the $WO^{\infty}$- or the $CO^{\infty}$-topologies.
\end{cl}

{\bf Proof of Proposition \ref{evcon}. }
Take $V \subset Y$ open and consider a pair $(f, x) \in \textrm{ev}^{-1}(V) \subset \mathscr{C}^{0}(X, Y) \times X$. Since $f$ is continuous, $f^{-1}(V)$ is an open neighborhood of $x$ in $X$ (observe that $f(x)=\textrm{ev}(f,x) \in V$). Since $X$ is Hausdorff and locally compact, there is an open neighborhood $U$ of $x$ in $X$ with compact closure $\overline{U}$ fully contained in $f^{-1}(V)$. That means that $\overline{U} \subset f^{-1}(V)$ and thus $f(\overline{U}) \subset V$, and further $f \in CO(\overline{U}, V)$. We claim that the open neighborhood $CO(\overline{U}, V) \times U$  of $(f, x)$ is fully contained in $\textrm{ev}^{-1}(V)$. Given $(f^{\prime},x^{\prime}) \in CO(\overline{U}, V) \times U$, $x^{\prime} \in U \subset \overline{U}$ holds, so that $\textrm{ev}(f^{\prime}, x^{\prime}) = f^{\prime} (x^{\prime}) \in f^{\prime} (\overline{U}) \subset V$.
This shows $\textrm{ev}$ is continuous with respect to the $CO$-topology, but since the $WO$-topology is finer, it also shows that it is continuous with respect to that other topology.
\qed

{\bf Proof of Corollary \ref{ary}. }
The second statement is immediate since we are working with the product topology on $\E \times M$. For the first statement, we only need to show it for $k = \infty$ since all the other cases follow from it ($j^k = \pi_{\infty}^k \circ j^{\infty}$ would be the composition of two continuous maps). But, following the argument above, $j^{\infty} = \textrm{ev} \circ \left( J^{\infty} \times \textrm{id}_M \right)$. The first map is continuous since the $CO^{\infty}$ and $WO^{\infty}$ are the initial topologies with respect to $J^{\infty}$. The evaluation map is continuous by Lemma \ref{evcon} since $M$ is Hausdorff and locally compact (it is a smooth manifold).
\qed

Observe that the fact that $j^{\infty} \colon \E \times M \rightarrow J^{\infty} E$ is continuous amounts to say that the $j^{\infty}$-topology is weaker than the product topology on $\E \times M$ with respect to the $CO^{\infty}$- or the $WO^{\infty}$-topologies on $\E$.  Actually the topology is strictly weaker as we will see in the next result. This means that, by working with the $CO^{\infty}$- and the $WO^{\infty}$- topologies, we might be loosing $j^k$ being open in the soft sheaf case.

\begin{pp} The $j^{\infty}$-topology is strictly weaker than the product topology on $\E \times M$ with the $CO^{\infty}$- or the $WO^{\infty}$-topologies on $\E$. 
\end{pp}

\dem The proof is very similar to the one of Proposition \ref{noco}, we need to replace the open subset in $E$ with an open subset in $J^{\infty} E$ not containing all the connected component of the fiber at $x^{\prime}$ and apply Whitney's extension theorem in our setting (Corollary \ref{WETc}). That shows that there are open sets in the product with the $CO^{\infty}$-topology (and hence on the one with the $WO^{\infty}$-topology) which are not open in the $j^{\infty}$-topology.
\qed

The $CO^{\infty}$- and the $WO^{\infty}$-topologies are the convenient ones to compare local maps and continuous maps. We fix two smooth fiber bundles $\pi \colon E \rightarrow M$ and $\rho \colon F \rightarrow N$ with corresponding spaces of smooth sections $\E$ and $\F$. We are going to relate local maps and continuous maps under certain assumptions. These assumptions will be satisfied in some cases in our future study, but not in all of them. There are weaker assumptions in which locality implies continuity, but we will explore them with more detail when working on the smooth structure on $\E$. In that case we will have a different result stating local implies smooth, and thus continuous.

\begin{tm}\label{maincon} Let $f=f_{\F} \times f_ N \colon \E \times M \rightarrow \F \times N$ be a local map such that $f_N$ is continuous and surjective. Then $f_{\F} \times f_N$ is continuous with respect to the $WO^{\infty}$-topologies on $\E$ and $\F$.
\end{tm}

\dem Consider $\mathcal{W} \subset \F \times N$ open and $(\varphi, x) \in f^{-1}(\mathcal{W}) \subset \E \times M$. We are going to construct an open neighborhood of $(\varphi, x)$ with respect to the $WO^{\infty}$-topology, fully contained in $f^{-1}(\mathcal{W})$. There exists $V$ open in $J^{\infty} F$ and $U$ open in $N$ such that $(f_{\F}(\varphi), f_{N}(x)) \in (J^{\infty})^{-1}\left(WO(V)\right) \times U \subset \mathcal{W}$. Consider $f^{\infty} \colon J^{\infty} E \rightarrow J^{\infty} F$ such that $f$ descends to. By Lemma \ref{pfsac} we know that $f^{\infty}$ is continuous and hence $(f^{\infty})^{-1}(V)$ is open in $J^{\infty} E$. We claim that $\mathcal{V} \defeq (J^{\infty})^{-1}\left(WO\left( (f^{\infty})^{-1}(V) \right) \right) \times f_N^{-1}(U)$ is the desired open neighborhood. Using the previous arguments and the fact that $f_N$ is continuous it follows that $\mathcal{V}$ is indeed open.

First observe that $\psi \in (J^{\infty})^{-1}  \left(WO\left( (f^{\infty})^{-1}(V) \right) \right) $ if and only if $j_{f_N (x^{\prime})}^{\infty} f_{\F}(\psi) \in V$ for every $x^{\prime} \in M$. This follows from locality:  $\psi \in (J^{\infty})^{-1} \left(WO\left( (f^{\infty})^{-1}(V) \right) \right)$ if and only if for each $x^{\prime} \in M$, $J^{\infty} \psi (x^{\prime}) = j_{x^{\prime}}^{\infty} \psi \in (f^{\infty})^{-1}(V)$ , and now

$$j_{x^{\prime}}^{\infty} \psi \in (f^{\infty})^{-1}(V) \Longleftrightarrow f^{\infty} (j_{x^{\prime}}^{\infty} \psi) \in V \Longleftrightarrow j_{f_N(x^{\prime})}^{\infty} f_{\F}(\psi) \in V .$$

Since $f_{\F} (\varphi) \in (J^{\infty})^{-1}(WO(V))$ we conclude that $\varphi \in (J^{\infty})^{-1} \left(WO\left( (f^{\infty})^{-1}(V) \right) \right) $ and it is clear that $x \in f_N^{-1}(U)$. So, indeed $\mathcal{V}$ is a neighborhood of $(\varphi, x)$. It remains to show that it is in the preimage of $\mathcal{W}$.

$$ \hspace{-5em} (J^{\infty})^{-1}\left(WO\left( (f^{\infty})^{-1}(V) \right) \right) \times f_N^{-1}(U) = \mathcal{V} \subset f^{-1}\left( \mathcal{W} \right) \Longleftrightarrow f \left( \mathcal{V} \right) \subset \mathcal{W} \Longleftrightarrow $$
\vspace{-2.5ex}
$$\hspace{9em}  \Longleftrightarrow f_{\F}\left( (J^{\infty})^{-1} \left(WO\left( (f^{\infty})^{-1}(V) \right) \right) \right) \times f_N (f_N^{-1} (U)) \subset \mathcal{W}.$$

From one side it is true that $f_N \circ f_N^{-1} (U) \subset U$ and from the other side, using the description above for elements of $(J^{\infty})^{-1}(WO((f^{\infty})^{-1})(V))$ we see that since $\psi$ is in $(J^{\infty})^{-1}(WO( (f^{\infty})^{-1}(V) ))$, we get that $j_{N}^{\infty} f_{\F}(\psi) \subset V$ using the surjectivity of $f_N$. This shows that $f$ is a local map.
\qed

We want to emphasize that the previous proof does not intrinsically depend on any features of the infinite jet bundles. The same arguments pulls through when the situation is as follows: we consider always $WO$-topologies on the space of smooth functions, $\E$ is topologized via $G_E^{-1}$ where $g_E \colon \E \times M \rightarrow X$ and $G_E \colon \E \rightarrow \mathscr{C}^{0}(M, X)$ maps $\varphi$ to the map sending $x \in M$ to $g_E(x, \varphi)$, $\F$ is topologized similarly for some map $g_F \colon \F \times N \rightarrow Y$ and $f$ descends to a continuous map $\widehat{f} \colon X \rightarrow Y$ commuting with $g_E$ and $g_F$. In that case $f$ will also be continuous.

Back into our setting, relevant choices for $g_E$ and $g_F$ are finite jet bundles evaluations such as $j_E^k$ and $j_F^l$. In that case, the topology on $\E$ induced by the $WO$-topology on $\mathscr{C}^0(M, J^k E)$ will be called the {\it $WO^k$-topology}. (The same goes for the {\it $CO^k$-topology} on $\E$.) Observe that the $WO^{\infty}$-topology is finer that any $WO^k$-topologies on $\E$. 

\begin{cl} Let $f=f_{\F} \times f_ N \colon \E \times M \rightarrow \F \times N$ be a map such that $f_N$ is continuous and surjective. If $f$ descends to a smooth map $f^l \colon J^k E \rightarrow J^l F$ then $f_{\F} \times f_N$ is continuous with respect to the $WO^{k}$-topology on $\E$ and the $WO^{l}$-topology on $\F$. It is also continuous with respect to the $WO^{\infty}$-topology on $\E$ and the $WO^{l}$-topology on $\F$.
\end{cl}

\begin{rk}\label{loconco}
In the case in which $M$ is compact, the $WO$-topologies and the $CO$-topologies agree on the spaces $\mathscr{C}^0(M, Y)$ for any topological space $Y$. In particular, under the assumption $M$ compact, local maps $f_{\F} \times f_N$ --with $f_N$ a continuous surjection-- are continuous with respect to the $CO^{\infty}$-topology.
\end{rk}

We want to include a result that can be found in the book of Kriegl and Michor \cite{KM} which is a particular case of Theorem \ref{maincon} due to Example \ref{secti}:

\begin{pp}[Kriegl and Michor \cite{KM}] Let $E, F \rightarrow M$ be two smooth fiber bundles over the same manifold. If $g \colon E \rightarrow F$ is a smooth bundle morphism (over the identity), then the push-forward $g_* \colon \E \rightarrow \F$ is continuous with respect to the $WO^{\infty}$-topologies.
\end{pp} 

We want to finish this section, and hence this chapter, with some concluding remarks about the different topologies on $\E$:

\begin{rk} We will explore in the following chapters, smooth structures on $\E$. The topologies arising from those structures will be finer than the $WO^{\infty}$-topology on $\E$. We will be able to show that under certain assumptions, more general than the ones in Theorem \ref{maincon}, that locality implies smoothness and thus continuity.
\end{rk}
\begin{rk} It is possible to explore the relations between the different topologies in $\E$ in a much more detailed way. That can be found certainly in the literature, showing the following chain of topologies
$$CO^{k-1} \hspace{-0.15ex} \subset \hspace{-0.15ex} CO^k \hspace{-0.15ex} \subset \hspace{-0.15ex} WO^{k} \hspace{-0.15ex} \subset \hspace{-0.15ex} WO^{k+1} \hspace{-0.15ex} \subset \hspace{-0.15ex} \lim_{k \in \mathbb{N}}{WO^k} \hspace{-0.15ex} \subset \hspace{-0.15ex} \E \rightarrow \lim_{k \in \mathbb{N}}{(\mathscr{C}^0(M, J^k E), WO)} \hspace{-0.15ex} \subset \hspace{-0.15ex} WO^{\infty}$$
In the case in which $M$ is compact all the limits agree and so do the $WO^k$- and $CO^k$-topologies for finite and infinite $k$.


\end{rk}


\newpage
\chapter{Smoothness and locality on \boldmath{\mbox{$\E \times M$}}}

The space of compactly supported smooth sections of a smooth vector bundle is a Fr\'echet space. For a general smooth fiber bundle, the space of smooth sections (not only the compactly supported ones, but all smooth sections) is a Fr\'echet manifold modeled in Fr\'echet spaces of the previous kind with local transition maps.\\

The locally convex topology agrees with the $WO^{\infty}$-topology in the case of a compact base. The topologies on $\E$ coming from the Fr\'echet manifold structure in a general fiber bundle are finer than the $WO^{\infty}$-topology.\\

Smooth maps involving the space of sections can be characterized in terms of smooth maps involving only finite dimensional manifolds. The infinite jet evaluations are smooth and hence continuous with respect to the locally convex topology.\\

The comparison between local maps and smooth maps cannot be done in full generality, since none of the two notions is weaker than the other. On the other hand there is a stronger notion of locality, insularity, for which insular maps  are smooth. \\

{\it This chapter collects different Fr\'echet smooth structures on the space of smooth sections of a smooth fiber bundle. It gives conditions under which maps involving such spaces are smooth and, in particular, includes the proof that jet evaluations are smooth. The chapter also addresses the topic of comparing local maps and smooth maps, stating that insular maps are smooth. The references in this chapter are Hamilton \cite{HAM} and Kriegl and Michor \cite{KM}.}


\section{Smooth structures on \boldmath{\mbox{$\E$}}}

{\it This section is a review about Fr\'echet manifold structures on $\E$. It is explained in full detail working from the easiest case of a vector bundle over a compact manifold to the general case of a fiber bundle over a non-compact base. It includes remarks about the comparison between the Fr\'echet manifold topology and the topologies investigated in the previous chapter. Besides reviewing the main references, we prove that the charts defining the  Fr\'echet manifold structure on $\E$ are local. The main references in this chapter are Hamilton \cite{HAM} and Kriegl and Michor \cite{KM}.}\\

The space of smooth sections of a smooth vector bundle $V \rightarrow M$ is a vector space and hence we can ask ourselves whether or not it can be given the structure of a Fr\'echet space. In the case of a general smooth fiber bundle $E \rightarrow M$, the strategy is to make use of $TE$ to give local charts and induce a Fr\'echet manifold structure on $\E$.

As in the study about the topological structure on $\E$, the case in which the base manifold is compact is easier, since many of the possible Fr\'echet structures on $\E$ agree. The results about the Fr\'echet manifold structures on spaces of smooth sections of a smooth bundle over a compact base is original to Hamilton \cite{HAM} and it studied in full generality by Kriegl and Michor \cite{KM}:

\begin{dfpp}\label{eman} Let $V \rightarrow M$ be a smooth fiber bundle. Let $\{K_{p}\}_{p \in \mathbb{N}}$ be an exhaustion by compact sets of $M$, (it can be taken to be countable). For each $p$, let $\{U^p_a \}_{a \in A^p}$ be a finite atlas of $K_p$ such that the closure of each open set is inside some trivializing neighborhood for the bundle. Consider a Riemannian metric on $M$. Then, the space of compactly supported sections of the bundle, $\mathcal{V}_c$ is a Fr\'echet space with seminorms $\{|\cdot |_{p,q}\}_{p, q \in \mathbb{N}}$ defined by:
$$ | \varphi |_{p,q} \defeq \sum_{a \in A^p} \,
\max_{   \begin{smallmatrix}
 x \in \overbar{U_{a}^p} \\ 
 \alpha \in \{ 1, \ldots, \textrm{rank}(V) \}
 \end{smallmatrix} } { \, \left( \sum_{i=0}^q \left| D^i \varphi_a^{\alpha} (x) \right| \right)} \, \textrm{ for all } \varphi \in \V_c,$$
where $\varphi_a^{\alpha}$ is the  $\alpha$-th component of the trivialization of $\varphi$ using the charts indexed by $a$ and $D^i$ denotes the iterative total derivative of order $i$. The structure does not depend on the compact exhaustion chosen nor on the Riemannian structure.
\end{dfpp}

For the proof, we refer to the work of Kriegl and Michor \cite[30.4]{KM}. The only subtlety is to observe that the topology is independent of the compact exhaustion chosen, otherwise the proof follows from the aforementioned result.\\

{\bf Proof of the independence of the compact exhaustion.} Consider two compact exhaustions of $M$, $\{K_{p}\}_{p \in \mathbb{N}}$ and $\{K^{\prime}_{p}\}_{p \in \mathbb{N}}$. Given any $p \in \mathbb{N}$ there exists $p^{\prime} \in \mathbb{N}$ such that $K_p \subset K_{p^{\prime}}$  so that if for some $\varepsilon > 0$ and some $\varphi, \psi \in \V$ we have that $| \varphi - \psi|_{p,k^{\prime}}^{\prime} < \epsilon$ then $| \varphi - \psi|_{p,n}^{} < \epsilon$ and also holds. Hence $V_{\varepsilon}^{\{(p,q^{\prime})\}} \subset U_{\varepsilon}^{\{(p,q)\}}$ showing that the topology indexed by the $K^{\prime}$'s is finer that the one indexed by the $K$'s. The same argument exchanging the exhaustions shows the independence on the exhaustion chosen.
\qed

\begin{rk}\label{km1} Kriegl and Michor \cite{KM} consider the more general case in which the fiber is not finite dimensional. In that case, the space of smooth sections over a compact base is not a locally convex space. They have to pass to the bornological topology (the finest locally convex topology with the same bounded sets as the topology of the space of sections).

In the case in which the fiber is finite dimensional that problem is no longer there (see the original result by Hamilton \cite{HAM}). They construct a convenient vector space structure on the space of compactly supported smooth sections that is the one here presented for our particular case. The only difference being that there is no need for bornologification.

On top of that, they consider base spaces which might not be $\sigma$-compact (as it is the case of smooth manifolds), since they consider manifolds modeled in more general vector spaces. This has as a consequence that the resulting space is actually Fr\'echet.
\end{rk}

Still, this does not equip the total space of sections with a Fr\'echet structure. The original result for the compact case, which replaces compactly supported sections with sections, is due to Hamilton \cite{HAM}. We mention it here as a corollary of the previous proposition. For a more modern approach of the proof of this corollary than the one by Hamilton we refer to Sharko \cite{SHA}.

\begin{cl}[Hamilton, {\cite[I. 1.1.5]{HAM}}]\label{eman4} Let $V \rightarrow K$ be a smooth fiber bundle. Let $\{U_{a}\}_{a \in A}$ be a finite atlas of $K$ such that the closure of each open set is inside some trivializing neighborhood for the bundle. Consider a Riemannian metric on $K$. The space of sections $\mathcal{V}$ is a Fr\'echet space with seminorms $\{|\cdot |_{q}\}_{q \in \mathbb{N}}$given by:
$$ | \varphi |_{q} \defeq \sum_{a \in A} \,
\max_{   \begin{smallmatrix}
	x \in \overbar{U_{a}} \\ 
	\alpha \in \{ 1, \ldots, \textrm{rank}(V) \}
	\end{smallmatrix} } { \, \left( \sum_{i=0}^q \left| D^i \varphi_a^{\alpha} (x) \right| \right)} \, \textrm{ for all } \varphi \in \V,$$
The structure does not depend on the Riemannian metric.
\end{cl}

Now we are ready to understand the Fr\'echet manifold structure on spaces of smooth sections. For general fiber bundles, the space of sections is no longer a vector space, hence it cannot be given the structure of a Fr\'echet space. It can, nevertheless, be given the structure of a Fr\'echet {\it manifold}. For simplicity in the argument, consider $E \rightarrow K$ a smooth fiber bundle over a compact manifold. Fixing a Riemannian metric on $E$ we can choose $e = (\textrm{exp}, \textrm{pr}_E) \colon TE \rightarrow E \times E$ a local diffeomorphism around the zero section of $TE$ and the diagonal on $E \times E$. Denote it by $e \colon V \stackrel{\cong}{\rightarrow} U$. Given any section $\varphi \in \E$ we consider the pullback bundle $\varphi^* TE$ over $K$. Define the following two sets:
$$ U_{\varphi} \defeq \{\psi \in \E \colon (\varphi, \psi)(K) \subset U \} \textrm{ and}$$
$$ V_{\varphi} \defeq \{\gamma \in \Gamma^{\infty}(K,\varphi^* TE) \colon \textrm{pr}_{TE} \circ \gamma (K) \subset V \}.$$

\begin{center}
\begin{tikzpicture}[description/.style={fill=white,inner sep=2pt}]
\matrix (m) [matrix of math nodes, row sep=2.9em,
column sep=2.9em, text height=1.5ex, text depth=0.25ex]
{ K & & \\
  & \varphi^* TE & TE \\
  & K & E \\};
\path[->,font=\scriptsize, color=darkdelion]
(m-1-1) edge [bend left=30] node[auto] {$\textrm{pr}_{TE} \circ \gamma$} (m-2-3);
\path[->,font=\scriptsize]
(m-2-2) edge node[auto] {$\textrm{pr}_M$} (m-3-2)
(m-1-1) edge [bend right=30] node[right=1mm] {$\textrm{id}_K$} (m-3-2)
(m-2-2) edge node[auto] {$\textrm{pr}_{TE}$} (m-2-3)
(m-3-2) edge node[auto] {$\varphi$} (m-3-3)
(m-2-3) edge node[auto] {$\textrm{pr}_E$} (m-3-3)
(m-1-1) edge[dashed] node[auto] {$\gamma$} (m-2-2);
\begin{scope}[shift=($(m-2-2)!.5!(m-3-3)$)]
\draw +(-.3,0) -- +(0,0)  -- +(0,.3);
\fill +(-.15,.15) circle (.05);
\end{scope}
\end{tikzpicture}
\end{center}

The two sets are open with respect to the $WO$-topologies on $\E$ and on $\Gamma^{\infty}(K,\varphi^* TE)$ respectively. Moreover, since the $WO$-topology is weaker than the $WO^{\infty}$-topology on $\Gamma^{\infty}(K,\varphi^* TE)$, $V_{\varphi}$ is also open with respect to the locally convex topology as a consequence of Proposition \ref{iguales}. The map 
\begin{eqnarray}\label{aefi}
    A_{\varphi}^e \colon U_{\varphi} & \longrightarrow & V_{\varphi} \nonumber \\
    \psi & \longmapsto & \left[ x \mapsto \left(x, e\left(\varphi(x), \psi(x) \right) \right) \right].
\end{eqnarray}
happens to be a homeomorphism (as we will see in Lemma \ref{tranloc}), and it is used to endow $\E$ with the structure of a Fr\'echet manifold.

\begin{pp}[Hamilton, {\cite[I. 4.1.2]{HAM}}]\label{eman2} Let $E \rightarrow K$ be a smooth fiber bundle over a compact manifold. The smooth space of sections $\E$ is a Fr\'echet manifold modeled on $\Gamma^{\infty}(K,\varphi^* TE)$ for every $\varphi \in \E$ with charts given by $A_{\varphi}^e$ as in equation \ref{aefi}. The structure is independent of the Riemannian metric chosen for $E$.
\end{pp} 

We refer to Hamilton \cite{HAM} or to the more general version by Kriegl and Michor \cite{KM} for the details, especially concerning the smoothness of the transition functions and the independence on the Riemannian metric. Nevertheless, we want to say something about the charts: $A_{\varphi}^e \times \textrm{id}_K$ are not only homeomorphisms, but also local maps. This is very interesting from our point of view.

\begin{lm}\label{tranloc} Let $\pi \colon E \rightarrow K$ be a smooth fiber bundle over a compact manifold. The charts $A_{\varphi}^e \times \textrm{id}_K$ as in equation \ref{aefi} are local and homeomorphisms.
\end{lm}

\dem $A_{\varphi}^e \times \textrm{id}_K$ descends to a map between the $0$-jet spaces of both bundles:
\begin{center}
\begin{tikzpicture}[description/.style={fill=white,inner sep=2pt}]
\matrix (m) [matrix of math nodes, row sep=3em,
column sep=3em, text height=1.5ex, text depth=0.25ex]
{ U_{\varphi} \times K & & V_{\varphi} \times K \\
E & E \times E & \varphi^* TE \\};
\path[->,font=\scriptsize]
(m-1-1) edge node[auto] {$A_{\varphi}^e \times \textrm{id}_K$} (m-1-3)
(m-1-3) edge node[auto] {$j^0$} (m-2-3)
(m-1-1) edge node[auto] {$j^0$} (m-2-1)
(m-2-1) edge node[auto] {$(\textrm{inc}, \varphi \circ \pi)$} (m-2-2)
(m-2-2) edge node[auto] {$e$} (m-2-3);
\end{tikzpicture}
\end{center}

Observe that the map indeed covers the identity on $K$ since on the lower row we have the composition of $e$ (a bundle map) with $(\textrm{inc}, \varphi \circ \pi)$ which is also a bundle map over the identity.

We can use the prolongations of $e \circ (\textrm{inc}, \varphi \circ \pi)$
as in Definition \ref{PRO1} to construct a pro-finite smooth map between the infinite jet bundles that $A_{\varphi}^e \times \textrm{id}_K$ descends to, showing that it is a local map:

\begin{center}
\begin{tikzpicture}[description/.style={fill=white,inner sep=2pt}]
\matrix (m) [matrix of math nodes, row sep=3em,
column sep=3em, text height=1.5ex, text depth=0.25ex]
{ U_{\varphi} \times K & & V_{\varphi} \times K \\
J^{\infty} E & & J^{\infty} (\varphi^* TE) \\
E & E \times E & \varphi^* TE \\};
\path[->,font=\scriptsize]
(m-1-1) edge node[auto] {$A_{\varphi}^e \times \textrm{id}_K$} (m-1-3)
(m-2-1) edge node[auto] {$j^{\infty} (e \circ (\textrm{inc}, \varphi \circ \pi))$} (m-2-3)
(m-1-3) edge node[auto] {$j^{\infty}$} (m-2-3)
(m-1-1) edge node[auto] {$j^{\infty}$} (m-2-1)
(m-2-3) edge node[auto] {} (m-3-3)
(m-2-1) edge node[auto] {} (m-3-1)
(m-3-1) edge node[auto] {$(\textrm{inc}, \varphi \circ \pi)$} (m-3-2)
(m-3-2) edge node[auto] {$e$} (m-3-3);
\end{tikzpicture}
\end{center}

Since the map is local, it is continuous by Theorem \ref{maincon}. The same arguments apply to show that $(A_{\varphi}^e)^{-1} \times \textrm{id}_K$ is local and continuous. This only shows that the charts are homeomorphisms with respect to the $WO^{\infty}$-topologies, but as we will see in Proposition \ref{iguales} those topologies agree with the locally convex ones, hence proving the result.
\qed

In the general fiber case over a non-compact base, it is possible to use the same technique as in Definition/Proposition \ref{eman} to define a smooth manifold structure on $\E$. The first step is to refine the $WO^{\infty}$-topology on $\E$ to include for all $\varphi \in \E$ the set:
$$ U_{c,\varphi} \defeq \{\psi \in \E \colon (\varphi, \psi)(M) \subset U \textrm{ and } \restrict{\varphi}{M \smallsetminus K} = \restrict{\psi}{M \smallsetminus K} \textrm{ for } K \subset M \textrm{ compact}\}.$$

Fix a Riemannian metric on $E$ and as before, and let $e \colon V \stackrel{\cong}{\rightarrow} U$ be the local diffeomorphism around the zero section of $TE$ and the diagonal on $E \times E$. Given any section $\varphi \in \E$ we consider the pullback bundle $\varphi^* TE$ over $M$. The charts $A_{\varphi}^e$ are as in equation \ref{aefi}, the only difference from before is that we have to take compactly supported sections instead of smooth sections in the pullback bundle. $U_{c,\varphi}$ is then homeomorphic to:
$$ V_{c, \varphi} \defeq \{\gamma \in \Gamma_c^{\infty}(M,\varphi^* TE) \colon \textrm{pr}_{TE} \circ \gamma (M) \subset V \}.$$

\begin{pp}[Michor \cite{MIC}]\label{eman3} Let $E \rightarrow M$ be a smooth fiber bundle. The smooth space of sections $\E$ is a Fr\'echet manifold modeled on $\Gamma_c^{\infty}(M,\varphi^* TE)$ for every $\varphi \in \E$ with charts given by
	\begin{eqnarray}\label{aefi2}
	A_{\varphi}^e \colon U_{c,\varphi} & \longrightarrow & V_{c,\varphi} \nonumber \\
	\psi & \longmapsto & \left[ x \mapsto \left(x, e\left(\varphi(x), \psi(x) \right) \right) \right].
	\end{eqnarray}
	The structure is independent of the Riemannian metric chosen for $E$.
\end{pp} 

\begin{rk}\label{km2}
Kriegl and Michor \cite[Theorem 42.1]{KM} consider the same charts, but they define the topological structure a posteriori, via the trivializations. They use the final topology with respect to all smooth curves in $\Gamma_c^{\infty}(M,\varphi^* TE)$. As pointed out in Remark \ref{km1}, in our case $\Gamma_c^{\infty}(M,\varphi^* TE)$ is a Fr\'echet space, and hence the locally convex topology coincides with the  final topology with respect to all smooth curves (see the book by Kriegl and Michor,\cite[4.11]{KM}).
\end{rk}

It is clear that the arguments in the proof of Lemma \ref{tranloc} still hold in this more general setting, we have:

\begin{lm}\label{tranloc2} Let $\pi \colon E \rightarrow M$ be a smooth fiber bundle over. The charts $A_{\varphi}^e \times \textrm{id}_M$ from equation \ref{aefi2} are local and homeomorphisms.
\end{lm}


\subsection{Locally convex topology on \boldmath{\mbox{$\E$}}}

In order to compare the Fr\'echet manifold topology to the ones in Chapter \ref{top} we need to make a comment about the Fr\'echet manifold charts on $J^{\infty} E$. We have used a family of charts on Definition \ref{above}, but a different one will be more useful this time around: one in which the seminorms on the target Fr\'echet space are similar to the ones in Definition/Proposition \ref{eman} for $\V_c$. This is exactly the approach by Kriegl and Michor \cite{KM}.

\begin{rk}\label{aboove}
Kriegl and Michor \cite{KM} consider an atlas on $J^{\infty} E$ given by the family $\{ \pi_{\infty}^{-1}(V_{a}) \}_{a \in A}$ where $\{V_{a}\}_{a \in A}$ is a cover of $M$ by trivial neighborhoods. Each of the $\pi_{\infty}^{-1}(V_{a})$ are trivialized using charts to the space $\RE^m \oplus \prod_{k \in \mathbb{N} \cup \{0\}} \textrm{Lin}_{\textrm{sym}}^k (\RE^m, \RE^e)$ given by the Taylor series, where $m$ is the dimension of $M$ and $e$ is the rank of $E$. That space is a sequential pro-finite normed space, hence a Fr\'echet space by Lemma \ref{lmfs}, where the seminorms are given as follows (see Lewis \cite{LW} for an explicit exposition of this matter):
$$| \chi |_n \defeq \max_{j \in \{ 1, \ldots n \}} \sup_{v \in \RE^m, |v|=1 } \, \left( \left. \frac{\partial^j \varphi}{\partial(v, \cdots, v)} \right|_x \right) \, \textrm{ where } \chi = [ (\varphi, x) ] .$$
Making use of Lemma \ref{saun1} one proves that the transition functions are smooth, hence $J^{\infty} E$ is also a Fr\'echet space with respect to that atlas. 
\end{rk}

\begin{lm}
The locally convex topologies coming from the seminorms in Remark \ref{aboove} and in Definition \ref{above} agree.
\end{lm}

\dem
Consider a cover of $E$ by trivial neighborhoods $\{U_{a}\}_{a \in A}$ where each $U_{a}$ is provided with a trivialization to $\pi(U_{a}) \times F$ for $F$ the fiber of the bundle. Given such a cover, we get a cover of $M$ by trivial neighborhoods $V_{a} = \pi(U_{a})$ as in Definition \ref{above}. The induced covers on $J^{\infty} E$ given by Definition \ref{above} with respect to the $U_{a}$'s and by Remark \ref{aboove} with respect to the $V_{a}$'s are the same since

$$\left( \pi_{\infty}^0 \right)^{-1} \left(U_{a} \right) = \left( \pi_{\infty}^0 \right)^{-1} \left(\pi^{-1}(V_{a})\right) = \pi_{\infty}^{-1}(V_{a}) \textrm{ for all } a \in A.$$
Observe that the later topology was given by the charts $u_{a}^{\infty}$ taking values in $\RE^{\infty}$ (and the components were given by $x^i, u^{\alpha}, u_I^{\alpha}$). For any $n \in \mathbb{N} \cup \{0\}$, the fiber over $\pi(U_{a})$ of that chart is precisely given by \mbox{ $\prod_{k \in \{0, \ldots, n\}} \textrm{Lin}_{\textrm{sym}}^k (\RE^m, \RE^e)$} and we could hence consider that operator norm; showing that indeed in every chart the locally convex topologies agree.
\qed

With this result in hand, we are able to compare the locally convex topology on $\V$ and the $WO^{\infty}$-topology for the case of a compact base:

\begin{pp}\label{iguales} Let $V \rightarrow K$ be a smooth vector bundle over a compact manifold $K$ with space of smooth sections denoted by $\V$. The locally convex topology on $\V$ from Definition/Proposition \ref{eman} agrees with the $WO^{\infty}$-topology (and hence with the $CO^{\infty}$-topology since $K$ is compact).
\end{pp}

We postpone the proof of this fact to the end of the section. First we want to gather similar results for the other Fr\'echet structures developed for $\E$ in this section. First of all, using the locality of the local charts, we can prove the analog of Proposition \ref{iguales} to fiber bundles over a compact base:

\begin{cl}\label{igualess} Let $E \rightarrow K$ be a smooth fiber bundle over a compact manifold $K$ with space of smooth sections denoted by $\E$. The locally convex topology on $\E$ from Proposition \ref{eman2} agrees with the $WO^{\infty}$-topology (and hence with the $CO^{\infty}$-topology since $K$ is compact).
\end{cl}

\dem For each $\varphi \in \E$, the charts $A_{\varphi}^e$ are local as seen in Lemma \ref{tranloc}, in particular they are covered by a continuous map between the infinite jet bundles. This shows that the $CO^{\infty}$-open sets in $\E$ are mapped to $CO^{\infty}$-open sets in $\Gamma^{\infty}(K, \varphi^* TE)$. Those are open with respect to the locally convex topology as seen in Proposition \ref{iguales} and are hence the open sets in the Fr\'echet manifold topology on $\E$. 
\qed

In the non-compact base, the topology, by construction, is strictly larger than the $WO^{\infty}$-topology (this is pointed out by Kriegl and Michor \cite{KM}). This means that all the structure maps out of $\E \times M$ are still continuous: $\E \times M \rightarrow M$, $\E \times M \rightarrow \E$, $\E \times M \rightarrow J^{\infty} E$, and $\E \times M \rightarrow J^k E$ for finite $k$ are all continuous. On the other hand, Proposition \ref{iguales} does not hold in the non-compact base generality. We finish this section with the proof of that result in the case of a compact base and a vector bundle:\\

{\bf Proof of Proposition \ref{iguales}. } Let $\mathcal{W} \subset \V$ be open with respect to the Fr\'echet space topology on $\V$ from Definition/Proposition \ref{eman}. We are going to show that $\mathcal{W}$ is open with respect to the $WO^{\infty}$-topology on $\V$ (which agrees with the $CO^{\infty}$-topology since $K$ is compact). In order to do so, we are going to find a $WO^{\infty}$-open neighborhood of every point in $\mathcal{W}$ fully contained in $\mathcal{W}$. We fix $\{U_{a} \}_{a \in A}$ a finite cover of $K$ by trivial neighborhoods as in Definition/Proposition \ref{eman}.

Let $\varphi \in \mathcal{W}$, there exist $I = ( i_1 , \cdots , i_{|I|} )$ a finite multi-index and $\varepsilon > 0$ such that $\varphi \in U_{\varepsilon}^I(\varphi) \subset \mathcal{W}$ where $U_{\varepsilon}^I(\varphi)$ denotes the basic local neighborhood of $\varphi$ with respect to the seminorms from Definition/Proposition \ref{eman} in the notation of Definition \ref{lctvs}. Consider for every $n \leqslant | I |$ the following set
$$W^n \defeq \bigcup_{a \in A} \bigcup_{x \in \overbar{U_{a}}} \alpha_{\frac{\varepsilon}{|A| \cdot (n+1)}}^{\{i_1, \cdots, i_n\}} \left( j_x^{\infty} \varphi \right),$$
where $\alpha_{\delta}^J$ denotes the local basic neighborhood on $J^{\infty} V$ around $j_x^{\infty} \varphi$ given by the Fr\'echet space structure on $\pi_{\infty}^{-1}(U_{a})$ defined in Remark \ref{aboove}. Clearly, $W^n$ is open with respect to the topology induced by that Fr\'echet manifold structure, and hence with respect to the limit topology as a consequence of the comments after that remark and after Definition \ref{above}. Since $I$ is finite $W \defeq \bigcap_{n =1}^{|I|} W^n$ is also open with respect to the limit topology on $J^{\infty} V$.

Consider $\psi \in \left( J^{\infty} \right)^{-1} \left( WO(W) \right)$, by a direct calculation, we will see that $|\varphi - \psi |_n < \varepsilon$ for all $n \leqslant | I |$. Thus $\varphi \in \left( J^{\infty} \right)^{-1} \left( WO(W) \right) \subset U_{\varepsilon}^I(\varphi) \subset \mathcal{W}$ showing that $\mathcal{W}$ is open with respect to the $WO^{\infty}$-topology on $\E$. Explicitly
$$ \left| D^i (\varphi  - \psi)_a^{\alpha} (x) \right| < \frac{\varepsilon}{|A| \cdot (n+1)}$$  for all $a \in A$, $x \in \overbar{U_{a}}$, $\alpha \in \{1, \dots, \textrm{rank} (V) \}$, $i \in \{ 0 , \cdots, n\}$ and $n \leqslant | I |$.
Using an appropriate norm on the space of matrices (the same argument as in Remark \ref{aboove}), the previous result means that
$$ \sum_{i=0}^n \left| D^i (\varphi  - \psi)_a^{\alpha} (x) \right| < \frac{\varepsilon}{|A|} \hspace{2em} \forall a \in A \, \forall x \in \overbar{U_{a}} \, \forall \alpha \in \{1, \dots, \textrm{rank} (V) \} \, \forall n \leqslant | I |;$$
$$\textrm{thus } \max_{   \begin{smallmatrix}
 x \in \overbar{U_{a}} \\ 
 \alpha \in \{ 1, \ldots, \textrm{rank} (V) \}
 \end{smallmatrix} } { \, \left(\sum_{j=0}^n \sum_{i=0}^n \left| D^i (\varphi  - \psi)_a^{\alpha} (x) \right| \right) } < \frac{\varepsilon}{|A|}  \hspace{3em} \forall a \in A \, \forall n \leqslant | I | ;$$
and hence $\forall n \leqslant | I |$
$$ | \varphi - \psi |_n = \sum_{a \in A}{\,
\max_{   \begin{smallmatrix}
	x \in \overbar{U_{a}} \\ 
	\alpha \in \{ 1, \ldots, \textrm{rank} (V) \}
	\end{smallmatrix} } { \, \left(\sum_{j=0}^n \sum_{i=0}^n \left| D^i (\varphi  - \psi)_a^{\alpha} (x) \right| \right) } } < \varepsilon .$$

To prove that every $\mathcal{W} \subset \V$ open in the $WO^{\infty}$-topology is open in the Fr\'echet space topology we need to use the same idea backwards. We fix $\varphi \in \mathcal{W}$, we can assume that $\varphi \in \left( J^{\infty}\right)^{-1} \left( WO(W) \right) \subset \mathcal{W}$ where $W$ is open in the Fr\'echet manifold topology on $J^{\infty} V$ introduced in Remark \ref{aboove}. We fix $a \in A$ and consider for every $x \in \overbar{U_{a}}$ an open neighborhood of $j_x^{\infty} \varphi$ in $W$ given by $\alpha_{\delta_x}^{J_{x}} (j_x^{\infty} \varphi)$. $\varphi(\overbar{U_{a}})$ is compact and covered by the union of the $\alpha_{\delta_x}^{J_{x}} (j_x^{\infty} \varphi)$. Hence, we can extract a finite subcover $\{ \alpha_{\delta_i}^{J_{i}} (j_{x_i}^{\infty} \varphi) \}_{i \in I_{a}}$. Consider $I \defeq \bigcup_{a \in A} I_{a}$ and $\varepsilon = \min_{i \in I} \delta_i$. For every $\psi \in U_{\varepsilon}^I (\varphi)$ we have, by construction and using Remark \ref{aboove}, that
$$ \psi (K) \subset \bigcup_{i \in I} \alpha_{\delta_i}^{J_{i}} (j_{x_i}^{\infty} \varphi) \subset W $$
showing that $U_{\varepsilon}^I (\varphi) \subset \left(J^{\infty}\right)^{-1}(WO(W)) \subset \mathcal{W}$ and thus that $\mathcal{W}$ is open in the Fr\'echet space topology.
\qed


\section{Smooth maps involving \boldmath{\mbox{$\E \times M$}}}\label{cinpudos}

{\it We characterize smooth maps taking values in $\E \times M$ both in the compact and the non-compact base cases. We also prove that jet evaluations are smooth. We compare locality and smoothness, concluding that no notion is stringer than the other, but that under certain assumptions locality implies smoothness. The principal reference in this chapter is Kriegl and Michor \cite{KM}.}\\

Given two smooth fiber bundles $\pi \colon E \rightarrow M$ and $\rho \colon F \rightarrow N$ with spaces of smooth sections $\E$ and $\F$ respectively, we can consider local maps $f \colon \E \times M \rightarrow \F \times N$. We have seen in the previous section that the spaces of smooth sections $\E$ and $\F$ carry the structure of a Fr\'echet manifold. This also allows us to talk about smooth maps $f \colon \E \times M \rightarrow \F \times N$. We want to explore what is the relation between these two concepts.

In order to understand what a smooth map $f \colon \E \times M \rightarrow \F \times N$ is we need to first observe that jet evaluations are smooth.

\begin{pp}\label{evsmo} Let $E \rightarrow M$ be a smooth fiber bundle with space of smooth sections $\E$. Then the jet evaluations $j^k \colon \E \times M \rightarrow J^k E$ are smooth for all $k \in \mathbb{N} \cup \{ \infty \}$.
\end{pp}

\dem Smoothness is to be checked in every coordinate chart, so that it is enough to do it for the case in which the bundle is a vector bundle and the base manifold is compact because local trivializations are local maps (see Lemma \ref{tranloc2}).

Jet evaluations are continuous because the Fr\'echet manifold topology agrees with the $WO^{\infty}$-topology (Corollary \ref{igualess}) and they are continuous with respect to that topology (Corollary \ref{ary}). It is then possible to talk about the G\^ateaux derivatives of $j^k$ (see Appendix \ref{app2} for the definition). We fix $k$ finite and we will argue for infinite $k$ later.

A straight forward calculation shows that the G\^ateaux derivative of $j^k$ at $(\varphi, x)$ in the direction of $(\psi, y)$ is simply given by 
$$ Dj^k((\varphi, x)){(\psi, y)} = j^k(\psi, x) + \frac{\partial j^k \varphi }{\partial y_i}\mid_{x} y^i .$$

We have seen in Corollary \ref{igualess} that the seminorms on $\E$ and $J^k E$ are compatible, this means that jet evaluations are continuous and that the first G\^ateaux derivative is also continuous. The same argument works for higher derivatives. This shows that all finite jet evaluations are continuous.

In the case $k = \infty$ the map $j^{\infty} \colon \E \times K \rightarrow J^{\infty} E$ is smooth if and only if it is smooth in every chart:
$$j_{\varphi, a}^{\infty} = j^{\infty} \circ \left( A_{\varphi}^e, \textrm{id}_K \right)^{-1} \colon V_{\varphi} \times U_{a} \rightarrow J^{\infty} E$$ for all $\varphi \in \E$ and all $a$ in $A$ where $\{U_{a}\}_{a \in A}$ is an atlas for $K$. Those are maps between Fr\'echet spaces as in Lemma \ref{saun1}, so that they are smooth if and only if the composition with any of the projections $\pi_{\infty}^l$ are smooth. But that happens if and only if all finite jet evaluations are smooth, which was showed earlier in this proof.
\qed

A map $f \colon \E \times M \rightarrow \F \times N$ is smooth if and only if the two components $f_{\F} \colon \E \times M \rightarrow \F$ and $f_N \colon \E \times M \rightarrow N$ are smooth (result by Kriegl-Michor \cite[27.3]{KM}). With this in mind, we want to give an idea of how apart a local map $f \colon \E \times M \rightarrow \F \times N$ is from being smooth. The key property is that infinite jet evaluations are smooth. If $f$ is local, with pro-finite smooth representative $f^{\infty} \colon J^{\infty} E \rightarrow J^{\infty} F$, by looking at the following commutative diagram

\begin{center}
\begin{tikzpicture}[description/.style={fill=white,inner sep=2pt}]
\matrix (m) [matrix of math nodes, row sep=3em,
column sep=3em, text height=1.5ex, text depth=0.25ex]
{ \E \times M & \F \times N & N\\
J^{\infty} E & J^{\infty} F \\};
\path[->,font=\scriptsize]
(m-1-1) edge [bend left=40] node[above=0.2em] {$f_N$} (m-1-3)
(m-1-1) edge node[auto] {$f$} (m-1-2)
(m-2-1) edge node[auto] {$f_{\infty}$} (m-2-2)
(m-1-2) edge node[auto] {$\textrm{pr}_N$} (m-1-3)
(m-1-1) edge node[auto] {$j^{\infty}$} (m-2-1)
(m-1-2) edge node[auto] {$j^{\infty}$} (m-2-2)
(m-2-2) edge node[below=0.2em, right=0.1em] {$\rho_{\infty}$} (m-1-3);
\end{tikzpicture}
\end{center}

{\noindent it is clear that $f_N$ is smooth since it is the composition of smooth maps:}

$$f_N = \textrm{pr}_N \circ f = \rho \circ j^0 \circ f = \rho_{\infty} \circ j^{\infty} \circ f = \rho_{\infty} \circ f^{\infty} \circ j^{\infty},$$

{\noindent because jet evaluations are smooth ($j^{\infty}$ is smooth by Proposition \ref{evsmo}), pro-finite smooth maps are smooth (Corollary \ref{prosmosmo}) and $\rho_{\infty}$ is smooth (Proposition \ref{proprosmo}). We conclude the following:}

\begin{lm} Let $E \rightarrow M$ and $F \rightarrow N$ be two smooth fiber bundles with spaces of smooth of sections $\E$ and $\F$ respectively. A local map $f \colon \E \times M \rightarrow \F \times N$ is smooth if and only if $f_{\F} \colon \E \times M \rightarrow \F$ is smooth.
\end{lm}

Kriegl and Michor \cite{KM} give characterizations of when such maps are smooth. Given two Fr\'echet manifolds $\M$ and $\N$, a map $f \colon \M \rightarrow \N$ is smooth if and only if $\gamma \circ f \colon \RE \rightarrow \N$ is smooth for all $\gamma \colon \RE \rightarrow \M$ smooth curve (see Kriegl and Michor \cite[27.2]{KM}). Again, using their work, this time \cite[27.3]{KM}, $\gamma \colon \RE \rightarrow \E \times M$ is smooth if and only if $\gamma_{\E}$ and $\gamma_M$ are smooth. We will hence have to deal with smooth maps $\gamma_{\E} \colon \RE \rightarrow \E$ and $f \circ \gamma \colon \RE \rightarrow \F$. The smooth curves in spaces of sections are different depending on whether the base manifold is compact or not.

\begin{lm}[Kriegl and Michor \cite{KM}]\label{lkm} Let $\E$ be the space of smooth sections of a smooth fiber bundle over a compact manifold $K$. A map $\gamma \colon \RE \rightarrow \E$ is smooth if and only if 
\begin{center}
    \begin{tabular}{rcl}
    $\gamma^{\smallsmile} \colon \RE \times K$ & $\longrightarrow$ & $F$ \\
    $(t,x)$ & $\longmapsto$ & $\gamma(t)(x)$ \\
    \end{tabular}
\end{center}
{\noindent is smooth.}\\

If $\E$ be the space of smooth sections of a smooth fiber bundle over a non-compact manifold $M$. A map $\gamma \colon \RE \rightarrow \E$ is smooth if and only if 
\begin{center}
    \begin{tabular}{rcl}
    $\gamma^{\smallsmile} \colon \RE \times M$ & $\longrightarrow$ & $F$ \\
    $(t,x)$ & $\longmapsto$ & $\gamma(t)(x)$ \\
    \end{tabular}
\end{center}
{\noindent is smooth and for all $[a,b] \subset \RE$ there exists $K_{[a,b]} \subset M$ compact such that  for all $t \in [a,b]$ we have that $\restrict{\gamma(t)}{K_{[a,b]}} = \restrict{\gamma(a)}{K_{[a,b]}}$.}
\end{lm}

The statement can be found in the book of Kriegl and Michor \cite[30.8]{KM}, for the vector bundle case, and \cite[42.5]{KM} for the general case.

The following two examples show that there are smooth maps which are not local and vice versa: there are local maps which are not smooth. For simplicity we will consider one dimensional trivial vector bundles over compact one dimensional manifolds: $\RE \times \S^1 \rightarrow \S^1$:

\begin{ej}[Smooth map which is not local]\label{sl}
Consider the trivial vector bundle $\RE \times \S^1 \rightarrow \S^1$. Its space of smooth sections can be identified with $\C(\S^1)$. The map
\begin{center}
    \begin{tabular}{rcl}
    $f \colon \C(\S^1) \times \S^1$ & $\longrightarrow$ & $\C(\S^1) \times \S^1$ \\
    $(\varphi,x)$ & $\longmapsto$ & $(\varphi,x + \pi)$ \\
    \end{tabular}
\end{center}
{\noindent is clearly not local. It is nevertheless smooth since given any smooth curve $\gamma \colon \RE \rightarrow \C(\S^1) \times \S^1$ the induced curve $f \circ \gamma \colon \RE \rightarrow \C(\S^1) \times \S^1$ is given by the smooth assignation $t \mapsto \left( \gamma_{\C(\S^1)}(t), \gamma_{\S^1}(t) + \pi \right)$ which is the composition of a smooth map with a translation in $\S^1$ (which is also smooth).}
\end{ej}

\begin{ej}[Local map which is not smooth]\label{ls}
As in the previous example, consider the trivial vector bundle $\RE \times \S^1 \rightarrow \S^1$. Recall its space of smooth sections can be identified with $\C(\S^1)$.
Consider a smooth bump function $b \in \C(\S^1)$ such that $\restrict{b}{[-\frac{\pi}{4}, \frac{\pi}{4}]} = 0$ and $\restrict{b}{[\frac{3 \pi}{4}, \frac{5 \pi}{4}]} = 1$. Using this function we are going to construct the following a local map:
\begin{center}
    \begin{tabular}{rcl}
    $f \colon \C(\S^1) \times \S^1$ & $\longrightarrow$ & $\C(\S^1) \times \S^1$ \\
    $(\varphi,x)$ & $\longmapsto$ & $(|\varphi(x)|\cdot b, 0).$ \\
    \end{tabular}
\end{center}
Clearly $\restrict{\left( |\varphi(x)|\cdot b \right)}{[-\frac{\pi}{4}, \frac{\pi}{4}]} = 0$ so that $j_0^{\infty} (|\varphi(x)|\cdot b) = (0,0)$. This shows that the map $f$ is local. On the other hand, $f$ is not smooth. Consider the following smooth curve:
\begin{center}
    \begin{tabular}{rcl}
    $\gamma \colon \RE$ & $\longrightarrow$ & $\C(\S^1) \times \S^1$ \\
    $t$ & $\longmapsto$ & $([x \mapsto t], 0).$ \\
    \end{tabular}
\end{center}
It is indeed smooth since $\gamma_{S^1} = 0$ is smooth and $\gamma_{\C(\S^1)}^{\smallsmile} \colon \RE \times \S^1 \rightarrow \RE \times \S^1$ sending $(t,x)$ to $(t,x)$ is also smooth. But now consider the map $f \circ \gamma$:
\begin{center}
    \begin{tabular}{rcl}
    $f \circ \gamma \colon \RE$ & $\longrightarrow$ & $\C(\S^1) \times \S^1$ \\
    $t$ & $\longmapsto$ & $([x \mapsto |t|\cdot b(x)], 0).$ \\
    \end{tabular}
\end{center}
It turns out that $(f \circ \gamma)_{\C(\S^1)}$ is not smooth since $(f \circ \gamma)_{\C(\S^1)}^{\smallsmile}$ maps $(t, x)$ to $(|t| b(x), x)$ and that means that around $\pi$, $\restrict{(f \circ \gamma)_{\C(\S^1)}^{\smallsmile}}{[\frac{3 \pi}{4}, \frac{5 \pi}{4}]}$ maps $(t,x)$ to $(|t|, x)$ which is not smooth.
\end{ej}

When considering local maps $f \colon \E \times M \rightarrow \F \times N$ or smooth maps, we have seen in the examples \ref{sl} and \ref{ls} that none of these two notions are stronger than the other. On the other hand, there are some cases in which local maps can be proven to be smooth:

\begin{pp}\label{lis1} Let $f_{\F} \times f_N \colon \E \times M \rightarrow \F \times N$ be a local map where $\E$ and $\F$ are the spaces of smooth sections of the smooth fiber bundles $\pi \colon E \rightarrow M$ and $\rho \colon F \rightarrow N$ respectively. If $f_N$ is a surjective submersion, then $f$ is smooth.
\end{pp}

\dem Let $(\varphi, x) \in \E \times M$. In order to show that $f$ is smooth we need to see that for all $\gamma \colon \RE \rightarrow \E \times M$ smooth the composition with $f$ is again smooth, $f \circ \gamma \colon \RE \rightarrow N$. Since $f$ is a product, it is enough to check that $f_{\F} \circ \gamma_{\E} \colon \RE \rightarrow \F$ is smooth for all $\gamma_{\E} \colon \RE \rightarrow \E$ smooth (the other component is the composition of two smooth maps always).

This amounts to check if $(f_{\F} \circ \gamma_{\E})^{\smallsmile} \colon \RE \times N \rightarrow F$ is smooth at every $(t, n) \in \RE \times N$. Consider $\iota_n$ a local smooth section of $f_N$ around $n$, to be precise $\iota_n \colon N \rightarrow M$ such that $\restrict{f_N \circ \iota_n}{V_n} = \textrm{id}_{V_n}$ in some open neighborhood $V_n$ of $n$. We can factor $(f_{\F} \circ \gamma_{\E})^{\smallsmile}$ in the following way:
\vspace{-4ex}
\begin{center}
\begin{tikzpicture}[description/.style={fill=white,inner sep=2pt}]
\matrix (m) [matrix of math nodes, row sep=3em,
column sep=3.5em, text height=1.5ex, text depth=0.25ex]
{ \RE \times V_n & \E \times M & \F \times V_n & \\
& J^{\infty} E & J^{\infty} F & F\\};
\path[->,font=\scriptsize]
(m-1-1) edge [bend left=40] node[below=0.25em] {$f_{\F} \circ \gamma_{\E} \times \textrm{id}_{V_n}$} (m-1-3)
(m-1-1) edge [bend left=70] node[right=3em] {$(f_{\F} \circ \gamma_{\E})^{\smallsmile}$} (m-2-4)
(m-1-1) edge node[auto] {$\gamma_{\E} \times \iota_n$} (m-1-2)
(m-2-2) edge node[auto] {$f_{\infty}$} (m-2-3)
(m-1-2) edge node[auto] {$f_{\F} \times f_N$} (m-1-3)
(m-1-3) edge node[left] {$j^{\infty}$} (m-2-3)
(m-1-2) edge node[auto] {$j^{\infty}$} (m-2-2)
(m-1-3) edge node[auto] {$j^0$} (m-2-4)
(m-2-3) edge node[auto] {$\rho_{\infty}^0$} (m-2-4);
\end{tikzpicture}
\end{center}
That shows that $(f_{\F} \circ \gamma_{\E})^{\smallsmile}$ is smooth as it is the composition of smooth maps.
\qed

Observe that the strategy used in the proof of Proposition \ref{lis1} could not have worked in example \ref{ls} because $f_{\F}$ was changing with the point. Being more precise, $f_{\F}$ was changing along the image of $\iota_n$ (as in the proof of that proposition): this should be reminiscent of the discussion about jet prolongations of maps respecting the Cartan distribution. As a preview of what is going to happen next, insular maps are a particular kind of local maps preserving the Cartan distribution, in which the locality is stronger. These will be defined in Part III, but we want to state now one of their main properties:

\begin{tm*}[Insular maps are smooth]\label{imas} Let $f \colon \E \times M \rightarrow \F \times N$ be an insular map with smooth local family of sections. If $N$ is compact, then $f$ is smooth.
\end{tm*}

This is Theorem \ref{t3}, and we postpone its proof to Part III.

Observe that besides the insularity assumption, that we are not able to understand yet, it is said that it has a smooth local family of sections: this family of sections are the equivalent to $\iota_n$ in the proof of Proposition \ref{lis1}. For a local map $f \colon \E \times M \rightarrow \F \times N$ we observed that $f_N$ is always smooth. Moreover, the map $\E \rightarrow \C(M,N)$ sending $\varphi$ to $f_N(\varphi, -)$ is also smooth (at least in the case $M$ is compact) since for all $\gamma \colon \RE \rightarrow \E$ smooth the map $\RE \times M \rightarrow N$ mapping $(t,m)$ to $f_N(\gamma(t), m)$ decomposes into $\gamma \times \textrm{id}_M$ and $f_N$ which are smooth. It is realistic then to be able to find a smooth map $\iota \colon \E \times M \rightarrow \C(N,M)$ giving sections of $f_N(\varphi, -)$.\\



\printbibliography

\setcounter{part}{2}
\setcounter{chapter}{5}


\newpage
\part{Insularity and Cartan-preserving maps}

\newpage
\tableofcontents

\chapter*{\color{darkdelion} Insularity and Cartan-preserving maps}

There are different notions of locality in the literature: from the most common ones of maps induced by morphisms of sheaves, or germ-local maps, passing through differential operators to maps preserving the Cartan foliation. All of them have their advantages and all of them are related to local maps as defined in Part II.\\

Differential operators such that the lower map $J^k E \rightarrow F$ uniquely determines the global map between the infinite dimensional Fr\'echet manifolds are very interesting because of that: all the information can be encoded in a map between finite dimensional manifolds. Local maps are differential operators, but the converse is not true.\\

Differential operators along the identity are local and also smooth. They form a category in which composition is given by taking infinite jet prolongations. These maps are the ones we encounter often in our research.\\

In the more general setting, there are ways of imposing conditions on the differential operators so that jet prolongations exist and the associated local map is smooth. Maps under those conditions are called insular and are the main tool in this part. Insular maps have associated foliations such that, when restricted to every leaf, the map towards the space of sections is constant, and the one to the base manifold is a diffeomorphism.\\

Finally, Peetre's theorem relates the concept of germ-local maps, maps of sheaves and differential operators. There exists also a version of Peetre's theorem in our insular setting. Germ-local maps along a submersion are locally like an insular map.\\

{\it This part explores the relations between the different notions of locality. It provides definitions of the main concepts: differential operators, Cartan-preserving maps, germ-local maps, etc. Two categories are produced: that of differential operators along the identity and that of insular manifolds, the first being a subcategory of the second. Insular differential operators are defined and proven to be local and smooth. To conclude, we revisit Peetre's theorem in our setting, showing that under certain assumptions all notions of locality agree at least locally. The main references in this part are Chetverikov \cite{CHE}, Kock \cite{K}, and Slov\'ak \cite{SLO}.}


\newpage
\chapter{Differential operators}

Differential operators are maps $\E \times M \rightarrow \F \times N$ that descend to a map $J^k E \rightarrow F$, where as usual, $\E$ and $\F$ denote the space of smooth sections of the smooth fiber bundles $E \rightarrow M$ and $F \rightarrow N$ respectively. Differential operators have the advantage that the information is encoded in a map between finite dimensional manifolds. They also have disadvantages, since, for example, they do not form a category.\\

Local maps are differential operators, but the converse is not true; jet prolongations do not exist for all differential operators nor are they always covered by the original map. The easiest case for which we can develop hypothesis to get a converse result is when the differential operator covers the identity. In that case, differential operators along the identity (products $f_{\F} \times \textrm{id}_M$) are the ones that give rise to extended local maps with many other good properties.\\

{\it This chapter focuses in the relation existing between differential operators and local maps. Moreover, it studies the category of differential operators along the identity, which is particularly well behaved. The main reference for this chapter is Kock \cite{K}.}


\section{Local maps or differential operators?}\label{question}

{\it This section introduces differential operators and states the three question necessary to compare them to local maps. It gives an example of a differential operator which is not a local map, nor it induces a map between the corresponding \'etal\'e spaces. The main references in this chapter are Mac Lane and Moerdijk \cite{McMd} and Kock \cite{K}.}\\

In this chapter we are interested in maps from $\E \times M$ to $\F \times N$ where $\E$ and $\F$ are the spaces of smooth sections of the smooth fiber bundles $\pi \colon E \rightarrow M$ and $\rho \colon F \rightarrow N$ respectively. In the study of pro-smooth maps between infinite jet bundles it is well known that the maps preserving the Cartan distribution are specially well behaved: given a Cartan distribution preserving map $f^0 \colon J^k E \rightarrow F$ there exists a unique pro-smooth map $j^{\infty} f^0 \colon J^{\infty} E \rightarrow J^{\infty} F$ covering $f^0$ and preserving the Cartan distribution (Proposition \ref{PR} and Corollary \ref{chocho}). On the side of maps involving $\E \times M$, if they cover a map $f^0 \colon J^k E \rightarrow F$ they have a name of their own: they are called differential operators. Differential operators are widely studied in the literature and they constitute a field of research on their own.

\begin{df}[Differential operator of order $k$] A map $f \colon \E \times M \rightarrow \F \times N$ is called a differential operator of order $k \in \mathbb{N}$ if it descends to a smooth map $f^0 \colon j^k (\E \times M) \subset J^k E \rightarrow F$:
\begin{center}
\begin{tikzpicture}[description/.style={fill=white,inner sep=2pt}]
\matrix (m) [matrix of math nodes, row sep=3em,
column sep=2.5em, text height=1.5ex, text depth=0.25ex]
{\E \times M & \F \times N \\
 j^k (\E \times M) \subset J^k E & F \\};
\path[->,font=\scriptsize]
(m-1-1) edge node[auto] {$f$} (m-1-2)
(m-1-1) edge node[auto] {$j^{k}$} (m-2-1)
(m-1-2) edge node[auto] {$j^{0}$} (m-2-2)
(m-2-1) edge node[auto] {$f^0$} (m-2-2);
\end{tikzpicture}
\end{center} 
\end{df}

This is a very well known term in the literature. We want to refer to Kock \cite{K} where its approach is similar to ours here. He gives almost the same Definition, where he calls these maps {\it sheaf theoretic differential operators} although he only considers the case in which $f_N = \textrm{pr}_{M=N}$ and replaces $j^k(\E \times M)$ by $J^k E$ (which holds if $\E$ is soft).

Local maps are differential operators since any pro-smooth map $f^{\infty} \colon J^{\infty} E \rightarrow J^{\infty} F$ is given by a family of maps smooth maps $\{f^l \colon J^{k(l)} E \rightarrow J^l F\}$ commuting with $\{\pi_{k(l^{\prime})}^{k(l)}\}$ and $\{\rho_{l^{\prime}}^{l}\}$ for all $l, l^{\prime} \in \mathbb{N}$. In particular if $l = 0$ we have the following diagrams (for simplicity of the diagrams, we are assuming $\E$ and $\F$ are soft):

\begin{center}
\begin{tikzpicture}[description/.style={fill=white,inner sep=2pt}]
\matrix (m) [matrix of math nodes, row sep=3em,
column sep=2.5em, text height=1.5ex, text depth=0.25ex]
{ \E \times M & \F \times N & & \E \times M & \F \times N \\
  J^{\infty} E & J^{\infty} F & & & \\
	J^{k} E & F & & J^{k} E & F \\};
\path[->,font=\scriptsize]
(m-1-1) edge node[auto] {$f$} (m-1-2)
(m-1-1) edge node[auto] {$j^{\infty}$} (m-2-1)
(m-2-1) edge node[auto] {$f^{\infty}$} (m-2-2)
(m-1-2) edge node[auto] {$j^{\infty}$} (m-2-2)
(m-2-1) edge node[auto] {$\pi_{\infty}^k$} (m-3-1)
(m-2-2) edge node[auto] {$\rho_{\infty}^0$} (m-3-2)
(m-3-1) edge node[auto] {$f^0$} (m-3-2)

(m-1-4) edge node[auto] {$f$} (m-1-5)
(m-1-4) edge node[auto] {$j^{k}$} (m-3-4)
(m-1-5) edge node[auto] {$j^{0}$} (m-3-5)
(m-3-4) edge node[auto] {$f^0$} (m-3-5);
\end{tikzpicture}
\end{center}

From our point of view, the key {\bf good property} that differential operators satisfy is that in order to study $f$, a map between infinite dimensional Fr\'echet space, we can pass to $f^0$ a much better behaved map between smooth bundles of {\it finite dimensional} rank over finite dimensional manifolds. 

The {\bf bad property} about differential operators is that they {\it do not form a category}. Let $\pi \colon E \rightarrow M$, $\rho \colon F \rightarrow N$ and $\sigma \colon G \rightarrow P$ be three smooth fiber bundles with corresponding spaces of smooth sections $\E$, $\F$ and $\G$. Given two differential operators $(f, f^0)$ from $\E \times M$ to $\F \times N$ and $(g, g^0)$ from $\F \times N$ to $\G \times P$ there is no natural way of composing $f^0 \colon J^{k} E \rightarrow F$ with $g^0 \colon J^l F \rightarrow G$ (again assume $\E$ and $\F$ are soft for simplicity):

\begin{center}
\begin{tikzpicture}[description/.style={fill=white,inner sep=2pt}]
\matrix (m) [matrix of math nodes, row sep=3em,
column sep=2.5em, text height=1.5ex, text depth=0.25ex]
{ \E \times M & \F \times N & \G \times P \\
   & J^{l} F & G  \\
   J^{k} E & F  &  \\};
\path[->,font=\scriptsize]
(m-1-1) edge node[auto] {$f$} (m-1-2)
(m-1-2) edge node[auto] {$g$} (m-1-3)

(m-1-1) edge node[auto] {$j^{k}$} (m-3-1)
(m-1-2) edge node[auto] {$j^{l}$} (m-2-2)
(m-2-2) edge node[auto] {$\rho_{l}^0$} (m-3-2)
(m-1-3) edge node[auto] {$j^{0}$} (m-2-3)

(m-3-1) edge node[auto] {$f^0$} (m-3-2)
(m-2-2) edge node[auto] {$g^0$} (m-2-3);
\end{tikzpicture}
\end{center} 

There are naturally arising questions at this point. We have studied in Section \ref{jpg} conditions under which infinite jet prolongations of $f^0$ exist and are unique, but we do not know how does this infinite jet prolongation behave, in particular:

\begin{enumerate}
	\item[DO.1] If $j^{\infty} f^0$ (the infinite jet prolongation of $f^0$) exists, is it covered by $f$?
	\item[DO.2] Is there a notion of Cartan-preserving local maps so that we can recover $f$ from $f^0$?
	\item[DO.3] Is the map sending $[(\varphi, x)]$ to $[f(\varphi, x)]$ well defined? (By $[-]$ we mean the germ class of the pair $(\varphi, x)$.)
\end{enumerate}

In the case in which we can develop conditions so that all questions can be answered positively, we would have the best of the two worlds: the finite dimensionality properties of the differential operators and the category structure properties of the local maps.

Recall that there is an important issue we had to deal with and it is that when $\E$ is soft, it is possible that there exists a map $f \colon \E \times M \rightarrow \F \times N$ covering different maps $f^{\infty} \colon J^{\infty} E \rightarrow J^{\infty} F$. We have solved this by working with extended local maps (see Definition \ref{elMan}). Recall that it amounts to consider maps $f$ covering pro-smooth maps $f^{\infty} \colon j^{\infty}(\E \times M) \rightarrow j^{\infty}(\F \times N)$:

\begin{center}
\begin{tikzpicture}[description/.style={fill=white,inner sep=2pt}]
\matrix (m) [matrix of math nodes, row sep=3em,
column sep=2.5em, text height=1.5ex, text depth=0.25ex]
{\E \times M & \F \times N \\
  j^{\infty}(\E \times M) & j^{\infty}(\F \times N) \\};
\path[->,font=\scriptsize]
(m-1-1) edge node[auto] {$f$} (m-1-2)
(m-1-1) edge node[auto] {$j^{\infty}$} (m-2-1)
(m-1-2) edge node[auto] {$j^{\infty}$} (m-2-2)
(m-2-1) edge node[auto] [swap]{$f^{\infty}$} (m-2-2);
\end{tikzpicture}
\end{center}

Observe that by working with extended local maps, we know that if a map $f$ is local, it covers a unique map $f^{\infty} \colon j^{\infty}(\E \times M) \rightarrow j^{\infty}(\F \times N)$: the one sending $j_x^{\infty} \varphi$ to $j_{f_N(\varphi, x)}^{\infty} (f_{\F} (\varphi, x))$. That is why question [DO.3] is relevant: in order to answer positively to [DO.1] we would need to develop conditions so that [DO.3] also holds.

We want to be more precise about [DO.3], in order to do so we need some notation. Given a sheaf $\E$ on $M$ (in our case it will always be the sheaf of sections of a smooth fiber bundle), the {\it \'etal\'e space} associated to it is the union of the stalks. The stalk of $\E$ at $x$, denoted by $\E_x$, is $\E_x \defeq  \underset{x \in U, \subset}{\textrm{colim }} \E(U)$. An element of the stalk is an equivalence class of elements $\varphi_U \in \mathcal \E(U)$, where the equivalence relation is called sharing the same {\it germ} at $x$. The \'etal\'e space associated to $\E$ is the disjoint union of all stalks: $\widehat{E} \defeq  \bigsqcup_{x \in M} \E_x$, with some natural topology that can be found for instance in the book of Mac Lane and Moerdijk, \cite[II.5]{McMd}. It receives a map from $\E \times M$, $g_E \colon \E \times M \rightarrow \widehat{E}$ sending a section and at point to its germ at that point. The finite jet bundles are quotients of the \'etal\'e space: $J^k E = \bigslant{\widehat{E}}{\sim}$ where two germs are equivalent if their $k$--th jets agree. That means that the jet evaluation factors though the \'etal\'e space, as we can see in the following commutative diagram for all $k$:

\begin{center}
\begin{tikzpicture}[description/.style={fill=white,inner sep=2pt}]
\matrix (m) [matrix of math nodes, row sep=3em,
column sep=2.5em, text height=1.5ex, text depth=0.25ex]
{\E \times M &  \\
 \widehat{E}  \hspace{0.8em}& J^{k}(E) \\};
\path[->,font=\scriptsize]
(m-1-1) edge node[auto] {$g_{E}$} (m-2-1)
(m-2-1) edge node[auto] {$\widehat{j}^{k}$} (m-2-2)
(m-1-1) edge node[auto] {$j^{k}$} (m-2-2);
\end{tikzpicture}
\end{center}

Consider a differential operator $(f,f^0) \colon \E \times M \rightarrow \F \times N$. By using $f^0$ we know the zero-th jet of $f_{\F}(\varphi, x)$ at $f_N(\varphi,x)$ for all $(\varphi, x) \in \E \times M$. But, a priori, we do not know the value of $f_{\F}(\varphi, x)$ in any neighborhood of $f_{N}(\varphi, x)$. This is so because as soon as we move away from $(\varphi, x)$, $f_{\F}(\varphi, x)$ might change and $f^0$ does not help. Even for local maps, where thanks to $f^{\infty}$ we know the infinite jet of $f_{\F}(\varphi, x)$ at $f_N(\varphi,x)$, that is still not enough to know the germ of $f_{\F}(\varphi, x)$ at $f_N(\varphi,x)$ since we are working with smooth maps and not analytic ones.\\

The examples in which [DO.1] and [DO.3] fail are not interesting and they are precisely the situations that we want to avoid: they are not natural, in the precise sense that $f$ does not induce a map between the corresponding \'etal\'e spaces. The following example illustrates the problems caused by the failure of [DO.1] and [DO.3], in the case of a differential operator of order $0$:

\begin{ej}\label{con} Consider the trivial one dimensional vector bundle over $\RE$. We write $\C(\RE)$ for its space of sections using the usual isomorphism. For every pair of real numbers $r$ and $x$ we can construct a polynomial (in particular a smooth map) whose value at $x$ is zero and whose derivative at $x$ is $r$. Explicitly:
$$p_{r,x}(y)\defeq  y^2 + (r -2x)y +(x - r)x.$$
Consider the following map:

\begin{center}
    \begin{tabular}{rcl}
    $f \colon \C(\RE) \times \RE$ & $\longrightarrow$ & $\C(\RE) \times \RE$ \\
    $(\varphi, x)$ & $\longmapsto$ & $(\varphi + p_{\varphi(x+1),x}, x)$. \\
    \end{tabular}
\end{center}

The $0$-th jet of $\varphi + p_{\varphi(x+1),x}$ at $x$ is simply given by $(\varphi(x), x) \in \mathbb{R} \times \RE$. In this way we see that $f$ is a differential operator of order $0$ (even an extended one since the bundle is a vector bundle, hence its sheaf of sections is soft):

\begin{center}
\begin{tikzpicture}[description/.style={fill=white,inner sep=2pt}]
\matrix (m) [matrix of math nodes, row sep=3em,
column sep=2.5em, text height=1.5ex, text depth=0.25ex]
{ \C(\RE) \times \RE & \C(\RE) \times \RE \\
  \RE \times \RE  & \RE \times \RE \\};
\path[->,font=\scriptsize]
(m-1-1) edge node[auto] {$f$} (m-1-2)
(m-1-1) edge node[auto] {$j^{0}$} (m-2-1)
(m-2-1) edge node[auto] {id} (m-2-2)
(m-1-2) edge node[auto] {$j^{0}$} (m-2-2);
\end{tikzpicture}
\end{center}

Clearly, we can see that $f$ does not cover a map between the associated \'etal\'e spaces since the fact that $\varphi$ and $\varphi^{\prime}$ have the same germ at $x$ tells us nothing about their value at $x+1$, which is used in the definition of $f$. This means that [DO.3] is answered negatively in this case. 

Although we can argue from the previous paragraph that $f$ is not a local map either, we can also calculate that explicitly. We can take jet prolongations of any order of the map $\textrm{id}_{\RE \times \RE}$ because it is a bundle map over $\textrm{id}_{\RE}$ (see Proposition \ref{pro}). As a mater of fact, all these prolongations are just the identity again so that $j^{\infty} \textrm{id}_{\RE \times \RE} = \textrm{id}_{J^{\infty}(\RE \times \RE)}$. But it is impossible to know the first jet of $\varphi + p_{\varphi(x+1),x}$ at $x$ just by knowing local information of $\varphi$ around $x$: $(\varphi + p_{\varphi(x+1),x})^{\prime}(x) = \varphi^{\prime}(x) + \varphi(x+1)$. This tells us that $f$ is not a local map and thus that [DO.1] is answered negatively.

Even thinking about it further, the identity on $\C(\RE) \times \RE$ is a local map and its degree $0$ part is again $\textrm{id}_{\RE \times \RE}$: this tells us that {\it two different differential operators can give rise to the same pro-smooth map between the infinite jet bundles by prolongation, and yet not share the property that they are local maps}.
\end{ej}

We finish this section with some conclusions, claiming that [DO.1], [DO.2] and [DO.3] are not answered positively in full generality:

\begin{itemize}
	\item There are differential operators which are not local maps.
	\item In an extended local map, the map between the infinite jets does not determine the map upstairs.
	\item There are local maps that do not induce a map between the associated \'etal\'e spaces.
\end{itemize}

Developing a right notion of Cartan-preserving local maps solves not only [DO.2] but also the other two issues. The rest of the chapter is devoted to do so and hence to find ways to overcome those problems.


\section{Local maps along the identity}

{\it In this section we study a special case of local maps and differential operators: those covering the identity on the base manifold. We show that in the case of products with the identity on the base manifold (maps along the identity), differential operators are equivalent to local maps (a result by Kock) and they satisfy some other properties. The main reference at this point is Kock \cite{K}.}\\

We want to put our focus on a special kind of local maps and differential operators: those covering the identity. Recall that a map $f \colon \E \times M \rightarrow \F \times M$ is called a local map covering the identity if $f = (f_{\F}, \textrm{pr}_M)$. Maps of this kind are the easiest case in which we can explore the relation among local maps and differential operators --and this has been done already in the literature--. This section does not mention at all the Cartan-preserving property that we are trying to investigate, but it gives a flavor of what is going to happen next. 

\begin{df}[Extended local maps covering the identity]\label{lmai} Consider the subcategory of extended local maps where we only allow objects such that the base manifold is a given smooth manifold $M$ and the morphisms are of the kind $(f_{\F}, \textrm{pr}_M) \colon \E \times M \rightarrow \F \times M$. It will be denoted by $\blMan_M$. 

\begin{center}
\begin{tikzpicture}[description/.style={fill=white,inner sep=2pt}]
\matrix (m) [matrix of math nodes, row sep=3em,
column sep=2.5em, text height=1.5ex, text depth=0.25ex]
{ \E \times M & \F \times M \\
  j^{\infty}(\E \times M)  & j^{\infty}(\F \times M) \\
 M & M\\};
\path[->,font=\scriptsize]
(m-1-1) edge node[auto] {$(f_{\F},\textrm{pr}_M)$} (m-1-2)
(m-1-1) edge node[auto] {$j^{\infty}$} (m-2-1)
(m-2-1) edge node[auto] {$f^{\infty}$} (m-2-2)
(m-1-2) edge node[auto] {$j^{\infty}$} (m-2-2)
(m-2-1) edge node[auto] {$\pi_{\infty}$} (m-3-1)
(m-3-1) edge node[auto] {$\textrm{id}_M$} (m-3-2)
(m-2-2) edge node[auto] {$\rho_{\infty}$} (m-3-2)
(m-1-1) edge [bend right=70] node[left] {$\textrm{pr}_M$} (m-3-1)
(m-1-2) edge [bend left=70] node[right] {$\textrm{pr}_M$} (m-3-2);
\end{tikzpicture}
\end{center}
\end{df}

It is a subcategory since the identity on $\E \times M$ is of that kind (it is actually a product) and the composition of two such maps has the same property. Observe that $\E \times M$ can be considered as a trivial bundle over $M$ and maps in $\blMan_M$ are bundle maps over $\textrm{id}_M$ and so they are all the intermediate maps $f^l \colon J^{k(l)} E \rightarrow J^l F$. The notation $\blMan_M$ refers to the fact that all objects and maps are not only local, but also bundles over $M$.

Most, but not all, of the local maps that we are interested can be thought of as local maps of this kind. We give an example:

\begin{ej} Consider $E \rightarrow M$ a smooth fiber bundle whose space of sections is provided with a representation on $\Diff(M)$ ($l \colon \E \rightarrow \Diff(M)$). Imagine $\E$ also acts on the space of smooth sections of another bundle $F \rightarrow M$. Local Lie group actions (we will study them in Part \ref{lla}) are local maps of the kind

\begin{center}
    \begin{tabular}{rcl}
    $f \colon \E \times \F \times M $ & $\longrightarrow$ & $\F \times M$ \\
    $(\varphi, \psi, x)$ & $\longmapsto$ & $\left( \varphi \cdot \psi, l(\varphi)(x) \right)$. \\
    \end{tabular}
\end{center}

Even if $f$ does not look like a map in $\blMan_M$, we can decompose $f$ into a map along the identity and an evaluation:

\begin{center}
    \begin{tabular}{rclcl}
    $\E \times \F \times M $ & $\stackrel{\widetilde{f}}{\longrightarrow}$ & $\F \times \C(M, M) \times M$ & $\stackrel{\textrm{ev}}{\longrightarrow}$ & $\F \times M$\\
    $(\varphi, \psi, x)$ & $\longmapsto$ & $\left( \varphi \cdot \psi, l(\varphi), x \right)$ & & \\
    & & \hfill $\left( \varphi^{\prime}, f, x \right)$ & $\longmapsto$ & $\left( \varphi^{\prime}, f(x) \right)$. \\
    \end{tabular}
\end{center}
\end{ej}

Recall that in the previous section we had identified some relevant questions regarding the relation between differential operators and local maps. We can define differential operators along the identity in the same way as we have defined local maps. Given $(f, f^0)$ a differential operator along the identity, recall that the questions that we are interested in are the following:

\begin{enumerate}
	\item[DO.1] If $j^{\infty} f^0$ (the infinite jet prolongation of $f^0$) exists, is it covered by $f$?
	\item[DO.2] Is there a notion of Cartan preserving local maps so that we can recover $f$ from $f^0$?
	\item[DO.3] Is the map sending $[(\varphi, x)]$ to $[f(\varphi, x)]$ well defined? (By $[-]$ we mean the germ class of the pair $(\varphi, x)$.)
\end{enumerate}

Proposition \ref{pro} precisely states that jet prolongations exist for bundle maps covering a diffeomorphism (holonomic-jet prolongations). Recall that in Example \ref{con} we looked at a differential operator covering the identity, which failed to satisfy all the three properties. That means that even for local maps covering the identity [DO.1], [DO.2] and [DO.3] cannot be always answered positively. The problem, once again, is that in a differential operator $(f, f^{0})$ we know $f_{\F}(\varphi, x)$ at $x$ thanks to $f^0$ but not at any other point: if $y$ is near $x$, $f^0$ only gives information at $y$ about $f_{\F}(\varphi, y)$ but not about $f_{\F}(\varphi, x)$. When $f = f_{\F} \times f_M$ this problem no longer shows up: 

\begin{pp}\label{cinco} Let $f = f_{\F} \times \textrm{id}_M \colon \E \times M \rightarrow \F \times M$ be a differential operator along the identity. Let $f^0 \colon J^k E \rightarrow F$ be the map making $f$ into a differential operator. Then the following statements hold
\begin{enumerate}
	\item Infinite jet prolongations of $f^0$ exist and are covered by $f$ (i.e., [DO.1] holds). 
	\item $j^{\infty} f^0$ preserves the Cartan distribution and we can recover $f$ from $f^0$, at least locally (i.e., [DO.2] holds).
	\item $f$ induces a map between the associated \'etal\'e spaces (i.e., [DO.3] holds).
	\item $f$ is a local map. 
	\item $f_{\F} \colon \E \rightarrow \F$ is smooth if $M$ is compact.
\end{enumerate}
\end{pp}

\dem $4$ follows from $1$, and $5$ will be discussed separately in Corollary \ref{underr}. As matter of fact, we will see that $2$ is also a consequence of $1$. We start then by showing that $f$ induces a map between the associated \'etal\'e spaces, which is $3$. Let then $(f,f^0)$ be a differential operator along the identity. Consider $[(\varphi, x)] \in \widehat{E}$ a germ of a section at a point. Let $\varphi, \varphi^{\prime} \in \E$ be two sections representing the germ (this means that there exists $U \subset M$ open neighborhood of $x$ such that $\varphi$ and $\varphi^{\prime}$ coincide in $U$). Then $f_{\F}(\varphi)$ and $f_{\F}(\varphi)$ have the same germ at $x$. This is so because at every point $y \in U$, the two sections have the same finite jet evaluations at $y$. This implies that: 
\begin{equation}\label{jj}
f_{\F}(\varphi)(y) = j^0 \circ f (\varphi, y) = f^0 \circ j^k (\varphi, y) = f^0 \circ j^k (\varphi^{\prime}, y) = f_{\F}(\varphi)(y).
\end{equation}
That means that the map sending $[(\varphi, x)]$ to $[f(\varphi, x)]$ is well defined (showing $3$).
Now, we want to prove $1$: applying Proposition \ref{pro}, $j^{\infty} f^0$ exists and it is given by 
\begin{center}
    \begin{tabular}{rcl}
    $j^{\infty} f^0 \colon J^{k+l} E$ & $\longrightarrow$ & $J^l F$ \\
    $j_x^{\infty} \varphi$ & $\longmapsto$ & $j_{x}^{\infty} (f^0 \circ j^k \varphi)$. \\
    \end{tabular}
\end{center}
Examining equation \ref{jj}, $f^0 \circ j^k \varphi$ has also the same germ as $f(\varphi)$ around $x$. Since jet evaluations factor through the germ evaluation, we have that $j_{x}^{\infty} (f^0 \circ j^k \varphi) = j_x^{\infty} (f_{\F}(\varphi))$ showing that $j^{\infty} f^0$ is covered by $f$ (claim $1$) and hence that $(f, j^{\infty} f^0)$ is an extended local map (claim $4$).

Now, in order to show $2$ we need to check if $(j^{\infty} f^0)^* \mathtt{C} \subset \mathtt{C}$. This is equivalent to $(j^{\infty} \varphi)^* (j^{\infty} f^0)^* \mathtt{C} = 0$ for all $\varphi \in \E$. Observe that $j^{\infty} f^0 \circ j^{\infty}\varphi \colon M \rightarrow J^{\infty} F$ maps $y$ into $j_{x}^{\infty} (f^0 \circ j^k \varphi)$ and by the previous calculation that equals $j_y^{\infty} f_{\F}(\varphi)$. In other words, we have shown that $j^{\infty} f^0 \circ j^{\infty}\varphi = j^{\infty} f_{\F}(\varphi)$ and by Lemma \ref{XI}, $(j^{\infty} f^0 \circ j^{\infty}\varphi)^* \mathtt{C} = 0$. We simply have to put this together with our previous equation to conclude that indeed $j^{\infty} f^0$ preserves the Cartan distribution.

This is not all there is to show in $2$, we also want to recover $f$ from $f^0$ locally. But this is ensured precisely by equation \ref{jj} in this proof. Locally $f_{\F}(\varphi) = f^0 \circ j^k \varphi$.
\qed

One of the conclusions we can draw from the previous proposition is that differential operators along the identity can be recovered from the bundle map along the identity $f^0 \colon J^k E \rightarrow F$. This is nothing new, Kock \cite{K} proves the same result but with a difference in language and assumptions. Kock distinguishes between {\it sheaf theoretic differential operators} (differential operators along the identity in our language) and {\it bundle theoretic differential operators} (bundle maps $f^0 \colon J^k E \rightarrow F$ along the identity). He shows that under the assumption of $\E$ being soft (so that $j^{\infty}(\E \times M) = J^{\infty} E$) these two notions agree. We simply have replaced the softness condition on $\E$ by working with the image of the jet evaluations.

\begin{rk}\label{ilcd}
Observe that the key hypothesis has been that the map $f$ is {\it along} the identity and not only {\it covering} the identity. This has to do precisely with the fact that $j^{\infty} f^0$ preserves the Cartan distribution. Recall that heuristically a map preserves the Cartan distribution if and only if it sends each jet prolongation $j^{\infty} \varphi$ to the union of several jet prolongations 
\begin{equation}\label{aalpha}
\bigcup_{a \in A} j^{\infty} \psi_{a}^{\varphi} \qquad \textrm{where } \varphi \in \E \textrm{ and } \psi_{a}^{\varphi} \in \F \quad \forall a \in A.
\end{equation}
In the case the map covers the identity there is only room for a single section on the right of equation \ref{aalpha}. That means that $j^{\infty} f^0$ should map $j^{\infty} \varphi$ to $j^{\infty} \psi^{\varphi}$. Imagine now that the map $j^{\infty} f^0$ covers a map between the total spaces $f \colon \E \times M \rightarrow \F \times M$. The most straightforward condition that such $f$ should satisfy in order to define a local map $(f, j^{\infty} f^0)$ is that $f_{\F}(\varphi,x) = \psi^{\varphi}$ for all $\varphi \in \E$. This is precisely that $f_{\F}$ is independent of $M$, in other words $f$ is a local map {\it along} the identity.
\end{rk}

Using those results, we can consider extended local maps along the identity to be part of a category (this result can be found, for instance in the book by Kock, \cite{K}):

\begin{dfpp}[Category of differential operators along the identity]\label{maybe} Given $M$ a smooth manifold, we define the category $\Diff_M$ whose objects are $\E$ spaces of sections of smooth fiber bundles over $M$ and morphisms are extended local maps along the identity $\E \rightarrow \F$ as in Definition \ref{lmai}. Given two maps $(f,f^0)$ and $(g, g^0)$ the composition $g \circ f$ is given by $(g \circ f, g^0 \circ j^l f^0)$ where $l$ is the degree of $g^0$.
\begin{center}
\begin{tikzpicture}[description/.style={fill=white,inner sep=2pt}]
\matrix (m) [matrix of math nodes, row sep=3em,
column sep=2.5em, text height=1.5ex, text depth=0.25ex]
{ \E \times M & \F \times N & \G \times P \\
   j^{k+l} (\E \times M) & j^{l} (\F \times N) & j^0 (\G \times P)  \\
   j^{k} (\E \times M) & j^0 (\F \times N)  &  \\};
\path[->,font=\scriptsize]
(m-1-1) edge node[auto] {$f$} (m-1-2)
(m-1-2) edge node[auto] {$g$} (m-1-3)

(m-1-1) edge node[auto] {$j^{k+l}$} (m-2-1)
(m-2-1) edge node[auto] {$\pi_{k+l}^{l}$} (m-3-1)
(m-1-2) edge node[auto] {$j^{l}$} (m-2-2)
(m-2-2) edge node[auto] {$\rho_{l}^0$} (m-3-2)
(m-1-3) edge node[auto] {$j^{0}$} (m-2-3)

(m-3-1) edge node[auto] {$f^0$} (m-3-2)
(m-2-1) edge node[auto] {$j^l f^0$} (m-2-2)
(m-2-2) edge node[auto] {$g^0$} (m-2-3);
\end{tikzpicture}
\end{center} 
\end{dfpp}

We can reinterpret some of the claims in Proposition \ref{cinco} to induce functors out of $\Diff_M$. The whole picture is the following:

\begin{center}
\begin{tikzpicture}[description/.style={fill=white,inner sep=2pt}]
\matrix (m) [matrix of math nodes, row sep=1em,
column sep=5em, text height=1.5ex, text depth=0.25ex]
{  & \blMan_M \\
  \Diff_M  & \Fr\Man \\
 & \Sh_M \\};
\path[->,font=\scriptsize]
(m-2-1) edge node[auto] {} (m-1-2)
(m-2-1) edge node[auto] {} (m-2-2)
(m-2-1) edge node[auto] {} (m-3-2);
\end{tikzpicture}
\end{center}

The top one is given by sending $(f, f^0)$ to $(f, j^{\infty} f^0)$ and it is fully faithful. The second one is given by forgetting about $f^0$ and will be justified by Corollary \ref{underr}. The third one is also given by forgetting about $f^0$ and the relation will only be explained in in chapter \ref{PT}.

The main idea in this part is to develop conditions under which we have the same behavior for other kinds of local maps, not only local maps covering the identity. As mentioned above, we are also interested in a question about smoothness:
\begin{itemize}
	\item{[S.1]} Are there reasonable assumptions so that some extended local maps are smooth?
\end{itemize}


\newpage
\chapter{Insularity}\label{ins}

The notion of Cartan-preserving maps is central for the discussion on infinite jet prolongations as we have see in Section \ref{jpg}. There is an analogous version of such concept for maps $f\colon \E \times M \rightarrow \F \times N$. It involves taking sections of the map $f_N(\varphi, -)$ for each field $\varphi$. As an example, differential operators along the identity are Cartan-preserving.\\

Insular maps $f \colon \E \times M \rightarrow \F \times N$ are provided with a local family of sections for the maps $f_N(\varphi, -)$ and are such that $f_{\F}(\varphi, -)$ is locally constant along the image of the sections for all $\varphi \in \E$. Insular maps look locally like a very specific kind of maps: globally insular maps. For globally insular maps both the source and the target are given by bundles over Euclidean spaces $M$ and $N$. The map $f_N$ is a projection to a Euclidean subspace and the map $f_{\F}$ is constant along the fibers.\\

Insular maps are Cartan-preserving maps with the extra property that the sections of $f_N(\varphi, -)$ can be put together into a local map. That is why they are called that way: insular maps are maps which are even more local than usual. Insular maps are smooth provided $N$ is compact.\\

{\it In this chapter we introduce the concept of Cartan-preserving maps $\E \times M \rightarrow \F \times N$. A local interpretation of the differential operators satisfying that property is given using globally insular differential operators. Insular maps are introduced, as a local version of Cartan-preserving maps. We give several examples and counterexamples of all the implications among the different concepts. We prove that insular maps are smooth provided $N$ is compact. There are no relevant references in the literature at this point. }


\section{Cartan-preserving maps}

{\it In this section we introduce Cartan-preserving maps $\E \times M \rightarrow \F \times N$, a related concept to that of pro-smooth maps preserving the Cartan distribution. We show that extended local maps $(f, f^{\infty})$ which are also Cartan-preserving are such that infinite jet prolongations of $f^{0}$ exist. We give examples, especially focusing on maps covering the identity which are also Cartan-preserving: those are actually maps along the identity.}\\

We want a class of extended local maps for which we give positive answers to questions [DO.1], [DO.2], [DO.3], and [S.1]. We first start by developing a notion of Cartan-preserving local maps so that question [DO.2] can be answered affirmatively. The starting point is Remark \ref{ilcd} were we discussed the basics for local maps along the identity. We fix $\pi \colon E \rightarrow M$ and $\pi \colon F \rightarrow N$ two smooth fiber bundles with spaces of sections $\E$ and $\F$ respectively.

Instead of directly dealing with the Cartan distribution, let us start with the Cartan foliation. Remember that the leaves of the Cartan foliation on $J^{\infty} E$ were of the kind $j^{\infty} \varphi (M)$. A map $f^{\infty} \colon J^{\infty} E \rightarrow J^{\infty} F$ was said to preserve the Cartan foliation if $f^{\infty} j^{\infty} \varphi (U) = \bigcup_{a} j^{\infty} \psi_{a} (V)$ for every local section (Remark \ref{genius}). Equivalently we will say that $f \colon \E \times M \rightarrow \F \times N$ {\it preserves the Cartan foliation} if for every section $\varphi \in \E$ and for every $U \subset M$ open we have that $f(\varphi,-) (U) = \bigcup_{a \in A} (\psi_{a}) \times V$ for some family of sections $\psi_{a} \in \F$, and some open set $V \subset N$. 

By exploring such maps we come to the conclusion that the easiest way in which that can happen is when $f_N(\varphi,- )(U)=V$ (in particular $f_N(\varphi,-)$ is open for all $\varphi \in \E$) and when there exists $\iota_{\varphi, \alpha} \colon V \rightarrow U$ such that $U_{\varphi}^{a} \defeq \iota_{\varphi, a}(V)$ foliate $U$ and $\restrict{f_{\F}(\varphi, -)}{U_{\varphi}^{a}}$ is the singleton $\psi_{a}$. Since we want everything to be smooth at the level of maps between $M$ and $N$ we can adapt the previous discussion and give the following definition:

\begin{df}[Cartan-preserving map]\label{cpm} A map $f \colon \E \times M \rightarrow \F \times N$ is said to be Cartan-preserving if:
\begin{itemize}
	\item $f_N(\varphi,-)$ is a submersion for each $\varphi \in \E$.
	\item There exist smooth local trivializations of $f_N(\varphi,-)$ around any point $x \in M$ such that $f_N(\varphi, -) \colon V \times A \rightarrow V$ and $f_{\F}(\varphi, -)$ is a singleton at every $V \times \{a\}$, for some transverse fiber $A$.
\end{itemize} 
\end{df}

To be precise, the second point asks for the existence of $U$ an open neighborhood of $x$ in $M$, $U$ to be foliated by $\bigcup_{a \in A} U_{\varphi}^{a}$ where each $U_{\varphi}^{a}$ is diffeomorphic to $V \defeq f_N(\varphi, -)(U)$ and such that for every $a$, $\restrict{f_{\F}(\varphi, -)}{U_{\varphi}^{a}}$ is a singleton $\psi_{a}$. Hence $f(\varphi,-) (U) = \left( \bigcup_{a \in A} \psi_{\alpha} \right) \times V$. In other other words, there exists a family of smooth maps of the kind $\iota_{\varphi, a} \colon V \rightarrow U$, sections of $f_N(\varphi,-)$ whose images foliate $U$ and such that $f_{\F}(\varphi, \iota_{\varphi, a}(V))$ is a singleton (to compare with the previous notation, $U_{\varphi}^{a} \defeq \iota_{\varphi, a}(V)$).

Fix $\varphi \in \E$. For simplicity of the following argument, let us assume that $M$ is connected and hence that we can glue the sets $\{U_{\varphi}^{a}\}$ into a global foliation of $M$ denoted by $\{M_{\varphi}^{a} \}_{a \in A}$ such that
\begin{itemize}
	\item $\restrict{f_{N}(\varphi, -)}{M_{\varphi}^{\alpha}} \colon M_{\varphi}^{\alpha} \rightarrow N$ is a local diffeomorphism for all $\alpha \in A$.
	\item $\restrict{f_{\F}(\varphi, -)}{M_{\varphi}^{\alpha}} \colon M_{\varphi}^{\alpha} \rightarrow \{\psi_{\varphi}^{\alpha}\} \in \F$ is constant for all $\alpha \in A$.
\end{itemize}

Such a foliation is called an {\it island} associated to $\varphi$. Giving an island for every $\varphi \in \E$ is equivalent to show that the map is Cartan preserving.


\begin{ej} We consider a Cartan-preserving map $f$ such that $M = N$ and that $f_N(\varphi,-)$ is a local diffeomorphism for every $\varphi \in \E$. The only possible local section of $f_N(\varphi, -)$ is $f_N(\varphi,-)^{-1}$. The fact that $f_{\F}$ is constant along the image of the section is equivalent to say that $f_{\F}$ is independent of the second variable.
\end{ej} 

\begin{ej}\label{e3}  For a Cartan preserving map $f$ such that $M = N$ and that $f_N = \textrm{pr}_M$ is of the previous kind, so that $f_{\F}$ is independent of the second variable and $f = f_{\F} \times \textrm{id}_M$ is a differential operator along the identity as in Definition \ref{maybe}. This was already pointed ou in Remark \ref{ilcd}. As a conclusion we have that:

Cartan-preserving maps covering the identity are precisely differential operators along the identity, that is: product maps $f = f_{\F} \times \textrm{id}_M \colon \E \times M \rightarrow \F \times M$. In particular, they are local.
\end{ej} 

But up to this moment, we do not have enough information about Cartan-preserving maps to decide whether or not they are local in general. That would be discussed in the following chapter. Up to now, what we can do instead is to assume that the map is already local and see if we can extract some interesting conclusions.

The existence of such sections $\iota_{\varphi, a}$ should be very reminiscent of what was happening for jet prolongations. If we define for every $x \in U$, $\iota_{\varphi, x} \defeq \iota_{\varphi, a}$ where $a \in A$ is the unique index such that $x \in U_{\varphi}^a$ we get that $\iota_{\varphi, x}(f_N(\varphi, x)) = x$ and that $\iota_{\varphi, \iota_{\varphi,x}(y)} = \iota_{\varphi, x}$ for all $y \in V$. Those were precisely the assumptions we developed in Remark \ref{genius} to say that a map $f^{\infty} \colon J^{\infty} E \rightarrow J^{\infty} F$ preserved the Cartan foliation. The idea is that if $(f, f^{\infty})$ is a local map, asking $f$ to be Cartan-preserving ensures $f^{\infty}$ to preserve the Cartan distribution. Moreover,

\begin{pp}\label{finhoy} If an extended local map $(f,f^{\infty})$ is Cartan-preserving then, given $\{f^l\}$ a representative of $f^{\infty}$, infinite jet prolongations of $f^0 \colon J^k E \rightarrow F$ in the sense of Definition/Proposition \ref{PRO1} exist, they are pro-smooth and $f^{\infty} = j^{\infty} f^0$:
\begin{center}
    \begin{tabular}{rcl}
    $f^{\infty} \colon J^{k+l} E$ & $\longrightarrow$ & $J^l F$ \\
    $j_x^{\infty} \varphi$ & $\longmapsto$ & $j_{\rho \circ f^0 \circ j^{k} \varphi (x)}^{\infty} (f^0 \circ j^k \varphi \circ \iota_{\varphi,x})$. \\
    \end{tabular}
\end{center}
\end{pp} 

\dem The conclusions are precisely the same as those in Proposition \ref{PR}. What we are going to do is to prove that the hypothesis in that proposition follow from the ones here, with the only difference that now we want everything to be defined on the image of the jet evaluations instead that on the whole jet bundles:
\begin{enumerate}
\item For all sections $\varphi \in \E$ the map $\tau_{\varphi} \defeq  \rho_{\infty} \circ f^{\infty} \circ j^{\infty} \varphi \colon M \rightarrow N$ is a submersion.
\item $f^* (\mathtt{C}(J^{\infty} F)) \subset \mathtt{C}(J^{\infty} E)$.
\end{enumerate}
$1$ is easy, since $(f ,f^{\infty})$ is extended local, for all $\varphi \in \E$ we have
$\rho_{\infty} \circ f^{\infty} \circ j^{\infty} = f_N(\varphi, -)$ which is a submersion by hypothesis. In order to prove $2$ first observe that for all $(\varphi, x) \in \E \times M$ there is $U$ open neighborhood of $x$ such that
$$f^{\infty} j^{\infty} \varphi (U) = f^{\infty} \circ j^{\infty} (\varphi, U) = j^{\infty} \circ f (\varphi, U) = j^{\infty} \left( \bigcup_{a \in A} (\psi_{a}) \times V \right) = \bigcup_{a \in A}  \psi_{a}( V) .$$
Now, by a general argument, since $f^{\infty}$ preserves the Cartan foliation, it also preserves the Cartan distribution, showing $2$. To be precise, for all $\omega \in \mathtt{C}$ we have that $$(j^{\infty} \varphi)^* \circ (f^{\infty})^* \omega = \sum_{a \in A} (j^{\infty} \psi_{a})^* \omega = 0.$$
\qed

We want to point out that the assumption of $f_N$ being a submersion might have important consequences:

\begin{ej}\label{e4} If $f$ is a product, $f = f_{\F} \times f_N$, the condition $f_N(\varphi, -)$ being a submersion for every $\varphi \in \E$ is equivalent to say that $f_N \colon M \rightarrow N$ is a submersion. We distinguish two cases:

\begin{enumerate}
	\item In the case $M = N$ and $\tau$ a diffeomorphism. This was treated in Example \ref{e3}.
	\item Consider $f_N$ to be a strict submersion, namely it locally looks like a projection $\RE^m \times \RE^n \rightarrow \RE^m$ with $n \geqslant 1$. In this case, differential operators are very limited (and thus local maps). For simplicity take the following $0$--degree differential operator between trivial vector bundles on Euclidean spaces:
	
\begin{center}
\begin{tikzpicture}[description/.style={fill=white,inner sep=2pt}]
\matrix (m) [matrix of math nodes, row sep=2.5em,
column sep=3.5em, text height=1.5ex, text depth=0.25ex]
{ \C(\RE^m \times\RE^n, \RE^p) \times \RE^m \times \RE^n & \C(\RE^m, \RE^q) \times \RE^m \\
  \RE^m \times\RE^n \times \RE^p & \RE^m \times \RE^q \\};
\path[->,font=\scriptsize]
(m-1-1) edge node[auto] {$A \times \textrm{pr}_{\RE^m}$} (m-1-2)
(m-2-1) edge node[auto] {$\widehat{A}= (\textrm{pr}_{\RE^m}, g)$} (m-2-2)
(m-1-1) edge node[auto] {$j^0$} (m-2-1)
(m-1-2) edge node[auto] {$j^0$} (m-2-2);
\end{tikzpicture}
\end{center}

We want to argue that $g$ depends only on $\RE^m$. Take any triple $(x,y,z)$ in $\RE^m \times\RE^n \times \RE^p$ and represent it as the $0$--th jet of the constant function with value $z$, call it $Z$. Now
\begin{eqnarray*}
g(x,y,z) &=& \textrm{pr}_{\RE^q} \circ j^0 \circ (A \times \textrm{pr}_{\RE^n}) (Z, x, y) = A(Z)(x) \\
&=& j^0 \circ (A \times \textrm{pr}_{\RE^n}) (Z, x, y^{\prime}) = g(x, y^{\prime}, z).
\end{eqnarray*}
This shows that $g$ is independent of $y$. To show that it is also independent of $z$ we should take some section depending on $y$. For instance, represent $(x,y,z)$ by the function $s_z$ sending $(x,y)$ to $(z_1, \cdots, y_1 \times z_i, \cdots, z_q)$ (this is possible since $n \geqslant 1$). Take $z \neq z^{\prime}$, assume $z_i \neq z_i \neq 0$. Now
\begin{eqnarray*}
g(x,y,z) &=& \textrm{pr}_{\RE^q} \circ j^0 \circ (A \times \textrm{pr}_{\RE^n}) (s_z, x, y) = A(s_z)(x) \\
&=& j^0 \circ (A \times \textrm{pr}_{\RE^n}) (s_{z}, x, y^{\prime}) = g(x, y^{\prime}, z^{\prime}) = g(x,y,z^{\prime})
\end{eqnarray*}
{\noindent where $y^{\prime} = (z_i^{\prime} / z_i, y_2, \cdots, y_n)$.}

This shows that $g$ is a function $\RE^m \rightarrow \RE^q$. In particular $A(\varphi) = g$ for all $\varphi \in \C(\RE^m \times\RE^n, \RE^p)$. We have shown the simplest version of the following fact: {\it any differential operator which is a product of a map $A$ and a strict submersion is such that $A$ is constant}.
\end{enumerate}
\end{ej}


\section{Globally insular differential operators: a local mo\-del}

{\it In this section we define globally insular differential operators. Those are local models for differential operators that preserve the Cartan foliation. We show that globally insular maps are Cartan-preserving and that any operator which looks locally like globally insular differential operators is Cartan-preserving (and vice versa). We consider examples of both kinds of maps.}\\

We expect local maps to satisfy several properties, such as being smooth and having a unique jet prolongation of $f^0$ still covered by $f$. We will define maps which satisfy these conditions: they will be Cartan-preserving maps with a stronger notion of locality, that is why we will call them {\it insular}. In particular, we will be interested in differential operators $(f, f^0)$ such that $f$ preserves the Cartan distribution. So far, the term Cartan-preserving map has been introduced using the local Cartan foliation. In this section we would like to give a local description of such differential operators, in the sense that Cartan-preserving differential operators have charts mapping to the local models. Those local models are, as a matter of fact, insular, as we will see in the following sections.

\begin{ej}\label{e2} Consider the local map 
\begin{center}
    \begin{tabular}{rcl}
    $f \colon \C(\RE^2) \times \RE^2$ & $\longrightarrow$ & $\C(\RE) \times \RE$ \\
    $\left(\varphi, (x,y)\right)$ & $\longmapsto$ & $\left( \left[ x^{\prime} \mapsto f(x^{\prime}, y) \right], x \right)$. \\
    \end{tabular}
\end{center}
The map is local with $f^l \colon J^l(\RE^2 \times \RE) \rightarrow J^l (\RE \times \RE)$ given by forgetting the derivatives in the $y$ direction. The map even descends to a morphism of pro-nanofolds in the sense of Blohmann \cite{B}.

Observe that $f_{\RE}(\varphi, -) \colon \RE^2 \rightarrow \RE$ mapping $(x,y) \mapsto x$ is a submersion and for any $y \in \RE$ we have a section $\iota_{\varphi, y} \colon \RE \rightarrow \RE^2$ sending $x^{\prime}$ to $(x^{\prime}, y)$ (in this case the index set $A$ is $\RE$). The section $f_{\C(\RE)}(\varphi, -)$ is simply the composition of $\iota_{\varphi, y}$ with $\varphi$. The map is clearly an extended local map which is also Cartan-preserving. It follows from the study done in the previous chapter that the higher maps $\{f^l\}$ can be recovered from the bottom map $f^0 \colon \RE^2 \times \RE \rightarrow \RE \times \RE$ (which is given by $(x,y,z) \mapsto (x,z)$):
\begin{center}
    \begin{tabular}{rcl}
    $j^l f \colon J^l E$ & $\longrightarrow$ & $J^l F$ \\
    $[(\varphi ,x)]$ & $\longmapsto$ & $[(f^0 ( \varphi \circ \iota_{\varphi, y}, x )] = [(\varphi \circ \iota_{\varphi, y}, x)]$.\\
    \end{tabular}
\end{center}
\end{ej}

We start by looking more closely to Example \ref{e2}: we have that $f_{\F}$ is a restriction of $\varphi$ to a submanifold of $M$. By making this a bit more involved, we can think of maps in which $f_{\varphi}$ is the restriction of $j^k \varphi$ to a submanifold of $M$.

\begin{df}[Globally insular differential operator]\label{gim} Let $E \rightarrow \RE^{n+m}$ and $F \rightarrow \RE^n$ be two smooth fiber bundles. Denote by $\textrm{pr}_n \colon \RE^{n+m} \rightarrow \RE^n$ the projection to the first $n$ components.

Given $\varphi \in \E$, a map $f(\varphi) \colon \RE^{n+m} \rightarrow \F \times \RE^n$ is called a globally insular at $\varphi$ if $f(\varphi)_{\RE^n} = \textrm{pr}_n$ and there exists a smooth map $f^0 \colon J^k E \rightarrow F$ such that $f(\varphi)_{\F}(x, y) = f^0 \circ j^k \varphi \circ \textrm{inc}_y(x)$ where $\textrm{inc}_y \colon \RE^n \rightarrow \RE^{n+m}$ maps $x^{\prime}$ to $(x^{\prime}, y)$ for all $(x,y) \in \RE^{n+m}$. Defining Calling $f (\varphi, (x,y)) \defeq f(\varphi)(x,y)$ we say that $(f, f^0)$ is a globally insular differential operator if it is a differential operator and globally insular at every section with respect to the same $f^0$.
\begin{center}
    \begin{tabular}{rcl}
    $f \colon \E \times \RE^{n+m}$ & $\longrightarrow$ & $\F \times \RE^n$ \\
    $(\varphi, (x,y))$ & $\longmapsto$ & $\left( \left[x^{\prime} \mapsto f^0 \left(j_{(x^{\prime}, y)}^k \varphi \right) \right], x \right)$. \\
    \end{tabular}
\end{center}
\end{df}

The first observation is that globally insular maps are local and Cartan-preserving. It would also be immediate to check that $f$ is smooth, but we postpone the proof to Theorem \ref{t3}.

\begin{pp} Let $f \colon \E \times \RE^{n+m} \rightarrow \F \times \RE^n$ be a globally insular differential operator. Then it is Cartan-preserving and local.
\end{pp}

\dem
For all $\varphi \in \E$ we have that $f_N(\varphi, -) = \textrm{pr}_n$ is a submersion. We have a foliation of $\RE^{n+m}$ given by the images of $\textrm{inc}_y(\RE^n)$. $f_{\F}(\varphi,-)$ is constant at every leave, to be precise $\restrict{f_{\F}(\varphi,-)}{\textrm{inc}_y(\RE^n)} = \{*\}$ is a singleton.

Since $f_{\F}$ depends directly on $f^0$ it is clear that jet prolongations exist and are covered by $f$. This means in particular that $f$ is also local.
\qed

Now we give an example of a map in which $f_{N}$ depends on $\varphi$ but gives still good properties such as [S.1] (remember that meant that the map is also smooth).

\begin{ej}\label{e1} Consider $\textrm{Diff}(\S^1) \times \C(\S^1) \subset \Gamma^{\infty}(\S^1, \S^1 \times \S^1 \times \RE)$. It is open since the space of diffeomorphisms $\textrm{Diff}(M)$ is open in $\Gamma(M, M \times M) = \C(M, M)$ with the Fr\'echet space topology for any smooth manifold $M$ (this can be found in the book by Kriegl and Michor \cite[43.1]{KM}). The map
\begin{center}
    \begin{tabular}{rcl}
    $f \colon \textrm{Diff}(\S^1) \times \C(\S^1) \times \S^1$ & $\longrightarrow$ & $\C(\S^1) \times \S^1$ \\
    $(\tau, g, x)$ & $\longmapsto$ & $\left( g \circ \tau^{-1}, \tau(x) \right)$ \\
    \end{tabular}
\end{center}
{\noindent is local (for more details about local maps of products of sections see Example \ref{fpb}) with $f^l \colon J^l(\S^1 \times \S^1) \times_{\S^1} J^l(\RE \times \S^1) \rightarrow J^l(\RE \times \S^1)$ given by fiber-wise multiplication.}

Any local map which is a diffeomorphism action on the base can be shown to be smooth (this will follow from Theorem \ref{t3}). But we can show that this one map in particular is smooth. In this case, since $f_{\C(\S^1)}$ is independent of $\S^1$ we can consider smooth maps $\gamma \colon \RE \rightarrow \textrm{Diff}(S^1) \times \C(\S^1)$, then the associated maps $\left(f_{\C(\S^1)} \circ \gamma \right)^{\smallsmile}$ are smooth because of being the composition of $\gamma_{\textrm{Diff}(\S^1)}^{\smallsmile}$ and $\gamma_{\C(\S^1)}^{\smallsmile}$ which are smooth by assumption:
\begin{center}
    \begin{tabular}{rcccl}
    $\RE \times \S^1$ & $\longrightarrow$ & $\RE \times \S^1$ & $\longrightarrow$ & $\RE$\\
    $(t,x)$ & $\longmapsto$ & $(t, \gamma_{\textrm{Diff}(\S^1)}^{\smallsmile}(t,x))$ & $\longmapsto$ & $\gamma_{\C(\S^1)}^{\smallsmile} \left(t, \gamma_{\textrm{Diff}(\S^1)}^{\smallsmile}(t,x)\right)$. \\
    \end{tabular}
\end{center}
\end{ej}

The idea is that the map from Example \ref{e1} is locally like the ones introduced in Definition \ref{gim}.

\begin{df}\label{loclike}
Let $\pi \colon E \rightarrow M$ and $\rho \colon F \rightarrow N$ be two smooth fiber bundles. A map $f \colon \E \times M \rightarrow \F \times N$ is said to look locally like a globally insular differential operator if for all $(\varphi, x) \in \E$ there exist $(u, U)$ chart around $x$ in $M$ and $(v, V)$ chart around $f_N(\varphi, x)$ in $N$ such that
\begin{eqnarray*}
g(u^* \varphi) \colon \RE^{n+m} & \longrightarrow & v^* \F \times \RE^n \\
(x,y) & \longmapsto & (v^* \times v)\left( \restrict{f}{U}(\varphi, u^{-1}(x,y) ) \right)
\end{eqnarray*}
{\noindent is globally insular at $u^* \varphi$ in the sense of Definition \ref{gim} and it has smooth transition functions for different $f^0$'s.}
\end{df}

\begin{lm}\label{lcp} Maps that locally look like globally insular differential operators are Cartan-preserving differential operators and vice versa.
\end{lm}

\dem In order to check that the map is Cartan-preserving we need to show that $f_N(\varphi,-)$ is a submersion for all $\varphi$. But that is a local property, and all submersions look like Euclidean projections as in the model. The second condition is that $f_{\F}(\varphi,-)$ is constant along the fibers of the submersion, but that is explicitly required in the local model.

The different maps $f^0$ are assumed to have smooth transition functions so that they define a smooth map $f^0 \colon J^k E \rightarrow F$  covered by $f$, showing that $(f, f^0)$ is indeed a differential operator.

In the other direction, if a map is Cartan preserving, locally it looks like the model (submersions are Euclidean projections locally) and the map $f^0$ is the same for all such local triviallizations, showing that the maps is under the assumptions of Definition \ref{loclike}.
\qed


\section{Insular maps}

{\it In this section we define insular maps and prove that they are Cartan-preserving. As a matter of fact they are simply Cartan-preserving maps with an extra locality assumption. Insular maps are examples of maps modeled in globally insular maps. We also show that insular maps are smooth when the base manifold is compact.}\\

The idea behind insular maps is the following: given a Cartan-preserving map we have identified for every $\varphi$ an island $\{M_{\varphi}^{a}\}_{a \in A_{\varphi}}$; we want all islands to depend locally on $\varphi$. Cartan-preserving differential operators with that extra property  happen to be local (among other properties) and that is why they are called insular, their locality is stronger than for general local maps: one can associate a family of local sections to $f_{N}$ (recall the islands are the images of those sections).

To be precise, we want to consider Cartan-preserving maps $f \colon \E \times M \rightarrow \F \times N$ such that there exists a smooth map $\iota \colon \E \times M \rightarrow \C(N,M)$ making the following diagram commute locally:
\begin{center}\label{LFSd}
\begin{tikzpicture}[description/.style={fill=white,inner sep=2pt}]
\matrix (m) [matrix of math nodes, row sep=4em,
column sep=8em, text height=1.5ex, text depth=0.25ex]
{ & \E \times M \\
\E \times M & \left(\E \times M\right) \times \left(\C(N,M) \times N\right) \\
& \C(N,N) \times N \\};
\path[->,font=\scriptsize]
(m-2-1) edge node[auto] {$\textrm{id}_{\E \times M}$} (m-1-2)
(m-2-1) edge node[above=0.6em, right=-2.2em] {$\textrm{id}_{\E \times M}, \, (\iota, f_N)$} (m-2-2)
(m-2-2) edge node[right] {$\textrm{pr}_{\E} \times \textrm{ev}$} (m-1-2)
(m-2-1) edge node[below=1em, left=0em] {$\textrm{ct}_{\textrm{id}_N}, f_N$} (m-3-2)
(m-2-2) edge node[auto] {} (m-3-2);

\node[font=\scriptsize] (i1) at (4,-0.5) {$(\varphi,m) \, (g, n)$};
\node[font=\scriptsize] (i2) at (4,-2.3) {$\left( [n^{\prime} \mapsto f_N(\varphi, g(n^{\prime}) )] , \, n \right)$};
\node[font=\bf] (label) at (6,0) {[LFS]};

\path[|->]
(i1) edge node[auto] {} (i2);

\end{tikzpicture}
\end{center}

We have said that the diagram has to commute {\it locally}, by this we mean that for all $(\varphi, x) \in \E \times M$ there exist $U$ an open neighborhood of $x$ in $M$ such that the diagram commutes when restricted to $\E \times U$.\footnote{We have been not so rigorous when considering $\left(\E \times M\right) \times \left(\C(N,M) \times N\right)$ as a product of $\Gamma^{\infty}(M \sqcup N, E \sqcup (M \times N)) \times (M \sqcup N)$ since we are suddenly allowing bundles over manifolds with no fixed dimension and no fixed rank. This is not an issue, we will not consider such maps later on, we have only done it that way in the diagrams {\bf [LFS]} and {\bf [cLFS]} in order to be able to summarize the information in a more compact fashion.}

The fact that the diagram commutes can be rephrased in the following terms:
\begin{itemize}
	\item The commutativity of the lower triangle expresses the fact that $\iota(\varphi, x)$ is a local section of $f_N(\varphi, -)$ for all $(\varphi, x) \in \E \times M$. That is $\restrict{f_N(\varphi,-)\circ \iota(\varphi,x)}{V} = \textrm{id}_V$ for some $V$ open neighborhood of $f_N(\varphi, x)$ in $N$.
	\item The commutativity of the upper triangle simply says that $\iota(\varphi, x)$ passes through $x$. That is $\iota(\varphi, x) (f_N(\varphi,x))=x$.
\end{itemize}

\begin{df}[Local family of sections]\label{LFS}
A map $f \colon \E \times M \rightarrow \F \times N$ is said to have a local family of sections for $f_N$ if $f_N(\varphi, -)$ is a submersion for all $\varphi \in \E$ and there exists a local map $I =(\iota, f_N) \colon \E \times M \rightarrow \C(N,M) \times N$ making the diagram {\bf [LFS]} locally commute.
\end{df}

In the case in which $f$ has a local family of sections, we are interested in the case in which $f_{\F}(\varphi, -)$ is constant along the image of the sections. Again we express this fact in another locally commutative diagram. Let $n_0 \in N$:
\begin{center}\label{cLFSd}
\begin{tikzpicture}[description/.style={fill=white,inner sep=2pt}]
\matrix (m) [matrix of math nodes, row sep=4em,
column sep=8em, text height=1.5ex, text depth=0.25ex]
{ \E \times M & \left(\E \times M\right) \times \left(\C(N,M) \times N\right) \\
& \F \times N \\};
\path[->,font=\scriptsize]
(m-1-1) edge node[above=0.6em, right=-2.2em] {$\textrm{id}_{\E \times M}, \, (\iota, f_N)$} (m-1-2)
(m-1-1) edge node[below=1em, left=0em] {$f$} (m-2-2)
(m-1-2) edge node[auto] {$f^{\prime}$} (m-2-2);

\node[font=\scriptsize] (i1) at (3, 0.6) {$(\varphi,m) \, (g, n)$};
\node[font=\scriptsize] (i2) at (3,-1.3) {$\left( f_{\F}(\varphi, g(n_0)), \, n \right)$};
\node[font=\bf] (label) at (6,0) {[cLFS]};

\path[|->]
(i1) edge node[auto] {} (i2);

\end{tikzpicture}
\end{center}

Again, we give an explicit interpretation of the fact that the diagram commutes: 
\begin{itemize}
	\item The diagram {\bf [cLFS]} says that $f_{\F}(\varphi, -)$ is constant along the image of the local sections, in other words $f_{\F}(\varphi, \iota(\varphi, x)(V))$ is a singleton for all $(\varphi, x) \in \E \times M$ ($V$ some open set since the diagram is supposed to commute locally).
\end{itemize} 

\begin{df}[Insular map]\label{cLFS}
A map $f \colon \E \times M \rightarrow \F \times N$ is said to be insular if there is $\iota \colon \E \times M \rightarrow \C(N,M)$ a local family of sections as in Definition \ref{LFS} making the diagram {\bf [cLFS]} commute.
\end{df}

The clear implication is the following:

\begin{pp}\label{icp} Insular maps are Cartan-preserving.
\end{pp}

\dem
We fix an insular map $f \colon \E \times M \rightarrow \F \times N$ and a point $(\varphi, x) \in \E \times M$. By assumptions $f_N(\varphi, -)$ is a submersion. The local section $\iota_{\varphi, x} \defeq \iota(\varphi,x) \in \C(N, M)$ is a section of $f_{N}(\varphi, -)$ such that $f_{\F}(\varphi, \iota_{\varphi,x}(V))$ is constant for some open neighborhood $V$ of $f_N(\varphi,x)$. 

Moreover, the fact that $f_{\F}(\varphi, -)$ is constant along the image of $\iota_{\varphi,x}$ tells us precisely that $f_{\F}(\varphi, -)$ is constant at every leaf, so that $\{M_{\varphi}^{a}\}_{a \in A}$ is an island associated to $\varphi$.
\qed

As a Corollary we have that insular maps look locally like globally insular maps (using Lemma \ref{lcp}):

\begin{cl} Insular maps look locally like globally insular maps in the sense of Definition \ref{loclike}.
\end{cl}

Insular maps satisfy [S.1], one of the key properties we wanted our maps to meet.

\begin{tm}[Insular maps are smooth]\label{t3} Let $f \colon \E \times M \rightarrow \F \times N$ be an insular map with smooth local family of sections. If $N$ is compact, then $f$ is smooth.
\end{tm}

\dem As discussed in Section \ref{cinpudos}, a local map is smooth if and only if for all $\gamma \colon \RE \rightarrow \E \times M$ the composition with $f_{\F}$ is again smooth, that is, if $(f_{\F} \circ \gamma)^{\smallsmile} \colon \RE \times N \rightarrow F$ sending $(t,n)$ to $f_{\F}(\gamma(t))(n)$ is smooth. In order to show so we are going to use the local family of sections $\iota \colon \E \times M \rightarrow \C(N,M) \times N$:

Consider the following map
\begin{center}
    \begin{tabular}{rcl}
    $\gamma^{\prime} \colon \RE \times N$ & $\longrightarrow$ & $\E \times M$ \\
    $(t,n)$ & $\longmapsto$ & $\left( \gamma_{\E}(t), \, \left( \iota \circ \gamma (t) \right) (n) \right)$ \\
    \end{tabular}
\end{center}

The map $\gamma^{\prime}$ is smooth since the first component is given by $\gamma_{\E}$ (which is smooth by assumption), and the second one is simply the $\left(\iota \circ \gamma \right)^{\smallsmile}$ which is smooth since we have assumed the local family of sections being smooth.

Using the fact that $\iota$ is a family of sections of $f_N$ we see that $f_N(\gamma_{\E}(t), \iota \circ \gamma(t)) = \textrm{id}_N$, in particular $f_N(\gamma_{\E}(t), \iota \circ \gamma(t)(n)) = n$ so that 
$$f_{N} \circ \gamma^{\prime} = \textrm{id}_N.$$
We also know that $\iota \left( \gamma_{\E}(t), \gamma_M(t) ) (f_N(\gamma(t)) \right) = \gamma_M(t)$ since the section passes through the $M$ component. Taking $n_0 = f_N \circ \gamma (t)$ and using the fact that $f_{\F}$ is constant along the image of the sections, we have that:
\begin{eqnarray*}
f_{\F} \circ \gamma^{\prime} (t,n) &=& f_{\F} \left( \gamma_{\E}(t), \iota \circ \gamma(t)(n)   \right) = f_{\F} \left( \gamma_{\E}(t), \iota \circ \gamma(t)(n_0)   \right)\\ 
&=& f_{\F} \left( \gamma_{\E}(t), \gamma_M(t) )   \right) = f_{\F} \circ \gamma (t).
\end{eqnarray*}

That shows that $f \circ \gamma^{\prime} = (f_{\F} \circ \gamma^{\prime}, f_N \circ \gamma^{\prime}) = (f_{\F} \circ \gamma, \textrm{id}_N)$. Now we are in the situation of Lemma \ref{lis1}: all the information can be put together into the following commutative diagram:
\vspace{-4ex}
\begin{center}
\begin{tikzpicture}[description/.style={fill=white,inner sep=2pt}]
\matrix (m) [matrix of math nodes, row sep=3em,
column sep=3.5em, text height=1.5ex, text depth=0.25ex]
{ \RE \times N & \E \times M & \F \times N & \\
& J^{\infty} E & J^{\infty} F & F\\};
\path[->,font=\scriptsize]
(m-1-1) edge [bend left=40] node[below=0.25em] {$f_{\F} \circ \gamma, \, \textrm{id}_{N}$} (m-1-3)
(m-1-1) edge [bend left=70] node[right=3em] {$(f_{\F} \circ \gamma)^{\smallsmile}$} (m-2-4)
(m-1-1) edge node[auto] {$\gamma^{\prime}$} (m-1-2)
(m-2-2) edge node[auto] {$f_{\infty}$} (m-2-3)
(m-1-2) edge node[auto] {$f$} (m-1-3)
(m-1-3) edge node[left] {$j^{\infty}$} (m-2-3)
(m-1-2) edge node[auto] {$j^{\infty}$} (m-2-2)
(m-1-3) edge node[auto] {$j^0$} (m-2-4)
(m-2-3) edge node[auto] {$\rho_{\infty}^0$} (m-2-4);
\end{tikzpicture}
\end{center}

We conclude that $(f_{F} \circ \gamma)^{\smallsmile}$ is smooth since it is the composition of smooth maps. \qed

We finish this section with the conclusion necessary to prove the last statement of Proposition \ref{cinco}.

\begin{cl}\label{underr} Differential operators along the identity of a compact manifold are smooth.
\end{cl}

\dem Differential operators along the identity are insular by Example \ref{e3} and those are smooth by Theorem \ref{t3}.
\qed


\newpage
\chapter{Differential operators revisited}


\section{Jet prolongations of insular differential operators}

{\it In this chapter we construct jet prolongations of insular differential operators: this shows in particular that insular differential operators are local. Their jet prolongations are unique and hence it is possible to define the category of insular manifolds where morphisms are given by insular maps. We will show that insular differential operators satisfy [DO.1], [DO.2], and [DO.3].}\\

We fix two smooth fiber bundles $\pi \colon E \rightarrow M$ and $\rho \colon F \rightarrow N$ over two possibly distinct manifolds. In section \ref{jpg} we were able to construct unique jet prolongations of bundle maps $f^0 \colon J^k E \rightarrow F$ covering different submersions depending on the field $\varphi \in \E$ provided they preserved the Cartan distribution. In proposition \ref{finhoy} we learned that local Cartan-preserving maps have unique jet prolongations:

\begin{lm}\label{tale} For an insular local map $(f,f^{\infty})$, infinite jet prolongations, in the sense of Definition/Proposition \ref{PRO1}, of $f^0 \colon J^k E \rightarrow F$  exist, they are pro-smooth, and preserve the Cartan distribution. Moreover $f^{\infty} = j^{\infty} f^0$:
\begin{center}
    \begin{tabular}{rcl}
    $f^{\infty} \colon J^{k+l} E$ & $\longrightarrow$ & $J^l F$ \\
    $j_x^{\infty} \varphi$ & $\longmapsto$ & $j_{\rho \circ f^0 \circ j^{k} \varphi (x)}^{\infty} (f^0 \circ j^k \varphi \circ \iota_{\varphi,x})$. \\
    \end{tabular}
\end{center}
\end{lm} 

\dem Insular maps are Cartan preserving by Proposition \ref{icp}. It follows from Proposition \ref{finhoy} that local maps which are also Cartan preserving satisfy the conclusions of this Lemma. 
\qed 

When working with general differential operators $(f,f^0)$, the map $f_{\F}(\varphi, x)$ could change as soon as we move a little bit away from $x$ (for $(\varphi, x) \in \E \times M$). In this way, we will not be able to know any information about the germ of $f_{\F}(\varphi, x)$ at $f_N(\varphi ,x)$ (equal to $\tau(x)$ in the previous example) using $f^0$. This does not happen for insular maps. 

\begin{rk}[Insular differential operator]\label{docLFS}
In general, we can consider differential operators $(f, f^0)$ such that $f$ is an insular map, those are called {\it insular differential operators}.
\end{rk}

It is possible to construct jet prolongations for insular differential operators (showing [DO.3]):

\begin{lm}\label{etale} Let $f \colon \E \times M \rightarrow \F \times N$ be an insular differential operator. Then $f$ covers a unique map between the \'etal\'e spaces $\widehat{f} \colon g_E(\E \times M) \subset \widehat{E} \rightarrow \widehat{F}$ which extends to the whole of $\widehat{E}$ if $\E$ is soft. The map is given by $\widehat{f}(g_E(\varphi, x)) = g_F(f(\varphi, x))$.
\end{lm}

\dem We need to show that $f_{\F}(\varphi, x)$ and $f_{\F}(\widetilde{\varphi}, x)$ have the same germ at $f_N(\varphi, x)$ provided $\varphi$ and $\widetilde{\varphi}$ have the same germ at $x$. Fix any two such sections.

Since $\varphi$ and $\widetilde{\varphi}$ agree on a full neighborhood $U$ of $x$ in $M$, for every point $x^{\prime} \in U$, $j_{x^{\prime}}^k \varphi = j_{x^{\prime}}^k \widetilde{\varphi}$. Hence $f_{N}(\varphi, x^{\prime}) = f_{N}(\widetilde{\varphi}, x^{\prime})$ since both of them agree also with $f^0 (j_{x^{\prime}}^k \varphi)$. This shows that $\iota_{\varphi, x}$ is a local section of both $f_{N}(\varphi, -)$ and $f_{N}(\widetilde{\varphi}, -)$ since they agree on some neighborhood of $x$. We take $V \subset N$ an open neighborhood of $f_N(\varphi, x)$ such that $f_{\F}(\varphi, \iota_{\varphi,x}(V))$ and $f_{\F}(\varphi, \iota_{\widetilde{\varphi},x}(V))$ are constant. $f_{\F}(\varphi,x)$ and $f_{\F}(\widetilde{\varphi},x)$ have the same germ at $f_N(\varphi, x)$ since they agree on $V$. Call $y_0 \defeq f_N(\varphi, x)$. Then, for any $y \in V$:
\begin{align*}
	f_{\F}(\varphi, x)(y) & = f_{\F}(\varphi, x) f_N(\varphi, \iota_{\varphi,x}(y)) = f_{\F}(\varphi, \iota_{\varphi,x}(y_0)) f_N(\varphi, \iota_{\varphi,x}(y)) =\\
	& = f_{\F}(\varphi, \iota_{\varphi,x}(y)) f_N(\varphi, \iota_{\varphi,x}(y)) = j^0 \circ f (\varphi, \iota_{\varphi,x}(y)) =\\
	& = f^0 \circ j^k (\varphi, \iota_{\varphi,x}(y)) = f^0 \circ j^k (\widetilde{\varphi}, \iota_{\varphi,x}(y)) = f_{\F}(\widetilde{\varphi}, x)(y)
\end{align*} 
\qed
  
The map between the corresponding \'etal\'e spaces can be recovered from the map $f^0 \colon J^k E \rightarrow F$. We have seen this fact in the proof of the Lemma \ref{etale}. The formula of such map is given by the following equation (where it does not matter which family of sections we get since we know the map is unique from Lemma \ref{etale}):
\begin{center}
    \begin{tabular}{rcl}
    $\widehat{f} \colon g_E(\E \times M) \subset \widehat{E}$ & $\longrightarrow$ & $\widehat{F}$ \\
    $g_E(\varphi, x)$ & $\longmapsto$ & $[(f^0 \circ j^k \varphi \circ \iota_{\varphi,x}, \, \rho \circ f^0 j_x^k \varphi )]$. \\
    \end{tabular}
\end{center}
In principle we will think of insular differential operators as maps where that local behavior at the germs is global, but it is important to keep in mind that this is only true at the germ level. We will consider maps of the kind:
\begin{center}
    \begin{tabular}{rcl}
    $f \colon \E \times M$ & $\longrightarrow$ & $\F \times N$ \\
    $(\varphi, x)$ & $\longmapsto$ & $(f^0 \circ j^k \varphi \circ \iota_{\varphi,x}, \, \rho \circ f^0 j^k \varphi (x) )$. \\
    \end{tabular}
\end{center}

The following step is to observe that the map between the \'etal\'e spaces actually descends to a pro-finite smooth map between the infinite jet bundles because $I = (\iota, f_N)$ is local:

\begin{center}
\begin{tikzpicture}[description/.style={fill=white,inner sep=2pt}]
\matrix (m) [matrix of math nodes, row sep=3em,
column sep=2.5em, text height=1.5ex, text depth=0.25ex]
{ \E \times M & \C(N, M) \times N & M \\
  j^{\infty} (\E \times M) & j^{\infty} (\C(N, M) \times N)  & \\
	j^{k} (\E \times M) & j^0 (\C(N, M) \times N) & M \\};
\path[->,font=\scriptsize]
(m-1-1) edge node[auto] {$(I, f_N)$} (m-1-2)
(m-1-1) edge node[auto] {$j^{\infty}$} (m-2-1)
(m-2-1) edge node[auto] {$I^{\infty}$} (m-2-2)
(m-1-2) edge node[auto] {$j^{\infty}$} (m-2-2)
(m-1-2) edge node[auto] {ev} (m-1-3)
(m-3-2) edge node[auto] {$(\textrm{pr}_M)_0$} (m-3-3)
(m-2-1) edge node[auto] {$j_{\infty}^k$} (m-3-1)
(m-2-2) edge node[auto] {$j_{\infty}^0$} (m-3-2)
(m-3-1) edge node[auto] {$I^0$} (m-3-2);
\end{tikzpicture}
\end{center}

%
%
%
%
%

\begin{lm}[Insular differential operators are local maps]\label{ketale}
Let $f \colon \E \times M \rightarrow \F \times N$ be an insular differential operator. Then $f$ is a local map, unique in the following sense: 

It covers a unique map between the \'etal\'e spaces $\widehat{f} \colon g_E(\E \times M) \subset \widehat{E} \rightarrow \widehat{F}$ and a unique map $f^{\infty} \colon j^{\infty}(\E \times M) \rightarrow J^{\infty} F$ which extend to the whole of $\widehat{E}$ and $J^{\infty} E$ if $\E$ is soft. The map between the infinite jet bundles is given by:

\begin{center}
    \begin{tabular}{rcl}
    $f^{l} \colon j^{k(l)\defeq  \max(I(l), k + l)}(\E \times M)$ & $\longrightarrow$ & $J^l F$ \\
     $j_x^{k(l)} \varphi$ & $\longmapsto$ & $j_{\rho \circ f^0 j^k (\varphi,x)}^{l} \left( f^0 \circ j^k \varphi \circ \iota_{\varphi,x} \right),$ \\
    \end{tabular}
\end{center}
where $I$ also denotes the index functor for $I^{\infty}$.
\end{lm} 

By the inverse function theorem, it is possible to choose $I(l)=l$ so that the jet prolongations are parallel ($k$ goes to $0$, $k+l$ goes to $l$).

\dem The statement about the \'etal\'e maps follows directly from Proposition \ref{etale}. Using the remark after Proposition \ref{etale} we can see that the map between the \'etal\'e spaces can be expressed in terms of $f^0$. Now, using $I^l$ we know the $l$-th jet of $\iota_{\varphi,x}$ at $f_N(\varphi,x)$. Knowing the $(k+l)$-th jet of $\varphi$ at $x$ is enough to know the $l$-th jet of $j^k \varphi$ at $\iota_{\varphi,x}(f_N(\varphi, x)) =x$. Finally, the derivatives of $f^0$ are given at $j_x^k(\varphi)$.  
\qed 

\begin{rk} By construction, the unique map between the infinite jet bundles is also given by:
	\begin{center}
    \begin{tabular}{rcl}
    $f^{\infty} \colon j^{\infty}(\E \times M)$ & $\longrightarrow$ & $j^{\infty}(\F \times N)$ \\
    $j^{\infty}(\varphi, x)$ & $\longmapsto$ & $j_{f_N(\varphi,x)}^{\infty} f_{\F}(\varphi, x)$. \\
    \end{tabular}
	\end{center}
\end{rk}

We have an analogue of Proposition \ref{cinco} to the general case of insular differential operators:

\begin{tm}  Let $(f,f^0)$ be an insular differential operator. Then the following statements hold
\begin{enumerate}
	\item Infinite jet prolongations of $f^0$ exist and are covered by $f$ (i.e., [DO.1] holds). 
	\item $j^{\infty} f^0$ preserves the Cartan distribution and we can recover $f$ from $f^0$, at least locally (i.e., [DO.2] holds).
	\item $f$ induces a map between the associated \'etal\'e spaces (i.e., [DO.3] holds).
	\item $f$ is an extended local map. 
	\item $f_{\F} \colon \E \rightarrow \F$ is smooth if $N$ is compact (i.e., [S.1] holds).
\end{enumerate}
\end{tm}

\dem The three previous lemmas are precisely the three first statements. Lemma \ref{ketale} gives $2$, Lemma \ref{etale} gives $3$ and Lemma \ref{tale} together with Lemma \ref{ketale} imply $1$. As it happened in Proposition \ref{cinco}, $4$ follows from $1$. Finally, $5$ is Theorem \ref{t3}.
\qed

Recall that we wanted [DO.1], [DO.2], and [DO.3] to hold so that we could have the best of the worlds of differential operators and local maps. This is true for insular differential operators, they form a category:

\begin{dfpp}[Category of insular manifolds] The category of insular manifolds $\iMan$ is the subcategory of $\elMan$ with the same objects and with insular differential operators as morphisms. Composition is given by composition of the unique local maps associated to the morphisms by Lemma \ref{ketale}.
\end{dfpp}

\dem By Lemma \ref{ketale}, insular differential operators are local maps in a unique way. The composition between local maps is well defined. The identity of every object is an insular map (product over the identity as in Example \ref{e3}). We only need to check that the composition of insular maps is again insular.

Consider two composable maps 
$$ W \subset \left( \E \times M \right) \xrightarrow{f} W^{\prime} \subset \left( \F \times N \right) \xrightarrow{g} W^{\prime \prime} \subset \left( \G \times P \right)$$

Given $(\varphi, x) \in W \subset \E \times M$, we get an associated island $\{M_{\varphi}^{a}\}_{a \in A}$ of an open neighborhood of $x$ in $M$ which lies inside of $\left( \{\varphi\} \times M \right) \cap W$ (we are going to use repeatedly the argument that everything can be taken to be inside of $W$, $W^{\prime}$, $W^{\prime \prime}$ since the topology on $\E \times M$ is the product one (equivalently for $\F \times N$ and $\G \times P$)).

Given any $a \in A$ denote by $\psi_{a} \defeq \restrict{f_{\F}(\varphi,-)}{M_{\varphi}^{a}} \in \F$. Consider the associated island to $\psi_{a}$: $\{N_{\psi_{a}}^{b}\}_{b \in B_{a}}$. Using $\left(\restrict{f_N(\varphi, -)}{M_{\varphi}^{a}}\right)^{-1}$ we can bring the foliations indexed by $B_{a}$ to $M$. Call $\widetilde{M}_{\varphi}^{a, b} \defeq \restrict{\left(\restrict{f_N(\varphi, -)}{M_{\varphi}^{a}}\right)^{-1}}{ N_{\psi_{a}}^{b}}$ for any $a \in A$ and $b \in B_{a}$. We get a foliation of $M$ given by $\{ \widetilde{M}_{\varphi}^{a, b} \}_{a \in A, \, b \in B_{a}}$.

\begin{figure}[h]
	\centering
	\includegraphics[width=0.6\textwidth]{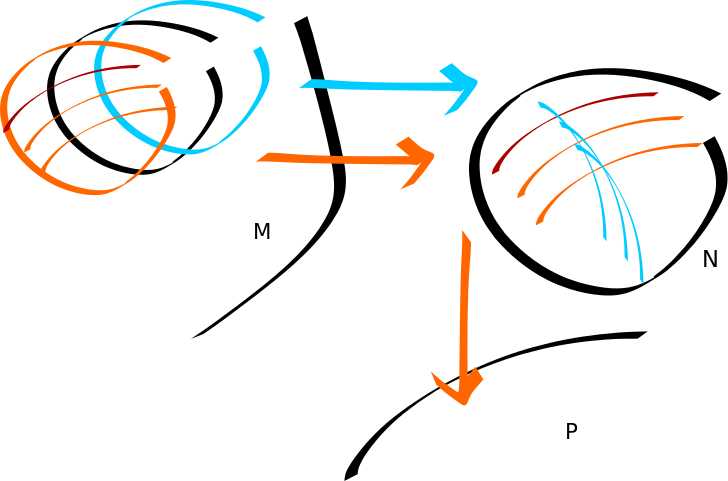}
	\caption{The island $\{ \widetilde{M}_{\varphi}^{a, b} \}_{a \in A, \, b \in B_{a}}$ associated to $\varphi$ depicted in dark red. In the image we can see other islands for $f$, $g$ and $g \circ f$.}
	\label{fig:hier}
\end{figure}

The map $(g \circ f)_P(\varphi, -)$ is a submersion because when restricted to $\widetilde{M}_{\varphi}^{a, b}$ it is a local diffeomorphism. Consider $x \in \widetilde{M}_{\varphi}^{a, b}$ and take local sections $\iota_{\varphi, x}$ and $\kappa_{\psi_{a}, f_{N}(\varphi, x)}$ of $f_N(\varphi, -)$ and $g_P(\psi_{a}, -)$ respectively. Their composition $\iota_{\varphi, x} \circ \kappa_{\psi_{a}, f_{N}(\varphi, x)}$ is an inverse of $(g \circ f)_P(\varphi, -)$ since it is given by selecting the leaf $\widetilde{M}_{\varphi}^{a, b}$, where $b$ is the leaf of $f_N(\varphi, x)$ with respect to $\psi_{a} = f_{\F}(\varphi, x)$. The composition is local since local maps form a category.

We only are left to show that $(g \circ f)_{\G}(\varphi, -)$ is constant along the image of each section $\iota_{\varphi, x} \circ \kappa_{\psi_{a}, f_{N}(\varphi, x)}$, this is $\widetilde{M}_{\varphi}^{a, b}$. This is immediate:
$$\restrict{(g \circ f)_{\G} (\varphi, -)}{\widetilde{M}_{\varphi}^{a, b}} =  \restrict{g_{\G}\left(f_{\F}(\varphi, -), f_N(\varphi, -)\right)}{\widetilde{M}_{\varphi}^{a, b}} = \restrict{g_{\G}(\psi_{a}, -)}{N_{\psi_{a}}^{b}}$$
{\noindent which is a singleton.}\qed

As a consequence of Example \ref{e3} we can conclude that the following functor is fully faithful:
$$\Diff_M \longrightarrow \iMan.$$

\begin{rk}
In the particular case in which $W = \E \times M$, an insular manifold morphism is given by a foliation of $M$ for every $\varphi$ (a collection of islands) and by a smooth map $f^0 \in \C(J^k E, F)$. If the number of islands is finite, all the information of an insular map can be encoded in a map between finite dimensional manifolds.
\end{rk}

\newpage
\chapter{Peetre's Theorem (maps of sheaves)}\label{PT}

The concept of locality is not unique to the world of jet bundles. In topology, a map between spaces of smooth sections over the same base manifold is called local if it comes from a morphism of sheaves. This concept can be generalized to the case in which the base manifolds are different, calling a map local in topology, or germ-local, if it descends to a map from the \'etal\'e space to the $0$-th jet bundle.\\

The relation between differential operators and germ local maps is well known. Peetre's theorem is the essential result for understanding the link between the two notions: it states that germ local maps are locally differential operators.\\

The notion of locality for product maps $A \times \tau \colon \E \times M \rightarrow \F \times N$ is thus very closely related to that of a local map in topology. Under insularity conditions, germ-local maps are locally like a local map. \\

{\it In this chapter we define the local maps in topology and relate them to local maps in our setting. We review some versions of Peetre's theorem and  apply it to our framework. We present proves of some results using the theory of insular maps developed in the previous chapter. The main references at this point are Kol\'a\v{r}, Michor, and Slov\'ak \cite{KMS}; and Slov\'ak \cite{SLO}.}


\section{Peetre's Theorem for maps over the identity}

{\it This section explores the relations between the different concepts of locality: germ-local, map of sheaves and local maps; for product maps $f_{\F} \times \textrm{id}_M \colon \E \times M \rightarrow \F \times M$ along the identity. Among other results, it states the main theorem in this area, known as Peetre's Theorem in terms of local maps and it uses insular maps to prove those results. The main references in this chapter are Kol\'a\v{r}, Michor, and Slov\'ak \cite{KMS}; and Slov\'ak \cite{SLO}.}\\

In order to address the comparison between maps of sheaves $f_{\F} \colon \E \rightarrow \F$ and local maps, we have to restrict ourselves to consider product maps $f_{\F} \times f_N \colon \E \times M \rightarrow \F \times N$. But recall that there are examples of local maps, even of insular maps, which are not products: the action of a local Lie groupoid would be an essential example of this kind as we will see in Part \ref{lla}.

In this first section, we focus on the case $M = N$ and $f_N = \textrm{id}_M$. We denote $A \defeq  f_{\F}$ to make clear the difference with the general case. Under these assumptions $\E$ and $\F$ are sheaves over the same manifold $M$. A map $A \colon \E \rightarrow \F$ is called in topology, local if it comes from a map of sheaves. At this point it is better to change the notation momentarily and denote $\Gamma^{\infty}(M,E)$ by $\E(M)$ and reserve $\E$ for the sheaf (the same goes for $\F$). So that $\E(U)$ where $U$ is open in $M$ denotes the space of sections $\Gamma^{\infty}(U,\restrict{E}{\pi^{-1}(U)})$ and $\F(U) = \Gamma^{\infty}(U,\restrict{F}{\rho^{-1}(U)})$ respectively. The question is clear: when does a local map $A(M) \colon \E(M) \rightarrow \F(M)$ come from a map of sheaves and vice versa?

As a first step into answering that question, we would like to point out the connection between a map of sheaves and a map which only depends on the germ of a section. 

\begin{df}[Germ-local map] A map $A(M) \colon \E(M) \rightarrow \F(M)$ is said to depend only on the germs or to be germ-local if $A(M)(\varphi)(x)=A(M)(\widetilde{\varphi})(x)$ provided $\varphi$ and $\widetilde{\varphi}$ have the same germ at $x$. This simply means that $A \times \textrm{id}_M$ descends to a map with source the \'etal\'e space associated to $\E$:

\begin{center}
\begin{tikzpicture}[description/.style={fill=white,inner sep=2pt}]
\matrix (m) [matrix of math nodes, row sep=3em,
column sep=2.5em, text height=1.5ex, text depth=0.25ex]
{\E(M) \times M & \F(M) \times M \\
 \widehat{E} \phantom{x} & F \\};
\path[->,font=\scriptsize]
(m-1-1) edge node[auto] {$A \times \textrm{id}_M$} (m-1-2)
(m-1-1) edge node[auto] {$g_E$} (m-2-1)
(m-1-2) edge node[auto] {$j^0$} (m-2-2)
(m-2-1) edge node[auto] {$\widehat{A}$} (m-2-2);
\end{tikzpicture}
\end{center} 
\end{df}

Germ-local maps are easier to work with than maps of sheaves when talking about the relationship between these concepts and local maps. This result is well known and it is used, without a proof, for instance in the book by Kol\'a\v{r}, Michor, and Slov\'ak \cite{KMS}.

\begin{pp}\label{soft}
Let $E, F \rightarrow M$ be two fiber bundles over the same base manifold $M$. Denote the associated sheaves of sections by $\E$ and $\F$ respectively. Let $A(M) \colon \E(M) \rightarrow \F(M)$ be a map.
\begin{itemize}
	\item If $A(M)$ comes from a map of sheaves $A \colon \E \rightarrow \F$, then $A(M)$ is germ local.
	\item Conversely, if $A(M)$ is germ local and $\E$ is soft, then $A(M)$ comes from a map of sheaves. 
\end{itemize}
\end{pp}

\dem If we have a map of sheaves $A \colon \E \rightarrow \F$ and two sections $\varphi$ and $\widetilde{\varphi}$ with the same germ at $x \in M$ we want to show that $A(M)(\varphi)(x) = A(M)(\widetilde{\varphi})(x)$ so that we would have proven the first point. Since $\varphi$ and $\widetilde{\varphi}$ have the same germ at $x$, there exists an open neighborhood $U$ of $x$ in $M$ such that $\restrict{\varphi}{U} = \restrict{\widetilde{\varphi}}{U}$. Now, since $A$ is a map of sheaves, we have that:
\begin{center}
\begin{tikzpicture}[description/.style={fill=white,inner sep=2pt}]
\matrix (m) [matrix of math nodes, row sep=3em,
column sep=2.5em, text height=1.5ex, text depth=0.25ex]
{ \E(M) & \F(M) \\
  \E(U) & \F(U) \\};
\path[->,font=\scriptsize]
(m-1-1) edge node[auto] {$A(M)$} (m-1-2)
(m-1-1) edge node[auto] {$\restrict{}{U}$} (m-2-1)
(m-2-1) edge node[auto] {$A(U)$} (m-2-2)
(m-1-2) edge node[auto] {$\restrict{}{U}$} (m-2-2);
\end{tikzpicture}
\end{center}
Because $\restrict{\varphi}{U} = \restrict{\widetilde{\varphi}}{U}$ it follows that
$$\restrict{\left(A(M)(\varphi)\right)}{U} = A(U) \circ \restrict{\varphi}{U} = A(U)\circ \restrict{\widetilde{\varphi}}{U} = \restrict{\left(A(M)(\widetilde{\varphi})\right)}{U}.$$
As $x \in \, U$, we conclude that $A(M)(\varphi)(x) = A(M)(\widetilde{\varphi})(x)$ and hence that $A(M)$ is germ-local.

To prove the other statement, let $A(M)$ be a germ-local map and $\E$ be a soft sheaf. We want to construct a map $A(U) \colon \E(U) \rightarrow \F(U)$ for every $U$ open in $X$. We fix such a $U$, for every point $x \in \, U$, take a neighborhood $U_x$ of $x$ in $U$ such that the closure of $U_x$, $\overbar{U_x}$ remains in $U$ (this is possible since $M$ is a smooth manifold). Given any section $\varphi \in \, \E(U)$ take $\varphi_x \defeq  \restrict{\varphi}{\overbar{U_x}}$. Since $\E$ is soft and $\overbar{U_x}$ is closed, we can extend $\varphi_x$ to a global section $\widetilde{\varphi}_x \in \E(M)$.

Because $M$ is a manifold, it is paracompact and so it is $U$. The family $\{U_x\}_{x \in U}$ is an open cover, so we can take a locally finite subcover of it: $\{ V_y \}$ where $y$ is indexed over some subset of points of $U$.
Define $A(U)(\varphi)(x)$ as $A(\widetilde{\varphi}_y)(x)$ for some $y$ such that $x \in \, V_y$.

If we will be taking any other $\widetilde{\varphi}_{y^{\prime}}$ such that $x \in V_{y^{\prime}}$, both sections will agree with $\varphi$ on $V_{y} \cap V_{y^{\prime}}$ which is an open neighborhood of $x$, and thus by germ-locality the definition of $A(U)(\varphi)(x)$ is independent of any of the choices made so far. Using the same idea, it is clear that $A(U)(\varphi)$ is smooth and it commutes with the projections maps, showing that $A(M)$ actually comes from a map of sheaves $A \colon \E \rightarrow \F$.
\qed

\begin{rk}
A flabby or flasque sheaf is one in which sections over open sets extend to global sections. It is possible to think that if instead of requiring $\E$ to be soft, we ask it to by flabby, we get an easier proof of the Proposition \ref{soft}. This is true, but flabby sheaves are not very common: trivial vector bundles are not flabby (and hence we are in deep trouble when finding a smooth fiber bundle whose sheaf of sections is flabby). Contrary, all sheaves of sections of vector bundles are soft. Still, there are examples of fiber bundles which are not soft (fiber bundles with no global sections are an example of this: the tangent bundle of $\mathbb{S}^ 2$ without the zero section is an explicit one).
\end{rk}

The following result also appears, for example, in the book by Kol\'a\v{r}, Michor, and Slov\'ak \cite{KMS}. We provide an independent prove using insular maps:

\begin{pp}\label{easy}
If $A(M) \colon \E(M) \rightarrow \F(M)$ is a differential operator along the identity, then it is a germ-local map. If $\E$ is soft, then $A(M)$ comes from a morphism of sheaves.
\end{pp}

\dem Differential operators along the identity which are products are insular by Example \ref{e3}. Insular maps are germ local by Proposition \ref{etale}. Applying Proposition \ref{soft} we get that in the case in which $\E$ is sof, the map comes from a morphism of sheaves. 
\qed

Observe that using the fact that such differential operators are insular and hence local (Proposition \ref{ketale}) we can conclude the following:

\begin{cl}\label{lts}
Let $E, F \rightarrow M$ be two fiber bundles over the same base manifold $M$, with associated sheaves of sections $\E$ and $\F$ respectively. If $A(M) \colon \E(M) \rightarrow \F(M)$ is a local map, then it is a germ-local map. If $\E$ is soft, then $A(M)$ comes from a morphism of sheaves.
\end{cl}

\dem The last statement follows from Proposition \ref{soft}. Imagine $\varphi$ and $\widetilde{\varphi}$ are two sections with the same germ at $x \in \, M$. It is clear that $j^{\infty}(\varphi,x) = j^{\infty}(\widetilde{\varphi},x)$ and since $A(M)$ is local it follows that $A(M)(\varphi)(x) = A(M)(\widetilde{\varphi})(x)$.
\qed

Observe that in the case of analytic manifolds, germ-local and local in our sense are the same. Any analytic function is fully determined in a neighborhood of a point by its Taylor polynomial. In smooth manifolds, this is no longer the case. Think for example of $e^{\frac{-1}{x^2}}$ and $0$ as functions on $\RE$ (or as sections of the trivial one dimensional vector bundle over $\RE$). Both functions have zero derivatives at every order at $x=0$ but they do not agree at any neighborhood of $0$: these two functions have the same $\infty$-jet, but they have different germs at $x=0$.

The previous paragraph tells us why one should not expect a germ-local map to be a local map in general. But Peetre's theorem tells us that a partially converse to Proposition \ref{easy} does hold. We include the statement by Slov\'ak \cite{SLO}:

\begin{tm}[Peetre's Theorem for bundles over the same manifold]\label{Pee1}
Consider $E, F \rightarrow M$ be two fiber bundles over the same base manifold $M$. Denote the associated sheaves of sections by $\E$ and $\F$ respectively. Let $A(M) \colon \E(M) \rightarrow \F(M)$ be a germ-local map.

For every $\varphi \in \, \E$ and every compact submanifold $K \subset M$ there exists $k$ a natural number, $\mathcal{V} \subset \E(M)$ an open neighborhood of $\varphi$ in the $CO^{k}$-topology (and hence in the $CO^{\infty}$-topology) such that $\restrict{\left( A(M) \times \textrm{id} \right)}{\mathcal{V} \times K}$ is a $k$-th order differential operator.

In other words, there exists a smooth map $A^0 \colon J^ k E \rightarrow F$ such that the following diagram commutes:
\begin{center}
\begin{tikzpicture}[description/.style={fill=white,inner sep=2pt}]
\matrix (m) [matrix of math nodes, row sep=3em,
column sep=3.5em, text height=1.5ex, text depth=0.25ex]
{ \mathcal{V} \times K & \F(M) \times M \\
  J^k E  & F \\};
\path[->,font=\scriptsize]
(m-1-1) edge node[auto] {$A(M) \times \textrm{id}$} (m-1-2)
(m-1-1) edge node[auto] {$j^ k$} (m-2-1)
(m-2-1) edge node[auto] {$A^0$} (m-2-2)
(m-1-2) edge node[auto] {$j^0$} (m-2-2);
\end{tikzpicture}
\end{center}
\end{tm}

The original proof is from Peetre, first \cite[Theorem 2]{PEE1} and later completed \cite{PEE2}. The proof is quite involved, and we refer to Kol\'a\v{r}, Michor, and Slov\'ak \cite{KMS}, for a linear (easier) version of the theorem and to Slov\'ak \cite{SLO}, for the general case in a more modern language.

In particular if $A \colon \E \rightarrow \F$ is a map of sheaves, then it is a local map (along compact submanifolds and open neighborhoods of a given section).

\begin{cl}[Peetre's Theorem for $\Diff_M$]\label{Peecl}
Let $E, F \rightarrow M$ be two fiber bundles over the same base manifold $M$. Denote the associated sheaves of sections by $\E$ and $\F$ respectively. Assume $\E$ is soft. Let $A \colon \E \rightarrow \F$ be a map of sheaves.

For every $\varphi \in \, \E$ and every compact submanifold $K \subset M$ there exists $\mathcal{V} \subset \E(M)$ an open neighborhood of $\varphi$ in the $CO^{k}$-topology (and hence in the $CO^{\infty}$-topology) such that $\restrict{\left( A(M) \times \textrm{id} \right)}{\mathcal{V} \times K}$ is in $\Diff_M$ (a local map along the identity).

In other words, there exists a pro-smooth map $A^{\infty} \colon J^{\infty} E \rightarrow F^{\infty}$ such that the following diagram commutes:
\begin{center}
\begin{tikzpicture}[description/.style={fill=white,inner sep=2pt}]
\matrix (m) [matrix of math nodes, row sep=3em,
column sep=3.5em, text height=1.5ex, text depth=0.25ex]
{ W \times K & \F(M) \times M \\
  J^{\infty} E  & J^{\infty} F \\};
\path[->,font=\scriptsize]
(m-1-1) edge node[auto] {$A(M) \times \textrm{id}$} (m-1-2)
(m-1-1) edge node[auto] {$j^{\infty}$} (m-2-1)
(m-2-1) edge node[auto] {$A^{\infty}$} (m-2-2)
(m-1-2) edge node[auto] {$j^{\infty}$} (m-2-2);
\end{tikzpicture}
\end{center}
\end{cl}

\dem The proof is a consequence of several of the results we have mentioned so far. Since $A$ is a map of sheaves, by Proposition \ref{soft} $A(M)$ is a germ-local map. Since $A(M)$ is germ-local map, there exist $\mathcal{V}$ and $K$ such that $\restrict{A(M) \times \textrm{id}}{\mathcal{V} \times K}$ is a differential operator along the identity, this is nothing else but Theorem \ref{Pee1}. Since the map is a product with the identity, it is in particular insular (Example \ref{e3}), and by Lemma \ref{ketale} we conclude that the map is local.
\qed


\section{Peetre's Theorem for products}

{\it This section continues to explore the relations between germ-locality and local maps, this time for product maps $f_{\F} \times \tau \colon \E \times M \rightarrow \F \times N$ where now $\tau$ is an arbitrary smooth map. The main theorem in this area, Peetre's Theorem, also holds and it can be stated in terms of insular maps. The starting point of this chapter is the book by Kol\'a\v{r}, Michor, and Slov\'ak \cite{KMS}.}\\

In the previous section we discussed the relationship between germ-locality and (jet)-locality or differentiability of an operator between global sections of bundles over the same manifold. Since we are working with $\E \times M$ and $\F \times N$ in general, we would like to have similar statements to the ones in the previous section for bundles over different base manifolds.

We abandon the notation $\E(M)$ because we will not have a map of sheaves this time. Similarly to the case of an operator between bundles over the same manifold, we can talk about a germ-local map $A \colon \E \rightarrow \F$ (along $\tau$):

\begin{df}[Germ-local along $\tau$]
Let $\pi \colon E \rightarrow M$ and $\rho \colon F \rightarrow N$ be two fiber bundles and $\tau \colon M \rightarrow N$ be a smooth map between the bases. A map $A \colon \E \rightarrow \F$ is said to be germ-local along $\tau$ if $A(\varphi)(\tau(x)) = A(\widetilde{\varphi})(\tau(x))$ provided $\varphi \in \, \E$ and $\widetilde{\varphi} \in \, \E$ have the same germ at  $x \in \, M$. This means that there exists a map $\widehat{A}$ such that the following diagram commutes:
\begin{center}
\begin{tikzpicture}[description/.style={fill=white,inner sep=2pt}]
\matrix (m) [matrix of math nodes, row sep=3em,
column sep=2.5em, text height=1.5ex, text depth=0.25ex]
{\E(M) \times M & \F(M) \times M \\
 \widehat{E} & F \\};
\path[->,font=\scriptsize]
(m-1-1) edge node[auto] {$A \times \textrm{id}_M$} (m-1-2)
(m-1-1) edge node[auto] {$g_E$} (m-2-1)
(m-1-2) edge node[auto] {$j^0$} (m-2-2)
(m-2-1) edge node[auto] {$\widehat{A}$} (m-2-2);
\end{tikzpicture}
\end{center} 
\end{df}

Using the same proof as in Proposition \ref{easy} we get the following result:

\begin{pp}\label{easy2}
Let $\pi \colon E \rightarrow M$ and $\rho \colon F \rightarrow N$ be two fiber bundles and $\tau \colon M \rightarrow N$ be a smooth map between the bases. Denote the associated spaces of global smooth sections by $\E$ and $\F$ respectively. If $A \times \tau \colon \E \times M \rightarrow \F \times N$ is a differential operator, then it is a germ-local map along $\tau$.
\end{pp}

Let $\pi \colon E \rightarrow M$ and $\rho \colon F \rightarrow N$ be two fiber bundles and $\tau \colon M \rightarrow N$ be a smooth map between the bases. The space of global sections of $\tau^* F$, will be denoted by $\tau^* \F$, since it is the sheafification of the pullback pre-sheaf of $\F$ along $\tau$.

Similarly, we can construct the {\it pullback of $A$ along $\tau$}: given $A \colon \E \rightarrow \F$, we define
\begin{eqnarray*}
\tau^* A \colon \E & \longrightarrow & \tau^* \F \\
(\varphi)(x) & \longmapsto & \left( x, A(\varphi)(\tau(x)) \right).
\end{eqnarray*}
It is clear that $A(\varphi)(\tau(x))$ is in the fiber over $x$: $\rho(A(\varphi)(\tau(x)))= \tau(x)$. Now we have two bundles over the same manifold and we are reduced to study the previous case provided we can find relations between the notions of jet-locality and germ-locality for both maps $A$ and $\tau^* A$. These are the objectives of the following two results:

\begin{pp}\label{gl}
Let $\pi \colon E \rightarrow M$ and $\rho \colon F \rightarrow N$ be two fiber bundles and $\tau \colon M \rightarrow N$ be a smooth map between the bases. If $A \colon \E \rightarrow \F$ is germ-local along $\tau$, then $\tau^* A \colon \E \rightarrow \tau^* \F$ is germ local.
\end{pp}

\begin{pp}\label{jl}
Let $\pi \colon E \rightarrow M$ and $\rho \colon F \rightarrow N$ be two fiber bundles and $\tau \colon M \rightarrow N$ be a smooth map between the bases. Let $A \colon \E \rightarrow \F$ be a map and $\tau^* A \colon \E \rightarrow \tau^* \F$ be the corresponding pull-back map. If $\tau^* A$ is a differential operator of order $k$, so it is $A$.
\end{pp}

{\bf Proof of Proposition \ref{gl}:}
We want to check that $\tau^* A$ is germ-local. In order to do so, we fix a point $x \in \, M$ and two sections $\varphi$, $\widetilde{\varphi}$ which agree on a neighborhood of $x$ in $M$. Since $A$ is germ-local along $\tau$ we know that $A(\varphi)(\tau(x)) = A(\widetilde{\varphi})(\tau(x))$, but this is precisely the definition of $\tau^* A$ applied to $\varphi$ and $\widetilde{\varphi}$ at $x$.
\qed

{\bf Proof of Proposition \ref{jl}:}
The proof uses the following fact: an operator is a differential operator of order $k$ if and only if the same diagram commutes when replacing the target space by the image of the operator. We express this condition for $A \times \tau$:
\begin{center}
\begin{tikzpicture}[description/.style={fill=white,inner sep=2pt}]
\matrix (m) [matrix of math nodes, row sep=3em,
column sep=2.5em, text height=1.5ex, text depth=0.25ex]
{ \E \times M & \restrict{A(\E)}{\tau(M)} \times \tau(M) & \F \times N \\
  J^k E & F & \\};
\path[->,font=\scriptsize]
(m-1-1) edge node[auto] {$A \times \tau$} (m-1-2)
(m-1-2) edge node[auto] {inc} (m-1-3)
(m-1-1) edge node[auto] {$j^ k$} (m-2-1)
(m-2-1) edge node[auto] {$A^0$} (m-2-2)
(m-1-2) edge node[auto] {ev} (m-2-2)
(m-1-3) edge node[auto] {$j^ 0$} (m-2-2);
\end{tikzpicture}
\end{center}

The statement is that the outer trapezoid commutes if and only of the inner square does. This is true because the triangle on the right is commutative.

In general, we do not have a map from $\tau^* \F \times M$ to $\F \times N$, which will make the proof work. But do have 
\begin{eqnarray*}
\underline{\tau} \colon \tau^* A (\E) \times M &\longrightarrow & \restrict{A(\E)}{\tau(M)} \times \tau(M)\\
(\tau^*A(\varphi), x) & \longmapsto & \left(A(\varphi), \tau(x)\right).
\end{eqnarray*}
Observe we do not need $A$ to be injective in order for this map to be well-defined, if $\tau^*A(\varphi) = \tau^*A(\widetilde{\varphi})$ then $A(\varphi) = A(\widetilde{\varphi})$ along $\tau(M)$.

Now, we assume that $\tau^* A$ is a differential operator of degree $k$, this means we have a map $\left(\tau^* A \right)^0 \colon J^k E \rightarrow \tau^* F$ such that the inner left square in the following diagram commutes:

\begin{center}
\begin{tikzpicture}[description/.style={fill=white,inner sep=2pt}]
\matrix (m) [matrix of math nodes, row sep=3em,
column sep=3.5em, text height=1.5ex, text depth=0.25ex]
{ \E \times M & \tau^*A(\E) \times M & \restrict{A(\E)}{\tau(M)} \times \tau(M) \\
  J^k E & \tau^* F & F \\};
\path[->,font=\scriptsize]
(m-1-1) edge node[auto] {$\tau^*A \times \textrm{id}$} (m-1-2)
(m-1-2) edge node[auto] {$\underline{\tau}$} (m-1-3)
(m-1-1) edge node[auto] {$j^ k$} (m-2-1)
(m-2-1) edge node[auto] {$\left(\tau^* A \right)^0$} (m-2-2)
(m-1-2) edge node[auto] {ev} (m-2-2)
(m-1-3) edge node[auto] {ev} (m-2-3)
(m-2-2) edge node[auto] {$\textrm{pr}_F$} (m-2-3);
\end{tikzpicture}
\end{center}

The right square is also commutative
$$\textrm{ev} \circ \underline{\tau}(\tau^*A(\varphi), x)= A(\varphi)(\tau(x)) =  \textrm{pr}_F(\textrm{ev}((A(\varphi))\tau(x))) = \textrm{pr}_F(\textrm{ev}(\tau^*A(\varphi), x)).$$

All there is left to see is that $\underline{\tau}(\tau^* A \times \textrm{id}) = A \times \tau$ but this is true by definition of $\underline{\tau}$ (observe we only care about what happens in the image of both maps):
$$\underline{\tau}(\tau^* A \times \textrm{id})(\varphi, x)= \underline{\tau}(\tau^* A(\varphi), x) = \left(A(\varphi), \tau(x)\right) = (A \times \tau)(\varphi, x).$$

This proves the Proposition, since $A \times \tau$ descends to $A^0 \defeq \textrm{pr}_F \circ \left(\tau^* A \right)^0 \colon J^k E \rightarrow F$.
\qed

Now we can use Peetre's theorem for bundles over the same manifold to get a similar result for bundles over different base manifolds:

\begin{tm}[Peetre's Theorem for products]\label{Peer}
Let $\pi \colon E \rightarrow M$ and $\rho \colon F \rightarrow N$ be two fiber bundles and $\tau \colon M \rightarrow N$ be a smooth map between the bases. Let $A \colon \E \rightarrow \F$ be a germ-local map.

For every $\varphi \in \, \E$ and every compact submanifold $K \subset M$ there exists $k$ a natural number, $\mathcal{V} \subset \E(M)$ an open neighborhood of $\varphi$ in the $CO^{k}$-topology (and hence on the $CO^{\infty}$-topology) such that $\restrict{\left( A \times \tau \right)}{\mathcal{V} \times K}$ is a $k$-th order differential operator:
\begin{center}
\begin{tikzpicture}[description/.style={fill=white,inner sep=2pt}]
\matrix (m) [matrix of math nodes, row sep=3em,
column sep=3.5em, text height=1.5ex, text depth=0.25ex]
{ \mathcal{V} \times K & \F \times N \\
  J^k E  & F \\};
\path[->,font=\scriptsize]
(m-1-1) edge node[auto] {$A \times \tau$} (m-1-2)
(m-1-1) edge node[auto] {$j^ k$} (m-2-1)
(m-2-1) edge node[auto] {$A^0$} (m-2-2)
(m-1-2) edge node[auto] {$j^ 0$} (m-2-2);
\end{tikzpicture}
\end{center}
\end{tm}

\dem Since $A$ is germ-local, so it is $\tau^* A$ by Proposition \ref{gl}. Now we can apply Peetre's theorem (\ref{Pee1}) for $\tau^* A$, $\varphi \in \, \E$ and $K \subset X$ compact to find $k$, $\mathcal{V}$ and $\left(\tau^* A \right)^0$ such that $\tau^* A$ is a local map of order $k$ along $\mathcal{V} \times K$. Applying Proposition \ref{jl} to $\tau^* A$ and restricting the maps to $\mathcal{V} \times K$ we find that the statement is true.
\qed

We can try to use our knowledge about insular maps to take jet prolongations of this differential operator and get a local map under certain circumstances:

\begin{cl}[Peetre's Theorem for local product maps]\label{Pee25}
Let $\pi \colon E \rightarrow M$ and $\rho \colon F \rightarrow N$ be two fiber bundles and $\tau \colon M \rightarrow N$ be a submersion between the bases. Assume $\E$ is soft. Let $A \colon \E \rightarrow \F$ be a germ-local map.

For every $\varphi \in \, \E$ and every compact submanifold $K \subset M$ there exists $\mathcal{V} \subset \E(M)$ an open neighborhood of $\varphi$ in the $CO^k$-topology (and hence on the $CO^{\infty}$) such that $\restrict{\left( A \times \tau \right)}{\mathcal{V} \times K}$ is a local map.

In other words, there exists a pro-smooth map $A^{\infty} \colon J^{\infty} E \rightarrow J^{\infty} F$ such that the following diagram commutes:
\begin{center}
\begin{tikzpicture}[description/.style={fill=white,inner sep=2pt}]
\matrix (m) [matrix of math nodes, row sep=3em,
column sep=3.5em, text height=1.5ex, text depth=0.25ex]
{ \mathcal{V} \times K & \F \times N \\
  J^{\infty} E  & J^{\infty} F \\};
\path[->,font=\scriptsize]
(m-1-1) edge node[auto] {$A \times \tau$} (m-1-2)
(m-1-1) edge node[auto] {$j^{\infty}$} (m-2-1)
(m-2-1) edge node[auto] {$A^{\infty}$} (m-2-2)
(m-1-2) edge node[auto] {$j^{\infty}$} (m-2-2);
\end{tikzpicture}
\end{center}
\end{cl}

\dem Using Peetre's Theorem for products \ref{Peer} we see that $A \times \tau$ is a differential operator which is a product with a submersion. Those are insular by Example \ref{e4} and hence local by Proposition \ref{ketale}.
\qed

\begin{rk} From $A \colon \E \rightarrow \F$ and $\tau \colon M \rightarrow N$ we can also construct the push-forward sheaf $\tau_* \E$ over $N$. This way of proceeding is not successful because $\tau_* \E$ is not the sheaf of sections of a smooth vector bundle in general. This is bad because we are no longer working with the same objects. But it is also bad because in this way, even though we can construct a map $\tau_* A$, this will not be in the assumptions of Peetre's Theorem (\ref{Pee1}) and will not be able to have a similar statement for $A$.

Observe that two sections in $\tau_* \E$ coincide at a point if and only if the have the same germ at that point (the only way of evaluating such a section at a point is talking about the stalk). This is definitely not what we are looking for.

Finally, even thought it does not have anything to do with field theory, there is also the notion of $\tau$-local map for a map $B \colon \F \rightarrow \E$ in the other direction. Kock \cite{K} studies differential operators of this kind. In this setting, Peetre's theorem also applies provided $\tau$ is locally non-constant. Slov\'ak's paper on the subject \cite{SLO} is a wonderful reference for minimal conditions under which Peetre's theorem still holds true. 
\end{rk}

All the relations among the different concepts of locality are summarized in the following diagram:

\begin{center}
\begin{framed}
\begin{tikzpicture}[transform shape,mylabel/.style={thin, draw=black, align=center, minimum width=0.5em, minimum height=0.5ex,fill=white}]
\matrix (m) [matrix of math nodes, row sep=2.5em,
column sep=2.5ex, text height=1.5ex, text depth=0.25ex]
{ & \textrm{Locality for products } A \times \tau & \\
 \textrm{Local} && \textrm{Differential} \\
 && \\
 \textrm{Sheaves} && \textrm{Germ-local} \\};
\path[->,font=\scriptsize]
(m-2-1) edge node[auto] {} (m-2-3)
(m-2-3) edge node[mylabel] {Proposition \ref{easy2}} (m-4-3)
(m-4-3) edge [bend left=25] node[mylabel] {Proposition \ref{soft} \\ if $\tau = \textrm{id}$ \\ $\E$ soft} (m-4-1)
(m-4-3) edge [bend right=90] node[mylabel, right=-1em] {Theorem \ref{Peer} \\ $M$ compact \\ locally} (m-2-3)
(m-2-3) edge [bend left=25] node[mylabel] {Proposition \ref{ketale} \\ if $\E$ soft \\ $\tau$ submersion } (m-2-1)
(m-4-1) edge node[mylabel] {Proposition \ref{soft} \\ if $\tau = \textrm{id}$} (m-4-3);
\end{tikzpicture}
\end{framed}
\end{center}


\printbibliography

\setcounter{part}{3}
\setcounter{chapter}{9}


\newpage
\part{{Lagrangian field theory}}\label{lft}

\newpage
\tableofcontents

\chapter*{\color{darkdelion} Lagrangian field theory}

Lagrangian field theory is a mathematical formulation of classical field theories in physics. The action is usually an integral over a base manifold $M$ of a local function on the fields. Fields are sections of a smooth fiber bundle over $M$ ($\E \defeq \Gamma(M, E)$). The adjective local is closely related to that of local maps. As a matter of fact, the Lagrangian is the pullback via the infinite jet evaluation of a bidegree $(0, \textsf{top})$ local form on the associated infinite jet bundle $J^{\infty} E$.\\

In order to properly talk about the pullback of a form on the infinite jet bundle to the insular manifold $\E \times M$ one needs to develop the concept of the tangent bundle to $\E \times M$. Once that definition has been established it is possible to talk about vector fields and forms on insular manifolds. The interesting ones from the Lagrangian field theory point of view are those coming from pro-smooth vector fields and ind-differential forms on the associated infinite jet bundle. The forms form a bicomplex, the pullback of the variational bicomplex, called the bicomplex of local forms.\\

The commutative relations among the graded endomorphisms of the bicomplex of local forms given by the differentials and the insertion of local vector fields are known as local Cartan calculus. The equations are more involved than usual due to the relation between $\E$ and $M$. One distinguishes among local, insular and decomposable vector fields -all of them with their different Cartan calculus expressions.\\

The derivation of the equations of motions, or Euler-Lagrange equations in field theories has a nice interpretation in Lagrangian field theory. It is no more than a consequence of the cohomology of the bicomplex of local forms, which has very interesting properties. The relation between symmetries and currents known as Noether's first theorem can also be proven using the same techniques.\\

{\it This part develops a theory of tangent objects in insular manifolds together with local differential forms and insular vector fields. The first chapter focuses on the commutation relations among insertion of insular vector fields and the action of the differentials. It later reviews the Lagrangian formalism developed by Zuckerman and exposes his main work: the fundamental formulae theorem and his version of Noether's first theorem. We also compare Zuckerman's paper to more recent approaches to Lagrangian field theories related to multisymplectic geometry. On the meantime, we introduce some notation and results that will be needed in the future, when talking about the \Linf { of} observables associated to a Lagrangian field theory. The basic references in this chapter are Anderson \cite{AND}, Deligne and Freed \cite{DEL}, Takens \cite{TAK}, and Zuckerman \cite{ZUC}.}


\newpage
\chapter{Local Cartan calculus}\label{lcc}

There is a notion of tangent objects for insular manifolds. Given a smooth fiber bundle $\pi \colon E \rightarrow M$ with space of smooth sections $\E$, the tangent to the insular manifold $\E \times M$ is $\mathrm{T}(\E \times M) = \Gamma(M, \ker T \pi) \times TM$. There is a map, called tangent bundle, from $\mathrm{T} (\E \times M)$ to $\E \times M$.\\

The theory of local forms and local vector fields on insular (and local manifolds) such as $\E \times M$ is of big relevance in this thesis. These concepts are defined in the same way they are defined in finite dimensional manifolds: local vector fields are special kind of sections of the tangent bundle and local forms are bundle maps from exterior powers of the tangent bundle to the trivial bundle (or the orientation line bundle in the case of $M$-twisted forms). What these maps have in special is that they cover or come from vector fields  and ind-differential forms  in the variational bicomplex respectively.\\

The pullback of a local form along a local map is again local. Insertion of local vector fields into local forms are again insular. The usual formulas known as Cartan calculus hold in the local setting. The interesting result is that once we consider the bigrading into account there are interesting non trivial commutators. For example, there are local vector fields which split into an evolutionary and a total part. The commutator of the insertion of a total vector field and the vertical differential is not trivial.\\

{\it In this chapter we define the tangent bundle to an insular manifold. We give an interpretation of it in terms of local maps. In particular we compare $j_{TE}^k$-local maps and $T (j_E^k)$-local maps. We define local vector fields and local forms and distinguish between insular and local vector fields (insular vector fields cover a decomposable vector field on the infinite jet bundle). We study the commutation relations among insertion of insular vector fields, differentials and Lie derivatives in the local setting. The basic references of this chapter are Anderson \cite{AND} and Deligne and Freed \cite{DEL}.}


\section{Tangent to \boldmath{\mbox{$\E \times M$}}}\label{fnote}

{\it This section introduces a field-theoretic version of the tangent bundle on $\E \times M$ given by $\Gamma(M, \ker{T \pi}) \times TM$. It defines the tangent map of the finite jet evaluations. It compares the insular manifold $\Gamma(M, TE) \times M$ to $\Gamma(M, \ker{T \pi}) \times TM$ by constructing a map covering a pro-smooth map between $J^{\infty} (TE)$ and $T (J^{\infty} E)$. The basic reference at this point is Deligne and Freed \cite{DEL}.}\\

We fix a smooth fiber bundle $\pi \colon E \rightarrow M$ with space of smooth section $\E$. We denote the vertical tangent bundle by $V E \defeq \ker{T \pi}$.

There are several ways of defining a tangent bundle to $\E$ and $\E \times M$. The first one is by considering them as Fr\'echet manifolds and studying their tangent spaces within the theory of Fr\'echet manifold. As a matter of fact, Fr\'echet manifolds are a subcategory of diffeological spaces (result original to Fr\"olicher \cite[Theorem 1]{FRO}, reformulated in terms of fully faithful functors by Losik \cite[Theorem 3.1.1]{LOS}), and the diffeology on $\E \times M$ and $\E$ is enough to define the concept of tangent spaces. On $\E \times M$ we the diffeology is given by local maps from any $U \subset \RE^n$ open ($n$ arbitrary) to $\E \times M$. There are two different ways of defining tangents at a point for diffeological spaces: the internal tangent space from Hector \cite{HEC} and the external tangent space from Christensen and Wu \cite{CW}. The external one is nothing else than smooth derivations at a point of the diffeological $\RE$-algebra of germs of smooth functions on $\E \times M$ (this concept is being defined by Christensen and Wu \cite{CW}). When talking about the tangent bundle, the internal tangent space is preferred, due to having a closer relation to the concept of diffeological space, Christensen and Wu \cite{CW} staudy two different diffeologies on the bundle. Our approach in this chapter is to give an insular definition of the tangent bundle of $\E \times M$. (Insular) sections of this bundle will be derivations of the algebra of local functions on $\E \times M$, the corresponding diffeological $\RE$-algebra of germs of smooth functions on $\E \times M$ --showing the relation to the diffeological exterior definition--. Since the objective of this thesis is to avoid talking about the smooth structure on $\E \times M$, including its diffeology, we will now focus on the insular approach to the tangent space and do not refer to the diffeological notion again. 

\begin{df}[Tangent bundle to $\E \times M$]\label{TEM1}
Given a smooth fiber bundle $E \rightarrow M$ we define $\textrm{T} \E := \Gamma(M, VE)$, $\textrm{T} (\E \times M) := \Gamma(M, VE) \times TM$ and the tangent bundle of $\E \times M$ to be the map:
$$\textrm{pr}_{\E \times M} \defeq (\textrm{pr}_E)_* \times \textrm{pr}_M \colon \textrm{T} (\E \times M) = \Gamma(M, V E) \times TM \rightarrow \E \times M.$$
\end{df}

This definition is not original from this thesis. It follows the idea of considering a variation of $\varphi \in \E$ to be given by a map associating to every $x \in M$ a vector $\partial \varphi (x) \in V_{\varphi(x)} E$, a variation in the vertical direction $V E = \ker{T \pi}$. In other words $\partial \varphi \in \Gamma(M, \varphi^* V E)$. The collection of all sections $\varphi$ and all variations at that section is precisely $\Gamma(M, VE)$: what has been defined as $\mathrm{T} \E$. This approach is the one followed by Deligne and Freed as can be seen \cite[page 158]{DEL}.

\begin{figure}[h]
	\centering
	\includegraphics[width=0.38\textwidth]{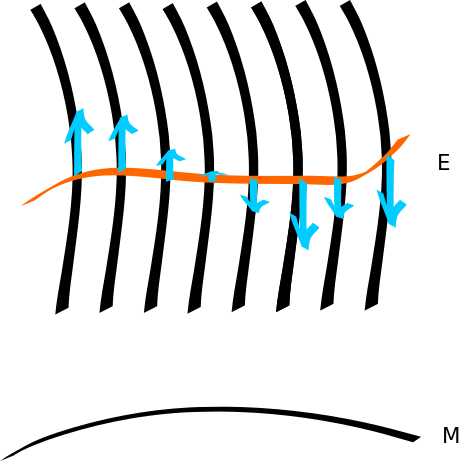}
	\caption{A section $\varphi$ (in orange), and a variation $\partial \varphi$ of it in blue.}
	\label{fig:isl}
\end{figure}

Now that we know what the tangent of an object in $\iMan$ is, we would like to know what the tangent of an insular map is. As a disclaimer, we would not do such a thing in full generality, simply because we do not need to. For us the tangent of an insular (or local for that matter) map is only relevant because we can use it to pullback local differential forms on the associated objects. These will be defined in the next section where we will also learn that in order to take the pullback of a local differential form it is enough to know the tangent of the finite jet evaluations: $j^k \colon \E \times M \rightarrow J^k E$. Observe that finite jet evaluations are also insular maps, so that we are going to define at least some tangents of insular maps. In order to define $\mathrm{T} j^k$ we need the following lemma: 

\begin{lm}\label{uk} The map $u^k \colon J^k (VE) \rightarrow V (J^k E)$ sending $\left(\frac{\partial}{\partial u_{\alpha}}\right)^{I}$ to $\frac{\partial}{\partial u_{\alpha}^I}$ is well defined, smooth and induces a pro-smooth map $u^{\infty} \colon J^{\infty} E \rightarrow T (J^{\infty} E)$.
\end{lm}

The proof of the lemma is postponed to the next subsection.

\begin{df}
Given a smooth fiber bundle $E \rightarrow M$ and an integer $k$ we define the tangent of the $k$-th jet evaluation to be the map:
\begin{eqnarray}\label{odin}
\textrm{T} j^k \colon \textrm{T} \E \times TM & \longrightarrow & T (J^k E) \nonumber \\
((\varphi,x), (\partial \varphi, v))& \longmapsto & T(j^k \varphi) (x, v) + u^k (j_x^k(\varphi, \partial \varphi)).
\end{eqnarray}
\end{df}

\begin{rk}\label{nlin} We want to stress out that the map is $\RE$-linear in the fibers since tangent maps and the prolongations are $\RE$-linear. But observe that fixing $(x,v) \in TM$ and $\varphi \in \E$, the map 
$$\textrm{T}_{(\varphi, x)} j^k (-,v) \colon \Gamma(M, \varphi^* VE) \rightarrow T(j_x^k \varphi) (J^k E) \textrm{ is not } \C(M)-\textrm{linear}$$
since $j_x^k (f \partial \varphi) \neq f(x) j_x^k (\partial \varphi)$ for a general smooth function $f \in \C(M)$.
\end{rk}

As a consequence of this definition, we have that the tangent bundle covers the pro-smooth map $\textrm{pr}_{J^{\infty} E} \colon T (J^{\infty} E) \rightarrow J^{\infty} E$ since $u^k$ is pro-smooth by Lemma \ref{uk} and the other summand clearly commutes with $T \pi_k^l$:

\begin{pp}\label{ff} The tangent bundle $\textrm{pr}_{\E \times M} \colon \Gamma(M, V E) \times TM \rightarrow \E \times M$ covers the pro-smooth map $\textrm{pr}_{J^{\infty} E} \colon T (J^{\infty} E) \rightarrow J^{\infty} E$.
\end{pp}

\begin{center}
\begin{tikzpicture}[description/.style={fill=white,inner sep=2pt}]
\matrix (m) [matrix of math nodes, row sep=3em,
column sep=4em, text height=1.5ex, text depth=0.25ex]
{ \textrm{T} \E \times TM & \E \times M\\
  T (J^{\infty} TE) & J^{\infty} E \\};
\path[->,font=\scriptsize]
(m-1-1) edge node[auto] {$\textrm{pr}_{\E \times M}$} (m-1-2)
(m-1-2) edge node[auto] {$j^{\infty}$} (m-2-2)
(m-1-1) edge node[auto] {$T j^{\infty}$} (m-2-1)
(m-2-1) edge node[auto] {$\textrm{pr}_{J^{\infty}E}$} (m-2-2);
\end{tikzpicture}
\end{center}

\begin{rk} Once again, this definition comes from the same spirit as Definition \ref{TEM1}. The tangent of $j^k$, $\textrm{T} j^k \colon \Gamma(M, VE) \times M \rightarrow T (J^k E)$, should be $\RE$-linear in the fibers. This means that
$$\textrm{T}_{(\varphi, x)} j^k (\partial \varphi, v) = \textrm{T}_{(\varphi, x)} j^k (0 , v) + \textrm{T}_{(\varphi, x)} j^k (\partial \varphi, 0),$$
{\noindent where $(\varphi, x) \in \E \times M$; $\partial \varphi, 0 \in \Gamma(M, \varphi^* VE)$; and $v, 0 \in T_x M$. We would like to recover the usual tangent when restricting to smooth manifolds so that $\textrm{T}_{(\varphi, x)} j^k (0 , v)$ has no other option but to be $T (j^k \varphi) (x,v) \in T(J^k E)$.}

To compute $\textrm{T}_{(\varphi, x)} j^k (\partial \varphi, 0)$ we simply need to think of $\partial \varphi$ as an infinitesimal translation of $\varphi$, so that 
$$\frac{\partial}{\partial x^I} (\varphi_{\alpha} + (\partial \varphi)_{\alpha}) = \varphi_{\alpha}^I + (\partial \varphi)_{\alpha}^I.$$ 
In this way  $\textrm{T}_{(\varphi, -)} j^k (-, 0) \colon \Gamma(M, \varphi^* VE) \times M \rightarrow T (J^k E)$ should send $(\partial \varphi)_{\alpha} \frac{\partial}{\partial u_{\alpha}}$ to $(\partial \varphi)_{\alpha}^I \frac{\partial}{\partial u_{\alpha}^I}$. These are the coefficients of $j_x^k(\partial \varphi)$, but observe that $j_x^k(\partial \varphi) \in J^k(V E)$ which is inside of $J^k(TE)$. That is not the same space as the desired $T(J^k E)$. In order to fix this we use the map $u^k \colon J^k (VE) \rightarrow V (J^k E)$ from Lemma \ref{uk}.
\end{rk}

\vspace{1em}

At this point we are ready to define local vector fields and local $1$-forms to be maps commuting with $\{j^k\}$ and $\{\textrm{T} j ^k\}$. We refer to the following two sections for the detailed definitions. But before we do that, we want to point out that such maps are not going to be local maps, since when working with $\Gamma(M, VE) \times TM$ we use $\textrm{T} j^k$ and not $j^k$ for any bundle $F \rightarrow N$. As a matter of fact, $VE \rightarrow M$ is not a bundle over $TM$ so that not even $\Gamma(M, VE) \times TM$ is an object in insular or local manifolds.

Nevertheless, there is another approach to this problem which keeps things in $\iMan$ and which is motivated from the the study of the smooth Fr\'echet structures on $\E \times M$. Recall that a trivialization of $\E$ was given for all $\varphi \in \E$ by a map to the Fr\'echet space $\Gamma(K, \varphi^* T E)$ where $K$ is a compact subset of $M$. $\Gamma(M, TE) \times M$ is an object in insular (and local) manifolds. Observe the two apparently different pairs $\Gamma(M, VE) \times TM$ and $\Gamma(M, TE) \times M$. It turns out that the two products are closely related. Let us study $\Gamma(M, TE)$ in more detail:

\begin{tm}\label{spl} Let $\pi \colon E \rightarrow M$ be a smooth vector bundle. We have the following decomposition:
$$\Gamma(M, TE) \cong \textrm{T} \E \times \mathfrak{X}(M).$$
\end{tm}

Before proving the theorem we would like to point out that at the end of Appendix \ref{tf} we mentioned that we needed a connection in order to split $\Gamma(E,TE)$ into vertical and horizontal parts. Theorem \ref{spl} precisely says that when the sections are taken over $M$ there is no need of a connection.

\dem Observe that 
$$\E \times \mathfrak{X}(M) = \Gamma(M, E) \times \Gamma(M, TM) \cong \Gamma(M, E \times_M TM) = \Gamma(M, \pi^* TM).$$ We have the following short exact sequence of vector bundles over $E$:
$$ 0 \longrightarrow VE \longrightarrow TE \longrightarrow \pi^* TM \longrightarrow 0.$$
Taking sections over $M$ we get a sequence of maps:
$$ 0 \longrightarrow \Gamma(M,VE) \longrightarrow \Gamma(M,TE) \longrightarrow \Gamma(M, \pi^* TM) \longrightarrow 0,$$
\noindent{ which we prefer to rewrite as:}
$$ 0 \longrightarrow \textrm{T} \E \stackrel{\textrm{inc}_*}{\xrightarrow{\hspace*{3em}}} \Gamma(M,TE) \stackrel{((\textrm{pr}_E)_* , (T \pi)_*)}{\xrightarrow{\hspace*{3em}}} \E \times \mathfrak{X}(M) \longrightarrow 0.$$
All maps in this sequence are local as a consequence of Example \ref{secti}. Moreover, the sequence is a short exact sequence in the appropriate category (the over category over $\E$ of sections of fiber bundles on $M$ with the extra property that the fibers on $\E$ are vector spaces --or vector bundles over $\E$ if we want to approach the problem that way--). We will actually show in the next lemma the exactness at every level by finding local one side inverses of both maps. This will show that the sequence splits and hence that
$$\Gamma(M, TE) \cong T \E \times_{\E} (\E \times \mathfrak{X}(M)) \cong T \E \times \mathfrak{X}(M).$$
\qed

The proof of Theorem \ref{spl} relies on the following lemma that will be proven in the next subsection.

\begin{lm}\label{rands} Let $E \rightarrow M$ be a smooth fiber bundle. The maps
\begin{center}
    \begin{tabular}{rcl}
	  $r \colon \Gamma(M, TE) $ & $\longrightarrow$ & $T \E$ \\
    $\chi$ & $\longmapsto$ & $\chi - \left(T(\textrm{pr}_E \circ \chi) \circ T \pi \circ \chi \right)$ \vspace{1em} \\
	  $s \colon \E \times \mathfrak{X}(M)$ & $\longrightarrow$ & $\Gamma(M, TE)$ \\
      $(\varphi, X)$ & $\longmapsto$ & $T \varphi \circ X$ \\
    \end{tabular}
\end{center}
{\noindent are well defined, $r$ is a retraction of $\mathrm{inc}_* \colon T \E \rightarrow \Gamma(M, TE)$, and $s$ is a section of $((\textrm{pr}_E)_* , (T \pi)_*) \colon \Gamma(M,TE) \rightarrow \E \times \mathfrak{X}(M)$. Moreover $r \times \textrm{id}_M$ and $s \times \textrm{id}_M$ are local.}
\end{lm}

In order to relate $\Gamma(M, TE) \times M$ and $\textrm{T} \E \times TM$ we need the previous isomorphism $\Gamma(M, TE) \cong \textrm{T} \E \times \mathfrak{X}(M)$, the map $u^k$ and one extra ingredient:
\begin{eqnarray*}
v^k \colon J^{k+1} (E \times_M TM) & \longrightarrow & T (J^k E)\\
(j_x^{k+1} \varphi, j_x^0 X) & \longmapsto & T(j^k \varphi) \circ X (x).
\end{eqnarray*}

\begin{pp}\label{tjvsj} Let $\pi \colon E \rightarrow M$ be a smooth vector bundle. While not being local, the map $\textrm{id}_{\textrm{T} \E} \times j^0 \colon \textrm{T} \E \times \mathfrak{X}(M) \times M \rightarrow \textrm{T} \E \times TM$ covers a pro-smooth map $\{u^k + v^k \colon J^{k+1}(TE) \rightarrow T (J^k E)\}_{k \in \mathbb{N}}$ in the sense that the following diagram commutes for all $k \geqslant l$:
\begin{center}
\begin{tikzpicture}[description/.style={fill=white,inner sep=2pt}]
\matrix (m) [matrix of math nodes, row sep=3em,
column sep=4em, text height=1.5ex, text depth=0.25ex]
{ \textrm{T} \E \times \mathfrak{X}(M) \times M & \textrm{T} \E \times TM\\
  J^{k+1} (TE) & T (J^{k} E) \\
  J^{l+1} (TE) & T (J^{l} E) \\};
\path[->,font=\scriptsize]
(m-1-1) edge node[auto] {$\textrm{id}_{\textrm{T} \E} \times j^0$} (m-1-2)
(m-1-2) edge node[auto] {$j^k$} (m-2-2)
(m-1-1) edge node[auto] [swap]{$\widetilde{j^{k+1}}$} (m-2-1)
(m-2-1) edge node[auto] {$u^k + v^k$} (m-2-2)
(m-2-1) edge node[auto] [swap]{$(\pi \circ \textrm{pr}_E)_{k+1}^{l+1}$} (m-3-1)
(m-2-2) edge node[auto] {$T \pi_k^l$} (m-3-2)
(m-3-1) edge node[auto] {$u^l + v^l$} (m-3-2);
\end{tikzpicture}
\end{center}
\end{pp}

We should be a bit careful in understanding the previous diagram cause we are using the isomorphism given by Theorem \ref{spl} heavily. This is explained in the proof of the result:

\dem Observe, first of all that the map $\widetilde{j^{k+1}} \colon \textrm{T} \E \times \mathfrak{X}(M) \times M \rightarrow J^{k+1} TE$ is not quite $j_{TE}^{k+1}$ but $j_{TE}^{k+1} \circ (\textrm{inc}^* \times s)$ due to Theorem \ref{spl}. To be explicit 
\begin{eqnarray}\label{zva}
\widetilde{j^{k+1}} \colon \E \times \mathfrak{X}(M) \times M & \longrightarrow & J^{k+1} (T E) \nonumber \\
(j_x^{k+1} \varphi, j_x^{k+1} X) & \longmapsto & j_x^{k+1}(T \varphi \circ X) + j_x^{k+1} \psi,
\end{eqnarray}

Comparing the maps \ref{odin} and \ref{zva} we can see that the first summand is taken care of by the map $u^{k+1}$ and the map $v^k$ should replace 
$j^{k+1}(T \varphi \circ X) = j^k(\varphi_* X)$ by $T (j^{k} \varphi) \circ X = (j^k \varphi)_* X$, but that is precisely how $v^k$ is defined above.

As a matter of fact, in order to apply $u_k$ and $v_k$ we prefer to take the explicit jet evaluations of the two parts $\E \times \mathfrak{X}(M)$ and $\mathrm{T} \E \times M$ respectively, because they help us to see that actually the jet degree does not change for $u^k$. Here we use once again Theorem \ref{spl}. The two diagrams are the following:

\begin{center}
\begin{tikzpicture}[description/.style={fill=white,inner sep=2pt}]
\matrix (m) [matrix of math nodes, row sep=3em,
column sep=4em, text height=1.5ex, text depth=0.25ex]
{ \E \times \mathfrak{X} \times M & \E \times TM & \textrm{T} \E \times M &  \textrm{T} \E \times M\\
  J^{k+1} (E \times_M TM ) & T (J^{k} E) & J^k (VE) & T (J^k E)\\};
\path[->,font=\scriptsize]
(m-1-1) edge node[auto] {$\textrm{id}_{\E} \times j^0$} (m-1-2)
(m-1-2) edge node[auto] {$T j^k(0,-)$} (m-2-2)
(m-1-1) edge node[auto] {$j^{k+1}$} (m-2-1)
(m-2-1) edge node[auto] {$v^k$} (m-2-2)

(m-1-3) edge node[auto] {$\textrm{id}$} (m-1-4)
(m-1-4) edge node[auto] {$T j^k(-,0)$} (m-2-4)
(m-1-3) edge node[auto] {$j^{k}$} (m-2-3)
(m-2-3) edge node[auto] {$u^k$} (m-2-4);
\end{tikzpicture}
\end{center}
\qed

They key point of the previous result is that we can jump from local maps to/from $\mathrm{T} \E \times \mathfrak{X}(M) \times M$ and maps commuting with $\{\textrm{T} j^k \}$ to/from $\mathrm{T} \E \times TM$. As a a first example of this, we will study local vector fields in the following section.


\subsection{Proof of the technical lemmas}

The proofs of two technical Lemmas have been postponed to this point in the section. The first of them Lemma \ref{uk} was:

\begin{lm1} The map $u^k \colon J^k (VE) \rightarrow V (J^k E)$ sending $\left(\frac{\partial}{\partial u_{\alpha}}\right)^{I}$ to $\frac{\partial}{\partial u_{\alpha}^I}$ is well defined, smooth and induces a pro-smooth map $u^{\infty} \colon J^{\infty} E \rightarrow T (J^{\infty E})$.
\end{lm1}

{\bf Proof of Lemma \ref{uk}.}
The proof for vector bundles is easier to understand, so we include it to provide better intuition into the problem.
 
If $E$ is a vector bundle, $(\varphi^* VE)_x \cong E_x$ so that $\partial \varphi$ can be understood as a section of $E$. Using this interpretation, $j^k(\partial \varphi)$ is seen as a section of $J^k E$, and once again since $\left((j^k \varphi)^* V(J^k E)\right)_x \cong (J^k E)_x$ we can view $j^k(\partial \varphi)$ as a section of $(j^k \varphi)^* V(J^k E)$. The map $u^k$ defined in the lemma maps $j_x^k(\varphi, \partial \varphi)$ to $(j_x^{k} \varphi, j_x^{k} \partial \varphi)$ using the equivalences explained above.

In the general case, where $E$ is not necessarily a vector bundle, we call $\psi \defeq (\varphi, \partial \varphi)$ for simplicity. We want to show that the map $u^k$ sends $j_x^k (\psi)$ to $\psi^k(x)$ where the map $\psi^k \colon M \rightarrow V(J^k E)$ will be a section defined shortly. First observe that $\psi \colon M \rightarrow VE$ can be represented by a map $f \colon M \times \RE \rightarrow E$ with the following properties:
\begin{itemize}
	\item[$a_0$)] $f(-,0) = \varphi$.
	\item[$b_0$)] $\restrict{\frac{\partial}{\partial t} f(x,-)}{t=0} = \psi(x)$ for all $x \in M$.
	\item[$c_0$)] $f(-,t) \in \E$ for all $t \in \RE$.
\end{itemize}

We want to construct a map $\psi^k \colon M \rightarrow V(J^k E)$ in a similar way from some $f^k$ with the property that $(\psi^k)_{\alpha}^I = \frac{\partial^I}{\partial x^I} (\partial \varphi)_{\alpha}$, showing that $u^k$ maps indeed $j_x^k (\psi)$ to $\psi^k(x)$. We define the map:
\begin{eqnarray*}
f^k \colon M \times \RE &\longrightarrow & J^k E\\
(x, t) & \longmapsto & j_x^k f(-,t).
\end{eqnarray*}

We need to check that $f^k$ does not depend on anything else than the $k$-th jet of $\psi$. It satisfies the following properties:
\begin{itemize}
	\item[$a_k$)] $f^k(-,0) = j^k(\varphi)$.
	\item[$b_k$)] $\left(\restrict{\frac{\partial}{\partial t} f^k(x,-)}{t=0}\right)_{\alpha}^I = \frac{\partial^I}{\partial x^I} (\partial \varphi)_{\alpha}$ for all $x \in M$.
	\item[$c_k$)] $f^k(-,t) \in \mathcal{J}^k \E$ for all $t \in \RE$.
\end{itemize}

$a_k$ follows from $a_0$ and $c_k$ from $c_0$ automatically. For $b_k$ just observe that $$f^k(x,-)_{\alpha}^I(t)= u_{\alpha}^I(f(-,t), x) = \frac{\partial^I }{\partial x^I} f_{\alpha}(-, t) \textrm{ so that } \frac{\partial}{\partial t}  f^k(x,-)_{\alpha}^I(t) = \frac{\partial^{I,t} }{\partial x^I \partial t} f_{\alpha}(-, t).$$ When evaluating at $t=0$, using $b_0$ we can show that:
$$\left(\restrict{\frac{\partial}{\partial t} f^k(x,-)}{t=0}\right)_{\alpha}^I =  \frac{\partial^{I} }{\partial x^I } \restrict{\frac{\partial}{\partial t} f(x,-)}{t=0} = \frac{\partial^I}{\partial x^I} (\partial \varphi)_{\alpha}.$$

In any case, $u_k$ clearly commutes with $\pi_k^l$ and $T \pi_k^l$ since it sends $\left(\frac{\partial}{\partial u_{\alpha}}\right)^{I}$ to $\frac{\partial}{\partial u_{\alpha}^I}$.
\qed

The second one, Lemma \ref{rands}, was the following:

\begin{lm2} Let $E \rightarrow M$ be a smooth fiber bundle. The maps
\begin{center}
    \begin{tabular}{rcl}
	  $r \colon \Gamma(M, TE) $ & $\longrightarrow$ & $T \E$ \\
    $\chi$ & $\longmapsto$ & $\chi - \left(T(\textrm{pr}_E \circ \chi) \circ T \pi \circ \chi \right)$ \vspace{1em} \\
	  $s \colon \E \times \mathfrak{X}(M)$ & $\longrightarrow$ & $\Gamma(M, TE)$ \\
      $(\varphi, X)$ & $\longmapsto$ & $T \varphi \circ X$ \\
    \end{tabular}
\end{center}
{\noindent are well defined, $r$ is a retraction of $\mathrm{inc}_* \colon T \E \rightarrow \Gamma(M, TE)$, and $s$ is a section of $((\textrm{pr}_E)_* , (T \pi)_*) \colon \Gamma(M,TE) \rightarrow \E \times \mathfrak{X}(M)$. Moreover $r \times \textrm{id}_M$ and $s \times \textrm{id}_M$ are local.}
\end{lm2}

{\bf Proof of Lemma \ref{rands}:}
We begin by pointing out that for all $\varphi \in \E$ and all $X \in \mathfrak{X}(M)$ we have that
$$\textrm{pr}_E \circ T \varphi \circ X = \varphi \circ \textrm{pr}_M \circ X = \varphi.$$
Further composing with $\pi$ we see that $\pi \circ \textrm{pr}_E \circ s(\varphi, X) = \pi \circ \varphi = \textrm{id}_M$ so that $s(\varphi, X)$ is indeed a section of $TE \rightarrow M$.

But we can also use the previous equation to show that $r(\chi)$ is a well defined smooth map from $M$ to $TE$ for all $\chi \in \Gamma(M, TE)$ since both $\chi$ and $\left(T(\textrm{pr}_E \circ \chi) \circ T \pi \circ \chi \right)$ are supported over the same section in $\E$, namely $\textrm{pr}_E \circ \chi$. This is so because when applying the previous equation to $\varphi = \textrm{pr}_E \circ \chi$ and $X = T \pi \circ \chi$ we have
$$\textrm{pr}_E \left( T(\textrm{pr}_E \circ \chi) \circ T \pi \circ \chi  \right) =  \textrm{pr}_E \circ \chi.$$

It is indeed a section of $TE \rightarrow M$ since $\chi$ was a section. We also have to see that $\chi - \left(T(\textrm{pr}_E \circ \chi) \circ T \pi \circ \chi \right)$ is in the kernel of $T \pi$. But $\pi \circ \textrm{pr}_E \circ \chi = \textrm{id}_M$ so that $T(\pi \circ \textrm{pr}_E \circ \chi )= \textrm{id}_{TM}$ and hence $T \pi \circ \chi = T \pi \circ \left(T(\textrm{pr}_E \circ \chi) \circ T \pi \circ \chi \right)$ showing what $T \pi \circ r(\chi) = 0$ and thus that $r(\chi) \in T \E$.

The next step is to show that $r$ is a retraction of $\textrm{inc}_*$. We want to show that for all $\xi \in T \E$, $r(\xi)=\xi$. But this is clear since $T \pi \circ \xi = 0$ and hence $r(\xi) = \xi - T(\textrm{pr}_E \circ \xi) \circ 0 = \xi$.

Now we check that $s$ is a section of $(\textrm{pr}_E)_* \times (T \pi)_* \colon \Gamma(M,TE) \rightarrow \E \times \mathfrak{X}(M)$; in other words given $(\varphi, X) \in \E \times \mathfrak{X}(M)$ we want to show that 
$$(\textrm{pr}_E \circ T \varphi \circ X, T \pi \circ T \varphi \circ X) = (\varphi, X)$$ but this is true since from one side
\begin{eqnarray*}
\textrm{pr}_E \circ T \varphi \circ X &=& \varphi \circ \textrm{pr}_M \circ X = \varphi \circ \textrm{id}_M = \varphi \textrm{ and from the other side} \\
T \pi \circ T \varphi \circ X &=& T(\pi \circ \varphi) \circ X = \textrm{id}_{TM} \circ X = X.
\end{eqnarray*}

The locality of the maps $(r \times \textrm{id}_M)$ and $(r \times \textrm{id}_M)$ is immediate. They descend to bundle maps over the identity, $J^1(TE) \rightarrow VE$ and $J^1 E \times_M TM \subset J^!(E \times_M TM) \rightarrow TE$ that by Proposition \ref{pro} give rise to the corresponding pro-smooth maps making $(r \times \textrm{id}_M)$ and $(r \times \textrm{id}_M)$ local.

Finally we also show the exactness in the middle of the sequence
$$ 0 \longrightarrow T \E \stackrel{\textrm{inc}_*}{\xrightarrow{\hspace*{3em}}} \Gamma(M,TE) \stackrel{(\textrm{pr}_E)_* \times (T \pi)_*}{\xrightarrow{\hspace*{3em}}} \E \times \mathfrak{X}(M) \longrightarrow 0.$$

This is a consequence of the original short exact sequence, but it can also be proven using the maps $r$ and $s$. Consider the composite 
$$s \circ ((\textrm{pr}_E)_* \times (T \pi)_*) \colon \Gamma(M, TE) \rightarrow \Gamma(M, TE)$$ sending $\chi \in \Gamma(M, TE)$ to $T(\textrm{pr}_E \circ \chi)\circ T \pi \circ \chi$. This shows the interesting equality $r = \textrm{id}_{\Gamma(M, TE)} - (s \circ ((\textrm{pr}_E)_* \times (T \pi)_*))$.

Now $((\textrm{pr}_E)_* \times (T \pi)_*)(\chi)=0$ if and only if $s \circ ((\textrm{pr}_E)_* \times (T \pi)_*) (\chi) = 0$ since $s$ is injective (it is a section). More over that is equivalent to $\chi = r(\chi)$ by the previous comment, or more precisely $\chi = \textrm{inc}_* r(\chi)$ which shows that the image of $\textrm{inc}_*$ is the kernel of $(\textrm{pr}_E)_* \times (T \pi)_*$. 
\qed


\section{Local vector fields}\label{lmlvf}

{\it In this section we give three equivalent definitions of local vector fields. The first one as sections of the tangent bundle covering a vector field on the infinite jet bundle. The second one as a derivation of the local algebra of functions depending on a vector field on the infinite jet bundle. The last one is as certain local sections of $\Gamma(M, TE) \times M \rightarrow \E \times M$. For local vector fields, being a product of a vertical and a local family of horizontal vector fields, being insular and preserving the Cartan distribution are equivalent notions. Our starting point for the definition of a local vector field is that of Deligne and Freed \cite{DEL}.}\\

We fix a smooth fiber bundle $\pi \colon E \rightarrow M$ with space of smooth sections $\E$. Recall that the tangent bundle of $\E \times M$ covers the pro-smooth map projecting $T(J^{\infty} E)$ to its base (Proposition \ref{ff}).

\begin{df}[$\mathfrak{X}_{\textrm{loc}}(\E)$]\label{lve}
A local vector field on $\E$ is a section $\xi$ of $\textrm{pr}_E^* \colon \textrm{T} \E \rightarrow \E$ such that $\xi \times \textrm{id}_M$ is covered by a vector field on $J^{\infty} E$ supported on $j^{\infty}(\E \times M)$.
\end{df}

This is the definition Deligne and Freed follow \cite[page 166]{DEL}. The way they talk about local vector fields on $\E$, it seems that they jump from this notion to the equivalent one from Definition/Proposition \ref{lvf2} that will be found at the end of the section. (See Remark \ref{notation} for a more detailed explanation of this.)

We are interested in more general vector fields, also involving the $M$ direction:

\begin{df}[$\mathfrak{X}_{\textrm{loc}}(\E \times M)$]\label{lvf}
A local vector field on $\E \times M$ is a section of the tangent bundle of $\E \times M$ covered by a vector field on $J^{\infty} E$ supported on $j^{\infty}(\E \times M)$.
\end{df}

There is no motivation at the moment for the use of the word local. Later on we will justify that term. A local vector field can be represented by the following diagram

\begin{center}
\begin{tikzpicture}[description/.style={fill=white,inner sep=2pt}]
\matrix (m) [matrix of math nodes, row sep=3em,
column sep=4em, text height=1.5ex, text depth=0.25ex]
{\E \times M & \textrm{T} \E \times TM & \E \times M\\
  J^{\infty} E & T (J^{\infty} E) & J^{\infty} E\\};
\path[->,font=\scriptsize]
(m-1-1) edge node[auto] {$(\chi, \textrm{id}_M)$} (m-1-2)
(m-1-1) edge node[auto] {$j_{E}^{\infty}$} (m-2-1)
(m-1-2) edge node[auto] {$T j^{\infty}$} (m-2-2)
(m-2-1) edge node[auto] [swap]{$\chi^{\infty}$} (m-2-2)
(m-1-2) edge node[auto] {$\textrm{pr}_{\E \times M}$} (m-1-3)
(m-1-3) edge node[auto] {$j_{E}^{\infty}$} (m-2-3)
(m-2-2) edge node[auto] [swap]{$\textrm{pr}_{J^{\infty}E}$} (m-2-3)
(m-1-1) edge [bend left=40] node[below=0.4em] {$\textrm{id}_{\E \times M}$} (m-1-3)
(m-2-1) edge [bend right=40] node[above=0.3em] {$\textrm{id}_{J^{\infty} E}$} (m-2-3);
\end{tikzpicture}
\end{center}

For simplicity we have assumed the bundle is soft. To be precise we should replace the infinite jet bundle and its tangent to their restriction to the open subset $j^{\infty}(\E \times M)$.

Recall that $\C_{\textrm{loc}}(\E \times M)$ is an algebra due to Proposition \ref{subal}.

\begin{dfpp}[Lie algebra of local vector fields] A derivation $D$ of $\C_{\textrm{loc}}(\E \times M)$ is called local if and only if there is a vector field $\chi^{\infty}$ on $J^{\infty} E$ such that $D f = (\chi^{\infty} \cdot f^{\infty}) \circ j^{\infty}$ for every local function $(f, f^{\infty})$. When $D$ and $\chi^{\infty}$ are related in that way we denote the local derivation as $(D ,\chi^{\infty})$. Then: 
\begin{itemize}
	\item Local vector fields $(\chi, \chi^{\infty})$ are in one to one correspondence with local derivations of $\C_{\textrm{loc}}(\E \times M)$, $(D, \chi^{\infty})$.
	\item Local derivations form a Lie subalgebra of derivations.
	\item Given two local derivations $(D_1, \chi_1^{\infty})$ and $(D_2, \chi_2^{\infty})$ we have that for every local function $(f, f^{\infty})$, $[D_1, D_2] f = ([\chi_1^{\infty}, \chi_2^{\infty}] \cdot f^{\infty}) \circ j^{\infty}$ so that $([D_1, D_2], [\chi_1^{\infty}, \chi_2^{\infty}])$ is the corresponding local derivation.
\end{itemize}
\end{dfpp}

\dem By Proposition \ref{vfeq} we know that vector fields on $J^{\infty} E$ are in one to one correspondence with ind-derivations of $\C(J^{\infty} E)$. By definition a, local derivation of $\C_{\textrm{loc}}(\E \times M)$ is one that comes from an ind-derivation. Given $(\chi, \chi^{\infty})$ we define $D f \defeq (\chi^{\infty} \cdot f^{\infty}) \circ j^{\infty}$ for every local function $(f, f^{\infty})$. Conversely, given $(D, \chi^{\infty})$ we define $\chi (\varphi, x) \defeq \left(\varphi, x, Dx^i(\varphi, x) \frac{\partial}{\partial x^i},  Du_{\alpha}(\varphi, x) \frac{\partial}{\partial u_{\alpha}} \right)$ in local coordinates. This proves the first item. A quick computation proves $3$:
\begin{eqnarray*}
	{[}D_1, D_2{]} f & =&  D_2 \circ D_1 (f) - D_1 \circ D_2 (f) \\
	&=& Q_2 ((\chi_1^{\infty} \cdot f^{\infty}) \circ j^{\infty}) - D_1 ((\chi_2^{\infty} \cdot f^{\infty}) \circ j^{\infty}) \\
	& = & (\chi_2^{\infty} \cdot (\chi_1^{\infty} \cdot f^{\infty}))\circ j^{\infty} - (\chi_1^{\infty} \cdot (\chi_2^{\infty} \cdot f^{\infty}))\circ j^{\infty} \\
	&=& ({[}\chi_1^{\infty}, \chi_2^{\infty}{]} \cdot f^{\infty}) \circ j^{\infty}.\\
\end{eqnarray*}
Item $2$ follows from $3$. \qed

There is a third equivalent way of defining local vector fields and it has to do with the map $\textrm{id}_{\textrm{T} \E} \times j^0 \colon \textrm{T} \E \times \mathfrak{X}(M) \times M \rightarrow \textrm{T} \E \times TM$ from Proposition \ref{tjvsj}.

\begin{lm}\label{oneto} We have a one to one correspondence between sections of the tangent bundle on $\E \times M$ and sections of $\textrm{T} \E \times \mathfrak{X}(M) \times M \rightarrow \E \times M$ which have the vector field part independent of $M$. The correspondence is the following
\begin{eqnarray*}
\Gamma_{\Set}(\E \times M, \textrm{T} \E \times TM) & \stackrel{1 \colon 1}{\longleftrightarrow} & \Gamma(\E \times M, \textrm{T} \E \times M) \times \textrm{Hom}_{\Set}(\E, \mathfrak{X}(M)).\\
\chi=(\chi_{\textrm{T} \E}, \chi_{TM}) & \longmapsto & \left( \chi_{\textrm{T} \E}, \overbar{\chi}_{\mathfrak{X}(M)}(\varphi) \defeq \chi_{TM}(\varphi, -) \right) \\
\left( \overbar{\chi}_{\textrm{T} \E}, \chi_{TM}(\varphi, x) \defeq \overbar{\chi}_{\mathfrak{X}(M)}(\varphi)(x) \right) & \longleftmapsto & \left( \overbar{\chi}_{\textrm{T} \E}, \overbar{\chi}_{\mathfrak{X}(M)} \right).
\end{eqnarray*}
\end{lm}

The proof is immediate. In short, we have said that $\mathfrak{X}(M)$ is the exponential object $TM^M$ in some category (at the moment a category very close to sets). We would like to extend this to a category that takes into account all the local and smooth structures of the maps. We have decided not to be explicit about the category theory construction, since developing the correct category where $\mathfrak{X}(M) = TM^M$ is more lengthy than the following lemma and proposition, which have the same effect.

Observe that $\chi = (\textrm{id}_{\textrm{T} \E} \times j^0) \circ \overbar{\chi}$. Given any local map $(\overbar{\chi}, \textrm{id}_M)$, by composition with the map from Proposition \ref{tjvsj}, we get a map which is a candidate of a local vector field:
\begin{center}
\begin{tikzpicture}[description/.style={fill=white,inner sep=2pt}]
\matrix (m) [matrix of math nodes, row sep=3em,
column sep=4em, text height=1.5ex, text depth=0.25ex]
{\E \times M & \textrm{T} \E \times \mathfrak{X}(M) \times M & \textrm{T} \E \times TM\\
  J^{\infty} E & J^{\infty} (TE) & T (J^{\infty} E)\\};
\path[->, font=\scriptsize]
(m-1-1) edge node[auto] {$(\overbar{\chi}, \textrm{id}_M)$} (m-1-2)
(m-1-1) edge node[auto] {$j_{E}^{\infty}$} (m-2-1)
(m-1-2) edge node[auto] {$j_{TE}^{\infty}$} (m-2-2)
(m-2-1) edge node[auto] {$\overbar{\chi}^{\infty}$} (m-2-2)
(m-2-2) edge node[auto] {$u^{\infty} + v^{\infty}$} (m-2-3)
(m-1-1) edge [bend left=30] node[below=0.4em] {$\chi$} (m-1-3)
(m-2-1) edge [bend right=20] node[above=0.4em] {$\chi^{\infty}$} (m-2-3);
\path[->,color=darkdelion, font=\scriptsize]
(m-1-2) edge node[auto] {$\textrm{id}_{\textrm{T} \E} \times j^0$} (m-1-3)
(m-1-3) edge node[auto] {$T j_E^{\infty}$} (m-2-3);
\end{tikzpicture}
\end{center}

\begin{rk}\label{dlin} Similarly to what was said in Remark \ref{nlin}, the composition depicted in orange, $\textrm{T} j_E^{k} \circ (\textrm{id}_{\textrm{T} \E} \times j^0) \colon \textrm{T} \E \times \mathfrak{X}(M) \times M \rightarrow T(J^k E)$ is not $\C(M)$-linear on the field direction. Nevertheless, it is $\C(M)$-linear in the vector field direction since by looking at the definition of $v^k$, a vector field $X$ only participates into that composition via $(j^k \varphi)_* X$, which is $\C(M)$-linear in $X$. We have that:
$$\textrm{T}_{(\varphi, x)} j_E^{k} \circ (\textrm{id}_{\textrm{T} \E} \times j^0) (\partial \varphi, f X) = f(x) \cdot \textrm{T}_{(\varphi, x)} j_E^{k} \circ (\textrm{id}_{\textrm{T} \E} \times j^0) (\partial \varphi, X).$$
\end{rk}

\begin{lm}\label{twoto} Given any local map $(\overbar{\chi}, \textrm{id}_M) \colon \E \times M \rightarrow \textrm{T} \E \times \mathfrak{X}(M) \times M$ covering the identity, which is a section of $\textrm{pr}_E^* \times \textrm{id}_M$ and such that $\overbar{\chi}_{\mathfrak{X}(M)}$ is independent of $M$, the associated section of the tangent bundle $\chi$ from Lemma \ref{oneto} is a local vector field.

Moreover, all local vector fields $\chi$ come from this way, although the map $\chi^{\infty} \mapsto \overbar{\chi}^{\infty}$ is not canonical.

\end{lm}

\dem The diagram before the Lemma shows how to construct $\chi^{\infty}$ from $\overbar{\chi}^{\infty}$. The composition of pro-smooth maps is pro-smooth, and the projections to $J^{\infty} E$ commute with $u^{\infty} + v^{\infty}$ so that $\chi^{\infty}$ is a vector field on $J^{\infty} E$. This, together with the definition of $\chi$ as $\chi \defeq (\textrm{id}_{\textrm{T} \E} \times j^0) \circ \overbar{\chi}$ as in Lemma \ref{oneto} proves that what we get is a local vector field.

Using a connection in $T(J^k E)$ for all $k$ and the map $u^k$ we see that 
$$T(J^k E) \cong V(J^k E) \times_M  TM \cong J^k (VE) \times_M TM  \subset J^k (VE \times_M TM) \cong J^k (TE).$$
Then we have a one side inverse of $u^k + v^k$, managing to define $\overbar{\chi}^{\infty}$ from $\chi^{\infty}$. Connections always exists, but there is not a canonical choice for them; thus there is no canonical way of constructing $\overbar{\chi}^{\infty}$ from $\chi^{\infty}$.
\qed

As a consequence, {\it all local vector fields split into an evolutionary an a total part, in the sense that $\chi = \xi +X$ ($\xi \colon \E \times M \rightarrow \mathrm{T} \E$, $X \colon \E \times M \rightarrow \mathfrak{X}(M)$)}. But there is no pro-smooth vector field associated to it in a canonical way which splits into the two parts (in particular there is no Cartan preserving vector field on $J^{\infty} E$ associated to it).

Observe that local vector fields $\chi = \xi + X$ such that $\xi = 0$ are local maps along the identity $X \colon \E \rightarrow \mathfrak{X}(M)$. In particular, they are insular and infinite jet prolongations exist. Nevertheless, if $\xi \neq 0$ the map is generally not insular. Only if $\xi$ is independent of $M$ the map will be insular. But observe that vector fields $\chi = \xi + X$ such that $X = 0$ and $\xi$ are independent of $M$ are precisely local vector fields on $\E$. In other words, sums of a local vector field on $\E$ and a local vector field on $M$ are insular and have unique prolongations. As a matter of fact these prolongations agree with the prolongation of a vector field as in equation \ref{pr2}.

We get to the third equivalent definition of a local vector field, which explains the term local in the definition:

\begin{pp}\label{lvf2}
Local vector fields on $\E \times M$ are equivalent to local maps covering the identity $(\overbar{\chi}, \textrm{id}_M) \colon \E \times M \rightarrow \textrm{T} \E \times \mathfrak{X}(M) \times M$ which are sections of $\textrm{pr}_E^* \times \textrm{id}_M$ and such that $\overbar{\chi}_{\mathfrak{X}(M)}$ is independent of $M$.

Infinite jet prolongations of the lowest map $\overbar{\chi}^{0} \colon J^{k} E \rightarrow TE$ exists and are the unique ones that preserve the contact ideal. Moreover, 
$$(u^{\infty} + v^{\infty}) \circ j^{\infty} \overbar{\chi}^0 = \textrm{pr}\left((u^{0} + v^{0}) \circ j^{1} \overbar{\chi}^0\right).$$ 
If $(\overbar{\chi}, \textrm{id}_M)$ is insular, $\chi$ covers the infinite jet prolongation of $\chi^0$.  
\end{pp}

The penultimate statement says in particular that the the Cartan preserving properties for $\overbar{\chi}^{\infty}$ and $\chi^{\infty}$ are equivalent, since $\textrm{pr}(\chi^0)$ is the unique vector field preserving the contact ideal covering $\chi^0$ and $\chi^0 = (u^{0} + v^{0}) \circ j^{1} \overbar{\chi}^0$ by Lemma \ref{twoto}.

In the insular case we have: 
\begin{center}
\begin{tikzpicture}[description/.style={fill=white,inner sep=2pt}]
\matrix (m) [matrix of math nodes, row sep=3em,
column sep=4em, text height=1.5ex, text depth=0.25ex]
{\E \times M & \textrm{T} \E \times \mathfrak{X}(M) \times M & \textrm{T} \E \times TM\\
  J^{k+1} E & J^{1} (TE) & T E\\
  J^{k} E & TE & \\};
\path[->,font=\scriptsize]
(m-1-1) edge node[auto] {$(\overbar{\chi}, \textrm{id}_M)$} (m-1-2)
(m-1-1) edge node[auto] {$j_{E}^{k+1}$} (m-2-1)
(m-1-2) edge node[auto] {$j_{TE}^{1}$} (m-2-2)
(m-2-1) edge node[auto] {$j^1 \overbar{\chi}^{0}$} (m-2-2)
(m-1-2) edge node[auto] {$\textrm{id}_{\textrm{T} \E} \times j^0$} (m-1-3)
(m-1-3) edge node[auto] {$T j_E^{0}$} (m-2-3)
(m-2-2) edge node[auto] {$u^{0} + v^{0}$} (m-2-3)
(m-1-1) edge [bend left=30] node[below=0.4em] {$\chi$} (m-1-3)
(m-3-1) edge node[auto] {$\overbar{\chi}^{0}$} (m-3-2)
(m-2-1) edge node[auto] {$\pi_{k+1}^k$} (m-3-1)
(m-2-2) edge node[auto] {$(\pi \circ \textrm{pr}_E)_{1}^{0}$} (m-3-2);
\end{tikzpicture}
\end{center}

\begin{cl} Local vector fields on $\E$ are equivalent to local maps along the identity $\overbar{\xi} \times \textrm{id}_M \colon \E \times M \rightarrow \textrm{T} \E \times M$ which are sections of $\textrm{pr}_E^*$.

Infinite jet prolongations of the lowest map $\overbar{\xi}^{0} \colon J^{k} E \rightarrow VE$ exists and the unique ones that preserve the contact ideal. They are covered by the original local vector field. Moreover, 
$$u^{\infty}  \circ j^{\infty} \overbar{\xi}^0 = \textrm{pr}\left(u^{0} \circ \overbar{\xi}^0 \right).$$ 
\end{cl}

The Corollary follows from Proposition \ref{lvf2} and the comments before the Proposition together with the fact that $v^k$ is the one taking care of the action of the vector field on $M$ and it is the responsible for the shift in jet degrees.

These last two results say that the tangent of $\E \times M$ at a point, defined in an insular fashion, actually agrees with the notion of exterior tangent for diffeological spaces as by Christensen and Wu \cite{CW} as noted at the beginning of Section \ref{fnote} (up to considering all derivations and not only smooth ones).\\

{\bf Proof of Proposition \ref{lvf2}.}
Given $(\overbar{\chi},\overbar{\chi}^{\infty})$ in the hypothesis of the proposition, the associated map $\chi^0 = (u^{0} + v^{0}) \circ j^{1} \overbar{\chi}^0$ by Lemma \ref{twoto} is given in local coordinates by
$$\chi^0 = \overbar{\chi}_i^0 \frac{\partial}{\partial x^i} + \left( \overbar{\chi}_{\alpha}^0 + \overbar{\chi}_i^0 u_i^{\alpha} \right) \frac{\partial}{\partial u^\alpha}.$$
We are working with the dual coordinates to $u_{I}^{\alpha}$ so that we need to use equation \ref{pr2} for the prolongation of a vector field. We get that
$$\textrm{pr}(\chi^0)_{\alpha}^I = D_I \left(\overbar{\chi}_{\alpha}^0 + \overbar{\chi}_i^0 u_i^{\alpha} - \overbar{\chi}_i^0 u_i^{\alpha} \right) + \overbar{\chi}_i^0 u_{I,i}^{\alpha} = D_I \overbar{\chi}_{\alpha}^0 + \overbar{\chi}_i^0 u_{I,i}^{\alpha}.$$
Similarly, $(u^k \circ j^{k+1} \overbar{\chi}^0)_{\alpha}^I = D_I(\overbar{\chi}_{\alpha}^0)$ and $(v^k \circ j^{k+1} \overbar{\chi}^0)_{\alpha}^I = \overbar{\chi}_i^0 u_{\alpha}^{I,i}$, showing the wanted result that for all $k$
$$\left((u^k+v^k) \circ j^{k+1} \overbar{\chi}^0\right)_{\alpha}^I = \textrm{pr}(\chi^0)_{\alpha}^I.$$
\qed

\begin{cl} Local vector fields form a $\C_{\textrm{loc}}(\E \times M)$-module given by point-wise multiplication.
\end{cl}

\dem The point-wise multiplication of a local map and a local function is clearly local, since it factors through the jet bundle of the maximum degree of the two maps. Since the fibers in the image are vector bundles, fiber-wise multiplication gives the structure of a module.
\qed

The notions of insularity, Cartan preserving and decomposability agree for vector fields.

\begin{pp}\label{insvf} Let $(\chi, \chi^{\infty})$ be a local vector field. The following conditions are equivalent:
\begin{enumerate}[(i)]
	\item The corresponding local map $\overbar{\chi}$ from Proposition \ref{lvf2} is insular.
	\item The local vector field $\chi$ covers the infinite jet prolongation of its lower map, $\textrm{pr}(\chi^ 0)$.
	\item $\chi^{\infty}$ is a Cartan preserving vector field, hence decomposable (Proposition \ref{last}).
\end{enumerate}
\end{pp}

\dem Local vector fields covering a Cartan preserving vector field are decomposable by Proposition \ref{last}. Local vector fields covering the unique Cartan preserving vector field are insular. The jet prolongations for $\chi$ and $\overbar{\chi}$ agree by Proposition \ref{lvf2}. This completes the proof.
\qed

\begin{df}[Insular vector field] Any local vector field satisfying any of the equivalent conditions from Proposition \ref{insvf} are called insular (or also Cartan preserving). Insular vector fields are denoted by $\mathfrak{X}_{\textrm{ins}}(\E \times M)$. 
\end{df}

\begin{rk}\label{notation}
We want to distinguish between {\it total} vector fields: insular vector fields $\E \rightarrow \mathfrak{X}(M)$ and {\it horizontal} vector fields, those aresimply vector fields on $M$ which can be thought of as constant total vector fields $\E \rightarrow \{*\} \rightarrow \mathfrak{X}(M)$. The best example to distinguish the two of them is the coordinate ones: while $D_i$ is a total vector field, $\frac {\partial}{\partial x^i}$ is horizontal.

Observe that the fact that a vector field on $J^{\infty} E$ is insular does not mean that $\chi$ splits into an insular vector field on $\E$, also called {\it evolutionary} vector field, and a horizontal vector field on $M$. In general the {\it $M$ direction of an insular field will depend on $\E$, and hence it will be total instead of horizontal}. Vector fields that split into an evolutionary vector field on $\E$ and a horizontal vector field are called decomposable by Deligne and Freed \cite[page 168]{DEL}. We want to follow this notion, although the notation is quite unfortunate by what has been said just above. We want to clarify that this notion of decomposability is much stronger than the one of insularity. We will see the differences when studying the local Cartan calculus (Cartan calculus on the bicomplex of local forms in Section \ref{lill}) and when working with symmetries versus families of symmetries.
\end{rk}

\begin{df}[Decomposable vector field]\label{decvf} A local vector field $(\chi, \chi^{\infty}) $ is called decomposable if it is the sum of an evolutionary and a horizontal vector field: $\chi = \xi + X$ where $\chi \in \mathfrak{X}_{\textrm{loc}}(\E)$ and $X \in \mathfrak{X}(M)$. 
\end{df}

Decomposable vector fields are insular vector fields such that the total part $\E \rightarrow \mathfrak{X}(M)$ is constant.


\section{Bicomplex of local forms}\label{blf}

{\it In this section we define local forms on $\E \times M$. Instead of using the defined smooth structures on $\E \times M$, we pullback $M$-twisted forms from the jet bundles of $E$. The resulting bicomplex, called the bicomplex of local forms is the main object of study in Lagrangian field theory. Zuckerman \cite{ZUC} gave the basic definitions and properties using the variational bicomplex for the non-$M$-twisted case. Besides the results from Zuckerman, we prove define pullbacks of local forms along local maps. We follow the $M$-twisted approach from Deligne and Freed \cite{DEL}, but replacing the bigrading they use by the bigrading of the variational bicomplex used by Anderson \cite{AND} and Zuckerman \cite{ZUC}, but using the ind/pro-language developed in Part I.}\\

We want to be able to talk about forms and $M$-twisted forms on $\E \times M$ pulled back from $\Omega^{\bullet, \bullet}(J^{\infty} E)$ the same way we did it for local functions (Definition \ref{locfun}). As an example, in field theory, it is common to say that the Lagrangian is a function of the fields $\E$, valued on integrable top-degree forms on the base manifold $M$ (on densities). On top of that it is said that the Lagrangian depends on finitely many derivatives of the filed. As we have seen in the previous part, we can make sense of this idea by using $M$-twisted ind-differential forms on $J^{\infty} E$ and pull them back through $j^{\infty} \colon \E \times M \rightarrow J^{\infty} E$.

We start by pulling back non-twisted, $1$-forms on a finite jet bundle:

\begin{df}[$(j^k)^*\Omega^1(J^k E)$] Given $\omega_{k} \in \Omega^1(J^{k} E)$ we define the pullback of $\omega_k$ by $j^k$ to be the bundle map over $\E \times M$:
$$(j^k)^* \omega_k \colon \textrm{T} \E \times TM \longrightarrow \E \times M \times \RE$$
given by the map $\omega_k \circ \textrm{T} j^k \colon \textrm{T} \E \times TM \rightarrow J^k E \times \RE$.
\end{df}

For pulling back higher forms we need to develop a notion of $\wedge_{\E \times M}^n (\textrm{T} \E \times TM)$. This is done thinking of the the tangent bundle as a vector bundle over $\E \times M$ (an $\RE$-vector bundle) and taking the anti-symmetric product of each fiber. We define:

\begin{eqnarray}\label{fiber}
\left(\wedge_{\E \times M}^n (\textrm{T} \E \times TM)\right)_{(\varphi,x)} &\defeq& \bigoplus_{p+q = n} \wedge_{\RE}^p \Gamma(M, \varphi^* VE) \otimes_{\RE} \wedge_{\RE}^q T_x M \nonumber \\
& \cong &  \bigoplus_{p+q = n} \wedge_{\RE}^p \textrm{T}_{\varphi} \E  \otimes \wedge_{\RE}^q T_x M.
\end{eqnarray}

Observe that the exterior product on the tangent to the field direction is not over $\C(M)$ so that
\begin{equation}\label{solito}
\wedge_{\RE}^p \Gamma(M, \varphi^* VE) \neq \Gamma(M, \wedge_E^p \varphi^* VE).
\end{equation}

In any case, we can regroup all the fibers together and define:

\begin{equation}\label{fiber2}
\wedge_{\E \times M}^n (\textrm{T} \E \times TM) \defeq \bigoplus_{p+q = n} \left( \wedge_{\E}^p \Gamma(M, VE) \times M \right) \otimes \left( \E \times \wedge_M^q TM\right)
\end{equation}

{\noindent where the tensor product in the middle is of $\RE$-vector bundles over $\E \times M$, so that actually the fiber over $(\varphi, x)$ is the one given by equation \ref{fiber}. We have emphasized over which base the anti-symmetric powers of each of the vector bundles are taken to avoid confusion.}

Now we define $\wedge^n \mathrm{T} j^k \colon \wedge_{\E \times M}^n (\textrm{T} \E \times TM) \rightarrow \wedge_{J^k E}^n T(J^k E)$ by $\RE$-multilinearity using the $\RE$-linearity of $\mathrm{T} j^k \colon \textrm{T} \E \times TM \rightarrow T (J^k E)$.
 
\begin{df}[$(j^k)^*\Omega^n(J^k E)$] Given $\omega_{k} \in \Omega^n(J^{k} E)$ we define the pullback of $\omega_k$ by $j^k$ to be the bundle map over $\E \times M$:
$$(j^k)^* \omega_k \colon \wedge_{\E \times M}^n (\textrm{T} \E \times TM) \longrightarrow \E \times M \times \RE$$
given by the map $\omega_k \circ \wedge^n  (\textrm{T} j^k) \colon \wedge_{\E \times M}^n (\textrm{T} \E \times TM) \rightarrow J^k E\times \RE$.
\end{df}

The next steps is to  generalize the pullback construction for $M$-twisted forms.

\begin{df}[$(j^k)^*\Omega_{\textrm{tw}}^n(J^k E)$] Given $\omega_{k} \in \Omega_{\textrm{tw}}^n(J^{k} E)$ we define the pullback of $\omega_k$ by $j^k$ to be the bundle map over $\E \times M$:
$$(j^k)^* \omega_k \colon \wedge_{\E \times M}^n (\textrm{T} \E \times TM) \longrightarrow \E \times \mathfrak{o}_M \cong (\E \times M) \times_M \mathfrak{o}_M$$
given by the map $\omega_k \circ \wedge^n  (\textrm{T} j^k) \colon \wedge_{\E \times M}^n (\textrm{T} \E \times TM) \rightarrow \pi_k^* \mathfrak{o}_M$.
\end{df}

\begin{center}
\begin{tikzpicture}[description/.style={fill=white,inner sep=2pt}]
\matrix (m) [matrix of math nodes, row sep=2em,
column sep=2.5em, text height=1.5ex, text depth=0.25ex]
{\wedge_{\E \times M}^n ( \textrm{T} \E \times T M ) & \wedge^n T (J^k E) & \pi_k^* \mathfrak{o}_M & \mathfrak{o}_M \\
 \E \times M & J^k E & J^k E & M\\};
\path[->,font=\scriptsize]
(m-1-1) edge [bend left=25] node[above=0.2em] {$\omega_k \circ \wedge^n \textrm{T} j^k$} (m-1-3)
(m-1-1) edge node[auto] {$\wedge^n (\textrm{T} j^k)$} (m-1-2)
(m-1-2) edge node[auto] {$\textrm{pr}_{J^k E}$} (m-2-2)
(m-1-1) edge node[auto] {$\textrm{pr}_{\E \times M}$} (m-2-1)
(m-2-1) edge node[auto] [swap]{$j^k$} (m-2-2)
(m-1-2) edge node[auto] {$\omega_k$} (m-1-3)
(m-1-3) edge node[auto] {} (m-2-3)
(m-2-2) edge node[auto] [swap]{$\textrm{id}_M$} (m-2-3)
(m-1-3) edge node[auto] {} (m-1-4)
(m-2-3) edge node[auto] {$\pi_k$} (m-2-4)
(m-1-4) edge node[auto] {} (m-2-4);
\begin{scope}[shift=($(m-1-3)!.5!(m-2-4)$)]
\draw +(-.3,0) -- +(0,0)  -- +(0,.3);
\fill +(-.15,.15) circle (.05);
\end{scope}
\end{tikzpicture}
\end{center}

\begin{rk}\label{conr} Using the splitting of equation \ref{fiber2}, we can observe that an element of $(j^k)^*\Omega_{\textrm{tw}}^n(J^k E)$ is a sum of special kinds of maps
$$\left( \wedge_{\E}^p \Gamma(M, VE) \times M \right) \otimes \left( \E \times \wedge_M^q TM\right) \longrightarrow \E \times \mathfrak{o}_M.$$
The maps are $\RE$-linear, so that we can dualize the part on $TM$ to get a map:
$$\wedge_{\E}^p \Gamma(M, VE)  \longrightarrow \textrm{Hom}_{VB}\left( \wedge_M^q TM, \mathfrak{o}_M\right) \cong \Gamma(M, \wedge_M^q T^*M \otimes \mathfrak{o}_M) \cong \textrm{Dens}^{\textsf{top}-q}(M).$$
The process of getting from a form to the corresponding density is already pointed out by Zuckermann \cite{ZUC}.
We will therefore denote any form $\alpha = (j^k)^*\alpha_k$ also as such a map, and we will jump from one interpretation to the other without making any change in the notation. To be explicit $\alpha$ will also be understood as a map:
\begin{eqnarray}\label{formi}
\alpha \colon \wedge_{\E}^p \Gamma(M, VE)  &\longrightarrow & \textrm{Dens}^{\textsf{top}-q}(M) \\
(\varphi, \partial \varphi_1 \wedge \cdots \wedge \partial \varphi_p) & \longmapsto &
\left( (j^k)^* \alpha_k \right) \left((\partial \varphi_1, 0), \cdots, (\partial \varphi_p, 0), (0,-), \cdots, (0,-)\right) \nonumber 
\end{eqnarray}
The very important remark is that the map \ref{formi} {\bf is not $\C(M)$-linear}. As a matter of fact the domain is $\wedge_{\E}^p \Gamma(M, VE) $ and not $\wedge_{\C(M)}^p \Gamma(M, VE) \cong  \Gamma(M, \wedge_{E}^p VE)$, as mentioned in Equation \ref{solito}. So that in particular 
$$(j^k)^*\Omega_{\textrm{tw}}^n(J^k E) \not\subset \bigoplus_{p+q= n}\Gamma(M, \wedge_{E}^p VE) \otimes_{\C(M)} \textrm{Dens}^{\textsf{top}-q}(M).$$
This is a very common source of confusion and we have wanted to be quite explicit about it.
\end{rk}

We can finally define the bicomplex of local forms using the twisted variational bicomplex from Definition \ref{TVB}. Similar notions have been studied before by Zuckerman \cite{ZUC}, Deligne and Freed \cite{DEL} or Anderson \cite{AND}. Our input is to combine the best of all the three worlds and add our way of talking about the variational bicomplex by using ind-differential forms. Zuckerman \cite{ZUC} uses the variational bicomplex, but only for the non-$M$-twisted case. Deligne and Freed \cite{DEL} work with $M$-twisted forms but they do not use the infinite jet bundle. In any case, neither Zuckerman \cite{ZUC}, nor Anderson \cite{AND} use the pro/ind-categorical approach we have developed in Part I.

\begin{df}[Bicomplex of ($M$-twisted) local forms]\label{LF}
Given a fiber bundle $\pi \colon E \rightarrow M$, we define a bicomplex  $\left( \Omega_{\textrm{loc}}^{\bullet, \bullet}(\E \times M), \delta, d \right)$, called the bicomplex of local forms on $\E \times M$ given by $$\Omega^{p,q}_{\mathrm{loc}}(\E \times M) \defeq (j^\infty)^* \left(\Omega^{p,q}_{\mathrm{tw}}(J^{\infty} E) \right)$$ 
or in short,
$$ \Omega_{\textrm{loc}}^{\bullet, \bullet}(\E \times M) =  (j^{\infty})^*\left( \Omega_{\textrm{tw}}^{\bullet, \bullet}(J^{\infty} E) \right);$$ equipped with differentials $\delta$ and $d$, $$\delta((j^{\infty})^*\alpha)\defeq(j^{\infty})^*(d_V \alpha) \textrm{ and }d((j^{\infty})^*\alpha)\defeq(j^{\infty})^*(d_H \alpha).$$
The splitting of $\wedge^n(\textrm{T} \E \times TM)$ agrees with the splitting coming from the variational bicomplex.
\end{df}

We should have called this definition a Definition/Proposition, because it comprises two hidden statements. The first one is that $\delta$ and $d$ are independent of the preimage chosen. But this is clear since $d_V \alpha$ and $d_H \alpha$ depend on at most one more jet than $\alpha$ and $(j^{\infty})^* \alpha$ fixes all jets. The second statement is that the bidegree decomposition coming from the variational bicomplex and the splitting of the exterior powers of the tangent bundle on $\E \times M$ agree. But this is also clear: if a form is the pullback of an element in $\Omega_H^{n,s}(J^{\infty} E)$ then it is annihilated by all $n-s+1$ vertical vectors (those coming from $\mathrm{T} \E$) and not from any $n-s$ ones. This shows that the form actually comes from a map $(\wedge^{n-s} \mathrm{T} \E \times M) \otimes (\E \times \wedge^s TM) \rightarrow \E \otimes \mathfrak{o}_M$. Observe that we have preferred now to use the letters $p$ and $q$ instead of $r$ and $s$ to distinguish from the variational bicomplex and also to match what is done by Deligne and Freed \cite{DEL}.

There is a theory of differential forms on $\E\ \times M$ from the Fr\'echet manifold point of view. We have seen that jet evaluations are smooth, and hence we can talk about pulling back differential forms to $\E \times M$ in that setting. From that point of view, all local forms are differential forms on $\E \times M$ and $\delta$ is the differential in the $\E$ direction while $d$ is the differential in the $M$ direction. This is the introduction to the concept of local forms given by Zuckerman \cite{ZUC}. The bicomplex of local forms is not only interesting because of the properties that it might have, being a colimit of complexes over finite dimensional manifolds, but also because the bigger bicomplex $\Omega^{\bullet}(\E \times M)$ might easily have trivial cohomology groups.

In the case in which $j^{\infty}$ is not surjective, $\Omega_{\textrm{loc}}^{\bullet, \bullet}(\E \times M)$ is not the same as $\Omega_{\textrm{tw}}^{\bullet}(J^{\infty} E)$. When we are in the case of a vector bundle over an oriented manifold $M$, we have that $\Omega_{{\textrm{loc}}}^{\bullet , \bullet}(\E \times M)$ is isomorphic to $\Omega^{\bullet} (J^{\infty} E)$. (We have the same result if $\E$ is soft, but replacing $\Omega^{\bullet} (J^{\infty} E)$ by $\Omega^{\bullet}_{\textrm{tw}}(J^{\infty} E)$.) We will be abusive in terms of vocabulary and call elements of $\C_{\textrm{loc}}(\E \times M) \defeq \Omega_{\textrm{loc}}^{0}(\E \times M)$ simply local functions although we defined those to be something else in Definition \ref{locfun} (the two definitions agree if $M$ is orientable). What local functions really are, are zero-degree non-$M$-twisted local forms:

\begin{df}[Bicomplex of non-$M$-twisted local forms]\label{nLF}
Given a fiber bundle $\pi \colon E \rightarrow M$, we define a bicomplex  $\left( \Omega_{\textrm{ntw-loc}}^{\bullet, \bullet}(\E \times M), \delta, d \right)$, called the bicomplex of non-$M$-twisted local forms on $\E \times M$ given by $$\Omega^{p,q}_{\mathrm{ntw-loc}}(\E \times M) \defeq (j^\infty)^* \left(\Omega^{p,q}(J^{\infty} E) \right)$$ 
equipped with differentials $\delta$ and $d$, $$\delta((j^{\infty})^*\alpha)\defeq(j^{\infty})^*(d_V \alpha) \textrm{ and }d((j^{\infty})^*\alpha)\defeq(j^{\infty})^*(d_H \alpha).$$
\end{df}

In general, we will use more often the $M$-twisted bicomplex, that is why we have reserved a special name and notation for the non-$M$-twisted case. The only remarkable property about non-$M$-twisted local forms in particular is that pullbacks respect them, while they do not respect $M$-twisted local ones. Observe this is not something characteristic of local forms, or ind-differential forms or even forms on a vector bundle: it is always the case. Given a smooth map $f \colon X \rightarrow Y$, the pullback along $f$ of an $n$-dimensional density on $Y$ gives a bundle map from $\wedge^n T X \rightarrow \mathfrak{o}_Y$ which is {\it not} a density on $X$.

\begin{pp}\label{1292} Let $f \colon \E \times M \rightarrow \F \times N$ be a local map. Let $\alpha \in \Omega^{n}_{\mathrm{ntw-loc}}(\F \times N)$, then $f^* \alpha \in \Omega^{n}_{\mathrm{ntw-loc}}(\E \times M)$ where the function is defined as $f^* \alpha \defeq (j^{\infty})^* (f^{\infty})^* \alpha_{\infty}$ where $\alpha = (j^{\infty})^*\alpha_{\infty}$. 
\end{pp}

\dem The idea is that we do not need to define what $\textrm{T} f$ is, it is enough to know how to pullback pro-smooth maps and jet prolongations. If $\alpha = (j^l)^* \alpha_l$ and $f^l \colon J^{k(l)} E \rightarrow J^l F$ we have that the defined form in the Proposition is:
\begin{center}
\begin{tikzpicture}[description/.style={fill=white,inner sep=2pt}]
\matrix (m) [matrix of math nodes, row sep=2em,
column sep=3.5em, text height=1.5ex, text depth=0.25ex]
{\wedge_{\E \times M}^n ( \textrm{T} \E \times T M ) & \wedge^n ( \textrm{T} \F \times T N ) &  \\
 \wedge^n T (J^{k(l)} E) & \wedge^n T (J^{l} F) & \RE \\};
\path[->,font=\scriptsize]
(m-1-1) edge [bend left=50] node[above=0.2em] {$f^* \alpha$} (m-2-3)
(m-2-1) edge [bend right=25] node[below=0.2em] {$(f^{\infty})^* \alpha_k$} (m-2-3)

(m-1-1) edge node[auto] {$\wedge^n \textrm{T} j^{k(l)}$} (m-2-1)
(m-1-2) edge node[auto] [swap]{$\wedge^n \textrm{T} j^{l}$} (m-2-2)
(m-2-1) edge node[auto] {$\wedge^n T f^k$} (m-2-2)
(m-2-2) edge node[auto] {$\alpha_l$} (m-2-3)
(m-1-2) edge node[auto] {$\alpha$} (m-2-3);
\end{tikzpicture}
\end{center}
This map is well defined since any other choice of $f^{\infty}$ or $\alpha_{\infty}$ will give raise to commutative diagrams so that the corresponding definitions of $f^* \alpha$ will still agree.
\qed

One should keep in mind from last proposition that the equality $j^{\infty} \circ f = f^{\infty} \circ j^{\infty}$ still holds when taking pullbacks: $$(j^{\infty} \circ f)^* \alpha_{\infty} = f^* (j^{\infty})^* \alpha_{\infty} \defeq (f^{\infty} \circ j^{\infty})^* \alpha_{\infty}.$$


\subsection{Diagrams and pictures}\label{1285}

We wish to make some comments on this diagram, which will soon become very common in this thesis.

\begin{center}
\begin{tikzpicture}[description/.style={fill=white,inner sep=2.5pt},scale=1.4]
\node (a) at (-1,4) {\color{cyan} $0$};
\node (b) at (-2,3) {\color{cyan} $1$};
\node (c) at (-3,2) {\color{cyan} $2$};
\node (d) at (-4,1)  {\color{cyan} $\iddots$};
\node (e) at (1,4) {\color{dandelion} $\textsf{top}$};
\node (f) at (2,3) {\color{dandelion} $\textsf{top}-1$};
\node (g) at (3,2) {\color{dandelion} $\textsf{top}-2$};
\node (h) at (4,1) {\color{dandelion} $\ddots$};
\node (i) at (5,0) {\color{dandelion} $0$};
\node (j) at (0,3) {$\Omega_{\textrm{loc}}^{0, \textsf{top}}(\E \times M)$};
\node (k) at (-1,2) {$\Omega_{\textrm{loc}}^{1, \textsf{top}}(\E \times M)$};
\node (l) at (1,2) {$\Omega_{\textrm{loc}}^{0, \textsf{top}-1}(\E \times M)$};
\node (m) at (-2,1) {$\Omega_{\textrm{loc}}^{2, \textsf{top}}(\E \times M)$};
\node (n) at(2,1) {$\Omega_{\textrm{loc}}^{0, \textsf{top}-2}(\E \times M)$};
\node (o) at(0,1) {$\Omega_{\textrm{loc}}^{1, \textsf{top}-1}(\E \times M)$};
\node (p) at (4,-1) {$\C_{\textrm{loc}}(\E \times M)$};
\node (q) at (3,-2) {$\Omega_{\textrm{loc}}^{1, 0}(\E \times M)$};
\node (r) at (2,-3) {$\Omega_{\textrm{loc}}^{2, 0}(\E \times M)$};
\node (s) at (1,-4) {$\iddots$};
\node (t) at (-3,0) {$\iddots$};
\node (u) at (3,0) {$\ddots$};
\node (v) at (-1,0) {$\vdots$};
\node (w) at (-1,-2) {$\vdots$};
\node (x) at (-1,-4) {$\vdots$};
\node (y) at (1,0) {$\vdots$};
\node (z) at (1,-2) {$\vdots$};
\node (aa) at (-3,-2) {$\vdots$};
\node (ab) at (-3,-4) {$\vdots$};
\draw (-4,0) -- (0,4);
\draw (5,-1) -- (0,4);
\draw (5,-1) -- (1,-5);
\path[->,font=\scriptsize,color=cyan]
(j) edge node[auto] {$\delta$} (k)
(k) edge node[auto] {$\delta$} (m)
(m) edge node[auto] {$\delta$} (t)
(l) edge node[auto] {$\delta$} (o)
(o) edge node[auto] {$\delta$} (v)
(n) edge node[auto] {$\delta$} (y)
(p) edge node[auto] {$\delta$} (q)
(q) edge node[auto] {$\delta$} (r)
(r) edge node[auto] {$\delta$} (s);
\path[->,font=\scriptsize,color=dandelion]
(l) edge node[auto] {$d$} (j)
(o) edge node[auto] {$d$} (k)
(v) edge node[auto] {$d$} (m)
(n) edge node[auto] {$d$} (l)
(y) edge node[auto] {$d$} (o)
(u) edge node[auto] {$d$} (n)
(p) edge node[auto] {$d$} (u)
(r) edge node[auto] {$d$} (z);
\end{tikzpicture}
\end{center}

\begin{itemize}
	\item The \color{BlueViolet}(blue) \color{black} index depicted on the left diagonal represents the vertical direction, in other words the $\E$ direction and it is the $p$-index in $\Omega_{\textrm{loc}}^{\color{BlueViolet}p\color{black},q}(\E \times M)$. $\delta$ is the differential in this direction and goes from top-right to bottom-left.
	\item Contrary, the  \color{darkdelion}(orange) \color{black} index depicted on the right diagonal represents the horizontal direction, in other words the $M$ direction and it is the $q$-index in the decomposition $\Omega_{\textrm{loc}}^{p, \color{darkdelion}q\color{black}}(\E \times M)$. $d$ is the differential in this direction and goes from bottom-right to top-left.
	\item The elements in the same column are on the same total degree. The total degree is the sum of the blue and orange indices associated to a certain node in the diagram. The total differential $D = \delta + d$ goes from right to left.
	\item Outside of the black rectangle the complex is zero. This means in particular that the total complex is bounded from the right. Since our differential goes to the left, this is usually called ``bounded from below''. It also implies that at each total degree, there are finitely many nodes. Actually there are at most $\textsf{top}$-many nodes.
	\item Usually we will write in the nodes elements instead of the total spaces. This, together with a similar diagram before the beginning of the next subsection, are the only occasions in which we will do the opposite.
	\item The variational bicomplex and the twisted one from Definitions \ref{JVB} and \ref{TVB} respectively can be depicted in a similar diagram where one has to replace the nodes by the corresponding ones with the same bidegrees.
\end{itemize}

Usually bicomplexes are depicted in a non-tilted way in which the horizontal differential has horizontal direction and the vertical one, vertical. This will correspond to tilting our diagram by $-\frac{3 \pi}{4}$. Another way of understanding the orange part will be to think of it as the density part. With this convention $\textsf{top} - p$ becomes $p$ as in $\Omega^{|-p|}(M) = \textrm{Dens}^p$. With this other bigrading, the tilting would be by $\frac{\pi}{4}$ and a reflection along the horizontal axis. We have decided not to do that since it is very relevant for us to consider the elements in the surface of the bicomplex.

Observe that we can refer to an element of the bicomplex of local forms not only by its bidegree $(p,q)$ but also by another combination of two integers:

\begin{df}[Surface forms]
Given a local form $\omega \in \Omega^{p,q}(\E \times M)$ we say that $\omega$ is of { total degree} $p+q$ and { depth} $\textrm{max}(p, \textsf{top}-q)$. The elements of zero depth are called called {surface} forms.
\end{df}

The surface forms are the most exterior nodes at every total degree: those with $p = 0$ or $q = \textsf{top}$. To emphasize the importance of depth and the fact that the outer most nodes are on the {\it surface} we depict the complex with the axes below the surface and in green (like a grass field). The total degrees can be though of as {\bf wells} that go into the ground. To further highlight this distinction, forms of a given total degree will be denoted by Greek letters: $\alpha$ and they decompose as the sum of their surface part denoted by capital Roman letter $A$ and the deeper parts indexed by depth: 
$$\alpha = (A, \alpha_1, \alpha_2, \cdots).$$

In the following example we have indicated with vertical stripes the wells: the forms that have same total degree. Observe that in the lower diagonal, the colors are swapped to emphasize that $B$ is in total degree $\textsf{top}-2$ and $\epsilon_2$ is in total degree $2$; which might be different (they are different provided $\textsf{top} \neq 4$).

\begin{center}
\begin{tikzpicture}[description/.style={fill=white,inner sep=2.5pt},scale=1]

\draw [fill=cyan!30, cyan!30] (-0.4,3.4) rectangle (0.4,-1.4);
\draw [fill=dandelion!30, dandelion!30] (-1.4,2.4) rectangle (-0.6,-1.4);
\draw [fill=cyan!30, cyan!30] (-2.4,1.4) rectangle (-1.6,-1.4);
\draw [fill=dandelion!30, dandelion!30] (-3.4,0.4) rectangle (-2.6,-1.4);
\draw [fill=dandelion!30, dandelion!30] (0.6,2.4) rectangle (1.4,-1.4);


\draw [fill=cyan!30, cyan!30] (1.6,1.4) rectangle (2.4,-1);
\draw [fill=dandelion!30, dandelion!30] (1.6,-2) rectangle (2.4,-3.4);
\draw [fill=cyan!30, cyan!30] (2.6,-1) rectangle (3.4,-2.4);
\draw [fill=dandelion!30, dandelion!30] (3.6,-0.6) rectangle (4.4,-1.4);

\node (a) at (-1,4) {\color{cyan} $0$};
\node (b) at (-2,3) {\color{cyan} $1$};
\node (c) at (-3,2) {\color{cyan} $2$};
\node (d) at (-4,1)  {\color{cyan} $\iddots$};
\node (e) at (1,4) {\color{dandelion} $\textsf{top}$};
\node (f) at (2,3) {\color{dandelion} $\textsf{top}-1$};
\node (g) at (3,2) {\color{dandelion} $\textsf{top}-2$};
\node (h) at (4,1) {\color{dandelion} $\ddots$};
\node (i) at (5,0) {\color{dandelion} $0$};

\node (j) at (0,3) {$L$};
\node (k) at (-1,2) {$EL$};
\node (l) at (1,2) {$A$};
\node (m) at (-2,1) {$0$};
\node (m2) at (-2,-1) {$0$};
\node (n) at(2,1) {$B$};
\node (o) at(0,1) {$\lambda_1$};
\node (p) at (4,-1) {$E$};
\node (q) at (3,-2) {$\epsilon_1$};
\node (r) at (2,-3) {$\epsilon_2$};
\node (s) at (1,-4) {$\iddots$};
\node (t) at (-3,0) {$\iddots$};
\node (u) at (3,0) {$\ddots$};
\node (v) at (-1,0) {$\omega_1$};
\node (w) at (-1,-2) {$\vdots$};
\node (x) at (-1,-4) {$\vdots$};
\node (y) at (1,0) {$\vdots$};
\node (z) at (1,-2) {$\vdots$};
\node (aa) at (-3,-2) {$\vdots$};
\node (ab) at (-3,-4) {$\vdots$};
\draw [rounded corners, dashed, thick, teal] (-2.4,-0.4) to[out=45,in=180] (0,1.5);
\draw [rounded corners, dashed, thick, teal] (-4,-2) to (-2.4,-0.4);
\draw [rounded corners, dashed, thick, teal] (0, 1.5) to[out=0,in=135] (2.4,-0.4);
\draw [dotted, thick, teal] (2.6,-0.6) -- (3,-1);
\draw [rounded corners, dashed, thick, teal] (3.2,-1.2) -- (4, -2) -- (1,-5);

%
\path[->,font=\scriptsize]
(j) edge node[auto] {} (k)
(k) edge node[auto] {} (m)
(v) edge node[auto] {} (m2)
(m) edge node[auto] {} (t)
(l) edge node[auto] {} (o)
(o) edge node[auto] {} (v)
(n) edge node[auto] {} (y)
(p) edge node[auto] {} (q)
(q) edge node[auto] {} (r)
(r) edge node[auto] {} (s);
\path[->,font=\scriptsize]
(l) edge node[auto] {} (j)
(o) edge node[auto] {} (k)
(v) edge node[auto] {} (m)
(n) edge node[auto] {} (l)
(y) edge node[auto] {} (o)
(u) edge node[auto] {} (n)
(p) edge node[auto] {} (u)
(r) edge node[auto] {} (z);
\end{tikzpicture}
\end{center}


We conclude this section with another example of some forms in the bicomplex of local forms.

\begin{ej}\label{clamech}
We can repeat the diagram above in the case in which $\textsf{top}=1$: in other words when the base manifold is one dimensional. The bicomplex in this case looks as follows:

\begin{center}
\begin{tikzpicture}[description/.style={fill=white,inner sep=2.5pt},scale=1]

\draw [fill=black!20, black!20] (-0.4,3.4) rectangle (0.4,0.6);
\draw [fill=black!20, black!20] (-1.4,2.4) rectangle (-0.6,-0.4);
\draw [fill=black!20, black!20] (-2.4,1.4) rectangle (-1.6,-1.4);
\draw [fill=black!20, black!20] (-3.4,0.4) rectangle (-2.6,-2.4);
\draw [fill=black!20, black!20] (0.5,2.4) rectangle (1.5,1.6);

\node (a) at (-1,4) {\color{cyan} $0$};
\node (b) at (-2,3) {\color{cyan} $1$};
\node (c) at (-3,2) {\color{cyan} $2$};
\node (d) at (-4,1)  {\color{cyan} $\iddots$};
\node (e) at (1,4) {\color{dandelion} $1$};
\node (f) at (2,3) {\color{dandelion} $0$};

\node (j) at (0,3) {$L$};
\node (k) at (-1,2) {$EL$};
\node (l) at (1,2) {$\alpha = A$};
\node (m) at (-2,1) {$0$};
\node (m2) at (-2,-1) {$0$};
\node (o) at(0,1) {$\lambda_1$};
\node (aa) at (-3,-2) {$\iddots$};
\node (t) at (-3,0) {$\iddots$};
\node (v) at (-1,0) {$\omega_1$};

\draw [rounded corners, dashed, thick, teal] (-4,-2) -- (0,2) -- (1,1) -- (-3,-3);
\path[->,font=\scriptsize,color=cyan]
(j) edge node[auto] {} (k)
(k) edge node[auto] {} (m)
(v) edge node[auto] {} (m2)
(m2) edge node[auto] {} (aa)
(m) edge node[auto] {} (t)
(l) edge node[auto] {} (o)
(o) edge node[auto] {} (v);
\path[->,font=\scriptsize,color=dandelion]
(l) edge node[auto] {} (j)
(o) edge node[auto] {} (k)
(v) edge node[auto] {} (m);
\end{tikzpicture}
\end{center}

Observe that the notation of Greek letters for the lower degrees and a Roman letter for the surface ones matches the usual conventions in Lagrangian field theory which will be discussed in the following sections (although usually $\lambda_1$ is called $-\gamma$).
\end{ej}

\begin{figure}[h]
	\centering
	\includegraphics[width=0.6\textwidth]{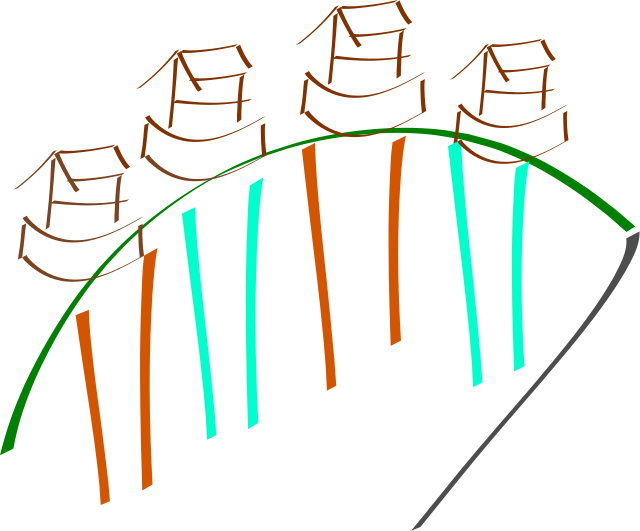}
	\caption{The bicomplex of local forms.}
	\label{fig:isl}
\end{figure}


\section{Local insertion operator and local Lie derivative}\label{lill}

{\it In this section we introduce the insertion of a local vector field into a local form. We specialize for the case in which the vector field is insular and study the graded Lie algebra generated by the total differential $D$ and the insertion of these vector fields, as a graded Lie subalgebra of graded endomorphisms of $\Omega_{\textrm{loc}}^{\bullet}(\E \times M)$. We study what happens when we also introduce the horizontal and vertical differentials. The vertical differential commutes with the insertion of horizontal vector fields, but it does not commute with the insertion of total vector fields, giving rise to an interesting theory. We give formulas for the most basic commutators involving all of those operators. In this section we also introduce local coordinates for local forms and local vector fields in analogy to what was done for the infinite jet bundle. The main references are Anderson \cite{AND} and Deligne and Freed \cite{DEL}.}\\

We have defined local forms and local vector fields on $\E \times M$. The next natural question is to try to replicate insertion operators and Lie derivatives in the local world.

\begin{df}[Insertion of a local vector field]
Given a local form of degree $n$, $\omega \in \Omega_{\textrm{loc}}^{n}(\E \times M)$ and a local vector field $\chi \in \mathfrak{X}_{\textrm{loc}}(\E \times M)$ the insertion of $\chi$ into $\omega$ denoted by $\iota_{\chi}\omega$ is the map 
$$\iota_{\chi}(\omega) \defeq \omega(\chi, -) \colon \wedge^{n-1} (\textrm{T} \E \times TM) \rightarrow \E \times \mathfrak{o}_M$$
\end{df}

This definition is used in the work of Deligne and Freed \cite{DEL} but only for the insertion of vertical vector fields. The theory at this point is complete: the insertion of a local vector field into a local form is again local. Moreover, we have the following result:

\begin{pp}
Given $\omega \in \Omega_{\textrm{loc}}^{n}(\E \times M)$ and $\chi$ a local vector field, the insertion of $\chi$ in $\omega$ is local: $\iota_{\chi} \omega \in \Omega_{\textrm{loc}}^{n-1}(\E \times M)$. Moreover if the bidegree of $\omega$ is $(p,q)$ and $\chi \in \mathfrak{X}_{\textrm{ins}}(\E\ \times M)$ with decomposition $\chi = \xi + X$ we have that $\iota_{\xi} \omega \in \Omega_{\textrm{loc}}^{p-1,q}(\E \times M)$ and $\iota_{X}\omega \in \Omega_{\textrm{loc}}^{p,q-1}(\E \times M)$.
\end{pp}

\dem
The proof goes by taking representatives both of the form and the vector field: $\omega$ is the pullback of $\omega_{\infty}$, an $M$-twisted $n$-degree ind-differential form on $J^{\infty} E$ and $\chi$ covers a pro-vector field $\chi^{\infty}$ on $J^{\infty} E$. The pullback along $j^{\infty}$ of the $M$-twisted $(n-1)$-degree ind-differential form $\iota_{\chi^{\infty}} \omega_{\infty}$ is $\iota_{\chi} \omega$. To see this observe that $\iota_{\chi^{\infty}} \omega_{\infty}$ is given by $\iota_{\chi^{l}}\omega_l$ where $l$ is the order of $\omega_{\infty}$. Letting $k(l)$ be the corresponding order of $\chi^{\infty}$ at $l$ (and assuming $k(l) \geqslant l$), we can see that $\iota_{\chi^{\infty}} \omega_{\infty}$ is given by $(\iota_{\chi^{l}}\omega_l) \in \Omega_{\textrm{tw}}^{n-1}(J^{k(l)} E)$. Explicitly, defining $\iota_{\chi^{l}}\omega_l \defeq \omega_l( \chi^{l}(-), T \pi_{k(l)}^l (-), \cdots, T \pi_{k(l)}^l (-))$, we have that:
\begin{align*}
\left(j^{k(l)}\right)^{*} \left(\iota_{\chi^{l}}\omega_l\right) &= \iota_{\chi^{l}}\omega_l (\textrm{T} j^{k(l)} (-), \ldots, \textrm{T} j^{k(l)} (-)) \\
&= \omega_l (\chi^l, T \pi_{k(l)}^l \textrm{T} j^{k(l)} (-), \ldots, T \pi_{k(l)}^l \textrm{T} j^{k(l)} (-)) \\
&= \omega_l (\chi^l, \textrm{T} j^{l} (-), \ldots, \textrm{T} j^{l} (-)) \\
&= \omega_l (\textrm{T} j^{l} \chi, \textrm{T} j^{l} (-), \ldots, \textrm{T} j^{l} (-)) = \omega(\chi, -) .
\end{align*}

The question about the bigrading of $\omega$ is very simple. It can be answered in local (variational) coordinates. Evolutionary vector fields have no $D_i$ components, so that the horizontal part of $\omega$ does not change when we plug in an evolutionary vector field. Similarly, total vector fields have no $\partial_{\alpha}^I$ component so that the vertical degree of a form remains unchanged under contraction with it.

We also present a coordinate free proof: recall that the bidegree of the bicomplex of local forms agrees with the splitting of $\wedge^n (\mathrm{T} \E \times TM)$ into vertical and horizontal parts. This means that $\omega \in \Omega_{\textrm{loc}}^{p,n-p}(\E \times M)$ can be thought as a map
$$\omega \colon  (\wedge^{p} \mathrm{T} \E \times M) \otimes (\E \times \wedge^{n-p} TM) \rightarrow \E \otimes \mathfrak{o}_M.$$
When we plug in a vertical vector field $\xi \colon \E \times M \rightarrow \mathrm{T} \E$ we clearly get a map 
$$\iota_{\xi }\omega \colon (\wedge^{p-1} \mathrm{T} \E \times M) \otimes (\E \times \wedge^{n-p} TM) \rightarrow \E \otimes \mathfrak{o}_M,$$
and hence $\iota_{\xi} \omega \in \Omega_{\textrm{loc}}^{p-1,n-p}(\E \times M)$. Contrary, if what we plug is a total vector field $X \colon \E \rightarrow \mathfrak{X}(M)$ we get a map:
$$\iota_{X}\omega \colon (\wedge^{p} \mathrm{T} \E \times M) \otimes (\E \times \wedge^{n-p-1} TM) \rightarrow \E \otimes \mathfrak{o}_M.$$
It follows that $\iota_{X} \omega \in \Omega_{\textrm{loc}}^{p,n-p-1}(\E \times M)$.
\qed

As a consequence of the previous proposition, the insertion operator of the local vector field $\chi = \xi + X$ has opposite directions of $\delta$ and $d$:

\begin{center}
\begin{tikzpicture}[description/.style={fill=white,inner sep=2.5pt},scale=1.4]
\node (a) at (-1,4) {\color{cyan} $0$};
\node (b) at (-2,3) {\color{cyan} $1$};
\node (c) at (-3,2) {\color{cyan} $2$};
\node (d) at (-4,1)  {\color{cyan} $\iddots$};
\node (e) at (1,4) {\color{dandelion} $\textsf{top}$};
\node (f) at (2,3) {\color{dandelion} $\textsf{top}-1$};
\node (g) at (3,2) {\color{dandelion} $\textsf{top}-2$};
\node (h) at (4,1) {\color{dandelion} $\ddots$};
\node (i) at (5,0) {\color{dandelion} $0$};
\node (j) at (0,3) {$\Omega_{\textrm{loc}}^{0, \textsf{top}}(\E \times M)$};
\node (k) at (-1,2) {$\Omega_{\textrm{loc}}^{1, \textsf{top}}(\E \times M)$};
\node (l) at (1,2) {$\Omega_{\textrm{loc}}^{0, \textsf{top}-1}(\E \times M)$};
\node (m) at (-2,1) {$\Omega_{\textrm{loc}}^{2, \textsf{top}}(\E \times M)$};
\node (n) at(2,1) {$\Omega_{\textrm{loc}}^{0, \textsf{top}-2}(\E \times M)$};
\node (o) at(0,1) {$\Omega_{\textrm{loc}}^{1, \textsf{top}-1}(\E \times M)$};
\node (p) at (4,-1) {$\C_{\textrm{loc}}(\E \times M)$};
\node (q) at (3,-2) {$\Omega_{\textrm{loc}}^{1, 0}(\E \times M)$};
\node (r) at (2,-3) {$\Omega_{\textrm{loc}}^{2, 0}(\E \times M)$};
\node (s) at (1,-4) {$\iddots$};
\node (t) at (-3,0) {$\iddots$};
\node (u) at (3,0) {$\ddots$};
\node (v) at (-1,0) {$\vdots$};
\node (w) at (-1,-2) {$\vdots$};
\node (x) at (-1,-4) {$\vdots$};
\node (y) at (1,0) {$\vdots$};
\node (z) at (1,-2) {$\vdots$};
\node (aa) at (-3,-2) {$\vdots$};
\node (ab) at (-3,-4) {$\vdots$};
\draw [rounded corners, dashed, thick, teal] (-2.4,-0.4) to[out=45,in=180] (0,1.5);
\draw [rounded corners, dashed, thick, teal] (-4,-2) to (-2.4,-0.4);
\draw [rounded corners, dashed, thick, teal] (0, 1.5) to[out=0,in=135] (2.4,-0.4);
\draw [dotted, thick, teal] (2.6,-0.6) -- (3,-1);
\draw [rounded corners, dashed, thick, teal] (3.2,-1.2) -- (4, -2) -- (1,-5);
\path[->,font=\scriptsize,color=cyan]
(k) edge node[auto] {$\iota_{\xi}$} (j)
(m) edge node[auto] {$\iota_{\xi}$} (k)
(t) edge node[auto] {$\iota_{\xi}$} (m)
(o) edge node[auto] {$\iota_{\xi}$} (l)
(v) edge node[auto] {$\iota_{\xi}$} (o)
(y) edge node[auto] {$\iota_{\xi}$} (n)
(q) edge node[auto] {$\iota_{\xi}$} (p)
(r) edge node[auto] {$\iota_{\xi}$} (q)
(s) edge node[auto] {$\iota_{\xi}$} (r);
\path[->,font=\scriptsize,color=dandelion]
(j) edge node[auto] {$\iota_X$} (l)
(k) edge node[auto] {$\iota_X$} (o)
(m) edge node[auto] {$\iota_X$} (v)
(l) edge node[auto] {$\iota_X$} (n)
(o) edge node[auto] {$\iota_X$} (y)
(n) edge node[auto] {$\iota_X$} (u)
(u) edge node[auto] {$\iota_X$} (p)
(z) edge node[auto] {$\iota_X$} (r);
\end{tikzpicture}
\end{center}


\subsection{Local coordinates}\label{lcooss}

We want to introduce coordinates for the local forms and the local vector fields. This is done in a mirrored way from that done for the corresponding notions in the infinite jet bundle:

The coordinate system in Definition \ref{coordinates} induces a coordinate system in the space of local forms on $\E \times M$. We are going to assume that $M$ is oriented for simplicity (otherwise every form will be a sum of the forms here described tensored with a section of the orientation line bundle). Denote $(j^{\infty})^*(x^i)$ simply by $x^i$ and $(j^{\infty})^*(u_I^{\alpha})$ by $u_I^{\alpha}$ (as local functions on $\E \times M$). Recall how these functions act on $\E \times M$:
\begin{eqnarray*}
	x^i (\varphi, x) & =& x^i(x) \textrm{ and } \\
    u_I^{\alpha} (\varphi, x) & =& {\left. \frac{\partial^{|I|}(u^{\alpha} \circ \varphi)}{\partial x^I} \right|}_x .
\end{eqnarray*}

We can reinterpret $d x^i$ as a form in $\Omega_{\textrm{loc}}^{0, 1}(\E \times M)$ and $\delta u_I^{\alpha} \in \, \Omega_{\textrm{loc}}^{1, 0}(\E \times M) $. They determine a basis of $\Omega_{\textrm{loc}}^{\bullet, \bullet}(\E \times M)$. A form $\omega \in \Oml$ is now of the kind:

$$ \omega = \omega_{{\alpha}_1, \ldots, {\alpha}_p; i_1, \ldots, i_{q}}^{I_1, \ldots, I_p} \delta u_{I_1}^{{\alpha}_1} \wedge \cdots \wedge \delta u_{I_p}^{{\alpha}_p} \wedge d x^{i_1} \wedge \cdots \wedge d x^{i_{q}}. $$

Observe that if $M$ was not oriented one should add $\epsilon \in \mathfrak{O}_M$ to the previous expression. The functions $\omega_{{\alpha}_1, \ldots, {\alpha}_p; i_1, \ldots, i_{q}}^{I_1, \ldots, I_p}$ are in $(j^{\infty})^*(\C(J^{\infty} E))$.

\begin{rk} It is very important to observe that $dx^i$ and $\delta u_I^{\alpha}$ anti-commute. We should keep in mind that these coordinates are {\it not} the coordinates coming from the decomposition
\begin{equation*}
\left(\wedge_{\E \times M}^n (\textrm{T} \E \times TM)\right)_{(\varphi,x)}  \cong   \bigoplus_{p+q = n} \wedge_{\RE}^p \textrm{T}_{\varphi} \E  \otimes \wedge_{\RE}^q T_x M.
\end{equation*}
This is so because $\delta u_I^{\alpha}$ mixes horizontal and vertical directions.
\end{rk}

Using equation \ref{ventiuno} we can see that for any local function $f \in \, \Omega_{\textrm{loc}}^{0,0}(\E \times M)$.
\begin{eqnarray*}
d f &=& D_i f \, dx^i  \\
\delta f & =& \partial_{\alpha}^I f \, \delta u_I^{\alpha},
\end{eqnarray*}

{\noindent where $D_i$ and $\partial_{\alpha}^I$ are like in Chapter 2: $(D_i f) (\varphi, x) \defeq  \frac{\partial}{\partial x^i} f^{\infty}(j_x^{\infty} \varphi) + u_{I, i}^{\alpha} \partial_{\alpha}^I f^{\infty}(j_x^{\infty} \varphi)$ and $(\partial_I^{\alpha} f)(\varphi, x) \defeq \frac{l_1! \cdots l_m! }{k!} \frac{\partial }{\partial u_I^{\alpha}} f^{\infty}(j_x^{\infty} \varphi)$ for $f = (j^{\infty})^* f^{\infty}$. The operator also extends for multi-indices $I = (i_1, \ldots, i_k)$ by setting $D_I \defeq D_{i_1} \cdots D_{i_k}$.}

Applying the previous formulas to the coordinate functions we get:
\begin{eqnarray*}
d x^i = d x^i & \phantom{aa} &  d u_I^{\alpha} = u_{I,i}^{\alpha} d x^i , \\
\delta x^i = 0 & \phantom{aa} & \delta u_I^{\alpha} = \delta u_I^{\alpha} , \\
D x^i = d x^i & \phantom{aa} & D u_I^{\alpha} \, = \delta u_I^{\alpha} + u_{I,i}^{\alpha} d x^i,\\
D_{i^{\prime}} x^i = \delta_i^{i^{\prime}} & \phantom{aa} & D_i u_I^{\alpha} = u_{I,i}^{\alpha}.
\end{eqnarray*} 

For local vector fields $\chi \in \mathfrak{X}_{\textrm{loc}}(\E \times M)$ we want to use the variational coordinates introduced in Remark \ref{IMPO} 
$$\chi^{\infty} = \chi_i^{\infty} D_i + (\chi^{\infty})_{\alpha}^I \partial_{\alpha}^I,$$
{\noindent in $T (J^{\infty} E)$. Now $\chi(f) = \iota_{\chi} (df) =  {\chi}_i D_i f + {\chi}_{\alpha}^I \partial_{\alpha}^I f$ which suggest to use as a notation for local vector fields the following:}
$$\chi = \chi_i D_i  + \chi_{\alpha}^I \partial_{\alpha}^I.$$

If $\chi$ is insular $\chi = \xi + X$ then by the prolongation formula \ref{pr} we have that
$$X = X_i \, D_i \qquad \xi = \xi_{\alpha}  \, \partial_{\alpha} + D_I (\xi_{\alpha})  \, \partial_{\alpha}^I.$$



\subsection{Cartan calculus}\label{lcoocc}

The commutation relations among the insertion operator of vector fields, the differential and the Lie derivatives are commonly referred to as Cartan calculus. When local forms are the space of forms all these operators act upon, we talk about local Cartan calculus. We shall differentiate among the calculus in three different cases:
\begin{itemize}
	\item When only total differentials and insertion of local vector fields are involved.
	\item When we consider differentials in both directions and only decomposable vector fields.
	\item When we consider differentials in both directions but insertion of all insular vector fields.
\end{itemize}

The formulas are different in each case.

We begin by considering the case dealing with total differentials and insertion of local vector fields. All the results from the interaction between vector fields and ind-differential forms on the infinite jet bundle are also true for the bicomplex of local forms. We gather the basic results from Section \ref{vf} in the following Proposition:

\begin{pp}[Local Cartan calculus]\label{CC3}
Let $\pi \colon E \rightarrow M$ be a smooth fiber bundle. Let $\omega \in \Omega_{\textrm{loc}}(\E \times M)$ be a local differential form of total degree $n$. Let ${\chi} \in \mathfrak{X}_{\textrm{loc}}(\E \times M)$ be a local vector field. Let $f$ be a local function. Then
\begin{itemize}
	\item $\chi(f)$ is defined to be the pro-smooth function $\iota_{\chi} (df)$.
	\item The Lie derivative of $\omega$ along $\chi$, denoted by $\mathcal{L}_{\chi} \omega$ is the local form of degree $n$ defined by Cartan's magic formula: $\mathcal{L}_{\chi} \omega = \iota_{\chi} D \omega + D \iota_{\chi} \omega = [D, \iota_{\chi}]$.
	\item The total derivative $D$, the insertion operators $\iota_{\chi}$ and the Lie derivatives $\mathcal{L}_{\chi}$ (for each local vector field ${\chi}$) form a closed subalgebra of the Lie algebra of graded endomorphisms of $\Omega_{\textrm{loc}}(\E \times M)$ with only nontrivial commutators (besides the definition of the Lie derivative):
	\begin{multicols}{2}
\begin{enumerate}
 	\item $[\mathcal{L}_{\chi}, \iota_{{\chi}^{\prime}}] = \iota_{[{\chi}, {\chi}^{\prime}]}$.
	\item $[\mathcal{L}_{\chi}, \mathcal{L}_{{\chi}^{\prime}}] = \mathcal{L}_{[{\chi}, {\chi}^{\prime}]}$.
\end{enumerate}
\end{multicols}
	\item The bracket of two insular vector fields is again insular.
	\item The bracket of two evolutionary vector fields is again an evolutionary vector field in $\E$.
	\item The bracket of two total vector fields is again total.
\end{itemize}
\end{pp}

\dem Definition \ref{CC2}, Proposition \ref{CC} and Proposition \ref{closed} are the corresponding results for the infinite jet bundle. Since the definition of the insertion operator and the Lie derivative are given in terms of the associated elements in the infinite jet bundle, all computations can be done there and brought back using the pullback of the infinite jet evaluation.
\qed

It is important to distinguish between total and horizontal vector fields. The following is a generalization of \cite[Proposition 1.21 ii)]{AND} by Anderson.

\begin{pp}\label{CC4}
Consider $E \rightarrow M$ a fiber bundle.
\begin{itemize}
	\item The bracket of two horizontal vector fields is again horizontal.
	\item The bracket of an evolutionary vector field and a horizontal vector field is zero.
\end{itemize}
\end{pp}

\dem Consider $X, Y \in \mathfrak{X}(M)$ two horizontal vector fields and $\xi \in \mathfrak{X}_{\textrm{loc}}(\E)$ and evolutionary vector field. For any local function $f \in \C_{\textrm{loc}}(\E \times M)$ we have that
$$X(f) = X_i D_i f \qquad Y(f) = Y_j D_j f,$$
$$Y (X(f)) = Y_j D_j (X_i D_i f) = Y_j D_j (X_i) D_i f + Y_j X_i D_{i,j} f,$$
$$X (Y(f)) = X_i D_i (X_j D_j f) = X_i D_i (Y_j) D_j f + X_i Y_j D_{i,j} f,$$
so that 
$$[X, Y](f) = X_i D_i (Y_j) D_j f - Y_j D_j (X_i) D_i f = \left( X_j D_j Y_i - Y_j D_j X_i \right) D_i f.$$
In order to see that $[X, Y]$ is horizontal, since it is insular by Proposition \ref{CC3}, it is enough to check that $\partial_{\alpha} \left( X_j D_j Y_i - Y_j D_j X_i \right) = 0$ for all $i$, where sum over $j$ is meant. For doing so we need to understand the difference between $\partial_{\alpha} D_j$ and $D_j \partial_{\alpha} $. They happen to be the same. In order to prove so, we are going to compare more generally $\partial_{\alpha}^I D_j$ and $D_j \partial_{\alpha}^I $. We take another test function $g \in \C_{\textrm{loc}}(\E \times M)$
\begin{eqnarray}\label{numerito}
D_j \partial_{\alpha}^I g &=& \frac{\partial^2 g}{\partial x^j \partial u_{\alpha}^I} + u_{\beta}^{J,j} \frac{\partial^2 g}{\partial u_{\beta}^{J} \partial u_{\alpha}^I} \textrm{ while,} \\
\partial_{\alpha}^I D_j g &=& \frac{\partial^2 g}{\partial x^j \partial u_{\alpha}^I} + u_{\beta}^{J,j} \frac{\partial^2 g}{\partial u_{\beta}^{J} \partial u_{\alpha}^I} + \frac{\partial g}{\partial u_{\alpha}^{I \smallsetminus j}} = D_j \partial_{\alpha}^I g + \frac{\partial g}{\partial u_{\alpha}^{I \smallsetminus j}}. \nonumber
\end{eqnarray}
Going back to the computation of $\partial_{\alpha} \left( X_j D_j Y_i - Y_j D_j X_i \right)$, since $\partial_{\alpha} X_k = \partial_{\alpha} Y_k = 0$ for all $k$, we conclude that $\partial_{\alpha} [X, Y]_i = 0$ for all $i$, which shows the bracket of two horizontal vector fields is again horizontal.

For the computation of the commutator of an evolutionary and a horizontal vector field we also need the following equations:
\begin{eqnarray*}
\xi (f) &=& (D_I \xi_{\alpha}) \partial_{\alpha}^I f.\\
\xi (X(f)) &=& (D_I \xi_{\alpha}) \partial_{\alpha}^I (X_i D_i f) = (D_I \xi_{\alpha}) X_i \partial_{\alpha}^I D_i f \\
&=& X_i (D_I \xi_{\alpha}) D_i \partial_{\alpha}^I f + X_i (D_{I, i} \xi_{\alpha}) \partial_{\alpha}^I f.\\
X(\xi(f)) &=& X_i D_i ((D_I \xi_{\alpha}) \partial_{\alpha}^I f) = X_i (D_I \xi_{\alpha}) D_i \partial_{\alpha}^I f + X_i (D_{I, i} \xi_{\alpha}) \partial_{\alpha}^I f.
\end{eqnarray*}

That shows that $[\xi,X] = 0$ proving the Proposition.
\qed

Now we can consider the Lie subalgebra generated by vertical and horizontal differentials, together with insertion of evolutionary and horizontal vector fields, this is decomposable vector fields. Decomposable vector fields are well behaved, and some of its fundamental commutative properties are studied by Anderson \cite[Proposition 1.17]{AND} (for the infinite jet bundle case).

\begin{pp}[Decomposable Cartan calculus]\label{lcc0}
We consider two decomposable vector fields $\chi = \xi + X$ and $\chi^{\prime} = \xi^{\prime} + X^{\prime}$. The only nontrivial commutators between any two of the following endomorphisms: $D, d, \delta, \iota_{\xi}, \iota_{\xi^{\prime}}, \iota_X, \iota_{X^{\prime}}, \mathcal{L}_{\xi}, \mathcal{L}_{\xi^{\prime}}, \mathcal{L}_X, \textrm{ and } \mathcal{L}_{X^{\prime}}$; besides the ones given by Proposition \ref{CC3}; are the following:
\begin{multicols}{2}
\begin{enumerate}
	\setcounter{enumi}{2}
	\item $[\delta, \iota_{\xi}] = [D, \iota_{\xi}] = \mathcal{L}_{\xi}$.
	\item $[d, \iota_{X}] = [D, \iota_{X}] = \mathcal{L}_{X}$.
\end{enumerate}
\end{multicols}
\end{pp}

\dem The anti-commutativity between all the different differentials is precisely the definition of a bicomplex. The commutators that do not include $d$ or $\delta$ follow from the Cartan calculus for the variational bicomplex (Proposition \ref{CC}). The commutators including a vertical vector field and no horizontal ones are proven by Anderson \cite[Proposition 1.17]{AND} (as a matter of fact Deligne and Freed also use this fact without giving an explicit proof, for example \cite[Equation 2.53]{DEL}). We are referring to equation $3$ and the proof can be done simply for the basic $1$-forms using derivation properties:
\begin{eqnarray*}
\iota_{\xi} d (f d x^i) & =& \iota_{\xi} (D_j f dx^j dx^i) = 0 = -d 0 = -d \iota_{\xi} (f  dx^i)\\
\iota_{\xi} d (f \delta u_I^{\alpha}) & = & \iota_{\xi} (df \delta u_I^{\alpha} + f d \delta u_I^{\alpha}) =\iota_{\xi} (D_i f dx^i \delta u_I^{\alpha} - f \delta u_{I, i}^{\alpha} dx^i) \\
& = & - D_I \xi_{\alpha} D_i f dx^i - D_{I, i} \xi_{\alpha} f dx^i = -d (f D_I \xi_{\alpha}) \\
&=& -d (\iota_{\xi} f \delta u_I^{\alpha})
\end{eqnarray*}

Finally, the counter part in the horizontal direction, equation $4$ is equivalent to $[\delta, \iota_X] = 0$. This is used in the work of Deligne and Freed \cite[page 168]{DEL}. The computations for the coordinate $1$-forms are the following:
\begin{eqnarray*}
\iota_X \delta (f \delta u_I^{\alpha}) &=& \iota_X (\partial_{\beta}^J f \delta u_J^{\beta} \delta u_I^{\alpha}) = 0 = - \delta 0 = - \delta \iota_X (f \delta u_I^{\alpha}) \\
\iota_X \delta (f dx^i) & = & \iota_X (\partial_{\alpha}^I f \delta u_I^{\alpha} dx^i) = -X_i \partial_{\alpha}^I f \delta u_I^{\alpha} = - \partial_{\alpha}^I (X_i f) \delta u_I^{\alpha} \\
& = & -\delta (X_i f) = - \delta \iota_X (f dx^i)
\end{eqnarray*}

\qed

The key observation at this point is that in the last proof, the second to last equality does not hold for $X$ total. In that case $\partial_{\alpha}^I X_i$ is not necessarily true. Hence, $[\delta, \iota_X] \neq 0$ if $X$ is total but not horizontal. {\it This means in particular that $\mathcal{L}_{X}$ does not respect the bigrading.} Thus, the Lie subalgebra generated by the horizontal and vertical differentials, together with insertion of insular vector fields is not as simple as one might think it is. 

\begin{tm}[Insular Cartan calculus]\label{lcc1} 
We consider two insular vector fields $\chi = \xi + X$ and $\chi^{\prime} = \xi^{\prime} + X^{\prime}$. The only nontrivial commutators between any two of the following endomorphisms: $D, d, \delta, \iota_{\xi}, \iota_{\xi^{\prime}}, \iota_X, \iota_{X^{\prime}}, \mathcal{L}_{\xi}, \mathcal{L}_{\xi^{\prime}}, \mathcal{L}_X, \textrm{ and } \mathcal{L}_{X^{\prime}}$ are the following:
\begin{multicols}{2}
\begin{enumerate}
	\item $\mathcal{L}_X = [D, \iota_{X}]$.
	\item $\mathcal{L}_{\xi}^{\delta} \defeq [\delta, \iota_{\xi}] = [D, \iota_{\xi}] = \mathcal{L}_{\xi}$.
	\item $\mathcal{L}_X^d \defeq [d, \iota_X]$.
	\setcounter{enumi}{4}
	\item $[\delta, \iota_X] = \mathcal{L}_X - \mathcal{L}_X^d$.
	\item $[\mathcal{L}_{\chi}, \iota_{\chi^{\prime}}] = \iota_{[\chi, \chi^{\prime}]}$.
	\item $[\mathcal{L}_{\chi}, \mathcal{L}_{\chi^{\prime}}] = \mathcal{L}_{[\chi, \chi^{\prime}]}$.
\end{enumerate}
\end{multicols}
\vspace{-0.8em}
\begin{enumerate}
\setcounter{enumi}{3}
	\item $\mathcal{M}_X \defeq [D, \mathcal{L}_X^d] = [\delta, \mathcal{L}_X^d] = [\delta, \mathcal{L}_X]  = -[d, \mathcal{L}_X] = -[d, [\delta, \iota_X]]$.
\end{enumerate}
Where items $5$, $6$ and $7$ hold also replacing $\chi$ by $\xi$ and $X$ respectively.
\end{tm}

Observe that the theorem does not say that those are the only nontrivial commutators in the Lie subalgebra generated by $D, d, \delta, \iota_{\xi}, \iota_{\xi^{\prime}}, \iota_X, \iota_{X^{\prime}}, \mathcal{L}_{\xi}, \mathcal{L}_{\xi^{\prime}}, \mathcal{L}_X, \textrm{ and } \mathcal{L}_{X^{\prime}}$. As a matter of fact there are many non-trivial commutators involving $\mathcal{L}_X^d$ and $\mathcal{M}_X$. We are not interested in these operators as we will focus on the Lie derivatives in the total and vertical directions, but we want to include a list of some of the commutators for the sake of completeness.

\begin{pp}[Insular Cartan calculus, part 2]\label{lcc2} 
We consider two insular vector field $\chi = \xi + X$ and $\chi^{\prime} = \xi^{\prime} + X^{\prime}$. Define $\mathcal{L}_{\chi}^d \defeq [d, \iota_{\chi}]$ and $\mathcal{M}_{\chi} \defeq [\delta, \mathcal{L}_{\chi}^d]$ in analogy to what was done in Theorem \ref{lcc1}. Then
\begin{multicols}{2}
\begin{enumerate}
\setcounter{enumi}{7}
	\item $[\iota_{\xi}, \mathcal{L}_X^d]=0$.
	\item $[\mathcal{L}_{X}^{d}, \iota_{X^{\prime}}] = \iota_{[X, X^{\prime}]}$.
	\item $[\mathcal{L}_{X}^{d}, \mathcal{L}_{X^{\prime}}^d] = \mathcal{L}_{[X, X^{\prime}]}^d$.
	\item $[\mathcal{M}_X, D] = [\mathcal{M}_X, d] = [\mathcal{M}_X, \delta] = 0$.
	\item $[\mathcal{L}_{\xi}, \mathcal{L}_X] = \mathcal{L}_{[X, \xi]}^d$.
	\item $[\mathcal{L}_{\xi}, \mathcal{M}_X] = \mathcal{M}_{[\xi, X]}$.
\end{enumerate}
\end{multicols}
\end{pp}

The proof of this Proposition can be found in Appendix \ref{Alcc}.

We conclude this subsection, and with it this section and this chapter, with the proof of the Cartan calculus equations for the insular case (as mentioned in the proof, most of the equations were already known and we have cited the corresponding references):

{\bf Proof of Theorem \ref{lcc1}.}
Once again, the anti-commutativity between all the different differentials is precisely the definition of a bicomplex. The commutators that do not include $d$ or $\delta$ follow from the Cartan calculus for the variational bicomplex (Proposition \ref{CC}). The commutators including a vertical vector field and no horizontal follow from the decomposable Cartan calculus, Proposition \ref{lcc0}. 

What remains to prove is simply equation $4$. The key fact is that
$$[d , \mathcal{L}_X^d] = d \mathcal{L}_X^d - \mathcal{L}_X^d d = d \iota_X d - d \iota_X d = 0,$$
and symmetrically $[\delta,[\delta, \iota_X]] = 0$. Now
\begin{eqnarray*}
\mathcal{M}_X &\defeq& [D, \mathcal{L}_X^d] = [d, \mathcal{L}_X^d] + [\delta, \mathcal{L}_X^d] \\
&=& [\delta, \mathcal{L}_X^d] = [\delta, \mathcal{L}_X - [\delta, \iota_X]] \\
&=& [\delta, \mathcal{L}_X]  = [D, \mathcal{L}_X] - [d, \mathcal{L}_X] \\
&=& -[d, \mathcal{L}_X] = -\left( [d, \mathcal{L}_X^d] + [d,[\delta, \iota_X]]\right) \\
&=& -[d, [\delta, \iota_X]].
\end{eqnarray*}
\qed

\begin{rk}\label{dpull} The evaluation of a form $\omega \in \Omega_{\textrm{loc}}^{0,q}(\E \times M)$ on a filed $\varphi \in \E$ gives rise to a $(\textsf{top}-q)$-density on $M$. In this remark we are going to assume $M$ is oriented to avoid any further problems. The evaluation commutes with the horizontal differential (this can be found in the paper by Zuckerman \cite[Remark 1.2]{ZUC}). We want to be explicit about this, since that result has a nice interpretation in terms of pullbacks. The local form $\omega$ is the pullback of a form $\omega_{\infty}$ on the infinite jet bundle: $\omega = (j^{\infty})^{*} \omega_{\infty}$.  Evaluation at $\varphi$ is no more than pulling back along $\varphi$:
	\begin{equation*}
	\omega (\varphi) = (\varphi, -)^{*} (j^{\infty})^{*} \omega_{\infty} = (j^{\infty} \varphi)^* \omega_{\infty}.
	\end{equation*}
	Observe that $T j^{\varphi} (\frac{\partial}{\partial x^i}) = D_i$ for all $i$ (as vector fields). In that case it is clear that $\iota_{\frac{\partial}{\partial x^i}}(j^{\infty} \varphi)^* = (j^{\infty} \varphi)^* \iota_{D_i}$ for all $i$. The objective of this remark is to compare $(d \omega) (\varphi)$ and $d (\omega (\varphi))$. In local coordinates $\omega = \omega_{i_1, \ldots , i_q} dx^{i_1} \wedge \cdots \wedge dx^{i_q}$ and $d \omega = D_i \omega_{i_1, \ldots , i_q} dx^i \wedge dx^{i_1} \wedge \cdots \wedge dx^{i_q}$. Observe that $D_i \omega_{i_1, \ldots , i_q} = \mathcal{L}_{D_i} \omega_{i_1, \ldots , i_q} = \iota_{D_i} D \omega_{i_1, \ldots , i_q}$. Now a simply computation yields:
	\begin{eqnarray}\label{prev}
	(d \omega )(\varphi) &=& (j^{\infty} \varphi)^* \iota_{D_i} D \omega_{i_1, \ldots , i_q} dx^i \wedge dx^{i_1} \wedge \cdots \wedge dx^{i_q} \nonumber \\ 
	&=& \iota_{\frac{\partial}{\partial x^i}} (j^{\infty} \varphi)^* D \omega_{i_1, \ldots , i_q} dx^i \wedge dx^{i_1} \wedge \cdots \wedge dx^{i_q} \nonumber \\
	&=& \iota_{\frac{\partial}{\partial x^i}} d (j^{\infty} \varphi)^* \omega_{i_1, \ldots , i_q} dx^i \wedge dx^{i_1} \wedge \cdots \wedge dx^{i_q} \nonumber \\
	&=& d (\omega (\varphi)).
	\end{eqnarray}
	
	We can do the same for forms $\alpha \in \Omega_{\textrm{loc}}^{p,q}(\E \times M)$ with higher $p$. This goes along the lines of Remark \ref{conr} and the result can be found again in the paper by Zuckerman \cite[Remark 1.3]{ZUC}. For us it is a consequence of the local Cartan calculus. Given $\delta \varphi_1 , \cdots , \delta \varphi_p \in \mathfrak{X}_{\textrm{loc}}(\E)$ and $\varphi \in \E$ the associated density $(\iota_{\delta \varphi_1 \cdots \delta \varphi_p} \alpha)(\varphi)$ commutes with the horizontal differentials:
	\begin{equation}\label{zucpull}
	(\iota_{\delta \varphi_1 \cdots \delta \varphi_p} d \alpha) (\varphi) = (d \iota_{\delta \varphi_1 \cdots \delta \varphi_p} \alpha) (\varphi) = d (\iota_{\delta \varphi_1 \cdots \delta \varphi_p} \alpha (\varphi)).
	\end{equation}
	In order to derive this equation we need Equation $2$ from Theorem \ref{lcc1} and the previous Equation \ref{prev}.
\end{rk}

\newpage
\chapter{Basics in Lagrangian Field Theory}\label{blft}

In Lagrangian Field Theory the basic ingredients are a space of fields $\E$ and a Lagrangian $L$. The space of fields is the space of smooth sections of some vector bundle $\pi \colon E \rightarrow M$ and it is denoted by $\E \defeq \Gamma^{\infty}(M,E)$. The Lagrangian is a $(0, \textsf{top})$-local form. In other words, it is a local function on the space of fields which can be integrated over the base manifold.\\

The physical interpretation of a Lagrangian is that it gives, after integration, an action. The fields for which this action are minimal (or extremal to be precise) are to be considered as the physically relevant ones. In order to study which fields are extremal, the calculus of variations is used. Knowing the cohomology of the bicomplex of local forms is of tremendous importance at this point.\\

The space of extrema are solutions to a partial differential equation known as the Euler-Lagrange equation. A classical result expresses the derivation of the Euler-Lagrange equations as a calculus on the bicomplex of local forms. This point of view is very effective, as it eases the calculus used to relate symmetries of the Lagrangian and conserved quantities along the space of extrema: what is known as Noether's first theorem.\\

{\it This chapter starts by reviewing the known results in the literature about the cohomology of the bicomplex of local forms. It gives an interpretations of those results in terms of an useful result called pulley theorem. We review the paper by Zuckerman in which the relation between symmetries and conserved currents is addressed in the Language of the variational bicomplex. Most of this chapter is a bibliographical review in which we re-express the classical results in a language which will be convenient for the future study of the observables of a lagrangian field theory. The basic references in this chapter are Anderson \cite{AND}, Takens \cite{TAK}, and Zuckerman \cite{ZUC}.}


\section{The pulley theorem}\label{at}

{\it The global cohomological properties of the variational bicomplex (with respect to $\delta$ and $d$) have been extensively studied by several authors. This section summarizes the most relevant results from our point of view, and we use them to prove a theorem about the $D$-cohomology which will become relevant in the following chapters: the pulley theorem. The main references in this section are Anderson \cite{AND} and Takens in\cite{TAK}.}\\

We studied in Part I some properties of the variational bicomplex. Most of what was said there applies to the bicomplex of local forms. The study of the cohomology of the variational bicomplex started with the work of Takens \cite{TAK} where he showed that the $d$-complex is acyclic under the surface\footnote{As mentioned in the first part, the original result by Takens \cite{TAK} only deals with the locally finite bounded jet case. In the variational bicomplex, and hence in the bicomplex of local forms, the finite bound on the jets is global (every local form is the pullback of a differential form on a finite jet bundle). The globally bounded version of Theorem \ref{TAT} is original to Bauderon \cite{BAU} and Anderson \cite{AND}. For a non bounded version (something more general than the scope of the variational bicomplex) we refer to the paper by Giachetta, Mangiarotti and Sardanashvily \cite{GMS}.}.

From now on we will denote the dimension of $M$ by $\textsf{top}$.

\begin{tm}[Takens' acyclicity theorem, Takens \cite{TAK}]\label{TAT}
For $p \geqslant 1$ the complex $\left( \Omega_{\textrm{loc}}^{p,\bullet}(\E \times M), d \right)$ is exact except in top degree $\bullet = \textsf{top}$.
\end{tm}

\begin{center}
\begin{tikzpicture}[description/.style={fill=white,inner sep=2.5pt},scale=1]

\draw [fill=black!20, black!20] (-4,-2) to (-2.4,-0.4) to[out=45,in=180] (0,1.5) to[out=0,in=135] (2.4,-0.4) -- (2.6,-0.6) -- (3,-1) -- (3.2,-1.2) -- (4, -2) -- (1,-5) -- (-4,-5);

\node (a) at (-1,4) {\color{cyan} $0$};
\node (b) at (-2,3) {\color{cyan} $1$};
\node (c) at (-3,2) {\color{cyan} $2$};
\node (d) at (-4,1)  {\color{cyan} $\iddots$};
\node (e) at (1,4) {\color{dandelion} $\textsf{top}$};
\node (f) at (2,3) {\color{dandelion} $\textsf{top}-1$};
\node (g) at (3,2) {\color{dandelion} $\textsf{top}-2$};
\node (h) at (4,1) {\color{dandelion} $\ddots$};
\node (i) at (5,0) {\color{dandelion} $0$};

\node (j) at (0,3) {$ $};
\node (k) at (-1,2) {$ $};
\node (l) at (1,2) {$ $};
\node (m) at (-2,1) {$ $};
\node (m2) at (-2,-1) {$ $};
\node (n) at(2,1) {$ $};
\node (o) at(0,1) {$ $};
\node (p) at (4,-1) {$ $};
\node (q) at (3,-2) {$ $};
\node (r) at (2,-3) {$ $};
\node (s) at (1,-4) {$\iddots$};
\node (t) at (-3,0) {$\iddots$};
\node (u) at (3,0) {$\ddots$};
\node (v) at (-1,0) {$ $};
\node (w) at (-1,-2) {$\vdots$};
\node (x) at (-1,-4) {$\vdots$};
\node (y) at (1,0) {$\vdots$};
\node (z) at (1,-2) {$\vdots$};
\node (aa) at (-3,-2) {$\vdots$};
\node (ab) at (-3,-4) {$\vdots$};
\draw [rounded corners, dashed, thick, teal] (-2.4,-0.4) to[out=45,in=180] (0,1.5);
\draw [rounded corners, dashed, thick, teal] (-4,-2) to (-2.4,-0.4);
\draw [rounded corners, dashed, thick, teal] (0, 1.5) to[out=0,in=135] (2.4,-0.4);
\draw [dotted, thick, teal] (2.6,-0.6) -- (3,-1);
\draw [rounded corners, dashed, thick, teal] (3.2,-1.2) -- (4, -2) -- (1,-5);

%
\path[->,font=\scriptsize, color= cyan]
(j) edge node[auto] {} (k)
(k) edge node[auto] {} (m)
(v) edge node[auto] {} (m2)
(m) edge node[auto] {} (t)
(l) edge node[auto] {} (o)
(o) edge node[auto] {} (v)
(n) edge node[auto] {} (y)
(p) edge node[auto] {} (q)
(q) edge node[auto] {} (r)
(r) edge node[auto] {} (s);
\path[->,font=\scriptsize, color=dandelion]
(l) edge node[auto] {} (j)
(o) edge node[auto] {} (k)
(v) edge node[auto] {} (m)
(n) edge node[auto] {} (l)
(y) edge node[auto] {} (o)
(u) edge node[auto] {} (n)
(p) edge node[auto] {} (u)
(r) edge node[auto] {} (z);
\end{tikzpicture}
\end{center}

Taken's acyclicity Theorem tells us that every element under the surface (the gray part in the previous diagram) which is $d$-closed, $d \alpha_{i \geqslant 1} = 0$, is actually $d$-exact ($\alpha_i = d \beta_{j \in \{i, i+1\} }$). Here we are using the notation introduced in Subsection \ref{1285} where any form of total degree $n$, $\alpha \in \Omega_{\textrm{loc}}^n(\E \times M)$ decomposes in terms indexed by depth $\alpha = (A, \alpha_1, \alpha_2, \ldots)$. Sometimes the surface part will also be denoted by $\alpha_0$ if that simplifies the notation. Observe that if the bidegree of A is $(p\geqslant 1,q)$, a form $\beta$ such that $d \beta = \alpha_i$ has depth $i+1$, while if the bidegree is $(p\leqslant 0,q)$, such $\beta$ has depth $i$.

The local $(1, \textsf{top})$-forms that depend only on $dx^i$ and $\delta u^{\alpha}$, and not on higher $\delta u_I^{\alpha}$, are called {\bf source} forms (these are $\pi_{\infty}^0$-vertical forms). There is a similar definition for $p \geqslant 1$ that we will give in a second. There is an operator, called the Interior Euler operator that works as a projection to the source forms. The following definitions and results are taken from Anderson's work on the variational bicomplex, \cite{AND}.

\begin{df}[Interior Euler operator]
The following family $\mathbf{I}=\sum_{p \geqslant 1} \mathbf{I}^p$ of operators is globally well defined and it is called the interior Euler operator in $(\E \times M)$. 
\begin{eqnarray*}
	\mathbf{I}^{p\geqslant 1} \colon \Omega_{\textrm{loc}}^{p,\textsf{top}}(\E \times M) & \longrightarrow & \Omega_{\textrm{loc}}^{p,\textsf{top}}(\E \times M) \\
    A & \longmapsto & \frac{1}{p} \delta u^{\alpha} \wedge (-1)^{|I|}D_I \left( \iota_{ \partial_I^{\alpha} } A \right).
\end{eqnarray*}
The image $\mathbf{I}(\Omega_{\textrm{loc}}^{p,q})$ is defined to be the space of source forms and it is denoted by $\Omega_{\textrm{source}}^{p,\textsf{top}}(\E \times M)$.
\end{df}

Observe that $I$ can be thought as being an everywhere defined automorphism of the bicomplex of local forms, which is extended by zero to other bidegrees. In this sense, $I$ is only non-zero on the left-hand side slope of the surface part of the bicomplex (this is done in the work by Musilov\'a, see \cite{MUS} for example). Pictorially, the area in which $I$ is non-trivially defined is the shaded region in blue:

\begin{center}
\begin{tikzpicture}[description/.style={fill=white,inner sep=2.5pt},scale=1]

\draw [fill=cyan!30, cyan!30] (-4, 0) to (-1,3) to (0,2) -- (-4,-2);

\draw [fill=black!20, black!20] (-4,-2) to (-2.4,-0.4) to[out=45,in=180] (0,1.5) to[out=0,in=135] (2.4,-0.4) -- (2.6,-0.6) -- (3,-1) -- (3.2,-1.2) -- (4, -2) -- (1,-5) -- (-4,-5);

\node (a) at (-1,4) {\color{cyan} $0$};
\node (b) at (-2,3) {\color{cyan} $1$};
\node (c) at (-3,2) {\color{cyan} $2$};
\node (d) at (-4,1)  {\color{cyan} $\iddots$};
\node (e) at (1,4) {\color{dandelion} $\textsf{top}$};
\node (f) at (2,3) {\color{dandelion} $\textsf{top}-1$};
\node (g) at (3,2) {\color{dandelion} $\textsf{top}-2$};
\node (h) at (4,1) {\color{dandelion} $\ddots$};
\node (i) at (5,0) {\color{dandelion} $0$};

\node (j) at (0,3) {$ $};
\node (k) at (-1,2) {$ $};
\node (l) at (1,2) {$ $};
\node (m) at (-2,1) {$ $};
\node (m2) at (-2,-1) {$ $};
\node (n) at(2,1) {$ $};
\node (o) at(0,1) {$ $};
\node (p) at (4,-1) {$ $};
\node (q) at (3,-2) {$ $};
\node (r) at (2,-3) {$ $};
\node (s) at (1,-4) {$\iddots$};
\node (t) at (-3,0) {$\iddots$};
\node (u) at (3,0) {$\ddots$};
\node (v) at (-1,0) {$ $};
\node (w) at (-1,-2) {$\vdots$};
\node (x) at (-1,-4) {$\vdots$};
\node (y) at (1,0) {$\vdots$};
\node (z) at (1,-2) {$\vdots$};
\node (aa) at (-3,-2) {$\vdots$};
\node (ab) at (-3,-4) {$\vdots$};
\draw [rounded corners, dashed, thick, teal] (-2.4,-0.4) to[out=45,in=180] (0,1.5);
\draw [rounded corners, dashed, thick, teal] (-4,-2) to (-2.4,-0.4);
\draw [rounded corners, dashed, thick, teal] (0, 1.5) to[out=0,in=135] (2.4,-0.4);
\draw [dotted, thick, teal] (2.6,-0.6) -- (3,-1);
\draw [rounded corners, dashed, thick, teal] (3.2,-1.2) -- (4, -2) -- (1,-5);

%
\path[->,font=\scriptsize]
(j) edge node[auto] {} (k)
(k) edge node[auto] {} (m)
(v) edge node[auto] {} (m2)
(m) edge node[auto] {} (t)
(l) edge node[auto] {} (o)
(o) edge node[auto] {} (v)
(n) edge node[auto] {} (y)
(p) edge node[auto] {} (q)
(q) edge node[auto] {} (r)
(r) edge node[auto] {} (s);
\path[->,font=\scriptsize]
(l) edge node[auto] {} (j)
(o) edge node[auto] {} (k)
(v) edge node[auto] {} (m)
(n) edge node[auto] {} (l)
(y) edge node[auto] {} (o)
(u) edge node[auto] {} (n)
(p) edge node[auto] {} (u)
(r) edge node[auto] {} (z);
\end{tikzpicture}
\end{center}

This operator, as mentioned before, is a projection and it satisfies some other properties.

\begin{tm}[Anderson {\cite[Theorems 2.12 and 5.1]{AND}}]\label{ieop}
The interior Euler operator has the following properties:
\begin{enumerate}
	\item $\mathbf{I} \circ \mathbf{I} = \mathbf{I}$,
	\item $d \circ \mathbf{I} = d = 0$ on $\Omega_{\textrm{loc}}^{p \geqslant 1,\textsf{top}}(\E \times M)$,
	\item ${(\mathbf{I} \circ \delta)}^2 = 0$ and
	\item $\textrm{Ker}\, \mathbf{I}^{p} = d \Omega_{\textrm{loc}}^{p,\textsf{top}-1}(\E \times M)$ for all $p \geqslant 1$.
\end{enumerate}
\end{tm}

The fourth statement was first proven by Takens \cite{TAK} for the case $p =1$. Anderson \cite{AND} extends the proof for global $p$, something which can also be found in the work of Musilov\'a and Krbek, although they do not use the language of the variational bicomplex (see \cite{MUS}). That precise statement tells us how the cohomology of the $d$-complex is in top degree. This together with the acyclicity Theorem \ref{TAT} gives us all the information of the $d$-cohomology in the shaded area (both blue and gray) of the previous diagram.

The operator $\mathbf{I} \circ \delta$ plays an important role on Lagrangian field theories, it is given a special name and symbol:

\begin{df}[Exterior Euler operator]
The exterior Euler operator is the composition $\mathbf{E} \defeq \mathbf{I} \circ \delta$. It squares to zero $\mathbf{E}^2 = 0$.
\end{df}

Recall that because of the definition of $\mathbf{I}$, the exterior Euler operator is only non-trivially defined on the left slope of the surface together with the summit:
\begin{equation*}
	\mathbf{E}^{p\geqslant 0} \colon \Omega_{\textrm{loc}}^{p,\textsf{top}}(\E \times M) \longrightarrow \Omega_{\textrm{source}}^{p+1,\textsf{top}}(\E \times M). 
\end{equation*}

The vertical cohomology is a bit more complex, but we will not be particularly interested in it throughout this thesis. For the sake of completeness we include the relevant result for the vertical cohomology treated by Anderson:

\begin{tm}[Anderson {\cite[Corollary 5.28]{AND}}]\label{AAT}
For any $0 \leqslant q \leqslant \textsf{top}$, the complex $\left( \Omega_{\textrm{loc}}^{\bullet,q}(\E \times M), \delta \right)$ is exact in all degrees $p \geqslant \textrm{rank}(E)$.
\end{tm}

This result tell us that all the total cohomology is in finitely many degrees besides the left surface $d$-cohomology, which is controlled by the interior Euler operator. The study of the general $D$-cohomology can be done using all the previous results. As a general rule: {\it the cohomology class of a form is determined by the cohomological properties of its surface part.} We distinguish between the forms on the right and on the left:

\begin{tm}[Right pulley theorem]\label{RPT}
Let $\alpha \in \Omega_{\textrm{loc}}^{m \leqslant \textsf{top}}(\E \times M)$ be a $D$-closed form, $D \alpha = 0$. If there exists $B \in \Omega_{\textrm{loc}}^{0, m-1}(\E \times M)$ such that 
$$d B = A, \qquad A = \alpha_0$$
 then there exists $\beta \in \Omega_{\textrm{loc}}^{m-1}(\E \times M)$ such that $B = \beta_0$ and 
$$D \beta = \alpha.$$
Moreover, such a $\beta$ is unique up to a $D$-exact term.
\end{tm}

\begin{tm}[Left pulley theorem]\label{LPT}
Let $\alpha \in \Omega_{\textrm{loc}}^{m \geqslant \textsf{top} + 1}(\E \times M)$ be a $D$-closed form, $D \alpha = 0$. If there exists $B \in \Omega_{\textrm{loc}}^{m-\textsf{top}-1,\textsf{top}}(\E \times M)$ such that 
$$\mathbf{E} B = \mathbf{I} A$$
then there exists $\beta \in \Omega_{\textrm{loc}}^{m-1}(\E \times M)$ such that $B = \beta_0$ and 
$$D \beta = \alpha.$$
Moreover, such a $\beta$ is unique up to a $D$-exact term.
\end{tm}

The relevant theorem for Lagrangian field theories is the left pulley theorem, which will be called simply {\it pulley theorem}.

The proof of both theorems is constructive. From the surface form $B$ one constructs $\beta_1$, and then $\beta_2$ and then $\beta_3$ all the way down to the end of the complex. Since we are picturing the total degrees of the bicomplex as water wells, this process can be thought of using a pulley to complete the surface form $B$ into a form $\beta$ of the same total degree with the specified property.\\

\begin{center}
\begin{tikzpicture}[description/.style={fill=white,inner sep=2.5pt},scale=1]


\node (m) at (-2,1) {$B$};
\node (m2) at (-2,-1) {$\beta_1$};
\node (m3) at (-2,-3) {$\beta_2$};
\node (m4) at (-2,-5) {$\vdots$};

\node (l) at (-3,0) {$A$};
\node (l2) at (-3,-2) {$\alpha_1$};
\node (l3) at (-3,-4) {$\alpha_2$};
\node (l4) at (-3,-6) {$\vdots$};

\node (n) at(2,1) {$A^{\prime}$};
\node (n2) at (2,-1) {$\alpha_1^{\prime}$};
\node (n3) at (2,-3) {$\alpha_2^{\prime}$};
\node (n4) at (2,-5) {$\vdots$};

\node (q1) at (3,0) {$B^{\prime}$};
\node (q2) at (3,-2) {$\beta_1^{\prime}$};
\node (q3) at (3,-4) {$\vdots$};

\node (v) at (-1,0) {$\vdots$};
\node (w) at (-1,-2) {$\vdots$};
\node (x) at (-1,-4) {$\vdots$};

\node (y) at (1,0) {$\vdots$};
\node (z) at (1,-2) {$\vdots$};
\node (s) at (1,-4) {$\vdots$};

\draw [rounded corners, dashed, thick, teal] (-4,-2) to (-2.4,-0.4);

\draw [rounded corners, dashed, thick, teal] (-2.4,-0.4) to[out=45,in=210] (-1.2,1);
\draw [rounded corners, dotted, thick, teal] (-1.2,1) to[out=30,in=180] (0,1.5);
\draw [rounded corners, dotted, thick, teal] (1.2,1) to[out=150,in=0] (0,1.5);
\draw [rounded corners, dashed, thick, teal] (2.4,-0.4) to[out=135,in=330] (1.2,1);
\draw [rounded corners, dashed, thick, teal] (4,-2) to (2.4,-0.4);
\draw [rounded corners, dotted, thick, teal] (4,-2) to (4.6,-2.6);
\draw [rounded corners, dashed, thick, teal] (4.6,-2.6) -- (5, -3) -- (2,-6);

\path[->, font=\scriptsize]
(m) edge[bend left=60] node[above=0.3em] {} (m2)
(m2) edge[bend left=60] node[above=0.3em] {} (m3)
(q1) edge[bend left=60] node[above=0.3em] {} (q2);

\path[->,font=\scriptsize, color=cyan]
(m) edge node[auto] {} (l)
(m2) edge node[auto] {} (l2)
(m3) edge node[auto] {} (l3)
(q1) edge node[auto] {} (n2)
(q2) edge node[auto] {} (n3);
\path[->,font=\scriptsize,color=darkdelion]
(m2) edge node[auto] {} (l)
(m3) edge node[auto] {} (l2)
(q1) edge node[auto] {} (n)
(q2) edge node[auto] {} (n2);

\end{tikzpicture}
\end{center}

{\bf Proof of Theorems \ref{RPT} and \ref{LPT}.}
The proof of the two theorems is the same after we have constructed $\beta_1$. We start by considering $\alpha$ and $B$ as in the right pulley Theorem \ref{RPT}. In this case 
$$d(\alpha_1 - \delta B) = d \alpha_1 + \delta A = (D \alpha)_1 = 0.$$
Since $\alpha_1 - \delta B$ is not on surface anymore, Taken's acyclicity theorem (\ref{TAT}) applies and we get a form $\beta_1$ such that $d \beta_1 = \alpha_1 - \delta B$, in other words 
\begin{equation}\label{osito}
(D (B + \beta_1))_1 = \alpha_1.
\end{equation}

In the case in which $\alpha$ and $B$ are like in the left pulley Theorem \ref{RPT} we proceed as follows.
$$\mathbf{I} (A - \delta B) = \mathbf{I} A - \mathbf{E} B = 0.$$
Applying the fourth item of Theorem \ref{ieop}, we have that there exists $\beta_1$ such that $d \beta_1 = A - \delta B$. Once again, this can be rewritten as  
\begin{equation}\label{dosito}
(D (B + \beta_1))_0 = \alpha_0.
\end{equation}

Observe that there is a difference in the depth of equations \ref{osito} and \ref{dosito}. In order to proceed in a unique way with the rest of the proof, we define $\widetilde{i}$ to be $i+1$ in the case we are working on the right (on the assumptions of the right pulley Theorem), and $i$ otherwise. With this notation, equations \ref{osito} and \ref{dosito} can be both written as:
\begin{equation}\label{cosito}
(D (B + \beta_1))_{\widetilde{0}} = \alpha_{\widetilde{0}}.
\end{equation}

Now we proceed by induction. Assume we have constructed $(\beta_0 =B, \ldots, \beta_i)$ such that
\begin{equation}\label{losito}
(D (\beta_j + \beta_{j+1}))_{\widetilde{j}} = \alpha_{\widetilde{j}},
\end{equation}
for all $j \in \{ 0, \ldots, i-1\}$. Now
$$d(\alpha_{\widetilde{i}} - \delta \beta_{i}) = d \alpha_{\widetilde{i}} + \delta (\alpha_{\widetilde{i-1}} - \delta \beta_{i-1} ) = d \alpha_{\widetilde{i}} + \delta \alpha_{\widetilde{i-1}} = 0.$$
Applying Taken's acyclicity Theorem \ref{TAT} we get $\beta_{i+1}$ such that $d \beta_{i+1} = \alpha_{\widetilde{i}} - \delta \beta_{i}$. In other words, equation \ref{losito} holds for $i$. Since the induction hypothesis is equation \ref{cosito} (where $B \defeq \beta_{0}$)we have proven that equation \ref{losito} holds for maximum depth, in other words, calling $\beta \defeq (B, \beta_1, \ldots)$ we have that $D \beta = \alpha$ as wanted.

For the uniqueness up to $D$-exact term, we also use the pulley theorem. Given $\beta$ and $\beta^{\prime}$ satisfying the conditions from the theorems, the form $\beta - \beta^{\prime}$ is $D$-closed: $D \beta - D \beta^{\prime} = \alpha - \alpha =0$. Consider $\Gamma \defeq \beta_0 - \beta_0^{\prime} = B - B = 0$. We have that $D \Gamma = \mathbf{E} \Gamma = 0$ so that $(\Gamma, \beta - \beta^{\prime})$ are in the hypothesis of the pulley theorem (the left or the right one, it does not matter). In any case, there exists a form $\gamma$ such that $D \gamma = \beta - \beta^{\prime}$ and $\gamma_{0} = \Gamma = 0$. This completes the proof.
\qed

Now that we have proved the two theorems, let us describe two of the possible uses of it:

\begin{cl}\label{clpt1} Let $B \in \Omega_{\textrm{loc}}^{p,\textsf{top}}(\E \times M)$ be such that 
$$\mathbf{E} B = 0$$
then there exists $\beta \in \Omega_{\textrm{loc}}^{p+ \textsf{top}}(\E \times M)$ such that $B = \beta_0$ and 
$$D \beta = 0.$$
Moreover, such a $\beta$ is unique up to a $D$-exact term.
\end{cl}

\dem Consider $\alpha = 0 \in \Omega_{\textrm{loc}}^{p + \textsf{top} + 1}(\E \times M)$, clearly $D \alpha = 0$. The form $B$ is such that $\mathbf{E} B = \mathbf{I} \alpha_0 = 0$. We can apply the pulley theorem (the left one \ref{LPT}) to get a unique up to $D$-exact term form $\beta$ such that $\beta_0 = B$ and $D \beta = 0$.
\qed

\begin{df} A local form $\beta \in  \Omega_{\textrm{loc}}^{p+ \textsf{top}}(\E \times M)$ is a Lepagean form for the surface form $B \in \Omega_{\textrm{loc}}^{p,\textsf{top}}(\E \times M)$ if $\beta_0 = B$ and $(D \beta)_0 \in \Omega_{\textrm{source}}^{p+1, \textsf{top}}(\E \times M)$.
\end{df}

The theorey of Lepagean forms is quite fruitful. We have taken this definition from Anderson \cite{AND}. It is related to higher analogues to the Poincar\'e-Cartan form in Classical mechanics when applied to the Lagrangian, as we will see in the next section. Our approach to the study of symmetries in Lagrangian field theory is heavily related to the existence of Lepagean forms, hence the importance of Corollary \ref{clpt2}.

\begin{cl}\label{clpt2} Any form $B \in \Omega_{\textrm{loc}}^{p,\textsf{top}}(\E \times M)$ admits a Lepagean, which is unique up to a $D$-exact term. Moreover, the Lepagean can be taken to be of maximum depth $1$, i.e. $\beta = B + \beta_1$.
\end{cl}

\dem The proof relies on the use of the pulley Theorem twice. Consider $A \defeq \mathbf{E} B$. In this case $\mathbf{E} A = \mathbf{E}^2  B = 0$ by Theorem \ref{ieop} so that we can apply Corollary \ref{clpt2} to get an $\alpha$ such that $D \alpha = 0$ and $\alpha_0 = A$. Now $\mathbf{E} B = \mathbf{I} \delta B = \mathbf{I}^2 \delta B = \mathbf{I} A$, again by Theorem \ref{ieop}. Now we can apply the pulley theorem to get a form $\beta$ such that $D \beta = \alpha$ and $\beta_0 = B$. Observe that $(D \beta)_0 = \alpha_0 = A = \mathbf{E} B = \mathbf{I} \delta B$ which is a source form by definition. Observe that $B + \beta_1$ is already a Lepagean for $B$ without the need to use all the lower $\beta$'s.
\qed

\begin{rk}\label{nontw1} All the results about the cohomology of the bicomplex of local forms also apply in the non-twisted case. The orientation line bundle remains stable under all the operations.
\end{rk}


\section{The fundamental formulae}\label{ffs}

{\it In this section we define Lagrangians and Lagrangian field theories for the first time in this thesis. A classical result by Zuckerman explains how the variation of a Lagrangian splits into the Euler-Lagrange equations and a boundary term. In this section we review this theorem and give new interpretation of it in terms of exact forms, Lepageans and the pulley theorem. We introduce the Poincar\'e-Cartan form associated to a Lagrangian, a special kind of local pre-$\mathsf{top}$-multisymplectic form on $\E \times M$. We study the case in which the Lagrangian is of first jet order and compare the construction given by the result of Zuckerman and the theory of first order Lagrangians common in the literature. The main references in this section are de Le\'on, Mart\'{i}n de Diego and Santamr\'{i}a-Merino \cite{dL} and Zuckerman \cite{ZUC}.}\\

Lagrangian Field Theory is a mathematical formulation of classical field theory in physics. All the integrals appearing in the theory are substituted by the integrands: those are local forms. A classical field theory is given by the space of fields, and the action, $S(\varphi)= \int_M L(\varphi)$ where $L$ is a function on the fields valued in $0$-densities on $M$, which is also local. We take these definitions from Blohmann \cite{B}.

\begin{df}[Lagrangian]
Given a smooth fiber bundle $\pi \colon E \rightarrow M$ and the cor\-re\-spond\-ing variational bicomplex $\Omega_{\textrm{loc}}^{\bullet, \bullet}(\E \times M)$ we call Lagrangian an element $L$ of $\Omega_{\textrm{loc}}^{0, \textsf{top}}(\E \times M)$.
\end{df}

\begin{df}[Lagrangian field theory]
A Lagrangian field theory is given by a smooth fiber bundle $\pi \colon E \rightarrow M$ and a Lagrangian $L \in \Omega_{\textrm{loc}}^{0, \textsf{top}}(\E \times M)$.
\end{df}

Anderson in his book about the variational bicomplex, which we are re-interpreting in terms of ind- and pro-categories, acknowledges that the subject was introduced by Vinogradov \cite{VINO}. The motivation to study such structures by Vinogradov was precisely to explore the geometric and algebraic foundations of field theory, what is now known as Lagrangian field theory today.

Using the ideas from the previous sections, we can consider $\pi, \rho \colon E, F \rightarrow M$ two smooth fiber bundles over the same orientable manifold $M$. We can pullback local forms of any degrees (Proposition \ref{1292}), in particular Lagrangians:

\begin{cl}
Let $M$ be an oriented manifold and consider two fiber bundles over it $E, F \rightarrow M$. If $(\F \times M, L)$ is a Lagrangian field theory, then the pullback of $L$ along any local map $f= f \colon \E \times M \rightarrow \F \times M$ is a Lagrangian for $E$.
\end{cl}

The idea behind classical field theory is to study the variation of the action $\delta \int_M L$ to get to the Euler-Lagrange equations: these are the equation whose solutions are the fields that minimize (extremalize, to be precise) the action. This is known as the {\it principle of least action}. In Lagrangian field theory this is done by studying $\delta L \in \Omega_{\textrm{loc}}^{1,\textsf{top}}(\E \times M)$. By performing integration by parts we get two different terms: one that only depends on $\pi_{\infty}^0$-vertical vectors (a source form) and a $d$-exact term (which is usually discarded by considering variations vanishing along the boundary of a submanifold with boundary inside of $M$). The derivation of the Euler--Lagrange equations and some other interesting features are summarized in the {\it Fundamental Formulae} theorem of Zuckerman \cite{ZUC} (Deligne reached independently part of the same result). We present the theorem in the way Zuckerman originally wrote the result:

\begin{tm}[Fundamental Formulae, Zuckerman \cite{ZUC}]\label{FF}
Given a Lagrangian $L \in \Omega_{\textrm{loc}}^{0,\textsf{top}}(\E \times M)$ there exist 
$$EL \in \, \Omega_{\textrm{source}}^{1,\textsf{top}}(\E \times M), \, \, \lambda_1 \in \, \Omega_{\textrm{loc}}^{1,\textsf{top}-1}(\E \times M) \textrm{ and }
 \omega_1 \in \, \Omega_{\textrm{loc}}^{2,\textsf{top} -1}(\E \times M) \textrm{ such that:}$$
\begin{multicols}{2}
 \begin{itemize}
 		\item $\delta L = EL - d\lambda_1$,
 		\item $\omega = \delta \lambda_1$,
 		\item $\delta \omega_1 = 0$ and
 		\item $d\omega_1 = -\delta EL$.
 \end{itemize}
 \end{multicols}
 Evenmore,
 \begin{enumerate}
 		\item $EL$ is uniquely determined by $L$,
 		\item $\lambda_1$ is determined by $L$ modulo the addition of $d \alpha_1$ and
 		\item $L$ modulo $d A$ does not change $\omega_1$ modulo $d \lambda_2$. Thus, we get a linear map:
 		$$ \widetilde{\omega} \colon \bigslant{\Omega_{\textrm{loc}}^{0,\textsf{top}}(\E \times M)}{\Omega_{\textrm{loc}}^{0,\textsf{top}-1}(\E \times M)} \longrightarrow  
 		\bigslant{\Omega_{\textrm{loc}}^{2,\textsf{top} -1}(\E \times M)}{\Omega_{\textrm{loc}}^{2,\textsf{top} -2}(\E \times M)} .$$
 \end{enumerate}
\end{tm}

One of purposes of this theorem is to find $\lambda_1$ such that $\delta L + d\lambda_1 = D(L - \lambda_1)_0$ is a source form. This is precisely the same as finding a Lepagean for $L$. In Corollary \ref{clpt2} we constructed a Lepagean for every surface form, such as $L$. Bringing the way of thinking of the complex of local forms that we explored during the previous section, we can reinterpret the statement of the Fundamental Formulae as follows:

\begin{tm}[Fundamental Formulae, pulley version]\label{FFP}
Given a Lagrangian $L$ in $\Omega_{\textrm{loc}}^{0,\textsf{top}}(\E \times M)$ there exist:
\begin{itemize} 
\item $\omega \in \Omega_{\textrm{loc}}^{\textsf{top}+1}(\E \times M)$ of maximum depth $1$ such that $D \omega = 0$ and $\omega_0 = \textbf{E}L$.
\item $\lambda \in \Omega_{\textrm{loc}}^{\textsf{top}}(\E \times M)$ a Lepagean form for $L$ of maximum depth $1$ such that $D \lambda = \omega$ (and $\lambda_0 = L$).
\end{itemize}
Moreover, the depth $1$ components $\lambda_1$ and $\omega_1$ are unique up to $d$-exact terms.
\end{tm}

\dem The proof now is an application of the pulley Theorem. As a matter of fact, it follows from Corollary \ref{clpt2}. We can construct $\omega$ and $\lambda$ an auxiliary form and a Lepagean form for $L$ ($\omega_0 = \mathbf{E} L$, $\lambda_0 = L$ and $D \lambda = \omega$).

The key observation is that we can change $\omega$ to $\omega - D(\lambda - L - \lambda_0)$ and still get a pulley for $\mathbf{E} L$. Observe that for all $i \geqslant 2$, $\omega_i = d \lambda_{i+1} + \delta \lambda_i$ so that $\left(\omega - D(\lambda - L - \lambda_0) \right)_i = 0$ for all $i \geqslant 2$. Now it is immediate to show that $L + \lambda_1$ is a Lepagean for $B$, and that $D(L + \lambda_1) = \delta L + d \lambda_1 + \delta \lambda_1 = \mathbf{E} L + \omega_1$. The uniqueness up to $d$-exact term follows from the corresponding statement of uniqueness up to $D$-exact term applied to depth $1$ forms.
\qed

\begin{rk}\label{nontw2} As mentioned in Remark \ref{nontw1}, all the cohomological properties of the bicomplex of local forms persist in the non-twisted case. In this way, an element of $\Omega_{\textrm{ntw-loc}}^{0,\textsf{top}}(\E \times M)$ is called a non-twisted Lagrangian.
\end{rk}

Observe that $\lambda = L + \lambda_1$ and $\omega = \textbf{E} L + \omega_1$. In the literature it is common to find $\gamma \defeq \lambda_1$ and $\omega_1 \defeq \delta \gamma$ (sometimes $\gamma \defeq -\lambda_1$). We prefer to use our notation to keep in mind that each couple is in a same total degree.\footnote{We are still violating our convention of denoting $\alpha = (A, \alpha_1, \alpha_2, \ldots)$ since $\omega = (\textbf{E} L, \omega_1)$. As a mnemonic, we can thing that the $\mathbf{E}$ stands for Euler (both as the exterior Euler operator applied to $L$ and as $\mathbf{E}L$ referring to the Euler-Lagrange equations). The phonetic transcription of Euler is /\textopeno \textsci l\textschwa r/ which in Greek could be transcribed as $\Omega \iota \lambda \epsilon \rho$, hence the $\omega$. As a matter of fact, this is only a mnemonic, because Euler is actually written $O \iota \lambda \epsilon \rho $ in Greek.}

\begin{center}
\begin{tikzpicture}[description/.style={fill=white,inner sep=2.5pt},scale=1]

\draw [fill=cyan!30, cyan!30] (-0.4,3.4) rectangle (0.4,0.6);
\draw [fill=dandelion!30, dandelion!30] (-1.4,2.4) rectangle (-0.6,-0.4);

\node (j) at (0,3) {$L$};
\node (k) at (-1,2) {$\mathbf{E}L$};
\node (m) at (-2,1) {$0$};
\node (m2) at (-2,-1) {$0$};
\node (o) at(0,1) {$\lambda_1$};
\node (v) at (-1,0) {$\omega_1$};
\node (w) at (-1,-2) {$\vdots$};
\node (y) at (1,0) {$\vdots$};
\node (z) at (1,-2) {$\vdots$};
\node (aa) at (-3,-2) {$\vdots$};
\draw [rounded corners, dashed, thick, teal] (-2.4,-0.4) to[out=45,in=180] (0,1.5);
\draw [rounded corners, dashed, thick, teal] (-4,-2) to (-2.4,-0.4);
\draw [rounded corners, dashed, thick, teal] (0, 1.5) to[out=0,in=135] (2.4,-0.4);
\draw [dotted, thick, teal] (2.6,-0.6) -- (3,-1);
\draw [rounded corners, dashed, thick, teal] (3.2,-1.2) -- (4, -2) -- (3.8,-2.2);

\path[->,font=\scriptsize]
(j) edge node[auto] {} (k)
(k) edge node[auto] {} (m)
(v) edge node[auto] {} (m2)
(o) edge node[auto] {} (v);
\path[->,font=\scriptsize]
(o) edge node[auto] {} (k)
(v) edge node[auto] {} (m);
\end{tikzpicture}
\end{center}

\begin{df}[Poincar\'e-Cartan form]\label{pcf} The $D$-closed form $\omega \in \Omega_{\textrm{loc}}^{\textsf{top}+1}(\E \times M)$ given by Theorem \ref{FFP} is called the Poincar\'e-Cartan local pre-$\textsf{top}$-multisymplectic form or simply Poincar\'e-Cartan form associated to the Lagrangian $L$.
\end{df}

A local pre-$m$-multisymplectic form is simply another word for a $D$-closed local form of total degree $m$. We will apply results of multisymplectic theory to $\omega$ in the following Parts, so that we want to already point out this aspect of the form. The proper definition appears as Definition \ref{lms}.

\begin{rk}\label{others}
The fundamental formulae theorem has been around during three decades now and our interpretation of it as a Lepagean form is not new. Somehow, the case in which $\lambda$ is the pullback of a form on the first jet bundle has been extensively treated. Observe that $(D \lambda)_0$ is a source form, and hence can be thought of as coming from a form on the first jet bundle, but this does not mean that $\lambda$ itself is a form on $J^1 E$.

We refer to the paper by de Le\'on, Mart\'{i}n de Diego and Santamr\'{i}a-Merino, \cite{dL} for a modern approach to the topic. In that reference, as well as in many other papers in the area, $\lambda$ is called the {\it Poincar\'e-Cartan $\mathbf{\textsf{top}}$ form} and $\omega$ is called the {\it Poincar\'e-Cartan $(\mathbf{\textsf{top}+1})$ form} (we will reserve the name Poincar\'e-Cartan for the higher degree form). Proposition \ref{fol} in the following subsection shows the equivalence of their $\lambda$ and the one from Zuckerman. They focus in the case in which $\lambda$ is the pullback of a form on $J^1 E$. De Le\'on, Mart\'{i}n de Diego and Santamr\'{i}a-Merino \cite{dL} observe that $\omega$ can be treated as a pre-multisymplectic form. They study the degeneracy conditions of this form and relate them to regular Lagrangians. The Legendre transformation relating the Lagrangian and Hamiltonian formalisms of field theories is available in their setup. They also study the symmetries of $\omega$ and $\lambda$, as we will do in this thesis and relate them to the symmetries of $L$.

The celebrated (but unpublished) GiMmsy project on Momentum maps and Classical Fields \cite{GIMMSY} also studies what is covered in the de Le\'on, Mart\'{i}n de Diego and Santamr\'{i}a-Merino paper \cite{dL}. Their study of momentum maps and reduction is more extensive and in some examples it agrees with our approach, but again, they only treat the $J^1 E$ case. 

The fundamental difference with those lines of research and our work is that we do not restrict ourselves to the case in which $\lambda$ is the pullback of a form on the first jet bundle. Many of their results fail for Lagrangians with higher jet dependencies and this is one of the starting points of this thesis.
\end{rk}

As mentioned before, and keeping in mind the example of classical mechanics, $\mathbf{E}L$ is going to give conditions on the extrema of the action. Being precise: suppose we have $N \subset \, M$ a codimension $0$ submanifold with smooth boundary $\partial N$. We fix a field $\varphi \in \, \E$ and a variation $\delta \varphi \in \, \mathrm{T}_{\varphi}{\E}$ vanishing along $\partial N$. Applying Stokes theorem and the previous theorem one easily gets:
\begin{eqnarray}\label{stokes}
\delta \int_{N}{L(\varphi), \delta \varphi)} &\defeq& \int_{N}{\delta L(\varphi, \delta_{\varphi})} = \int_{N}{\mathbf{E}L(\varphi, \delta \varphi)} + \int_{N}{d \lambda_1 (\varphi, \delta \varphi)} \nonumber \\
&=& \int_{N}{\mathbf{E}L(\varphi, \delta \varphi)} + \int_{\partial N}{\lambda_1 (\varphi, \delta \varphi)} = \int_{N}{\mathbf{E}L(\varphi, \delta \varphi)}.
\end{eqnarray}

\begin{df}[Extremal]\label{extremal}
An element $\varphi \in \, \E$ is an extremal for the Lagrangian field theory determined by $L$ if $\delta \int_{N}{L(\varphi), \delta \varphi)} = 0$ for all codimension $0$ $N \subset \, M$ relatively compact and $\delta \varphi$ vanishing on $\partial N$.
\end{df}

\begin{pp}
A field $\varphi \in \, \E$ is an extremal for $L$ if and only if $\left( \iota_{\xi}\mathbf{E}L \right)(\varphi) = 0$ for every evolutionary vector field $\xi$.
\end{pp}

This result can be found in the paper by Zuckerman \cite{ZUC}. The argument nevertheless has been known for a very long time. It combines Stokes theorem as in Equation \ref{stokes} and the Bois-Reymond Lemma\footnote{The result says that if $\int f \cdot g = 0$ for arbitrary $g$, then $f = 0$. In this case we apply to $g$ the coefficients of the arbitrary vector field and $f$ the Euler-Lagrange equations evaluated at $\varphi$. Kosmann-Schwarzbach \cite{YVE} attributes this result to Paul Du Bois-Reymond without any particular citation.}.

\begin{rk}\label{shell}
	We denote by $\E_{L}$ the variety of extrema of the Lagrangian field theory determined by $L$. It is not a smooth manifold in general, but we will denote by $\Tg_{\varphi} \E_L$ the set of variations of $\varphi$ satisfying the Jacobi-linearization around $\varphi$ of the Euler-Lagrange equations. We distinguish between what happens {\it off-shell} (on $\E$ and $\textrm{T}\E$) and {\it on-shell}, this is on $\E_L$ and on $\textrm{T}\E_L$.
\end{rk}

\begin{lm}[Zuckerman {\cite[Lemma 8]{ZUC}}]\label{lm8} For all $\xi_1, \xi_2$ evolutionary vector fields $\iota_{\xi_1} \iota_{\xi_1} \delta \mathbf{E} L = 0$ on shell.
\end{lm}

Observe that this means that $\iota_{\xi_1} \iota_{\xi_1} \delta \mathbf{E} L(\varphi) = 0$ for all $\varphi \in \E_L$ provided that $\xi_1(\varphi), \xi_1(\varphi) \in \mathrm{T}_{\varphi} \E$.

An advantage of the pulley version of the Fundamental Formulae Theorem \ref{FFP} is that we can apply it to any other (left) surface forms.

\begin{cl}\label{cucu}
Given a local form $L \in \Omega_{\textrm{loc}}^{p,\textsf{top}}(\E \times M)$ there exist:
\begin{itemize} 
\item $\omega \in \Omega_{\textrm{loc}}^{p+1}(\E \times M)$ of maximum depth $1$ such that $D \omega = 0$ and $\omega_0 = \textbf{E}L$.
\item $\lambda \in \Omega_{\textrm{loc}}^{p}(\E \times M)$ a Lepagean form for $L$ of maximum  depth $1$ such that $D \lambda = \omega$ (and $\lambda_0 = L$). 
\end{itemize}
Moreover, the depth $1$ components $\lambda_1$ and $\omega_1$ are unique up to $d$-exact term.
\end{cl}

\begin{rk} There is not a clear physical motivation for {\it Lagrangians} of the form $L \in \Omega_{\textrm{loc}}^{p,\textsf{top}}(\E \times M)$ for $p \geqslant 1$. We would like to do the same for forms of any bidegree $(p,q)$. The extra dimensions, departing from $(0, \textsf{top})$ might mean that we are considering functions on the fields which integrable in dimension $q$-submanifolds of $M$ where we fix some variations of the field to be vanishing along that submanifold. This might be the case if we are embedding the space--time $X$ into a larger submanifold $M$, and it is very reminiscent of (although not quite the same as) the actions in String Theory (integrals over dimension $2$--submanifolds of the space--time, called world--sheets). Possibly some formulation of field theories will make sense in this context making use of Corollary \ref{cucu}.
\end{rk}


\subsection{Fundamental formulae in local coordinates and first order Lagrangians}

In this subsection we give explicit formulas for the forms $\lambda_1$ and $\omega_1$ coming from the Fundamental Formulae Theorem \ref{FFP}. This calculations come from \cite{TAK} by Takens. We already know that
\begin{equation}\label{FFE}
\mathbf{E} L = \mathbf{I} \delta L = \delta u^{\alpha} \wedge (-1)^{|I|}D_I \left( \iota_{ \partial_I^{\alpha} } \delta L \right).
\end{equation}

Now we will show that $\lambda_1$ is given by the following formula (where sum over $\alpha$ is understood by the Einstein summation convention).
\begin{equation}\label{FFlambda}
\lambda_1 \defeq \sum_I \sum_{j=1}^{|I|} \delta u_{J_I (j)}^{\alpha} (-1)^{|K_I (j)|}D_{K_I (j)} \left( (-1)^{\textrm{top}} \iota_{\partial_I^\alpha} \iota_{d_{i_j}} \delta L\right).
\end{equation}

The symbols $K_I (j)$ and $J_I (j)$ are multi-indices coming from $I$. If $I = (i_1, \ldots, i_{|I|})$, $\left( J_I(j), i_j, K_I (j) \right) = I$, that is $J_I(j) \defeq (i_1, \ldots, i_{j-1})$ and $K_I(j) \defeq (i_{j+1}, \ldots, i_{|I|})$. Observe that for every $\alpha$, $I$ and $j$ we have that:
\begin{eqnarray*}
& & d \left( \delta u_{J_I (j)}^{\alpha} (-1)^{|K_I (j)|}D_{K_I (j)} \left( (-1)^{\textrm{top}} \iota_{\partial_I^\alpha} \iota_{d_{i_j}} \delta L \right) \right) = \\ 
& & \delta u_{J_I (j+1)}^{\alpha} (-1)^{|K_I (j)|+1} D_{K_I (j)} \left( \iota_{\partial_I^\alpha} \delta L\right) + \delta u_{J_I (j)}^{\alpha} (-1)^{|K_I (j)|+1} D_{K_I (j-1)} \left(  \iota_{\partial_I^\alpha} \delta L\right)
\end{eqnarray*}

Now $d \lambda_1$ can be expressed as:
\begin{eqnarray*}
d \lambda_1 &=& \sum_I \sum_{j=1}^{|I|} d \left( \delta u_{J_I (j)}^{\alpha} (-1)^{|K_I (j)|}D_{K_I (j)} \left( (-1)^{\textrm{top}} \iota_{\partial_I^\alpha} \iota_{d_{i_j}} \delta L \right) \right) \\ 
&=& \sum_I \left( \sum_{j=1}^{|I|} \delta u_{J_I (j+1)}^{\alpha} (-1)^{|K_I (j)|+1} D_{K_I (j)} \left( \iota_{\partial_I^\alpha} \delta L\right) \right. +\\
& & \left. \sum_{j=1}^{|I|} \delta u_{J_I (j)}^{\alpha} (-1)^{|K_I (j)|+1} D_{K_I (j-1)} \left(  \iota_{\partial_I^\alpha} \delta L\right) \right) \\
&=& \sum_I \left( \sum_{j=1}^{|I|} \delta u_{J_I (j+1)}^{\alpha} (-1)^{|K_I (j)|+1} D_{K_I (j)} \left( \iota_{\partial_I^\alpha} \delta L\right) \right. +\\
& & \left. \sum_{j=0}^{|I|-1} \delta u_{J_I (j+1)}^{\alpha} (-1)^{|K_I (j+1)|+1} D_{K_I (j)} \left(  \iota_{\partial_I^\alpha} \delta L\right) \right) \\
&=& \sum_I \left(  \delta u_{J_I (|I|+1)}^{\alpha} (-1)^{|K_I (|I|)|+1} D_{K_I (|I|)} \left( \iota_{\partial_I^\alpha} \delta L\right) \right. +\\
& & \left. \delta u_{J_I (1)}^{\alpha} (-1)^{|K_I (1)|+1} D_{K_I (0)} \left(  \iota_{\partial_I^\alpha} \delta L\right) \right)  \\
&=& \sum_I \left(  \delta u_{|I|}^{\alpha} (-1) \iota_{\partial_I^\alpha} \delta L  + \delta u^{\alpha} (-1)^{|I|} D_{I} \left(\iota_{\partial_I^\alpha} \delta L\right) \right) \\
&=& -\delta L + \mathbf{I}\delta L = -\delta L + \mathbf{E} L.
\end{eqnarray*}

This computation shows that $\lambda \defeq L + \lambda_1$ satisfies that $D(\lambda)_0 = \mathbf{E}L$. Moreover, by taking the $\delta$ differential on both sides of Equation \ref{FFlambda} we get an expression for $\omega_1$:
\begin{equation}\label{FFomega}
\omega_1 \defeq \delta \lambda_1 = \sum_I \sum_{j=1}^{|I|} \delta u_{J_I (j)}^{\alpha} (-1)^{|K_I (j)|} \delta D_{K_I (j)} \left( (-1)^{\textrm{top}} \iota_{\partial_I^\alpha} \iota_{d_{i_j}} \delta L\right).
\end{equation}

Now we study the case in which $L$ is the pullback of a form on $J^1 E$. Moreover we assume $M$ is oriented and $L = f \, \textrm{Vol}$ where $\textrm{Vol} \defeq  dx^1 \wedge \cdots \wedge dx^{\textsf{top}}$. We use the notation $\textrm{Vol}^i \defeq dx^1 \wedge \cdots \wedge dx^{i-1} \wedge dx^{i+1} \wedge \cdots \wedge dx^{\textsf{top}}$. In this case the result is as follows:

\begin{pp}[First order oriented Lagrangians]\label{fol} For $L = f \, \textrm{Vol}$ a Lagrangian over an oriented manifold which comes from the pullback of a differential form on the first jet bundle, the form $\lambda = L + \lambda_1$ coming from the Fundamental Formulae Theorem \ref{FFP} has the same expression as the Poincar\'e-Cartan $\textsf{top}$-form from de Le\'on, Mart\'{i}n de Diego and Santamr\'{i}a-Merino \cite{dL} up to a sign. Explicitly:
$$\lambda_1 = (-1)^{\textrm{top}+i-1} \left( \frac{\partial f}{\partial u_{\alpha}^i} d u_{\alpha} \textrm{Vol}^i - u_{\alpha}^i \frac{\partial f}{\partial u_{\alpha}^i} \textrm{Vol} \right).$$
\end{pp}

This first order $\lambda$ is called the Poincar\'e-Cartan $\textsf{top}$-form by de Le\'on, Mart\'{i}n de Diego and Santamr\'{i}a-Merino \cite[Definition 2.4]{dL} and the Cartan $\textsf{top}$-form \cite[Equation 3B.3]{GIMMSY} in the GiMmsy papers.

\dem If $L = f \textrm{Vol}$ where $\textrm{Vol} \defeq  dx^1 \wedge \cdots \wedge dx^{\textsf{top}}$ and $f$ is a function on $J^1 E$ we have that
$$\delta L = \frac{\partial f}{\partial u_{\alpha}} \delta u_{\alpha} \textrm{Vol} + \frac{\partial f}{\partial u_{\alpha}^i} \delta u_{\alpha}^{i}.$$
Observe the difference $\frac{\partial f}{\partial u_{\alpha}^i}$ versus $\partial_{\alpha}^i f$. Using equation \ref{FFlambda} we get:
\begin{equation*}\label{FFlambda2}
\lambda_1 = (-1)^{\textrm{top}+i-1}\frac{\partial f}{\partial u_{\alpha}^i} \delta u_{\alpha} \textrm{Vol}^i = (-1)^{\textrm{top}+i-1}\left( \frac{\partial f}{\partial u_{\alpha}^i} d u_{\alpha} \textrm{Vol}^i - u_{\alpha}^i \frac{\partial f}{\partial u_{\alpha}^i} \textrm{Vol} \right).
\end{equation*}
\qed

As mentioned in Remark \ref{others}, first order Lagrangians have been extensively studied in the literature. Projects similar to the one of this thesis relating the symmetries of $\delta \lambda_1$ or $D(L + \lambda_1)$ are available (the already mentioned de Le\'on, Mart\'{i}n de Diego and Santamar\'{i}a-Merino \cite{dL} and GiMmsy \cite{GIMMSY} are good examples of them). We will not treat the first order Lagrangians in any particular detail. Instead, we will focus on the properties and conclusions that we can achieve by looking at $\lambda$ and $\omega = D \lambda$ with full generality.

\begin{rk}\label{alalepage}
	The work on Poincar\'e-Cartan forms, even in higher order jets, was productive in the 1980s and 1990s. The approach was, nevertheless, different to the one presented here, and we will refer to it as ``Lepagean''. The general idea in Lepagean theories was to construct (multi)-symplectic structures on some finite jet bundle tensored with a vertical bundle and to bring that form to $\E \times M$. This procedure does not only work for first order Lagrangians, as pointed out here, but also for second order Lagrangians. In 1983, several authors published that there is no universal Poincar\'e-Cartan form, constructed in a Lepagean way, for Lagrangian theories of order greater or equal than three (this means there is not such a uniquely defined form Poincar\'e-Cartan \'a la Lepage, even modulo $d$-exact terms). The paper of Kol\'a\v{r} on the topic \cite{KOL1}, summarizes these ideas and it is a good reference to understand this way of constructing multi-symplectic structures from field theories in a Lepagean way. Kol\'a\v{r} himself points out that the original work comes from Ferraris \cite{FER1}, Garc\'ia and Mu\~noz Masqu\'e \cite{GM1}, Hor\'ak and Kol\'a\v{r} \cite{HK1}, and Krupka \cite{KRU1}. Mu\~noz in his PhD thesis and later in the article \cite{MUN1} gives a detailed definition of Poincar\'e-Cartan forms for Lagrangians of higher orders, making the dependency explicit on the choice of certain connections on the bundles involved. Our approach is different, we do not try to see if the form $\omega$ also defines a multi--symplectic structure on the lowest jet bundle in which it is defined, avoiding then the problem of choosing connections. Our $\omega$ is universal, and well defined from the Lagrangian modulo a $d$-exact term (that comes from granted from Zuckerman's result --Theorem \ref{FF} above-- \cite{ZUC}). Another way of interpreting our form $\omega$ is a coordinate-free, universal version of the Poincar\'e-Cartan form \'a la Lepage of Mu\~noz \cite{MUN1}.
\end{rk}


\section{Symmetries, conserved currents and charges}

{\it A symmetry of a Lagrangian field theory is an evolutionary vector field which does not change the Lagrangian up to a $d$-exact term. Symmetries are directly related to currents and charges: the conserved forms and quantities on shell. The classical theorem of Noether relating them has an easy presentation using the fundamental formulae theorem as it was done by Zuckerman. In this section we collect the relevant definitions and results for the study of symmetries and conserved currents and we discuss how other symmetries could be considered. The main references in this section are Noether \cite{NOE} and Zuckerman \cite{ZUC}.}\\

The concept of symmetry in physics is one of the most important tools in theoretical physics. Due to Noether's theorems, symmetry means invariance: the fact that a certain property is conserved. From this point of view, we can view all laws of nature to be encoded by symmetries.

We fix a Lagrangian field theory $(L, \pi \colon E \rightarrow M)$. We adopt the conventions coming from Deligne and Freed \cite{DEL}, where the following definitions are taken from. 

\begin{df}[Symmetry]\label{syme}
An evolutionary vector field $\xi \in \mathfrak{X}_{\textrm{loc}}(\E)$ is called a symmetry of the Lagrangian field theory given by $L$ if there exists a local form $A_{\xi} \in \Omega_{\textrm{loc}}^{0, \textsf{top}-1}$ such that $\mathcal{L}_{\xi} = d A_{\xi}$.
\end{df}

Usually these are called infinitesimal symmetries, since it is the infinitesimal analogue of a local function respecting the Lagrangian. Since we are more interested in the (infinitesimal) symmetries, we reserve the term symmetry without any further adjective to the ones as in Definition \ref{syme}.

\begin{df}[Integral Symmetry]
A local map $g \colon \E \times M \rightarrow \E \times M$ such that $g^* L = L$ is called an integral symmetry of the Lagrangian field theory given by $L$
\end{df}

One concept related to symmetries of the theory is the one of conserved currents.

\begin{df}[Current]\label{curr} A local form in $\Omega_{\textrm{loc}}^{0,\textsf{top}-1}(\E \times M)$ is called a current. Local forms of bidegree $(p, \textsf{top} -1)$ are called $p$-currents. \end{df}

\begin{df}[Conserved current]\label{ccrr}
In a Lagrangian Field Theory given by $L$, a $p$-current $\zeta \in \Omega_{\textrm{loc}}^{p,\textsf{top}-1}(\E \times M)$ is said to be conserved if  for all $\xi_1 \ldots \xi_p$ evolutionary vector fields on $\E$
$$ d \left( \iota_{\xi_p} \cdots \iota_{\xi_1} \zeta \right)  = 0 \quad \textrm{ on shell. }$$
\end{df}

For the most classical case of currents, $p = 0$, the condition is $d Z(\varphi) = 0$  whenever $\varphi \in \E_L$. A particular case is when $d Z = \iota_{\xi} \textbf{E} L$ for some evolutionary vector field $\xi \in \mathfrak{X}_{\textrm{loc}}(\E)$. But this is not the only case. Stasheff \cite{STA} gives a detailed explanation about the concept of ``vanishing on shell'' which we reproduce in the following Remark:

\begin{rk}[Following Stasheff \cite{STA}]\label{stasheff} We assume $M$ to be oriented and the bundles to be trivial for simplicity of the argument. In this case, the fact that $\varphi$ is on shell is equivalent to $\varphi$ being a solution of the Euler-Lagrange equations $\textbf{E} L_{\alpha} (\varphi) = 0$ for all $\alpha$ where $\textbf{E} L = \textbf{E} L_{\alpha} \delta u^{\alpha} \textrm{Vol}$ for a given volume form $\textrm{Vol}$ on $M$. The stationary surface $\Sigma \defeq \{ j_x^{\infty} \varphi \colon \textbf{E} L_{\alpha} (\varphi) = 0 \textrm{ for all } \alpha \}$ defines a differential ideal 
	$$\mathtt{I} \defeq \{ (f, f^{\infty}) \in \C_{\textrm{loc}}(\E \times M) \colon \restrict{f^{\infty}}{\Sigma} = 0\}$$
	 which precisely takes care of the functions that vanish on shell (along the pullback of the stationary surface in this language). The term differential ideal in this case in used in the sense of differential algebra and rings. The Lie derivatives $\mathcal{L}_{D_i}$ are understood as derivations of $\C_{\textrm{loc}}(\E \times M)$ and $\mathtt{I}$ is the ideal generated by  $\textbf{E} L_{\alpha}$ for all $\alpha$ closed under the set of derivations $\{\mathcal{L}_{D_i}\}_{i=1}^{\textsf{top}}$. A generic element of $\mathtt{I}$ is then $f = f_{\alpha}^I D_I \textbf{E} L_{\alpha}$. A conserved current as in Definition \ref{ccrr} is a form $\zeta \in \Omega_{\textrm{loc}}^{p,\textsf{top}-1}(\E \times M)$ such that for all $\xi_1 \ldots \xi_p$ evolutionary vector fields on $\E$ there exists $f = f_{\alpha}^I D_I \textbf{E} L_{\alpha}$ such that $d \left( \iota_{\xi_p} \cdots \iota_{\xi_1} \zeta \right)  = f$.
\end{rk}

The relation between symmetries and conserved currents is given by Noether's theorem. It was first proved by Noether  \cite[Equation 12]{NOE}. We cite the proof given by Zuckerman \cite[Theorem 12]{ZUC}:

\begin{tm}[Noether's first theorem]\label{NFT} Let $\xi$ be a symmetry of the Lagrangian field theory given by $L$, that is there exists $A_{\xi} \in \Omega_{\textrm{loc}}^{0, \textsf{top}-1}(\E \times M)$ such that $\mathcal{L}_{\xi} L = d A_{\xi}$. Then the Noether current associated to $\xi$ is defined as $$Z_{\xi} \defeq A_{\xi} - \iota_{\xi} \lambda_1$$
is conserved. 
\end{tm}

\dem We include the proof for its simplicity:
\begin{equation}\label{ahora}
dZ_{\xi} = d A_{\xi} - d \iota_{\xi} \lambda_1 = \mathcal{L}_{\xi} L + \iota_{\xi} d \lambda_1 = \iota_{\xi} (\delta L + d \lambda_1) = \iota_{\xi} \mathbf{E} L.
\end{equation}
If we evaluate at any $\varphi \in \E_L$, the right hand side vanishes by definition of $\E_L$.
\qed

\begin{df}\label{NPair}
	A pair $(\xi, Z_{\xi}) \in \mathfrak{X}_{\textrm{ins}}(\E \times M) \times \Omega_{\textrm{loc}}^{0,\textsf{top}-1}(\E \times M)$ such that Equation \ref{ahora} holds, that is $dZ_{\xi} = \iota_{\xi} \mathbf{E} L$, is called a Noether pair.
\end{df}

There is a Noether's second theorem, which relates families of symmetries with families of conserved currents. We will talk about them in more detail in Part \ref{lla}.

\begin{rk}\label{modulo} The relation between symmetries and conserved currents is bidirectional. Given a conserved current of the kind $d Z = \iota_{\xi} \mathbf{E} L$ it is clear by the previous Theorem that $\xi$ is a symmetry with associated form $Z + \iota_{\xi} \lambda_1$. For a general conserved current as in Remark \ref{stasheff}, $d Z  = f_{\alpha}^I D_I \textbf{E} L_{\alpha}$, it is possible to see (see for example Barnich and Brandt \cite[Section 2.3]{BB}) that $\xi_{\alpha} \defeq (-1)^{|I|} D_I (f_{\alpha}^I)$ is a symmetry of the Lagrangian. The idea behind this result is that $f_{\alpha}^I D_I \textbf{E} L_{\alpha} = (-1)^{|I|} D_I (f_{\alpha}^I) \mathbf{E} L_{\alpha}$ modulo a $d$-exact term as a consequence the product rule for derivatives. We actually want to give an interpretation of this result using simply Taken's acyclicity Theorem \ref{TAT}, and the formula for $\lambda_1$ from Equation \ref{FFlambda}.
In general, given two local functions $f$ and $g$ (we assume $M$ is oriented with volume form $\textrm{Vol}$) and a multi-index $I$ the following equation holds:
\begin{equation}\label{modeq}
(-1)^{|I|} D_I g \cdot f - g D_I f = \sum_{j = 1}^{|I|} (-1)^{|I|-j+\textsf{top}} D_{i_j} \left( D_{i_1 \cdots i_{j-1}} f \cdot D_{i_{j+1} \cdots i_{|I|}} g \right).
\end{equation}
\dem
Defining $\omega \defeq g \delta u_{I}^{\alpha} \, \textrm{Vol}$ and $\xi \defeq f \partial_{\alpha}$ we can calculate
\begin{eqnarray*}
\iota_{\xi} \omega &=& D_I f \cdot g \, \textrm{Vol} \\
\iota_{\xi} \mathbf{I} \omega &=& (-1)^{|I|} D_I g \cdot f \, \textrm{Vol}.
\end{eqnarray*}
By Taken's acyclicity Theorem \ref{TAT} there exists a form $\lambda_1$ such that $\mathbf{I} \omega - \omega = d \lambda_1$. From equation \ref{FFlambda} we know that $\lambda_1$ can be taken to be:
$$ \lambda_1 = \sum_{j = 1}^{|I|} \delta_{i_1, \ldots, i_{j-1}} (-1)^{|I|-j} D_{i_{j+1}, \ldots, i_{|I|}} g \, \textrm{Vol}^{i_j},$$
where $\textrm{Vol}^{i_j} \defeq (-1)^{\textsf{top}} \iota_{D_{i_j}} \textrm{Vol}$. Now $$d \iota_{\xi} \lambda_1 = \iota_{\xi} d \lambda_1 = \sum_{j = 1}^{|I|} (-1)^{|I|-j+\textsf{top}} D_{i_1, \ldots, i_{j-1}} f  D_{i_{j+1}, \ldots, i_{|I|}} g \, \textrm{Vol}.$$ Comparing all equations:
\begin{eqnarray*}
\left( (-1)^{|I|} D_I g \cdot f \textrm{Vol} - D_I f \cdot g \right) \textrm{Vol} &=&  \iota_{\xi} (\mathbf{I} \omega - \omega) = \iota_{\xi} d \lambda_1 \\
&=& \left( (-1)^{|I|-j+\textsf{top}} D_{i_1, \ldots, i_{j-1}} f  D_{i_{j+1}, \ldots, i_{|I|}} g \right) \, \textrm{Vol},
\end{eqnarray*}
which after getting rid of the volume form gives precisely equation \ref{modeq}
\qed
\end{rk}

\begin{df}[Charge]
In a Lagrangian Field Theory given by $L$, a charge is the integral of a conserved current. If $Z \in \Omega_{\textrm{loc}}^{0,\textsf{top}-1}(\E \times M)$ is conserved, we define the charge associated to it to be for $N \subset M$ of codimension $1$:
$$ Q_{Z} \defeq \int_{N} Z.$$
\end{df}

The idea behind charges is that they are conserved. Imagine we have a submersion $N \times [0,1] \hookrightarrow M$. Denote by $N_t$ the image of $N \times \{t\}$. Now $\partial N = N_0 - N_1$ after fixing an orientation. This is to be thought as $N$ being the space and $[0,1]$ representing the time direction. In that case, fixing $\varphi \in \E_L$ and using Stokes theorem:
$$0 = \int_{N \times  [0,1]} dZ (\varphi)= \int_{N_0} Z(\varphi) - \int_{N_1} Z(\varphi).$$
In this way, viewing $Q_{Z}$ as a function on $[0,1]$, $Q_{Z} \colon [0,1] \rightarrow \RE$ sending $t$ to $\int_{N_t} Z(\varphi)$, we can see that $Q_{Z}$ is constant. In other words the quantity $Q_{Z}$ is conserved over time. In the Physics bibliography one usually finds $\frac{d}{dt} Q_{A} = 0$, which is to be understood in the way just described.

Theorem \ref{FF} provides a current, that Zuckerman proved \cite{ZUC} to be conserved.

\begin{tm}[Universal conserved current]\label{cc}
The current $\omega_1$ given in the theorem \ref{FF} is conserved and it is called the universal conserved current associated to $L$.
\end{tm}

\dem  Once again, we replicate the proof, since it uses some of the equations we have discussed so far ($[d, \iota_{\xi}] = 0$ and Lemma \ref{lm8}):
$$d \iota_{\xi_2} \iota_{\xi_1} \omega_1 = \iota_{\xi_2} \iota_{\xi_1} d \omega_1 = - \iota_{\xi_2} \iota_{\xi_1} \delta \textbf{E} L = 0.$$
\qed

\begin{rk}
The Universal conserved current gives rise to a $2$--form on $\E$ which depends on the choice of a compact hypersurface $N$ in $M$ (up to homology):
$$\omega_{[N]} \defeq \int_N \omega_1$$
which is $\delta$-exact ($\omega_{[N]} = \delta \int_N \lambda_1$). Moreover, it defines a symplectic structure on $\E^{\prime}_L \times M$ (the smooth locus of the space of extrema) because the form is also $d$-closed when restricted to the space of extrema. There is no canonical choice for the $2$--form unless all compact hypersurfaces in $M$ are cohomologous (for instance, the case $M = \RE$ gives a well defined symplectic structure on the phase space).
\end{rk}


\subsection{Total and horizontal symmetries}

To conclude this section we would like to point out why only evolutionary vector fields have been considered as symmetries. In principle we could have ask the symmetry equation given by Definition \ref{syme} to hold for all insular or decomposable vector fields $\chi$, that is $\mathcal{L}_{\chi} L = D A_{\chi}$ for some $A_{\chi} \in \Omega_{\textrm{loc}}^{0, \textsf{top} -1}(\E \times M)$.

Horizontal vector fields are usually excluded from the study of symmetries because any horizontal vector field is a symmetry (we can find this argument in the work of Deligne and Freed \cite{DEL}). Consider $X \in \mathfrak{X}(M)$, using the decomposable Cartan calculus (Proposition \ref{lcc0}) we get:
\begin{equation}\label{horr}
\mathcal{L}_{X} L = [d, \iota_X] L = d \iota_X L.
\end{equation}
So that, defining $A_X \defeq \iota_X L$ we have proven that $X$ is a symmetry.

Total vector fields are completely different. In the case in which $X \in \mathfrak{X}_{\textrm{loc}}(M)$, this is $X$ is total (recall that meant $X \colon \E \rightarrow \mathfrak{X}(M)$ is a local map along the identity) we have that $\mathcal{L}_{X} \neq [d, \iota_X]$ (see the formulas for the insular Cartan calculus from Proposition \ref{lcc1}). Moreover $\mathcal{L}_{X}$ is not only in bidegree $(0, \textsf{top})$, but it also includes $[\delta, \iota_X] L \in \Omega_{\textrm{loc}}^{1, \textsf{top}-1}(\E \times M)$.


In this case it makes more sense to consider $X$ total vector fields such that $\mathcal{L}_{X} L = D {\alpha}_{X}$ for some $\alpha_{X} \in \Omega_{\textrm{loc}}^{\textsf{top} -1}(\E \times M)$. Still, the physical interpretation of this equation is not straightforward since we do also have the first term in the previous equation.

If we take decomposable vector fields, Noether's first theorem still holds due to equation \ref{horr}. Nevertheless, if we consider insular vector fields which are not decomposable, the theorem does not hold anymore. But observe we are in a very different setting now: the total part of an insular vector filed can  be thought as a family of horizontal vector fields  indexed by $\E$. In other words, we are dealing with families of decomposable symmetries. This is the territory of Noether's second theorem. The correct notion of the family of vector fields is that of a family indexed by a Lie algebroid. The next part is devoted to local Lie group and local Lie algebra actions where we will learn about these families of symmetries.


\printbibliography

\setcounter{part}{4}
\setcounter{chapter}{11}


\newpage
\part{Higher Lie groups and higher Lie algebras}\label{lla}

\newpage
\tableofcontents

\chapter*{\color{darkdelion} Higher Lie groups and higher Lie algebras}

In the category of insular manifolds there are notions of group-like and (Lie) algebra-like insular manifolds and insular actions. From the group side, one considers bisection groups of Lie groupoids. (Fibered) Lie groupoids are smooth manifolds $G$ provided with two fiber bundle structures over another base manifold $M$. Maps $M \rightarrow G$ which are transverse to both sets of fibers are called bisections. All bisections of a Lie groupoid (denoted by $\G^r$) constitute a, generally infinite dimensional, Lie group.  After some adjustments regarding the topology of $\G^r$, it is possible to talk about local maps defined on it: it is possible to define jet groupoids and to show that the operations involved in the bisection group are all local.\\

Every Lie groupoid $G \rightrightarrows M$ has an associated linearized structure, which is a Lie algebroid. Those are vector bundles $A \rightarrow M$, whose spaces of sections $\A$ have the structure of a Lie algebra satisfying a version of the Leibniz rule. Lie algebras are Lie algebroids over a point. The Lie bracket on the space of sections of the Lie algebroid of a Lie groupoid is local. It is well behaved with respect to taking finite jets in the following sense: $A(J^k G) = J^k (A(G))$ for every $k$ (where $A(G)$ is the Lie algebroid of the Lie groupoid $G$.)\\

Bisection groups act on insular manifolds. When the action is local (it is immediately insular) we talk about local Lie group actions. The same happens in the Lie algebroid world. Using the machinery of insular differential operators, it is possible to encode the conditions of a local Lie group action, and those of a local Lie algebra action, in terms of the most lower map between finite dimensional manifolds. Symmetries parametrized by a local Lie algebra action are usually called gauge symmetries. Noether's second theorem relates these symmetries with certain differential equations satisfied by the Euler-Lagrange term of a field theory.\\

Another higher analogue of Lie algebras is that of \Linf s. An {\Linf} is a graded vector space $L$ together with a family of totally antisymmetric multilinear maps (no with more than two entries, generally) satisfying certain higher Jacobi equation. Under certain finite dimensionality assumptions, an {\Linf} structure on $L$ is equivalent to a degree $1$, square to zero, derivation of the algebra $S([-1]L^*)$. The finite dimensionality hypothesis can be lowered to pro-finite dimensionality.\\

The two generalizations of Lie algebras are related through higher analogues of the a classical result by Vaintorb relating Lie algebroid structures on $A \rightarrow M$ and $\C(M)$ differential graded Lie algebra structures on $[-1]\A^*$. These ideas also apply in the pro-finite dimensional case. Peetre's theorem in its linear version, relates linear insular differential operators to pro-linear maps. Using that result, we define local \Linf s and compare them to local \Linf s in the sense of Costello using polydifferential operators.\\

{\it This part reviews and fixes notations for the concepts of groupoid, Lie groupoid, bisection group, local bisections, jet groupoids, Lie alebroids and jet Lie algebroids, graded vector spaces, and \Linf s. It shows that the bisection group is local and that the bracket in the Lie algebroid of a Lie groupoid is also local. Using the previously developped theory of insular differential operators, it revisits the result of Blohmann \cite{B} about local Lie group actions and proves an equivalent one for local Lie algebra actions. Furthermore, it includes a version of Noether's second theorem using the cohomology of the bicomplex of local forms and it studies Lie pseudogroups from an insular manifold perspective. Finally, creating a theory of duals in pro- and ind-graded vector spaces, it defines local \Linf s showing that the different approaches to that definition are equivalent. The main references in this part include Lada and Stasheff \cite{LS}, Moerdijk and Mr\v{c}un \cite{MM}, Noether \cite{NOE}, Vaintrob \cite{VAI}, and Yudilevich \cite{ORI}.}


\newpage
\chapter{Lie groupoids}

In the category of insular manifolds there are notions of group-like insular manifolds and insular actions. The basic example of such objects is the bisection group of a Lie groupoid. Lie groupoids are fiber bundle analogues of Lie groups. In particular there are two smooth maps $G \rightrightarrows M$ called the moment maps of a Lie groupoid. Maps $M \rightarrow G$ which induce diffeomorphisms on $M$ after post-composition with the moment maps are called bisections. All bisections of a Lie groupoid (denoted by $\G^r$) constitute a, generally infinite dimensional, Lie group.\\

Under the assumption that both moment maps are fiber bundles, we can approach Lie groupoids from an insular point of view. $\G^r$ is not open in the $WO$-topology so that we cannot talk about local maps defined on them. The solution to this problem is to consider local bisections. The jet spaces of local bisections are smooth manifolds and the general theory of ind-/pro-objects applies to them. Using local bisections, it is possible to define jet groupoids and to show that the operations involved in the bisection group are local.\\

Bisection groups act on insular manifolds. When the action is local (it is immediately insular) we talk about local Lie group actions. Using the machinery of insular differential operators, it is possible to encode the conditions of a local Lie group action in terms of the most lower map between finite dimensional manifolds.\\

{\it This chapter reviews the general notions of groupoid, Lie groupoid, bisection group, local bisections and jet groupoids. It shows that the bisection group is local, even insular. Moreover, using the techniques developed in previous parts, it reinterprets the result of Blohmann \cite{B} about local Lie group actions. The main references in this chapter are Blohmann \cite{B}, Moerdijk and Mr\v{c}un \cite{MM} and Yudilevich \cite{ORI}.}


\section{Groupoids and Lie groupoids}

{\it Lie groupoids are groupoids inner to the category of smooth manifolds satisfying some other conditions. Lie groupoids act on manifolds over its base in the same way that Lie groups act on manifolds. Given any Lie groupoid we can construct its associated group of bisections. In this section we review the relevant definitions available in the literature to talk about Lie groupoids. This is a very well known topic in the literature. All examples and definitions of this section could also be found in the book \cite{MM} by Moerdijk and Mr\v{c}un.}\\

Groupoids are objects that generalize the concept of groups. We can understand groups as a categories with only one element in which all morphisms are invertible. A morphism of groups is nothing more than a functor between two groups. This is an easy approach to the concept of groupoid.

\begin{df}[Groupoid] A groupoid $G$ is a small category in which all the arrows are invertible.
\end{df} 

In a groupoid, the set of objects will be denoted by $G_0$ and the set of morphisms by $G_1$. There are two maps, usually denoted by $r, l \colon G_1 \rightrightarrows G_0$ giving the source and the target objects of a morphism. The identity morphism can be thought as a map $1 \colon G_0 \longrightarrow G_1$. By constructing the pullback 
$$G_1 \times_{G_0} G_1 \defeq  \{ (f,g) \in \, G_1 \times G_1 \colon r(f) = l(g) \}$$
we can encode the composition by a map $\dia \colon G_1 \times_{G_0} G_1 \rightarrow G_1$ sending $(f, g)$ to $f \circ g$, which will be denoted by $f \dia g$ to avoid confusion with composition of other maps which are not elements of $G_1$. The inversion is simply a map $\iota \colon G_1 \rightarrow G_1$.

The maps $l$, $r$, $1$, $\dia$ and $\iota$ are called the {\it structure maps} of the groupoid. Explicitly, a pair of sets $G_0$ and $G_1$ together with maps $l$, $r$, $1$, $\dia$ and $\iota$ form a groupoid if and only if for all $f$, $g$, $h$ in $G_1$:
\begin{align*}
	\textrm{Composition} & \quad \textrm{If } l(g) = r(f) \textrm{ then } f \dia g \textrm{ exists.}\\
	\textrm{Compatibility} & \quad r(f \dia g) = r(g) \textrm{ and } l(f \dia g) = l(f).\\
	\textrm{Associativity} & \quad f \dia (g \dia h) = (f \dia g) \dia h.\\
	\textrm{Identity element} & \quad 1(l(g)) \dia g = g = g \dia 1(r(g)).\\
	\textrm{Inverse element} & \quad r(\iota g) = l(g), \, l(\iota g) = r(g), \, \iota(g) \dia g = 1(r(g)) \textrm{ and } g \dia \iota(g) = 1(l(g)).
\end{align*}

These are very similar to the group axioms, with the different that not all elements of $G_1$ can be composed (multiplied) only the ones with matching source and target. This is usually summarized as saying than the equations hold {\it whenever defined}.

\begin{rk}\label{aqui} There are several key properties that we would like to point out about groupoids. The maps $l$ and $r$ are surjective. It is easy to show from the axioms that $r(1(x)) = x = l(1(x))$ for every $x$ in $G_0$. Moreover, given any $x \in G_0$, the set of endomorphimsms of $x$ in the category has the structure of a group. A groupoid can be seen as a doubly fibered space where the identity is seen as a section of both $l$ and $r$ and in which every node is a group.

\begin{figure}[h]
	\centering
	\includegraphics[width=0.6\textwidth]{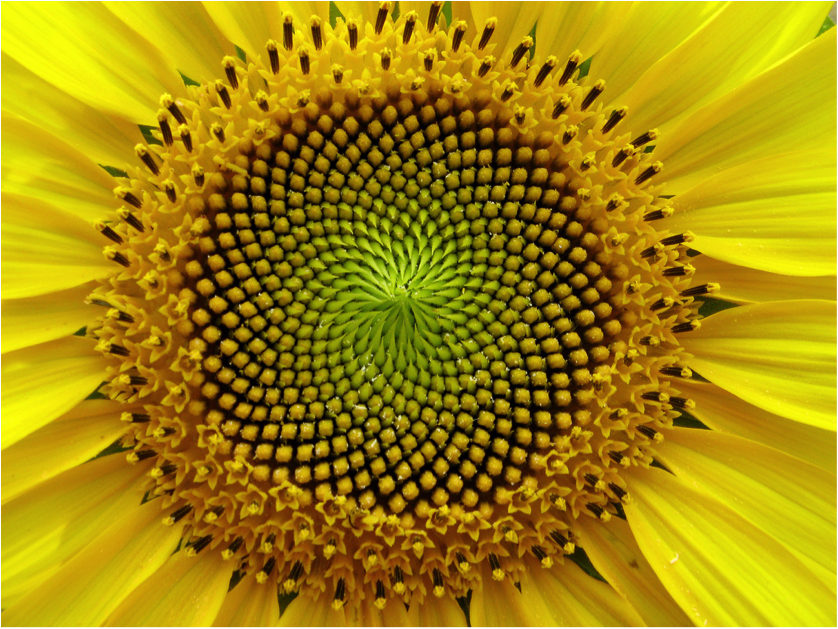}
	\caption{A Lie groupoid. The outer circle of the sunflower represents the identity morphism $M \rightarrow G$. The spirals arriving at that section with an angle smaller that $\frac{\pi}{2}$ are the $l$-sections, while the ones arriving with a greater angle are the $r$-sections. The intersection of two spirals is thick: it can be given the structure of a group modulo the choice of an identity element.}
	\label{fig:isl}
\end{figure}

\end{rk}

A Lie groupoid will be a groupoid in which $G_0$ is a smooth manifold, $G_1$ is a possibly non-Hausdorff, possibly non-second countable manifold such that $G_1 \times_{G_0} G_1$ is smooth and all the structure maps are smooth. The source and target maps are called {\it right} and {\it left} moment maps and denoted by $r$ and $l$ respectively. The condition $G_1 \times_{G_0} G_1$ being smooth is assured when $s$ is a submersion with Hausdorff fibers. It is possible to show that $s$ is a submersion if and only if $t$ is a submersion, if and only if $(s,t)$ is transverse to the diagonal map in $G_0$.

\begin{df}[Lie groupoid] A Lie groupoid is a groupoid $G_1 \rightrightarrows G_0$ in the category of non-Hausdorff, non-second countable smooth manifolds with moment maps ($l$ and $r$) which are submersions and such that $G_0$ is Hausdorff and second countable. A morphism of Lie groupoids is a smooth map that is a functor between the corresponding categories.
\end{df}

Morphisms of Lie groupoids respect all the structure maps. We will denote a morphism from $G_1 \rightrightarrows G_0$ to $H_1 \rightrightarrows H_0$ by the following diagram:

\begin{center}
\begin{tikzpicture}[description/.style={fill=white,inner sep=2pt}]
\matrix (m) [matrix of math nodes, row sep=3em,
column sep=3em, text height=1.5ex, text depth=0.25ex]
{ G_1 & H_1 \\
  G_0 & H_0 \\};
\path[->,font=\scriptsize]
(m-1-1) edge (m-1-2)
(m-2-1) edge (m-2-2);
\draw[transform canvas={xshift=0.5ex},->] (m-1-1) --(m-2-1) node[right,midway] {\scriptsize $r_G$};
\draw[transform canvas={xshift=-0.5ex},->] (m-1-1) --(m-2-1) node[left,midway] {\scriptsize $l_G$};
\draw[transform canvas={xshift=0.5ex},->] (m-1-2) --(m-2-2) node[right,midway] {\scriptsize $r_H$};
\draw[transform canvas={xshift=-0.5ex},->] (m-1-2) --(m-2-2) node[left,midway] {\scriptsize $l_H$}; 
\end{tikzpicture}
\end{center}

A Lie groupoid $G = (G_1 \rightrightarrows G_0)$ can act on manifolds over $G_0$, by this we mean on a smooth map: $l_E \colon E \rightarrow G_0$.

\begin{df}[Lie groupoid action]
A left action of a Lie Groupoid $G_1 \rightrightarrows G_0$ on manifold over $G_0$, $l_E \colon E \rightarrow G_0$ is a smooth map $\rho \colon G_1 \times_{G_0}^{r,l_E} E \rightarrow E$ denoted by $\rho(g,e) \defeq  g \cdot e$ such that $g^{\prime} \cdot (g \cdot e) = (g^{\prime} \dia g) \cdot e$ and $1(l_E(e)) \cdot e = e$ whenever defined.
\end{df}

Observe that we are using the notation explained prior to Remark \ref{aqui}, in which ``whenever defined'' means that if one of the sides of the equation is defined, so it is the other and they have the same values. In particular if $(g^{\prime} \dia g) \cdot e$ is defined, then $l(g) = r(g^{\prime})$. In that case $g^{\prime} \cdot (g \cdot e)$ should be also defined, so that $r(g^{\prime})=l_E(g \cdot e)$. We conclude that $l_E(g \cdot e) = l(g)$.

Our study so far has focused on spaces of smooth sections over fiber bundles. In the case of a groupoid, we have two candidates for a fiber bundle over $G_0$: the left and the right moment maps. The concept of a bisection encodes the idea of sections of both moment maps:

\begin{df}[Bisection group]\label{bg} The bisection group of a Lie groupoid $G = \left( G_1 \rightrightarrows G_0 \right)$ is given by $$\G^r \defeq  \{ \sigma \colon G_0 \rightarrow G_1 \colon r \circ \sigma = \textrm{id}_{G_0} \textrm{ and } l \circ \sigma \in \, \textrm{Diff}(G_0) \}.$$ The product of two bisections $\sigma$ and $\sigma^{\prime}$ in $\G^r$ is given at any $x$ of $G_0$ by:
$$\left(\sigma \sigma^{\prime} \right)(x) \defeq  \sigma \left( l \sigma^{\prime} (x) \right) \dia \sigma^{\prime}(x).$$
Sometimes we will use the dot notation $\sigma \cdot \sigma^{\prime}$ in case there is risk of confusion. The identity bisection is $1 \colon G_0 \rightarrow G_1$.
The inverse map $(-)^{-1} \colon \G^r \rightarrow \G^r$ is given by:
$$\sigma^{-1}(x) \defeq  \iota \sigma \left( (l \sigma)^{-1}(x) \right).$$
\end{df}

The notation chosen for the bisection group, $\G^r$, is so that it is easy to remember that it is a subset of sections of $\G \defeq  \Gamma^{\infty}(G_0,G_1,r)$ where we have included the map $r$ to clarify with respect to which map the sections are taken. It is not only a subset but also a topological subspace, moreover:

\begin{pp}\label{lie} The bisection group of a Lie groupoid is a Lie group.
\end{pp}

\dem
In the first place, $\G^r$ is a smooth open submanifold of $\C(G_0,G_1)$. This is true because $l_* \colon \C(G_0,G_1) \rightarrow \C(G_0,G_0)$ is continuous and $\mathrm{Diff}(G_0)$ is an open submanifold of $\C(G_0,G_0)$ (see the book of Kriegl and Michor, \cite[Theorem 43.1]{KM}).

Since $r (\sigma \left( l \sigma^{\prime} (x) \right)) = l \sigma^{\prime} (x)$ we conclude that $\sigma \left( l \sigma^{\prime} (x) \right)$ and $\sigma^{\prime}(x)$ are composable. The product is well defined and smooth. Similar calculations will show that $\G^r$ is indeed a Lie group with respect to that smooth structure (associativity and inversion can be checked).

As an example, let us compute $\sigma \sigma^{-1}$ (note that $l \iota \sigma = r \sigma = \textrm{id}_m$):
\begin{eqnarray*}
	\sigma \sigma^{-1} (x) & = & \sigma (l \sigma^{-1} (x)) \dia \sigma^{-1}(x) \\
	& = & \sigma (l \iota \sigma ((l \sigma)^{-1}(x))) \dia \iota\sigma((l\sigma)^{-1}(x)) \\
	& = & \sigma ((l \sigma)^{-1}(x)) \dia \iota\sigma((l\sigma)^{-1}(x)) \\
	& = & 1 (l \sigma (l \sigma^{-1} (x))) = 1(x).
\end{eqnarray*}\qed

\subsection{Examples}

There are two basic examples: the pair groupoid and Lie groups.

\begin{ej}[The pair groupoid]
Given a smooth manifold $M$ consider the Lie groupoid $M \times M \rightrightarrows M$ given by left and right projections as moment maps. The composition is given by forgetting a common middle element: $(x,y) \dia (y,z) = (x,z)$. This is called the pair groupoid associated to $M$.

Given a smooth map $l_E \colon E \rightarrow M \neq \emptyset$ consider an action of the pair groupoid $M \times M \rightrightarrows M$ on it. We have that $(M \times M) \times_M E$ is simply $M \times E$. The groupoid action conditions translate into $l_E(e) \cdot e = e$ and $x \cdot (x^{\prime} \cdot e) = x \cdot e$. Fix $x \in \, M$ and consider the following pair of maps:
	\begin{eqnarray*}
		M \times l_E^{-1}(x) & \longleftrightarrow & E \\
		(y,e) & \longmapsto & y \cdot e\\
		(l_E(f), x \cdot f) & \longleftmapsto & f
	\end{eqnarray*}	
	The map to the right is well defined because $l_E(x \cdot f)=x$ (this follows from the observation made after the definition of an action). Both maps are smooth, and they are actually inverse of one another: $l_E(f) \cdot (x \cdot f) = l_E(f) \cdot f = f$ and on the other direction for $e \in l_E^{-1}(x)$, $(l_E(y \cdot e), x \cdot (y \cdot e)) = (y, x \cdot e) = (y, l_E(e) \cdot e) = (y, e)$.\\
	This is then a trivialization of $l_E \colon E \rightarrow M$. As a matter of fact, any trivialization $\Phi \colon M \times F \rightarrow E$ will give a unique Lie groupoid action: $x \cdot e \defeq  \Phi(x, \textrm{pr}_F \Phi^{-1}(e))$. Checking the three axioms is trivial: 
	\begin{align*}
		l_E(x \cdot e) & = \textrm{pr}_M \Phi^{-1}(x \cdot e) = \textrm{pr}_M (x, \textrm{pr}_F \Phi^{-1}(e)) = x.\\
		l_E(e) \cdot e & = \textrm{pr}_X(\Phi^{-1}(e)) \cdot e = \Phi(\textrm{pr}_M(\Phi^{-1}(e)), \textrm{pr}_F(\Phi^{-1}(e))) = e. \\
		x \cdot (x^{\prime} \cdot e) & = x \cdot (\Phi(x^{\prime}, \textrm{pr}_F \Phi^{-1}(e))) = \Phi\left(x, \textrm{pr}_F \left( \Phi^{-1} \circ \Phi \left(x^{\prime}, \textrm{pr}_F \Phi^{-1}(e) \right) \right) \right) = \\
		& = \Phi(x, \textrm{pr}_F \Phi^{-1}(e)) = x \cdot e.
 	\end{align*}

The bisection group in this case is $\Gamma^{\infty}(M,  M \times M, r)^{r} \cong \mathrm{Diff}(M)$, the group of diffeomorphisms of $M$. Any map $M \rightarrow M \times M$ which is a section of the right moment map is of the kind $\sigma = (s, \textrm{id}_M)$, where $s \in \, \C(M, M)$. The fact that $\sigma$ is a bisection amounts to saying that $s$ is a diffeomorphism.
\end{ej}

\begin{ej}[Lie group]
Any Lie group $G$ can be seen as a Lie groupoid over a point: $G \rightrightarrows \{*\}$ with composition given by the group product. We denote this groupoid again by $G$.

Lie groups act on fiber bundles over a point, that is on manifolds. In this case $G \times_{ \{*\} } M = G \times M$ and a Lie groupoid action is nothing else but a left group action on the manifold $M$.

The bisection group is again the original group: $\G^r \cong G$. A smooth map $\{*\} \rightarrow G$ is just a point $g \in \, G$. Any such a map is a section of the right moment map and gives a diffeomorphism of $\{*\}$ when pulled-back along the right moment map. Even more, $\left(g g^{\prime} \right)(\{*\}) \defeq  g \left( l g^{\prime} (\{*\}) \right) \dia g^{\prime}(\{*\}) = g \cdot g^{\prime}$ where $\cdot$ denotes the group multiplication.
\end{ej}

By looking at the two previous examples we can think that a general Lie groupoid action is something between a flat connection and the action of some Lie group.

\begin{ej}\label{actionbabe}
Given any Lie groupoid $G_1 \rightrightarrows G_0$ we get a homomorphism of Lie groupoids to the pair groupoid $G_0 \times G_0 \rightrightarrows G_0$ given by $(l,r)$.

\begin{center}
\begin{tikzpicture}[description/.style={fill=white,inner sep=2pt}]
\matrix (m) [matrix of math nodes, row sep=3em,
column sep=3em, text height=1.5ex, text depth=0.25ex]
{ G_1 & G_0 \times G_0 \\
  G_0 & G_0 \\};
\path[->,font=\scriptsize]
(m-1-1) edge node[auto] {$(l, r)$} (m-1-2)
(m-2-1) edge node[auto] {id} (m-2-2);
\draw[transform canvas={xshift=0.5ex},->] (m-1-1) --(m-2-1) node[right,midway] {\scriptsize $r$};
\draw[transform canvas={xshift=-0.5ex},->] (m-1-1) --(m-2-1) node[left,midway] {\scriptsize $l$};
\draw[transform canvas={xshift=0.5ex},->] (m-1-2) --(m-2-2) node[right,midway] {\scriptsize $\textrm{pr}_r$};
\draw[transform canvas={xshift=-0.5ex},->] (m-1-2) --(m-2-2) node[left,midway] {\scriptsize $\textrm{pr}_l$}; 
\end{tikzpicture}
\end{center}

The induced map on bisections $l_* \colon \G^r \rightarrow \textrm{Diff}(G_0) \subset \Gamma(G_0, G_0 \times G_0)$ is a group homomorphism (as a matter of fact it is a Lie group homomorphism with the smooth structure from Proposition \ref{lie}). The fact that it is a group homomorphism tells us for instance that $l_*(1)(x)=l1(x)=x$ so that $1$ is sent to $\textrm{id}_{G_0}$.

Yet another notation for the map $l_*$: the evaluation at a point is going to be represented simply by a dot (for each $x \in G_0$ and every $\sigma$ in $\G^r$):
\begin{equation}\label{moruna}
\sigma \cdot x \defeq  l_*(\sigma)(x) = (l \circ \sigma)(x).
\end{equation}
\end{ej}

\begin{ej}[The action groupoid]
Given an action of a Lie group $G$ on a manifold $M$ we can consider the action groupoid $G \times M \rightrightarrows M$ with right moment map being the action and left moment map the projection to $M$. The inverse of $(g, x) \in G \times M$ is $(g^{-1}, g x)$, the identity $1(x)= (e, x)$ and the composition $(g, x) \dia (h, g \cdot x) = (h g, x)$ for all $x \in M$ and all $g, h \in G$. 
\end{ej}


\section{Local bisections}\label{mps}

{\it The Lie group of bisections of a Lie groupoid is not an open subset of $\G$ in the $WO$-topology. This represents a challenge when working with Lie groupoids and local maps. The solution is to consider the infinite jet bundle of local bisections. The main references in this section are Moerdijk and Mr\v{c}un \cite{MM} and Yudilevich \cite{ORI}.}\\

In this section we would like to start linking the world of Lie groupoids and the one of local maps. We already know that the way to do this is to consider the Lie group of bisections as pointed out in Definition \ref{bg}. In the world of local manifolds we always start by considering fiber bundles $G \rightarrow M$, which in particular are submersions. We would like to consider the case in which the previous map is part of a Lie groupoid $G \rightrightarrows M$ as the right moment map. In this way bisections will be denoted by $\varphi$ instead of $\sigma$ to match the general notation for insular manifolds. The fact that the right moment map is a fiber bundle implies that the left one is also a fiber bundle. This is so because of the compatibility relations among the structure maps, in particular between the moment maps and the inverse. \color{black}

\begin{df}[Fibered Lie groupoid]
A Lie groupoid is called fibered if both its moment maps are fiber bundles.
\end{df}

We are going to work with such fibered Lie groupoids $G \rightrightarrows M$. In the following section we will show that all its structure maps fit in diagrams of a local map. For example, it can be seen that the left moment map is part of a differential operator of order $0$:

\begin{center}
\begin{tikzpicture}[description/.style={fill=white,inner sep=2pt}]
\matrix (m) [matrix of math nodes, row sep=3em,
column sep=3.5em, text height=1.5ex, text depth=0.25ex]
{ \G^r \times M & \{\textrm{id}_M\} \times M \\
  G  & M \\};
\path[->,font=\scriptsize]
(m-1-1) edge node[auto] {$( *, \cdot )$} (m-1-2)
(m-1-1) edge node[auto] {$j^0$} (m-2-1)
(m-2-1) edge node[auto] {$l$} (m-2-2)
(m-1-2) edge node[auto] {$\textrm{pr}_M$} (m-2-2);
\end{tikzpicture}
\end{center}

Recall the space of sections of $\textrm{id}_M \colon M \rightarrow M$ is given just by $\{\textrm{id}\}$. The map $*$ is the unique map to $\{\textrm{id}_M\}$. We are using the notation $\cdot$ for the action of a bisection on a point as in Equation \ref{moruna}.

As usual, we should deal with the problem of the finite jet evaluations not being surjective. In this case the problem is even more serious since we are not even working with $\G$, but with the subset of bisections $\G^r$. We have studied in Part 3 ways of getting around this problem by replacing $J^k G$ by $j^k (\mathcal{W})$ for any $\mathcal{W}$ open in $\E \times M$. In Proposition \ref{lie} we learned that the bisection group is an open submanifold of $\G$. Now we could try to replace $J^k G$ by $j^k (\G^r \times M)$ and get back on track. However, this does not work. Proposition \ref{lie} states that $\G^r$ is an open {\it submanifold} of $\G$, so that it is an open subset with respect to the locally convex topology. Michor \cite[Corollary 5.7]{MIC} shows that $\textrm{Diff}(M)$ is open with respect to the $WO^1$-topology (which using the same argument as in Proposition \ref{lie}, will allow us to show that $\G^r$ is open in $\G$ with respect to the $WO^1$-topology). But the jet evaluation $j^k$ is not open with respect to those topologies, it is only open with respect to the $WO$-topology. In general, $\G^r$ is not open in that topology  and the argument fails.

We seem to have reached a dead end. Let us focus on the problem of extending the differential operator above to a local map. In order to do so we need to be able to take jet prolongations of $l$. That would require $l \circ \varphi$ to be invertible for any $j_x^k \varphi$ in some open set of $J^k G$ containing $j^k (\G^r \times M)$. The word {\it containing} is the relevant one here. In order for the jet prolongations to exist we do not need $j^k (\G^r \times M)$ to be open, but only to be inside of an open set where all $l\circ \varphi$ are invertible (for any $j_x^k \varphi$, $x \in M$).

Moerdijk and Mr\v{c}un \cite{MM} define a Lie groupoid of germs of bisections, where a local bisection is a local section of $r$ such that the pullback via $l$ is an open embedding. We have decided to treat this notion using a different language. We want to introduce at this point the monopresheaf of local diffeomorphisms: let $U$ be open in $M$ and define
$$\D(U) \defeq  \{ f \colon U \longrightarrow M | f \colon U \longrightarrow f(U) \textrm{ is a diffeomorphism} \}.$$

In other words, we consider embeddings of $U$ in $M$. Observe that the usual restriction map from $U$ to $V \subset U$ sends $\D(U)$ to $\D(V)$. As a matter of fact, $\D$ is a monopresheaf. This means that $\D$ is a presheaf in which the locality axiom is satisfied, but not the gluing axiom.

\begin{df}[Monopresheaf of local bisections] Given a fibered Lie groupoid $G \rightrightarrows M$, the monopresheaf of local bisections of $G$ is given by associating to every $U \subset M$ open set the following:
$$\G^r(U) \defeq  \{ \varphi \in \G(U) \colon l \circ \varphi \colon U \longrightarrow (l \circ \varphi)(U) \textrm{ is a diffeomorphism} \}.$$
\end{df}

Observe that this notation is compatible with that of the bisection group, that is $\G^r(M) = \G^r$ as defined in Definition \ref{bg}. We could define germs of $\G^r$ (as a monopresheaf) the same way it is done for sheaves. It is also possible to construct the \'etal\'e bundle associated to it by considering the union of all the germs. By taking sections of this bundle we do not recover the original monopresheaf. In oder words, the big difference with the sheaf case \color{black} is that there is no adjunction with the functor ``taking-sections-of'' since $\G^r$ is indeed not a sheaf. We denote the space of germs as usual by $\widehat{\G^r}$. As mentioned above, Moerdijk and Mr\v{c}un \cite{MM} talk about the Lie groupoid of germs of local bisections. We are more interested in its quotient by the equivalence relation of having the same $k$-th jet. This can be found for instance \cite{ORI}, the thesis of Yudilevich, where he also states that $\widehat{\G^r}$ is a possibly non-Hausdorff smooth manifold:

\begin{pp}[Yudilevich, {\cite[Section 2.2]{ORI}}]\label{openyo} Let $G \rightarrow M$ be a fibered groupoid. For every $k$ natural number denote the equivalence relation of having the same $k$-th jet by $\sim^k$. Then
$$J^k G^r \defeq \bigslant{\widehat{\G^r}}{\sim^k}$$
is an open submanifold of $J^k G$ and $G = J^0 \G^r$. For all $k \geqslant 1$ the map $\restrict{r_k^{k-1}}{J^k G^r}$ takes values in $J^{k-1} G^r$ and it is an affine bundle for all $k \neq 1$.
\end{pp}

We will simply denote $\restrict{r_k^{k-1}}{J^k G^r}$ by $r_k^{k-1}$. Observe that $r_1^0  \colon J^1 G^r \rightarrow G$ is not an affine bundle in general. Now we can define the corresponding pro-smooth manifold:

\begin{df} Let $G \rightarrow M$ be a fibered groupoid, the infinite jet bundle of local bisections, denoted by $J^{\infty} G^r$ is the pro-smooth manifold
$$\cdots \rightarrow J^k G^r \rightarrow J^{k-1} G^R \rightarrow \cdots \rightarrow J^1 G^r \rightarrow G.$$
\end{df}

It is important to keep in mind that $j^k (\G^r \times M) \neq J^k G^r$. And the reason is not only $\G$ not being soft as a sheaf. As a matter of fact, $j^k (\G^r \times M) \neq J^k G^r \cap j^k(\G \times M)$ \cite{ORI} (if the equality would hold, $\G^r \times M$ would be $WO$-open and we know that is not the case in general). In order to be able to talk about jet bundles associated to Lie groupoids we have had to relax our expectations about $j^k$ being surjective, since that does not hold in general.

\begin{rk}\label{estrella}
Observe that there are local bisections that do not extend to global ones even for the pair groupoid $G = M \times M \rightrightarrows M$. Consider $[(\tau, x)] \in \, J^k (M \times M)^r$ represented by a local diffeomorphism $\tau$ sending $x$ to $x$ and reversing a given orientation in some chart around $x$. $\tau$ can only be extended to a global diffeomorphism of $M$ if $M$ admits a global orientation-reversing map (assume $M$ is connected). But not every manifold has that property, for instance all $4k$-dimensional manifolds with odd ($2k$)-th Betti number admit no self-map of degree $-1$ (M\"ullner \cite[Corollary 5]{MUE}). 

In the case of the pair groupoid we could think of replacing $\G^r = \D$ by $\D^0$, where:
$$\D^0(U) \defeq  \{ f \colon U \rightarrow M \colon f \colon U \rightarrow f(U) \textrm{ is an orientation preserving diffeomorphism} \}.$$

This might be too restrictive since $M$ might have global orientation reversing diffeomorphisms. We could then work with $\D^0$ or $\D$ depending on $M$. We would be tempted to say that this solves all the problems for the pair groupoid. But for the general case, the bundle structure gives more obstructions:

Consider $G \rightrightarrows M$ a fibered Lie groupoid with soft sheaf of sections $\G$. Consider any local section $\varphi$ of $r \colon G \rightarrow M$ which induces an orientation preserving local diffeomorphism $\tau \defeq l \circ \varphi$ of the base $M$. Since $\G$ is soft we can define $\tau$ globally by extending $\varphi$.  Since the manifold $M$ is connected, the group of orientation preserving diffeomorphisms (the connected component of the identity in $\textrm{Diff}(M)$) acts transitively on $M$. That means that  we can find a global orientation preserving diffeomorphism $\sigma$ sending $\tau(x)$ to $x$. The map $\sigma \circ \tau$ is a local orientation preserving diffeomorphism fixing $x$. The idea is now that the boundary of a ball around $x$ is mapped (preserving the orientation) to some manifold diffeomorphic to $\mathbb{S}^{n-1}$ (where $n$ is the dimension of $M$). We can find an isotopy between that map and the identity and use it to extend $\sigma \circ \tau$ locally to a map that agrees eventually with the identity starting from the boundary of some ball around $x$. It is possible to extend, thus, $\sigma \circ \tau$ to a global orientation preserving diffeomorphism of $M$, call it $\iota$. The map $g^{-1} \circ \iota$ has the same germ as $\tau$ around $x$ and it is a global diffeomorphism. If we were now able to invert $l$ we could find a global bisection extending $\varphi$. This restricts our study to the case in which $l$ is a local diffeomorphism. These groupoids are called {\it \'etale}. The groupoid of local bisections is \'etale, but it is not fibered. That is definitely something we are not willing to give up in our study.

For more about \'etale Lie groupoids we refer to Moerdijk and Mr\v{c}un \cite{MM} or to Yudilevich \cite{ORI}. We will come back to the relation between local maps involving Lie groupoids and the concept of effective \'etale Lie groupoids from Yudilevich at a later point in this thesis (Section \ref{lpg}).
\end{rk}


\section{Jet groupoids and bisection groupoid}

{\it Lie groupoids whose moment maps are smooth fiber bundles can be treated in a local way. Given such a Lie groupoid it is possible to construct associated Lie groupoids on each finite jet bundle. The bisection group becomes local using these groupoids, and it is even possible to construct the local bisection groupoid as an action groupoid. The main reference is Yudilevich's thesis \cite{ORI}.}\\

Let $G \rightrightarrows M$ be a fibered groupoid. All the structure maps of its bisection group can be written as local maps as we will see in this section. We will start by constructing the infinite jet prolongations of all structure maps of $G$, and these will serve as the associated maps in the infinite jet bundle for the structure maps on $\G^r$. The two moment maps $l$ and $r$, together with the identity $1$ can be seen to be insular differential operators of order $0$ along the identity. The inversion is also an insular differential operator of order $0$.

\begin{center}
\begin{tikzpicture}[description/.style={fill=white,inner sep=2pt}]
\matrix (m) [matrix of math nodes, row sep=3em,
column sep=3.5em, text height=1.5ex, text depth=0.25ex]
{ \G^r \times M & \{\textrm{id}_M\} \times M & & \G^r \times M & \{\textrm{id}\} \times M\\
  G  & M && G  & M \\};
\path[->,font=\scriptsize]
(m-1-1) edge node[auto] {$( *, \cdot )$} (m-1-2)
(m-1-1) edge node[auto] {$j^0$} (m-2-1)
(m-2-1) edge node[auto] {$l$} (m-2-2)
(m-1-2) edge node[auto] {$\textrm{pr}_M$} (m-2-2)
(m-1-4) edge node[auto] {$( *, \textrm{id} )$} (m-1-5)
(m-1-4) edge node[auto] {$j^0$} (m-2-4)
(m-2-4) edge node[auto] {$r$} (m-2-5)
(m-1-5) edge node[auto] {$\textrm{pr}_M$} (m-2-5);
\end{tikzpicture}
\end{center}

We are using the notation $\cdot$ for the action of a bisection on a point as in Equation \ref{moruna}. For the identity element, the maps go in the other direction. For the inversion, we do not get an insular differential operator along the identity, but simply an insular operator:
\begin{center}
\begin{tikzpicture}[description/.style={fill=white,inner sep=2pt}]
\matrix (m) [matrix of math nodes, row sep=3em,
column sep=4.5em, text height=1.5ex, text depth=0.25ex]
{ \{\textrm{id}\} \times M & \G^r \times M & & \G^r \times M & \G^r \times M \\
  M  & G & & G & G \\};
\path[->,font=\scriptsize]
(m-1-1) edge node[auto] {$( 1, \textrm{id} ) \circ \textrm{pr}_M$} (m-1-2)
(m-1-1) edge node[auto] {$\textrm{pr}_M$} (m-2-1)
(m-2-1) edge node[auto] {$1$} (m-2-2)
(m-1-2) edge node[auto] {$j^0$} (m-2-2)

(m-1-4) edge node[auto] {$((-)^{-1}, \cdot)$} (m-1-5)
(m-2-4) edge node[auto] {$\iota$} (m-2-5)
(m-1-4) edge node[auto] {$j^0$} (m-2-4)
(m-1-5) edge node[auto] {$j^0$} (m-2-5);
\end{tikzpicture}
\end{center}

All jet prolongations exist as we are working with insular differential operators. It is important to observe that the prolongations only exists in $J^k G^r$ and not in $J^k G$ in general, since the conditions for insularity do not hold for $[(\varphi, x)]$ with $\varphi$ a local section which is not a bisection. We get the following explicit formulas (observe that $j^k r = r^k$):
\begin{eqnarray}\label{jg1}
	j^k l \colon J^k G^r & \longrightarrow & M \\
	{[} (\varphi, x) {]} & \longmapsto & (l \varphi) (x) \nonumber
\end{eqnarray}
\begin{eqnarray}\label{jg2}
	r_k = j^k r \colon J^k G^r & \longrightarrow & M \\
	{[}(\varphi, x){]} & \longmapsto &x \nonumber
\end{eqnarray}
\begin{eqnarray}\label{jg3}
	j^k 1 \colon M & \longrightarrow &J^k G^r \\
	m & \longmapsto & {[}(1,m){]}. \nonumber
\end{eqnarray}
\begin{eqnarray}\label{jg4}
	j^k \iota \colon J^k G^r & \longrightarrow & J^k G^r \\
	{[}(\varphi, x){]} & \longmapsto & {[}( \iota \varphi (l \varphi)^{-1}, (l \varphi)x ){]}. \nonumber
\end{eqnarray}

The composition is more problematic. First of all there is the issue of working with two bundles. Observe that $G \times_M G$ is the fiber product of $l \colon G \rightarrow M$ and $r \colon G \rightarrow M$. Since there are two different maps, we will specify the maps in the pullbacks. Using this notation the domain of the composition is $G \times_M^{r,l} G$.  Moreover, we will distinguish the jet evaluation and the jet bundles over $l$ by adding subscript $l$: $j_l^k$ and $J_l^k G$ to avoid confusion. A second issue is that $J^k (G \times_M^{r,l} G)$ and $J^k G \times_M^{j^k r, j^k l} J^k G$ are not the same space a priori.

To start addressing this problem, we observe that we could have chosen the other moment map to define the bisection group, in that case we were to consider the following set:
$$\G^{l} \defeq  \{ \psi \colon G \rightarrow M \colon l \circ \psi = \textrm{id}_{M} \textrm{ and } r \circ \psi \in \, \textrm{Diff}(M) \}.$$ 

We can extend the definition to a monopresheaf and to the jet bundle of local bisections with respect to $l$, denoted by $J^k G^l$. We can see that $\G^l$ and $\G^r$ are in one-to-one correspondence:
\begin{eqnarray*}
	RL \colon \G^r & \longleftrightarrow & \G^l \colon LR \\
	\varphi & \longmapsto & \varphi (l \varphi)^{-1} \\
	\psi (r \psi)^{-1} & \longleftmapsto & \psi
\end{eqnarray*}

As a matter of fact, $RL$ fits into is a degree $0$ differential operator:
\begin{center}
\begin{tikzpicture}[description/.style={fill=white,inner sep=2pt}]
\matrix (m) [matrix of math nodes, row sep=3em,
column sep=3.5em, text height=1.5ex, text depth=0.25ex]
{ \G^r \times M & \G^l \times M  & & (\varphi, x) & (\varphi(l\varphi)^{-1}, l\varphi(x))\\
  G  & G & & \varphi(x) & \varphi(x) \\};
\path[->,font=\scriptsize]
(m-1-1) edge node[auto] {$(RL, \cdot)$} (m-1-2)
(m-1-1) edge node[auto] {$j^{0}$} (m-2-1)
(m-2-1) edge node[auto] {id} (m-2-2)
(m-1-2) edge node[auto] {$j_l^0$} (m-2-2);
\path[|->]
(m-1-4) edge (m-1-5)
(m-1-4) edge (m-2-4)
(m-2-4) edge (m-2-5)
(m-1-5) edge (m-2-5);
\end{tikzpicture}
\end{center}

This map is once again insular (along $J^{\infty} G^r$), so that we can construct, for every $k$, the jet prolongations:
\begin{eqnarray*}
	j^k \textrm{id} \colon J^k G^r & \longrightarrow & J^k G^l \\
	{[}(\varphi, x){]} & \longmapsto & {[}( \varphi (l \varphi)^{-1}, l\varphi (x) ){]}
\end{eqnarray*}

Using the maps above, we can build a diagram as follows:

\begin{center}
\begin{tikzpicture}[description/.style={fill=white,inner sep=2pt}]
\matrix (m) [matrix of math nodes, row sep=3em,
column sep=3.5em, text height=1.5ex, text depth=0.25ex]
{ \G^r \times M & \G^l \times M  & & (\varphi, x) & (\varphi(l\varphi)^{-1}, l\varphi(x))\\
  J^k G^r  & J^k G^l & & \mbox{$[( \varphi, x )]_r$} &  \mbox{$[( \varphi(l\varphi)^{-1}, (l \varphi)x )]_l$} \\
	M & M & & l\varphi(x) & l\varphi(x) \\};
\path[->,font=\scriptsize]
(m-1-1) edge node[auto] {$(RL, \cdot)$} (m-1-2)
(m-1-1) edge node[auto] {$j^{k}$} (m-2-1)
(m-2-1) edge node[auto] {$j^k id$} (m-2-2)
(m-1-2) edge node[auto] {$j_l^k$} (m-2-2)
(m-2-1) edge node[auto] {$j^k l$} (m-3-1)
(m-2-2) edge node[auto] {$l_k$} (m-3-2)
(m-3-1) edge node[auto] {id} (m-3-2);
\path[|->]
(m-1-4) edge (m-1-5)
(m-1-4) edge (m-2-4)
(m-2-4) edge (m-2-5)
(m-1-5) edge (m-2-5)
(m-2-4) edge (m-3-4)
(m-3-4) edge (m-3-5)
(m-2-5) edge (m-3-5);
\end{tikzpicture}
\end{center}

The top square just comes from the fact that we are taking the jet prolongation of the pair $\left( (RL, \cdot), \textrm{id} \right)$, while the bottom square compares the projections $j^k l$ and $l_k$. This second square is crucial in order to understand the associated diffeomorphism we get for the pullbacks involving the jet prolongations of $r$ and $l$ and the one involving $r_k$ and $l_k$:
\begin{eqnarray}\label{pbs}
	\textrm{id}_{J^k G^r} \times_M j^k \textrm{id} \colon J^k G^r \times_M^{j^k r, j^k l} J^k G^l & \longrightarrow & J^k G^r \times_M^{r_k, l_k} J^k G^r \\ 
	 {[}(\varphi, x){]}, \, {[}(\varphi^{\prime},x){]} & \longmapsto & {[}(\varphi, x){]}, \, {[}( \varphi^{\prime}(l\varphi^{\prime})^{-1}, \left( l \varphi^{\prime} \right) x ){]} . \nonumber
\end{eqnarray}

We are going through all this comparisons to be able to construct the jet prolongation of the composition in $G$. Denote the multiplication among bisections by $m$, now consider the following map:

\begin{center}
\begin{tikzpicture}[description/.style={fill=white,inner sep=2pt}]
\matrix (m) [matrix of math nodes, row sep=3em,
column sep=2.5em, text height=1.5ex, text depth=0.25ex]
{ \G^r \times \G^l \times M & \G^r \times M  & & (\varphi, \psi, x) & (\varphi \cdot \psi(r \psi)^{-1}, (r \psi) x )\\
  G \times_M^{r,l} G  & G & & \left( \varphi(x), \psi(x) \right) & \varphi(x) \dia \psi(x) \\};
\path[->,font=\scriptsize]
(m-1-1) edge node[above=5] {$(m(\textrm{pr}_1, LR \circ \textrm{pr}_2), \cdot_l )$} (m-1-2)
(m-2-1) edge node[auto] {$\dia$} (m-2-2)
(m-1-1) edge node[auto] {$j^0 \times_M j_l^0$} (m-2-1)
(m-1-2) edge node[auto] {$j^0$} (m-2-2);
\path[|->]
(m-1-4) edge (m-1-5)
(m-1-4) edge (m-2-4)
(m-2-4) edge (m-2-5)
(m-1-5) edge (m-2-5);
\end{tikzpicture}
\end{center}

Observe that at this point we are using no symbol for composition, the dot for the bisection product and the diamond as the groupoid composition. \color{black} The diagram is commutative because 
$$\varphi \cdot \psi(r \psi)^{-1} (r \psi (x)) = \varphi (l \psi (r\psi)^{-1} (r\psi(x)) ) \dia \psi(r\psi)^{-1} (r\psi(x)) = \varphi (x) \dia \psi(x).$$

Once again, the map above is an insular differential operator of order $0$ (along $J^{\infty}(G^r \times_M G^l)$) and we can take jet prolongations of $\dia$. 
\begin{align*}
	j^k \dia \colon J^k G^r \times_M^{r_k, l_k} J^k G^l & \longrightarrow J^k G^r \\
	\left( [(\varphi, x)],[(\psi,x)] \right) & \longmapsto [( (\varphi \dia \psi) (r\psi)^{-1}  , r\psi(x))].
\end{align*}

We are going to use the isomorphism between the pullbacks of the jet bundles as in Equation \ref{pbs} to construct the following map:
\begin{eqnarray}\label{jg5}
	j^k \blacklozenge \colon J^k G^r \times_M^{j^k r, j^k l} J^k G^r & \longrightarrow & J^k G^r \\
	{[}(\varphi, x){]}, \, {[}(\varphi^{\prime},x) {]} & \longmapsto & {[}( (\varphi \dia \varphi^{\prime} (l\varphi^{\prime})^{-1}) (l\varphi^{\prime})  , (l\varphi^{\prime})^{-1} (x)){]}. \nonumber
\end{eqnarray}


\subsection{Definitions}

We have constructed something that resembles a Lie groupoid $J^k G^r \rightrightarrows M$: we have moment maps $j^k r, j^k l \colon J^k G^r \rightarrow M$ a composition $j^k \blacklozenge \colon J^k G^r \times_M^{j^k r, j^k l} J^k G^r$, an identity $j^k 1 \colon M \rightarrow J^k G^r$, and an inverse map $j^k \iota \colon J^k G^r \rightarrow J^k G^r$. As a matter of fact, it is indeed a Lie groupoid. 

\begin{dfpp}[$k$-th jet groupoid, following Yudilevich {\cite{ORI}}]\label{rfg} Let $G \rightrightarrows M$ be a fibered Lie groupoid. Then for every non-negative integer $k$ the maps \ref{jg1}, \ref{jg2}, \ref{jg3}, \ref{jg4} and \ref{jg5} define the structure of a fibered Lie groupoid on $J^k G^r \rightrightarrows M$ called the $k$-th jet groupoid.
\end{dfpp}

\dem
The collection of each structure map for different $k$'s is an infinite jet prolongation of an insular differential operators of order $0$. That means that the composition is well defined (the fact that pro-categories are categories) and also that the compositions keep the jet degree fixed. The Lie groupoid axioms are trivially satisfied. As an example we check one of the identities concerning the inverse element axiom:
$$ j^k r j^k \iota = j^k (r \iota) = j^k l.$$
The Lie groupoid we obtain in this way is fibered since $j^k r = r_k$ and $j^k l = l_k \circ j^k \textrm{id}_G$ are fiber bundles. 
\qed

It follows from the proof of the previous Definition/Proposition that the infinite jet prolongation of the structure maps of a fibered Lie groupoid is again a groupoid. As a matter of fact it is a groupoid inner to the category of pro-Lie groupoids since $r_k^{k^{\prime}} \colon J^k G^r \rightarrow J^{k^{\prime}} G^{r}$ is a morphism of Lie groupoids (see Yudilevich \cite{ORI} for example). The most relevant bit of information is that by looking at the equations \ref{jg3}, \ref{jg4} and \ref{jg5} we can see that they agree with the equations of the bisection group.

\begin{df}[Local Lie groupoid]
Let $r \colon G \rightarrow M$ be a smooth fiber bundle and let $\mathcal{H} \subset \G \defeq \Gamma(M, G)$ be an open submanifold of $\G$. $\mathcal{H} \times M \rightrightarrows N$ is said to be a local Lie groupoid if it is a Lie groupoid and all its structure maps are local. If the Lie groupoid is a Lie group the local Lie groupoid is called a local Lie group.
\end{df}

\begin{pp}\label{fibras} The bisection group of a fibered Lie groupoid is a local Lie group. A morphism of fibered Lie groupoids induces a local map between the bisection groups.
\end{pp}

\dem
This is an immediate consequence of Definition/Proposition \ref{rfg}. We want to be explicit about how to see the multiplication as a local map: we need to use the isomorphism given by $(RL, j^k \textrm{id})$:
\begin{center}
\begin{tikzpicture}[description/.style={fill=white,inner sep=2pt}]
\matrix (m) [matrix of math nodes, row sep=3em,
column sep=2em, text height=1.5ex, text depth=0.25ex]
{ \G^r \times \G^r \times M & \G^r \times M  & & (\varphi, \varphi^{\prime}, x) & (\varphi \cdot \varphi^{\prime}, (l \varphi^{\prime})^{-1} x )\\
  G \times_M^{r,l} G  & G & & \left( \varphi(x), \varphi((l\varphi)^{-1}x) \right) & \varphi(x) \dia \varphi((l\varphi)^{-1}x \\};
\path[->,font=\scriptsize]
(m-1-1) edge node[above=5] {$(m, (\cdot_2)^{-1} )$} (m-1-2)
(m-2-1) edge node[auto] {$\dia$} (m-2-2)
(m-1-1) edge node[auto] {$j_{RL}^0$} (m-2-1)
(m-1-2) edge node[auto] {$j^0$} (m-2-2);
\path[|->]
(m-1-4) edge (m-1-5)
(m-1-4) edge (m-2-4)
(m-2-4) edge (m-2-5)
(m-1-5) edge (m-2-5);
\end{tikzpicture}
\end{center}

Since all maps are local, so they are the composition between them. The corresponding maps between the infinite jet bundles will be given by the composition of the corresponding jet prolongations (we are just gluing together rectangles). 
%
%
%
%
%
%
A morphism of Lie groupoids induces a local map between the bisection groups because pullbacks of bundle maps are local as we saw in Example \ref{secti}.
\qed

The bisection group of $G$ maps to the bisection group of the pair groupoid: $\textrm{Diff}(M)$ (Example \ref{actionbabe}) and hence it acts on $M$. We get an induced action groupoid which in this case is local

\begin{dfpp}[Bisection groupoid]\label{defprop} Let $G \rightrightarrows M$ be a fibered Lie groupoid. Then following set of local maps defines a local Lie groupoid structure $\G^r \times M \rightrightarrows M$ called the local Lie groupoid associated to $G \rightrightarrows M$:
\vspace{2ex}

\begin{tabular}{ll}
	Left moment map: & $(*, \cdot) \colon \G^r \times M \longrightarrow M$ \\
	Right moment map: & $(*, \textrm{id}) \colon \G^r \times M \longrightarrow M$ \\
	Identity: &$(1, \textrm{id}) \colon M \longrightarrow \G^r \times M$ \\
	Inverse: &$((-)^{-1}, \cdot) \colon \G^r \times M \longrightarrow \G^r \times M$ \\
	Multiplication: &$(m, (\cdot_2)^{-1}) \colon \G^r \times \G^r \times M \longrightarrow \G^r \times M$ \\
\end{tabular}

\vspace{2ex}
The corresponding maps between the infinite jet bundles $J^{\infty} G \rightrightarrows M$ are given in the Definition/Proposition \ref{rfg}.
\end{dfpp}

\begin{rk} It is very important to understand what is the difference between working with pro-smooth maps involving $J^{\infty} G$ or $J^{\infty} G^r$. Going back to section \ref{question} we had three major questions regarding the comparison between differential operators $(f, f^0)$ and local maps.
\begin{enumerate}
	\item[DO.1] If $j^{\infty} f^0$ (the infinite jet prolongation of $f^0$) exists, is it covered by $f$?
	\item[DO.2] Is there a notion of Cartan-preserving local maps so that we can recover $f$ from $f^0$?
	\item[DO.3] Is the map sending $[(\varphi, x)]$ to $[f(\varphi, x)]$ well defined? (By $[-]$ we mean the germ class of the pair $(\varphi, x)$.)
\end{enumerate}
The key point is to replace the insularity conditions by insularity with respect to $J^{\infty} G^r$. In this way there exists unique jet prolongations preserving the Cartan distribution of bundle maps involving $J^k G^r$ (the argument used to prove the existence of jet prolongations in Proposition \ref{PR} is local).

We have seen that replacing $J^{\infty} G$ by $J^{\infty} G^r$ has the effect of turning the maps involving the action of the associated group of bisections into an insular map (it is possible to invert the action of local bisections). That means that $DO.1$ and $DO.3$ are positively answered also in the case in which the map is insular with respect to $J^{\infty} G^r$.

Nevertheless, the maps $\restrict{j^k}{\G^r \times M}$ are not surjective. This means that given a prosmooth map involving the jet groupoid of bisections we get a unique $f$ covering it. But on the other hand, there might be several pro-smooth maps covering the same $f$. This should not worry us, since we are interested in encoding all information from $f$ in a map between two finite dimensional manifolds alone.
\end{rk}


\section{Local Lie group actions}\label{llga}

{\it The bisection group $\G^r$ of a Lie groupoid can act on a space of sections $\E$ of a fiber bundle over the same common manifold. The action is said to be local if it can be encoded in a local map. Local Lie group actions are insular, in this way we present another conceptual way of understanding the result by Blohmann which states that all local Lie group actions can be recovered from a map $J^k G^r \times_M J^{k^{\prime}} E \rightarrow E$ satisfying some group-like axioms. The result this section revolves about is from Blohmann in  \cite{B}.}\\

We have already studied the action of a bisection group on the underlying base manifold. But from our point of view, we are also interested in the action of a Lie groupoid on the fields. Remember that what we are trying to do is to define what a family of symmetries on a Lagrangian field theory is. In this section we fix a fibered Lie groupoid $G \rightrightarrows M$ with group of bisections $\G^r$ and a smooth fiber bundle $\pi \colon E \rightarrow M$ over the same manifold. We use the notation of $\sigma$ for the bisections of $G$ and we denote the sections of $E$ by $\varphi$. 

Viewing $\G^r$ as a group and $\E$ as a set we can talk about (left-) actions of $\G^r$ on $\E$. Those are maps of sets $\G^r \times \E \rightarrow \E$, usually denoted by a dot such that $1 \cdot \varphi = \varphi$ and $(\sigma \sigma^{\prime}) \cdot \varphi = \sigma \cdot (\sigma^{\prime} \cdot \varphi)$ for all $\sigma, \sigma^{\prime} \in \G^r$ and all $\varphi \in \E$.

\begin{df}[Local Lie group action]\label{map4} A local Lie group action on $\E \times M$ where $\E$ is the space of smooth sections of a smooth fiber bundle is the action of a group of bisections $\G^r$ of a fibered Lie groupoid $G \rightrightarrows M$ on $\E$ such that the combined action of $\G^r$ on $\E$ and $M$ is insular with respect to the infinite jet bundles of bisections.
\end{df}

To unravel the previous definition: an action $\G^r \times \E \rightarrow \E$ is local if there exists $a^{\infty} \colon J^{\infty} G^r \times_M^{r_{\infty}, \pi_{\infty}} J^{\infty} E \rightarrow J^{\infty} E$ a pro-smooth map covering the action $a(\sigma, \varphi, x) \defeq (\sigma \cdot \varphi, \sigma \cdot x)$ for all bisections, sections and points.

\begin{center}
\begin{tikzpicture}[description/.style={fill=white,inner sep=2pt}]
\matrix (m) [matrix of math nodes, row sep=3em,
column sep=2.5em, text height=1.5ex, text depth=0.25ex]
{ \G^r \times \E \times M & \E \times M \\
  J^{\infty} G^r \times_M^{r_{\infty}, \pi_{\infty}} J^{\infty} E  & J^{\infty} E \\};
\path[->,font=\scriptsize]
(m-1-1) edge node[auto] {$a$} (m-1-2)
(m-1-1) edge node[left] {$j^{\infty}$} (m-2-1)
(m-2-1) edge node[auto] {$a^{\infty}$} (m-2-2)
(m-1-2) edge node[auto] {$j^{\infty}$} (m-2-2);
\end{tikzpicture}
\end{center}

As an example of a local Lie group action we have the action of $\G^r$ on $\C(M, M)$ which is local since it comes from a morphism of groupoids as we have seen in Proposition \ref{fibras}.

\begin{rk}\label{milkyway} We could of course encode the action of $\G^r$ on $M$ as a map of bisections. In this way we could look at maps 
\begin{equation}\label{otherway}
(l^*, \cdot) \colon \G^r \times \E \rightarrow \textrm{Diff}(M) \times \E
\end{equation}
 which are local along the identity (that is, insular). Observe here that the locality condition is different from the one on Definition \ref{map4}. On \ref{map4} we are asking $(\sigma \cdot \varphi) (l \circ \sigma x)$ to be known from $(j_x^k \sigma, j_x^k \varphi)$, while on \ref{otherway} we are asking $(\sigma \cdot \varphi)$ at $x$ to be known from the same initial information.
\end{rk}

Let us have a look at the conditions for $a$, a local Lie group action, to be insular.On the one hand $a_{M} (\sigma, \varphi)$ is always a diffeomorphism. On the other hand $a_{\E}$ depends only on $\G^r \times \E$ so that it is independent of the image of the inverse of $a_{M} (\sigma, \varphi)$. The map indeed satisfies the Cartan-like condition for insular maps. That means that all local Lie group actions are insular as well. We have decided to keep the word local not to make those actions sound more complicated than what they actually are.

Since local Lie group actions are insular we are able to recover $a$ from the lowest map $a^0 \colon J^k G^r \times_M J^{k^{\prime}} E \rightarrow E$. The natural question at this point is, what conditions does such a map have to satisfy to induce a local Lie group action? First we have to study which properties $a^0$ satisfy if coming from a local Lie group action. Let $\sigma, \sigma^{\prime}$ be local bisections of $G$ and $\varphi$ be a local section of $E$. Let $x$ be a point in $M$, we have that:
\begin{align*}
	\textrm{Bundle map} & \quad \pi \circ a^0 \left(j_x^k \sigma,j_x^{k^{\prime}} \varphi \right) = \sigma \cdot x. \\
	\textrm{Locality} & \quad a^0 \left(j_x^k \sigma,j_x^{k^{\prime}} \varphi \right) = (\sigma \cdot \varphi)(\sigma \cdot x). \\
	\textrm{Group action, identity} & \quad a^0 \left(j_x^k 1,j_x^{k^{\prime}} \varphi \right) = \varphi(x). \\
	\textrm{Group action, iteration} & \quad a^0 \left(j_x^k(\sigma \sigma^{\prime}),j_x^{k^{\prime}} \varphi \right) = a^0\left(j_{\sigma^{\prime} \cdot x}^k(\sigma),j_{\sigma^{\prime} \cdot x}^{k^{\prime}}(\sigma^{\prime} \cdot \varphi)\right).
\end{align*}

The unique insular map covering such a smooth map $a^0$ is actually a local Lie group action:

\begin{pp}[Blohmann {\cite[Proposition 3.2]{B}}]\label{lgaction} Let $\G^r \circlearrowright \E$ be a local Lie group action on the space of sections of a fiber bundle $\pi \colon E \rightarrow M$. Let $a^0 \colon J^k G^r \times_M J^{k^{\prime}} E \rightarrow E$ be a smooth map that the action descends to. Then $a^0$ satisfies for any $x \in \, M$, $\sigma$ and $\sigma^{\prime}$ in $\G^r$ and $\varphi \in \, \E$ the following conditions:
\begin{enumerate}
	\item[(i)] $\pi \circ a^0 \left(j_x^k \sigma,j_x^{k^{\prime}} \varphi \right) = \sigma \cdot x$.
	\item[(ii)] $a^0 \left(j_x^k 1,j_x^{k^{\prime}} \varphi \right) = \varphi(x)$.
	\item[(iii)] $a^0 \left(j_x^k(\sigma \sigma^{\prime}),j_x^{k^{\prime}} \varphi \right) = a^0 \left(j_{\sigma^{\prime} \cdot x}^k(\sigma), j^{k^{\prime}} a^0 \left( j_{x}^{k + k^{\prime}}\sigma^{\prime},  j_{x}^{2 k^{\prime}}\sigma^{\prime} \varphi \right) \right)$.
\end{enumerate}
Conversely, a smooth map $a^0 \colon J^k G^r \times_M J^{k^{\prime}} E \rightarrow E$ satisfying $(i)$ to $(iii)$ above comes from a unique local Lie group action.
\end{pp}

\dem The relation between $a^0$ and $a$ is a consequence of the fact that all local Lie group actions are insular. The conditions that $a^0$ satisfy if $a$ is a local Lie group action were already discussed before the proposition. The only difference being here that
$$j^{k^{\prime}} a^0 \left( j^{k + k^{\prime}}\sigma^{\prime},  j^{2 k^{\prime}}\sigma^{\prime} \varphi \right) = j^{k^{\prime}} (\sigma^{\prime} \cdot \varphi),$$
but that follows directly from the locality condition and the definition of jet prolongations.
The only thing left to show is that the unique insular map covering $a^0$ is actually an action. But that follows from the condition {\it ``locality''} from the list before the proposition. Given any $\sigma, \sigma^{\prime} \in \G^r$, $\varphi \in \E$ and $x \in M$ we have that:
$$(1 \cdot \varphi) (x) = (1 \cdot \varphi) (1 \cdot x) = a^0 \left(j_x^k 1, j_x^{k^{\prime}} \varphi \right) = \varphi (x).$$
This shows that $1 \cdot \varphi = \varphi$ for all sections. Moreover
$$(\sigma \sigma^{\prime} \cdot \varphi)(\sigma \sigma^{\prime} x) = a^0 \left(j_x^k(\sigma \sigma^{\prime}),j_x^{k^{\prime}} \varphi \right) = a^0 \left(j_{\sigma^{\prime} \cdot x}^k(\sigma),j_{\sigma^{\prime} \cdot x}^{k^{\prime}}(\sigma^{\prime} \cdot \varphi)\right) = \sigma \cdot (\sigma^{\prime} \cdot \varphi) (\sigma \sigma^{\prime} x).$$
This also shows $\sigma \sigma^{\prime} \cdot \varphi = \sigma \cdot (\sigma^{\prime} \cdot \varphi)$ since $\sigma \sigma^{\prime}$ acts as a diffeomorphism on $M$.
\qed


\begin{rk}[Local group actions and groupoid actions] If for a local Lie group action we can take $k^{\prime}=0$, then the map $a^0 \colon J^k G^r \times_M E \rightarrow E$ is a groupoid action. This simply follows from conditions $(ii)$ and $(iii)$ in Proposition \ref{lgaction} (denote $a^0 \left(j_x^k \sigma, \varphi(x) \right))$ by $j_x^k \sigma \cdot \varphi(x)$):
\begin{enumerate}
	\item[(ii)] $j_x^k 1 \cdot \varphi(x) = \varphi(x)$.
	\item[(iii)] $j_x^k(\sigma \sigma^{\prime}) \cdot \varphi(x) =  j_{\sigma^{\prime} \cdot x}^k(\sigma) \cdot \left( j_{x}^k \sigma^{\prime} \cdot \varphi(x) \right)$.
\end{enumerate}
By remembering how the identity and the multiplication were defined for the jet groupoid it is possible to conclude what we stated. In particular if $k=k^{\prime}=0$, local Lie group actions of $\G^r$ are groupoid actions of $G$.

For $k^{\prime} > 0$, $a^0$ does not define an action of $J^k G^r$ because the jet degree on $G$ changes all the time we do a new composition (see equation $(iii)$). Nevertheless, $a^{\infty}$, the unique Cartan-preserving map between the infinite jet spaces defines a groupoid action $J^{\infty} G^r \circlearrowright J^{\infty} E$ in the category of pro--finite dimensional manifolds. 
\end{rk}

\newpage
\chapter{Lie algebroids}

A Lie algebroid is a vector bundle over a manifold endowed with a bundle map to the tangent bundle of the base, called the anchor. Moreover, its space of sections has a Lie algebra structure, such that the anchor becomes a morphism of Lie algebras and that another relation, called the Leibniz rule, is satisfied.\\

Every Lie groupoid has an associated Lie algebroid in analogy to what happens in the Lie group/algebra case. The Lie bracket on the space of sections of the Lie algebroid of a Lie groupoid is local. The Lie algebroid of a jet groupoid is the jet bundle of the Lie algebroid of the underlying groupoid. Denoting by $A(G)$ the Lie algebroid of the Lie groupoid $G$, the statement is that $A(J^k G) = J^k (A(G))$.\\

The Lie algebra of sections of a Lie algebroid can act on spaces of sections of another fiber bundle over the same underlying base manifold. These actions give rise to vector fields on the space of sections which could be asked to be local or insular. All local Lie algebra actions are actually insular, and hence all the information can be encoded in a map between finite jet bundles.\\

Local Lie algebra actions can parametrize symmetries of a Lagrangian field theory. Noether's second theorem relates these families of symmetries with certain differential equations satisfied by the Euler-Lagrange term of a field theory. This statement can be understood using the cohomology of the bicomplex of local forms.\\

Lie pseudogroups encode group-like symmetries on a manifold. This notion is related to the concept of locality and insularity by the help of jet groupoids and representations on their Lie algebroids.\\

{\it This chapter starts with a bibliographical review of the concepts of Lie algebroids and jet Lie algebroids. It later studies local Lie algebra actions, providing an infinitesimal analogue to Proposition \ref{lgaction}. It gives a statement and proof of Noether's second theorem \`a la Zuckerman, using the cohomology of the bicomplex of local forms. The last section is devoted to find a connection between the theory of insular manifolds and Lie pseudogroups. The main references are Moerdijk and Mr\v{c}un \cite{MM}, Noether \cite{NOE} and Yudilevich \cite{ORI}.}


\section{Lie algebroids and the Lie algebroid of a Lie groupoid}

{\it The infinitesimal analogue of a Lie groupoid is a Lie algebroid. Given any Lie groupoid $G$ we get an associated Lie algebroid $A(G)$. The Lie algbroid of a jet groupoid $J^k G$ is the $k$-th jet bundle of the Lie algebroid of the underlying groupoid. In other words $A(J^k G) = J^k (A(G))$. We prove that the space of sections of the Lie algebroid of a Lie groupoid is a local Lie algebra. We have followed the study by Moerdijk and Mr\v{c}un \cite{MM} for the general notions and Yudilevich \cite{ORI} for the jet Lie algebroids.}\\

In Lagrangian field theories we are concerned about {\it infinitesimal symmetries} as seen in the previous part. When we go from Lie groups to Lie groupoids, the associated infinitesimal object is not a Lie algebra any more, but a Lie algebroid.

\begin{df}[Lie algebroid] A Lie algebroid over a smooth manifold $M$ is a vector bundle map $\rho \colon A \rightarrow TM$ over the identity, where the space of sections of $A \rightarrow M$ ais endowed with a Lie bracket $[-,-] \colon \A \times \A \rightarrow \A$. The bundle map, called the anchor, satisfies the following relation, called the Leibniz rule:
$$[\psi, f \psi^{\prime}] = \left( \rho(\psi) \cdot f \right) \psi^{\prime} + f[\psi,\psi^{\prime}] \textrm{ for every } f \in \, \C(M) \textrm{ and every } \psi, \, \psi^{\prime} \in \A.$$
\end{df}

The anchor map induces a map of Lie algebras between the spaces of sections of $A$ and $TM$ respectively. This fact follows from the Leibniz rule and the Jacobi identity for $[-,-]$.

We can observe the two extremes of this definition: the tangent bundle and a Lie algebra.

\begin{ej} Consider a smooth manifold $M$ and $A = TM \rightarrow M$ the tangent bundle. The usual Lie bracket of vector fields and the identity $TM \rightarrow TM$ as an anchor give $TM$ the structure of a Lie algebroid:
$$[X, fY] = \left( X \cdot f \right) Y + f[X,Y]$$
\noindent{is simply the Leibniz rule.}
\end{ej}

\begin{ej}  Lie algebroids over a point. The vector bundle $\mathfrak{g} \rightarrow \{*\}$ is simply a vector space $\mathfrak{g}$. Since its space of sections is again $\mathfrak{g}$ we have that the bracket on sections $[-,-] \colon \mathfrak{g} \times \mathfrak{g} \rightarrow \mathfrak{g}$, gives $\mathfrak{g}$ the structure of a Lie algebra. There is a unique bundle map to $T\{*\} \cong \{*\}$ so that $\rho$ gives no extra information. We conclude that Lie algebroids over a point are Lie algebras.
\end{ej}

Remember that the vector space of the Lie algebra of a Lie group is given by the tangent space at the identity or equivalently, by right-invariant vector fields on the Lie group. For the Lie groupoid/algebroid case one does the same: the identity {\it element} is replaced by the identity {\it map} and the concept of right invariant vector field changes accordingly from right {\it multiplication} to right {\it partial multiplication}.

Given a Lie groupoid $G \rightrightarrows M$, the right fibers constitute a foliation on $G$, the corresponding distribution will be denoted by $T^r G \defeq  \bigcup_{x \in \, M} T\left( r^{-1}(x) \right) \subset TG$. Since $r 1(x) = 0$ we can take the pullback of $T^r G$ along $1$:

\begin{center}
\begin{tikzpicture}[description/.style={fill=white,inner sep=2pt}]
\matrix (m) [matrix of math nodes, row sep=3em,
column sep=2.5em, text height=1.5ex, text depth=0.25ex]
{ 1^*\left( T^r G \right) & T^r G \\
  M  & G \\};
\path[->,font=\scriptsize]
(m-1-1) edge node[auto] {} (m-1-2)
(m-1-1) edge node[left] {} (m-2-1)
(m-2-1) edge node[auto] {$1$} (m-2-2)
(m-1-2) edge node[auto] {} (m-2-2);
\begin{scope}[shift=($(m-1-1)!.5!(m-2-2)$)]
\draw +(-.3,0) -- +(0,0)  -- +(0,.3);
\fill +(-.15,.15) circle (.05);
\end{scope}
\end{tikzpicture}
\end{center}

The bundle $1^*\left( T^r G \right)$ will be simply denoted by $A$ or by $A(G)$ depending on the occasion. It is a Lie algebroid over $M$, but in order to define the anchor and the bracket, we need a different interpretation of $\Gamma(M, A)$.

Right multiplication by $g \in G$ is only defined along $r^{-1}(l(g)) \subset G$.
\begin{eqnarray*}
R_g \colon r^{-1}(l(g)) \subset G &\longrightarrow &r^{-1}(r(g)) \subset G\\
h & \longmapsto &  h \dia r
\end{eqnarray*}

\begin{df}[Right invariant vector field]
A vector field $X \in \mathfrak{X}(G)$ is called right invariant if for all $g \in G$ and all $h \in r^{-1}(l(g))$ we have that
$$X(h) \in T_h r^{-1}(r(h)) \quad \textrm{ and} \quad (TR_g) X_h = X_{h \dia g}.$$ The set of right invariant vector fields is denoted by $\mathfrak{X}^r(G)$.
\end{df}

A right invariant vector field $X$ is determined by its values along $1(M)$. For every $g \in G$:
\begin{equation}\label{formula}
X_g = X_{1(l(g)) \dia g} = (TR_g) X_{1(l(g))}.
\end{equation}

This gives us a nice interpretation of $A = 1^* \left( T^r G \right)$.

\begin{dfpp}[see for instance Moerdijk and Mr\v{c}un in{\cite[Section 6.1]{MM}}]\label{lofl} Let $G \rightrightarrows M$ be a Lie groupoid. Denote by $A$ the vector bundle over $M$ given by $1^*(T^r G)$. Then
\begin{itemize}
	\item The space of right invariant vector fields is a Lie subalgebra of $\mathfrak{X}(G)$. 
	\item The space of sections $\Gamma^{\infty}(M, A)$ is in one to one correspondence with right invariant vector fields using the formula \ref{formula}.
	\item The map $\rho = Tl \circ \textrm{pr}_{T^r G} \colon 1^*(T^r G) \cong A \rightarrow TM$ satisfies the Leibniz rule.
\end{itemize}
The corresponding Lie algebroid $Tl \colon A \rightarrow TG$ is called the Lie algebroid of the Lie groupoid $G$.
\end{dfpp}


\begin{rk} The anchor, which is composition with $\restrict{Tl}{1(M)}$ gives the infinitesimal version of the action of a bisection: composition with $l$.
\end{rk}

\begin{ej} The Lie algebroid of the pair groupoid is the tangent bundle of the base manifold.
\end{ej}

\begin{ej} The Lie algebroid of a group is its Lie algebra.
\end{ej}


\subsection{Prolongations and jet algebroids}

The space of sections of a Lie algebroid $\alpha \colon A \rightarrow M$ is a Lie algebra by assumption. Hence, contrary to the groupoid case, we do not have to worry about this Lie algebra being open in $\A$. Moreover, since $A$ is a vector bundle over $M$ and not any general fiber bundle, $\A$ is soft as a sheaf. In conclusion we have that $j^k( \A \times M) = J^k A$ for all $k$.

The anchor of $A$ can be used to define an anchor on every jet bundle:  $\rho^k \defeq \rho \circ \alpha_k^0$. As a matter of fact, the jet bundles $J^k A$ inherit from $A$ the structure of a Lie algebroid since all elements of $J^k A$ can be represented by global sections:
$$[j_x^k \psi, j_x^k \psi^{\prime}] \defeq j_x^k [\psi, \psi^{\prime}].$$

When the Lie algebroid is the Lie algebroid of a fibered Lie groupoid $G \rightrightarrows M$, the right multiplication on $J^k G^r$ gives an isomorphism between $J^k (A(G))$ and $A(J^k G^r)$. That algebroid is called the {\it $k$-th jet algebroid} of $G$, or of $A(G)$. These notations and results can be found in the thesis of Yudilevich \cite{ORI}.

As in the group case, we can talk about local Lie algebras:

\begin{df}[Local Lie algebra]\label{sn}
Let $\alpha \colon A \rightarrow M$ be a smooth vector bundle. The space of sections $\A \defeq \Gamma(M, A)$ is said to be a local Lie algebra if it is a Lie algebra and its bracket is a local map along the identity.
\end{df}

\begin{pp}\label{fibras2} The Lie algebra of sections of the Lie algebroid of a Lie groupoid is local. A morphism of such Lie algebroids induces a local map between the  algebras of sections.
\end{pp}

\dem Since the bracket on a Lie algebroid of a Lie groupoid is the bracket of right invariant vector fields, we can simply show that the bracket of vector fields is local. But in local coordinates that result is clear, the bracket depends on the first jet of the original vector fields:
$$\left[ X_i \frac{\partial}{\partial x^ i}, Y_j \frac{\partial}{\partial x^ j} \right] (x) = \left( X_j(x) \frac{\partial Y_i}{\partial x^ j}(x) - Y_j(x) \frac{\partial X_i}{\partial x^j}(x) \right) \frac{\partial}{\partial x^i} .$$
\qed

The infinitesimal action of the space of sections of a Lie algebroid on its base manifold is always local because it comes from the anchor (see Example \ref{secti}).

Not all Lie algebroids are the Lie algebroid of a Lie groupoid, so that the argument used in \color{black} Proposition \ref{fibras2} does not hold for general Lie algebroids.


\section{Local Lie algebra actions}

{\it The Lie algebra of sections of a Lie algebroid can act infinitesimally on the space of sections of a fiber bundle over the same common manifold. Extending this action to the base manifold via the anchor we can ask when these maps are local or insular. The two notions are equivalent. That means that all such local Lie algebra actions are fully determined by the lower-most map involving only finite dimensional manifolds. In this section we provide a characterization of local Lie algebra actions, complementing the work of Blohmann \cite{B} for local Lie group actions.}\\

In analogy to what was done in Section \ref{llga}, we want to consider the infinitesimal actions of the Lie algebra $\A$ of sections of a Lie algebroid on the space of sections $\E$ of a smooth fiber bundle over the same manifold. This should be any bundle map $\xi \colon \A \times \E \rightarrow \mathrm{T} \E$ such that the associated map $\A \rightarrow \mathfrak{X}(\E)$ is a morphism of Lie algebras. In order to impose locality we want to make this compatible with the infinitesimal action of $\A$ on $M$ via the anchor $\rho$. This will give rise to a map 
\begin{equation}\label{yousee}
\A \times \E \times M \rightarrow \mathrm{T} \E \times TM \cong \mathrm{T} (\E \times M).
\end{equation}
We should use locality in the more generalized way of local vector fields developed in Part IV, using $T (J^{\infty} E)$ instead of $J^{\infty}(VE)$. Observe that, by simply reorganizing the brackets in equation \ref{yousee}, we can look at infinitesimal actions on $\E$ to be:
\begin{equation}\label{yousaw}
\A \times \left( \E \times M \right) \rightarrow \mathrm{T} \left( \E \times M \right) \times M,
\end{equation}
infinitesimal actions on $\E \times M$ instead. This is more clear when talking about locality. Observe that actions as in equation \ref{yousaw} could involve different symmetries of $M$ labeled by $\E$ unless we specify that the action on $M$ is the original one.

\begin{df}[Local Lie algebra action]\label{map5} A local Lie algebra action on $\E = \Gamma^{\infty}(M,E)$ is the action of a local Lie algebra of sections $\A$ of a Lie algebroid $A \rightrightarrows M$ on $\E$ that when extended to an action on $\E \times M$ using the anchor map (as in equation \ref{yousaw}), covers a pro-smooth map $J^{\infty} (A \times_M E) \rightarrow T (J^{\infty} E)$.
\end{df}

\begin{rk}\label{milkyway2} We could, also in this case, encode the infinitesimal action of $\A$ on $M$ as a map of bisections (this is similar to what we did in Remark \ref{milkyway}). In this way we could look at maps 
\begin{equation}\label{otherway2}
(\rho, \xi) \colon \A \times \E \rightarrow \mathfrak{X}(M) \times \E
\end{equation}
which are local along the identity (that is, insular). Observe here that the locality condition is different from the one on Definition \ref{map5}. On \ref{map5} we are asking $\xi(\varphi)(\rho(\alpha,x))$ to be known from $(j_x^k \alpha, j_x^k \varphi)$, while on \ref{otherway2} we are asking $\xi(\varphi)$ at $x$ to be known.
\end{rk}

Local Lie algebra actions give rise to maps $(\xi, \rho) \colon \A \rightarrow \mathfrak{X}_{\textrm{loc}}(\E \times M)$. These maps are morphisms of Lie algebras, since local vector fields constitute a Lie subalgebra of all vector fields. Actually, the maps land on insular vector fields $(\xi, \rho) \colon \A \rightarrow \mathfrak{X}_{\textrm{ins}}(\E \times M)$ since $\xi$ does not depend on $M$. That means that all local Lie algebra actions are insular in this sense. We have decided to keep the word local not to make those infinitesimal actions sound more complicated than what they actually are.

Similarly to the theory of local Lie group actions we have a finite dimensional description of local Lie group actions:

\begin{pp}\label{laaction} Let $\xi \colon \A \times \E \rightarrow \textrm{T} \E$ be a local Lie algebra action on the space of sections of the fiber bundle $\pi \colon E \rightarrow M$. Let $\xi^0 \colon J^k A \times_M J^{l} E \rightarrow T E$ be a smooth map that the action descends to. Then $\xi^0$ satisfies for any $x \in \, M$, $\psi$ and $\psi^{\prime}$ in $\A$ and $\varphi \in \E$ the following conditions:
\begin{enumerate}
	\item[(i)] $T \pi \circ \xi^0 \left(j_x^k \psi, j_x^{l} \varphi \right) = \rho \circ \psi (x)$, where $\rho$ is the anchor of $A$.
	\item[(ii)] $\xi^0 \left(j_x^k [ \psi, \psi^{\prime} ] , j_x^{l} \varphi \right) = [ \widetilde{\xi}(\psi), \widetilde{\xi}(\psi^{\prime}) ](\varphi)(x)$, where $\widetilde{\xi} \colon \A \rightarrow \mathfrak{X}_{\textrm{ins}}(\E)$ is given by the equation $\widetilde{\xi}(\psi)(\varphi) = \xi^{0} (j^k \psi, j^l \varphi)$ (and similarly for $\psi^{\prime}$).
\end{enumerate}
Conversely, any smooth map $\xi^0 \colon J^k A \times_M J^{l} E \rightarrow TE$ satisfying $(i)$ and $(ii)$ above comes from a unique local Lie group action.
\end{pp}

Since the Lie bracket on $\A$ is local, there is a map $b^0 \colon J^{k^{\prime}} A \times_M J^{k^{\prime}} A \rightarrow A$ covered by the bracket such that $j_x^k [ \psi, \psi^{\prime} ] = j^{k^{\prime}} b^0 (j_x^{k+ k^{\prime}} \psi, j_x^{k+ k^{\prime}} \psi^{\prime})$. In this way, the left hand side of equation $(ii)$ above can be written entirely in terms of finite jet bundles.

\dem Since all local Lie algebra actions are insular, $\xi^{0}$ fully determines $\xi$, so that $\widetilde{\xi} = \xi$. Now the two conditions are precisely those in the Definition of a local Lie algebra action \ref{map5}.
\qed

\begin{rk}[Local Lie algebra actions and algebroid anchors] If on a local Lie algebra action we can take $l=0$ we have that the map $\xi^0 \colon J^k A \times_M E \rightarrow TE$ is a bundle map over $E$. It is important to observe that sections of $J^k A \times_M E \rightarrow E$ do not inherit a Lie bracket from $\A$. Hence,  we cannot conclude that $\xi^0$ is an anchor of a Lie algebroid over $E$.
\end{rk}

\begin{rk} In the mirrored proposition for local Lie groups, Proposition \ref{lgaction}, we where able to encode the group action conditions only in terms of the map between finite jet bundles. In the Lie algebra case it is also possible to encode the fact that $\xi$ induces a map of Lie algebras by simply using $\xi^0$. We would like to express the right hand side of Equation $(ii)$ in Proposition \ref{laaction} only in terms of maps involving the infinite jet bundles. The first point to make is that in order to take a commutator of two vector fields we need to treat the coefficients of a vector field as functions. This is true in general. In our case we should consider local vector fields and local functions. For the vertical part, the equations are easy: 
\begin{eqnarray}\label{iii}
 \xi^0 \left(j_x^k  \psi,  j_x^{l} \varphi \right)_{\beta} \partial_{\beta} j^1 \xi^0 \left(j_x^{k+1}  \psi^{\prime},  j_x^{l+1} \varphi \right)_{\alpha} &-& \xi^0 \left(j_x^k  \psi^{\prime},  j_x^{l} \varphi \right)_{\beta} \partial_{\beta} j^1 \xi^0 \left(j_x^{k+1}  \psi,  j_x^{l+1} \varphi \right)_{\alpha} \nonumber\\
  &=& \xi^0 \left(j_x^k {[} \psi, \psi^{\prime} {]}, j_x^{l} \varphi \right)_{\alpha} \textrm{ for all } \alpha.
\end{eqnarray}
Observe that in order to ensure that $\partial_{\beta}$ of the $\alpha$ component of $\xi^0$ is well defined, we need to end up in $T (J^1 E)$, hence the need of the first jet prolongation.
Equation \ref{iii} is not equivalent to $(ii)$, in Proposition \ref{laaction}, since it does not take the horizontal part into account. It is important to be aware that the fact that the projection to $TM$ is $\rho$, according to $(i)$, does not give all the information since horizontal and vertical parts interact. We would like to point out one of the terms that do appear when considering commutators of local vector fields $\xi^1$ and $\xi^2$: since $D_i$ and $\partial_{\alpha}^I$ do not anti-commute (as seen in Equation \ref{numerito}), the expression for the bracket of two local vector fields, projected to $TE$ involve mixed terms of the kind 
$$\left( D_i \xi_{\alpha}^1 \partial_{\alpha}^i \xi_i^2 - D_i \xi_{\alpha}^2 \partial_{\alpha}^i \xi_i^1 \right) D_i .$$
Taking care of all expressions and writing them down explicitly is much less clear than just defining a vector field from $\xi^0$ and state properties about it as it was done in $(ii)$, Proposition \ref{laaction}. 
\end{rk}


\section{Noether's second theorem}\label{nono}

{\it Noether's second theorem relates families of symmetries of a Lagrangian field theory with certain relations satisfied by the Euler-Lagrange term of the theory and its derivatives. In this section we revisit this classical theorem with the objective of providing a statement and proof in terms of the bicomplex of local forms and its cohomology. The original result is from Noether in  \cite{NOE}, although we also refer to Sardanashvily \cite{SAR} as a recent inspiration.}\\

Noether's second theorem \cite[Equation 16]{NOE} is concerned with infinite dimensional groups infinitesimally parametrizing families of symmetries of a Lagrangian field theory. In a modern language, the infinite dimensional group is going to be the bisection group $\G^r$ of a fibered Lie groupoid $G \rightrightarrows M$, the infinitesimal parametrization is going to be given by a map from the Lie algebra of sections $\A$ of its associated Lie algebroid $A = A(G) \rightarrow M$. We know from Noether's first theorem (\cite[Equation 12]{NOE} or Theorem \ref{NFT}) that each of the symmetries gives a conserved current. The second theorem says that we do get more information from the fact that we have a {\it family} of symmetries than just the collection of Noether currents. This extra piece of information is usually referred to as a relation between the Euler-Lagrange term ($\mathbf{E} L$) and its derivatives and it encodes the degeneracies of the Euler-Lagrange equations.

Noether's theorems have been around for almost a century now, and many adaptations of the original results have been proposed in different mathematical languages. Nevertheless, Deligne and Freed do not study this result \cite{DEL}. The 2011 book of Kosmann-Schwarzbach on the topic \cite{YVE} provides a modern translation of the original paper of Noether and gives an overview of the two Theorems in later texts, she includes the aforementioned (in Remark \ref{stasheff}) paper of Stasheff \cite{STA}. More recently, the monograph of Sardanshvily on Noether's theorems \cite{SAR} provides a modern perspective on Noether's work. He focuses on BRST theories, and proves the relation between the Euler-Lagrange term and its derivatives \cite[Theorem 7.10]{SAR}. Sardanshvily uses the Euler operators to derive the result: and this is the same approach we have taken in this thesis. The difference between Sardanshvily's theorem and ours is the following: he works on $\Omega_{\textrm{loc}}^{\bullet, \bullet}(\A[1] \times \E \times M)$ (restricting to BRST theories) while we work on $\Omega_{\textrm{loc}}^{\bullet, \bullet}(\G^r \times M)$ with no restriction on the class of field theories and not passing to the graded world. It also differs in the presentation of the result. Our objective is to prove a version of Noether's second theorem \`a la Zuckerman: by this we mean using the cohomological properties of the bicomplex of local forms in the way Zuckerman proved Noether's first theorem \cite[Theorem 12]{ZUC}. In particular we do not want to argue that variations are arbitrary and can be taken to vanish along a boundary of a submanifold of $M$ in order to get an equation out of it. For us it suffices to express the result of Noether's second theorem in a $d$-cohomology fashion. The key idea used by Noether is to use the product rule for multiple derivatives. As we saw in Remark \ref{modulo} we can express that equation in terms of the interior Euler operator: that is precisely our approach in this section.

We begin by defining local forms and right-invariant forms on $\G^r \times M$. This is done the same way as it was done for $\G \times M$, even when $\G^r$ is not a Fr\'echet manifold, since $J^{\infty} G^r$ is a sub-pro-smooth manifold of $J^{\infty} G$ (Proposition \ref{openyo}) and that is the relevant space in order to define local forms.

\begin{df}[Bicomplex of local forms on $\G^r \times M$]\label{lfg}
	The bicomplex of (non-twisted) local forms on $\G^r \times M$, denoted by $\left( \Omega_{\textrm{ntw-loc}}^{\bullet, \bullet}(\G^r \times M), \delta, d \right)$, where $\G^r$ is the bisection group of a fibered Lie groupoid $G \rightrightarrows M$ is given by
	$$ \Omega_{\textrm{ntw-loc}}^{\bullet, \bullet}(\G^r \times M) =  (j^{\infty})^*\left( \Omega^{\bullet, \bullet}(J^{\infty} G^r) \right);$$ equipped with differentials $\delta$ and $d$, $$\delta((j^{\infty})^*\alpha)\defeq(j^{\infty})^*(d_V \alpha) \textrm{ and }d((j^{\infty})^*\alpha)\defeq(j^{\infty})^*(d_H \alpha).$$
\end{df}

Observe that we are working with non-twisted local forms since we are interested in right-invariant ones, and in order to define them we need to make sense of pullbacks (Proposition \ref{1292}). For every finite $k$, $J^k G^r$ is a groupoid and we can talk about right-invariant forms on it. The composition on $\G^r$ descends to the partial multiplication on each of these jet groupoids (Definition/Proposition \ref{defprop}). This enables us to talk about right-invariant local forms on $\G^r \times M$:

\begin{df}[Right invariant local forms on $\G^r \times M$]\label{rilf}
	Let $G \rightrightarrows M$ be a fibered Lie groupoid. Given a bisection $\sigma^{\prime} \in \G^r$ we define right multiplication by $\sigma^{\prime}$ to be the insular map:
\begin{eqnarray*}
	R_{\sigma^{\prime}} \colon \G^r \times M & \longrightarrow & \G^r \times M \\
(\sigma, x) & \longmapsto & \left( \sigma \sigma^{\prime} , (l\sigma^{\prime})^{-1} (x) \right).
\end{eqnarray*}
	A local form $\alpha \in \Omega_{\textrm{ntw-loc}}^{\bullet, \bullet}(\G^r \times M)$ is called right invariant if $R_{\sigma^{\prime}}^* \alpha = \alpha$ for all $\sigma^{\prime} \in \G^r$. Non-twisted, local, right invariant forms on $\G^r \times M$ will be denoted by $\Omega_{\textrm{r-inv}}^{\bullet, \bullet}(\G^r \times M)$. 
\end{df}

This right multiplication that we have just defined comes from the group operation of the bisection group (Definition \ref{bg}). By the previous observation, if a local form on $\G^r \times M$ is of order $k$, the associated form $\alpha_k \in \Omega^{\bullet}(J^k G^r)$ is right invariant with respect to $j_x^k \varphi$ for all $(\varphi, x) \in \G^r \times M$ (compare to Equation \ref{jg5}).

Since pullbacks commute with differentials, and insertion of right invariant vector fields on right-invariant forms are again right-invariant, we have that the Cartan calculus operations leave the subcomplex of right invariant local forms steady (with respect to the insertion and Lie derivatives of right invariant vector fields).

We distinguish between vertical and horizontal directions when talking about local forms. In a local form $\alpha$ of bidegree $(p,q)$ on $\G^r \times M$, the vertical direction depends on $\wedge_{\G^r}^p \Gamma (M, V G = \ker (T r))$. That kernel is nothing else than $T^r G$ the distribution associated to the foliation given by the $r$-fibers. Remember that the pullback to $M$ via the identity of the groupoid is simply $A$, the Lie algebroid associated to $G$. If $\alpha$ is right invariant, the vertical component only depends on $\wedge^p \A$, where now the exterior product is over $\C(M)$. In this way \color{black} a local form of bidegree $(p,q)$, supported on $\{1\} \times M \subset \G^r \times M$ is the same thing as a map from $p$ copies of $A$ to degree $q$ forms on $M$ (with the appropriate linearity conditions from Remark \ref{conr}). Given any local form $\alpha$ supported on $\{1\} \times M \subset \G^r \times M$, we get a unique extension of $\alpha$ to a right invariant local form on $\G^r \times M$ which we will also call $\alpha$. 

Assume now that we have a fiber bundle $\pi \colon E \rightarrow M$ over the same manifold and that we are given a local Lie algebra action $\xi \colon \A \times \E \rightarrow \mathrm{T} \E$ (as in Definition \ref{map5}). Following the argument in the previous paragraph, local forms on $\G^r \times \E \times M$ have a vertical and a horizontal direction. The vertical one depends on sections of $V(G \times_M E) \cong V G \times_M V E$. We can therefore split the vertical degree into two. We will talk about local forms of degree $(p_G, p_E, q)$ where $p_G$ refers to the multiplicity in the $V G$ direction, $p_E$ on the $V E$ direction, and $q$ in the $TM$ direction as usual. Moreover, we can talk about right invariant forms on the $\G^r$ direction alone, we will denote those as $\Omega_{\textrm{r-inv}}^{\bullet, \bullet}(\G^r \times \E \times M)$. In this way, given any local form $\alpha$ supported on $\{1\} \times \E \times M \subset \G^r \times \E \times M$, we get a unique extension of $\alpha$ to a right invariant local form on $\G^r \times \E \times M$ which we will also call $\alpha$. Observe that if $p_G = 0$, a right invariant form does not even depend on $\G^r$ since we are only interested on what happens along $1 \in \G^r$, therefore $(0, p_E, q)$ right-invariant local forms are in one to one correspondence with $(p_E, q)$ local forms on $\E \times M$.\color{black}

Since $\xi$ induces a morphism of Lie algebras, it is $\RE$-linear. Therefore it induces an endomorphism for each $p_E \geqslant 1$:
$$\iota_{\xi} \colon \Omega_{\textrm{r-inv}}^{p_G, p_E,q}(\G^r \times \E \times M) \rightarrow \Omega_{\textrm{r-inv}}^{p_G +1, p_E -1, q}(\G^r \times \E \times M).$$
given for each $\psi \in \A$ by $\iota_{\xi(\psi)} \alpha$. In particular, taking $p_G = 0$ we ca use the one to one correspondence explained in the previous paragraph to show that\color{black} any form $\alpha \in \Omega_{\textrm{ntw-loc}}^{p \geqslant 1,q}(\E \times M)$ defines a form 
\begin{equation}\label{unos}
\iota_{\xi} \alpha \in \Omega_{\textrm{ntw-loc}}^{1,p-1,q}(\G^r \times \E \times M)
\end{equation}
Conversely, given any form $\beta \in \Omega_{\textrm{ntw-loc}}^{1,p-1,q}(\G^r \times \E \times M)$ which is right invariant on $\G^r$, the insertion of $\psi \in \A$ and the evaluation at $1 \in \G^r$ give rise to a form 
\begin{equation}\label{doss}
\iota_{\psi} \beta (1) \in \Omega_{\textrm{ntw-loc}}^{p-1,q}(\E \times M).
\end{equation}

\begin{tm}[Noether's second theorem]\label{n2tt} Let $L$ be a non-twisted Lagrangian on the bicomplex associated to the smooth fiber bundle $E \rightarrow M$. Let $\G^r$ be the bisection group of a fibered Lie groupoid over the same manifold. Denote by $\A$ the Lie algebra of sections of its associated Lie algebroid. If $\A$ acts on $\E$ by symmetries via a local Lie algebra action $\xi$ then for all $\psi \in \A$ there exists a form $B_{\psi} \in \Omega_{\textrm{ntw-loc}}^{0,\textsf{top}-1}(\E \times M)$ such that
\begin{equation}\label{n2t}
\iota_{\psi} (\mathbf{I} \iota_{\xi} \mathbf{E} L) (1) = d (B_{\psi} + Z_{\xi(\psi)})
\end{equation}
where $Z_{\xi(\psi)}$ is the Noether current associated to the symmetry $\xi(\psi)$.
\end{tm}

Observe that Equation \ref{n2t} happens on $\Omega_{\textrm{ntw-loc}}^{0,\textsf{top}}(\E \times M)$.

\dem By equation \ref{unos} we get a form $\iota_{\xi} \mathbf{E} L \in \Omega_{\textrm{ntw-loc}}^{1,0,\textsf{top}}(\G^r \times \E \times M)$ which is right invariant on the bisection groupoid side. Using points $1$ and $4$ of Theorem \ref{ieop}, applied to that form, we know there exist a form $\beta \in \Omega_{\textrm{ntw-loc}}^{1,\textsf{top}-1}(\G^r \times \E \times M)$ such that
$$\mathbf{I} \iota_{\xi} \mathbf{E} L - \iota_{\xi} \mathbf{E} L = d \beta.$$
Since the left hand side is right invariant, so it is the right hand side (by the same argument it has multi-degree $(1,0, \textsf{top})$). Now we are going to insert a given $\psi \in \A$ and evaluate at $1 \in \G^r$ to obtain, via Equation \ref{doss}, an equality on $\Omega_{\textrm{ntw-loc}}^{0,\textsf{top}}(\E \times M)$.
\begin{equation}\label{est}
\left( \iota_{\psi} \left( \mathbf{I} \iota_{\xi} \mathbf{E} L - \iota_{\xi} \mathbf{E} L \right) \right) (1) = \left( \iota_{\psi} d \beta \right) (1)
\end{equation}
On the left hand side we have the term $\left( \iota_{\psi} \iota_{\xi} \mathbf{E} L \right) (1) = \iota_{\xi(\psi)} \mathbf{E} L$ by definition (in equation \ref{unos}). This term is equal to $d Z_{\xi(\psi)}$ where $Z_{\xi(\psi)}$ is the Noether current associated to the symmetry $\xi(\psi)$.
On the right hand side of Equation \ref{est} we have $\iota_{\psi} d \beta = -d \iota_{\psi} \beta$ because $\psi$ is evolutionary. By virtue of Equation \ref{prev} on Remark \ref{dpull} we also have that 
$$\left( \iota_{\psi} d \beta \right) (1) = \left( -d \iota_{\psi} \beta \right) (1) = d \left( -\iota_{\psi} \beta (1) \right).$$
We define $B_{\psi} \defeq -\iota_{\psi} \beta (1)$. Equation \ref{est} becomes simply equation \ref{n2t} proving the Theorem.
\qed

In order to give an interpretation of this theorem in the terms in which Noether's theorem is usually phrased we need some extra arguments: Stokes theorem and the Du Bois-Reymond lemma. We introduce local coordinates, where $v_{\beta}$ are the fiber coordinates of $A$. We assume $M$ is oriented with a given volume form $\textrm{Vol}$. Now 
$$\iota_{\xi} \textbf{E} L = \xi_{\alpha, \beta}^I \textbf{E} L_{\alpha} \delta v_{\beta}^I \, \textrm{Vol}.$$
Any $\psi \in \A$ is given (as a right invariant vector field) in evolutionary coordinates by $\psi = \psi_{\beta} \partial_{\beta} + D_I(\psi_{\beta}) \partial_{\beta}^I$. That means that  
$$\iota_{\psi} (\mathbf{I} \iota_{\xi} \mathbf{E} L) (1) = (-1)^{|I|} D_I \left( \xi_{\alpha, \beta}^I \textbf{E} L_{\alpha} \right) \psi_{\beta} \, \textrm{Vol}.$$
Now we can use the two results mentioned above to conclude that 
\begin{equation}\label{n2t2}
(-1)^{|I|} D_I \left( \xi_{\alpha, \beta}^I \textbf{E} L_{\alpha} \right) = 0.
\end{equation}
This is the usual relation between the Euler-Lagrange term ($\mathbf{E} L$) and its derivatives which encodes the degeneracies of the Euler-Lagrange equations.

Looking at equation \ref{n2t2} we can see that $\iota_{\psi} (\mathbf{I} \iota_{\xi} \mathbf{E} L) (1) = d (B_{\psi} + Z_{\xi(\psi)}) \in \mathtt{I}$ the differential ideal of local functions vanishing on shell (from Remark \ref{stasheff}). We would like to conclude from here that $Z_{\xi(\psi)}$ is $d$-exact modulo elements on $\mathtt{I}$. This is actually true and it is the other important result about gauge symmetries (that is, given by local Lie algebra actions). The statement from the mathematical point of view is usually phrased in terms of the {\it horizontal cohomology} (see the book by Krasil'shchik and Verbovetsky \cite[Chapter 3]{KV}) and its equivalence to the cohomology of the Koszul-Tate resolution for symmetries. The result can be stated as follows:

\begin{tm}[Barnich, Brandt, and Henneaux {\cite{BBH}}]
Let $L$ be a non-twisted La\-grangian on the bicomplex associated to the smooth fiber bundle $E \rightarrow M$. Let $\G^r$ be the bisection group of a fibered Lie groupoid over the same manifold. Denote by $\A$ the Lie algebra of sections of its associated Lie algebroid. If $\A$ acts on $\E$ by symmetries via a local Lie algebra action $\xi$ then for all $\psi \in \A$ the Noether current $Z_{\xi(\psi)}$ associated to the symmetry $\xi(\psi)$ is $d$-exact on shell, in other words, there exists a form $Y_{\xi(\psi)} \in \Omega_{\textrm{ntw-loc}}^{0,\textsf{top}-2}(\E \times M)$ such that $Z_{\xi(\psi)} - d Y_{\xi(\psi)}$ vanishes on shell.
\end{tm}

The proof uses Vinogradov's $\mathtt{C}$-spectral sequence (introduced by Vinogradov \cite{VINO}) and goes beyond the scope of this thesis. The closest result to this in terms of the bicomplex of local forms is \cite[Propositon 2.96]{DEL} by Deligne and Freed. A reformulation of this theorem and of the cohomology of the horizontal complex are possible future lines of work in Lagrangian field theories.


\section{Lie pseudogroups}\label{lpg}

{\it Lie pseudogroups are submonopresheaves of $\D$ (local diffeomorphisms) that encode group-like symmetries of the base manifold depending on finite jets. This notion is clearly close to our notion of local Lie groups and groupoids. This section provides a first link between these two worlds. The main reference is Yudilevich \cite{ORI}.}\\

The theory of pseudogroups, in particular Lie pseudogroups, has a long history. It starts with the works of Cartan \cite{Car3}, \cite{Car1} and \cite{Car2}; and it has been widely studied by the research group of Marius Crainic, the adviser of Yudilevich. Cartan's theorems on pseudogroups give conditions for some Lie algebroids to be integrable (contrary to the theory of Lie groups and algebras, not all Lie algebroids come from a Lie groupoid). In his thesis \cite{ORI} Yudilevich gives a modern approach to Cartan's formulation of the theory, including reduction techniques. Part of the purpose of this thesis is to connect the world of Lie pseudogroups to that of local Lie groupoids. This section tries to do the comparison.

We start by coming back to our discussion about fibered Lie groupoids, $G \rightrightarrows M$. Remember that in order to define the jet groupoids we had to use the monopresheaf of local bisections of $G$, which we had denoted by $\G^r$. Since $\G^r$ is not a sheaf, its space of germs $\widehat{\G^r}$ is not diffeomorphic to $G$ itself. The main issue is that $\widehat{\G^r}$ is often not Hausdorff. What it is, is an \'etale groupoid (see Moerdijk and Mr\v{c}un \cite{MM} and Remark \ref{estrella}). This new groupoid lies beyond the scope the study of this thesis since it is not fibered in general; but it has some other very interesting properties. Yudilevich \cite{ORI} gives an idea of what the local bisections of this new groupoid might look like, and we refer to him for the full discussion. 

Consider an \'etale groupoid which is also effective: that means that given any two local bisections $\varphi, \varphi^{\prime} \in \G^r(U)$ such that $r \circ \varphi = r \circ \varphi^{\prime}$ then $\varphi = \varphi^{\prime}$. In that case, Haefliger proved \cite{HAE} that the monopresheaf of local diffeomorphisms induced by the left moment map determines completely the groupoid. As a matter of fact, this monopresheaf has some group-like properties that classify all monopresheaves coming from an effective \'etale groupoid.

\begin{df}[Generalized pseudogroup, following Yudilevich \cite{ORI}] A generalized pseudogroup on a Lie groupoid $G \rightrightarrows M$ is a sub-monopresheaf $\mathcal{H}$ of $\G^r$ such that:
\begin{enumerate}
	\item If $\sigma^{\prime} \in \mathcal{H} (U)$ and $\sigma \in \mathcal{H}((l\circ \sigma^{\prime})(U))$, then $\sigma \sigma^{\prime} \in \mathcal{H} (U)$.
	\item If $\sigma \in \mathcal{H} (U)$ then $\sigma^{-1} \in \mathcal{H}((l\circ \sigma)(U))$.
	\item $1_{U} \in \mathcal{H}(U)$ for all open $U$.
\end{enumerate}
Moreover, if $\sigma \in \G^r (U)$ and if there exist $\{U_i\}_{i \in I}$ an open cover of $U$ and $\{ \sigma_i \in \mathcal{H}(U_i) \}_{i \in I}$ agreeing on the overlaps, then $\sigma \in \mathcal{H}(U)$.
A generalized pseudogroup on the pair groupoid $M \times M$ is simply called a pseudogroup on $M$. 
\end{df}

The last condition is a relaxation of the gluing axiom for sheaves. $\G^r$ is a generalized pseudogroup on $G$. From any groupoid we can pass to its generalized pseudogroup of local bisections $\G^r$ and from it to a pseudogroup using the left moment map. From any given pseudogroup, we can consider its space of germs. It is actually a groupoid. The result from Haefliger \cite{HAE} that we mentioned earlier is that these two constructions are inverse of one another, that means that all pseudogroups $\mathcal{H}$ are $\G^r$ for some $G$ effective and \'etale:

\begin{pp}[Haefliger \cite{HAE}] There is a one-to-one correspondence between effective \'etale groupoids and pseudogroups. 
\end{pp}

This has served as an invitation to the world of pseudogroups. The objects that we are more interested in are {\it Lie} pseudogroups. We want to give an introduction to them from the perspective of local maps. Consider $E, F \rightarrow M$ two smooth fiber bundles over the same base. Given a pro-smooth map between the associated infinite jet bundles, we have always focused on taking jet prolongations of the lowest map $J^k E \rightarrow F$. Actually, if the map between the infinite jet bundles preserves the Cartan foliation, each of the higher maps, say for example $J^{k+l} E \rightarrow J^ l F$, sends leaves to leaves of that foliation. In other words, we could extend the definition of jet prolongations for such maps, under the assumptions that leaves of the Cartan foliation are mapped to leaves of the Cartan foliation, just by taking the prolongation of the lowest map $J^k E \rightarrow F$.

There is a catch to the previous argument and it is that we do not have, yet, a way of isolating holonomic sections of $J^{k+l} E$ and $J^ l F$. Recall that Lemma \ref{XI}, established that holonomic sections of $J^{\infty} E$ are precisely those that take the contact ideal to zero via pullback. But now we are working with a {\it finite} jet bundle instead. As a matter of fact, there is a way around it: contact forms also exist for finite jet bundles.

\begin{df}[Cartan contact forms, following Yudilevich \cite{ORI}] Let $\pi \colon E \rightarrow M$ be a smooth fiber bundle and $k \geqslant 1$ a natural number. Denote the vertical tangent bundle of $J^{k-1} E$ by $V(J^{k-1} E)$ (that is $\ker T \pi_{k-1}$). The Cartan contact form on $J^k E$ is the form $\theta^k \in \Omega^1 \left(J^k E, (\pi_k^{k-1})^* V(J^{k-1} E) \right)$ given at every $[(\varphi, x)] \in J^k E$ by:
$${\theta^k}_{[(\varphi, x)]} =\left(T \pi_k^{k-1} - T j^{k-1} \varphi \circ T \pi_k \right)_{[(\varphi, x)]}.$$
\end{df}

Consider $\chi \in J^{k-1} E$. The kernel of the Cartan contact form $\theta^k$ at a given preimage $\chi^{\prime} \in (\pi_k^{k-1})^{-1}(\chi)$ is $T j^{k-1} \varphi (T_x M)$ where $\chi^{\prime} = j_{x = \pi_{k-1}(\chi)}^k \varphi$. This is so because $T \pi_k^{k-1}$ is surjective. Comparing this result to the Definition \ref{cd} of the Cartan distribution $C$ we can see that
$$C_{\chi}^{k-1} = \bigcup_{\chi^{\prime} \in (\pi_k^{k-1})^{-1}(\chi)} \ker(\theta^k)_{\chi^{\prime}}.$$

As a matter of fact, $C_{\chi}^{k-1} \subset \ker(\theta^{k-1})_{\chi}$ for all $\chi$ (this follows from the fact that $\theta^{k-1} \circ T \pi_{k}^{k-1} = T \pi_{k-1}^{k-2} \circ \theta^k$, done by Yudilevich \cite{ORI}). That shows that the associated distributions on $J^{\infty} E$ coming from $\ker(\theta)$ and the Cartan distribution as in Definition \ref{cd} are the same. Now we can introduce the analogue of Lemma \ref{XI} for finite jet bundles

\begin{pp}[Yudilevich {\cite[Proposition 1.3.3]{ORI}}]\label{XI2} Let $E \rightarrow M$ be a smooth fiber bundle with associated finite jet bundle $J^{k} E$ for some natural number $k \geqslant 1$. A local section $\Xi$ of $\pi_{k} \colon J^{k} E \rightarrow M$ is holonomic (this is, it is the prolongation of a local section $\varphi$ of $E \rightarrow M$) if and only if $\Xi^* \theta^k = 0$.
\end{pp}

\begin{rk}\label{rara}
Coming back to the previous line of argument. A map $J^{k+l} E \rightarrow J^ l F$ could be asked to preserve the Cartan contact form in order to use it to get infinite jet prolongations of the map $J^k E \rightarrow F$, even if we are not given explicitly that map.
\end{rk}

This is extremely convenient: sometimes we encounter monopresheaves $\mathcal{H} \subset \E$ such that $j^k (\mathcal{H} \times M)$ is only smooth starting from $k \geqslant k_0$ (to be precise we should work on the germs and say $\widehat{j}^k (\widehat{\mathcal{H}})$). Using Lemma \ref{XI2}, we are able to identify the elements of $j^0(\mathcal{H} \times M)$ from the smooth manifold $j^{k_0}(\mathcal{H} \times M)$ by asking which sections are holonomic. This is precisely the setup for Lie pseudogroups:

\begin{df}[Lie pseudogroup, following Yudilevich \cite{ORI}] A pseudogroup $\mathcal{H}$ on $M$ is called a Lie pseudogroup of order $k_0 + 1 \geqslant 0$ if for $k \in \{k_0, k_0 + 1\}$ the set
$$J^k \mathcal{H} \defeq \{ j_x^k \tau \colon \tau \in \mathcal{H}(U), x \in U\}$$
is a Lie subgroupoid of $J^k (M \times M)^r$, $\restrict{\pi_{k_0 +1}^{k_0}}{J^{k_0 +1} \mathcal{H}}$ is a submersion and all holonomic local sections of $J^{k_0 +1} \mathcal{H}$ are elements of $\mathcal{H}$.
\end{df}

The last condition means that if $\tau \in \D(U)$ for some $U \subset M$ open and $j_x^{k_0 +1} \tau \in J^k \mathcal{H}$ for all $x \in U$, then $\tau \in \mathcal{H}$. Observe that we also need $J^{k_0} \mathcal{H}$ to be defined, so that the Cartan contact form makes sense when restricted to $J^{k_0 +1} \mathcal{H}$

The Cartan contact form on a Lie groupoid can actually be asked to take values in the pullback via $l$ of its Lie algebroid, since $V G \cong l^* A(G)$. This is what is done by Yudilevich \cite{ORI}.

Now we are ready to understand the classification of Lie pseudogroups as holonomic sections given in \cite{ORI}:

\begin{pp}[Yudilevich {\cite[Proposition 3.4.1]{ORI}}] Lie pseudogroups of order $k$ are in one to one correspondence with subgroupoids of $J^k (M \times M)^r$.
\end{pp}

The correspondence is given by sending $\mathcal{H}$ to $J^k \mathcal{H}$ and $G \subset J^k (M \times M)^r$ to its holonomic bisections as explained in Proposition \ref{XI2}.

As a matter of fact, there is a way of talking about the groupoids $\mathcal{H}$ and its holonomic bisections without viewing them as a subgroupoid of $J^k (M \times M)^r$. It is possible to talk about {\it Lie-Pfaffian} Lie groupoids. Those are Lie groupoids endowed with a form $\theta$ with certain distribution properties, that allow to talk about holonomic bisections of $G$.

The conclusion is that Lie pseudogroups can be treated as Lie-Pfaffian groupoids. We could change the Cartan preserving conditions for both $J^{\infty} G$ and $\G^r$ to ensure insular maps exist from an insular differential operator involving such a groupoid, by always asking the prolongations to be taken using holonomic sections as explained in Remark \ref{rara}. This turns Lie psuedogroups into a category, and one can start exploring the concepts of Cartan equivalence for Lie pseudogroups and realizations using this category.


\newpage
\chapter{\boldmath{\Linf s} and locality}

The theory of Lie algebras up to homotopy has deep roots into mathematical physics. In modern language, they are referred to as \Linf s. They provide higher generalizations of Lie algebras in which the underlying vector space is now graded and brackets with a higher number of entries are also allowed. In particular, differential graded Lie algebras are examples of \Linf s.\\

There are different notions of \Linf s in the literature, all of which are equivalent under certain finite dimensionality, and non-positive concentration assumptions. This is so because the proof of the equivalence of the definitions relies on the fact that finite dimensional vector spaces are reflexive. The finite dimensionality hypothesis can be lowered to pro-finite dimensionality, since such spaces are also reflexive in the appropriate sense.\\

A classical result by Vaintorb relates Lie algebroid structures on $A \rightarrow M$ with differential graded Lie algebra structures on the non-reduced algebra $S_{\C(M)}^-([-1]\A^*)$. The same result applies in higher cases, getting to the notion of \Linfoid s. Replacing vector bundles by pro-vector bundles the result remains true. Peetre's theorem in its linear version, relates linear insular differential operators with pro-linear maps. In this way, local \Linf s are defined and related to pro-\Linfoid s by the pro-version of Vaintrob's result. The higher brackets in a local {\Linf} are examples of what is sometimes called a polydifferential operator in the literature.\\

{\it This chapter starts by fixing our conventions for graded vector spaces and \Linf s, and by presenting the classical results relating the different notions of \Linf s and \Linfoid s. It develops the theory of duals in pro- and ind-graded vector spaces and shows that pro-finite dimensionality is enough to compare the different definitions of (pro-)\Linf s. Finally, the notion of local \Linf s is presented. Using the linear version of Peetre's theorem, we show that our notion agrees with that using $D_M$-modules of Costello. The main references in the chapter are Costello and Gwilliam \cite{CG1}, Lada and Stasheff \cite{LS}, and Vaintrob \cite{VAI}.}


\section{\boldmath{\Linf s}}

{\it \mbox{$L_{\infty}$}--algebras are higher generalizations of Lie algebras which are stable under homotopies and which are useful in deformation theory and deformation quantization. Without entering into those details, this section provides the basic definitions and results about \mbox{$L_{\infty}$}--algebras that will be needed in this thesis. We fix our conventions regarding graded vector spaces and \mbox{$L_{\infty}$}--algebras. The main references are Lada and Stasheff \cite{LS} and Vaintrob \cite{VAI}.}\\

We work with vector spaces over the field $\mathbb{R}$. Most of what is said in this section also works for modules over rings, in particular we have in mind the case of $\C(M)$-modules.

Recall that the category of $G$-graded vector spaces, given $G$ an abelian group, has as objects families of vector spaces $V := \{ V_g \}_{g \in G}$ and as morphisms, families of linear maps $\{ f_g \colon V_g \rightarrow W_g\}_{g \in G}$. In our case, the grading is usually over $\mathbb{Z}$. It is a monoidal category with the tensor product gathering the vector spaces of same added degree:

$$(V \otimes W)_g := \bigoplus_{g = h + h^{\prime}}{V_h \otimes W_{h^{\prime}}}.$$

$\RE$ as a graded vector space concentrated in degree $0$ (the identity element on $G$) is the tensor unit. From now on we fix $G = \mathbb{Z}$. We follow the usual sign convention, where the linear maps 
\begin{eqnarray}\label{srule}
	\tau_{V, W} \colon V \otimes W &\longrightarrow& W \otimes V  \nonumber \\
	(v \otimes w ) &\mapsto& (-1)^{|v||w|} (w \otimes v) 
\end{eqnarray} 
define a symmetric structure on $\mathbb{Z}$-graded vector spaces. We can consider the $n$-th tensor product $W_1 \otimes \cdots \otimes W_n$ where each $W_n$ is a $\mathbb{Z}$-graded vector space. A permutation $\sigma \in \mathcal{S}_n$ acts in such tensor products by decomposing $\sigma$ into transpositions and let them act in order using the sign convention \ref{srule}: 
\begin{eqnarray*}
	\sigma_{W_1 \otimes \cdots \otimes W_n} \colon W_1 \otimes \cdots \otimes W_n &\longrightarrow& W_{\sigma(1)} \otimes \cdots \otimes W_{\sigma(n)} \\
	(w_1 \otimes \cdots \otimes w_n ) &\mapsto& \epsilon(\sigma, w) (w_{\sigma(1)} \otimes \cdots \otimes w_{\sigma(n)})
\end{eqnarray*} 
$\epsilon(\sigma, w)$ is called the Koszul sign of the permutation.

Degree shifting is an important tool in graded vector spaces. We adopt here a, somehow uncommon, convention: if $V$ is a $\mathbb{Z}$-graded vector space and $n$ an element in $\mathbb{Z}$, the $\mathbb{Z}$-graded vector space $[n]V$ (called $V$ shifted by $n$)  is defined degree-wise by $[n]V_m := V_{n+m}$ for all $m \in \, \mathbb{Z}$. The reason we write the degree shift $n$ on the left (instead on the right, what is usually found in the literature) is to emphasize how $[n]V$ is truly defined using the monoidal structure.
\begin{equation}\label{cshift}
[n]V := [n]\RE \otimes V
\end{equation}

This is very convenient when examining how degree shifting behaves with respect to tensor products. The tool used to study that behavior is called the d\'ecalage isomorphism:
\begin{eqnarray*}
	\dec \colon ([-1]V)^{\otimes n} & \longrightarrow & [-n](V^{\otimes n}) \\
	\uparrow v_1 \otimes \cdots \otimes \uparrow v_n & \mapsto & (-1)^{\sum_{i=1}^n {(n-i)|v_i|}} \uparrow \stackrel{n}{\cdots}\uparrow (v_1 \otimes \cdots \otimes v_n).
\end{eqnarray*}

We have used the arrow notation from Lada and Stasheff \cite{LS} where the unit $1 \in [-1]\RE$ is denoted by $\uparrow$ and the unit $1 \in [-1]\RE$ is denoted by $\downarrow$ (the same isomorphism works for down arrows as well). The d\'ecalage isomorphism is nothing else that a consequence of the symmetric structure on graded vector spaces given by the sign rule \ref{srule} and the convention for shifts \ref{cshift}. Observe for example that with this convention
$$[-n][n]V \cong (-1)^{\frac{n(n-1)}{2}}V.$$

Considering the action of the symmetric groups we can take the symmetrization $S(V)$ of $T(V)$. The anti-symmetrization of the tensor algebra is denoted by $\wedge V$. As in the non-graded world, those algebras are degree-wise isomorphic to the co-invariants of the canonical action of $\mathcal{S}_n$ on $V^{\otimes n}$ and of the action given by $(-1)^{\sigma} \sigma$ respectively. The tensor degree is called {\it polynomial degree} in $S(V) := \bigoplus_{n \geqslant 1} S^ n (V)$.
$$S(V) := T(V) / <v \otimes w - (-1)^{|x||y|}y \otimes x> \textrm{ and}$$
$$\wedge(V) := T(V) / <v \otimes w + (-1)^{|x||y|}y \otimes x>.$$
$S(V)$ is also a coalgebra, both as an algebra and as a coalgebra it is free (respectively cofree) on $V$. The symmetric algebra is really the {\it reduced} symmetric algebra, where we have discarded $\RE = S^0 V$. The notation is usually $S^+$ for the reduced symmetric algebra and $S$ for the symmetric one. We have decided to reserve the special symbol for the less common one in this thesis: $S^-(V) \defeq \bigoplus_{n \geqslant 0} S^ n (V) = \RE \oplus S(V)$ is the non-reduced symmetric algebra on $V$.
 
The d\'eclage isomorphism induces an isomorphism on the symmetric and anti-symmetric powers just by considering the invariants under the action of the symmetric groups on both sides:
$$\dec \colon S^n([-1]V) \stackrel{\scriptscriptstyle{\cong}}{\longrightarrow} [-n](\wedge^{n} V).$$

Given two $\mathbb{Z}$-graded vector spaces $V$ and $W$, we can view them as vector spaces by taking the direct sum of all the vector spaces in each degree; $V^{\oplus} \defeq \bigoplus_{n \in \mathbb{Z}} V_n$. The hom--space $\underline{\mathrm{Hom}}(V, W) := \mathrm{Hom}_{{\mathcal{V}ec}}(V^{\oplus}, W^\oplus)$ carries a $\mathbb{Z}$ grading given by 
$$\underline{\mathrm{Hom}}(V, W)_n = \prod_{m \in G}{\underline{\mathrm{Hom}}_{\mathcal{V}ec}(V_m, W_{n+m})}.$$

Elements of $\underline{\mathrm{Hom}}(V, W)_n$ are called morphisms of degree $n$. We could be tempted to say that the morphisms between $S^n([-1]V)$ and $[-1]V$ are isomorphic to the ones between $[-n](\wedge^{n} V)$ and $V$ through the d\'ecalage isomorphism. But one has to be a little bit careful: The shifting degree functor interacts with graded morphisms in a non-trivial way:
$$\ihom(V, W)_j \cong \ihom(V, [j]W)_0 \stackrel{\scriptscriptstyle{(-1)^{ij}}}{\cong} \ihom([i]V, [i]W)_j.$$

The first isomorphisms in trivial but the second one is not. We can see that the following diagram is the commutative one, not the one given by $[i]f$ (read as composition $[i] \circ f$).

\begin{center}
	\begin{tikzpicture}[description/.style={fill=white,inner sep=2pt}]
	\matrix (m) [matrix of math nodes, row sep=1.5em,
	column sep=5em, text height=1.5ex, text depth=0.25ex]
	{   & V & \phantom{.} [j]W \\
		\phantom{.} [i]V &   &\\
		& \phantom{.} [j]([i]W) & \phantom{.} [i]([j]W) \\};
	\path[->,font=\scriptsize]
	(m-1-2) edge node[auto] {$f$} (m-1-3)
	(m-2-1) edge node[auto] {$[-i]$} (m-1-2)
	(m-2-1) edge node[auto] {$(-1)^{ij} [i]f$} (m-3-2)
	(m-3-3) edge node[auto] {$\tau_{[-i]\mathbb{K}, [j]\mathbb{K}}$} (m-3-2)
	(m-1-3) edge node[auto] {$[i]$} (m-3-3);
	\end{tikzpicture}
\end{center}

Now we get isomorphisms (which we also call d\'{e}calage, abusing the notation):
\begin{equation}\label{dec2}
\dec \colon \ihom(\wedge^n W, W)_{-n-j} \longrightarrow \ihom(S^n([1]W), [1]W)_{-1 -j}.
\end{equation}

We follow the original definition of \Linf s by Lada and Staheff \cite{LS}. They attribute the definition to Jones \cite{JON}. We will later see how this generalizes the notion of Lie algebra.

\begin{df}[Jones \cite{JON}] An {\Linf} is a $\mathbb{Z}$-graded vector space $L$ together with a family of maps $\{l_n \colon \wedge^n L \rightarrow L\}_{n\in \mathbb{N}}$ of degree $2-n$ such that for every $n \in \mathbb{N}$, $n \geqslant 1$ and every $x = x_1 \otimes \cdots \otimes x_n$ in $L^{\otimes n}$:
	\begin{equation}\label{jacojon}
	\sum_{i+j=n+1}\sum_{\sigma \in \mathcal{S}h_{j-1}^i} (-1)^{j(i-1)} (-1)^{\sigma} \epsilon(\sigma) l_j(l_i(x_{\sigma(1)} \wedge \cdots \wedge x_{\sigma(i)}) \wedge x_{\sigma(i+1)} \wedge \cdots \wedge x_{\sigma(n)})=0.
	\end{equation}
\end{df}

The $(i,j-1)$-unshuffles, denoted by $\mathcal{S}h_{j-1}^i$, are the permutations $\sigma \in \mathcal{S}_{i+j-1}$ that leave the first $i$ entries and the last $j-1$ entries ordered:
$$\sigma(1) < \sigma(2) < \cdots < \sigma(i), \quad \sigma(i+1) < \sigma(i+2) < \cdots < \sigma(i+j-1).$$
The left hand side of equation \ref{jacojon} is called the $n$-th Jacobiator. The map $l_n$ is called the $n$--ary bracket and it will usually be referred to as a totally antisymmetric map $l_n \colon T^n L = L^{\otimes n} \rightarrow L$ instead of a map from the exterior algebra. According to that convention the elements will not be separated by wedges, but simply by commas.

\begin{lm}\label{lee}
	Let $V$ be a $\mathbb{Z}$-graded vector space and $i \in \mathbb{Z}$. Then we have the following isomorphisms:
	\begin{enumerate}
		\item $\textrm{Coder}^i\left(S\left([1]V\right)\right) \cong \ihom\left(S\left([1]V\right), [1]V\right)_{i}$.
		\item $\textrm{Der}^i\left(S\left([-1]V^*\right)\right) \cong \ihom\left([-1]V^*, S\left([-1]V^*\right)\right)_{i}$.
		\item If $V$ is finite dimensional and concentrated in non-positive degrees, then $$\ihom\left([-1]V^*, S\left([-1]V^*\right)\right)_{i} \cong \ihom\left(S\left([1]V\right), [1]V\right)_i.$$
		\item If $V$ is finite dimensional and concentrated in non-positive degrees, then $$\textrm{Der}^i\left(S\left([-1]V^*\right)\right) \cong \textrm{Coder}^i\left(S\left([1]V\right)\right).$$
	\end{enumerate} 
\end{lm}

In the previous lemma $\textrm{Der}$ and $\textrm{Coder}$ denote graded derivations and graded co\-der\-i\-va\-tions respectively. The proof is a simple calculation, the interesting result is the specialization to degree one, square to zero maps (sometimes called differentials and codifferentials respectively).

\begin{tm}[Lada and Stasheff \cite{LS}]\label{tml}
	Let $V$ be a $\mathbb{Z}$-graded vector space. Then we have the following one to one correspondences:
	\begin{enumerate}
		\item Degree $1$, square to zero coderivations of  $S\left([1]V\right)$ correspond to {\Linf} structures on $V$.
		\item If $V$ is finite dimensional and concentrated in non-positive degrees; then degree 1, square to zero derivations of $S \left([-1]V^*\right)$ correspond to families of maps \mbox{$\{q_n \in \ihom\left(S^n\left([1]V \right), [1]V \right)_{1} \}_{n \in \mathbb{N}}$} satisfying for every $n \in \mathbb{N}$, $n \geqslant 1$ and every $x = x_1 \otimes \cdots \otimes x_n$ in $V^{\otimes n}$:
			\begin{equation}\label{jaco2}
			\sum_{i+j=n+1}\sum_{\sigma \in \mathcal{S}h_{j-1}^i} \epsilon(\sigma) q_j(q_i(x_{\sigma(1)} \otimes \cdots \otimes x_{\sigma(i)}) \otimes x_{\sigma(i+1)} \otimes \cdots \otimes x_{\sigma(n)}).
			\end{equation}
		\item If $V$ is finite dimensional and concentrated in non-positive degrees, collections of maps $\{q_n \in \ihom\left(S^n\left([1]V \right), [1]V \right)_{1} \}_{n \in \mathbb{N}}$ satisfying \ref{jaco2} and collections of maps \mbox{$\{l_n \in \ihom\left( \wedge^n V, V \right)_{2-n} \}_{n\in \mathbb{N}}$} satisfying \ref{jacojon} are in one to one correspondence.
		\item If $V$ is finite dimensional and concentrated in non-positive degrees, then degree 1, square to zero derivations of $S \left([-1]V^*\right)$ are in one to one correspondence with degree $1$, square to zero coderivations of  $S\left([1]V\right)$.
	\end{enumerate} 
\end{tm}

\begin{rk}\label{nonred} At this point it becomes relevant whether or not the index category $\mathbb{N}$ includes $0$ or not. Derivations of $S \left([1]V\right)$, the {\it reduced} free algebra on $[1] V$, starting at polynomial degree $1$, correspond to families of maps indexed by $\{1, 2, \ldots \}$; while derivations of $S^- \left([1]V\right)$, with polynomial degrees starting at $0$, also include a map $q_1 \colon V \rightarrow \RE$.
\end{rk}

All the isomorphisms that involve changing symmetric powers by antisymmetric ones are d\'ecalages, from Equation \ref{dec2} ($\ihom(\wedge^n W, W)_{2-n} \rightarrow \ihom(S^n([1]W), [1]W)_{1}$). Any of the other equivalent notions from Theorem \ref{tml} are found as definitions of $L_{\infty}$--algebras in the literature. Using those equivalences it is clear that if $l_{n \neq 2} = 0$, $(L, \{l_n\}) = (L, l_2)$ is a (graded) Lie algebra, and if we also have $l_1$ we get differential graded Lie algebras.

When $l_2 \colon L \otimes L \rightarrow L$ is a Lie algebra bracket, the complex $(S\left([1]L\right), d)$ with the associated coderivation $d$ is called the {\it Chevalley-Eilenberg complex} of the Lie algebra $L$. We adopt the same notation for a general {\Linf} L.

Morphisms of \Linf s are defined as morphisms of coalgebras with coderivations following the identification from Theorem \ref{tml}. The definition can also be found in the paper by Lada and Stasheff \cite{LS}:

\begin{df}[${L_{\infty}}$-morphism]\label{morphi}
An \mbox{$L_{\infty}$} morphism between two \Linf s $(L, \{ l_n\})$ and $(V, \{ v_n\})$, $f \colon (L, \{ l_n\}) \rightarrow (V, \{ v_n\})$ is a collection of symmetric morphisms (of degree $0$) $f:=\{f_k \colon L^{\otimes k} \rightarrow V\}_{k \in \mathbb{N}}$ such that for every $n \in \mathbb{N}$, $n \geqslant 1$ and every $x = x_1 \otimes \cdots \otimes x_n$ in $L^{\otimes n}$
$$\sum_{i+j =n +1} \sum_{\sigma \in \mathcal{S}{h}_{j-1}^i} f_j (l_i, -) (\sigma \cdot x) = \sum_{p_1 + \ldots p_q = n}^{p_i \geqslant 1} \sum_{\sigma \in \mathcal{S}{h}(p_1,\ldots, p_q)} \frac{1}{q!} (v_q \circ (f_{p_1} \otimes \cdots \otimes f_{p_q})) \, (\sigma \cdot x).$$
\end{df}

As mentioned at the beginning of the section, we are also interested in graded $\C(M)$-modules and in {\Linf} structures inner to that category. The modules that we will be interested in are sections of graded vector bundles over $M$. A graded vector bundle is a vector bundle $V \rightarrow M$ in which the fibers are given a grading, compatible with the transition functions. It is important to observe that we view $\C(M)$ as in degree $0$. The non-reduced algebra $S_{\C(M)}^-([-1]\V^*) = \Gamma(M, S^-([-1]V^*))$ has always $\C(M)$ in degree $0$, even if asking $V$ to be concentrated in non-positive degrees. $S_{\C(M)}^-([-1]\V^*)$ can be understood as the space of functions on a graded manifold with non-zero coordinates given by $V$. Usually such graded manifold is denoted by $\mathcal{M}$ (as in Kontsevich's paper \cite{KON}).

The two generalizations to the notion of Lie group explored in this part are connected by a classical result by Vaintrob:

\begin{tm}[Vaintrob \cite{VAI}]
	Let $A \rightarrow M$ be a finite rank smooth vector bundle. Lie algebroid structures on $A$ are in one to one correspondence to degree $1$, square to zero derivations of $S_{\C(M)}^-([-1]\A^*) = \Gamma(M, S^-([-1]A^*))$ as a $\C(M)$-algebra.
\end{tm}

Since the algebra is the non-reduced one, we get an extra map involving $\C(M)$ (see Remark \ref{nonred}). That map induces the anchor. The substraction of that anchor to the $\C(M)$-linear map $l_2$, associated to the derivation from Theorem \ref{tml}, gives the bracket $\{-,-\}$ which is not $\C(M)$-linear anymore. The structure of a degree $1$, square to zero derivation of $S_{\C(M)}^-([-1]\A^*)$ is viewed in the supermanifold $\mathcal{M}$ as a cohomological vector field. Is in those terms that Vaintrob phrases his result \cite{VAI}. 

The result generalizes to the case in which $A$ is not a vector bundle, but a $\mathbb{Z}$-graded vector bundle to begin with. In that case we get higher brackets and higher anchors. We will not enter into the bracket definition of an {\Linfoid} having the following as a definition instead:

\begin{df}[Bruce \cite{BRU}]\label{oid}
	Let $A \rightarrow M$ be a total finite rank $\mathbb{Z}$-graded vector bundle concentrated in non-positive degrees. An {\Linfoid} structure on $A$ is a degree 1, square to zero derivation of the $\C(M)$-algebra $S_{\C(M)}^-([-1]\A^*) = \Gamma(M, S^-([-1]A^*))$.
\end{df}

Bruce shows that this definition is equivalent to that including higher brackets and anchors available in the literature. That equivalence is in essence, a generalization of Vaintrob's result. Observe that we need the requirement about total finite rank, since otherwise Theorem \ref{tml} will not hold. The study by Bruce might not be the first one in the direction of generalizing Vaintrob's result in a graded setting, see for example the paper by Khudaverdian and Th. Th. Voronov \cite{KhV}.


\section{Pro-\boldmath{\Linf s}}

{\it The different definitions of \Linf s are only equivalent for reflexive vector spaces. Working with pro-finite dimensional vector spaces and defining the duals in an ind-pro way, makes pro-finite dimensional vector spaces reflexive. This section introduces the concept of duality in pro-categories in which all objects are reflexive in the original category. It also defines and proves the equivalence between different definitions of \Linf s inner to those categories.}\\

As seen in Theorem \ref{tml}, the equivalence between the different definitions of \Linf s is only possible when the underlying vector space in finite dimensional. Pro and ind-finite dimensional spaces are one step away from finite dimensionality and that is enough to ensure the equivalence of definitions for such spaces. We work with the category $\mathbb{Z}$-$\Vect$ of $\mathbb{Z}$-graded, total finite dimensional vector spaces over $\RE$. We will refer to it as {``graded vector spaces''} for simplicity. 

\begin{df}[$\mathbb{Z}$-$\Vect$]\label{grvc}
	The category of graded vector spaces, denoted by $\mathbb{Z}$-$\Vect$, is the category of $\mathbb{Z}$-graded, total finite dimensional vector spaces over $\RE$.
\end{df}

Observe that, contrary to what was done in the previous chapter, now we work with {\it finite} dimensional vector spaces. In our notation a $\mathbb{Z}$-graded vector space can be infinite dimensional, but not a graded vector space. The category of graded vector spaces has duals and all its objects are reflexive ($(V^*)^* \cong V$ for all $V$ graded vector space).

\begin{dfpp}[Duality of pro- and ind-graded vector spaces]\label{duality}
Let $X \colon \mathcal{I} \rightarrow \mathbb{Z}$-$\Vect$ be a pro-graded vector space. The functor $X^{*} \colon \mathcal{I}^{\textsf{op}} \rightarrow \mathbb{Z}$-$\Vect$ given by $X^*(i) \defeq X(i)^*$ and $X^{*}(i^{\prime}, i) \defeq (X(i,i^{\prime}))^*$ is an ind-graded vector space called the dual of $X$.

Conversely, given $Y \colon \mathcal{J} \rightarrow \mathbb{Z}$-$\Vect$ be an ind-graded vector space. The functor $Y^{*} \colon \mathcal{J}^{\textsf{op}} \rightarrow \mathbb{Z}$-$\Vect$ given by $Y^*(j) \defeq X(j)^*$ and $Y^{*}(j^{\prime}, j) \defeq (Y(j,j^{\prime}))^*$ is a pro-graded vector space called the dual of $Y$.

All pro-graded vector spaces and all ind-graded vector spaces are reflexive in the sense that $(X^*)^* = X$ and $(Y^*)^* = Y$.
\end{dfpp}

\dem The opposite category of an essentially small cofiltered category is essentially small filtered and vice-versa. Since $\mathbb{Z}$-$\Vect$ has all duals, the maps $(X(i,i^{\prime}))^*$ and $(Y(j,j^{\prime}))^*$ are again maps in the same category. Dual of commutative diagrams are commutative again. This proves that $X^*$ and $Y^*$ are ind-graded vector spaces and pro-graded vector spaces respectively, provided $X \colon \mathcal{I} \rightarrow \mathbb{Z}$-$\Vect$ is a pro- and \mbox{$Y \colon \mathcal{J} \rightarrow \mathbb{Z}$-$\Vect$} is an ind-graded vector space. The opposite of an opposite category is again the same category; the functor  $(X^*)^*  \colon \mathcal{I} = (\mathcal{I}^{\textsf{op}})^{\textsf{op}} \rightarrow \mathbb{Z}$-$\Vect$ is the same as $X$ since $(X^*)^*(i) \defeq (X(i)^*)^* \cong X(i)$ and $(X^{*})^*(i, i^{\prime}) \defeq ((X(i,i^{\prime}))^*)^*$. The same argument works for ind-graded vector spaces.
\qed

As usual, we will work with sequential pro- and  ind-objects. If $X$ is a sequential pro-graded vector space
$$X_0 \stackrel{\scriptscriptstyle{X_1^0}}{\longleftarrow} X_1 \longleftarrow  \cdots \longleftarrow X_n \stackrel{\scriptscriptstyle{X_{n+1}^n}}{\longleftarrow} X_{n+1} \longleftarrow \cdots$$ 
The dual sequential ind-graded vector space $X^*$ is given by
$$X_0^* \stackrel{\scriptscriptstyle{(X_1^0)^*}}{\longrightarrow} X_1^* \longrightarrow  \cdots \longrightarrow X_n^* \stackrel{\scriptscriptstyle{(X_{n+1}^n)^*}}{\longrightarrow} X_{n+1}^* \longrightarrow \cdots$$

Ind- and pro-graded linear maps have the same notation as before, but specifying the category in which we are working every time. Degree shifts work in ind- and pro-graded vector spaces in the obvious way: $([n]X)(i) \defeq [n](X(i))$.  Observe that $\mathbb{Z}$-$\Vect$ has finite products, but not all infinite ones. While $S^n(V)$ and $\wedge^n(V)$ are in $\mathbb{Z}$-$\Vect$ for all graded vector space $V$ and all finite $n$; $S(V)$ and $\wedge(V)$ are not. The d\'ecalage isomorphism (which is defined at every finite polynomial degree) passes to pro- and ind-graded vector spaces with no further do.

\begin{lm}\label{lee2}
	Let $X$ be a pro-graded vector space concentrated in non-positive degrees (each $X(i)$ is concentrated in non-positive degrees for all $i$). We have the following isomorphisms for each $n \in \mathbb{N}$:
\begin{eqnarray*}
	\ihom_{\textrm{Ind}(\mathbb{Z}\textrm{-}\Vect)}\left([-1]X^*, S^n \left([-1]X^*\right)\right)_{1} &\cong& \ihom_{\textrm{Pro}(\mathbb{Z}\textrm{-}\Vect)}\left(S^n \left([1]X\right), [1]X\right)_1 \\
	&\cong& \ihom_{\textrm{Pro}(\mathbb{Z}\textrm{-}\Vect)}(\wedge^n X, X)_{2-n}.
\end{eqnarray*}
\end{lm}

\dem In graded vector spaces, finite tensor products and duals commute in the sense that we have natural isomorphisms 
$$(V \otimes \stackrel{\scriptscriptstyle{n}}{\cdots} \otimes V )^* \cong V^* \otimes \stackrel{\scriptscriptstyle{n}}{\cdots} \otimes V^*$$
(recall that graded vector spaces are finite dimensional for us, following Definition \ref{grvc}). Now it is possible to apply the concept of duality from Definition/Proposition \ref{duality} to get the first equality. For the second one we only need to use the d\'ecalage isomorphism for linear maps from Equation \ref{dec2}.
\qed

If $X$ is a pro-graded vector space, the comultiplications $S^{n-1} X \rightarrow S^n X$ are all pro-linear. Similarly, if $Y$ is an ind-graded vector space, the multiplications $S^{n+1} Y \rightarrow S^n Y$ are all ind-linear. (In both cases the operations happen at every $X(i)$ or $Y(j)$ respectively.) We can treat $S(X)$ and $S(Y)$ as $\mathbb{Z}$-graded vector spaces since that category, without the finite dimensionality condition, has all products. For every finite polynomial degree $S^n(X)$ and $S^n(Y)$ are cocompact and compact on $\mathbb{Z}$-$\Vect$ respectively. In that way, we can view pro- and ind-graded linear maps as special cases of graded linear maps. Following that spirit, we can define pro-coderivations and ind-coderivations of $S(X)$ and $S(Y)$ respectively using Lemma \ref{lee}.

\begin{df}\label{proco}
Let $X$ be a pro-graded vector space and $Y$ an ind-graded vector space. Treating $S(X)$ and $S(Y)$ as $\mathbb{Z}$-graded vector spaces, we define a degree $i$ pro-coderivation of $S(X)$ to be a degree $i$ coderivation of $S(X)$ such that the associated maps in $\ihom\left(S\left(X\right), X\right)_{i}$ are pro-graded linear.

Similarly, a degree $i$ ind-derivation of $S(Y)$ is a degree $i$ derivation of $S(Y)$ such that the associated maps in $\ihom\left(Y, S\left(Y\right)\right)_{i}$ are ind-graded linear. 
\end{df}

The equivalent of Theorem \ref{tml} for pro- and ind-graded vector spaces is now immediate, where the two first statements are true by definition.

\begin{tm}\label{tmm}
	Let $X$ be a pro-graded vector space concentrated in non-positive degrees (each $X(i)$ is concentrated in non-positive degrees for all $i$). Then we have the following one to one correspondences:
	\begin{enumerate}
		\item Degree $1$, square to zero pro-coderivations of  $S\left([1]X\right)$ correspond to {\Linf} structures on $X$ inner to \mbox{$\textrm{Pro}\left(\mathbb{Z}\right.$-$\left.\Vect\right)$}, in other words to collections of pro-graded linear maps $\{l_n \in \ihom_{\textrm{Pro}(\mathbb{Z}\textrm{-}\Vect)}\left( \wedge^n X, X \right)_{2-n} \}_{n\in \mathbb{N}}$ satisfying \ref{jacojon}.
		\item Degree 1, square to zero ind-derivations of $S \left([-1]X^*\right)$ correspond to families of pro-graded linear maps $\{q_n \in \ihom_{\textrm{Pro}(\mathbb{Z}\textrm{-}\Vect)}\left(S^n\left(X \right), X \right)_{1} \}_{n \in \mathbb{N}}$ satisfying equation \ref{jaco2}.
		\item Families of pro-graded linear maps $\{q_n \in \ihom_{\textrm{Pro}(\mathbb{Z}\textrm{-}\Vect)}\left(S^n\left([1]X \right), [1]X \right)_{1} \}_{n \in \mathbb{N}}$ satisfying \ref{jaco2} and $\{l_n \in \ihom_{\textrm{Pro}(\mathbb{Z}\textrm{-}\Vect)}\left( \wedge^n X, X \right)_{2-n} \}_{n\in \mathbb{N}}$ satisfying \ref{jacojon} are in one to one correspondence.
		\item Degree 1, square to zero ind-derivations of $S \left([-1]X^*\right)$ are in one to one correspondence with degree $1$, square to zero pro-coderivations of  $S\left([1]X\right)$.
	\end{enumerate} 
\end{tm}

\dem Item $3$ is a direct consequence of Lemma \ref{lee2} and the third statement of Theorem \ref{tml} for $\mathbb{Z}$-graded vector spaces. Item $1$ is simply the definition of pro-coderivation (Definition \ref{proco}) together with the formula of the Jacobiators from Theorem \ref{tml}. Similarly, the same argument applies for item $2$, but now again we have to use the duality results from Definition/Proposition \ref{duality}. Finally, item $4$ is a corollary of all the other three statements together.
\qed 

This is the main theorem of this section. It basically says that the different definitions of \Linf s agree on pro-graded vector spaces, lowering the usual condition of finite dimensionality.

\begin{df}
	A pro-{\Linf} is a pro-graded vector space $X$ provided with a family of pro-linear maps $\{l_n \colon \wedge^n X \rightarrow X\}_{n\in \mathbb{N}}$ of degree $2-n$ such that for every $n \in \mathbb{N}$, $n \geqslant 1$ and every $x = x_1 \otimes \cdots \otimes x_n$ in $L^{\otimes n}$
	\begin{equation}\label{jacolinf}
	\sum_{i+j=n+1}\sum_{\sigma \in \mathcal{S}h_{j-1}^i} (-1)^{j(i-1)} (-1)^{\sigma} \epsilon(\sigma) l_j(l_i(x_{\sigma(1)} \wedge \cdots \wedge x_{\sigma(i)}) \wedge x_{\sigma(i+1)} \wedge \cdots \wedge x_{\sigma(n)})=0.
	\end{equation}
\end{df}

\begin{rk}
	The only condition for pro-{\Linf s} to be equivalent to all the other notions from Theorem \ref{tmm} is the concentration in non-positive degrees.
\end{rk}

\subsection{Pro- and ind-graded vector bundles}

The same arguments used in the previous section apply when we replace $\mathbb{Z}$-$\Vect$ by $\mathbb{Z}$-$\VBun$ the category of total finite dimensional, $\mathbb{Z}$-graded smooth vector bundles. Similarly to what was discussed on the previous section, sections of such vector bundles are both graded vector spaces and graded $\C(M)$-modules, where $M$ is the common base to all the bundles. Since there is no limitation to take duals due to Theorem \ref{tmm}, we have a new version of Vaintrob's result in the ind/pro-setting:

\begin{df}
Let $A \rightarrow M$ be a pro-finite dimensional graded smooth vector bundle concentrated in non-positive degrees. A pro-{\Linfoid} structure on $A$ is a degree 1, square to zero ind-derivation of the $\C(M)$-algebra $S_{\C(M)}^-([-1]\A^*) = \Gamma(M, S^-([-1]A^*))$.
\end{df}

In our case this is not written as a theorem but as a definition since we have not introduced higher anchors and higher brackets on \Linfoid s (see Definition \ref{oid}). The closest we can get to a result, without entering into any further details is the following: 

\begin{tm}[Pro-Vaintrob]
	Let $A \rightarrow M$ be a pro-finite dimensional graded smooth vector bundle concentrated in non-positive degrees. Pro-{\Linfoid} structures on $A$ are in one to one correspondence to pro-graded version of the concept of {\Linfoid} structures on $A$ in the sense of higher brackets and higher anchors from Bruce \cite{BRU}.
\end{tm}


\section{Local \boldmath{\Linf s}}

{\it Finite jet bundles of a vector bundle are vector bundles again. Pro-\Linf s on the infinite jet bundle of a vector bundle are called local. A linear version of Peetre's theorem characterizes which insular differential operators are linear. Using that result it is possible to compare local-\Linf s to polydifferential operators. We finish the section by comparing local \Linf s and \Linf s made up from local forms. The main references in this section are Costello and Gwilliam \cite{CG1} and \cite{CG2}; and K\'ola\v{r}, Michor, and Slov\'ak \cite{KMS}.}\\

We have considered so far fiber bundles $\pi \colon E \rightarrow M$, its associated jet bundles and results involving locality on $\E \times M$ where $\E$ is the set of smooth sections of $\pi$. When talking about \Linf s we need a linear and graded version of those concepts. As a mater of fact, everything becomes linear in the appropriate way when we start with a smooth graded-{\it vector} bundle $\pi \colon V \rightarrow M$ to begin with.

\begin{pp}[Michor \cite{MIC}]
	The finite jet bundles of a smooth graded-vector bundle $\pi \colon V \rightarrow M$ are smooth graded-vector bundles over $M$. The sumbersions $\pi_k^l \colon J^k V \rightarrow J^l V$ are vector bundle morphisms of degree $0$ for every pair $k \geqslant l$.
\end{pp}

\begin{rk}
Now, $J^{\infty} V$ can be viewed as a {\it pro-graded vector bundle} instead of as a pro-smooth manifold. Observe that now we are restricting to bundle maps as morphisms. We will still denote the infinite jet bundle in the same way, but we should keep in mind in this section the change of pro-category that we are working with.
\end{rk}

Peetre's theorem in a linear fashion also holds and has interesting consequences for us at this moment:

\begin{tm}[Linear version of Peetre's Theorem]\label{Pee8}
	Let $E, F \rightarrow M$ be two graded vector bundles. Denote the associated (graded) sheaves of sections by $\E$ and $\F$ respectively. Let $A(M) \colon \E(M) \rightarrow \F(M)$ be an $\RE$-linear, germ-local map.
	
Then, for every compact submanifold $K \subset M$ there exists $k$ a natural number such that $\restrict{\left( A(M) \times \textrm{id} \right)}{\E(M) \times K}$ is a $k$-th order insular differential operator map. That means, there exists a vector bundle map $A^0 \colon J^ k E \rightarrow F$ such that the following diagram commutes:
	\begin{center}
		\begin{tikzpicture}[description/.style={fill=white,inner sep=2pt}]
		\matrix (m) [matrix of math nodes, row sep=3em,
		column sep=3.5em, text height=1.5ex, text depth=0.25ex]
		{ \E(M) \times K & \F(M) \times K \\
			J^k E  & F \\};
		\path[->,font=\scriptsize]
		(m-1-1) edge node[auto] {$A(M) \times \textrm{id}$} (m-1-2)
		(m-1-1) edge node[auto] {$j^k$} (m-2-1)
		(m-2-1) edge node[auto] {$A^0$} (m-2-2)
		(m-1-2) edge node[auto] {$j^0$} (m-2-2);
		\end{tikzpicture}
	\end{center}
\end{tm}

The proof can be found \cite{KMS}, the book by K\'ola\v{r}, Michor, and Slov\'ak. They do not consider the graded case, but that is irrelevant for the proof of the theorem. It can be seen as a consequence of the non-linear Peetre's theorem discussed earlier in this Thesis (Corollary \ref{Pee25}), by observing that finite jet evaluations are $\RE$-linear (and not $\C(M)$-linear).

A map in which the source is $\V_1 \times \cdots \times \V_n$ and which satisfies the commutativity diagram in Theorem \ref{Pee8} is called a {\it polydifferential operator} for many authors. We mention in particular Costello and Gwilliam \cite{CG2} and Costello \cite{COS}.

\begin{lm}\label{je}
	Given $V_1, \ldots, V_n \rightarrow M$ smooth vector bundles, multilinear insular maps along the identity $\V_1 \times \cdots \times \V_n \rightarrow \V_{n+1}$ are polydifferential operators in the sense of Costello \cite{COS}.
\end{lm}

\dem The proof is a consequence of the linear version of Peetre's theorem, \ref{Pee8}.
\qed

The interesting consequence is now that insular maps which are $\RE$-linear are covered by morphisms $J^{\infty} E \rightarrow J^{\infty} F$ in the category of pro-graded vector bundles. Now we can apply the machinery of pro-\Linf s and pro-\Linfoid s developed in the previous section. Even more, duals are particularly well behaved with respect to finite jet bundles: $J^k (V^*) \cong (J^k V)^*$ for all $k$. This allow us to look at maps $\V^* \times M \rightarrow (J^k V)^*$.

\begin{df}\label{llinf}
	Let $V \rightarrow M$ be a smooth graded vector bundle. An {\Linf} structure on $\V \defeq \Gamma(M, V)$ over $\RE$ is called local if all the structure maps are local.
\end{df}

Due to Peetre's linear Theorem \ref{Pee8} these notions are equivalent to having pro-{\Linf} structures on the associated infinite jet bundle. This is just simply by definition.

\begin{pp}\label{sus}
	Let $V \rightarrow M$ be a smooth graded vector bundle. If $V$ is concentrated in non-negative degrees, a family of local maps $\{l_n \colon \wedge^n \V \rightarrow \V \}_{n\in \mathbb{N}}$ of degree $2-n$ is a local {\Linf} if the associated maps between the infinite jet bundles induce the structure of a pro-{\Linf} on $J^{\infty} V$.
\end{pp} 

\begin{rk} Similarly, if in a local {\Linf} all maps are $\C(M)$-linear, and we also include maps to account for the non-reduced symmetric algebra $S_{\C(M)}^-([-1]\V^*)$, we get a pro-{\Linfoid} structure on the infinite jet bundle by definition.
\end{rk}

 Costello and Gwilliam define \cite{CG2} a local {\Linf} as an {\Linf} in which all maps are polydifferential operators.

\begin{pp}
	A local {\Linf} in the sense of Costello and Gwilliam \cite{CG2} is equivalent to a local {\Linf} in the sense of Definition \ref{llinf}.
\end{pp}

\dem The proof is a consequence of Lemma \ref{je} and Proposition \ref{sus} above.
\qed

In particular, the passage from derivations to families of polydifferential operators in the work of Costello and Gwilliam is also justified due to Theorem \ref{tmm}. Costello developed \cite{COS} a different symmetric additive category (that of differential operators along the identity) which he takes as a basis to construct local \Linf s. The previous proposition is simply a consequence of the theory of insular differential operators along the identity: the two categories are the same.
 
Local \Linf s do not only have multilinear and local properties, but also sheaf-like properties. More interestingly, if we replace $\V$ by $\V_{c}$, the cosheaf of compactly supported sections, such local \Linf s are factorization algebras. Factorization algebras have become a very important tool in quantum field theories after the work of Costello and Gwilliam \cite{CG1} (in particular perturbative quantum field theories). The result about the compactly supported sections can be found in the second volume of their book, \cite{CG2}.

\subsection{Two \boldmath{\Linf s}: local \boldmath{\Linf} of forms versus \boldmath{\Linf} of local forms}\label{lall}

In the following part we will study how Lie and \Linf s on $\Omega^{\bullet}(X)$ for a finite dimensional manifold $X$ can be generalized to Lie- and \Linf s on $\Omega_{\textrm{loc}}^{\bullet, \bullet}(\E \times M)$. We want to point out that, even when the vector space underlying an {\Linf} has the adjective {\it local}, as in {\it local forms}, it does not mean necessarily that the {\Linf} is local. By this we mean: $\E \times M$ is not a finite dimensional smooth manifold, so that $\Omega^{p, q}(\E \times M)$ cannot be seen as the space of smooth sections of a finite rank smooth vector bundle over it. We are not in the hypothesis of the definition of local \Linf s, Definition \ref{llinf}. One could argue that we can consider $\E \times M$ as a Fr\'echet manifold and we can try to replicate the study done in this chapter for Fr\'echet bundles and Fr\'echet vector spaces. This is also not a good solution: all Fr\'echet spaces are not reflexive: the dual of a Fr\'echet space is only Fr\'echet when the space was Banach to begin with (see Dodson, Galanis, and Vassiliou \cite{DGV} for example). The \Linf s where the vector spaces are local forms are called {\it \Linf s of local forms}. On the other hand if we are given a local {\Linf} structure on $\V^{\bullet} = \Omega^{\bullet}(X)$, it will be called a {\it local {\Linf} of forms}.

We could still think about $\Omega^{p, q}(\E \times M)$ as a $\C_{\textrm{loc}}(\E \times M)$-module. As such, we could think of \Linf s of local forms as \Linf s on $\wedge^{\bullet, \bullet} \mathrm{T} (\E \times M)$. As mentioned before, due to the problems with taking duals, Vaintrob's theorem will not apply in this setting, making the compared two notions inequivalent. But, as a matter of fact, we will not have to worry about this issue at all. We are {\it not} going to be working with such \Linfoid s since we will have vector subspaces of $\Omega^{p, q}(\E \times M)$ which are not $\C_{\textrm{loc}}(\E \times M)$-modules. Hence, we will not have an associated Fr\'echet bundle in which we wish to consider {\Linfoid} structures. This topic will be discussed when defining the corresponding \Linf s in the next part, but we wanted to be explicit about it now since we have spent some time talking about \Linfoid s and their relation to polydifferential operators.


\printbibliography

\setcounter{part}{5}
\setcounter{chapter}{14}


\newpage
\part{\boldmath{\Linf} of local observables}\label{pmim}

\newpage
\tableofcontents

\chapter*{\color{darkdelion} \boldmath{\Linf} of local observables}

A pre-$m$-symplectic manifold is a finite dimensional smooth manifold endowed with a closed differential form of degree $m$. The study of the dynamics generated by that form is done via the study of the Hamiltonian vector fields and forms in a similar way to what is done is pre-symplectic geometry. There is a higher generalization of the Poisson bracket on Hamiltonian forms in higher pre-multisymplectic manifolds. The construction is original to Rogers \cite{ROG} and provides an {\Linf} of Hamiltonian forms.\\

In the world of insular manifolds it is possible to talk about local pre-multi\-symplectic forms, insular Hamiltonian vector fields and local Hamiltonian functions. Since the cohomology of the variational bicomplex is concentrated on surface, Hamiltonian pairs are easily studied by what occurs on surface.\\

The main example of a local pre-multisymplectic form will be the closed form $\omega$ extracted via the fundamental formulae theorem out of a Lagrangian field theory. We will call that form the Poincar\'e-Cartan form. In this case, the surface part of a Hamiltonian pair is a Noether pair. The Poincar\'e-Cartan form simplifies on shell, where it can be further integrated along a Cauchy hyper-surface of the base to get a $2$ form on the space of extrema of the action. That space together with that $2$ form is precisely the covariant phase space of the original Lagrangian field theory.\\

The {\Linf} of Hamiltonian pairs associated to the Poincar\'e-Cartan form is called the {\Linf} of local observables. By adding certain $D$-exact term to the $2$-bracket of two local observables we get a bilinear map which splits by depth into well defined brackets. On surface, we recover the usual bracket on Noether currents and on depth one, the Poisson bracket on Hamiltonian functions on the covariant phase space (after the choice of a Cauchy hyper-surface).\\

{\it This part reviews the theory of \Linf s coming out of finite dimensional pre-multisymplectic manifolds. We present a generalization of that construction to the category of insular manifolds, showing explicit formulas for symplectic and Hamiltonian vector fields. These formulas are simple due to the particular cohomology of the bicomplex of local forms. Furthermore, we present the Poincar\'e-Cartan form, a universal pre-$\textsf{top}$-symplectic form associated to any Lagrangian. We study how that form restricts on shell, after integration along a Cauchy hyper-surface, to the pre-symplectic structure on the phase space. Finally, we present the {\Linf} of Hamiltonian pairs for the Poincar\'e-Cartan form, called the {\Linf} of local observables and we study how to interpret the $2$-bracket both as a bracket on currents and as a bracket on Hamiltonian functions on the phase space. The basic references in this part are Deligne and Freed \cite{DEL}; GiMmsy \cite{GIMMSY}; Rogers \cite{ROG} and together with Callies, Fr\'egier, and Zambon \cite{FRZ}; and Zuckerman \cite{ZUC}.}


\newpage
\chapter{Finite dimensional pre-multisymplectic geometry}

A symplectic manifold is a finite dimensional smooth manifold provided with a $2$ form which is closed and non-degenerate. The form is said to be symplectic, and if we drop the non-degeneracy condition, pre-symplectic. A way of getting higher analogues to that construction is to let the degree of the form be different than $2$: that is the notion of a pre-multisymplectic manifold.\\

The study of the dynamics generated by a pre-symplectic form is done through the concepts of Hamiltonian vector fields and Hamiltonian functions. There is an associated bracket on the later, known as the Poisson bracket, which endows the space of Hamiltonian functions with the structure of a Lie algebra. The higher generalization of the form degree on the pre-symplectic side gives a higher generalization, in terms of higher brackets, on the Poisson side. The corresponding structure is an {\Linf}, called of Hamiltonian forms.\\

{\it In this short chapter we review the general definitions and results in pre-multi\-symplectic geometry regarding Hamiltonian actions and co-momentum maps. We will follow Rogers \cite{ROG}, his work with Callies, Fr\'egier and Zambon \cite{FRZ}, and my previous work on the topic \cite{YO}.}


\section{Pre-multisymplectic forms}


Pre-multisymplectic manifolds are higher analogues to presymplectic manifolds, where higher means that the form $\omega$ can be of a degree higher than $2$. We are interested in Hamiltonian actions of a Lie groupoid in a pre-multisymplectic manifold. For that we need to define symplectic and Hamiltonian vector fields, higher analogues of the Poisson bracket on Hamiltonian functions, and comomentum maps. There are different approaches to these notions, for example the work of GiMmsy \cite{GIMMSY}. We follow the ideas from Rogers, first developed in his PhD thesis \cite{ROG} and later in a joint paper with Callies, Fr\'egier, and Zambon \cite{FRZ}. My Master thesis \cite{YO} continued that approach, by generalizing all the construction to multi-vector fields and higher \Linf s acting on a pre-multisymplectic manifold. As it will be mentioned later, we do not need the full generality of that thesis in the present document. That is why we summarize the important results and notations for comomentum maps in pre-multisymplectic manifolds following Callies, Fr\'egier, Rogers, and Zambon \cite{FRZ}. In this section, we have decided to denote our basic smooth manifold by the letters $Y$. The reason to do so is that, later, we will be interested in generalizing these concepts to the world of local forms on $\E \times M$, hence replacing $Y$ by $\E \times M$ (replacing $M$ by $\E \times M$ could potentially cause some confusion).

\begin{df}[Pre-multi-symplectic manifold]
	A differential form $\omega$ of degree $(m+1)$ on a manifold $Y$ is said to be pre-$m$-symplectic if it is closed. The pair $(Y,\omega)$ is called a pre-$m$-symplectic manifold.
\end{df}

We recover pre-symplectic manifolds if $m=1$. The step from pre-multi-symplectic to multi-symplectic structures is like the one from pre-symplectic to multi-symplectic manifolds.

\begin{df}[Multi-symplectic manifold]
	A form $\omega \in \Omega^{m+1}(Y)$ is called non-degenerate if $\widetilde{\omega} : \mathfrak{X}(Y) \rightarrow \Omega^{m}(Y)$ given by $\widetilde{\omega}(X):=\iota_X \omega$ is injective. A form $\omega$ in $\Omega^{n+1}(Y)$ is called $m$-symplectic if it is both closed and non-degenerate. The pair $(Y, \omega)$ is then called an $m$-symplectic manifold.
\end{df}

It is possible to extend the definition of symplectic and Hamiltonian vector fields to the multi-symplectic case.

\begin{df}[Multi-symplectic vector field]\label{zz}
	Let $(Y,\omega)$ be a pre-$m$-symplectic manifold. A vector field $X \in\, \mathfrak{X}(Y)$ is called $m$-symplectic, or simply symplectic, if $\mathcal{L}_{X}\omega=0$.
\end{df}

Since $\omega$ is closed $\mathcal{L}_X \omega = d\iota_X \omega$, and hence $m$-symplectic vector fields are precisely those for which $\widetilde{\omega}(X)$ is closed. The set of $m$-symplectic vector fields is an $\RE$-vector subspace of $\mathfrak{X}(Y)$ and it is denoted by $\mathfrak{X}_{\textrm{sym}}(Y)$.

\begin{df}[Hamiltonian vector fields, forms and pairs]\label{zzz}
	Let $(Y,\omega)$ be a pre-$m$-symplectic manifold.
	\begin{itemize}
		\item A vector field $X$ in $\mathfrak{X}^(Y)$, is called Hamiltonian when there exists $\beta \in \, \Omega^{m-1}(Y)$ such that $\iota_{X}\omega=d\beta$.
		\item A differential form $\beta \in \Omega^{m-1}(Y)$ is called Hamiltonian if there exists a Hamiltonian  vector field $X$ such that $\iota_X \omega = d \beta$.
		\item  A pair $(X, \beta) \in \mathfrak{X}(Y) \times \Omega^{m-1}(Y)$ such that $\iota_X \omega = d \beta$ is called a Hamiltonian pair.
	\end{itemize}
	The $\RE$-vector space of Hamiltonian vector fields is denoted by $\mathfrak{X}_{\textrm{ham}}(Y)$, that of Hamiltonian forms by $\Omega_{\textrm{ham}}^{m-1}(Y)$, and that of Hamiltonian pairs by $P_1(Y, \omega)$. 
\end{df}

All these four last definitions have followed Rogers \cite{ROG}.

Hamiltonian vector fields are precisely those such that $\widetilde{\omega}(X)$ is exact. Thus, Hamiltonian vector fields are in particular symplectic. In the case $m=1$, and $\omega$ non-degenerate (that is, in the symplectic case), $\mathfrak{X}_{\textrm{ham}}(Y)$ is also a $\C(Y)$-submodule of $\mathfrak{X}(Y)$. By the Cartan calculus formulas, $\mathfrak{X}_{\textrm{sym}}(Y)$ and $\mathfrak{X}_{\textrm{ham}}(Y)$ are closed under the bracket, and hence they are Lie sub-algebras of $\mathfrak{X}(Y)$. See the computations for $X_1$ and $X_2$ Hamiltonian:
\begin{equation}\label{l2}
\iota_{[X_1, X_2]} \omega = [\mathcal{L}_{X_1}, \iota_{X_2}] \omega = d \iota_{X_1} \iota_{X_2} \omega.
\end{equation}

In the right hand side of the previous equation, we can treat $X_1 \wedge X_2$ as a bivector field on $Y$. More generally, we can talk about multi-vector fields. Those are sections of the $n$-th exterior power of the tangent bundle of $Y$ for every natural number $n$:
$$\mathfrak{X}^{n}(Y) \defeq \Gamma(Y, \wedge^n TY) \cong \wedge_{\mathcal{C}^{\infty}(Y)}^n \mathfrak{X}(Y).$$

Our convention for Lie algebras and \Linf s has been to concentrate the spaces into non-positive degrees. We would like to view $\mathfrak{X}^{n}(Y)$ in degree $-n$. The corresponding graded vector space will be denoted by $\mathfrak{X}^{\bullet}(Y)$. This corresponds to shift $TY$ to degree $-1$. In other words, as graded $\C(Y)$-modules, $\mathfrak{X}^{n}(Y) = S_{\mathcal{C}^{\infty}(Y)}^n \left( [1] \mathfrak{X}(Y) \right)$ (observe the change from exterior products to symmetric products due to the d\'ecalage isomorphism). Since the usual commutator of vector fields endows $\mathfrak{X}(Y)$ with the structure of a Lie algebra, via Theorem \ref{tml} we get an associated Chevalley-Eilenberg differential on multi-vector fields, $\left( S_{\mathcal{C}^{\infty}(Y)} \left( [1] \mathfrak{X}(Y) \right), d_{\textrm{CE}} \right) $:
\begin{equation}\label{CEd}
d_{\textrm{CE}}^n (X_1, \ldots, X_n) = \sum_{\sigma \in \mathcal{S}h_2^{n-2}} \epsilon(\sigma) [ X_{\sigma(1)}, X_{\sigma(2)} ] \wedge X_{\sigma(3)} \wedge \cdots X_{\sigma(n)}.\footnote{Multi-vector fields can be endowed with a Gerstenhaber algebra structure and even an {\Linf} structure (see N.L.D. \cite{YO}). The higher brackets in that {\Linf}, when restricted to vector fields $X_i \in \mathfrak{X}(Y)$ agree with the Chevalley-Eilenberg differential from equation \ref{CEd}. Actually, a pencil of \Linf s originated by that {\Linf} from \cite{YO} has been created by Azimi, Laurent-Gengoux, and Nunes da Costa \cite{ALG}.
}
\end{equation}

Observe that we have decided to drop the up-arrow/down-arrow notation in Equation \ref{CEd} to make the equation more readable. By doing so, it seems that the map is not graded symmetric, but graded antisymmetric. It is important to keep in mind that this is not the case and that we can simply think of $X_i$ to be in degree $-1$ for all $i$. Another observation regarding Equation \ref{CEd}, when using a multi-vector field as the input of a map we will drop the wedge sign, not to cause any confusion with the symmetric or antisymmetric interpretation of the Chevalley-Eilenberg complex of multi-vector fields.

\begin{rk}\label{insertme} We use the following convention for the insertion of a multivector field on a form:
	let $X=X_1 \wedge \cdots \wedge X_n \in \mathfrak{X}^{n}(Y)$ be a multi-vector field and $\beta \in \Omega^{\bullet}(Y)$. The insertion of $X$ into $\beta$ is defined by:
	$$\iota_{X_1 \cdots X_n} \beta := \iota_{X_1} \cdots \iota_{X_n} \beta = \beta(X_n, \ldots, X_1, -).$$
	 Usually the contraction operator is defined to be $\hat{\iota}_{X_1 \cdots X_n} \beta := \beta(X_1, \ldots, X_n, -)$ (see for example the paper of Forger, Paufler, and R\"{o}mer \cite{FPR}). We have chosen a different convention, following the previous work \cite{YO}, since it has some advantages as we will see in Proposition \ref{maincl}. 
\end{rk}

Similarly, the Lie derivative along a multi-vector field $X \in \, \mathfrak{X}^{ n}(M)$ is defined through Cartan's magic formula:
$\mathcal{L}_X := d \iota_X + \iota_X d$. Cartan calculus generalizes to multi-vector fields as done by Forger, Paufler and R\"{o}mer \cite[Proposition A.3]{FPR}. The relevant Corollary out of those formulas for the theory of pre-multisymplectic manifolds is the following: 

\begin{lm}[Fiorenza, Rogers, and Schreiber {\cite[Lemma 3.2.1]{FRS}}]\label{FR}
	Let $X = X_1 \wedge \cdots \wedge X_n$ be a multi-vector field. Then
	$$\mathcal{L}_X - \iota_{d_{\textrm{CE}}^n(X_1, \ldots, X_n)} = (-1)^{n +1}\sum_{\sigma \in \mathcal{S}h_1^{n-1}} \iota_{X_{\sigma(1)} \cdots X_{\sigma(n-1)}} \mathcal{L}_{X_{\sigma(n)}}.$$ 
\end{lm}

As a consequence of the previous lemma, when applied to pre-$m$-symplectic manifolds we get the following:

\begin{pp}[N.L.D. \cite{YO}]\label{maincl}
	Let $(Y, \omega)$ be a pre-$m$-symplectic manifold. For all $X_1, \cdots, X_n$ symplectic vector fields on $Y$ we have that
	$$d \iota_{X_1, \cdots, X_n} \omega - \iota_{d_{\textrm{CE}}^n(X_1,\ldots, X_n)} \omega = 0.$$ In other words, the map
	$$\iota_{-} \omega \colon S_{\mathcal{C}^{\infty}(Y)} \left( [1] \mathfrak{X}_{\textrm{sym}}(Y) \right) \longrightarrow [m]\mathsf{tr}_m(\Omega^{\bullet}(Y))$$
	is a cochain map, where $\mathsf{tr}_m$ denotes truncation in degrees smaller of equal than $m$.
\end{pp}

Callies, Fr\'egier, Rogers, and Zambon {\cite{FRZ}} also have such an result, but using a slight weaker language and in less generality that what is discussed \cite{YO}.


\subsection{Poisson multibrackets}

In symplectic geometry there is a Lie bracket, the Poisson bracket defined on Hamiltonian forms (in that case Hamiltonian functions). If $(Y, \omega)$ is a pre-symplectic manifold, the bracket is given by $\{f, g\} = -\iota_{X_f  X_g} \omega$ where $(X_f, f)$ and $(X_g, g)$ are Hamiltonian pairs. Due to the fact that $\omega$ is antisymmetric, the bracket is antisymmetric as well. If $(Y, \omega)$ is now a pre-$m$-symplectic manifold, the same bracket on Hamiltonian forms, $\{\alpha, \beta\}:= \iota_{X_{\beta} X_{\alpha}} \omega$ does not satisfy the Jacobi identity. The way to address this issue is through an {\Linf}. The idea is original to Rogers \cite{ROG}:

\begin{tm}[{\Linf} of Hamiltonian forms. Rogers \cite{ROG}]\label{rog1}
	Let $(Y, \omega)$ be a pre-$m$-symplectic manifold. Define the following graded vector space
	\[
	L_{n}(Y, \omega) =
	\begin{cases} 
	\hfill \Omega_{ham}^{m-1}(Y)   \hfill & \text{ if $n=1$ } \\
	\hfill \Omega^{m-n}(Y) \hfill & \text{ if $n \in \, \{2, \ldots, m\}$ } \\
	\hfill \{0\} \hfill & \text{ else. }
	\end{cases}
	\]
	together with the following family of brackets is an {\Linf}: $l_1 = -d$ and for every $n \geqslant 2$
	\[
	l_{n}(\beta_1 \otimes \cdots \otimes \beta_{n}) =
	\begin{cases} 
	\hfill \iota_{X_1 \cdots X_{n}} \omega    \hfill & \text{ if $d \beta_i = \iota_{X_i} \omega$ for every $1 \leqslant i \leqslant n$ } \\
	\hfill 0 \hfill & \text{ else. } \\
	\end{cases}
	\]
\end{tm}

\begin{center}
	\begin{tikzpicture}[description/.style={fill=white,inner sep=2pt},descr/.style={fill=white,inner sep=2.5pt}]
	\matrix (m) [matrix of math nodes, row sep=1.5em,
	column sep=2.2em, text height=1.5ex, text depth=0.25ex]
	{ \Omega_{\textrm{ham}}(Y)&\Omega^{m-2}(Y)&\cdots&\Omega^{1}(Y)&\Omega^{0}(Y)\\
		\color{darkdelion}1&\color{darkdelion}2&\color{darkdelion}\ldots&\color{darkdelion}m-1&\color{darkdelion}m\\};
	\path[->,font=\scriptsize]
	(m-1-2) edge node[auto] {$-d$} (m-1-1)
	(m-1-5) edge node[auto] {$-d$} (m-1-4)
	(m-1-4) edge node[auto] {$-d$} (m-1-3)
	(m-1-3) edge node[auto] {$-d$} (m-1-2);
	\end{tikzpicture}
\end{center}

Actually, the {\Linf} in the thesis of Rogers \cite{ROG} differs from this one in a sign: while $l_1$ is $-d$, the higher brackets ($n \geqslant 2$) are defined to be $-\iota_{X_1 \cdots X_n}\omega$. This occurs because of a different convention in defining what a Hamiltonian pair is and also because we are taking the opposite of the Poisson bracket (recall Remark \ref{insertme}). It is possible to replace Hamiltonian forms with Hamiltonian pairs to get another {\Linf}:

\begin{center}
	\begin{tikzpicture}[description/.style={fill=white,inner sep=2pt},descr/.style={fill=white,inner sep=2.5pt}]
	\matrix (m) [matrix of math nodes, row sep=1.5em,
	column sep=2.2em, text height=1.5ex, text depth=0.25ex]
	{ P_1(Y,\omega)&\Omega^{m-2}(Y)&\cdots&\Omega^{1}(Y)&\Omega^{0}(Y)\\
		\color{darkdelion}1&\color{darkdelion}2&\color{darkdelion}\ldots&\color{darkdelion}m-1&\color{darkdelion}m\\};
	\path[->,font=\scriptsize]
	(m-1-2) edge node[auto] {$-d$} (m-1-1)
	(m-1-5) edge node[auto] {$-d$} (m-1-4)
	(m-1-4) edge node[auto] {$-d$} (m-1-3)
	(m-1-3) edge node[auto] {$-d$} (m-1-2);
	\end{tikzpicture}
\end{center}

\begin{tm}[{\Linf} of Hamiltonian pairs. Callies, Fr\'egier, Rogers and Zambon {\cite[Theorem 4.7]{FRZ}}]\label{rog2}
	Let $(Y, \omega)$ be a pre-$m$-symplectic manifold. Define the following graded vector space
	\[
	P_{n}(Y, \omega) =
	\begin{cases} 
	\hfill P_1(Y, \omega)   \hfill & \text{ if $n=1$ } \\
	\hfill \Omega^{m-n}(Y) \hfill & \text{ if $n \in \, \{2, \ldots, m\}$ } \\
	\hfill \{0\} \hfill & \text{ else. }
	\end{cases}
	\]
	together with the following family of brackets is an {\Linf}: \[
	l_{1}(v) =
	\begin{cases} 
	\hfill (0, -d \beta)   \hfill & \text{ if $v=(X,\beta)$ } \\
	\hfill -d v \hfill & \text{ if $|v| \in \, \{2, \ldots, m\}$ } 
	\end{cases}
	\]
	
	\[
	l_{2}(v_1 \otimes v_2) =
	\begin{cases} 
	\hfill ([X_1, X_2], l_2(\beta_1 \otimes \beta_2))   \hfill & \text{ if $v_i=(X_i, \beta_i)$ for $i\in\,\{1,2\}$} \\
	\hfill 0 \hfill & \text{ else. }
	\end{cases}
	\]
	
	\[
	l_{n \geqslant 3}(v_1 \otimes \cdots \otimes v_n) =
	\begin{cases} 
	\hfill l_n(\beta_1 \otimes \cdots \otimes \beta_n) \hfill & \text{ if $v_i=(x_i, \beta_i)$ $\forall i\in\,\{1,\ldots,n\}$} \\
	\hfill 0 \hfill & \text{ else. }
	\end{cases}
	\]
\end{tm}

In the aforementioned reference, this {\Linf} is called Poisson Lie-m-algebra (up two a minus sign on the brackets $l_{n\geqslant 2}$). There is a projection from $P(Y, \omega)$ to $P_1(Y, \omega)$ and further to $\mathfrak{X}_{\textrm{ham}}(Y)$ denoted by $\pi_v$.

\begin{rk}\label{aboutme}
In \cite{YO}, I generalized the above constructions to include lower degree Hamiltonian forms: those are $\beta \in \Omega^{m-n}(Y)$ ($n\geqslant 1$) such that $d \beta = \iota_{X_1 \cdots X_n} \omega$ for $n$ vector fields $X_i \in \mathfrak{X}(Y)$. That {\Linf} encodes higher symmetries of $\omega$. In the Lagrangian field theory setting, it would include not only Lie algebroid symmetries (gauge symmetries) but also {\Linfoid} symmetries. Since we do not known of a physical theory with such families of symmetries, and for the sake of conciseness and clarity in our discussion, we have preferred not to enter into the theory developed \cite{YO}. From the sheer mathematical point of view, studying such \Linf s in the insular world is perfectly plausible and it is an open line of research.
\end{rk}

The last bit of information about multi-symplectic geometry that we want to address is that of symplectic and Hamiltonian actions of Lie algebras on a pre-$m$-symplectic manifold. Let $\mathfrak{g}$ be a Lie algebra acting by symplectic vector fields on the symplectic manifold $(Y, \omega)$:
$$\mathsf{a} \colon \mathfrak{g} \longrightarrow \mathfrak{X}_{\mathrm{sym}}(Y).$$
In symplectic geometry a comomentum map (see the book of Cannas da Silva \cite[Section 18.1]{DS}, for example) is a morphism of Lie algebras $\mu \colon \mathfrak{g} \rightarrow \C(Y)$ such that $\iota_{\mathsf{a}x} \omega = d \mu(x)$ for all $x \in \, \mathfrak{g}$ (that condition is called the {\it generating function condition}). In particular, the image of $\mathsf{a}$ lies within the space of Hamiltonian vector fields: $\mathsf{a} \colon \mathfrak{g} \rightarrow \mathfrak{X}_{\mathrm{ham}}(Y)$. A symplectic action is said to be {\it Hamiltonian} if there exists a comomentum map for the action.

Similarly, by looking at the equation $\iota_{\mathsf{a}x} \omega = d \mu(x)$ and remembering that the insertion of vector fields is a cochain map (Proposition \ref{maincl}), in multisymplectic geometry one defines:

\begin{df}[Homotopy momentum map, Callies, Fr\'egier, Rogers, and Zambon \cite{FRZ}]\label{dee} Let $\mathfrak{g}$ be a Lie algebra acting by ${\sf a}$ on the pre-$m$-symplectic manifold $(Y,\omega)$ by symplectic vector fields. Denote the associated cochain map using the Chevalley-Eilenberg complexes of both $\mathfrak{g}$ and $\mathfrak{X}(Y)$ by $\mathsf{f} \colon S^{\bullet} \left([1]\mathfrak{g}\right) \rightarrow [m]\mathsf{tr}_m(\Omega^{\bullet}(Y))$. A map $\mathsf{h}$ is defined to be a homotopy momentum map for the action if it defines a cochain null-homotopy
	$$(\mathsf{h},\mathsf{f}) \colon E \longrightarrow [m]\mathsf{tr}_m(\Omega^{\bullet}(Y)).$$
	We say that the action is Hamiltonian if there exists a homotopy momentum map for it.
\end{df}

The interesting fact about this definition is that it is related to the {\Linf} of Hamiltonian pairs defined above:

\begin{tm}[Callies, Fr\'egier, Rogers, and Zambon \cite{FRZ}]\label{tee}
	Let $\mathfrak{g}$ be a Lie algebra acting by ${\sf a}$ on the pre-$m$-symplectic manifold $(Y,\omega)$ by symplectic vector fields. A homotopy moment map on $\mathfrak{g}$ is equivalent to a lift of the action $\mathsf{a}$ to $P(Y, \omega)$ in the category of \Linf s:
	\begin{center}
		\begin{tikzpicture}[description/.style={fill=white,inner sep=2pt}]
		\matrix (m) [matrix of math nodes, row sep=1.5em,
		column sep=3.5em, text height=1.5ex, text depth=0.25ex]
		{ & P(Y, \omega) \\
			\mathfrak{g} & \mathfrak{X}(Y) \\};
		\path[->,font=\scriptsize]
		(m-2-1) edge node[auto] {${\sf a}$} (m-2-2)
		(m-1-2) edge node[auto] {$\pi_v$} (m-2-2)
		(m-2-1) edge node[auto] {$\tilde{\mathsf{h}}$} (m-1-2);
		\end{tikzpicture}
	\end{center}
\end{tm}

The work of Callies, Fr\'egier, Rogers, and Zambon \cite{FRZ} also includes the study of obstructions for the existence of homotopy momentum maps.

In \cite{YO}, I extended the results by Callies, Fr\'egier, Rogers and Zambon \cite{FRZ}, by including both Hamiltonian and multi-Hamiltonian {\Linf} actions. As mentioned in Remark \ref{aboutme} we are not interested in {\Linf} actions in field theories, so that we have decided not to include those definitions and results in the present document.


\newpage
\chapter{Local pre-multisymplectic geometry}

The theory of pre-multisymplectic manifolds easily generalizes to the category of insular manifolds $\E \times M$. The relevant complex of forms is that of local forms: a pre-multisymplectic form is a closed form in that complex. Symmetries of the theory (both symplectic and Hamiltonian vector fields) are required to be insular. The particular cohomology of that complex, which is concentrated in surface, turns the conditions for a pair to be Hamiltonian into a condition involving the surface part of the form only.\\

When an insular manifold is endowed with a Lagrangian, the fundamental formulae theorem provides a closed form: $\omega$, the sum of the Euler-Lagrange term and the universal conserved current. It is possible to view it as a local pre-multisymplectic form. The surface part of a Hamiltonian pair in this case is a Noether pair. There are explicit formulas and diagrams both for symplectic and for Hamiltonian vector fields with respect to this form.\\

Since the local form $\omega$ agrees with the Poincar\'e-Cartan form for first order Lagrangians, we adopt that name both for $\omega$ and for the pre-multi-symplectic insular manifold $(\E \times M, \omega)$ (also in any arbitrary jet degree of the Lagrangian). The Poincar\'e-Cartan form simplifies on shell, where it can be further integrated along a codimension $1$-submanifold of the base to get a $2$ form on the space of extrema of the action. That space together with that $2$ form is precisely the covariant phase space of the original Lagrangian field theory.\\

{\it This chapter translates the theory of pre-multi-symplectic structures into the world of insular manifolds. It explains how to go from a Lagrangian to a local pre-multi-symplectic form and how that form, called the Poincar\'e-Cartan form, restricts on shell, after integration along a Cauchy hyper-surface, to the pre-symplectic structure on the phase space. The basic references in this section are GiMmsy \cite{GIMMSY}, Rogers \cite{ROG}, and Zuckerman \cite{ZUC}.}


\section{The Poincar\'e-Cartan pre-multisymplectic insular manifold}\label{df}

{\it Pre-multisymplectic manifolds in the world of insular geometry are insular manifolds endowed with a local form whose total differential is zero. Symplectic and Hamiltonian vector fields are defined in a similar way to their finite dimensional counterparts, only asking the extra condition that the vector fields are insular to begin with. All the relevant information about Hamiltonian vector fields is encoded in the surface parts of the bicomplex of local forms. The Euler-Lagrange term plus the universal conserved current, viewed as a local pre-multisymplectic form has Noether pairs as Hamiltonian pairs on surface. The associated pre-multisymplectic insular manifold is called the Poincar\'e-Cartan insular manifold and it is the main object of study in what follows in this thesis. There are no relevant bibliographical references at this point.}\\

We fix a smooth fiber bundle $\pi \colon E \rightarrow M$ with space of smooth sections $\E$. The following definition is the translation to local forms of the theory of pre-multisymplectic manifolds described in the previous section.

\begin{df}[Local pre-$m$-multisymplectic form]\label{lms} A local form of total degree $m+1$, $\omega \in \Omega_{\textrm{loc}}^{m+1}(\E \times M)$, is called a local pre-$m$-multisymplectic form if $D \omega = 0$. The pair $(\E \times M, \omega)$ is called a pre-$m$-symplectic insular manifold.
\end{df}

As a remark, we want to point out that the non-degeneracy condition does not usually hold in local pre-multisymplectic geometry. In other words, local $m$-symplectic insular manifolds do not exist for $m \neq 0$. The argument for $m \geqslant \textsf{top}$ is the following: even if $\iota_{-} \colon \mathfrak{X}_{\textrm{ins}}(\E) \rightarrow \Omega_{\textrm{loc}}^{0, m}(\E \times M)$ were non-degenerate, in the next step we have $\iota_{\xi} \omega_0 = - \iota_{X} \omega_1$ which cannot be possibly solved on $X$ for all $\xi$ by dimension reasons (we are indexing $\omega$ by depth). The same argument actually works for $m \geqslant 1$, varying the depth index in the proof.

The ideal in local pre-multisymplectic geometry is to consider everything to be insular. Thus, we do not only consider local forms, but also insular vector fields as multisymplectic or Hamiltonian vector fields in a pre-$m$-symplectic insular manifold. The following is the insular version of Definitions \ref{zz} and \ref{zzz}:

 \begin{df}
	Let $(\E \times M, \omega)$ be a pre-$m$-symplectic insular manifold. 
	\begin{itemize}
		\item An insular vector field $\chi \in \mathfrak{X}_{\textrm{ins}}(\E \times M)$ is called symplectic if $\mathcal{L}_{\chi}\omega = D \iota_{\chi} \omega =0$.
		\item An insular vector field $\chi \in \mathfrak{X}_{\textrm{ins}}(\E \times M)$, is called Hamiltonian when there exists $\beta \in \Omega_{\textrm{loc}}^{m-1}(\E \times M)$ such that $\iota_{\chi}\omega=D \beta$.
		\item A local differential form $\beta \in \Omega_{\textrm{loc}}^{m-1}(\E \times M)$ is called Hamiltonian if there exists a Hamiltonian vector field $\chi$ such that $\iota_{\chi} \omega = D \beta$.
		\item  A pair $(\chi, \beta) \in \mathfrak{X}_{\textrm{ins}}(\E \times M) \times \Omega_{\textrm{loc}}^{m-1}(\E \times M)$ such that $\iota_{\chi} \omega = D \beta$ is called a Hamiltonian pair.
	\end{itemize}
	The $\RE$-vector space of symplectic vector fields is denoted by $\mathfrak{X}_{\textrm{sym}}(\E \times M)$, the one of Hamiltonian vector fields by $\mathfrak{X}_{\textrm{ham}}(\E \times M)$, that of Hamiltonian forms by $\Omega_{\textrm{ham}}(\E \times M)$, and that of Hamiltonian pairs by $P_1(\E \times M, \omega)$. 
\end{df}

Making use of the pulley's theorems, we can characterize Hamiltonian forms by their surface parts. Observe that if $\chi$ is a Hamiltonian vector field, it is in particular symplectic, so that $\alpha \defeq \iota_{\chi} \omega \in \Omega_{\textrm{loc}}^{m}(\E \times M)$ is $D$-closed and thus in the hypothesis of the pulley's theorems. We will only work with pre-$m$-symplectic insular manifolds for which $m = \textsf{top}$, so that we are in the hypothesis of the right pulley Theorem \ref{RPT}. Moreover, the form $\iota_{\chi} \omega$ in that case has surface part equal to $\iota_{\xi} \omega_0 \in \Omega_{\textrm{loc}}^{0, \textsf{top}}(\E \times M)$ where $\chi = \xi + X$ is the splitting into evolutionary and total parts.

\begin{cl}\label{ahorita}
	Let $\chi = \xi + X$ be a symplectic vector field in the pre-$\textsf{top}$-symplectic insular manifold $(\E \times M, \omega)$. If there exists $B \in \Omega_{\textrm{loc}}^{0, \textsf{top}-1}(\E \times M)$ such that 
	$$d B = \iota_{\xi} \omega_0$$
	then there exists $\beta \in \Omega_{\textrm{loc}}^{\textsf{top}-1}(\E \times M)$ such that $B = \beta_0$ and $(\chi, \beta)$ is a Hamiltonian pair. In other words, $D \beta = \iota_{\chi} \omega$. Moreover, such a $\beta$ is unique up to a $D$-exact term.
\end{cl}

\dem The proof is a consequence of the right pulley Theorem for $\alpha \defeq \iota_{\chi} \omega$.
\qed

The previous Corollary basically says that the important information about Hamiltonian forms lies on surface. Moreover, it says that, provided $\chi$ is symplectic, the only obstructions for $\chi$ to be Hamiltonian are in the evolutionary part $\xi$ and not on $X$.

In section \ref{ffs} we already talked about a local pre-$\textsf{top}$-multisymplectic form: the Poincar\'e-Cartan form. We recover here its definition (Definition \ref{pcf}):

\begin{df}[Poincar\'e-Cartan manifold] Let $(\E \times M, L)$ be a Lagrangian field theory. The $D$-closed form $\omega \in \Omega_{\textrm{loc}}^{\textsf{top}+1}(\E \times M)$ given by Theorem \ref{FFP} is called the Poincar\'e-Cartan local pre-$\textsf{top}$-multisymplectic form or simply Poincar\'e-Cartan form associated to the Lagrangian $L$. $(\E \times M, \omega)$ is called the Poincar\'e-Cartan pre-$\textsf{top}$-multisymplectic insular manifold associated to the Lagrangian $L$, or simply the Poincar\'e-Cartan insular manifold associated to $L$.
\end{df}

By looking at the equation for Noether's currents (equation \ref{ahora}) and at Corollary \ref{ahorita} it seems clear that conserved currents and Hamiltonian vector fields on the Poincar\'e-Cartan insular manifold are related:

\begin{tm}\label{corr}
	Let $(\E \times M, \omega)$ be the Poincar\'e-Cartan pre-$\textsf{top}$-multisymplectic insular manifold associated to a Lagrangian $L$.
	\begin{enumerate}
		\item If $\chi = \xi + X$ is a Hamiltonian vector field, then $\xi$ is a symmetry of the Lagrangian.
		\item If $(\chi, \zeta)$ is a Hamiltonian pair, then $Z = \zeta_0$ is the Noether current associated to $\xi$, the evolutionary part of $\chi$.
		\item If $\chi = \xi + X$ is a symplectic vector field and $\xi$ is a symmetry, then $\xi$ is a Hamiltonian vector field. 
	\end{enumerate}
\end{tm}

Because of this theorem, we will prefer the notation $\zeta = (Z, \zeta_1, \ldots )$ for Hamiltonian forms, to remind us of the Noether current notation from Noether's first Theorem \ref{NFT}.

\dem Let $\chi = \xi + X$ be an insular vector field.
\begin{enumerate}
	\item If $\chi$ is Hamiltonian, there exists $\zeta \in \Omega_{\textrm{loc}}^{\textsf{top-1}}(\E \times M)$ such that $\iota_{\chi} \omega = D \zeta$. Let $Z = \zeta_0$. The equation on surface is 
	$$\iota_{\xi} \textbf{E} L = (\iota_{\chi} \omega)_0 = (D \zeta)_0 = d Z.$$
	We define $A_{\xi} \defeq Z + \iota_{\xi} \lambda_1$. By working backwards from the proof of Noether's first Theorem \ref{NFT}, we apply equation \ref{ahora} to get $\iota_{\xi} \delta L = d A_{\xi}$ showing that $\xi$ is a symmetry of the Lagrangian.
	\item Similarly to what was done in the previous point, the associated Noether's current to the symmetry $\iota_{\xi} \delta L = d (A_{\xi}) = d (Z + \iota_{\xi} \lambda_1)$ is $Z = Z + \iota_{\xi} \lambda_1 - \iota_{\xi} \lambda_1$ (again using Equation \ref{ahora}).
	\item This last statement is an application of the right pulley theorem once again. If $\chi$ is symplectic, $D (\iota_{\chi} \omega) = 0$. If $\xi$ is a symmetry with associated Noether's current $Z_{\xi}$ we have that $(\iota_{\chi} \omega)_0 = (\iota_{\xi} \textbf{E} L)_0 = d Z_{\xi}$. By the right pulley theorem, Theorem \ref{RPT}, there exists a form $\zeta \in \Omega_{\textrm{loc}}^{\textsf{top}-1}(\E \times M)$ such that $\zeta_0 = Z$ and $D Z = \iota_{\chi} \omega$.
\end{enumerate}
\qed

The second point  of Theorem \ref{corr} says that the surface part of a Hamiltonian pair is a Noether pair (Definition \ref{NPair}). Moreover, the third point says that provided the vector field is symplectic, all Noether pairs appear. We will make a more detailed comment about which Noether pairs actually appear later in Section \ref{random}, but first we want to give explicit formulas for what does it mean for an insular vector field to be symplectic and Hamiltonian respectively:

\begin{figure}[h]\label{besos}
	\centering
	\begin{tikzpicture}[description/.style={fill=white,inner sep=2.5pt},scale=1]
	
	\node (j) at (0,3) {$\iota_{\xi} \mathbf{E} L$};
	\node (k) at (-1,2) {$0$};
	\node (o) at (0,1) {$\iota_{\xi} \omega_1 + \iota_X \mathbf{E} L$};
	\node (oo) at (0,-1) {$\iota_{X} \omega_1$};
	\node (v) at (-1,0) {$0$};
	\node (w) at (-1,-2) {$0$};
	\draw [rounded corners, dashed, thick, teal] (-2.4,-0.4) to[out=45,in=180] (0,1.8);
	\draw [rounded corners, dashed, thick, teal] (-4,-2) to (-2.4,-0.4);
	\draw [rounded corners, dashed, thick, teal] (0, 1.8) to[out=0,in=135] (2.4,-0.4);
	\draw [dotted, thick, teal] (2.6,-0.6) -- (3,-1);
	\draw [rounded corners, dashed, thick, teal] (3.2,-1.2) -- (4, -2) -- (3.8,-2.2);
	
	\path[->,font=\scriptsize]
	(j) edge node[auto] {} (k)
	(oo) edge node[auto] {} (w)
	(o) edge node[auto] {} (v);
	\path[->,font=\scriptsize]
	(o) edge node[auto] {} (k)
	(oo) edge node[auto] {} (v);
	\end{tikzpicture}
	\caption{$\chi = \xi + X$ is a symplectic vector field for the Poincar\'e-Cartan insular manifold.}
\end{figure}
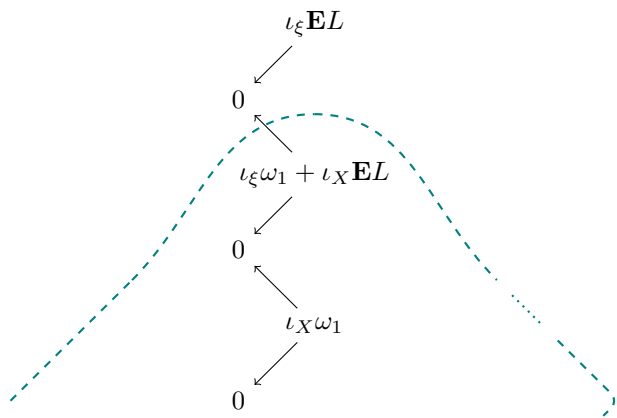

\begin{pp} Let $(\E \times M, \omega)$ be the Poincar\'e-Cartan pre-$\textsf{top}$-multisymplectic insular manifold associated to a Lagrangian $L$. An insular vector field $\chi = \xi + X$ is symplectic if and only if the following equations hold:
\begin{eqnarray}\label{svf}
\mathcal{L}_{\xi} \mathbf{E} L &=& -d \iota_X \mathbf{E} L \nonumber \\
\delta \iota_{\xi} \omega_1 &=& d \iota_X \omega_1 - \delta \iota_X \mathbf{E} L \nonumber \\
\delta \iota_X \omega_1 &=& 0.
\end{eqnarray}
\end{pp}

\dem The equations are the depth by depth translations of $D \iota_{\chi} \omega = 0$ (see Figure \ref{besos}). The only further transformation that we have used is that 
$$d \iota_{\xi} \omega_1 = - \iota_{\xi} d \omega_1 = \iota_{\xi} \delta \mathbf{E} L. $$
That follows from the insular Cartan calculus formulas (Theorem \ref{lcc1}) and the fundamental formulae theorem, Theorem \ref{FF}.
\qed

Similarly for Hamiltonian pairs:

\begin{pp} Let $(\E \times M, \omega)$ be the Poincar\'e-Cartan pre-$\textsf{top}$-multisymplectic insular manifold associated to a Lagrangian $L$. An insular vector field $\chi = \xi + X$ is Hamiltonian with associated Hamiltonian form $\zeta$ if and only if the following equations hold:
\begin{eqnarray}\label{hvf}
\iota_{\xi} \mathbf{E} L &=& d Z \nonumber \\
\iota_{\xi} \omega_1 &=& \delta Z + d \zeta_1 - \iota_X \mathbf{E} L \nonumber \\
\iota_X \omega_1 &=& \delta \zeta_1 + d \zeta_2 \nonumber \\
0 &=& \delta \zeta_i + d \zeta_{i+1} \textrm{ for all } i \geqslant 2
\end{eqnarray}
\end{pp}

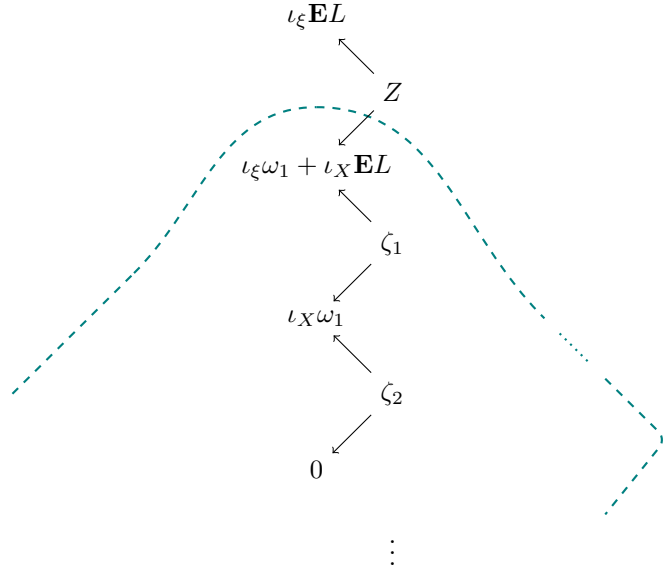
\begin{figure}[h]\label{rambla}
	\centering
	\begin{tikzpicture}[description/.style={fill=white,inner sep=2.5pt},scale=1]
	
	\node (j) at (0,3) {$\iota_{\xi} \mathbf{E} L$};
	\node (z) at (1,2) {$Z$};
	\node (z1) at (1,0) {$\zeta_1$};
	\node (z2) at (1,-2) {$\zeta_2$};
	\node (z3) at (1,-4) {$\vdots$};
	\node (o3) at (0,-3) {$0$};
	\node (o) at (0,1) {$\iota_{\xi} \omega_1 + \iota_X \mathbf{E} L$};
	\node (oo) at (0,-1) {$\iota_{X} \omega_1$};
	\draw [rounded corners, dashed, thick, teal] (-2.4,-0.4) to[out=45,in=180] (0,1.8);
	\draw [rounded corners, dashed, thick, teal] (-4,-2) to (-2.4,-0.4);
	\draw [rounded corners, dashed, thick, teal] (0, 1.8) to[out=0,in=135] (3,-1);
	\draw [dotted, thick, teal] (3.2,-1.2) -- (3.6,-1.6);
	\draw [rounded corners, dashed, thick, teal] (3.8,-1.8) -- (4.6, -2.6) -- (3.8,-3.6);
	
	\path[->,font=\scriptsize]
	(z) edge node[auto] {} (o)
	(z1) edge node[auto] {} (oo)
	(z2) edge node[auto] {} (o3);
	\path[->,font=\scriptsize]
	(z) edge node[auto] {} (j)
	(z1) edge node[auto] {} (o)
	(z2) edge node[auto] {} (oo);
	\end{tikzpicture}
	\caption{$\chi = \xi + X$ and $\zeta = (Z, \zeta_1, \zeta_2, \ldots )$ form a Hamiltonian pair for the Poincar\'e-Cartan insular manifold.}
\end{figure}

\dem In this case, no extra equalities have been changed from the original one $\iota_{\chi} \omega = D \zeta$. The Proposition is a splitting of that equation by depth (see Figure \ref{rambla}). \qed

Observe that by Takens's acyclicity Theorem (Theorem \ref{TAT}), the existence of $\zeta_{i}$ for each $i \geqslant 1$ is assured by asking the remaining terms in each equation to have the same $d$-derivative. For example, for the second equation in \ref{svf}, $\zeta_1$ exists if $d \iota_{\xi} \omega_1 = d \delta Z - d \iota_X \mathbf{E} L$. In particular this means that if $\delta \zeta_2 + d \zeta_{3} = 0$, then $d (\delta \zeta_3) = 0$. Thus there exists $\zeta_4$ such that $\delta \zeta_3 + d \zeta_4 = 0$ and so on. This means that once the forth equation in \ref{svf} holds for $i= 2$, there exists $\{\zeta_i\}_{i \geqslant 4}$ making the equation hold for higher $i$. Again, we refer to the following section, Section \ref{random} for a further interpretation of these equations.


\section{The Poincar\'e-Cartan insular manifold: some interpretations}\label{random}

{\it The Poincar\'e-Cartan local pre-$\textsf{top}$-multisymplectic form restricts on shell to the universal conserved current, seen as a pre-symplectic structure, on shell. Moreover, after fixing a Cauchy hyper-surface, the Poincar\'e-Cartan form restricts to the pre-symplectic structure on the covariant phase space associated to the theory and the hyper-surface. Symplectic and Hamiltonian vector fields of the Poincar\'e-Cartan form give, after restriction, symplectic and Hamiltonian vector fields on the phase space respectively. Noether pairs are symplectic vector fields on shell. The existence of symplectic vector fields is briefly discussed, comparing it to a similar discussion from the Lepagean school. The basic references in this sections are GiMmsy \cite{GIMMSY} and Zuckerman \cite{ZUC}.}\\

\subsection{On-shell versus off-shell}

As mentioned in Remark \ref{shell}, in Lagrangian field theories we are also interested in what happens on shell: this is on the variety of extrema (see Definition \ref{extremal}) of the action. It is often argued that there is a (pre-)symplectic structure on shell, on the so-called, covariant phase space of the theory (see Khavkine \cite{KHA}, for example). This pre-symplectic structure depends on the choice of a Cauchy hyper-surface (a codimension one sumbanifold of $M$, often compact and oriented), see \cite{CAR} for a definition in General relativity or \cite{ONE} for a definition in pseudo-Riemannian geometry. We want to mention a very productive language for talking about the covariant phase space: the BV-BRST formalism. See for example Cattaneo, Mn\"{e}v, and Reshetikhin \cite{CMR}; Costello and Gwilliam \cite{CG2}; or Cattaneo and Schiavina \cite{CS}). The other way of defining such pre-symplectic structures in a general Lagrangian field theory is through the universal conserved current (Theorem \ref{cc}) of Zuckerman \cite{ZUC}.

\begin{df}[Zuckerman {\cite[Definition 11]{ZUC}}]\label{o1N}
	Let $(\E \times M, L)$ be a Lagrangian field theory with associated universal conserved current $\omega_1 = \delta \lambda_1$ coming from the Fundamental Formulae Theorem \ref{FF}. For any given $N \subset M$, compact, oriented, codimension $1$ submanifold we define $\lambda_1^{[N]} \defeq \int_N \lambda_1$ and $\omega_1^{[N]} \defeq \int_N \omega_1$. $\omega_1^{[N]}$ is called the pre-symplectic form on the smooth locus of $\E_{L}$, the space of extrema. The pair is called the covariant phase space. The manifold $N$ is called a Cauchy hyper-surface.
\end{df}

The idea is to define on $\E_L$ a theory of differential forms with respect to the vertical differential $\delta$. The first problem to avoid is the lack of smoothness of this space, and to restrict to the smooth locus of it. The second one is to define which differential forms are we interested in (again, some notion of locality is needed):

\begin{df}[$Q$-local form]\label{qloc}
	Let $\E \times M$ be an insular manifold and $Q \subset M$ be a compact, oriented, dimension $q$ submanifold of $M$. $Q$-local forms of degree $p$ on $\W \subset \E$ are defined as $$\Omega_{Q-\textrm{loc}}^p (\W \times M) \defeq \{ \alpha^{[Q]} \defeq \restrict{\int_Q \alpha}{\W} \colon \alpha \in \Omega_{\textrm{loc}}^{p,q}(\W \times M) \}.$$
	The vertical differential on $Q$-local forms is defined by 
	\begin{eqnarray*}
		\delta \colon \Omega_{Q-\textrm{loc}}^p(\E \times M)  & \longrightarrow & \Omega_{Q\textrm{loc}}^{p+1}(\E \times M) \\
		\alpha^{[Q]} & \longmapsto & (\delta \alpha)^{[Q]}.
	\end{eqnarray*}
	It endows the graded space of $Q$-local forms with the structure of a cochain complex.
\end{df}

When $N=M$ and $M$ is compact, we will call $M$-local forms, int-local.\footnote{The concept of int-local maps first came across our study of the version of the BV-BRST formalism from Costello and Gwilliam \cite{CG2}, where the relevant algebra of functions seems to be that of int-local functions. The terminology is original to Blohmann (private communication).}

\begin{lm}\label{zuzu}
	Given a Lagrangian field theory and a Cauchy hyper-surface, the pre-symplectic form on the phase space $\omega_1^{[N]}$ is closed in the complex of $N$-local forms on the smooth locus of $\E_L$, which depends only on the homology class of the Cauchy hyper-surface $N$. 
\end{lm}

\dem The form is $\delta$-exact by definition, hence $\delta$-closed. The dependence on the homology class of $N$ is clear. It does not have in further dependencies because $\omega_1$ was unique up to a $d$-exact term and since $N$ is compact, the integral of a $d$-exact term vanishes.
\qed

Lemma \ref{zuzu} above is nothing more than \cite[Corollary of Theorem 10]{ZUC} in the paper by Zuckerman. All we have done is shrank the complex of differential forms on the smooth locus of $\E_L$ and, by doing so, avoiding to talk about the smooth structure on it. The term {\it pre-symplectic} in the definition of $\omega_1^{[N]}$ (Definition \ref{o1N}) is now justified. The dependency on the homology class of $N$ is irrelevant in classical mechanics and other theories with $\textrm{top}=1$, but not so on higher dimensional theories. The universal conserved current of Zuckerman, $\omega_1$ in our language (Theorem \ref{cc}), is a resolution of $\omega_1^{[N]}$ for all homology classes (we are using the term {\it resolution} in a vague way). Moreover, $\left( \Omega_{\textrm{loc}}^{\bullet, \textsf{top} -1 }(\E \times M), \delta \right)$ is also a cochain complex and $\omega_1$ can be seen as $2$-cocycle. The fact that $\omega_1$ is not $d$-closed, but only on shell makes it only possible to talk about a universal pre-symplectic structure on $\E_L \times M$ (rather, on its smooth locus).

\begin{pp}
	Let $(\E \times M, L)$ be a Lagrangian field theory. The associated Poincar\'e-Cartan local pre-$\textsf{top}$-multisymplectic form restricts to $\omega_1$ on shell and to $\omega_1^{[N]}$ up to integration along any Cauchy hyper-surface $N$ of $M$.
\end{pp}

\dem Since $\omega = \mathbf{E} L + \omega_1$ and on-shell $\mathbf{E} L = 0$, the statement is clear.
\qed

By the previous Proposition, the Poincar\'e-Cartan local form is also a resolution of all the pre-symplectic structures on the smooth locus of the space of extrema which also resolves the lack of smoothness of $\E_L$. This is so because, contrary to $\omega_1$ as seen by Zuckerman, $\omega$ is closed with respect to the total differential. Moreover, from the equations for symplectic and Hamiltonian vector fields for the Poincar\'e-Cartan local form, we can see how the dynamics of the different multi-symplectic structures are interconnected:

\begin{pp}
	Let $(\E \times M, \omega)$ be the Poincar\'e-Cartan pre-$\textsf{top}$-multisymplectic insular manifold associated to a Lagrangian $L$. If $\chi = \xi + X$ is a symplectic vector field for $\omega$, then $\xi$ is a symplectic vector field for $\omega_1^{[N]}$ for any Cauchy hyper-surface $N$.
\end{pp}

In other words, symplectic vector fields of $\omega$ are symplectic vector fields of $\omega_1$ on shell up to a $d$-exact term.

\dem Equations \ref{svf} for a symplectic vector field of $\omega$ reduce on shell to:
\begin{eqnarray*}
\delta \iota_{\xi} \omega_1 &=& d \iota_X \omega_1 \\
\delta \iota_X \omega_1 &=& 0
\end{eqnarray*}
The second equation is unimportant from the phase space point of view. The first one, after integration against a compact codimension $1$ submanifold of $N$, such as a Cauchy hyper-surface, gives $\delta \iota_{\xi} \omega_1^{[N]} = 0$ as a consequence of Stoke's theorem.
\qed

\begin{pp}\label{pis2}
	Let $(\E \times M, \omega)$ be the Poincar\'e-Cartan pre-$\textsf{top}$-multisymplectic insular manifold associated to a Lagrangian $L$. If $\chi = \xi + X$ is a Hamiltonian vector field for $\omega$ with associated Hamiltonian form $\zeta$, then $\xi$ is a Hamiltonian vector field for $\omega_1^{[N]}$ for any Cauchy hyper-surface $N$, with associated Hamiltonian form $\zeta_1^{[N]}$.
\end{pp}

\dem Equations \ref{hvf}, for Hamiltonian pairs reduce on shell to
\begin{eqnarray*}
0 &=& d Z \\
\iota_{\xi} \omega_1 &=& \delta Z + d \zeta_1  \\
\iota_X \omega_1 &=& \delta \zeta_1 + d \zeta_2  \\
0 &=& \delta \zeta_i + d \zeta_{i+1} \textrm{ for all } i \geqslant 2
\end{eqnarray*}
Furthermore, the reduction to the complex of $N$-local forms (for any Cauchy hyper-surface $N$)
\begin{equation*}
	\iota_{\xi} \omega_1^{[N]} = \delta \zeta_1^{[N]}
\end{equation*}
by the application of Stoke's theorem. This proves the proposition.
\qed

\begin{rk}
	As a conclusion, putting together Theorem \ref{corr} and Proposition \ref{pis2}, given a Hamiltonian pair $(\chi, \zeta)$ for the Poincar\'e-Cartan insular manifold, $(\xi, Z)$ is a Noether pair and $(\xi, \zeta_1^{[N]})$ is a Hamiltonian pair on the phase space associated to any Cauchy hyper-surface $N$.
\end{rk}

\subsection{Which symmetries do occur?}

So far we have said that any Noether pair $(\xi, Z)$ such that $\xi$ can be completed into a symplectic vector field $\chi = \xi + X$ (by adding a total vector field $X$) is Hamiltonian (Theorem \ref{corr}). But we have not clarified when that completion is possible. Deligne and Freed \cite{DEL} have a result that gives a first hint into answering this question:

\begin{pp}[Deligne and Freed {\cite[Proposition 2.76]{DEL}}]
	Let $(\E \times M, L)$ be a Lagrangian field theory. If $\xi$ is a symmetry of $L$ with associated Noether current $Z$, then there exists a form $\zeta_1 \in \Omega_{\textrm{loc}}^{1, \textsf{top}-1}(\E \times M)$ such that 
	$$\iota_{\xi} \omega_1 = \delta Z + d \zeta_1 \quad \textrm{ holds on-shell.}$$
\end{pp}

Comparing this to the second equation in \ref{hvf}, we are simply asking the deviation from the equation off-shell to be measured by $\iota_X \mathbf{E} L$. This seems reasonable, although there might be complications depending on the particular field theory we are considering.

First of all, we would like to point out that symplectic and Hamiltonian symmetries of $\omega$ are clearly related to symmetries of $\lambda$, the Lepagean equivalent of the Lagrangian $L$ given by the Fundamental Formulae Theorem \ref{FFP}. This is so because $\omega = D \lambda$. Symplectic vector fields are insular vector fields $\chi$ such that 
$$0 = D \iota_{\chi} \omega = \mathcal{L}_{\chi} \omega = \mathcal{L}_{\chi} D \lambda = D \mathcal{L}_{\chi} \lambda.$$
Similarly, an insular vector field $\chi$ is Hamiltonian with associated Hamiltonian form $\zeta$ if $\iota_{\chi} \omega = D \zeta$, which is equivalent to 
$$\mathcal{L}_{\chi} \lambda = D \iota_{\chi} \lambda + \iota_{\chi} \omega = D(\iota_{\chi} \lambda + \zeta).$$
This equation is the total degree analogue to Equation \ref{ahora}. These kind of symmetries of $\lambda$ are mentioned, but not studied in the work of Deligne and Freed \cite{DEL} (called symmetries of $\mathcal{L}$ in their language).

A similar question to the existence of symplectic vector fields, that is, insular vector fields such that $D \mathcal{L}_{\chi} \lambda = 0$, is the weaker question, when do insular vector fields $\chi$ such that $\mathcal{L}_{\chi} \lambda = 0$ exist? This is the approach taken by the Lepagean school (see Remark \ref{alalepage}). Recall that we showed in Proposition \ref{fol} that our Poincar\'e-Cartan form and theirs agrees for first order Lagrangians, and that as mentioned in Remark \ref{alalepage}, Poincar\'e-Cartan forms for higher order Lagrangian field theories also exist. GiMmsy \cite{GIMMSY} consider insular vector fields such that  $\mathcal{L}_{\chi} \lambda = 0$, but only evolutionary ones, calling them ``special covariant canonical transformations''. They argue that, in first and second order Lagrangians, those vector fields are in one to one correspondence with symmetries of the Lagrangian in the sense $\mathcal{L}_{\chi=\xi} L = 0$. In a more extensive result, Mu\~noz \cite{MUN1} shows that (for the same kind of special covariant canonical transformations), even if the Lagrangian is of a higher order, the one to one correspondence still holds true when restricted to insular vector fields of order $0$ (that is, with functions on $E$ as coefficients). GiMmsy \cite{GIMMSY} observe that for higher order Lagrangians, the existence of families of such kind of symmetries of the Poincar\'e-Cartan form is not known in general and has to be worked out in any given case.
	
Coming back to our notion of symplectic vector fields, we also allow horizontal vector fields and we do not focus on the study about ``special covariant canonical transformations''. This means that we have more flexibility to find symplectic vector fields. Nevertheless, we do not provide a general existence result. By working out the computations of equations \ref{svf} in local coordinates we discover that $\chi = \xi + X$ is a symplectic vector field if and only if:
	 \begin{eqnarray}\label{svf2}
	 \textbf{E} L_{\beta} \partial_{\alpha}^I \xi_{\beta} &=& 0 \quad \textrm{ for all } \alpha \textrm{ and all } I, |I| \geqslant 2 \nonumber \\
	 \textbf{E} L_{\beta} \partial_{\alpha}^i \xi_{\beta} &=& \textbf{E} L_{\alpha} X_i \quad \textrm{ for all } \alpha \textrm{ and all } i \nonumber \\
	 \left( \partial_{\beta}^I \mathbf{E} L_{\alpha}  \right) \left( D_I \xi_{\beta} \right) + \partial_{\alpha} \xi_{\beta} \mathbf{E} L_{\beta} &=& -D_i \left( \mathbf{E} L_{\alpha} X_i \right) \quad \textrm{ for all } \alpha .
	 \end{eqnarray}
 
As a matter of fact, we are only interested in symplectic vector fields $\chi = \xi + X$ such that $\xi$ is a symmetry of the Lagrangian. In that case $\textbf{E} L_{\beta} \xi_{\beta} = D_i Z_{\xi}^i$ where $Z_{\xi}$ is the associated Noether current to $\xi$. That equality further simplifies equation \ref{svf2}. Note that that in the case of first or second order Lagrangians, the first Equation in \ref{svf2} trivially holds, in agreement with the statement of GiMmsy \cite{GIMMSY} previously mentioned.


\newpage
\chapter{Local observables}

Multi-symplectic insular manifolds have associated \Linf s of Hamiltonian forms and pairs which are, by definition, local and insular respectively. In the case of the Poincar\'e-Cartan pre-multisymplectic insular manifold, Hamiltonian forms have a nice interpretation in terms of local observables of the underlying  Lagrangian field theory.\\

By adding certain $D$-exact term to the $2$-bracket on Hamiltonian pairs of the Poincar\'e-Cartan pre-multisymplectic insular manifold one gets a bilinear map which splits by depth into well defined brackets. On surface, that bracket agrees with the usual bracket on Noether pairs. On depth one, after the choice of a Cauchy hyper-surface, the bracket gives rise to a multiple of the Poisson bracket on Hamiltonian functions on the covariant phase space.\\

Homotopy moment maps for local Lie algebra actions which are symplectic for the Poincar\'e-Cartan pre-multisymplectic insular manifold are special kinds of Hamiltonian moment maps in the sense of Blohmann and Weinstein.\\

{\it This chapter discusses the \Linf s associated to the Poincar\'e-Cartan pre-multisymplectic insular manifold following the theory of Rogers \cite{ROG}. It also includes the relation of that {\Linf} to two Lie algebras known for Lagrangian field theories: that of Noether pairs and that of Hamiltonian functions on the phase space. Finally, it interprets homotopy moment maps coming from local Lie algebra actions in terms of Hamiltonian moment maps. The basic reference in this chapter is Deligne and Freed \cite{DEL}.}


\section{\boldmath{\Linf} of local observables}

{\it The term (classical) observable is ofter reserved to functions in the symplectic covariant phase space. The associated pre-symplectic structure is often degenerate, hence we are forced to work with Hamiltonian functions. Moreover, that space is only well defined up to homology class of the Cauchy hyper-surface and up to assuming the space of extrema is smooth. Instead, we define local observables as Hamiltonian forms of the Poincar\'e-Cartan pre-$\textsf{top}$-multisymplectic insular manifold. The {\Linf} of Hamiltonian forms of Rogers \cite{ROG} becomes an {\Linf} of local observables in this setting. Landsman \cite{LAN} is the main reference in this chapter.}\\

Given any local pre-multisymplectic structure we can construct the associated \Linf s of Hamiltonian vector fields and Hamiltonian pairs mimicking what Rogers in \cite{ROG} does for finite dimensional manifolds (Theorems \ref{rog1} and \ref{rog2}). Since the Cartan calculus is closed with respect to local vector fields and forms, so it is the {\Linf} of Hamiltonian forms: we get an {\Linf} of local forms as discussed in subsection \ref{lall}.

Fixing a Lagrangian field theory $(\E \times M, L)$, we are interested in the Poincar\'e-Cartan local pre-$\mathsf{top}$-symplectic form explained in the previous section. In that case, the associated {\Linf} from Theorem \ref{rog1} is going to be called the {\Linf} of local observables. We want to explain this notation.

An observable in a (classical) physical theory is sometimes thought as a function on the covariant phase space of the action (see Landsman \cite{LAN}). Observables form a Poisson algebra, which is a particular case of a Lie algebra. If the pre-symplectic structure on the phase space is non-degenerate, the bracket is the Poisson bracket. In the degenerate case we need to work with Hamiltonian functions on the phase space instead. Following what was explained in the last section, Hamiltonian forms of the Poincar\'e-Cartan pre-$\textsf{top}$-symplectic form give rise to Hamiltonian functions on the phase space, after restriction on shell and integration along a Cauchy hyper-surface (Proposition \ref{pis2}).

By working with Hamiltonian forms of the Poincar\'e-Cartan pre-$\textsf{top}$-symplectic form instead of with Hamiltonian functions of the covariant phase space we do not have to worry about the extrema of solutions of the action, $\E_L$, being non-smooth or about the choice of the Cauchy hyper-surface. (The problem of the lack of smoothness of $\E_L$ is also present in the definition of local observables followed by some other authors, like Khavkine \cite{KHA}, as int-local functions on the smooth locus of the space of extrema --recall Definition \ref{qloc}--.)

\begin{df}[Local observable] A local observable in a Lagrangian field theory is a Hamiltonian form for the associated Poincar\'e-Cartan pre-$\textsf{top}$-symplectic form.
\end{df}

We have justified the term local observable to refer to Hamiltonian forms of the Poincar\'e-Cartan insular manifold. In this way it is clear why the {\Linf} of Hamiltonian forms associated to that manifold deserve the name of {\Linf} of local observables.

\begin{dfpp}[{\Linf} of local observables]\label{fin1}
	Let $(\E \times M, \omega)$ be the Poincar\'e-Cartan pre-$\textsf{top}$-multisymplectic insular manifold associated to a Lagrangian $L$. The {\Linf} of local observables of the Lagrangian field theory $(\E \times M, L)$ is the graded vector space
	\[
	L_{n}(\E \times M, \omega) =
	\begin{cases} 
	\hfill \Omega_{ham}^{\textsf{top}-1}(\E \times M)   \hfill & \text{ if $n=1$ } \\
	\hfill \Omega_{\textrm{loc}}^{\textsf{top}-n}(\E \times M) \hfill & \text{ if $n \in \, \{2, \ldots, \textsf{top}\}$ } \\
	\hfill \{0\} \hfill & \text{ else. }
	\end{cases}
	\]
	together with the following family of brackets: $l_1 = -D$ and for every $n \geqslant 2$
	\[
	l_{n}(\zeta^1 \otimes \cdots \otimes \zeta^{n}) =
	\begin{cases} 
	\hfill \iota_{\chi^1 \cdots \chi^{n}} \omega    \hfill & \text{ if $D \zeta^i = \iota_{\chi^i} \omega$ for every $1 \leqslant i \leqslant n$ } \\
	\hfill 0 \hfill & \text{ else. } \\
	\end{cases}
	\]
\end{dfpp}

\dem The fact that the structure is an {\Linf} follows from Theorem \ref{rog1} and the fact that the Cartan calculus restricts to insular forms and insular vector fields (Theorem \ref{lcc1}).\qed

\begin{center}
\begin{tikzpicture}[description/.style={fill=white,inner sep=2.5pt},scale=1.4]
\draw [fill=black!20, black!20] (-4, -5.2) to (-0.5,-5.2) to (-0.5,3) -- (-4,-1.5);

\node (e) at (1,4) {$1$};
\node (f) at (2,3) {$2$};
\node (g) at (3,2) {$\ddots$};
\node (h) at (4,1) {$\textsf{top}$};
\node (l) at (0.9,2) {$\Omega_{\textrm{ham}}^{\textsf{top} -1}(\E \times M)^{0, \textsf{top}-1}$};
\node (n) at (2,1) {$\Omega_{\textrm{loc}}^{0, \textsf{top}-2}(\E \times M)$};
\node (o) at (0.75,0) {$\Omega_{\textrm{ham}}^{\textsf{top} -1}(\E \times M)^{1, \textsf{top}-2}$};
\node (p) at (4,-1) {$\C_{\textrm{loc}}(\E \times M)$};
\node (q) at (3,-2) {$\iddots$};
\node (r) at (2,-3) {$\Omega_{\textrm{loc}}^{2, 0}(\E \times M)$};
\node (s) at (0.9,-4) {$\Omega_{\textrm{ham}}^{\textsf{top} -1}(\E \times M)^{\textsf{top}-1,0}$};
\node (u) at (3,0) {$\ddots$};
\node (z) at (0.9,-2) {$\vdots$};
\draw [rounded corners, dashed, thick, teal] (-2.4,-0.8) to[out=45,in=180] (0,1.1);
\draw [rounded corners, dashed, thick, teal] (-4,-2.4) to (-2.4,-0.8);
\draw [rounded corners, dashed, thick, teal] (0, 1.1) to[out=0,in=135] (2.6,-0.6);
\draw [dotted, thick, teal] (2.8,-0.8) -- (3.2,-1.2);
\draw [rounded corners, dashed, thick, teal] (3.4,-1.4) -- (4.2, -2.2) -- (1.2,-5.2);
\path[->,font=\scriptsize,color=cyan]
(n) edge node[below] {$-\delta$} (o)
(p) edge node[auto] {$-\delta$} (q)
(q) edge node[auto] {$-\delta$} (r)
(r) edge node[auto] {$-\delta$} (s);
\path[->,font=\scriptsize,color=dandelion]
(n) edge node[auto] {$-d$} (l)
(u) edge node[auto] {$-d$} (n)
(p) edge node[auto] {$-d$} (u)
(r) edge node[auto] {$-d$} (z);
\end{tikzpicture}
\end{center}

Observe that in the diagram we cannot write $\Omega_{\textrm{ham}}^{p-1, \textsf{top}-p}(\E \times M)$ since that is not defined. Instead we should talk about the bidegree $(p-1, \textsf{top}-p)$ part of $\Omega_{\textrm{ham}}^{\textsf{top} -1}(\E \times M)$, that we have denoted by $\Omega_{\textrm{ham}}^{\textsf{top} -1}(\E \times M)^{p-1, \textsf{top}-p}$ (for all $p$ between $1$ and $\textsf{top}$). The grey part means the complex is zero at those degrees.

Similarly, we can talk about the {\Linf} of Noether pairs:

\begin{dfpp}[{\Linf} of Noether pairs]
	Let $(\E \times M, \omega)$ be the Poincar\'e-Cartan pre-$\textsf{top}$-multisymplectic insular manifold associated to a Lagrangian $L$. The {\Linf} of Noether pairs of the Lagrangian field theory $(\E \times M, L)$ is the graded vector space
	\[
	P_{n}(\E \times M, \omega) =
	\begin{cases} 
	\hfill P_1(\E \times M, \omega)   \hfill & \text{ if $n=1$ } \\
	\hfill \Omega_{\textrm{loc}}^{\textsf{top}-n}(\E \times M) \hfill & \text{ if $n \in \, \{2, \ldots, \textsf{top}\}$ } \\
	\hfill \{0\} \hfill & \text{ else. }
	\end{cases}
	\]
	together with the following family of brackets is an {\Linf}: \[
	l_{1}(v) =
	\begin{cases} 
	\hfill (0, -D \zeta)   \hfill & \text{ if $v=(\chi,\zeta)$ } \\
	\hfill -D v \hfill & \text{ if $|v| \in \, \{2, \ldots, \textsf{top}\}$ } 
	\end{cases}
	\]
	
	\[
	l_{2}(v^1 \otimes v^2) =
	\begin{cases} 
	\hfill ([\chi_1, \chi_2], l_2(\zeta^1 \otimes \zeta^2))   \hfill & \text{ if $v^i=(\chi^i, \zeta^i)$ for $i\in\,\{1,2\}$} \\
	\hfill 0 \hfill & \text{ else. }
	\end{cases}
	\]
	
	\[
	l_{n \geqslant 3}(v^1 \otimes \cdots \otimes v^n) =
	\begin{cases} 
	\hfill l_n(\zeta^1 \otimes \cdots \otimes \zeta^n) \hfill & \text{ if $v^i=(\chi^i, \zeta^i)$ $\forall i\in\,\{1,\ldots,n\}$} \\
	\hfill 0 \hfill & \text{ else. }
	\end{cases}
	\]
\end{dfpp}

\dem Once again, the fact that the structure is an {\Linf} follows from Theorem \ref{rog2} and the fact that the Cartan calculus restricts to insular forms and insular vector fields (Theorem \ref{lcc1}).\qed

To be completely formal, we should talk about the \Linf s of symplectic observables and of symplectic Noether pairs, as a consequence of Theorem \ref{corr}.

We conclude this section looking at what happens when $\textsf{top}=1$.

\begin{ej} In a Lagrangian field theory over a manifold $M$ of dimension $1$, a Hamiltonian pair $(\chi, Z)$ consists of an insular vector field $\chi = \xi + X$ and a local function $Z \in \C_{\textrm{loc}}(\E \times M)$ such that $\iota_{\xi} \mathbf{E} L = d Z$ and $\iota_X \mathbf{E} L + \iota_{\xi} \omega_1 =\delta Z$. The {\Linf} of local observables is concentrated in degree $1$ and consists of all Hamiltonian functions $Z$ as above. The only non-trivial bracket is the $2$-bracket, giving $\C_{\textrm{ham}}(\E \times M)$ the structure of a Lie algebra. The bracket is simply the Poisson bracket of two Hamiltonian functions translated to the insular world.
\begin{center}
\begin{tikzpicture}[description/.style={fill=white,inner sep=2.5pt},scale=1]

\node (a) at (-1,4) {\color{cyan} $0$};
\node (b) at (-2,3) {\color{cyan} $1$};
\node (c) at (-3,2) {\color{cyan} $2$};
\node (d) at (-4,1)  {\color{cyan} $\iddots$};
\node (e) at (1,4) {\color{dandelion} $1$};
\node (f) at (2,3) {\color{dandelion} $0$};

\node (j) at (0,3) {$\iota_{\xi} \mathbf{E}L$};
\node (k) at (-1,2) {$EL$};
\node (l) at (1,2) {$Z$};
\node (m) at (-2,1) {$0$};
\node (m2) at (-2,-1) {$0$};
\node (o) at(0,1) {$\iota_{\xi} \omega_1 + \iota_{X} \mathbf{E}L$};
\node (aa) at (-3,-2) {$\iddots$};
\node (t) at (-3,0) {$\iddots$};
\node (v) at (-1,0) {$\omega_1$};

\draw [rounded corners, dashed, thick, teal] (-4,-1.7) -- (0,2.3) -- (1.3,1) -- (-2.7,-3);
\path[->,font=\scriptsize,color=cyan]
(k) edge node[auto] {} (j)
(k) edge node[auto] {} (m)
(v) edge node[auto] {} (m2)
(m2) edge node[auto] {} (aa)
(m) edge node[auto] {} (t)
(l) edge node[auto] {} (o)
(v) edge node[auto] {} (o);
\path[->,font=\scriptsize,color=dandelion]
(l) edge node[auto] {} (j)
(k) edge node[auto] {} (o)
(v) edge node[auto] {} (m);
\end{tikzpicture}
\end{center}

\end{ej}


\section{Lie--algebras related to local observables}

{\it The {\Linf} of local observables provides a Hamiltonian form associated to the bracket of two Hamiltonian vector fields on the Poincar\'e-Cartan pre-multisymplectic insular manifold. There are other choices for such a Hamiltonian form. In this section we present an alternative choice, that when restricted on surface gives the usual bracket of Noether pairs, and when restricted to depth one, agrees with the Poisson bracket on the covariant phase space. The main reference in this section is Deligne and Freed \cite{DEL}.}\\

We start by fixing a Poincar\'e-Cartan pre-multisymplectic insular manifold $(\E \times M, \omega)$. In this section we are going to focus on the $2$-bracket of two local observables, the objective being giving another related bracket based on it, which has interesting properties. We fix two Hamiltonian pairs for the Poincar\'e-Cartan pre-multisymplectic insular manifold:

$$\left( \zeta=(Z, \zeta_1, \ldots), \, \xi + X \right) \quad \textrm{such that } D \zeta = \iota_{\xi + X} \omega \textrm{ and}$$
$$\left( \eta=(H, \eta_1, \ldots), \, \upsilon + Y \right) \quad \textrm{such that } D \eta = \iota_{\upsilon + Y} \omega .\phantom{ and}$$

Observe that, to have less variables around, we have decided not to give a specific name to the sum of the evolutionary and the total parts of the vector field. The expression for the $2$ bracket, $l_2(\zeta, \eta) = \iota_{\xi + X} \iota_{\upsilon + Y} \omega$, coming from Definition/Proposition \ref{fin1} does not have a nice decomposition by depth. By that we mean, for example in depth $1$
$$l_2((\zeta, \eta))_1 = \iota_{\xi} \iota_{Y} \mathbf{E} L - \iota_{\upsilon} \iota_{X} \mathbf{E} L + \iota_{\xi} \iota_{\upsilon} \omega_1$$
is not expressible only in terms of $\zeta_1$ and $\eta_1$. We propose a different bracket.

\begin{lm}\label{lemita} Let $(\zeta, \xi + X)$ and $(\eta, \upsilon + Y)$ be two Hamiltonian pairs of the Poincar\'e-Cartan pre-multisymplectic insular manifold $(\E \times M, \omega)$. Then,
$$D(\mathcal{L}_{\xi} \eta - \mathcal{L}_{\upsilon} \zeta - \iota_{\xi} \iota_{\upsilon} \omega + \iota_{X} \iota_{Y} \omega) = D(l_2(\zeta, \eta)).$$
\end{lm}

\dem In the notation of the Lemma
\begin{eqnarray*}
D(\mathcal{L}_{\xi} \eta - \mathcal{L}_{\upsilon} \zeta) &=& D \iota_{\xi} D \eta - D \iota_{\upsilon} D \zeta = D \iota_{\xi} \iota_{\upsilon + Y} \omega - D \iota_{\upsilon} \iota_{\xi + X} \omega \\
&=& D(\iota_{\chi + X} \iota_{\upsilon + Y} \omega) + D (\iota_{\xi} \iota_{\upsilon} \omega - \iota_{X} \iota_{Y} \omega)).
\end{eqnarray*}
After rearranging the two sides of the equation and using the definition of the $l_2$ bracket from Definition/Proposition \ref{fin1}, we get the desired result.
\qed

We define:

\begin{equation}\label{ham8}
[\zeta, \eta] \defeq \mathcal{L}_{\xi} \eta - \mathcal{L}_{\upsilon} \zeta - \iota_{\xi} \iota_{\upsilon} \omega + \iota_{X} \iota_{Y} \omega.
\end{equation}

The previous Lemma says that $[\xi + X, \upsilon + Y]$ is a Hamiltonian form (a local observable) associated to the Hamiltonian vector field $[\xi + X, \upsilon + Y]$. That bracket is explicitly antisymmetric, and it has a very nice decomposition by depth:
\begin{eqnarray}
	{[} \zeta, \eta {]}_0 &=& \mathcal{L}_{\xi} H - \mathcal{L}_{\upsilon} Z - \iota_{\xi} \iota_{\upsilon} \omega_1 \label{a}\\
	{[} \zeta, \eta  {]}_1 &=& \mathcal{L}_{\xi} \eta_1 - \mathcal{L}_{\upsilon} \zeta_1 + \iota_{X} \iota_{Y} \mathbf{E} L \label{b}\\
	{[} \zeta, \eta  {]}_2 &=& \mathcal{L}_{\xi} \eta_2 - \mathcal{L}_{\upsilon} \zeta_2 + \iota_{X} \iota_{Y} \omega_1 \label{c}\\
	{[} \zeta, \eta  {]}_i &=& \mathcal{L}_{\xi} \eta_i - \mathcal{L}_{\upsilon} \zeta_i \textrm{ for all } i \geqslant 3 . \nonumber
\end{eqnarray}

As a matter of fact, the two brackets differ by a $D$-exact term:

\begin{pp}\label{goodforlater}
Let $(\zeta, \xi + X)$ and $(\eta, \upsilon + Y)$ be two Hamiltonian pairs of the Poincar\'e-Cartan pre-multisymplectic insular manifold $(\E \times M, \omega)$. Then, there exists $\alpha \in \Omega_{\textrm{loc}}^{\textsf{top}-2}(\E \times M)$ such that $D \alpha = l_2(\zeta, \eta) - [\zeta, \eta]$.
\end{pp}

\dem By the pulley Theorem \ref{RPT}, it is enough to see that the difference $l_2(\zeta, \eta)_0 - {[} \zeta, \eta {]}_0$ is $d$-exact. But we can show that fact by a direct calculation using equation \ref{a}:
\begin{eqnarray}\label{esa}
{[} \zeta, \eta {]}_0 &=& \mathcal{L}_{\xi} H - \mathcal{L}_{\upsilon} Z - \iota_{\xi} \iota_{\upsilon} \omega_1 \nonumber \\
&=& \iota_{\xi} \iota_{\upsilon} \omega_1 - \iota_{\xi} d \eta_1 -  \iota_{\upsilon} \iota_{\xi} \omega_1 + \iota_{\upsilon} d \zeta_1 - \iota_{\xi} \iota_{\upsilon} \omega_1 \nonumber \\
&=& \iota_{\xi} \iota_{\upsilon} \omega_1 + d (\iota_{\xi} \eta_1 - \iota_{\upsilon} \zeta_1) = l_2(\zeta, \eta)_0 + d (\iota_{\xi} \eta_1 - \iota_{\upsilon} \zeta_1).
\end{eqnarray}
\qed

A disadvantage of this bracket is that it is not independent of the insular vector field chosen as a Hamiltonian pair for the forms. The reason for that is that $\iota_{X} \mathbf{E} L$ cannot be expressed only in terms of $\zeta$. Since $[\zeta, \eta]$ is a Hamiltonian form for the insular vector field $[\xi + X, \upsilon + Y]$ we can define a bracket on Hamiltonian pairs:

\begin{equation*}
	{[} (\zeta, \xi + X), (\eta, \upsilon + Y) {]} \defeq \left( {[}\zeta, \eta{]}, {[}\xi + X, \upsilon + Y{]}  \right).
\end{equation*}

If we try to write down the Jacobi identity for any of the two brackets we get into trouble very easily, because the evolutionary part of the bracket is not the bracket of the evolutionary parts of insular vector fields. The same goes for the total parts. Recall that when the total vector field is horizontal (independent of $\E$) the bracket does split (Proposition \ref{CC4}). Even in that case, the expression of the $3$ Jacobiator is not trivial:
\begin{equation}\label{boli}
\sum_{\sigma \in \mathcal{S}h_{1}^2}[\zeta^{\sigma(1)}, [\zeta^{\sigma(2)}, \zeta^{\sigma(3)}]] = D	(\iota_{\xi^1} - \iota_{X^1}) \iota_{X^2} \iota_{X^3} \omega_1.
\end{equation}
Nevertheless, if we look at this equation at every depth level we get interesting consequences. First of all, observe that the depth zero part of equation \ref{boli} is zero. In particular, if we restrict ourselves to the bracket between surface Hamiltonian pairs: that is Noether pairs, we get a Lie algebra:

\begin{dfpp}[Lie algebra of Noether pairs]\label{lanp}
Given a Lagrangian field theory $(\E \times M, L)$, the bracket \ref{a} on Noether pairs is a Lie algebra bracket. Explicitly:
\begin{equation}\label{bili}
	{[} (Z, \xi), (H, \upsilon) {]} \defeq \left( \mathcal{L}_{\xi} H - \mathcal{L}_{\upsilon} Z - \iota_{\xi} \iota_{\upsilon} \omega_1, {[}\xi, \upsilon{]}  \right).
\end{equation}
\end{dfpp}

This Lie algebra is already known in the literature. It appears in the work of Deligne and Freed \cite[equation 2.100]{DEL}. 

\dem One way of showing the result is using equation \ref{boli}. Nevertheless, we have not given explicit computations for it. A different way of proving it is to infer it from the Lie algebra of symmetries of the Lagrangian field theory. If $d A = \iota_{\xi} \delta L$ and $d B = \iota_{\upsilon} \delta L$ are two symmetries, then the bracket
$$\{A, B\} \defeq \mathcal{L}_{\xi} B - \mathcal{L}_{\upsilon} A$$
is a Lie algebra bracket. The associated symmetry to $\{A, B\}$ is $[\xi, \upsilon]$. Now, by twisting $A$ by adding $-\iota_{\xi} \lambda_1$ and $B$ accordingly, we get a bracket on the Noether currents $Z=A-\iota_{\xi} \lambda_1$
$$\{A-\iota_{\xi} \lambda_1, B -\iota_{\upsilon} \lambda_1\} \defeq \mathcal{L}_{\xi} B - \mathcal{L}_{\upsilon} A -\iota_{[\xi, \upsilon]} \lambda_1.$$
After using the commutation relations of the Cartan calculus we get that:
$$\{A-\iota_{\xi} \lambda_1, B -\iota_{\upsilon} \lambda_1\} = \mathcal{L}_{\xi} H - \mathcal{L}_{\upsilon} Z - \iota_{\xi} \iota_{\upsilon} \omega_1,$$
where $Z \defeq A-\iota_{\xi} \lambda_1$ and $H \defeq B -\iota_{\upsilon} \lambda_1$. Since the first bracket satisfies the Jacobi identity, so it does the second. That equation shows that the bracket \ref{bili} satisfies the Jacobi identity since the bracket of vector fields in indeed a Lie bracket. 
\qed

\begin{dfpp}[Lie algebra of Hamiltonian pairs on the phase space]
Given a Lagrangian field theory $(\E \times M, L)$, the bracket \ref{b}  restricted to shell and integrated along a Cauchy hyper-surface agrees with the Poisson bracket of Hamiltonian functions on the phase space associated to that Cauchy surface (up to a factor of $2$).
\end{dfpp}

\dem  The bracket \ref{b} on shell reduces to 
$${[} \zeta, \eta  {]}_1 = \mathcal{L}_{\xi} \eta_1 - \mathcal{L}_{\upsilon} \zeta_1.$$
Integrating along a Cauchy hyper-surface $N$:
$$\left[ \zeta_1^{[N]}, \eta_1^{[N]}  \right] = \mathcal{L}_{\xi} \eta_1^{[N]} - \mathcal{L}_{\upsilon} \zeta_1^{[N]}.$$
Hamiltonian pairs of the Poincar\'e-Cartan form give rise to Hamiltonian pairs of the pre-symplectic form on the phase space associated to $N$ (Proposition \ref{pis2}). It is well known that the bracket in a pre-symplectic manifold can also be written as
$$\left[ \zeta_1^{[N]}, \eta_1^{[N]}  \right] = \frac{1}{2} \left( \mathcal{L}_{\xi} \eta_1^{[N]} - \mathcal{L}_{\upsilon} \zeta_1^{[N]} \right),$$
which agrees with the previous equation (up to the factor of $2$).
\qed

We want to finish this section with some interpretations of the results here presented. The $l_2$ bracket on Hamiltonian pairs for the Poincar\'e-Cartan pre-$\textsf{top}$-symplectic insular manifold does not satisfy the Jacobi identity (observe what the $3$-Jacobiator in the {\Linf} of local observables is). We can deform\footnote{We are using the word deform in an informal way, not explicitly taking deformations of \Linf s.} that $l_2$ bracket by adding a $D$-exact term to get another bracket $[-,-]$ (see equation \ref{ham8}) which also gives a Hamiltonian form for commutator of the associated vector fields. This new bracket has a nice decomposition formula by depth. On depth zero, it agrees with the bracket of Noether pairs from Deligne and Freed \cite{DEL}. On depth one, after restriction on shell and integration along a Cauchy hyper-surface, it agrees with the Poisson bracket on the phase space. The non-deformed bracket $l_2$ does not have the nice {\it Lie} properties that the $[-,-]$ has, but it has the advantage that is part of an {\Linf} involving both surface and first order depth elements.


\section{Homotopy current maps}

{\it Homotopy moment maps for local Lie algebra actions by symplectomorphisms on a Poincar\'e-Cartan pre-multisymplectic insular manifold are here studied. They are particular case of Hamiltonian moment maps, an interesting tool in Lagrangian field theories motivated by the relation between Noether charges and the constraints for the initial data of the theory.}\\

In this section we come back to the setting that we left when talking about Noether's second theorem in section \ref{nono}. Assume we are given a local Lie algebra action on the insular manifold $\E \times M$ by the space of sections $\A$ of a Lie algebroid over $M$. Assume, moreover, that the action is by symmetries of a Lagrangian $L$ on $\E \times M$. By virtue of Noether's first theorem, we know that for every section $\psi \in \A$ there is an associated conserved current $Z_{\xi(\psi)}$ where $\xi \colon \A \rightarrow \mathfrak{X}_{\textrm{ins}}(\E)$ is the action. A local choice of conserved currents is called a current map:

\begin{df}[Current map, Blohmann and Weinstein\footnote{Private communication.}]
Let $\xi \colon \A \rightarrow \mathfrak{X}_{\textrm{ins}}(\E)$ be a local Lie algebra action on the insular manifold $\E \times M$ by symmetries of a Lagrangian $L$ on $\E \times M$. A current map is a map
$$Z \colon \A \longrightarrow \Omega_{\textrm{loc}}^{0, \textsf{top}-1}(\E \times M)$$
such that $d Z(\psi) = \iota_{\xi(\psi)} \mathbf{E} L$ for all $\psi \in \A$. 
\end{df}

Blohmann and Weinstein also define Hamiltonian current maps:

\begin{df}[Hamiltonian current map, Blohmann and Weinstein\footnote{Private communication.}]
A current map $Z \colon \A \rightarrow \Omega_{\textrm{loc}}^{0, \textsf{top}-1}(\E \times M)$ is Hamiltonian if the extension of $Z$ to
$$v \defeq (Z, \xi) \colon \A \longrightarrow \Omega_{\textrm{loc}}^{0, \textsf{top}-1}(\E \times M) \times \mathfrak{X}_{\textrm{ins}}(\E)$$
is such that $[v(\psi), v(\psi^{\prime})] - v([\psi, \psi^{\prime}])$ is $d$-exact, where the bracket on Noether pairs is the one from Definition \ref{lanp}.
\end{df}

Since Noether pairs are the surface part of Hamiltonian pairs for the Poincar\'e-Cartan pre-symplectic insular manifold associated to the Lagrangian $L$, we can use the theory of homotopy moment maps to define current maps by restriction to the surface:

\begin{df}[Homotopy current map]
Let $(A, \rho)$ be a Lie algebroid over $M$. Consider $\xi \colon \A \rightarrow \mathfrak{X}_{\textrm{ins}}(\E)$ a local Lie algebra action on the insular manifold $\E \times M$. Assume $(\xi, \rho^*)$ is an action on $\E \times M$ by symplectic vector fields of the Poincar\'e-Cartan form associated to a Lagrangian $L$ on $\E \times M$.
A homotopy current map on $\A$ is a lift $\tilde{h}$ of the action $\mathsf{a}$ to $P(Y, \omega)$ in the category of local \Linf s:
	\begin{center}
		\begin{tikzpicture}[description/.style={fill=white,inner sep=2pt}]
		\matrix (m) [matrix of math nodes, row sep=2.5em,
		column sep=3.5em, text height=1.5ex, text depth=0.25ex]
		{ & P(\E \times M, \omega) \\
			\A & \mathfrak{X}_{sym}(\E \times M) \\};
		\path[->,font=\scriptsize]
		(m-2-1) edge node[below] {$(\xi, \rho^*)$} (m-2-2)
		(m-1-2) edge node[auto] {$\pi_v$} (m-2-2)
		(m-2-1) edge node[auto] {$\tilde{h}$} (m-1-2);
		\end{tikzpicture}
	\end{center}
\end{df}

This definition is the translation to the category of insular manifolds of the Definition of a homotopy moment map \ref{dee}, via Theorem \ref{tee}, for the Poincar\'e-Cartan form associated to the Lagrangian $L$.

\begin{lm} Let $(A, \rho)$ be a Lie algebroid over $M$. Consider $\xi \colon \A \rightarrow \mathfrak{X}_{\textrm{ins}}(\E)$ a local Lie algebra action on the insular manifold $\E \times M$. Assume $(\xi, \rho^*)$ is an action on $\E \times M$ by symplectic vector fields of the Poincar\'e-Cartan form associated to a Lagrangian $L$ on $\E \times M$. Then, $\A$ acts by symmetries of the Lagrangian. Moreover, $\tilde{h}_1 \colon \A \rightarrow P_1(\E \times M, \omega)$ is a current map.
\end{lm}

This is a rephrasing of the second point of Theorem \ref{corr}, which said that Hamiltonian pairs on surface are Noether pairs when talking about the Poincar\'e-Cartan form associated to a Lagrangian $L$.

\begin{pp} Let $\tilde{h} \colon \A \rightarrow P(\E \times M, \omega)$ be a homotopy current map. Then $\tilde{h}_1 \colon \A \rightarrow P_1(\E \times M, \omega)$ is a Hamiltonian current map.
\end{pp}

\dem The map $\tilde{h}$, as a morphism of \Linf s is a collection of degree zero maps $f:=\{f_k \colon \left([-1]\A\right)^{\otimes k} \rightarrow P(\E \times M, \omega)\}$ such that certain compatibility relations hold (see Definition \ref{morphi})\footnote{Observe the shift by $-1$ of $\A$ so that it is an {\Linf} in our convention.}. In particular, given $\psi$ and $\psi^{\prime}$ two sections of $A$ we have that
$$\tilde{h}_1 ([\psi, \psi^{\prime}]) = l_2(\tilde{h}_1(\psi), \tilde{h}_1(\psi^{\prime})) + l_1(\tilde{h}_2(\psi, \psi^{\prime})).$$
That equation happens on Noether pairs, that is inside of $P_1(\E \times M, \omega)$. We denote by $\alpha \defeq \tilde{h}_2(\psi, \psi^{\prime})_0$ the surface part of the last element on its right hand side. On surface, the previous equation reads as follows:
$$\tilde{h}_1 ([\psi, \psi^{\prime}]) = l_2(\tilde{h}_1(\psi), \tilde{h}_1(\psi^{\prime}))_0 - d \alpha.$$
By Proposition \ref{goodforlater}, the difference between $l_2$ and $[-,-]$ on surface is d-exact, in particular, using equation \ref{esa} from the proof of that proposition:
$$\tilde{h}_1 ([\psi, \psi^{\prime}]) = [\tilde{h}_1(\psi), \tilde{h}_1(\psi^{\prime})]_0 -d(\iota_{\xi} \eta_1 - \iota_{\upsilon} \zeta_1 + \alpha).$$
This shows that $\tilde{h}_1$ is a Hamiltonian current map in the sense of Blohmann and Weinstein.
\qed



\printbibliography

\appendix

\newpage
\part{Appendices}


\chapter{Some details about the infinite jet bundle}


\section{Category structure of a pro-category}\label{app1}

{\it In this appendix we explain how is a pro-category a category. We will use repeatedly during this section a technique using final functors (see Mac Lane \cite{MAC}) without mentioning it explicitly.}\\

We defined a pro-category in terms of maps and diagrams in $\Cat$. In this section we answer the following questions: what is an identity morphism in a pro-category? How is composition defined in a pro-category? In other words, how is a pro-category a category?

A map $f$  from $X \colon \mathcal{I} \rightarrow \Cat$ to $Y \colon \mathcal{J} \rightarrow \Cat$ is given by a function $i \colon \mathcal{J} \rightarrow \mathcal{I}$ and maps $f^{j} \colon X_{i(j)} \rightarrow Y_j$ which are {\it compatible with all the structure maps of $X$ and $Y$}. We need to make sense of this sentence. If $j \rightarrow j^{\prime}$ in $\mathcal{J}$ then there exists an element $i(j\to j^{\prime}) \in \, \mathcal{I}$ and maps $i(j\to j^{\prime}) \rightarrow i(j)$, $i(j\to j^{\prime}) \rightarrow i(j^{\prime})$ in $\mathcal{I}$ such that the following diagram commutes:

\begin{center}
\begin{tikzpicture}[description/.style={fill=white,inner sep=2pt}]
\matrix (m) [matrix of math nodes, row sep=1.2em,
column sep=2.5em, text height=1.5ex, text depth=0.25ex]
{ & X_{i(j)} & & Y_{j} \\
 X_{i(j\to j^{\prime})} & & & \\
  & X_{i(j^{\prime})} & & Y_{j^{\prime}} \\ };
\path[-> ,font=\scriptsize]
(m-2-1) edge node[auto] {} (m-1-2)
(m-2-1) edge node[auto] {} (m-3-2)
(m-3-2) edge node[below] {$f^{j^{\prime}}$} (m-3-4)
(m-1-2) edge node[auto] {$f^j$} (m-1-4)
(m-1-4) edge node[auto] {} (m-3-4);
\end{tikzpicture}
\end{center}

The process of finding $X_{i(j\to j^{\prime})}$ such that the previous diagram commutes will be referred to as {\it closing the pentagon}.

Observe that in general there is no map between $X_{i(j)}$ and $X_{i(j^{\prime})}$. This is not the case when $\mathcal{I} = \mathbb{N}$ or $\mathbb{N} \times \mathbb{N}$. As a matter of fact, if $\mathcal{I}$ is a poset, the compatibility conditions for the maps $f^j$ are simply that for each $j \rightarrow j$ in $\mathcal{J}$, the following diagram commutes:

\begin{center}
\begin{tikzpicture}[description/.style={fill=white,inner sep=2pt}]
\matrix (m) [matrix of math nodes, row sep=1.2em,
column sep=2.5em, text height=1.5ex, text depth=0.25ex]
{ & X_{i(j)} & & Y_{j} \\
&&&\\
  & X_{i(j^{\prime})} & & Y_{j^{\prime}} \\ };
\path[-> ,font=\scriptsize]
(m-3-2) edge node[auto] {} (m-1-2)
(m-1-2) edge node[auto] {} (m-3-2)
(m-3-2) edge node[below] {$f^{j^{\prime}}$} (m-3-4)
(m-1-2) edge node[auto] {$f^j$} (m-1-4)
(m-1-4) edge node[auto] {} (m-3-4);
\end{tikzpicture}
\end{center}

Where it could be that the arrow on the left goes in one direction or in the other.

The map $f$ could have been represented by some other $\widetilde{i} \colon \mathcal{J} \rightarrow \mathcal{I}$ and other family of maps $\{ \widetilde{f}^{j} \colon X_{\widetilde{i}(j)} \rightarrow Y_j \}$. They define the same map in the pro-category when they are {\it compatible with all the structure maps of $X$ and $Y$}. This means, for every $j$ in $\mathcal{J}$ there exists an element $i^{\prime}(j) \in \, \mathcal{I}$ and maps $i^{\prime}(j) \rightarrow i(j)$, $i^{\prime}(j) \rightarrow \widetilde{i}(j)$ in $\mathcal{I}$ such that the following diagram commutes:

\begin{center}
\begin{tikzpicture}[description/.style={fill=white,inner sep=2pt}]
\matrix (m) [matrix of math nodes, row sep=1.2em,
column sep=2.5em, text height=1.5ex, text depth=0.25ex]
{ & X_{i(j)} &  \\
 X_{i^{\prime}(j)} & &  Y_{j} \\
  & X_{\widetilde{i}(j)} &  \\ };
\path[->,font=\scriptsize]
(m-2-1) edge node[auto] {} (m-1-2)
(m-2-1) edge node[auto] {} (m-3-2)
(m-3-2) edge node[below] {$\widetilde{f}^{j}$} (m-2-3)
(m-1-2) edge node[auto] {$f^j$} (m-2-3);
\end{tikzpicture}
\end{center}

Again, observe that in the case $\mathcal{I}$ is a poset, the two families represent the same $f$ if for every $j$ the following diagram commutes:

\begin{center}
\begin{tikzpicture}[description/.style={fill=white,inner sep=2pt}]
\matrix (m) [matrix of math nodes, row sep=3em,
column sep=3.5em, text height=1.5ex, text depth=0.25ex]
{X_{i(j)} & Y_j\\
 X_{\widetilde{i}(j)} &  \\};
\path[->,font=\scriptsize]
(m-1-1) edge node[auto] {$f^{j}$} (m-1-2)
(m-2-1) edge node[auto] {$\widetilde{f}^{j}$} (m-1-2)
(m-1-1) edge node[left] {} (m-2-1)
(m-2-1) edge node[left] {} (m-1-1);;
\end{tikzpicture}
\end{center}

Now we give the notion of the identity morphism from $X \colon \mathcal{I} \rightarrow \Cat$ to itself. Consider the identity maps $1_i \colon X_i \rightarrow X_i$ in $\Cat$, these induce elements of $\underset{j \in \mathcal{I}}{\textrm{colim}} \, \, \textrm{Hom}_{\Cat}(X_j, X_i)$. For every $i \to i^{\prime}$ take $X_{i\to i^{\prime}}$ to be $X_{i}$ and complete the pentagon in the following way:

\begin{center}
\begin{tikzpicture}[description/.style={fill=white,inner sep=2pt}]
\matrix (m) [matrix of math nodes, row sep=1.2em,
column sep=2.5em, text height=1.5ex, text depth=0.25ex]
{ & X_{i} & & X_{i} \\
 X_{i} & & & \\
  & X_{i^{\prime}} & & X_{i^{\prime}} \\ };
\path[-> ,font=\scriptsize]
(m-2-1) edge node[auto] {$1_{i}$} (m-1-2)
(m-2-1) edge node[auto] {} (m-3-2)
(m-3-2) edge node[below] {$1_{i^{\prime}}$} (m-3-4)
(m-1-2) edge node[auto] {$1_i$} (m-1-4)
(m-1-4) edge node[auto] {} (m-3-4);
\end{tikzpicture}
\end{center}

We see that this defines a map $1_X \in \, \textrm{Hom}_{\textrm{Pro}(\Cat)}(X, X)$. This map is called the {\bf identity} morphism from $X$ to $X$. We will later check that it has all the properties of an identity.

Given two maps $f$ from $X \colon \mathcal{I} \rightarrow \Cat$ to $Y \colon \mathcal{J} \rightarrow \Cat$ and $g$ from $Y \colon \mathcal{J} \rightarrow \Cat$ to $Z \colon \mathcal{K} \rightarrow \Cat$ we can easily construct a map $g \circ f$ from $X \colon \mathcal{I} \rightarrow \Cat$ to $Z \colon \mathcal{K} \rightarrow \Cat$. Let $i$ and $j$ denote the functions associated to $f$ and $g$ respectively, consider the composition $ij = i \circ j \colon \mathcal{K} \rightarrow \mathcal{I}$. Now consider the {\it composition} $g^{k} \circ f^{j(k)} \colon X_{ij(k)} \rightarrow Y_{j(k)} \rightarrow Z_k$. These maps satisfy the compatibility relations necessary to define a map from $X$ to $Z$:

\begin{center}
\begin{tikzpicture}[description/.style={fill=white,inner sep=2pt}]
\matrix (m) [matrix of math nodes, row sep=1.2em,
column sep=0.9em, text height=1.5ex, text depth=0.25ex]
{ & \color{violet} X_{i(j(k\to k^{\prime}) \to  j(k))} &  & \mathbf{X_{ij(k)}} & & Y_{j(k)} & & \mathbf{Z_k} \\
 \color{cyan} \mathbf{X_{i(k\to k^{\prime})}} & & X_{ij(k\to k^{\prime})} & & \color{dandelion} Y_{j(k\to k^{\prime})} & & & \\
  & \color{violet} X_{i(j(k\to k^{\prime}) \to  j(k)^{\prime})} &  & \mathbf{X_{ij(k^{\prime})}} & & Y_{j(k^{\prime})} & & \mathbf{Z_{k^{\prime}}} \\ };
\path[->,font=\scriptsize, color=cyan]
(m-2-1) edge node[auto] {} (m-1-2)
(m-2-1) edge node[auto] {} (m-3-2);
\path[->,font=\scriptsize, color=violet]
(m-3-2) edge node[below] {} (m-3-4)
(m-1-2) edge node[auto] {} (m-1-4)
(m-1-2) edge node[auto] {} (m-2-3)
(m-3-2) edge node[auto] {} (m-2-3);
\path[->,font=\scriptsize, color=darkdelion]
(m-2-5) edge node[auto] {} (m-1-6)
(m-2-5) edge node[auto] {} (m-3-6);
\path[->,font=\scriptsize]
(m-3-4) edge node[below] {$f^{j(k^{\prime})}$} (m-3-6)
(m-1-4) edge node[auto] {$f^{j(k)}$} (m-1-6)
(m-3-6) edge node[below] {$g^k$} (m-3-8)
(m-1-6) edge node[auto] {$g^{k^{\prime}}$} (m-1-8)
(m-2-3) edge node[auto] {$f^{j(k\to k^{\prime})}$} (m-2-5)
(m-1-8) edge node[auto] {} (m-3-8);
\end{tikzpicture}
\end{center}

In orange we have completed the pentagon for the map $g$, in violet we have completed two pentagons for the map $f$ and in blue we have used that pullbacks exists in cofiltered categories. The relevant pentagon is the one with the vertices in bold.

In the case where all the index categories are posets the composition is simply given by the following diagram:

\begin{center}
\begin{tikzpicture}[description/.style={fill=white,inner sep=2pt}]
\matrix (m) [matrix of math nodes, row sep=2em,
column sep=2.5em, text height=1.5ex, text depth=0.25ex]
{X_{ij(k)} & Y_{j(k)} & Z_k \\
X_{ij(k^{\prime})} & Y_{j(k^{\prime})} & Z_{k^{\prime}}\\};
\path[->,font=\scriptsize]
(m-1-1) edge node[auto] {$f^{j(k)}$} (m-1-2)
(m-1-2) edge node[auto] {$g^{k}$} (m-1-3)
(m-2-1) edge node[auto] {$f^{j(k^{\prime})}$} (m-2-2)
(m-2-2) edge node[auto] {$g^{k^{\prime}}$} (m-2-3)
(m-1-1) edge node[auto] {} (m-2-1)
(m-1-2) edge node[auto] {} (m-2-2)
(m-1-3) edge node[auto] {} (m-2-3);
\end{tikzpicture}
\end{center}

Observe that the composition is {\it associative} just because the composition of functors between the index categories is associative.

In order to establish that the pro-category is indeed a category, we just need to show that pre- and post-composition with the identity is the same as doing nothing. We have said that the identity is given by the identity function from $\mathcal{I}$ to $\mathcal{I}$ and all the maps are identities. It is clear then than if $f \colon Y \rightarrow X$ and $g \colon X \rightarrow Z$ are given by the functions $g^i \colon Y_{j(i)} \rightarrow X_i$ and $f^k \colon X_{i(k)} \rightarrow Z_k$ the compositions are just given by $(g \circ \textrm{id})^i \defeq  g^i \circ \textrm{id}^{i} = g^i$ and $(\textrm{id} \circ f)^k \defeq  \textrm{id}^{i(k)} \circ f^k = f^k$.

\begin{center}
\begin{tikzpicture}[description/.style={fill=white,inner sep=2pt}]
\matrix (m) [matrix of math nodes, row sep=2em,
column sep=2.5em, text height=1.5ex, text depth=0.25ex]
{Y_{j(i)} & X_{i} & X_{i} & X_{i(k)} & X_{i(k)} & Z_{k}\\
 Y_{j(i)} & & X_{i} & X_{i(k)} & & Z_{k}\\};
\path[->,font=\scriptsize]
(m-1-1) edge node[auto] {$g^{i}$} (m-1-2)
(m-1-2) edge node[auto] {id} (m-1-3)
(m-2-1) edge node[auto] {$g^{i}$} (m-2-3)
(m-1-1) edge node[auto] {id} (m-2-1)
(m-1-3) edge node[auto] {id} (m-2-3)
(m-1-4) edge node[auto] {id} (m-1-5)
(m-1-5) edge node[auto] {$f^{k}$} (m-1-6)
(m-2-4) edge node[auto] {$f^{k}$} (m-2-6)
(m-1-4) edge node[auto] {id} (m-2-4)
(m-1-6) edge node[auto] {id} (m-2-6);
\end{tikzpicture}
\end{center}

$f \colon X \rightarrow Y$ is an {\it isomorphism} in the pro-category if the associated function $i$ between the index categories is a bijection with two sided inverse $i^{-1}$ and such that there exists a map $g \colon Y \rightarrow X$ given by $g^{i} \colon Y_{i^{-1}(i)} \rightarrow X_i$ is such that $f^{j} \circ g^{i(j)} = \textrm{id}_{Y_{j}}$ and $g^i \circ f^{i^{-1}(i)} = \textrm{id}_{X_i}$.


\section{Fr\'echet spaces and manifolds}\label{app2}

{\it In this appendix we will define Fr\'echet spaces and manifolds. The main references in this chapter are Dodson, Galanis, and Vassiliou \cite{DGV} and Schaefer \cite{SCH}.}\\

As mentioned in Chapter \ref{jif}, a Fr\'echet space is a vector space equipped with a Hausdorff topology coming from a countable family of seminorms. All definitions in this chapter follow Dodson, Galanis, and Vassiliou \cite{DGV}.

\begin{df}[Seminorm] A seminorm on a vector space $V$ is a map $| \cdot | \colon V \rightarrow \RE$ such that
\begin{enumerate}
	\item $| v | \geqslant 0$ for all $v \in V$.
	\item $| v+ w | \leqslant | v | + | w |$ for all $v$ and $w$ in $V$.
	\item $| a \cdot v | = |a| \cdot | v |$ for all $v \in V$ and all $a \in \RE$.
\end{enumerate}
\end{df}

Observe that a seminorm such that $| v | = 0 \Leftrightarrow v = 0$ is a norm on $V$.

\begin{df}[Locally convex topological vector space]\label{lctvs} A locally convex topological vector space is a vector space $V$ endowed with a topology coming from a family of seminorms $\{| \cdot |_i\}_{i \in I}$.  A neighborhood basis of $v \in V$ is given by the sets $\{U_{\varepsilon}^J (v)\}_{\varepsilon > 0, J \subset I \textrm{ finite}}$ where
$$U_{\varepsilon}^J (v) \defeq  \{w \in V \colon | w-v |_j < \varepsilon \, \forall j \in J\}$$
\end{df}

A locally convex topological vector space is a topological space, hence we can look at some topological properties such as being Hausdorff or metrizable. The following is a well known characterization of those properties.

\begin{pp}[Dodson, Galanis, and Vassiliou {\cite[2.1.3]{DGV}}]\label{haumet} Given a locally convex topological vector space $\left( V, \{| \cdot |_i\}_{i \in I}\right)$ we have that
\begin{itemize}
	\item $V$ is Hausdorff if and only if $v=0 \Leftrightarrow \{|v|_i = 0 \, \forall i \in I\}$.
		\item $V$ is metrizable if and only if the topology can be defined by using only a countable subset of $I$.
\end{itemize}
\end{pp}

In particular, when working with a metrizable locally convex topological vector space we will consider that the index set for the seminorms to be $\mathbb{N}$ (repeating the seminorms if they are finite). In that case the norm is given by:
$$\textrm{d}(v,w) \defeq  \sum_{i=1}^{\infty} \frac{1}{2^i} \frac{|v-w|_i}{1+|v-w|_i}.$$

\begin{df}[Fr\'echet space]\label{fs} A Fr\'echet space is a sequentially complete Hausdorff metrizable locally convex topological vector space.
\end{df}

Recall that a metric space is sequentially complete if every Cauchy sequence converges. Fr\'echet spaces are projective limits of Banach spaces, we also recall that definition.

\begin{df}[Banach Space] A Banach space is a normed vector space which is sequentially complete.
\end{df}

Observe that Banach spaces are Fr\'echet spaces with a single seminorm $|\cdot |$. In that case the Hausdorff assumption means that the seminorm is a norm. As we said before, to define the metric we consider $|\cdot |_i = |\cdot |$ for all $i \in \mathbb{N}$ so that the metric is $\textrm{d}(v,w) \defeq  \frac{|v-w|}{1+|v-w|}$ which is equivalent to $|v-w|$.

\begin{tm}[Dodson, Galanis, and Vassiliou {\cite[2.3.8]{DGV}}]\label{probanach}
Every Fr\'echet space is isomorphic as vector spaces to a sequential projective limit of Banach spaces.
\end{tm}

Actually, the converse of Theorem \ref{probanach} is true: all sequential projective limits of Banach spaces are Fr\'echet. This can be found in the book by Dodson, Galanis, and Vassiliou \cite[Proposition 2.3.7]{DGV}.

\begin{rk}\label{probanachtop}In particular, if we restrict both sides to the subjacent topological structures this theorem says that the locally convex topology on a Fr\'echet space coincides with the projective limit topology of the associated Banach spaces.
\end{rk}

By sequential projective limit we mean a pro-object indexed by $\mathbb{N}$ as a cofiltered category. Observe that this theorem does not tell us which projective limits of Banach spaces are Fr\'echet spaces. Since our first examples of Fr\'echet structures have been sequential projective limits of finite dimensional normed spaces ($J^{\infty} E$) we focus on that case. 

The result is also proven in more detail by Schaefer \cite{SCH}. Inspecting this proof we can get some insight into the converse of the statement: which Fr\'echet spaces are sequential pro-finite dimensional normed spaces. That is the starting point of Corollary \ref{chus}.

One can talk about derivatives of continuous maps between Fr\'echet spaces.

\begin{df}[G\^ateaux derivative]\label{gtd} Let $U \subset V$ be an open set of a Fr\'echet space $V$ and $W$ be another Fr\'echet space.  The G\^ateaux derivative of a continuous map $f \colon U \rightarrow W$ at $u \in U$ in the direction of $v \in V$ is the limit
$$\textrm{D}f(u){v} \defeq  \lim_{t \rightarrow 0} \frac{f(u +tv) - f(u)}{t}.$$
\end{df}

\begin{df}[Smooth maps between Fr\'echet spaces]\label{FrSm}
Let $f \colon U \subset V \rightarrow W$ be a continuous map from an open subset $U$ of a Fr\'echet space $V$ to another Fr\'echet space $W$.
\begin{itemize}
	\item $f$ is said to be differentiable at $u$ in the direction of $v$ if the G\^ateaux derivative of $f$ at $u$ in the direction of $v$ exists.
	\item $f$ is said to be continuously differentiable, or differentiable of class $\mathscr{C}^1$ if it is differentiable at all points of $U$ in all directions and if the differential $\textrm{D}f$ is continuous, where
		\begin{center}
    		\begin{tabular}{rcl}
    			$\textrm{D}f \colon U \times V$ & $\longrightarrow$ & $W$ \\
    			$(u, v)$ & $\longmapsto$ & $\textrm{D}f(u){v}.$ \\
    		\end{tabular}
		\end{center}
	\item The higher differentials are defined inductively for all $k \in \mathbb{N}$,
			\begin{center}
    		\begin{tabular}{rcl}
    			$\textrm{D}^{k+1}f \colon U \times V \times \stackrel{k+1}{\ldots} \times V$ & $\longrightarrow$ & $W$ \\
    			$(u, v_1, \cdots, v_{k+1})$ & $\longmapsto$ & $\textrm{D}\left(\textrm{D}^k f (u){v_1, \cdots, v_{k}} \right)v_{k+1}.$ \\
    		\end{tabular}
		\end{center}
	\item $f$ is said to be of class $\mathscr{C}^k$ if $\textrm{D}^k f$ is continuous.
	\item $f$ is smooth if it is of class $\mathscr{C}^k$ for all $k \in \mathbb{N}$.
\end{itemize}
\end{df}

Now the definition of a Fr\'echet manifold makes sense: it is possible to define smooth manifolds with charts in Fr\'echet spaces. We can also extend the definition of smooth fiber bundle to the Fr\'echet case.

\begin{df}[Fr\'echet manifold] Hausdorff topological space $\M$ is a Fr\'echet manifold if it is provided with an atlas of homeomorphisms to open sets of a Fr\'echet space $V$ such that the transition functions are smooth in the sense of Definition \ref{FrSm}.
\end{df}


\section{Twisted forms}\label{tf}

{\it In this appendix we consider an analogue of the complex of ind-differential forms in the infinite jet bundle, in which the forms are valued on densitites instead. These are ind-differential $M$-twisted forms which are thereof defined. A compairson between the different splittings in the vertical and horizontal directions found in the literature is also discussed. The basic references in this appendix are Abraham, Marsden, and Ratiu \cite{AMR}; and Deligne and Freed \cite{DEL}.}\\

We fix a smooth fiber bundle $\pi \colon E \longrightarrow M$. Since $M$ is a priory not oriented, nor orientable; compactly supported differential forms are not integrable: we need to consider twisted forms, also called densities. We follow the definitions in the book by Abraham, Marsden, and Ratiu \cite{AMR}. 

\begin{df}[Orientation line bundle, following Abraham, Marsden, and Ratiu \cite{AMR}] Given a smooth manifold $M$, the orientation line bundle on $M$ is $$\mathfrak{o}_M \defeq \bigslant{ \{ (x, \mu, a) \colon x \in M, a \in \RE, \mu \textrm{ is an orientation of } T_x M \} }{(x, \mu, a) \sim (x, -\mu, -a)}.$$
It is a smooth vector bundle of rank one with locally constant transition functions. Its space of smooth sections is denoted by $\mathfrak{O}_M \defeq \Gamma^{\infty}(M, \mathfrak{o}_M)$.
\end{df}

\begin{df}[Densities, following Abraham, Marsden, and Ratiu \cite{AMR}] Given a smooth manifold $M$, the bundle of densities on $M$ is $\textrm{Dens} (M) \defeq \wedge^{\textrm{dim}(M)} T^* M \otimes \mathfrak{o}_M$. It is a smooth line bundle.
\end{df}

\begin{df}[Complex of twisted forms, following Abraham, Marsden, and Ratiu \cite{AMR}] Given a smooth manifold $M$, the complex of twisted forms on $M$, denoted by $\left( \Omega^{|\bullet|}(M), d\right)$ is defined as follows: 
$$\textrm{Dens}^p (M) = \Omega^{|-p|}(M) \defeq \Gamma^{\infty}\left(M, \, \wedge^p T M \, \otimes \, \textrm{Dens} (M) \right).$$
This can be rewritten as $\Gamma^{\infty}(M, \wedge^{\textsf{top}-p} T^* M \otimes \mathfrak{o}_M) = \Omega^{\textsf{top}-p}(M) \otimes_{\C(M)} \mathfrak{O}_M$ where (denote $\textsf{top} = \textrm{dim}(M)$). Using this interpretation, the de Rham differential on $\Omega^\bullet(M)$ induces a square zero map $d$ on $\Omega^{|\bullet|}(M)$ given by $d(\alpha \otimes \epsilon) = d\alpha \otimes \epsilon$ for all $\alpha \in \Omega^{\textsf{top}-p}(M)$ and $\epsilon \in \mathfrak{O}_M$. 
\end{df}

For convenience, it is possible to think of $M$ to be an oriented manifold. In this case $\mathfrak{o}_M$ is the trivial line bundle, and hence $\Omega^{|-p|} (M) \cong \Omega^{\textsf{top} -p} (M)$ for all $p$.

At this point we would like to talk about twisted forms in $J^k E$ for all $k \in \mathbb{N}$, but where the twist only happens at the level of $M$. 

\begin{df}[$M$-twisted sections] Given a smooth fiber bundle $\rho \colon F \rightarrow M$ and a vector bundle $\beta \colon V \rightarrow F$ we can consider $M$-twisted sections of $\beta$: 
$$\Gamma_{\textrm{tw}}(F, V) \defeq \Gamma(F, V) \otimes \Gamma(F, \rho^* \mathfrak{o}_M).$$
\end{df}

Observe that the tensor product is over $\C(F)$, hence it only works for sections of {\it vector} bundles over $F$. The pullback bundle $\rho^* \mathfrak{o}_M \rightarrow F$ is a line bundle with the same transition functions as $\mathfrak{o}_M$. It is trivial if $\mathfrak{o}_M \rightarrow M$ is trivial. This implies that if $M$ is orientable, $M$-twisted sections are the same as ordinary sections. Also observe that if $W \rightarrow M$ is any fiber bundle over $M$, the sections $\Gamma(F, \rho^* W) \cong \textrm{Bun}_M(F, W)$ are simply bundle maps over $M$. In particular $\Gamma(F, \rho^* \mathfrak{o}_M) \cong \textrm{Bun}_M(F, \mathfrak{o}_M)$.

We can repeat this construction for $\rho = \pi_k \colon F = J^k E \rightarrow M$ and for $\wedge^n T^* F \rightarrow F$

\begin{df}[($\Omega_{\textrm{tw}}^{\bullet}(J^k E), d)$] Given a smooth fiber bundle $\pi \colon E \rightarrow M$ and $k$ a non-negative integer we can define the complex of $M$-twisted forms on $J^k E$ by the following construction. For each $n \in \mathbb{N}$, the $n$-th degree $M$-twisted differential forms on $J^k E$ are the $M$-twisted sections of $\wedge^n T^* (J^k E) \rightarrow J^k E$:
$$\Omega_{\textrm{tw}}^n(J^k E) \defeq \Omega^n(J^k E) \otimes \Gamma(J^k E, \pi_k^* \mathfrak{o}_M).$$
The de Rham differential on $J^k E$ induces a square-zero map $d \defeq d \otimes \textrm{id}_{\Gamma(J^k E, \pi_k^* \mathfrak{o}_M)}$ since the transition functions of the bundle on the right are locally constant. The same applies to the wedge product which extends to $\wedge \defeq \wedge \otimes \cdot$ a product in $\Omega_{\textrm{tw}}^n(J^k E)$ where $\cdot$ is the usual product in $\mathbb{R}$. The resulting complex is a graded algebra with this product and $d$ is a derivation of the product.
\end{df}

In order to define the $M$-twisted differential forms on $J^{\infty} E$ we need maps between the $M$-twisted differential forms on different jet bundles. Consider $\pi_k^l \colon J^k E \rightarrow J^l E$ for $k \geqslant l$. Recall that $\Gamma(J^k E, \pi_k^* \mathfrak{o}_M) \cong \textrm{Bun}_M(J^k E, \mathfrak{o}_M)$ and similarly for $J^l E$. This shows that precomposition with $\pi_k^l$ induces a map $(\pi_k^l)^* \colon \Gamma(J^l E, \pi_l^* \mathfrak{o}_M) \rightarrow \Gamma(J^k E, \pi_k^* \mathfrak{o}_M)$. Using such a map we get:

$$(\pi_k^l)^* \defeq (\pi_k^l)^* \otimes (\pi_k^l)^* \colon \Omega_{\textrm{tw}}^{n}(J^l E) \rightarrow \Omega_{\textrm{tw}}^{n}(J^k E).$$

We can now define the ind-complex of $M$-twisted forms on $J^{\infty} E$ as follows:

\begin{df}[$M$-twisted ind-differential forms, $\Omega_{\textrm{tw}}^{\bullet}(J^{\infty}  E )$] 
Given a fiber bundle $\pi \colon E \longrightarrow M$. Ind-twisted forms on $J^{\infty}  E$ are defined to be the ind-object in the category of differential complexes given by the diagram $$\Omega_{\textrm{tw}}^{\bullet}(E) = \Omega_{\textrm{tw}}^{\bullet}(J^0 E ) \rightarrow \Omega_{\textrm{tw}}^{\bullet}(J^1 E ) \rightarrow \Omega_{\textrm{tw}}^{\bullet}(J^2 E) \rightarrow \cdots.$$
\end{df}

\begin{rk} The same remarks given for the ind-complex of differential forms can be repeated for the ind-complex of $M$-twisted forms:
\begin{itemize} 
	\item The differentials $d_k \colon \Omega_{\textrm{tw}}^{\bullet}(J^k E ) \rightarrow \Omega_{\textrm{tw}}^{\bullet + 1}(J^k E )$ can be interpreted as a morphism of ind-graded vector spaces $d \colon \Omega_{\textrm{tw}}^{\bullet}(J^{\infty}  E ) \rightarrow \Omega_{\textrm{tw}}^{\bullet + 1}(J^{\infty}  E )$ given by $\{d_k\}_{k \in \mathbb{N}}$. This map squares to zero in the sense that given any $\omega$ in $\Omega_{\textrm{tw}}^{\bullet}(J^{\infty} E )$ (that is a map $\mathbb{R} \rightarrow \Omega_{\textrm{tw}}^{\bullet}(J^{\infty} E )$ or equivalently $\omega_k \in \Omega^{\bullet}(J^k E)$ for some $k$) the ind-form $d \circ d (\omega)$ is the zero form of the appropriate degree and index.  
	\item Since the tensor product of two differential complexes is again a differential complex, we can talk about the ind-differential complex $\Omega_{\textrm{tw}}^{\bullet}(J^{\infty}  E) \otimes \Omega_{\textrm{tw}}^{\bullet}(J^{\infty}  E)$ which will be indexed by $\mathbb{N} \times \mathbb{N}$.
	\item The $M$-twisted ind-differential forms on the infinite jet bundle $\Omega_{\textrm{tw}}^{\bullet}(J^{\infty}  E)$ are equipped with an ind-morphism $\wedge \colon \Omega_{\textrm{tw}}^{\bullet}(J^{\infty}  E) \otimes \Omega_{\textrm{tw}}^{\bullet}(J^{\infty}  E) \rightarrow \Omega_{\textrm{tw}}^{\bullet}(J^{\infty}  E)$ given by the maps $\{ \wedge_{k,l} \colon \Omega_{\textrm{tw}}^{\bullet}(J^k  E) \otimes \Omega_{\textrm{tw}}^{\bullet}(J^l  E) \rightarrow \Omega_{\textrm{tw}}^{\bullet}(J^{\textrm{max}(k,l)}  E) \}_{(k,l)}$ defined as $\wedge_{k,l} = \wedge \circ (\pi_{k}^{\textrm{max}(k,l)} \otimes \pi_{l}^{\textrm{max}(k,l)})$. 
	\item In particular this means that when applied to two $M$-twisted ind-differential forms $\alpha_1$ and $\alpha_2$ the differential is, once again, a derivation of the product: $d(\alpha_1 \wedge \alpha_2) = d_1(\alpha_1) \wedge \alpha_2 + (-1)^{|\alpha_1|} \alpha_1 \wedge d_2(\alpha_2)$.
\end{itemize}
\end{rk}

If $M$ is orientable each of the pieces of $\Omega_{\textrm{tw}}^{\bullet}(J^{\infty}  E )$ is isomorphic to the non-twisted version. Moreover, $d$, $\wedge$ and $(\pi_k^l)^*$ reduce to the usual maps $d$, $\wedge$ and $(\pi_k^l)^*$. This means that in the case in which $M$ is orientable the two ind-graded algebras $\Omega_{\textrm{tw}}^{\bullet}(J^{\infty} E)$ and $\Omega^{\bullet}(J^{\infty} E)$ agree.

Remember that the ind-complex of differential forms on $J^{\infty} E$ could be considered as a bicomplex (called the variational bicomplex, see Definition \ref{JVB}). The same exact procedure applies for $M$-twisted ind-differential forms:

\begin{df}[$M$-twisted variational bicomplex]\label{TVB} 
Let $E \rightarrow M$ be a smooth fiber bundle and let $J^{\infty} E$ denote its infinite jet bundle. The $M$-twisted variational bicomplex associated to $E \rightarrow M$ is $(\Omega_{\textrm{tw}}^{r,s}(J^{\infty} E), d_H, d_V)$ where $\Omega_{\textrm{tw}}^{r,s}(J^{\infty} E)$ is the twisted analogue to $\Omega^{r,s}(J^{\infty} E) \defeq \Omega_V^{s+r,r} (J^{\infty} E) \cap \Omega_H^{s+r,s} (J^{\infty} E)$.
\end{df}

Before finishing this appendix, we want to point out how our definition of the complex of twisted forms differs from the one given by Deligne and Freed \cite{DEL}. This is relevant even for the non-twisted case since it departs from our convention (following Anderson \cite{AND}) on how to find vertical and horizontal forms. We discuss only the case $J^0 E = E$:

The map $T \pi \colon TE \rightarrow TM$ induces a map $TE \rightarrow \pi^* TM$ of vector bundles over $E$ which we will still denote by $T \pi$. The kernel of this map is a vector bundle over $E$ denoted by $\textrm{ker} \, T \pi$ or often $VE$. We have the following short exact sequence of vector bundles over $E$:

$$ 0 \rightarrow VE \rightarrow TE \rightarrow \pi^* TM \rightarrow 0.$$

A section of $T \pi$ induces a splitting $TE = VE \oplus \pi^* TM$: this is precisely an Ehresmann connection. Using a connection, differential forms of degree $n$ on $E$ split in the following way: 

$$\Omega^n(E) = \Gamma\left(E,  \bigoplus_{p+q = n} \wedge^p \left( VE \right)^* \otimes \pi^* \left( \wedge^q T M \right)^*  \right).$$

This is the way vertical and horizontal forms are defined in Deligne-Freed's book \cite{DEL} \cite{DEL}. Observe that using this splitting twisted differential forms can also be written in the following way:

\begin{eqnarray*}
\Omega_{\textrm{tw}}^n(E) & = & \Omega^n(E) \otimes \Gamma(E, \pi^* \mathfrak{o}_M) = \Gamma\left(E,  \bigoplus_{p+q = n} \wedge^p \left( VE \right)^* \otimes \pi^* \left( \wedge^q T M \right)^* \otimes \pi^* \mathfrak{o}_M \right) \\
& \cong & \Gamma\left(E,  \bigoplus_{p+q = n} \wedge^p \left( VE \right)^* \otimes \pi^* \left( \wedge^q T^* M \otimes \mathfrak{o}_M \right) \right).
\end{eqnarray*}

The later is the definition of twisted differential forms in the book by Deligne and Freed \cite{DEL}. Such definition does not use any connection in particular, but in order to recover a differential form on $E$ one needs to use a connection. This is not the case at all in our discussion.


\chapter{Extended local Cartan calculus}\label{Alcc}


{\it In this appendix we prove some equalities involving the commutators of the vertical and horizontal differentials; and insertion and Lie derivatives of horizontal and vertical insular vector fields.}\\

The proof of the second statement about local Cartan calculus has been postponed until this point. Here we present the proof.

\begin{pp*}[{\bf \ref{lcc2}} Local Cartan calculus, Part 2] 
We consider two insular vector fields $\chi = \xi + X$ and $\chi^{\prime} = \xi^{\prime} + X^{\prime}$. Define $\mathcal{L}_{\chi}^d \defeq [d, \iota_{\chi}]$ and $\mathcal{M}_{\chi} \defeq [\delta, \mathcal{L}_{\chi}^d]$ in analogy to what was done in Theorem \ref{lcc1}.  Then
\begin{multicols}{2}
\begin{enumerate}
\setcounter{enumi}{7}
	\item $[\iota_{\xi}, \mathcal{L}_X^d]=0$.
	\item $[\mathcal{L}_{X}^{d}, \iota_{X^{\prime}}] = \iota_{[X, X^{\prime}]}$.
	\item $[\mathcal{L}_{X}^{d}, \mathcal{L}_{X^{\prime}}^d] = \mathcal{L}_{[X, X^{\prime}]}^d$.
	\item $[\mathcal{M}_X, D] = [\mathcal{M}_X, d] = [\mathcal{M}_X, \delta] = 0$,
	\item $[\mathcal{L}_{\xi}, \mathcal{L}_X] = \mathcal{L}_{[X, \xi]}^d$.
	\item $[\mathcal{L}_{\xi}, \mathcal{M}_X] = \mathcal{M}_{[\xi, X]}$.
\end{enumerate}
\end{multicols}
\end{pp*}

In the proof, we will use extensively the other equations of the local Cartan calculus (Theorem \ref{lcc1}) that were proven in Chapter \ref{lill}. We include here the statement to make the reading and referencing easier:

\begin{tm*}[{\bf \ref{lcc1}} Local Cartan calculus] 
We consider two insular vector fields $\chi = \xi + X$ and $\chi^{\prime} = \xi^{\prime} + X^{\prime}$. The only nontrivial commutators between any two of the following endomorphisms: $D, d, \delta, \iota_{\xi}, \iota_{\xi^{\prime}}, \iota_X, \iota_{X^{\prime}}, \mathcal{L}_{\xi}, \mathcal{L}_{\xi^{\prime}}, \mathcal{L}_X, \textrm{ and } \mathcal{L}_{X^{\prime}}$ are the following:
\begin{multicols}{2}
\begin{enumerate}
	\item $\mathcal{L}_X = [D, \iota_{X}]$.
	\item $\mathcal{L}_{\xi}^{\delta} \defeq [\delta, \iota_{\xi}] = [D, \iota_{\xi}] = \mathcal{L}_{\xi}$.
	\item $\mathcal{L}_X^d \defeq [d, \iota_X]$.
	\setcounter{enumi}{4}
	\item $[\delta, \iota_X] = \mathcal{L}_X - \mathcal{L}_X^d$.
	\item $[\mathcal{L}_{\chi}, \iota_{\chi^{\prime}}] = \iota_{[\chi, \chi^{\prime}]}$.
	\item $[\mathcal{L}_{\chi}, \mathcal{L}_{\chi^{\prime}}] = \mathcal{L}_{[\chi, \chi^{\prime}]}$.
\end{enumerate}
\end{multicols}
\vspace{-1.8em}
\begin{enumerate}
\setcounter{enumi}{3}
	\item $\mathcal{M}_X \defeq [D, \mathcal{L}_X^d] = [\delta, \mathcal{L}_X^d] = [\delta, \mathcal{L}_X]  = -[d, \mathcal{L}_X] = -[d, [\delta, \iota_X]]$.
\end{enumerate}
Where items $5$, $6$ and $7$ hold also replacing $\chi$ by $\xi$ and $X$ respectively.
\end{tm*}

{\bf Proof of Proposition \ref{lcc2}.}
We start with the simplest, that is equation $8$. Using the graded Jacobi identity (even without paying attention to the signs) we can see that $[\iota_{\xi}, \mathcal{L}_X^d]$ is equal to a certain signed sum of terms including $[\iota_{\xi}, d]$ and $[\iota_{\xi}, \iota_X]$ which are both equal to zero by equation $2$ in Theorem \ref{lcc1} and the usual Cartan calculus equation for the commutation of two insertion operators which we have included in Proposition \ref{CC3}.

In order to prove $9$ we are going to show the equivalent relation $[[\delta, \iota_X], \iota_{X^{\prime}}] = 0$. Once that is proven, using $5$ and $6$ from Theorem \ref{lcc1} we would have
$$[\mathcal{L}_X^d, \iota_{X^{\prime}}] = [\mathcal{L}_X, \iota_{X^{\prime}}] - [[\delta, \iota_X], \iota_{X^{\prime}}] = [\mathcal{L}_X, \iota_{X^{\prime}}] = \iota_{[X, X^{\prime}]}.$$
In order to prove $[[\delta, \iota_X], \iota_{X^{\prime}}] = 0$, it is enough to check the equation evaluated at $(0,2)$ forms, by derivation properties. We take a homogeneous $(0,2)$ form $\omega = f d x^i \wedge d x^j$.
\begin{eqnarray*}
[\delta, \iota_X] \omega &=& \delta (f X_i d x^j - f X_j dx^i) + \iota_X (\delta f dx^i \wedge dx^j) \\
&=& \delta(f X_i) dx^j - \delta(f X_j) dx^i -X_i \delta f dx^j + X_j \delta f dx^i\\
&=& \left( \delta(f X_i) - X_i \delta f \right)dx^j - \left( \delta(f X_j) - X_j \delta f \right)dx^i\\
&=& f \delta X_i dx^j - f \delta X_j dx^i.
\end{eqnarray*}

And $\left(\iota_{X^{\prime}} \circ [\delta, \iota_X] \right) \omega = - X_j^{\prime} f \delta X_i + X_i^{\prime} f \delta X_i$.

On the other hand $\iota_{X^{\prime}} \omega = f X_i^{\prime} d x^j - f X_j^{\prime} dx^i$ and thus
\begin{eqnarray*}
[\delta, \iota_X] \circ \iota_{X^{\prime}} \omega &=& \delta (f X_i^{\prime} X_j - f X_j^{\prime} X_i) + \iota_X (\delta (f X_i^{\prime}) dx^j - \delta (f X_j^{\prime}) dx^i) \\
&=& \delta (f X_i^{\prime} X_j) - \delta(f X_j^{\prime} X_i) - X_j \delta (f X_i^{\prime}) + X_i \delta (f X_j^{\prime})\\
&=& f X_i^{\prime} \delta X_j - f X_j^{\prime} \delta X_i.
\end{eqnarray*}

This shows that $[\delta, \iota_X] \circ \iota_{X^{\prime}} \omega = \left(\iota_{X^{\prime}} \circ [\delta, \iota_X] \right) \omega$ and hence
$[[\delta, \iota_X], \iota_{X^{\prime}}] \omega = 0$ as wanted. This proves $9$.

The next equation, number $10$, follows directly from the fact that $[d , \mathcal{L}_X^d] = 0$ which was shown in the proof of Theorem \ref{lcc1} and equation $9$. Since using the graded Jacobi identity we have:
\begin{eqnarray*}
[\mathcal{L}_X^d , \mathcal{L}_{X^{\prime}}^d] &=& [\mathcal{L}_X^d , [d, \iota_{X^{\prime}}]] = [\iota_{X^{\prime}}, [\mathcal{L}_X^ d, d]] - [d, [\iota_{X^{\prime}}, \mathcal{L}_X^d]] \\
&=& - [d, [\iota_{X^{\prime}}, \mathcal{L}_X^d]] = [d, \iota_{[X, X^{\prime}]}] = \mathcal{L}_{[X, X^{\prime}]}^d .
\end{eqnarray*}

Now we start with the commutators involving $\mathcal{M}_X$. The first part of equation $11$ follows from the other two parts. In order to prove them we are going to use two different expressions for $\mathcal{M}_X$ which were proven to be equivalent in equation $7$, Theorem \ref{lcc1}.

For the commutativity with $d$, observe that $[d, \mathcal{M}_X] = [d, [D, \mathcal{L}_X^d]]$. Using the graded Jacobi identity, since $d$ graded commutes with both $D$ and $\mathcal{L}_X^d$ we conclude that it also graded commutes with $\mathcal{M}_X$. Similarly, using that $\mathcal{M}_X = -[d, [\delta, \iota_X]]$ we can see that $\delta$ graded-commutes with $\mathcal{M}_X$ since it graded-commutes both with $d$ and with $[\delta, \iota_X]$.

Now we focus on equation $12$,
$$\mathcal{L}_{[X, \xi]}^d = \mathcal{L}_{[X, \xi]} - [\delta, \iota_{[X, \xi]}] = [\mathcal{L}_{X}, \mathcal{L}_{\xi}] + [\delta, \iota_{[\chi, X]}] = [\mathcal{L}_{\xi}, \mathcal{L}_{X}] + [\delta, \iota_{[\chi, X]}].$$
We expand the last term:
$$[\delta, \iota_{[\chi, X]}] = [\delta, [\mathcal{L}_{\xi}, \iota_X]] = [\mathcal{L}_{\xi}, [\iota_X, \delta]] + [\iota_X, [\delta, \mathcal{L}_{xi}]] = [\mathcal{L}_{\xi}, [\iota_X, \delta]] = [\mathcal{L}_{\xi}, -[\delta, \iota_X]].$$
Putting the two last equations together we have that:
$$\mathcal{L}_{[X, \xi]}^d = [\mathcal{L}_{\xi}, \mathcal{L}_{X}] + [\mathcal{L}_{\xi}, -[\delta, \iota_X]] = [\mathcal{L}_{\xi}, \mathcal{L}_X^d].$$

That proves equation $12$, which we can now use for equation $13$:
\begin{eqnarray*}
\mathcal{M}_{[\xi, X]} &=& - \mathcal{M}_{[X, \xi]} = -[\delta, \mathcal{L}_{[X, \xi]}^d]= -[\delta, [\mathcal{L}_{\xi}, \mathcal{L}_X^d]] \\
&=&  [\mathcal{L}_{\xi}, [\mathcal{L}_X^d, \delta]] + [\mathcal{L}_X^d, [\delta, \mathcal{L}_{\xi}]]   =  [\mathcal{L}_{\xi}, [\mathcal{L}_X^d, \delta]] = [\mathcal{L}_{\xi}, \mathcal{M}_X].
\end{eqnarray*}

This completes the proof of the proposition.
\qed


\printbibliography



\nocite{*}
\newpage
\pagenumbering{gobble}
\printbibliography
\addcontentsline{toc}{part}{Bibliography}

\end{document}